\documentclass[twocolumn,twocolappendix]{aastex63} 

\usepackage{soul}
\usepackage[T1]{fontenc}
\usepackage{subfiles}
\usepackage{threeparttable}
\usepackage{xcolor}
\usepackage{amsmath}
\usepackage{comment}

\received{} 
\revised{}
\accepted{}

\submitjournal{ApJS}

\shorttitle{VLT/KMOS Survey of YSOs in Canis Major} 
\shortauthors{Itrich et al.}

\begin{document}

\title{Investigating the Impact of Metallicity on Star Formation in the Outer Galaxy.
 \\ I. VLT/KMOS Survey of Young Stellar Objects in Canis Major} 
\correspondingauthor{Agata Karska}
\email{agata.karska@umk.pl}

\author[0000-0002-9051-1781]{Dominika Itrich}
\affiliation{Institute of Astronomy, Faculty of Physics, Astronomy and Informatics, Nicolaus Copernicus University, Grudzi{\k a}dzka 5, 87-100 Toru{\'n}, Poland} 
\affiliation{European Southern Observatory, Karl-Schwarzschild-Str. 2, 85748 Garching bei M\"{u}nchen, Germany}
\affiliation{Universit\"{a}ts-Sternwarte, Ludwig-Maximilians-Universit\"{a}t, Scheinerstrasse 1, 81679 M\"{u}nchen, Germany}

\author[0000-0001-8913-925X]{Agata Karska}
\affiliation{Institute of Astronomy, Faculty of Physics, Astronomy and Informatics, Nicolaus Copernicus University, Grudzi{\k a}dzka 5, 87-100 Toru{\'n}, Poland} 
\affiliation{Max-Planck-Institut für Radioastronomie, Auf dem Hügel 69, 53121, Bonn, Germany}

\author[0000-0003-2248-6032]{Marta Sewi{\l}o}
\affiliation{Exoplanets and Stellar Astrophysics Laboratory, NASA Goddard Space Flight Center, Greenbelt, MD 20771, USA}
\affiliation{Center for Research and Exploration in Space Science and Technology, NASA Goddard Space Flight Center, Greenbelt, MD 20771}
\affiliation{Department of Astronomy, University of Maryland, College Park, MD 20742, USA}

\author{Lars E. Kristensen}
\affiliation{Niels Bohr Institute, Centre for Star and Planet Formation, University of Copenhagen, \O ster Voldgade 5-7, 1350 Copenhagen K, Denmark} 

\author{Gregory J. Herczeg}
\affiliation{Kavli Institute for Astronomy and Astrophysics, Peking University, Yiheyuan 5, Haidian Qu, 100871 Beijing, China}
\affiliation{Department of Astronomy, Peking University, Yiheyuan 5, Haidian Qu, 100871 Beijing, China}

\author{Suzanne Ramsay}
\affiliation{European Southern Observatory, Karl-Schwarzschild-Str. 2, 85748 Garching bei M\"{u}nchen, Germany}

\author[0000-0002-3747-2496]{William J. Fischer}
\affiliation{Space Telescope Science Institute, 3700 San Martin Dr., Baltimore, MD 21218, USA}

\author{Beno\^{i}t Tabone}
\affiliation{Leiden Observatory,Leiden University, PO Box 9513, 2300 RA Leiden, The Netherlands}
\affiliation{Institut d'Astrophysique Spatiale, Universit\'e Paris-Saclay, CNRS,  B$\hat{a}$timent 121, 91405 Orsay Cedex, France}

\author{Will R. M. Rocha}
\affiliation{Laboratory for Astrophysics, Leiden Observatory,Leiden University, PO Box 9513, 2300 RA Leiden, The Netherlands}

\author[0000-0001-5785-1154]{Maciej Koprowski}
\affiliation{Institute of Astronomy, Faculty of Physics, Astronomy and Informatics, Nicolaus Copernicus University, Grudzi{\k a}dzka 5, 87-100 Toru{\'n}, Poland} 

\author{Ng\^{a}n L\^{e}}
\affiliation{Institute of Astronomy, Faculty of Physics, Astronomy and Informatics, Nicolaus Copernicus University, Grudzi{\k a}dzka 5, 87-100 Toru{\'n}, Poland}

\author{Beata Deka-Szymankiewicz}
\affiliation{Institute of Astronomy, Faculty of Physics, Astronomy and Informatics, Nicolaus Copernicus University, Grudzi{\k a}dzka 5, 87-100 Toru{\'n}, Poland}  

\begin{abstract}

{The effects of metallicity on the evolution of protoplanetary disks may be studied in the outer Galaxy where the metallicity is lower than in the solar neighborhood.} We present the VLT/KMOS integral field spectroscopy in the near-infrared of $\sim$120 candidate {young stellar objects (YSOs)} in the CMa-$\ell$224 star-forming region located at a Galactocentric distance of 9.1 kpc. 
We characterise the YSO accretion luminosities and accretion 
rates using the hydrogen Br$\gamma$ 
emission {and find the median accretion luminosity of $\log{(L_{\rm acc})} = -0.82^{+0.80}_{-0.82} L_\odot$}.  
{Based on the measured accretion luminosities, we investigate the hypothesis of star formation history in the CMa-$\ell$224. Their median values suggest that Cluster C, where most of YSO candidates have been identified, might be the most evolved part of the region}. 
The accretion luminosities are {similar to} those observed toward low-mass YSOs in the Perseus and Orion molecular clouds, {and do not reveal the impact of lower metallicity.}
Similar studies in other outer Galaxy clouds covering a wide range of metallicities are critical to gain a complete picture of star formation in the Galaxy. 

\end{abstract}


\vspace{0.5cm}
\section{Introduction}
\label{sec:intro}
Stars form as a result of complex physico-chemical processes initiated by the gravitational collapse of a dense and cold molecular cloud. The formation of a rotating envelope and the embedded disk is associated with the ejection of jets and disk winds, together responsible for the removal of the angular momentum \citep[e.g.,][]{fr14}. The interaction between jets/winds and the envelope leads to the formation of outflow cavities and the dispersion of some mass reservoir \citep{vK10,visser12}. { Shock waves at the outflow/envelope interface compress and heat the envelope material to hundreds or thousands of K, even around low-mass protostars \citep{krist17,kar18}.} 

The net mass growth of a young star is a balance between the mass accretion from the envelope-disk system and the mass ejection by jets and winds. The main accretion phase occurs during the earliest evolutionary stages of a young stellar object (YSO; Class 0 and Class I), {accompanied by collimated H$_2$ jets \citep{davis95,stanke02,kristensen2007} and more extended molecular outflows \citep{marel2013,tobin2016,mottram2017}. Once the protostar evolves into Class II, the envelope mass reservoir is mostly depleted, the accretion rate decreases, and the outflow opening angle widens \citep{offner11,agra2014}.} The gradual decrease of accretion rates { \citep{manara2012,ansdell2017,testi2022}} is accompanied by decreasing mass loss rates and a transition from mostly molecular to atomic/ionic outflows \citep{ni15,bally16}.
\begin{figure*}
\includegraphics[width=\textwidth]{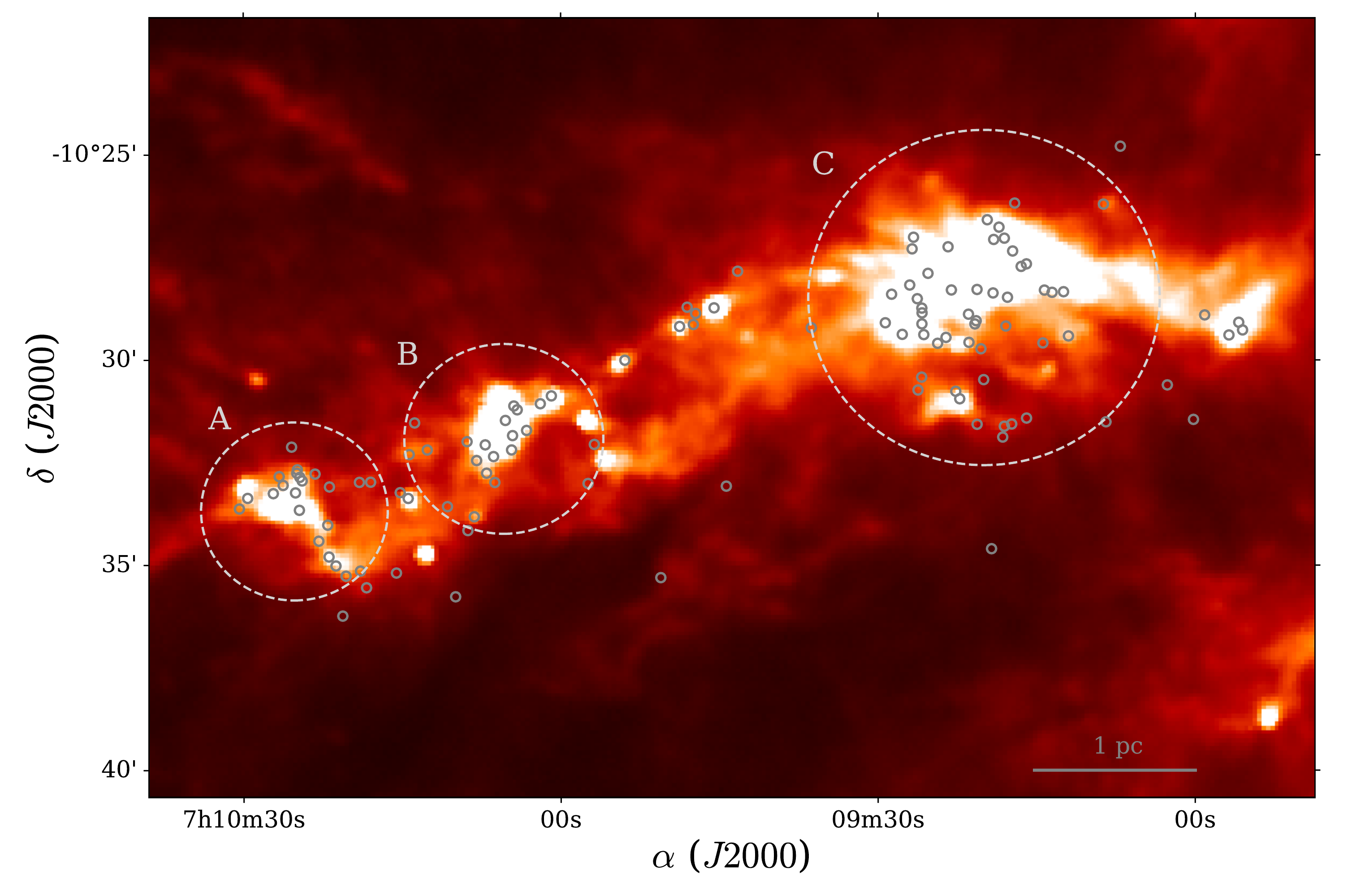}
\caption{\textit{Herschel}/SPIRE 250~$\mu$m image of the main filament in the CMa-$\ell$224 star-forming region. Grey circles show the positions of YSO candidates observed with VLT/KMOS, while light-grey dashed ellipses indicate regions hosting clusters of sources with the extended 4.5 $\mu$m emission (most likely tracing outflows) defined in \cite[Clusters A, B, and C]{sewilo2019}. North is up, east to the left.} 
\label{fig:pointing}
\end{figure*}

The location of a star-forming region in the Galaxy may influence the process of mass assembly. In the outer Galaxy, the gas surface density in molecular clouds is known to be lower than those in the solar neighborhood \citep{roman2010}. The decrease of metallicity with Galactocentric radius, traced by Fe or O abundance gradients, also translates to lower dust and molecular gas abundances { \citep{sodroski1997,lepine2011,hawkins2022}}. Additionally, lower cosmic-ray fluxes and the interstellar UV radiation field reduce the amount of gas and dust heating \citep{bloemen84}. These factors may affect physical and chemical conditions in star forming regions and result in \textit{globally} lower star formation rates and efficiencies in the outer Galaxy \citep{ke12,hd15,dj19}. 

Recent infrared studies revealed significant star formation activity in the Canis Major star-forming region in the outer Galaxy \citep{fischer16,sewilo2019}.
The 2.$^{\circ}5\times1.^{\circ}4$ region in Canis Major dubbed CMa-$\ell224$ centered on ($l$, $b$)=(224.$^\circ$5, $-$0.$^\circ$65) is particularly interesting because it hosts clusters of sources with extended 4.5 $\mu$m emission (likely tracing outflows), many of which were identified as YSOs \citep{sewilo2019}. CMa-$\ell224$ is located at a distance of 0.92~kpc from the Sun and 9.1 kpc from the Galactic Center (e.g., \citealt{claria1974}), 
where a subsolar metallicity is predicted by the O/H Galactocentric radial gradients determined based on the observations of H\,{\sc ii} regions ($Z$$\sim$$0.55-0.73$ $Z_\odot$: \citealt{balser11,fernandez2017,esteban18}). CMa-$\ell224$ corresponds to a { local} peak of the H$_2$ column density and the lowest { dust} temperatures in the far-infrared observations of the outer Galaxy by the \lq\lq Herschel infrared Galactic Plane Survey'' (Hi-GAL; \citealt{molinari10}). { The region} contains a  complex network of star-forming filaments \citep[Figure \ref{fig:pointing},][]{elia2013,schisano214} and a widespread emission in CO and { its} isotopologues, { tracing} dense, molecular gas \citep{olmi16,benedet20,lin21}.

\cite{sewilo2019} identified 294 YSO candidates in CMa-$\ell$224 based on the photometric data from the near- and mid-infrared catalogs: \textit{Spitzer}'s \lq\lq GLIMPSE360: Completing the Spitzer Galactic Plane Survey'' (PI: B. Whitney), the Two Micron All Sky Survey (2MASS; \citealt{skrutskie06}), and the AllWISE catalog that combines the \lq\lq Wide-field Infrared Survey Explorer'' (WISE; \citealt{wright2010}) and NEOWISE \citep{mainzer11} data. Some YSO candidates in CMa-$\ell$224 are associated with the extended 4.5~$\mu$m emission (Extended Green Objects, EGOs; \citealt{cyganowski08}), likely dominated by the H$_2$ line emission from outflow shocks (e.g., \citealt{cyganowski2011}). The Spectral Energy Distribution (SED) fitting with the \cite{robit17} YSO models identified 37 sources with envelopes, i.e., Class 0/I YSOs. Thus,  CMa-$\ell$224 offers an opportunity to evaluate how metallicity affects the ongoing star formation in a region that is significantly closer than, for example, the Magellanic Clouds. 

In this paper, we address the following questions. 
What are the mass accretion  
rates in YSOs in CMa-$\ell$224? How are they connected to the star formation scenarios in this region? Is there any impact of the reduced gas metallicity on the accretion 
properties of YSOs in CMa-$\ell$224?

To this end, we present the results of the { integral field} spectroscopy of 124 YSO candidates { in the CMa-$\ell$224 star-forming region using the $K$-band Multi Object Spectrograph (KMOS; \citealt{sharples2013}) on the Very Large Telescope (VLT). VLT/KMOS provides $2\rlap.{''}8\times2\rlap.{''}8$ spectral maps in ro-vibrational H$_2$ lines, hydrogen Br$\gamma$, and CO bandhead at 2.3~$\mu$m. At the distance of CMa-$\ell$224, we obtained $\sim$$2580\times2580$~au maps at the physical resolution of 184 au.}

The paper is organized as follows. In Section 2, we describe our sample selection, KMOS observations, and data reduction. In Section 3, we present the continuum and line emission maps, provide the statistics on line detections, discuss gas spatial distribution and describe calculations of the 
mass accretion rates. In Section 4, we discuss the results in the context of the star formation scenarios in CMa-$\ell$224, and in Section 5, we present the summary and conclusions. 

{This paper is the first in a series presenting the multi-wavelength spectroscopy of YSO candidates in the CMa-$\ell$224 star-forming region. {Two forthcoming} papers will discuss the $^{13}$CO and C$^{18}$O 2-1 observations of YSO candidates in the main filament in CMa-$\ell$224 with the {Atacama Large Millimeter/submillimeter Array} (ALMA; Koprowski et al., in prep.), and the spectral types and excess continuum measurements toward YSO candidates in the second brightest filament in the region using the SpeX instrument at the NASA Infrared Telescope Facility (IRTF; Le et al., in prep.).}

\newpage
\section{KMOS Observations and Data Reduction} \label{sec:obs}

We used VLT/KMOS to observe YSO candidates in the CMa-$\ell$224 star-forming region as part of the ESO programme 0102.C-0914(A) (PI:~A.~Karska). The sources were selected from the catalogue of YSO candidates from \cite{sewilo2019} based { on their 2MASS $K$-band brightness} and location within two KMOS patrol fields.  

Targets were selected from the $K$-band magnitude range of 9.5–15 to ensure a sufficient signal-to-noise ratio (SNR) without saturation for a single integration time applied to all exposures. 

KMOS is a second-generation instrument operating in the near-infrared (near-IR) at the B Nasmyth focus of the VLT Unit Telescope 1 on Paranal in Chile, in operation since November 2012 \citep{sharples2013}. It performs Integral Field Spectroscopy (IFS) for up to 24 targets simultaneously. Each of the 24 pick-off arms is connected to one of three identical spectrographs and detectors. KMOS arms are allocated in two planes (\lq\lq top'' and \lq\lq bottom'') to avoid interference between them. The patrol field is 7$\rlap.{'}$2 in diameter, the size of each IFU is $2\rlap.{''}8\times2\rlap.{''}8$, and the spatial sampling (spaxel size) is $0\rlap.{''}2\times0\rlap.{''}2$. 

Observations were prepared using the KMOS Arm Allocator (KARMA\footnote{\url{https://www.eso.org/sci/observing/phase2/SMGuidelines/KARMA.html}}) and {\tt p2}\footnote{\url{https://www.eso.org/sci/observing/phase2/p2intro.html}}, a web-based tool for preparation of the Phase 2 material. The observations were performed in October and December 2018. 

We used the K grating with the spectral range from 1.93 to 2.50 $\mu$m and { resolving power $R$ of $\sim$4000, with a spectral sampling $\Delta\lambda$ of $\sim$2.8~\AA~per spectral element}, corresponding to $\sim$39.7~km~s$^{-1}$ at 2.1218~$\mu$m. The total integration time for each pointing was 1300~s with 5 single exposures of 260~s, dithered by 0$\rlap.{''}$2. 

The Nod-to-Sky mode was used to observe sources in both the \lq science\rq~and \lq sky\rq~pointings. { The arms originally observing targets were next directed to the sky area, while those observing the sky were capturing science objects in the \lq sky\rq~pointing, usually offset by a few arcmin.} One of the pick-off arms was broken during semester 102, which resulted in a lack of observations of 6 sources. 

{Table \ref{tab:kmos:ob} shows the summary of the atmospheric conditions during our KMOS observations. The amount of precipitatable water vapour (PWV) was in the range from $\sim$2.5 to 4~mm, and the sky was covered with thin cirrus (TN). Seeing measured at the observatory site was typically below 1$''$ with the point spread function (PSF) of the data in the 0$\rlap.{''}$4--0$\rlap.{''}$7 range. }

\begin{table}[h!]
\caption{Atmospheric Conditions} 
\begin{center}
\scriptsize
\label{tab:kmos:ob} 
\begin{tabular}{ccccccccc}
\hline
OB&   Date    & Seeing & \multicolumn{2}{c}{Airmass\tablenotemark{a}} & PWV & Sky & Grade\\
    &         & ('')   & SCI & STD & (mm) &     & \\ 
\hline
 1 & 23.10.2018 & $\sim$1.5 & 1.20 & 1.70 & $<$2.5    & TN & B \\ 
 2 & 24.10.2018 & $\sim$0.5 & 1.18 & 1.15 & $\sim$2.5 & TN & B \\ 
 3 & 25.10.2018 & $\sim$0.6 & 1.25 & 1.83 & $\sim$2.5 & TN & B \\
 4 & 11.12.2018 & $\sim$1.0 & 1.34 & 1.20 & $\sim$4.0 & TN & B \\
 5 & 11.12.2018 & $<$0.6    & 1.13 & 1.20 & $\sim$4.0 & TN & B \\
 6 & 11.12.2018 & $\sim$0.5 & 1.05 & 1.20 & $>$3.0    & TN & B \\
\hline
\end{tabular}
\end{center}
\tablenotetext{a}{A mean airmass during the integration made for one of the targets per OB (SCI) and for a standard star used for telluric correction (STD).}
\end{table}

Data reduction was performed using the ESO Recipe Execution Tool (\texttt{esorex}, version 2.0.2 of the KMOS pipeline), the terminal-based software. The standard procedure included the processing of the raw calibration data: dark, flat-field, illumination, { and sky subtraction (a single sky observation per science target). The wavelength calibration was done using the Argon -- Neon lamp. } 

{We used a standard procedure for the telluric correction and flux calibration by observing a standard star, one per each of the three detectors, and comparing observations to the stellar model.  
The standard pipeline does not account for the difference in the airmass between the science and standard star observations, which might lead to under- or over-estimation of the telluric lines of up to 10\%. Additionally, each of the 24 arms have a slightly different spectral resolution resulting in different line shapes. }
{ It is possible to account for these effects by modeling the telluric lines using the science or the standard star spectrum \citep{coccato2019}. However, for our science data with the median SNR measured on the featureless parts of the spectra of less than 50, the differences between the standard method we used and modelling are negligible since the noise dominates any uncertainty coming from the telluric standard \citep{coccato2019}.} 

{
The imperfect telluric correction is a source of the unknown uncertainty, especially in the red part of the $K$-band (above 2.4~$\mu$m), where some of the H$_2$ lines are located. Also, other telluric features are poorly corrected, e.g., near 2~$\mu$m. All spectral regions particularly affected by the telluric lines are marked on the figures.}

The reduced { single exposures were} collapsed into the final data cubes. The entire process of creating the 3D KMOS science data cubes is described in \cite{davies2013}. For the analysis of sources observed multiple times, we used the data cubes combined with the KMOS pipeline task {\tt combine}.

The $K$-band continuum fluxes are calculated by fitting a 7th order Chebyshev polynomial to the spectrum in each spaxel (spatial pixel) of the data cube. The continuum level is defined as the value of the fit at 2.12~$\mu$m, near the middle of the Johnson $K$-band and the rest wavelength of the 1-0 S(1) H$_2$ line. The noise of the spectrum is calculated as the root mean square (RMS) of deviations from the continuum in the line-free parts of the spectrum: 2.076--2.106~$\mu$m, 2.125--2.143~$\mu$m, 2.179--2.187~$\mu$m, 2.198--2.201~$\mu$m, and 2.268--2.274~$\mu$m. In four cases of faint objects (No. 11, 14, 26, 69), we integrate the continuum emission over the full spectral range in each spaxel to detect these sources at a $>$3$\sigma$ level and verify their coordinates; the spectral analysis remained unchanged. 

\startlongtable
\begin{deluxetable*}{lcccccll}
\tablewidth{0pt}
\tablecaption{Catalog of Sources in CMa-$\ell$224 Detected with VLT/KMOS\label{tab:coordinates}}
\tablehead{
\colhead{No.} &
\colhead{IRAC Designation\tablenotemark{a}} &
\colhead{RA} & 
\colhead{DEC} & 
\colhead{Continuum near 2.12~$\mu$m} &
\colhead{Class} &
\colhead{YSO\tablenotemark{b}} &
\colhead{Remarks} \\
\colhead{~} & 
\colhead{SSTGLMA} &
\colhead{($^h$ $^m$ $^s$)} &
\colhead{($^o$ $\arcmin$ $\arcsec$)} &
\colhead{(10$^{-17}$ erg s$^{-1}$ cm$^{-2}$ \AA$^{-1}$)} &
\colhead{~} & \colhead{~} & \colhead{~}
}
\startdata
1   &	G224.2025-00.8569	&	 07 09 07.09 	&	 -10 24 47.50 	&	 08.27 	$\pm$	 0.12 &	 II 	&	 d 	&	 \\
2  	&	G224.2265-00.8620	&	 07 09 08.69 	&	 -10 26 12.22 	&	 26.40 	$\pm$	 0.13 &	 II 	&	 e+d 	&	 ext. H$_2$ emission\\
3  	&	G224.2420-00.8313	&	 07 09 17.08 	&	 -10 26 10.67 	&	 13.05 	$\pm$	 0.14 &	 II 	&	 d 	&	 \\
4  	&	G224.2449-00.9307	&	 07 08 55.89 	&	 -10 29 04.97 	&	 07.02 	$\pm$	 0.21 &	 I 	&	 null 	&	 ext. H$_2$ emission\\
5  	&	G224.2470-00.9334	&	 07 08 55.53 	&	 -10 29 16.20 	&	 38.74 	$\pm$	 0.60 &	 I 	&	 e 	&	 \\
6  	&	G224.2483-00.9176	&	 07 08 59.12 	&	 -10 28 54.25 	&	 11.22 	$\pm$	 0.14 &	 II 	&	 d 	&	 \\
7  	&	G224.2512-00.9297	&	 07 08 56.82 	&	 -10 29 23.57 	&	 07.49 	$\pm$	 0.29 &	 I 	&	 null 	&	 \\
8  	&	G224.2530-00.8250	&	 07 09 19.67 	&	 -10 26 35.06 	&	 07.84 	$\pm$	 0.15 &	 II 	&	 d 	&	 \\
9  	&	G224.2535-00.8306	&	 07 09 18.56 	&	 -10 26 45.89 	&	 10.17 	$\pm$	 0.10 &	 II 	&	 d 	&	 \\
10 	&	G224.2567-00.8346	&	 07 09 18.06 	&	 -10 27 02.03 	&	 11.21 	$\pm$	 0.13 &	 II 	&	 d 	&	 near edge\tablenotemark{d}  \\
11\tablenotemark{c} 	&	G224.2591-00.8305	&	 07 09 19.19 	&	 -10 27 03.71 	&	 2.52 	$\pm$	 0.57	&	 I 	&	 null 	&	 integrated cont.\\
12  	&	G224.2598-00.8398	&	 07 09 17.27 	&	 -10 27 20.77 	&	 28.52 	$\pm$	 0.13	&	 II 	&	 d 	&	 near edge \\
13 	&	G224.2621-00.8470	&	 07 09 15.97 	&	 -10 27 39.63 	&	 13.70 	$\pm$	 0.08 &	 I 	&	 d 	&	 near edge\\
14\tablenotemark{c}   	&	G224.2638-00.8456	&	 07 09 16.47 	&	 -10 27 43.32 	&	 2.88 	$\pm$	 0.38 &	 I 	&	 null 	&	 integrated cont.\\
15  	&	G224.2652-00.8647	&	 07 09 12.46 	&	 -10 28 20.32 	&	 13.27 	$\pm$	 0.12 &	 I 	&	 d 	&	 ext. H$_2$ emission\\
16  	&	G224.2674-00.8609	&	 07 09 13.54 	&	 -10 28 21.20 	&	 31.76 	$\pm$	 0.17 &	 II 	&	 d 	&	 \\
17  	&	G224.2680-00.8578	&	 07 09 14.27 	&	 -10 28 17.95 	&	 3.29 	$\pm$	 0.12 &	 II 	&	 null 	&	 \\
18  	&	G224.2697-00.8168	&	 07 09 23.38 	&	 -10 27 14.62 	&	 5.29 	$\pm$	 0.07 &	 II 	&	 null 	&	 \\
19  	&	G224.2724-00.8032	&	 07 09 26.64 	&	 -10 27 00.62 	&	 3.11 	$\pm$	 0.11 &	 I 	&	 d 	&	near edge, ext. H$_2$ emission\\
20  	&	G224.2769-00.8049	&	 07 09 26.78 	&	 -10 27 17.65 	&	 20.18 	$\pm$	 0.10 &	 II 	&	 d 	&	 near edge \\
21A 	&	G224.2778-00.8465	&	 07 09 17.76 	&	 -10 28 28.62 	&	 0.85 	$\pm$	 0.13 &	 I 	&	 null 	&	 on edge \\
21B 	&	G224.2778-00.8465	&	 07 09 17.83 	&	 -10 28 29.02 	&	 0.44 	$\pm$	 0.12 &  	&	  	&	 separation 1073 au \\
22  	&	G224.2784-00.8408	&	 07 09 19.14 	&	 -10 28 21.54 	&	 11.88 	$\pm$	 0.06 &	 II 	&	 null 	&	near edge \\
23  	&	G224.2800-00.8346	&	 07 09 20.64 	&	 -10 28 17.05 	&	 25.24 	$\pm$	 0.14 &	 I 	&	 null 	&	 \\
24  	&	G224.2802-00.8747	&	 07 09 12.00 	&	 -10 29 25.00 	&	 15.79 	$\pm$	 0.35 &	 II 	&	 d 	&	 \\
25  	&	G224.2802-00.9179	&	 07 09 02.64 	&	 -10 30 36.58 	&	 13.07 	$\pm$	 0.31 &	 II 	&	 null 	&	 \\
26\tablenotemark{c}  	&	G224.2828-00.8148	&	 07 09 25.28 	&	 -10 27 53.42 	&	 1.07 	$\pm$	 0.08 &	 I 	&	 d 	&	 integrated cont.\\
28  	&	G224.2872-00.8673	&	 07 09 14.41 	&	 -10 29 35.21 	&	 24.28 	$\pm$	 0.40 &	 II 	&	 d 	&	 \\
29  	&	G224.2881-00.9333	&	 07 09 00.17 	&	 -10 31 27.03 	&	 20.35 	$\pm$	 0.28 &	 II 	&	 null 	&	 \\
30  	&	G224.2882-00.8515	&	 07 09 17.92 	&	 -10 29 10.51 	&	 13.91 	$\pm$	 0.10 &	 II 	&	 d 	&	 near edge \\
31  	&	G224.2903-00.8108	&	 07 09 27.00 	&	 -10 28 10.52 	&	 2.08 	$\pm$	 0.14 &	 II 	&	 null 	&	 \\
32  	&	G224.2905-00.8364	&	 07 09 21.45 	&	 -10 28 53.19 	&	 68.62 	$\pm$	 0.10 &	 II 	&	 d 	&	 \\
33  	&	G224.2914-00.8402	&	 07 09 20.72 	&	 -10 29 02.78 	&	 6.02 	$\pm$	 0.09 &	 II 	&	 d 	&	 \\
34  	&	G224.2928-00.8403	&	 07 09 20.84 	&	 -10 29 07.42 	&	 11.50 	$\pm$	 0.07 &	 II 	&	 null 	&	 \\
35  	&	G224.2939-00.8157	&	 07 09 26.29 	&	 -10 28 30.56 	&	 13.85 	$\pm$	 0.21 &	 II 	&	 e+d 	&	 \\
36A 	&	G224.2965-00.8191	&	 07 09 25.86 	&	 -10 28 44.29 	&	 98.54 	$\pm$	 0.41 &	II	&	d	&	separation 1086 au \\
36B 	&	G224.2965-00.8191	&	 07 09 25.94 	&	 -10 28 44.29 	&	 6.51 	$\pm$	 0.04 &	 	&	  	&	on edge \\
38A 	&	G224.2983-00.8203	&	 07 09 25.84 	&	 -10 28 51.25 	&	 12.54 	$\pm$	 0.10 &	 I 	&	 null 	&	 \\
38B 	&	G224.2983-00.8203	&	 07 09 25.78 	&	 -10 28 52.25 	&	 2.84 	$\pm$	 0.09 &  	&	 	&	 separation 1178 au\\
39  	&	G224.3005-00.8417	&	 07 09 21.41 	&	 -10 29 34.38 	&	 193.49 	$\pm$	 0.67 &	 II 	&	 d 	&	 ext. H$_2$ emission \\
40  	&	G224.3006-00.8470	&	 07 09 20.27 	&	 -10 29 43.74 	&	 680.28 	$\pm$	 15.78	&	 II 	&	 d 	&	 \\
41  	&	G224.3022-00.8221	&	 07 09 25.85 	&	 -10 29 07.35 	&	 23.01 	$\pm$	 0.10 &	 I 	&	 null 	&	 \\
42  	&	G224.3028-00.8329	&	 07 09 23.58 	&	 -10 29 27.12 	&	 14.65 	$\pm$	 0.29 &	 II 	&	 d 	&	 \\
43  	&	G224.3044-00.9038	&	 07 09 08.45 	&	 -10 31 30.57 	&	 14.57 	$\pm$	 0.17 &	 II 	&	 d 	&	 \\
44  	&	G224.3057-00.8247	&	 07 09 25.68 	&	 -10 29 23.07 	&	 131.32 	$\pm$	 0.67	&	 I 	&	 null 	&	 \\
45  	&	G224.3063-00.8310	&	 07 09 24.37 	&	 -10 29 35.76 	&	 17.46 	$\pm$	 0.55 &	 II 	&	 d 	&	 \\
46  	&	G224.3083-00.8093	&	 07 09 29.31 	&	 -10 29 05.72 	&	 113.91 	$\pm$	 0.13	&	 II 	&	 d 	&	 ext. H$_2$ emission\\
47  	&	G224.3095-00.8172	&	 07 09 27.71 	&	 -10 29 22.66 	&	 3.73 	$\pm$	 0.08 &	 I 	&	 e 	&	 ext. H$_2$ emission\\
48  	&	G224.3112-00.8536	&	 07 09 20.01 	&	 -10 30 28.89 	&	 41.21 	$\pm$	 0.05 &	 II 	&	 d 	&	 \\
49  	&	G224.3159-00.7490	&	 07 09 43.27 	&	 -10 27 50.24 	&	 13.38 	$\pm$	 0.12 &	 I 	&	 d 	&	 near edge \\
50  	&	G224.3173-00.8757	&	 07 09 15.96 	&	 -10 31 25.09 	&	 12.63 	$\pm$	 0.07 &	 II 	&	 d 	&	 \\
51  	&	G224.3204-00.8462	&	 07 09 22.65 	&	 -10 30 45.91 	&	 29.36 	$\pm$	 0.50 &	 II 	&	 null 	&	 \\
52A 	&	G224.3215-00.8319	&	 07 09 25.88 	&	 -10 30 25.24 	&	 08.71 	$\pm$	 0.02 &	 II 	&	 null 	&	 \\
52B 	&	G224.3215-00.8319	&	 07 09 25.83 	&	 -10 30 25.24 	&	 6.85 	$\pm$	 0.02 &  	&	 	&	 separation 552 au \\
53  	&	G224.3222-00.8717	&	 07 09 17.36 	&	 -10 31 33.78 	&	 19.90 	$\pm$	 0.46 &	 II 	&	 d 	&	 \\
54  	&	G224.3225-00.8490	&	 07 09 22.28 	&	 -10 30 56.90 	&	 702.81 	$\pm$	 1.86	&	 I 	&	 null 	&	 ext. H$_2$ emission \\
55A 	&	G224.3233-00.7848	&	 07 09 36.31 	&	 -10 29 12.99 	&	 33.38 	$\pm$	 0.13 &	 II 	&	 d 	&	 \\
55B 	&	G224.3233-00.7848	&	 07 09 36.27 	&	 -10 29 13.19 	&	 27.61 	$\pm$	 0.08 &  	&	 	&	 separation 582 au\\
56  	&	G224.3243-00.8696	&	 07 09 18.06 	&	 -10 31 37.05 	&	 11.24 	$\pm$	 0.22 &	 II 	&	 d 	&	 \\
57  	&	G224.3268-00.8331	&	 07 09 26.22 	&	 -10 30 44.05 	&	 10.33 	$\pm$	 0.06 &	 II 	&	 null 	&	 \\
58  	&	G224.3285-00.8597	&	 07 09 20.64 	&	 -10 31 34.20 	&	 134.72 	$\pm$	 1.35	&	 II 	&	 d 	&	  \\
59  	&	G224.3286-00.8709	&	 07 09 18.21 	&	 -10 31 52.88 	&	 21.39 	$\pm$	 0.09 &	 II 	&	 d 	&	 \\
60  	&	G224.3333-00.7476	&	 07 09 45.51 	&	 -10 28 43.81 	&	 33.59 	$\pm$	 0.13 &	 I 	&	 e 	&	 ext. H$_2$ emission\\
61  	&	G224.3379-00.7384	&	 07 09 48.05 	&	 -10 28 42.87 	&	 29.73 	$\pm$	 0.31 &	 II 	&	 d 	&	 near edge \\
62  	&	G224.3386-00.7425	&	 07 09 47.25 	&	 -10 28 51.88 	&	 08.02 	$\pm$	 0.27 &	 II 	&	 d 	&	 near edge \\
63  	&	G224.3430-00.7437	&	 07 09 47.47 	&	 -10 29 08.11 	&	 18.24 	$\pm$	 0.30 &	 II 	&	 d 	&	 near edge \\
64  	&	G224.3462-00.7394	&	 07 09 48.76 	&	 -10 29 10.99 	&	 5.18 	$\pm$	 0.12 &	 I 	&	 e+d 	&	near edge, ext. H$_2$ emission\\
65  	&	G224.3684-00.7267	&	 07 09 53.94 	&	 -10 30 00.37 	&	 152.96 	$\pm$	 1.71	&	 II 	&	 e+d 	&	  \\
66  	&	G224.3710-00.8882	&	 07 09 19.27 	&	 -10 34 36.14 	&	 07.37 	$\pm$	 0.14	&	 II 	&	 null 	&	 near edge \\
68  	&	G224.3955-00.7851	&	 07 09 44.35 	&	 -10 33 04.55 	&	 67.11 	$\pm$	 0.99	&	 II 	&	 d 	&	 \\
69\tablenotemark{c}   	&	G224.3986-00.7060	&	 07 10 01.84 	&	 -10 31 02.49 	&	 00.08 	$\pm$	 0.14	&	 I 	&	 null 	&	 integrated cont.\\
70  	&	G224.4042-00.7319	&	 07 09 56.84 	&	 -10 32 03.40 	&	 08.36 	$\pm$	 0.08	&	 II 	&	 d 	&	 \\
71  	&	G224.4047-00.6971	&	 07 10 04.45 	&	 -10 31 07.18 	&	 16.38 	$\pm$	 0.26	&	 II 	&	 d 	&	 \\
72  	&	G224.4055-00.6989	&	 07 10 04.13 	&	 -10 31 12.79 	&	 14.10 	$\pm$	 0.34	&	 II 	&	 null 	&	uncorr. H$_2$ emission\tablenotemark{e} \\
73  	&	G224.4111-00.7059	&	 07 10 03.25 	&	 -10 31 42.93 	&	 10.08 	$\pm$	 0.04	&	 I 	&	 d 	&	 \\
74  	&	G224.4114-00.6969	&	 07 10 05.25 	&	 -10 31 28.33 	&	 3.29 	$\pm$	 0.15	&	 II 	&	 null 	&	 \\
75  	&	G224.4157-00.7023	&	 07 10 04.57 	&	 -10 31 50.47 	&	 6.83 	$\pm$	 0.15	&	 II 	&	 d 	&	 \\
76  	&	G224.4195-00.7372	&	 07 09 57.43 	&	 -10 33 00.55 	&	 11.10 	$\pm$	 0.33	&	 II 	&	 d 	&	 \\
77  	&	G224.4210-00.7045	&	 07 10 04.67 	&	 -10 32 11.44 	&	 2.83 	$\pm$	 0.03	&	 I 	&	 e+d 	&	 \\
78  	&	G224.4238-00.6944	&	 07 10 07.16 	&	 -10 32 04.00 	&	 13.27 	$\pm$	 0.05	&	 II 	&	 null 	&	uncorr. H$_2$ emission\\
79  	&	G224.4259-00.6877	&	 07 10 08.87 	&	 -10 31 59.31 	&	 6.82 	$\pm$	 0.16	&	 II 	&	 d 	&	 \\
80  	&	G224.4266-00.6996	&	 07 10 06.37 	&	 -10 32 21.11 	&	 35.91 	$\pm$	 0.22	&	 II 	&	 null 	&	uncorr. H$_2$ emission\\
81  	&	G224.4285-00.6662	&	 07 10 13.83 	&	 -10 31 31.86 	&	 25.17 	$\pm$	 0.75	&	 II 	&	 d 	&	 \\
82A 	&	G224.4309-00.6943	&	 07 10 07.97 	&	 -10 32 26.94 	&	 08.97 	$\pm$	 0.11	&	 II 	&	 e+d 	&	 \\
82B 	&	G224.4309-00.6943	&	 07 10 08.00 	&	 -10 32 26.34 	&	 07.96 	$\pm$	 0.25	&	  	&	 	&	 separation 663 au \\
83  	&	G224.4338-00.7003	&	 07 10 07.03 	&	 -10 32 45.39 	&	 4.11 	$\pm$	 0.21	&	 I 	&	 e+d 	&	 ext. H$_2$ emission? \\
84  	&	G224.4356-00.7049	&	 07 10 06.25 	&	 -10 32 59.03 	&	 67.02 	$\pm$	 1.28	&	 II 	&	 e 	&	 \\
85  	&	G224.4361-00.6756	&	 07 10 12.64 	&	 -10 32 11.47 	&	 26.07 	$\pm$	 0.07	&	 II 	&	 null 	&	 \\
86A 	&	G224.4404-00.7799	&	 07 09 50.55 	&	 -10 35 18.30 	&	 08.66 	$\pm$	 0.08	&	 II 	&	 d 	&	 \\
86B 	&	G224.4404-00.7799	&	 07 09 50.55 	&	 -10 35 19.30 	&	 5.73 	$\pm$	 0.08	&	  	&	 	&	 separation 920 au \\
87A 	&	G224.4410-00.6702	&	 07 10 14.35 	&	 -10 32 18.18 	&	 1.89 	$\pm$	 0.11	&	 I 	&	 d 	&	 ext. H$_2$ emission? \\
87B 	&	G224.4410-00.6702	&	 07 10 14.42 	&	 -10 32 18.38 	&	 1.58 	$\pm$	 0.17	&	  	&	 	&	 separation 938 au\\
88	&	G224.4514-00.7054	&		&		&		&	 	&	 	&	not det., uncorr. H$_2$ emission\\
89A 	&	G224.4528-00.6930	&	 07 10 10.73 	&	 -10 33 34.32 	&	 4.21 	$\pm$	 0.03	&	 II 	&	 null 	&	 \\
89B 	&	G224.4528-00.6930	&	 07 10 10.76 	&	 -10 33 34.32 	&	 4.27 	$\pm$	 0.05	&  	&	 	&	hardly resolved, sep. 368 au\\
90  	&	G224.4564-00.6744	&	 07 10 15.20 	&	 -10 33 13.83 	&	 09.90 	$\pm$	 0.15 &	 I 	&	 null 	&	 \\
91A 	&	G224.4567-00.6781	&	 07 10 14.44 	&	 -10 33 22.29 	&	 91.33 	$\pm$	 1.19	&	 I 	&	 null 	&	 ext. H$_2$ emission \\
91B 	&	G224.4567-00.6781	&	 07 10 14.43 	&	 -10 33 21.49 	&	 22.22 	$\pm$	 0.10 &  	&	  	&	ext. H$_2$ emission, sep. 759 au\\
92  	&	G224.4577-00.7046	&	 07 10 08.80 	&	 -10 34 09.34 	&	 17.79 	$\pm$	 0.57 &	 null 	&	 d 	&	 \\
93  	&	G224.4578-00.6621	&	 07 10 18.00 	&	 -10 32 58.47 	&	 16.30 	$\pm$	 0.06 &	 II 	&	 d 	&	 \\
94  	&	G224.4593-00.6284	&	 07 10 25.47 	&	 -10 32 07.09 	&	 17.32 	$\pm$	 0.36 &	 II 	&	 d 	&	 \\
95  	&	G224.4599-00.6582	&	 07 10 19.06 	&	 -10 32 58.73 	&	 6.80 	$\pm$	 0.14 &	 II 	&	 d 	&	 \\
96  	&	G224.4649-00.6415	&	 07 10 23.25 	&	 -10 32 46.55 	&	 42.58 	$\pm$	 1.03 &	 II 	&	 e 	&	 \\
97  	&	G224.4664-00.6346	&	 07 10 24.93 	&	 -10 32 40.28 	&	 78.36 	$\pm$	 0.87	&	 II 	&	 d 	&	 \\
98  	&	G224.4669-00.6489	&	 07 10 21.89 	&	 -10 33 05.39 	&	 14.29 	$\pm$	 0.16	&	 II 	&	 d 	&	 \\
99  	&	G224.4677-00.6351	&	 07 10 24.98 	&	 -10 32 45.19 	&	 12.05 	$\pm$	 0.10	&	 I 	&	 null 	&	 \\
100 	&	G224.4687-00.6368	&	 07 10 24.70 	&	 -10 32 51.26 	&	 12.85 	$\pm$	 0.37	&	 I 	&	 null 	&	 ext. H$_2$ emission\\
101 	&	G224.4697-00.6386	&	 07 10 24.48 	&	 -10 32 56.56 	&	 10.41 	$\pm$	 0.11	&	 II 	&	 null 	&	 near edge \\
102 	&	G224.4721-00.6296	&	 07 10 26.65 	&	 -10 32 50.22 	&	 94.01 	$\pm$	 0.93	&	 II 	&	 d 	&	 \\
103 	&	G224.4746-00.6328	&	 07 10 26.27 	&	 -10 33 03.10 	&	 16.35 	$\pm$	 0.16	&	 I 	&	 null 	&	 \\
104 	&	G224.4751-00.6383	&	 07 10 25.10 	&	 -10 33 14.00 	&	 54.74 	$\pm$	 0.42	&	 II 	&	 d 	&	  \\
105 	&	G224.4793-00.6309	&	 07 10 27.20 	&	 -10 33 15.10 	&	 3.60 	$\pm$	 0.05	&	 II 	&	 d 	&	 \\
106 	&	G224.4806-00.6430	&	 07 10 24.74 	&	 -10 33 39.50 	&	 1.94 	$\pm$	 0.23	&	 I 	&	 null 	&	 \\
108 	&	G224.4838-00.7128	&	 07 10 09.97 	&	 -10 35 46.15 	&	 14.17 	$\pm$	 0.04	&	 II 	&	 d 	&	 \\
109 	&	G224.4856-00.6229	&	 07 10 29.63 	&	 -10 33 22.03 	&	 08.77 	$\pm$	 0.20	&	 I 	&	 e+d 	&	 ext. H$_2$ emission \\
110 	&	G224.4859-00.6880	&	 07 10 15.56 	&	 -10 35 11.36 	&	 07.41 	$\pm$	 0.07	&	 II 	&	 d 	&	 \\
111 	&	G224.4883-00.6555	&	 07 10 22.90 	&	 -10 34 24.70 	&	 33.94 	$\pm$	 0.12	&	 II 	&	 d 	&	ext. H$_2$ emission \\
112A 	&	G224.4908-00.6223	&	 07 10 30.41 	&	 -10 33 37.50 	&	 2.25 	$\pm$	 0.06	&	 I 	&	 d 	&	 \\
112B 	&	G224.4908-00.6223	&	 07 10 30.36 	&	 -10 33 37.50 	&	 2.14 	$\pm$	 0.06 &	  	&	 	&	 separation 736 au \\
113 	&	G224.4916-00.6754	&	 07 10 18.97 	&	 -10 35 08.47 	&	 07.11 	$\pm$	 0.13 &	 II 	&	 d 	&	 \\
114 	&	G224.4921-00.6620	&	 07 10 21.94 	&	 -10 34 47.82 	&	 9.08 	$\pm$	 0.04 &	 II 	&	 d 	&	 ext. H$_2$ emission \\
115 	&	G224.4942-00.6660	&	 07 10 21.28 	&	 -10 35 00.91 	&	 19.05 	$\pm$	 0.21 &	 II 	&	 e+d 	&	uncorr. H$_2$ emission\tablenotemark{e}  \\
116 	&	G224.4959-00.6715	&	 07 10 20.33 	&	 -10 35 15.87 	&	 23.58 	$\pm$	 0.15 &	 II 	&	 d 	&	 \\
117 	&	G224.4965-00.6805	&	 07 10 18.39 	&	 -10 35 32.83 	&	 47.41 	$\pm$	 0.58 &	 II 	&	 d 	&	 \\
118 	&	G224.5109-00.6777	&	 07 10 20.64 	&	 -10 36 14.23 	&	 13.32 	$\pm$	 0.09 &	 II 	&	 d 	&	 \\
\enddata
\tablenotetext{a}{GLIMPSE360 IRAC Designations are \lq SSTGLMA' followed by the names listed in this column; IRAC Designations are based on Galactic coordinates \citep{meade14}.} 
\tablenotetext{b}{The YSO classification from \cite{sewilo2019}: the YSO Class and components (an envelope and/or a disk) identified based on the SED fitting with the \cite{robit17} YSO models (e - envelope, e+d - envelope and disk, d - disk-only); \lq null' indicates that the SED fitting results are not provided in \cite{sewilo2019}.}
\tablenotetext{c}{
Equatorial coordinates obtained from the integrated continuum maps in units of 10$^{-18}$ erg s$^{-1}$ cm$^{-2}$. See text for details.}
\tablenotetext{d}{The \lq near edge\rq~label is given to sources located off-center (less than 0.8\arcsec~from the map's edge).}
\tablenotetext{e}{The \lq uncorr. H$_2$ emission\rq~ note indicates the presence of the extended H$_2$ emission in the field that is unrelated to the studied source.}
\tablecomments{Resolved binary candidates are marked with \lq A\rq~and \lq B\rq~for the brighter and fainter source, respectively. The separation between the two binary components is given in au in the Remarks for component \lq B\rq. The pixel size (0.2$\arcsec$) corresponds to 184 au.}
\end{deluxetable*}


{The spectra of single sources were extracted using an aperture centered on the continuum peak with a radius of 3~pixels,  typically containing $\sim$70\% of the total flux. For the binary candidates, different aperture sizes were used depending on the separation between the stars. The spectra were subsequently aperture-corrected using the ratio of the continuum emission within the aperture to that in the entire field of view; see Appendix~\ref{app:spec} for details.}

{Since KMOS is a moderate-resolution spectrograph, we assume a simplified shape of a spectral line and calculate integrated fluxes using Gaussian fits.} The spectral line maps are constructed based on line fluxes estimated separately for each spaxel by fitting a Gaussian function to the emission line 50 times with randomly generated input parameters and choosing the fit with the lowest relative error. { All fits were visually confirmed.} The same approach is applied to fitting the spectral lines in the extracted spectra. 

\begin{figure*}[th!]
\includegraphics[width=0.253\textwidth]{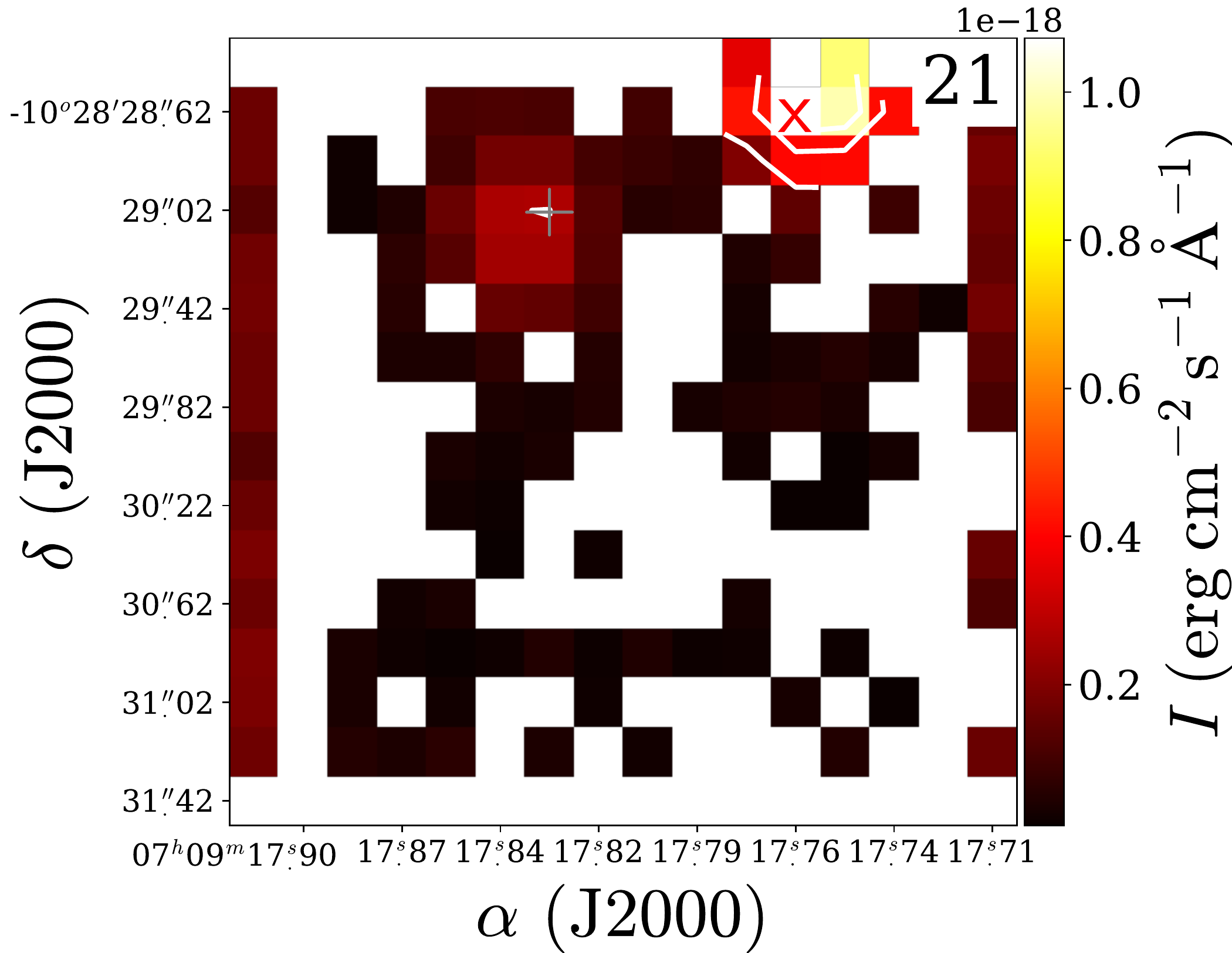}\hspace{-0.15cm}
\includegraphics[width=0.253\textwidth]{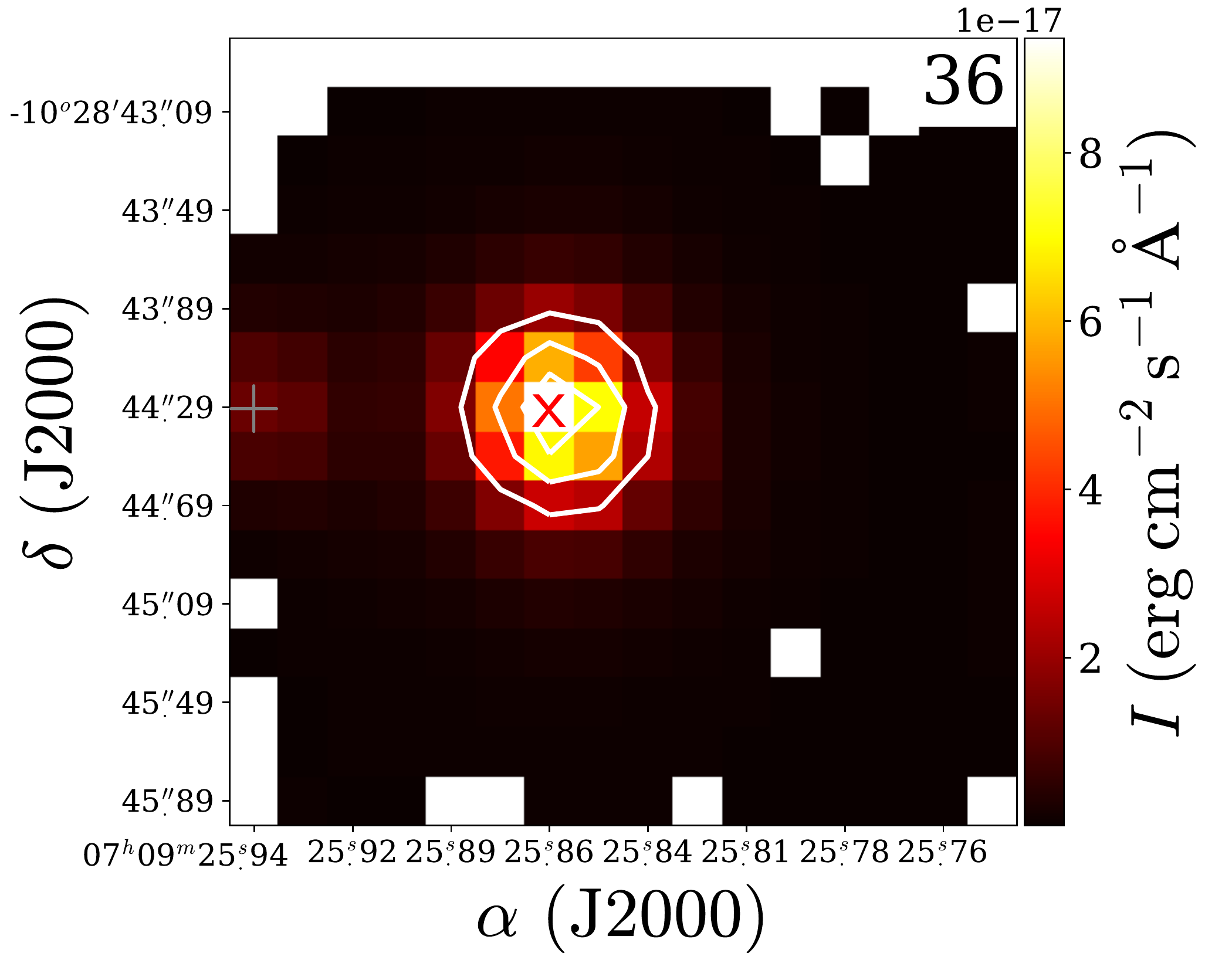}\hspace{-0.15cm}
\includegraphics[width=0.253\textwidth]{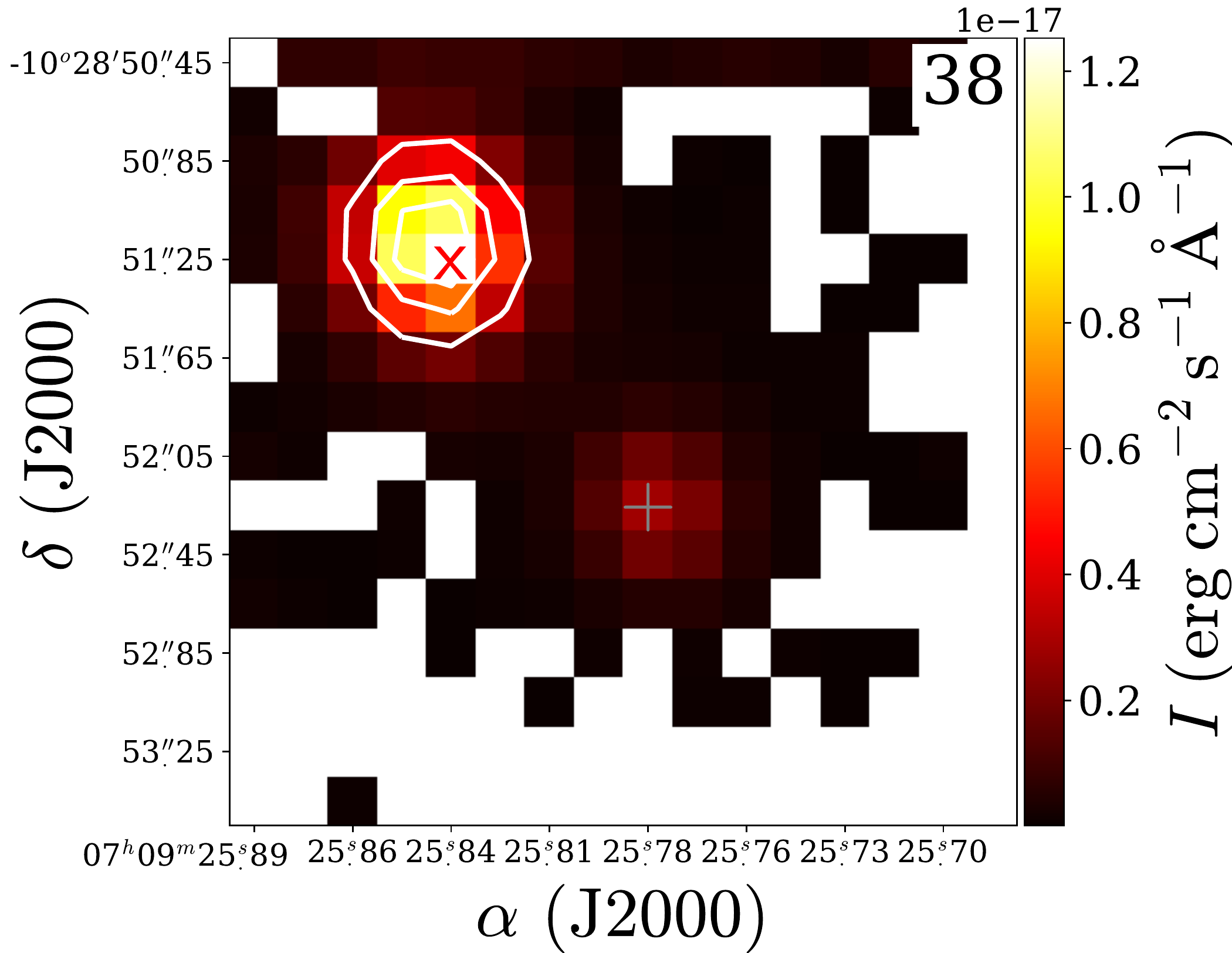}\hspace{-0.15cm}
\includegraphics[width=0.253\textwidth]{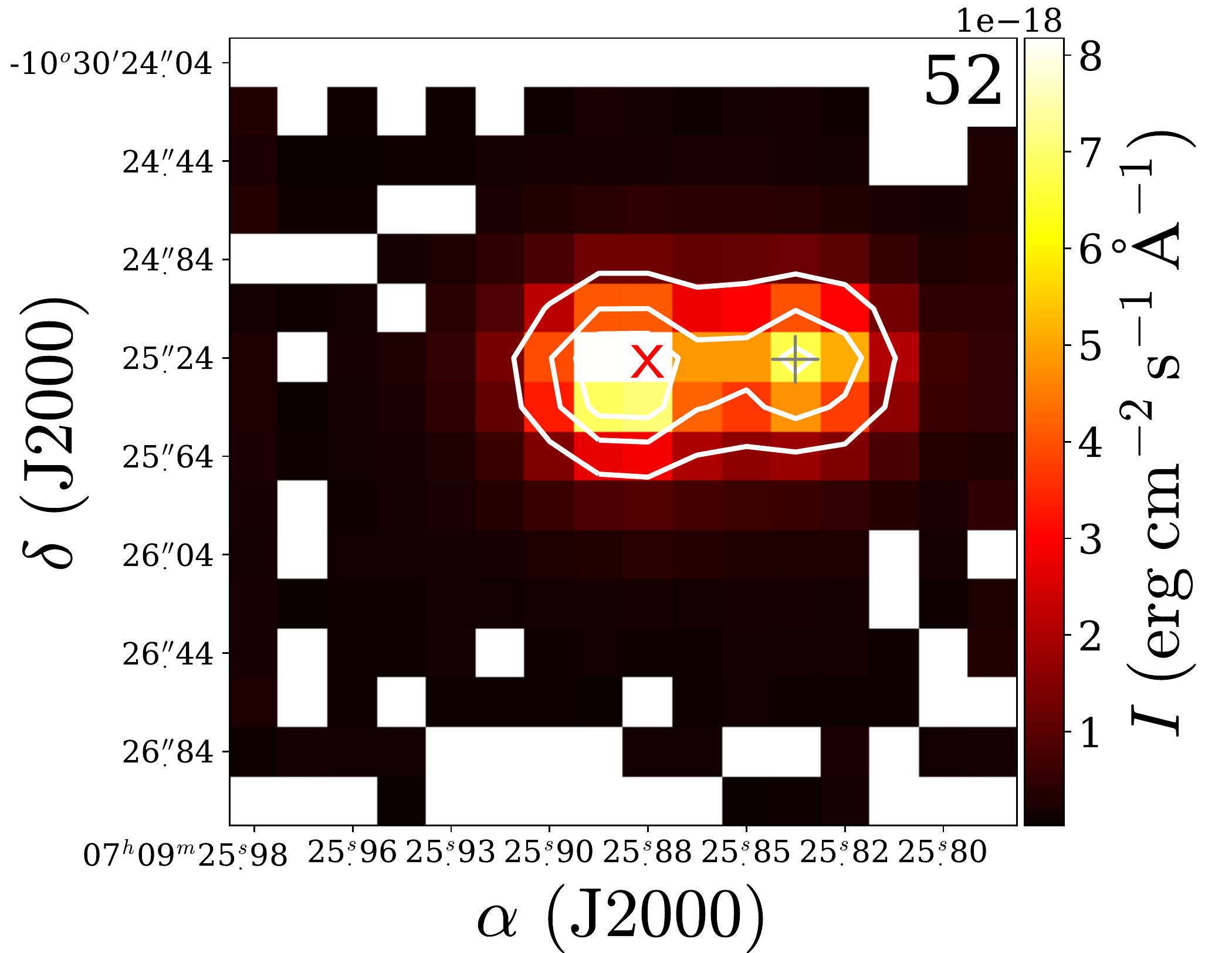}\hspace{-0.15cm}	\includegraphics[width=0.253\textwidth]{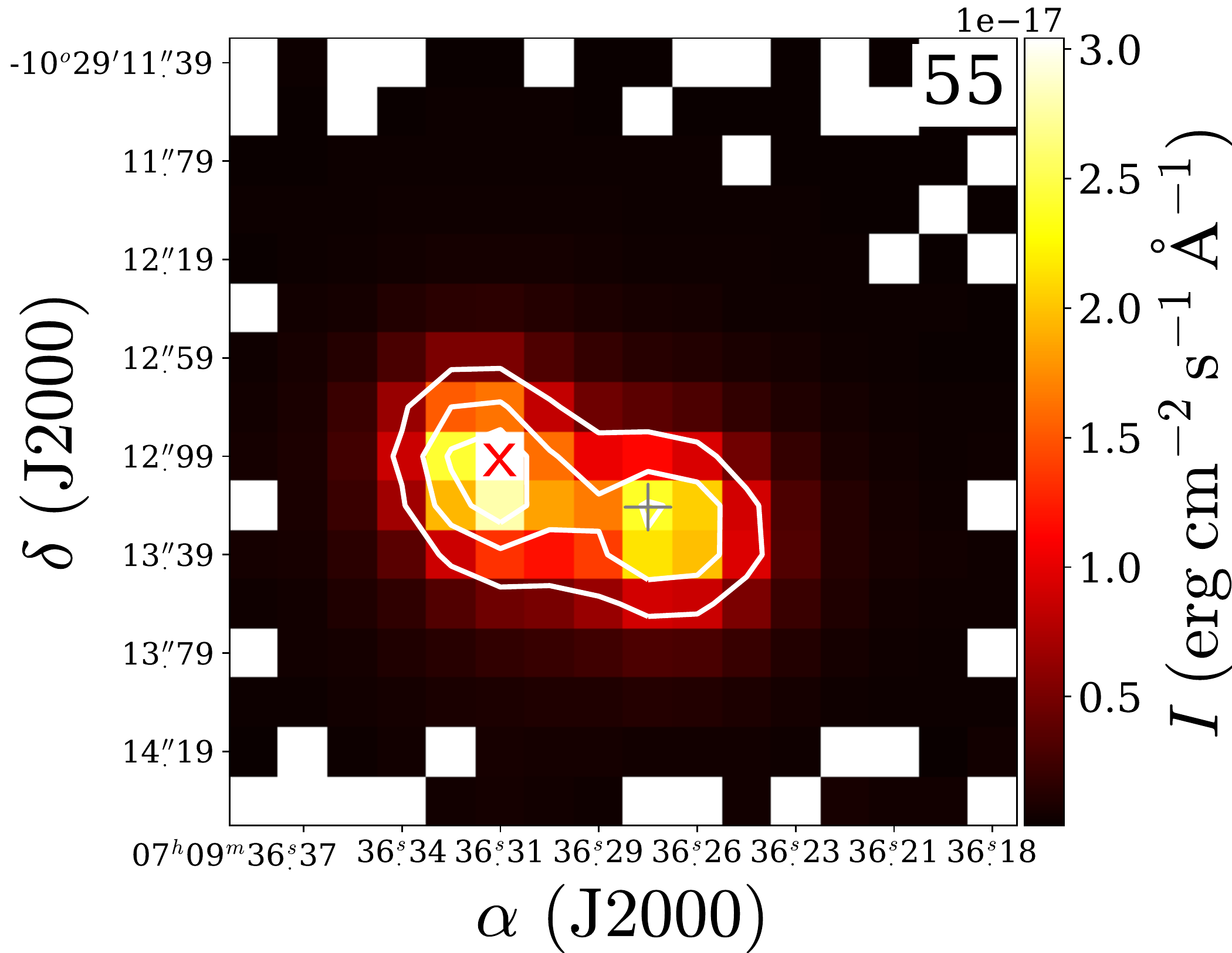}\hspace{-0.15cm}
\includegraphics[width=0.253\textwidth]{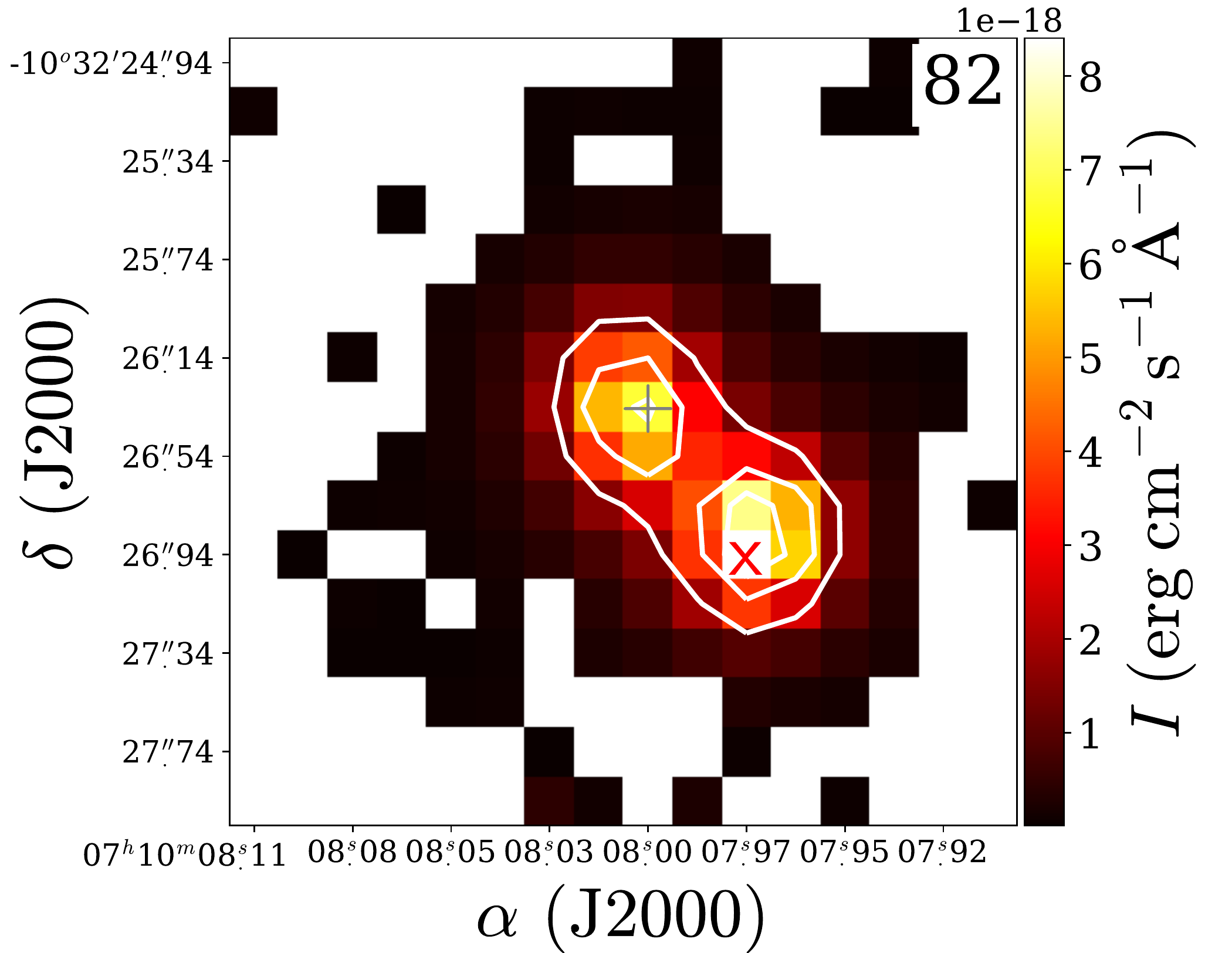}\hspace{-0.15cm}	\includegraphics[width=0.253\textwidth]{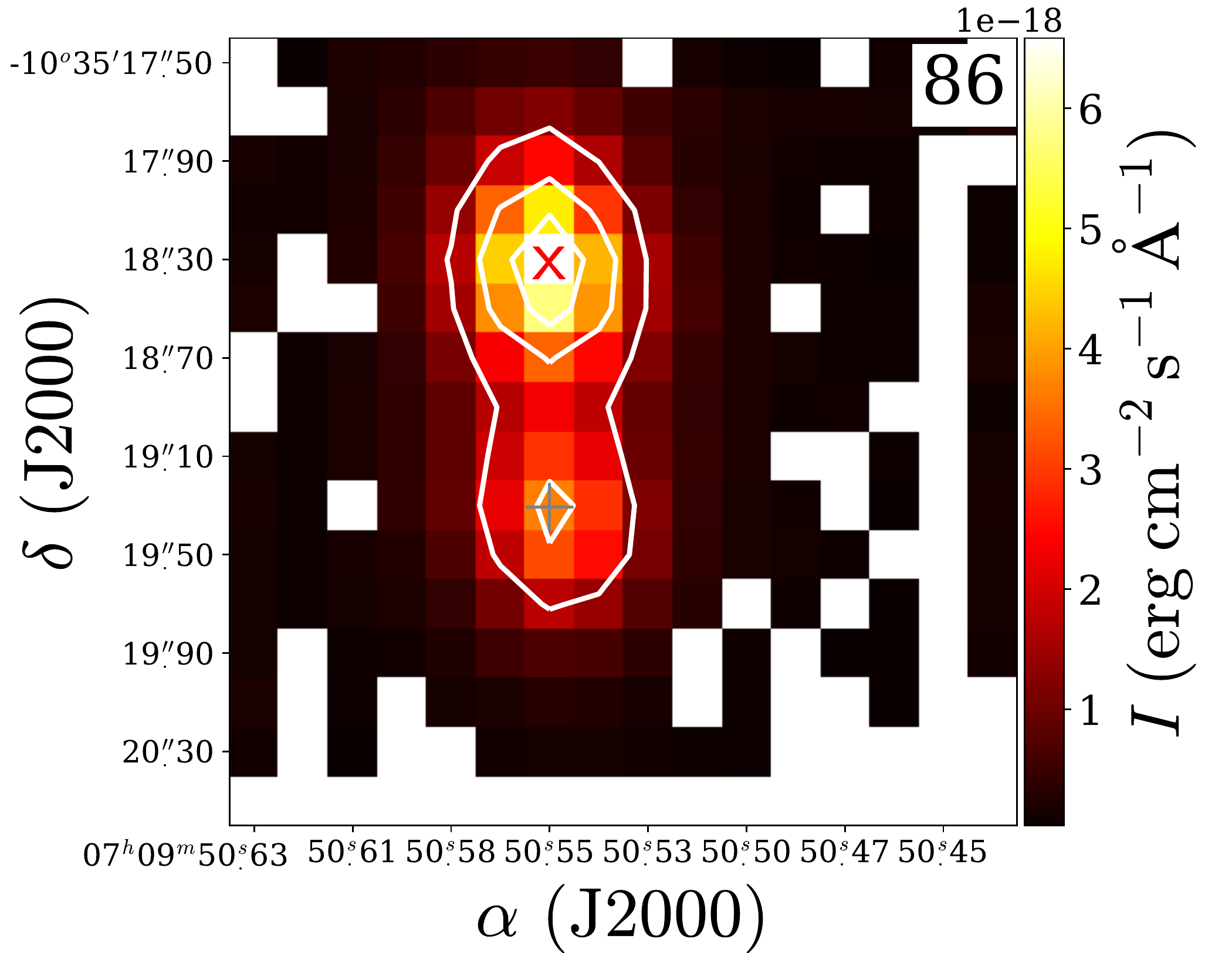}\hspace{-0.15cm}
\includegraphics[width=0.253\textwidth]{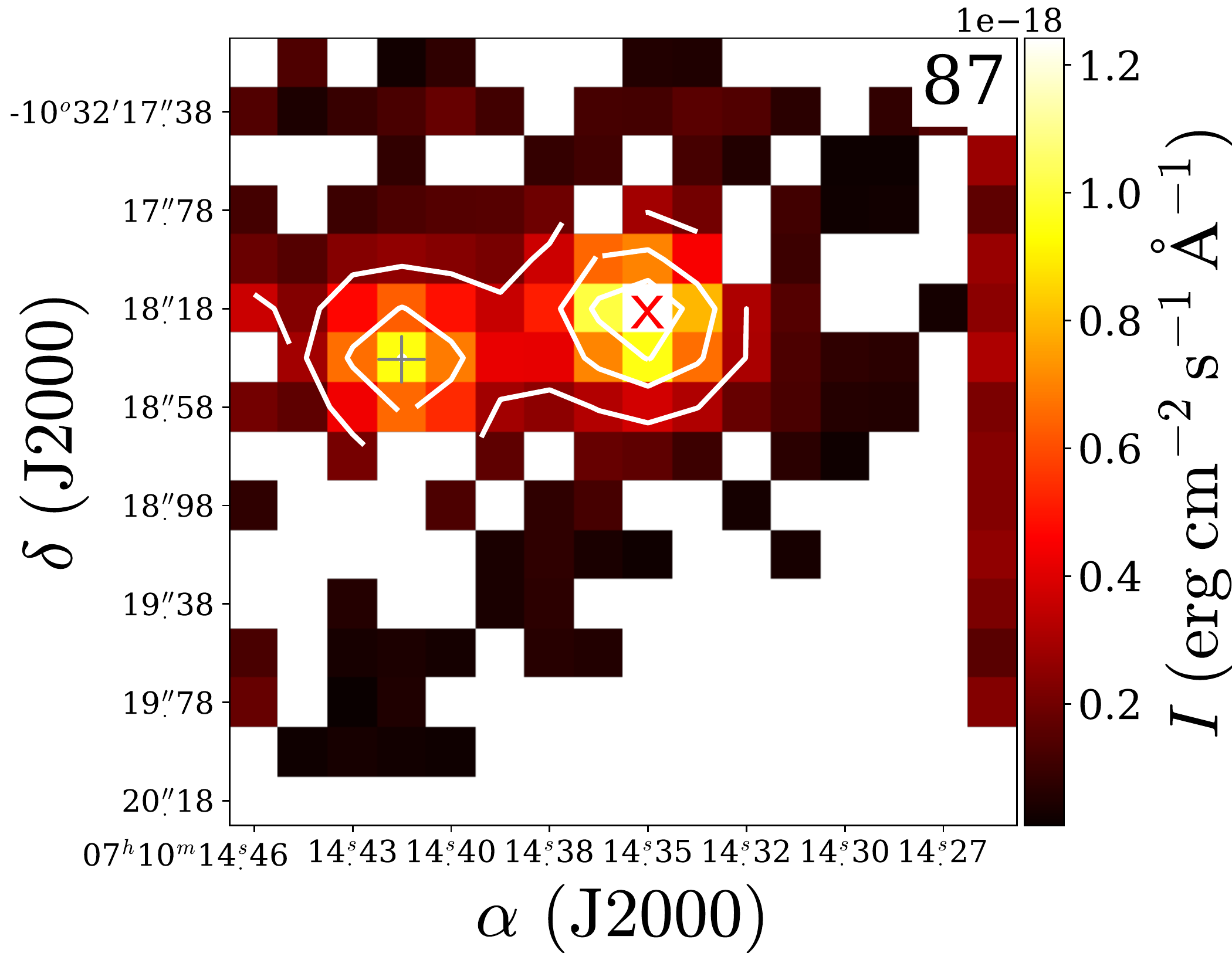}\hspace{-0.15cm}
\includegraphics[width=0.253\textwidth]{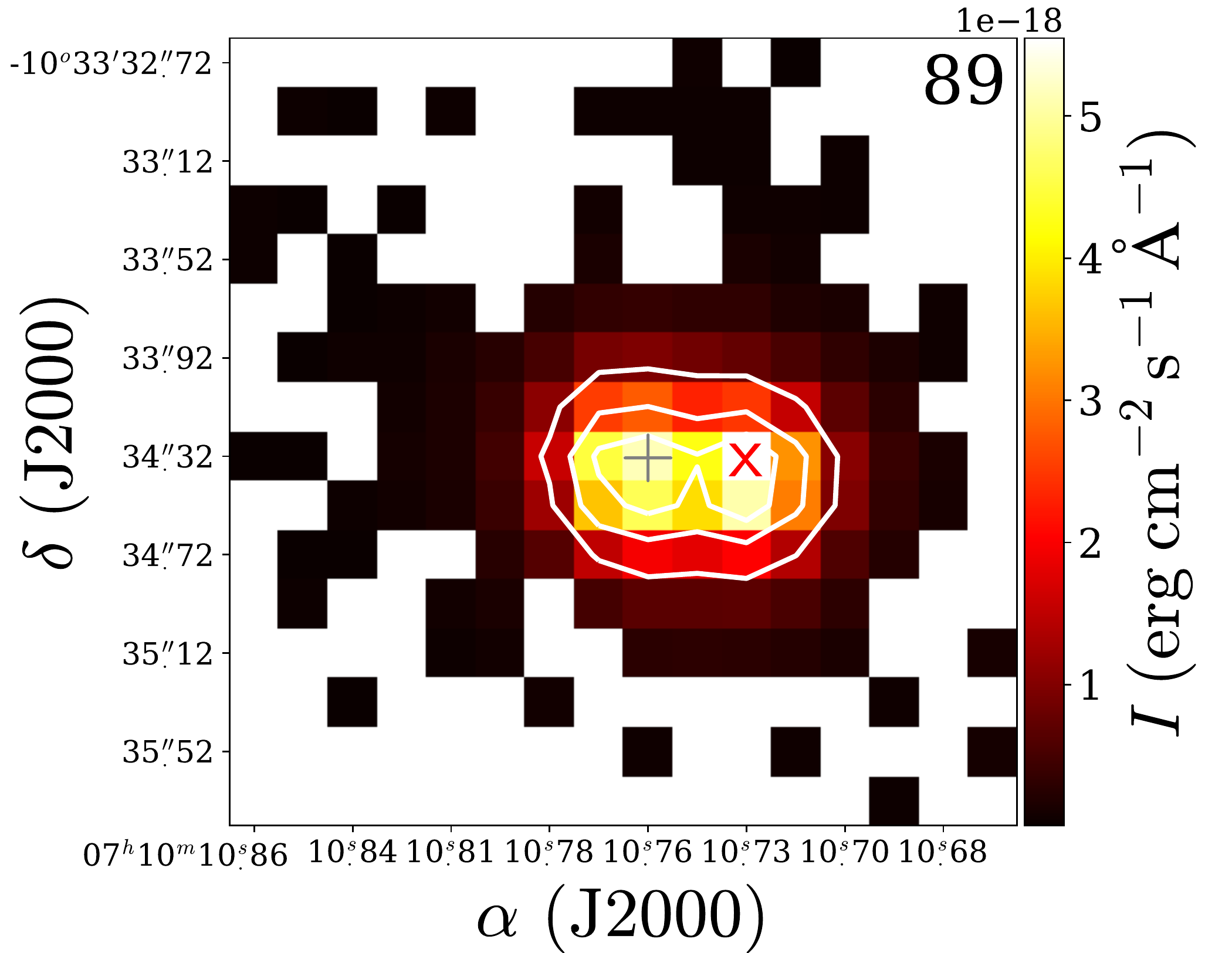}\hspace{-0.15cm}
\includegraphics[width=0.253\textwidth]{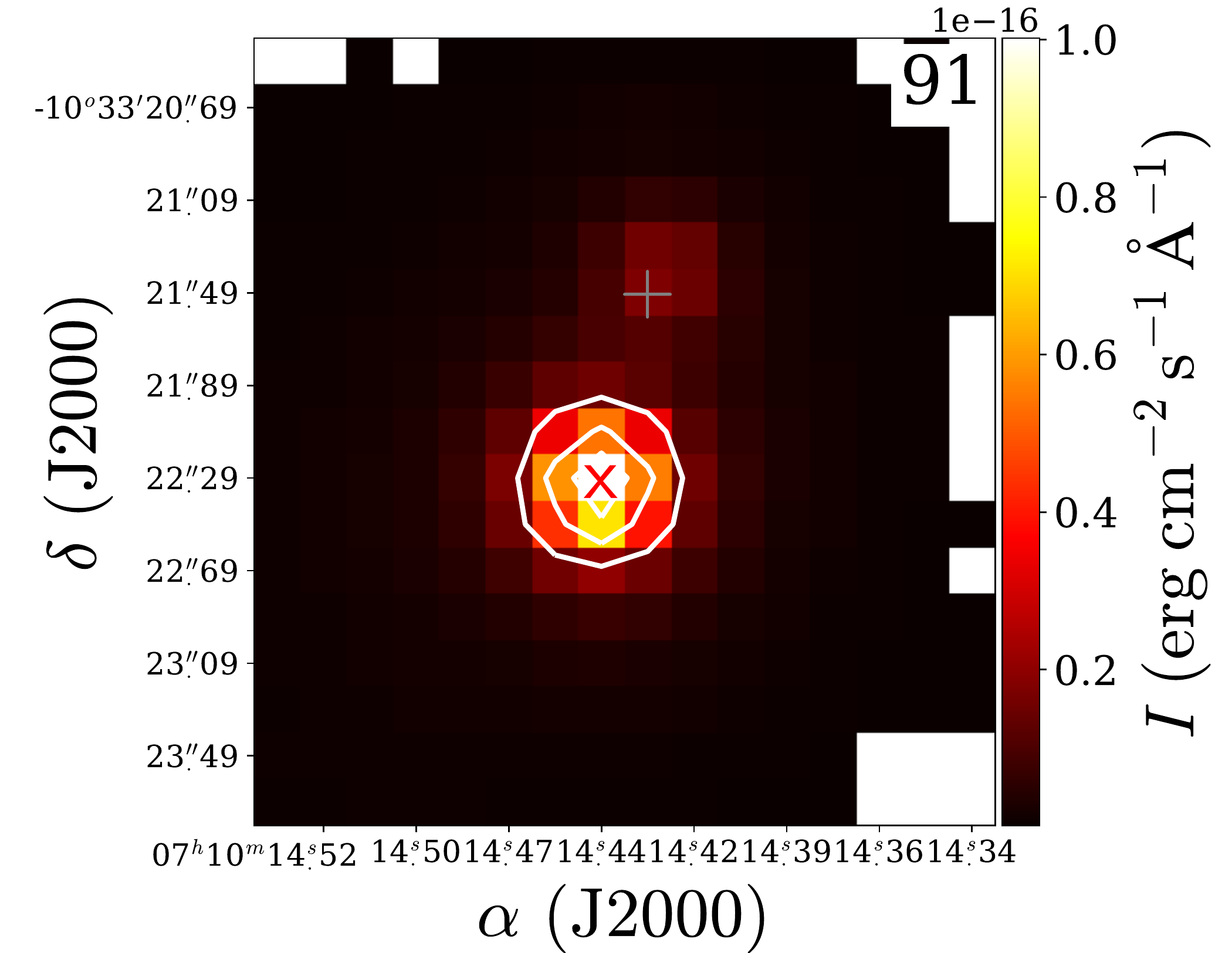}\hspace{-0.15cm}
\includegraphics[width=0.253\textwidth]{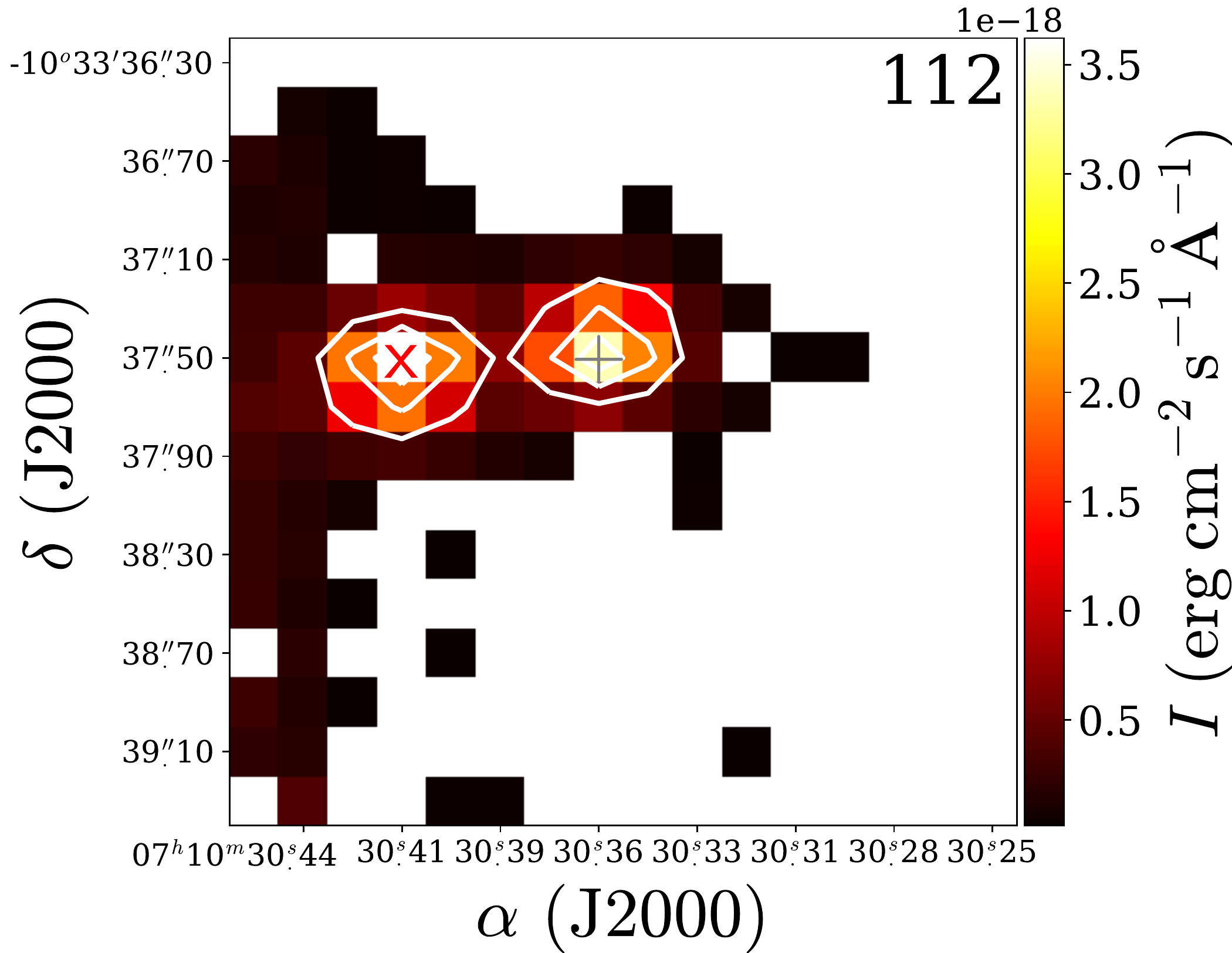}\hspace{-0.15cm}	
\caption{\label{fig:cont-double} KMOS $K$-band continuum maps of YSO candidates with possible stellar companions. Only spaxels with the continuum flux above 3$\sigma$ are shown. White contours correspond to 25\%, 50\%, and 75\% of the $K$-band continuum peak. The position of the brighter source in each map is indicated with a red '$\times$' symbol (the \lq\lq A\rq\rq~component in Table \ref{tab:coordinates}), and the fainter one with a grey '$+$' symbol (the \lq\lq B\rq\rq~component). The size of the pixel (0.2$\arcsec$) corresponds to 184 au.  
}
\end{figure*}

\section{Data Analysis and Results} \label{sec:res}

Out of 118 \textit{Spitzer} YSO candidates observed with KMOS, 5 were not detected and 11 were resolved into two near-IR sources. In total, our KMOS observations provide the $K$-band data for 124 sources. 

The IFU observations deliver both the spectral and spatial information. The $K$-band continuum emission allows us to identify all the continuum components and determine their positions and $K$-band fluxes. The spatial extent of the atomic and molecular emission hints at the underlying physical mechanisms and their characteristics. The hydrogen Br$\gamma$ emission is correlated with the UV continuum excess diagnostic of mass accretion. The CO ro-vibrational band head traces the inner disk. The H$_2$ emission lines, including the bright 1-0 S(1) line at 2.12 $\mu$m,  trace jets and shocks.

In this section, we discuss the spatial distribution of gas and dust, and summarize the line detections and kinematic information. 

\subsection{$K$-Band Continuum Emission}
\label{subsec:cont}
\begin{figure*}[t]
\includegraphics[width=\textwidth]{{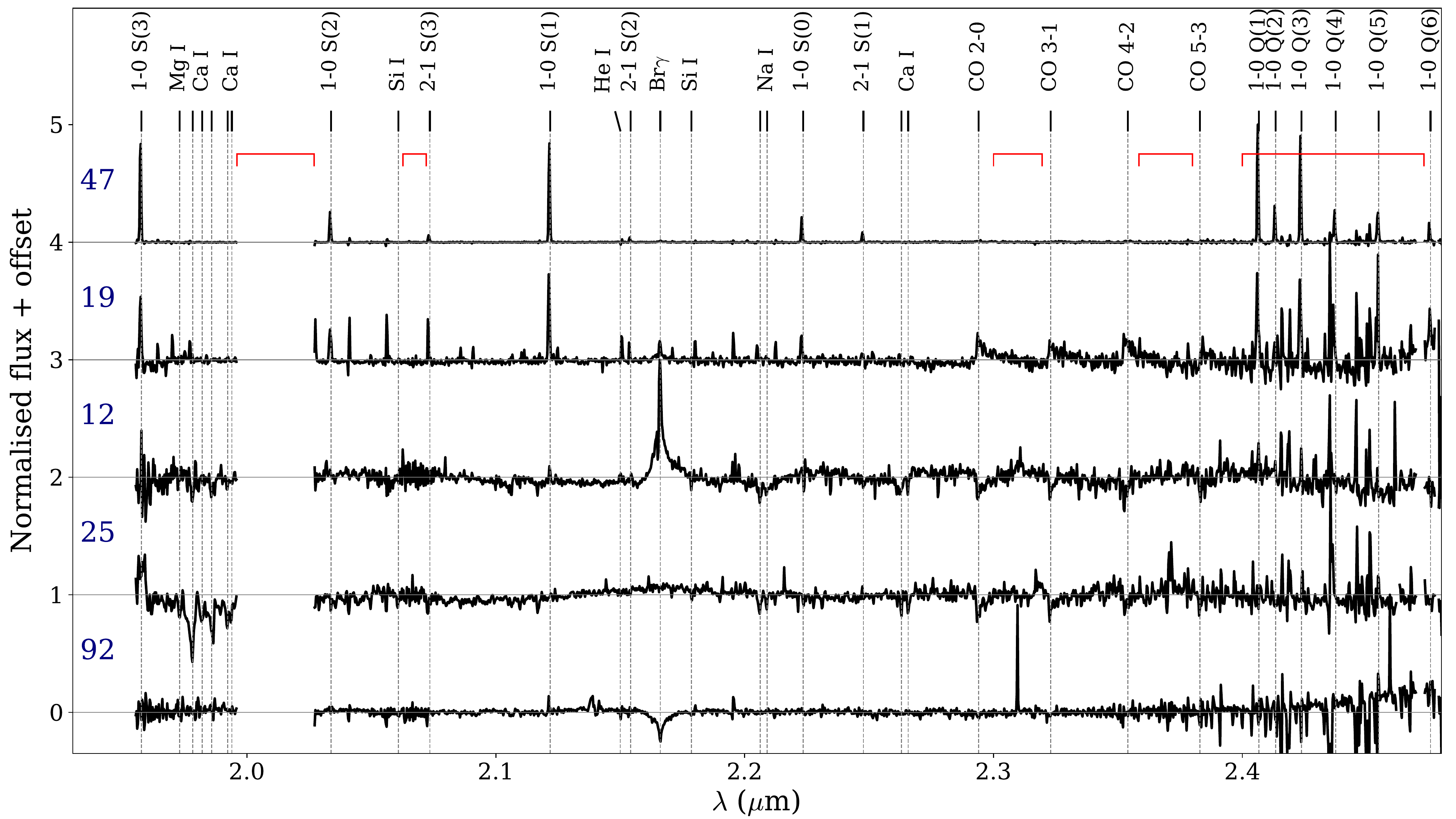}}
\caption{\label{fig:spec:ex} Example spectra of YSOs in CMa-$\ell$224. Source names from Table \ref{tab:coordinates} (No.) are provided on the left side of the respective spectra. All spectra are normalised to the continuum peak flux density in the wavelength range from 1.97 to 2.47~$\mu$m. The red horizontal lines show spectral ranges most affected by telluric lines; the wavelength range around 2.0 $\mu$m has been removed for clarity of the plot.}
\end{figure*}

{All the $K$-band continuum sources detected with VLT/KMOS in CMa-$\ell$224 are listed in Table~\ref{tab:coordinates}. The table provides the sources' GLIMPSE360 catalog names and assigned KMOS IDs, their equatorial coordinates corresponding to the position of the $K$-band continuum peak, the $K$-band continuum fluxes estimated near 2.12 $\mu$m, the YSO classification from \cite{sewilo2019}, and remarks on pointing, multiplicity, and the association with extended H$_2$ emission. Two $K$-band continuum components have been detected toward eleven fields centered on single \textit{Spitzer} YSO candidates (see Fig.~\ref{fig:cont-double}); the brighter source is listed in Table~\ref{tab:coordinates} as source \lq\lq A'' and the fainter one as source \lq\lq B.''}

Selected $K$-band continuum maps are shown in Figure~\ref{fig:cont-double} (11 fields with the detection of two sources) and Appendix~\ref{app:cont} (the remaining fields). 

{We consider a possibility that two sources detected in a single KMOS field are two components of a binary star/YSO. The median separation between the two components of the ``binary star candidates'' in CMa-$\ell$224 is 760~au (see Table \ref{tab:coordinates}), with a range from 370~au (source No. 89) to 1180~au (source No. 38).} { All but three pairs of visual binary star candidates have spectra typical of accreting young stars: emission of Br$\gamma$ and/or H$_2$. Two more pairs have CO bandhead in absorption typical for cool stars.} Given the proximity of the sources, they may be physically related. If they are binary systems, the multiplicity fraction defined as the ratio of the number of multiple systems to the total number of systems (both single and multiple), is $0.09\pm0.01$\footnote{We follow \cite{tobin2016} to estimate the uncertainty of derived multiplicity and adopt a bimodal statistic: $\sigma_{\rm{multi}}$=($N_{\rm{multi}}\cdot(1-N_{\rm{multi}}/N_{\rm{sys}})$)$^{-0.5}$ / $N_{\rm{sys}}$, where $N_{\rm{multi}}$ is a number of multiple systems and $N_{\rm{sys}}$ is a total number of systems.}. For comparison, { the multiplicity fraction for separations between 100 and 1000 au is $0.15\pm0.04$ for protostars in Perseus \citep{tobin2016}, $0.14\pm0.01$ for protostars and $0.13\pm0.01$ for Pre-Main Sequence (PMS) stars in the Orion molecular clouds \citep{kounkel2016}.} Adopting the approximate range recovered by KMOS (180-1180 au), the multiplicity fraction of protostars would be $\sim$0.09 in Perseus and 0.12 in Orion.

In summary, the binary star candidates in CMa-$\ell$224 have the multiplicity fraction that is consistent with YSOs in other molecular clouds, once we account for close binaries that are unresolved with KMOS. A more robust comparison requires an unbiased survey of YSOs at their earliest evolutionary stages and will be presented in the second paper in this series.

\newpage
\subsection{Spectral Line Detections}
\label{subsec:spec}

KMOS spectra contain several atomic and molecular lines emerging as a result of mass accretion and/or ejection in YSOs. Additionally, transitions of many atomic absorption lines which originate from a stellar photosphere are also detected and can be used to estimate spectral types and luminosity classes of YSOs \citep[e.g., ][]{Ni05,luhman1998}. 

Figure~\ref{fig:spec:ex} shows the $K$-band spectra toward selected YSO candidates with diverse characteristics to illustrate the differences between the sources. {For example, sources No. 47 and 19 show a strong H$_2$ emission, with only the latter showing also clear Br$\gamma$ and CO detections. Sources No. 12 and 25 both show CO bandhead in absorption. Yet, only source No. 12 is characterized also by a broad emission line profile in Br$\gamma$. Source No. 92 shows very weak H$_2$ emission, and Br$\gamma$ line in absorption. The line detections are discussed further in the subsequent sections.}

\subsubsection{Br$\gamma$}
\label{subseubsec:brg}

The most commonly detected line is the 2.1655 $\mu$m Br$\gamma$ hydrogen line, seen in emission in 58 sources (47\%) and in absorption in 16 sources (13\%). Twenty six sources with the Br$\gamma$ emission also show the CO lines in absorption. A similar pattern but higher detection rates of the Br$\gamma$ emission ($\sim75$\%), was found toward a sample of 19 Class I protostars \citep{connelley2010,connelley2014}. In a recent KMOS survey of YSOs in Perseus, the Br$\gamma$ was indeed more commonly detected in emission toward Class I (59\%) than Class II YSOs \citep[36\%, ][]{fiorellino2021}.

\begin{figure*}[ht!]
\includegraphics[width=0.2\textwidth]{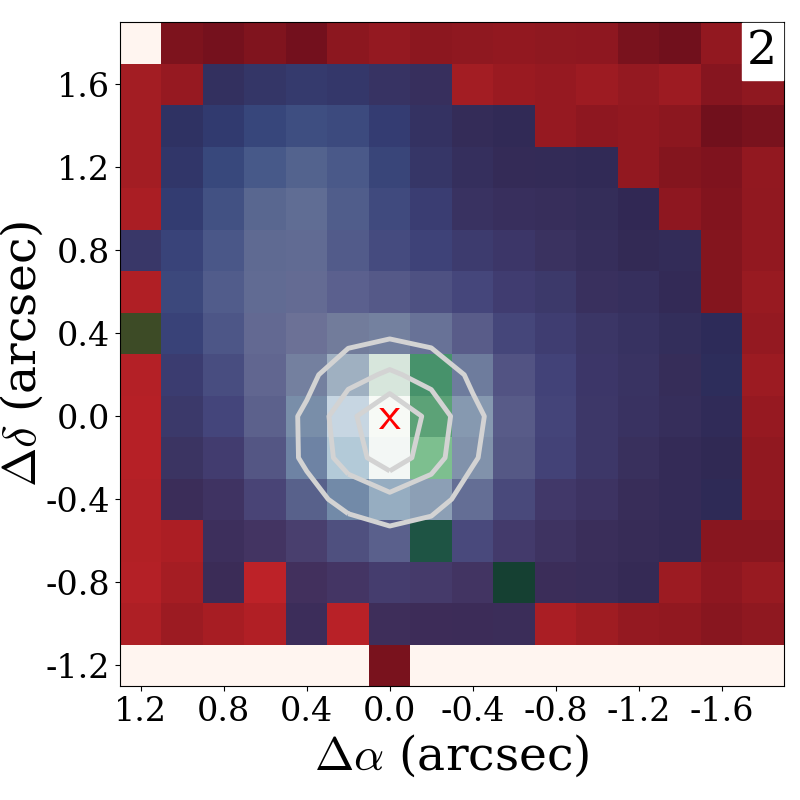}\hspace{-0.1cm}
\includegraphics[width=0.2\textwidth]{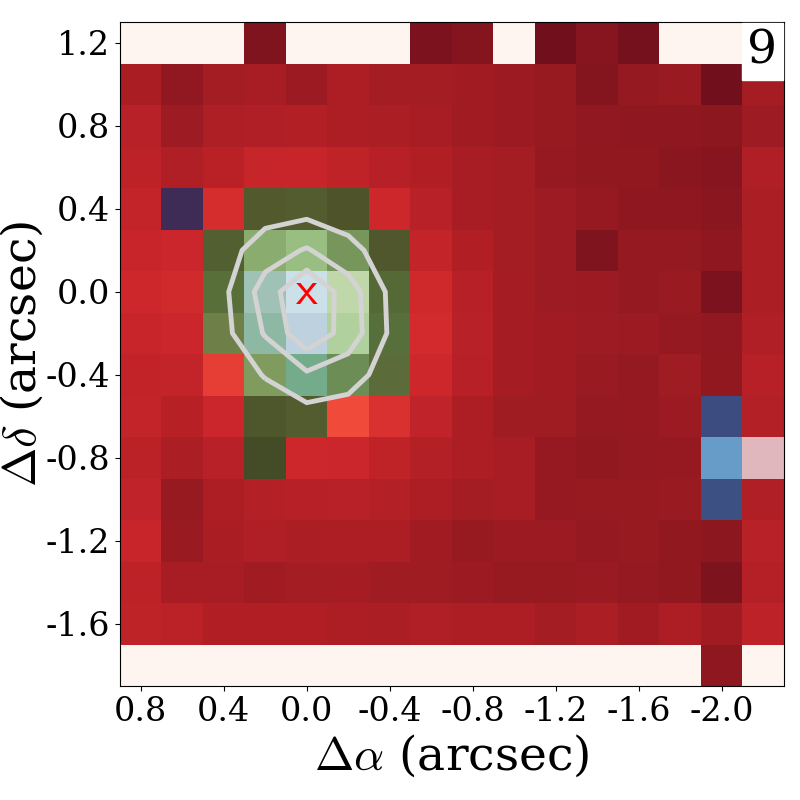}\hspace{-0.1cm}
\includegraphics[width=0.2\textwidth]{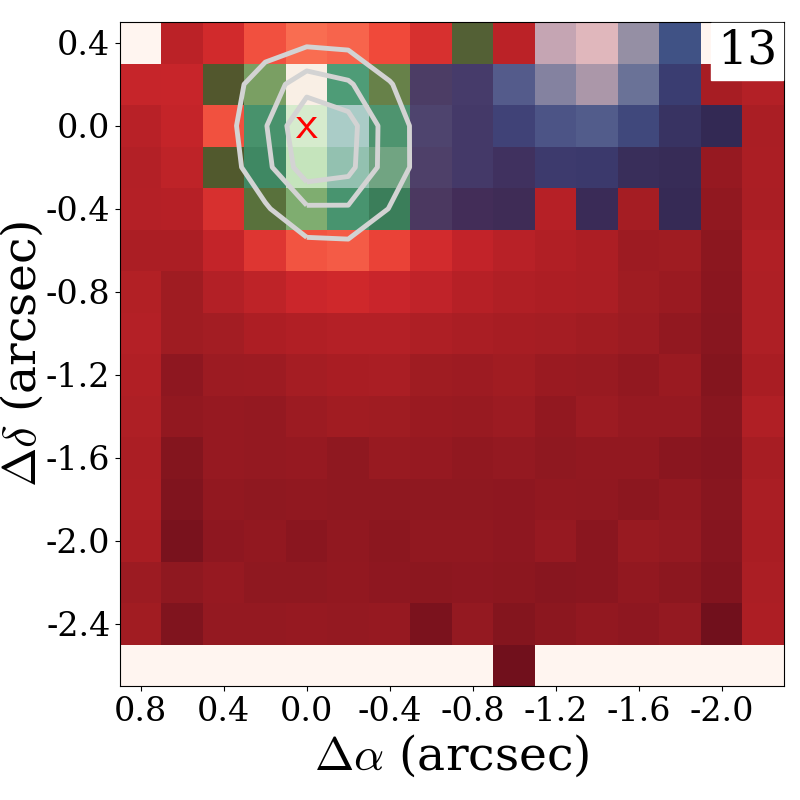}\hspace{-0.1cm}
\includegraphics[width=0.2\textwidth]{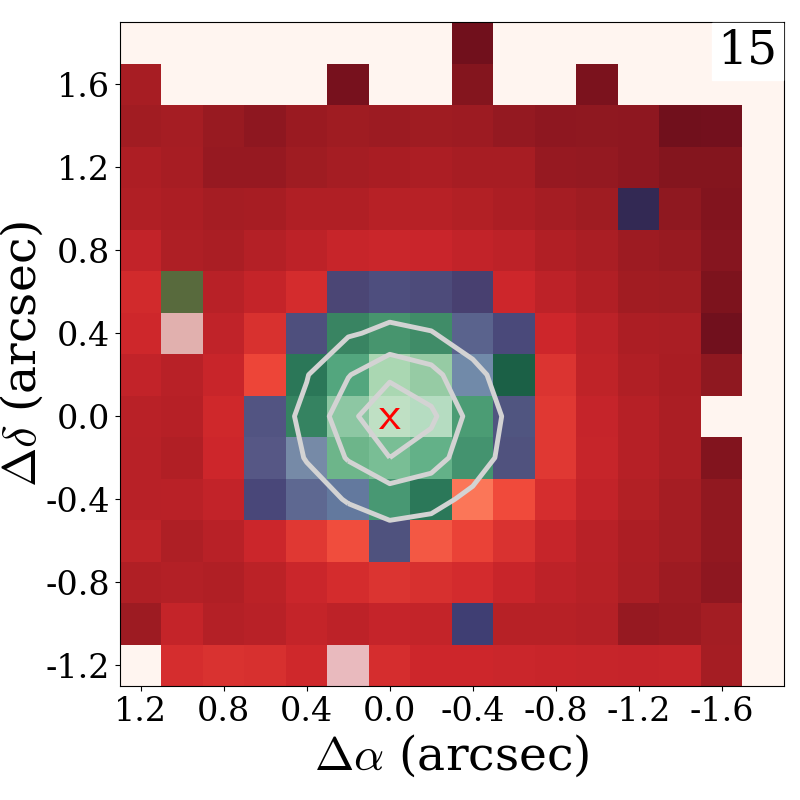}\hspace{-0.1cm}
\includegraphics[width=0.2\textwidth]{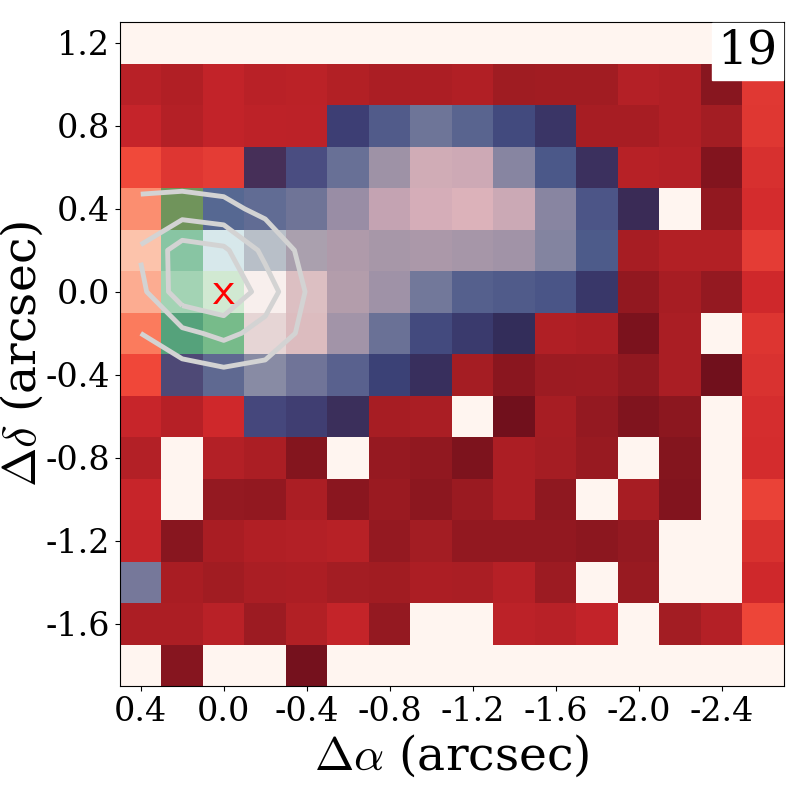}\hspace{-0.1cm}
\includegraphics[width=0.2\textwidth]{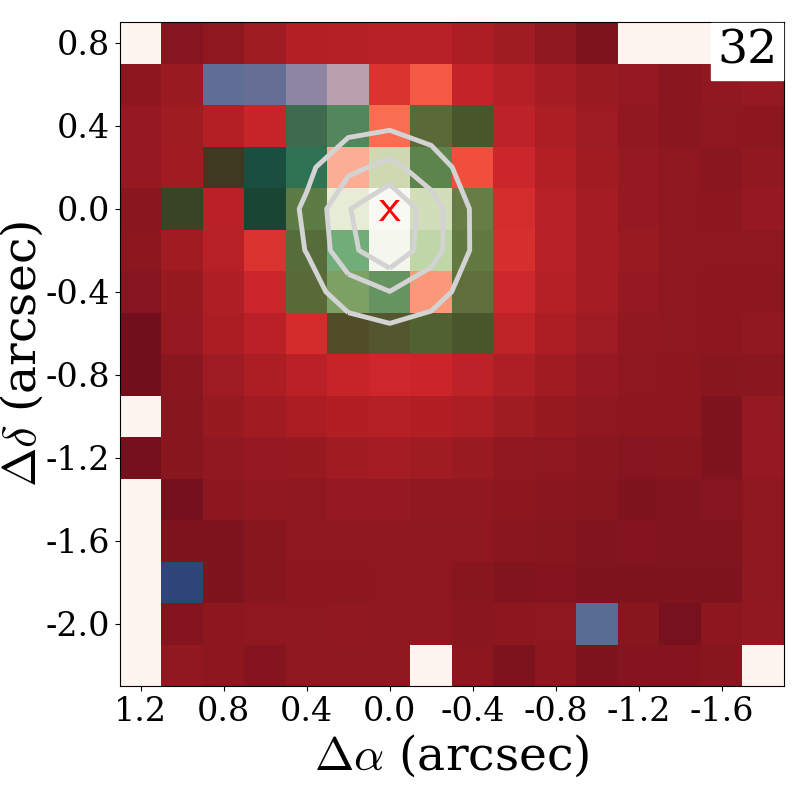}\hspace{-0.1cm}
\includegraphics[width=0.2\textwidth]{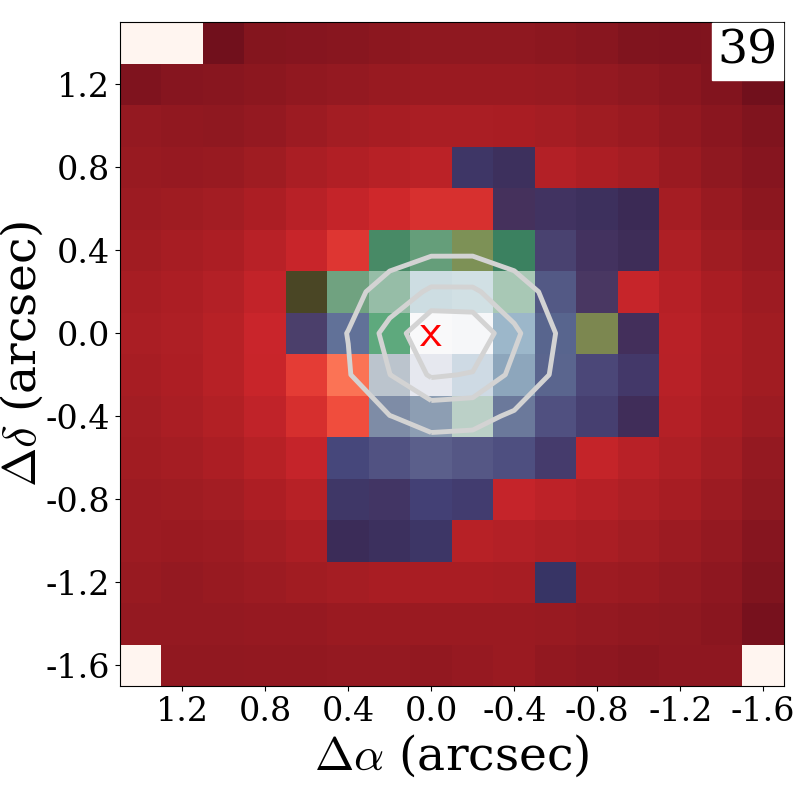}\hspace{-0.1cm}
\includegraphics[width=0.2\textwidth]{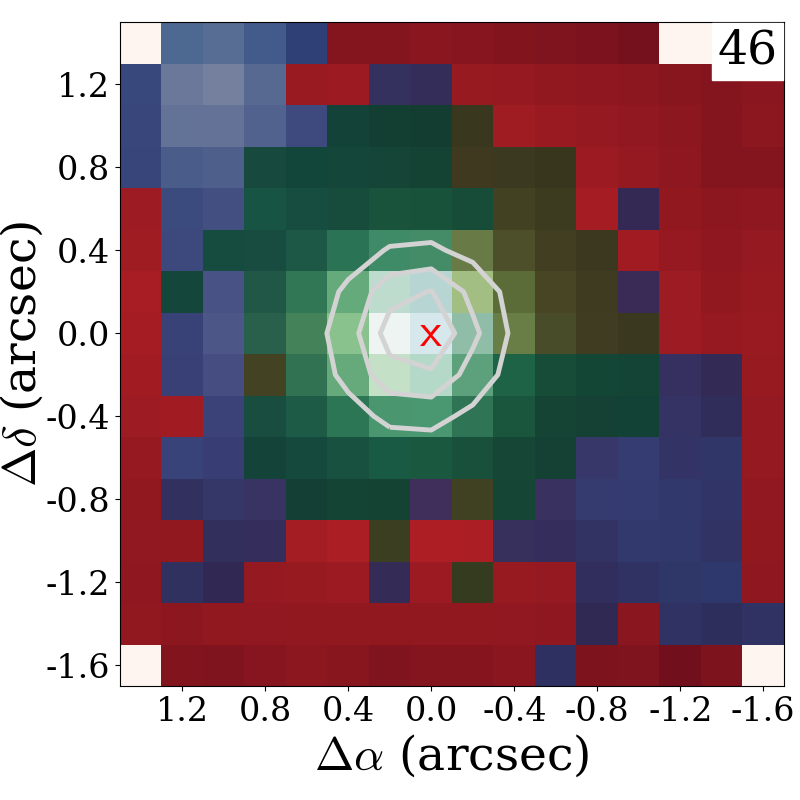}\hspace{-0.1cm}
\includegraphics[width=0.2\textwidth]{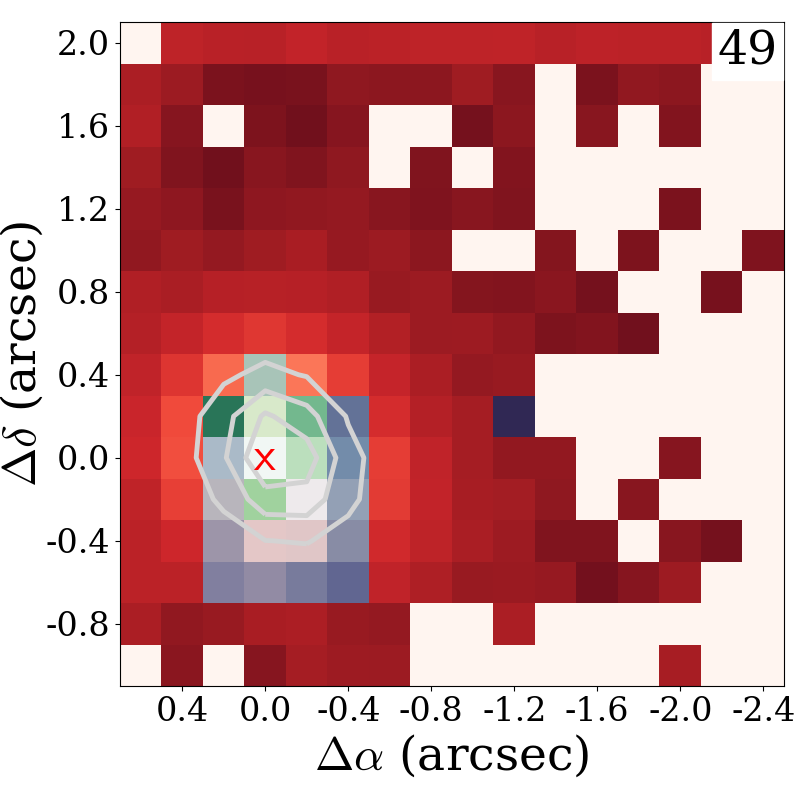}\hspace{-0.1cm}
\includegraphics[width=0.2\textwidth]{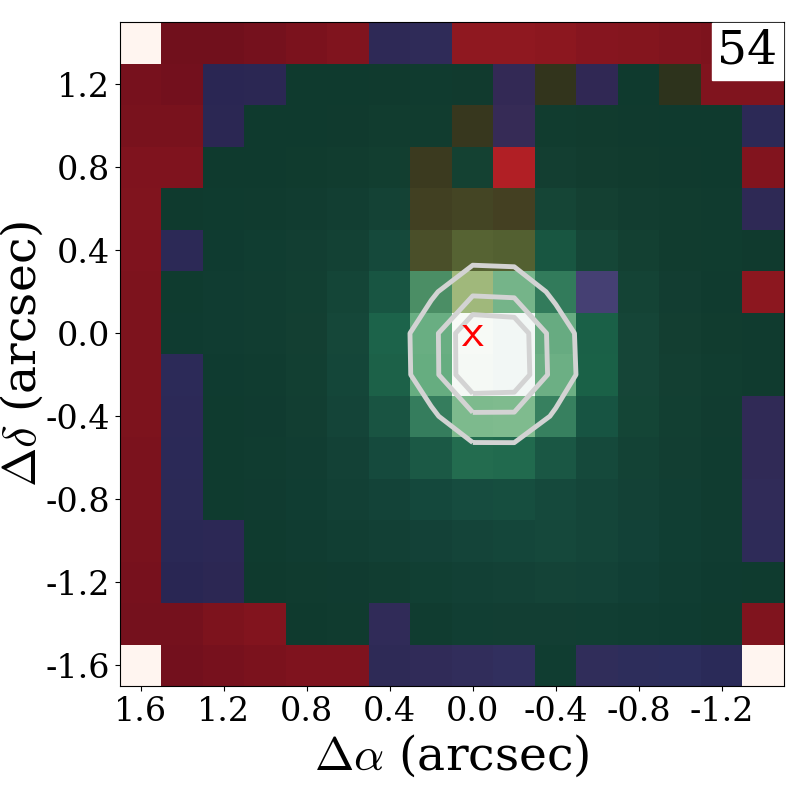}\hspace{-0.1cm}
\includegraphics[width=0.2\textwidth]{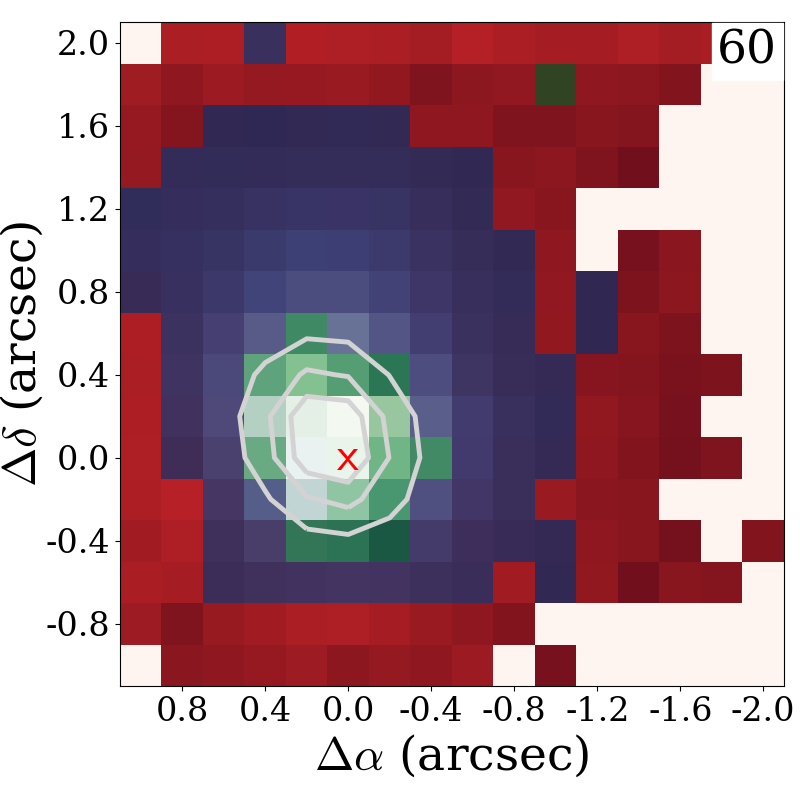}\hspace{-0.1cm}
\includegraphics[width=0.2\textwidth]{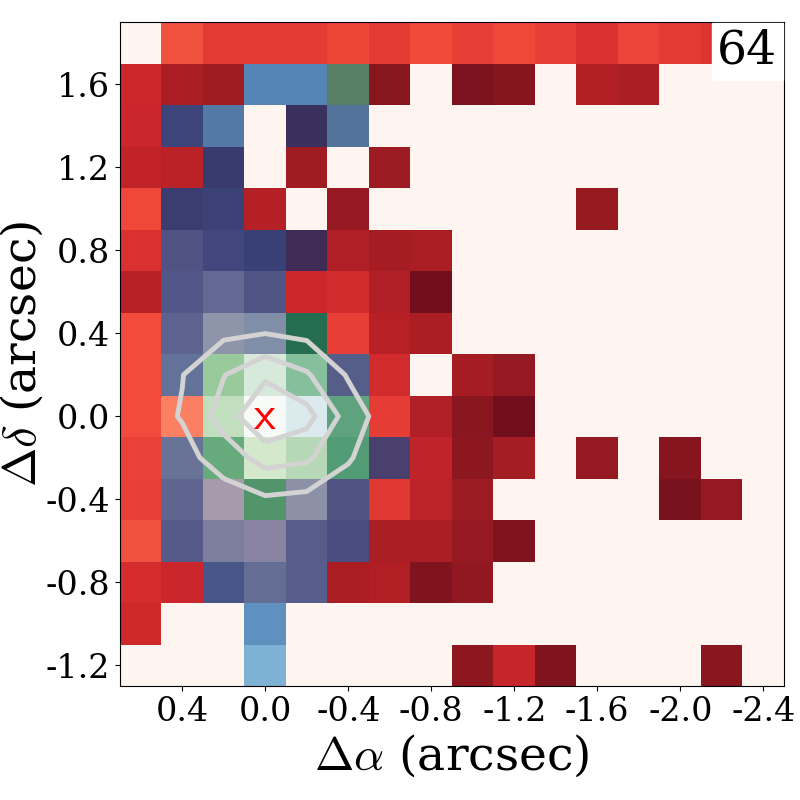}\hspace{-0.1cm}
\includegraphics[width=0.2\textwidth]{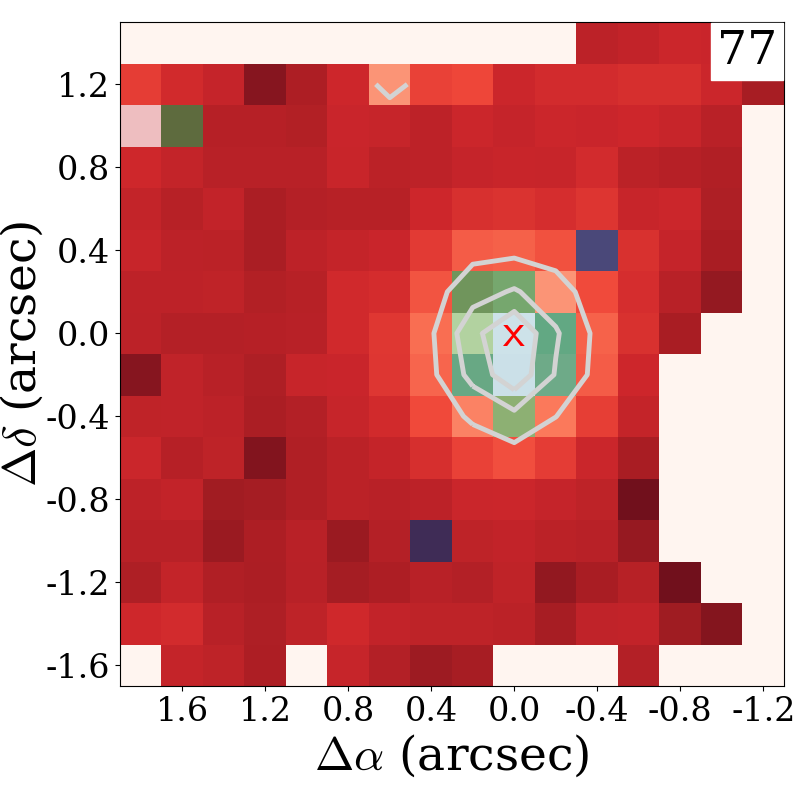}\hspace{-0.1cm}
\includegraphics[width=0.2\textwidth]{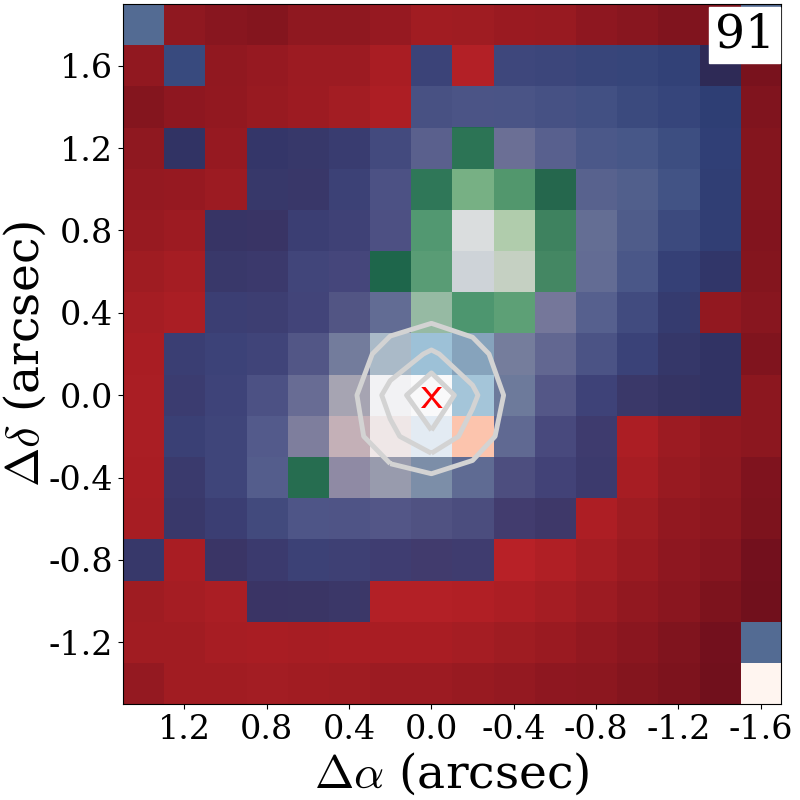}
\caption{Three-color composite images of YSO candidates with the $\geq$3$\sigma$ detection in the H$_2$ line at 2.12 $\mu$m (in blue), Br$\gamma$ line at 2.17~$\mu$m (in green), and the $K$-band continuum (in red). In each image, white contours correspond to the $K$-band continuum emission with contour levels of 25\%, 50\%, and 75\% of the continuum peak marked with the red '$\times$' symbol. Source IDs are indicated in the upper right corners.}
\label{fig:rgb}
\end{figure*}

The Br$\gamma$ line emission { in our sources} is compact and associated with the continuum emission (Section \ref{subsec:cont}, see also Fig. \ref{fig:rgb} and Appendix \ref{app:emiss}). In low-mass YSOs, Br$\gamma$ emission originates from the close vicinity of the young star and usually traces magnetospheric accretion columns, 
{ with possible contributions from} winds \citep{kraus2008}.  
{Some emission may also arise along the outflows \citep{beck10}}.

The spatial extent of atomic emission of  Br$\gamma$-emitting gas is {often more compact} than molecular gas, with the exception of source No. 54, a bright star with the $K$-band magnitude of 9.9 (Fig.~\ref{fig:rgb}). Extended emission around a young star can be partially explained by an emerging \ion{H}{2} region, but a broad line profile suggests that other physical processes { might also contribute \citep{cooper2013}. The physical extent of $\sim$2350~au {of Br$_\gamma$ emission from the source No. 54} cannot be explained with disk winds. In the subsequent analysis, the line fluxes are extracted from the area covering the continuum source, where 
Br$\gamma$ emission is likely dominated by accretion.}

\subsubsection{H$_2$}

H$_2$ is a homonuclear molecule that lacks a permanent dipole moment, thus only quadrupole transitions may occur ($\Delta J = 0, \pm2$). The extended distribution of the H$_2$ emission and its velocity structure indicate  the origin in bipolar outflows (Section \ref{subsec:linemap}). H$_2$ emission is detected in 33 sources in CMa-$\ell$224 (27\%), including 14 that also show a detection of the Br$\gamma$ line in emission, and only 3 with the CO ro-vibrational lines in absorption. A similar H$_2$ detection rate of 23\% was reported in a pioneering survey of \citet{carr1990}. Larger rates have been found toward Class I protostars \citep[43\%, ][]{connelley2014} and higher-mass YSOs (56\%, \citealt{cooper2013}; 76\%, \citealt{varricatt2010}). Line fluxes are reported in Appendix \ref{app:spec} in Table \ref{tab:H2flux-ex}.

\subsubsection{CO}

The CO bandhead in absorption traces the atmosphere of a star or a disk. Among YSO candidates in CMa-$\ell$224, the CO first overtone is detected in absorption in 60\% of YSO candidates, predominately (78\%) in those classified as Class II \citep[][see also Table \ref{tab:coordinates}]{sewilo2019}. Among Class I YSOs, the CO absorption is detected in 43\% of the sources, similar to the 57\% detection rate in \cite{connelley2010}, and $>$2.5 times more than in the \cite{cooper2013} survey. 
\begin{figure*}[t!]
\includegraphics[width=0.21\textwidth, trim={0 0 4cm 0}, clip]{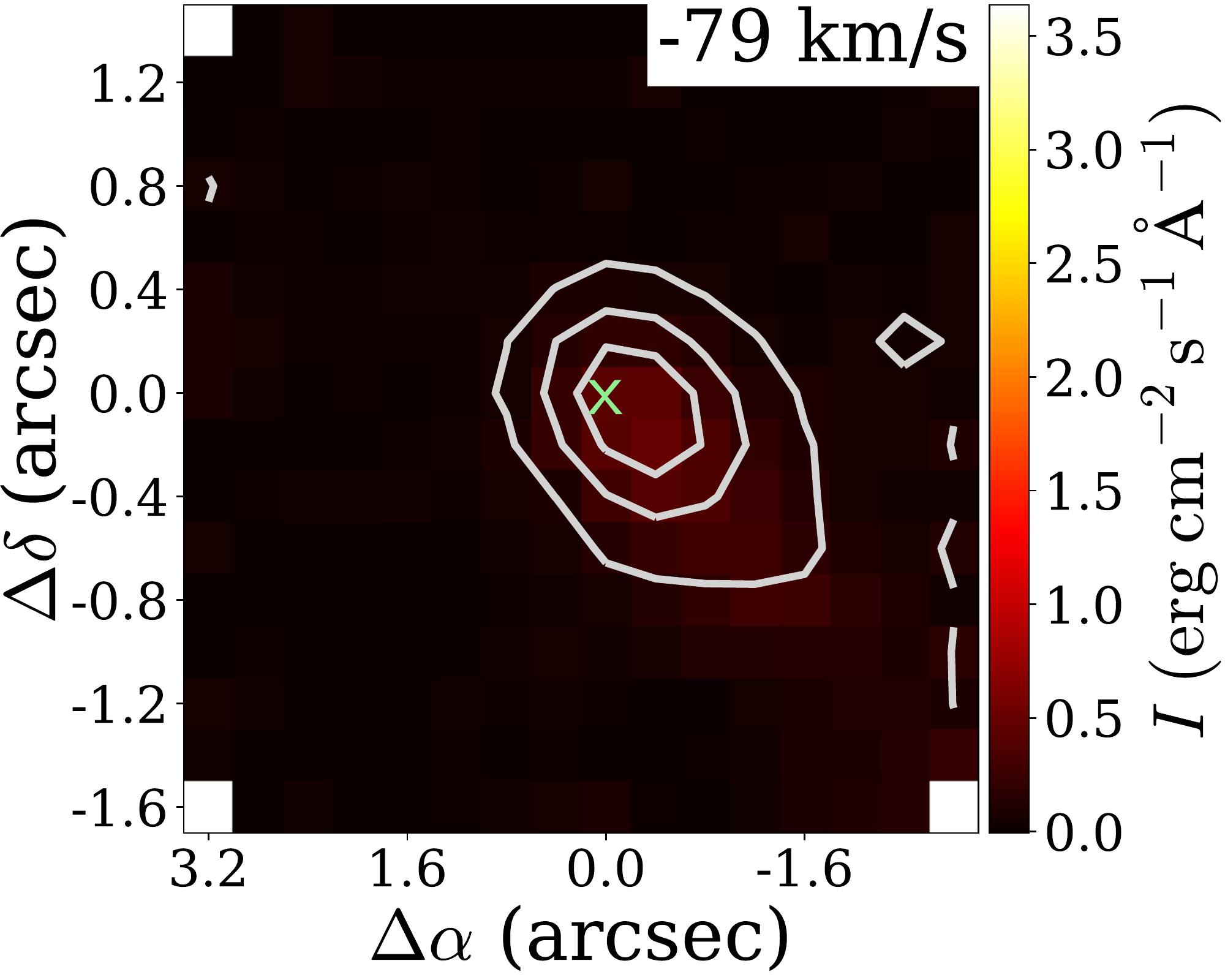}\hspace{-0.07cm}
\includegraphics[width=0.174\textwidth, trim={2.9cm 0 4cm 0}, clip]{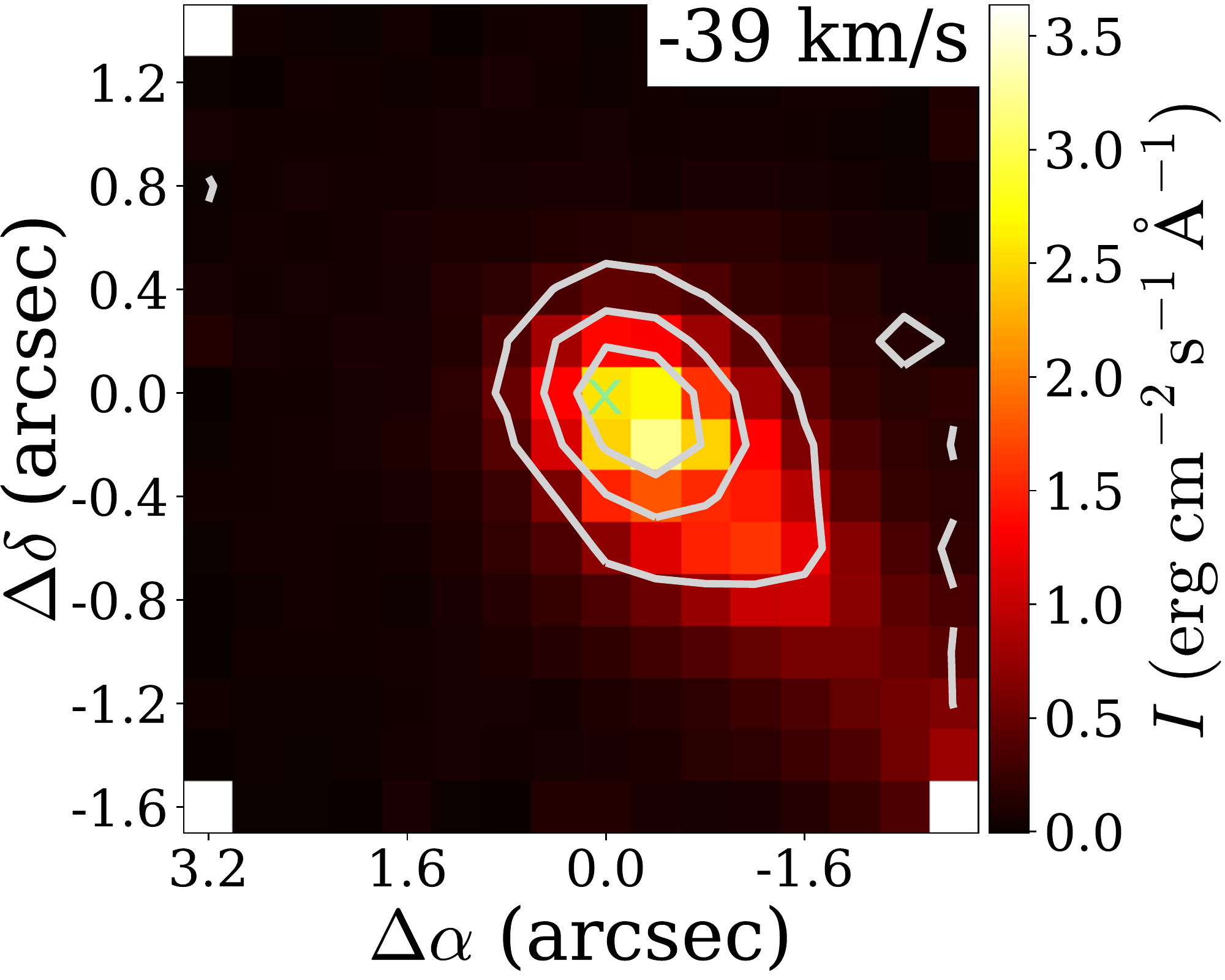}\hspace{-0.07cm}
\includegraphics[width=0.174\textwidth, trim={2.9cm 0 4cm 0}, clip]{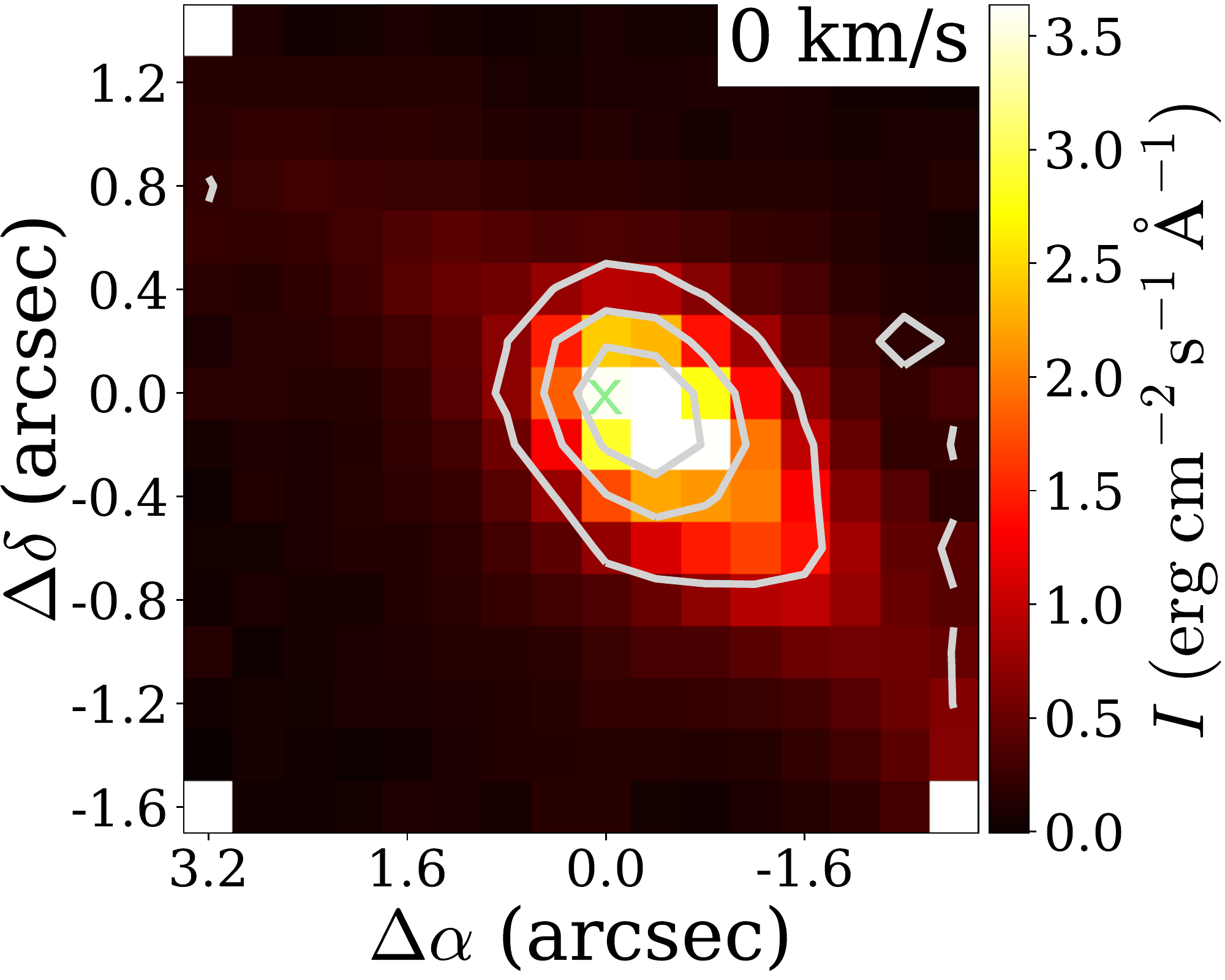}\hspace{-0.07cm}
\includegraphics[width=0.174\textwidth, trim={2.9cm 0 4cm 0}, clip]{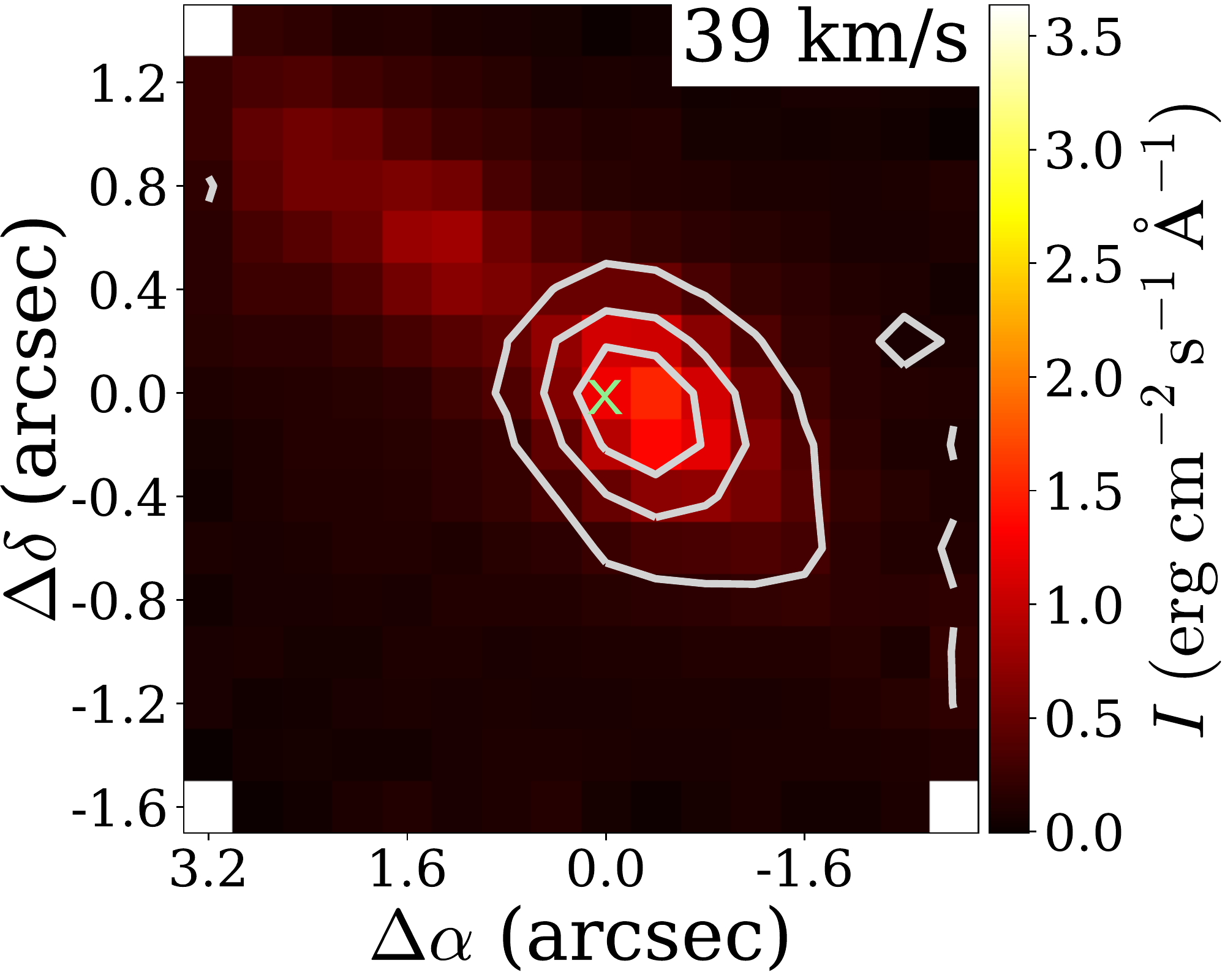}\hspace{-0.07cm}
\includegraphics[width=0.225\textwidth, trim={2.9cm 0 0 0}, clip]{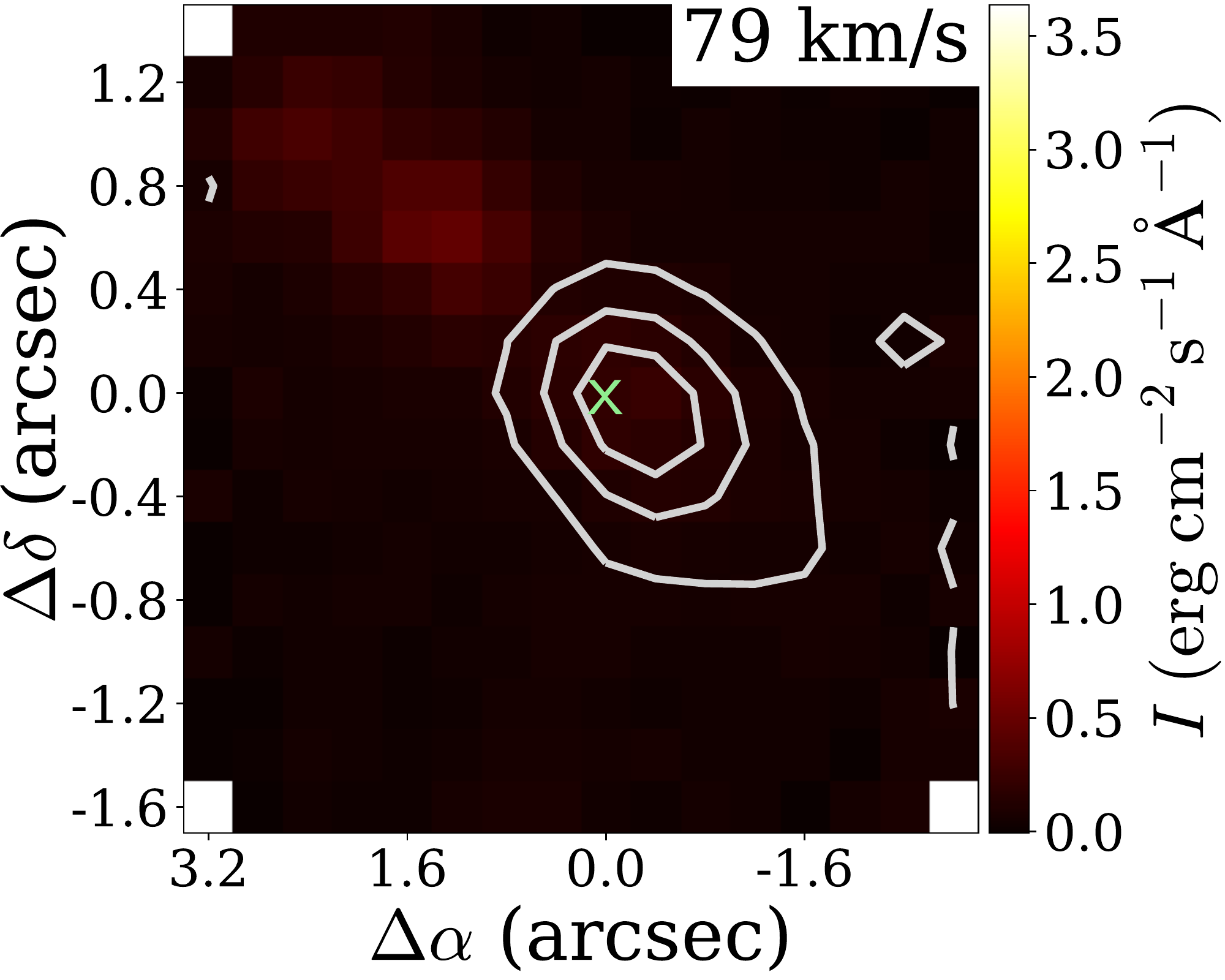}\hspace{-0.07cm}
\caption{The H$_2$ 1-0 S(1) channel maps for source No. 47. The velocities relative to the rest velocity of the H$_2$ line are shown in the upper right corners. The green '$\times$` symbol shows the position of the $K$-band continuum peak.}
\label{fig:chan47}
\end{figure*}

The CO bandhead in emission indicates high accretion rates onto the YSO disk, possibly also associated with an accretion burst { \citep[e.g.,][]{lorenzetti2009, guo2021}}.
The CO bandhead emission requires high temperatures {($2500-4000$~K)} and densities {($n_{\rm{H}}>10^{10}$~cm$^{-3}$)} to excite the upper levels \citep[e.g.,][]{carr1989,casali1996}. In CMa-$\ell$224, only 5\% of the sources detected with KMOS show the CO bandhead in emission (4 Class I and 2 Class II YSOs). This detection rate is a factor of 4-5 lower than in the \cite{carr1989} and \cite{connelley2010} surveys, likely due their focus on Class I protostars. All of our sources with the CO bandhead emission also show the H$_2$ and Br$\gamma$ emission, and in four { out of six} sources the H$_2$ emission is extended, suggesting intensive mass accretion and subsequent ejection.

\subsubsection{Other Lines}

Spectra of eleven YSO candidates show multiple atomic lines e.g., He~I, Mg~I, Ca~I, Na~I, K~I, or Si~I { typical for late spectral types}  \citep[K--M,][]{Ni05}; however, they do not reveal the presence of the Br$\gamma$, H$_2$, and CO lines. The confirmation of their YSO status would require careful spectral typing and further analysis, which will be presented in the future work in this series. Appendix~\ref{app:spec} shows spectra of all those sources and the information about the detection of the key spectral lines.

\begin{figure*}
\includegraphics[width=0.25\textwidth]{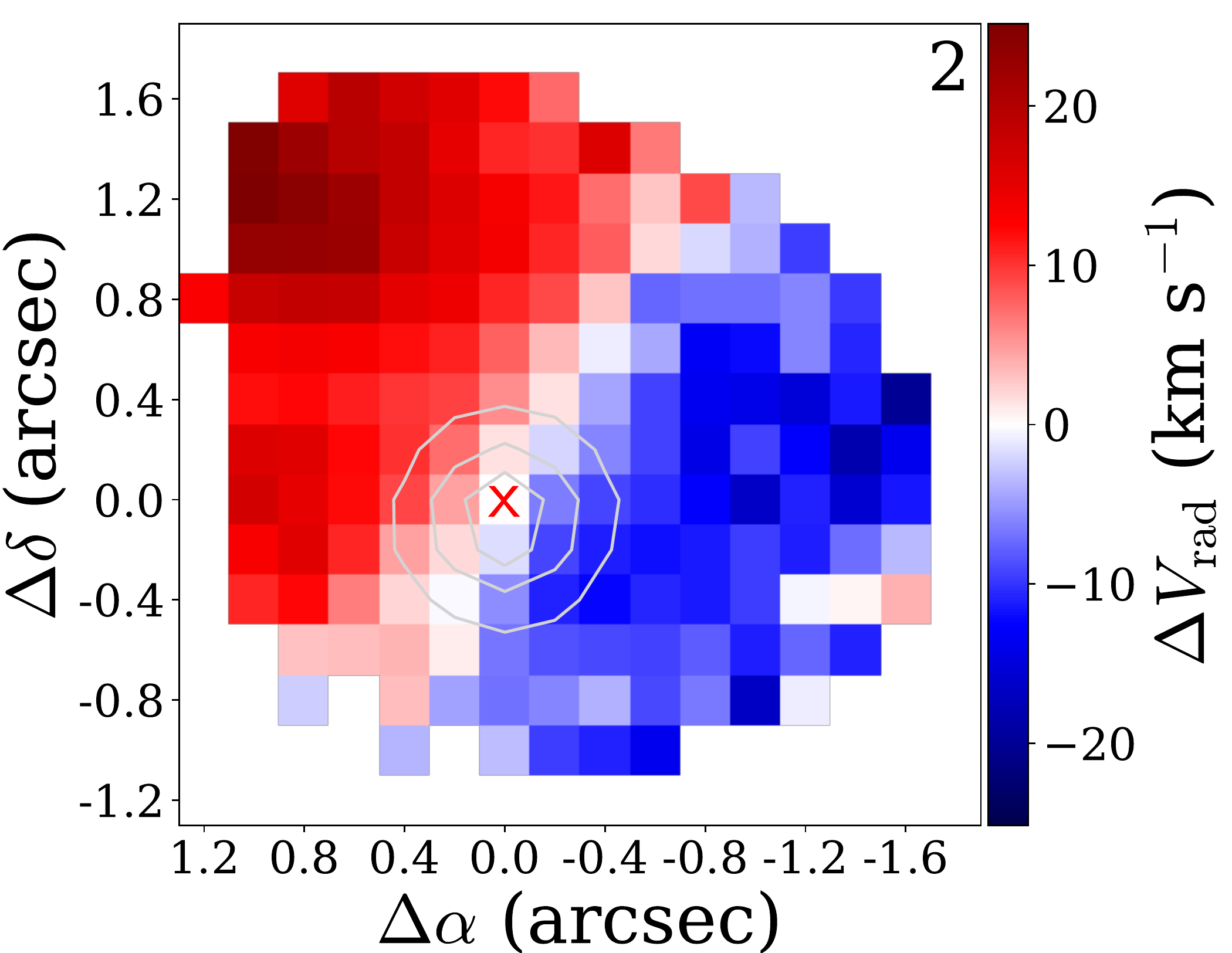}\hspace{-0.1cm}
\includegraphics[width=0.25\textwidth]{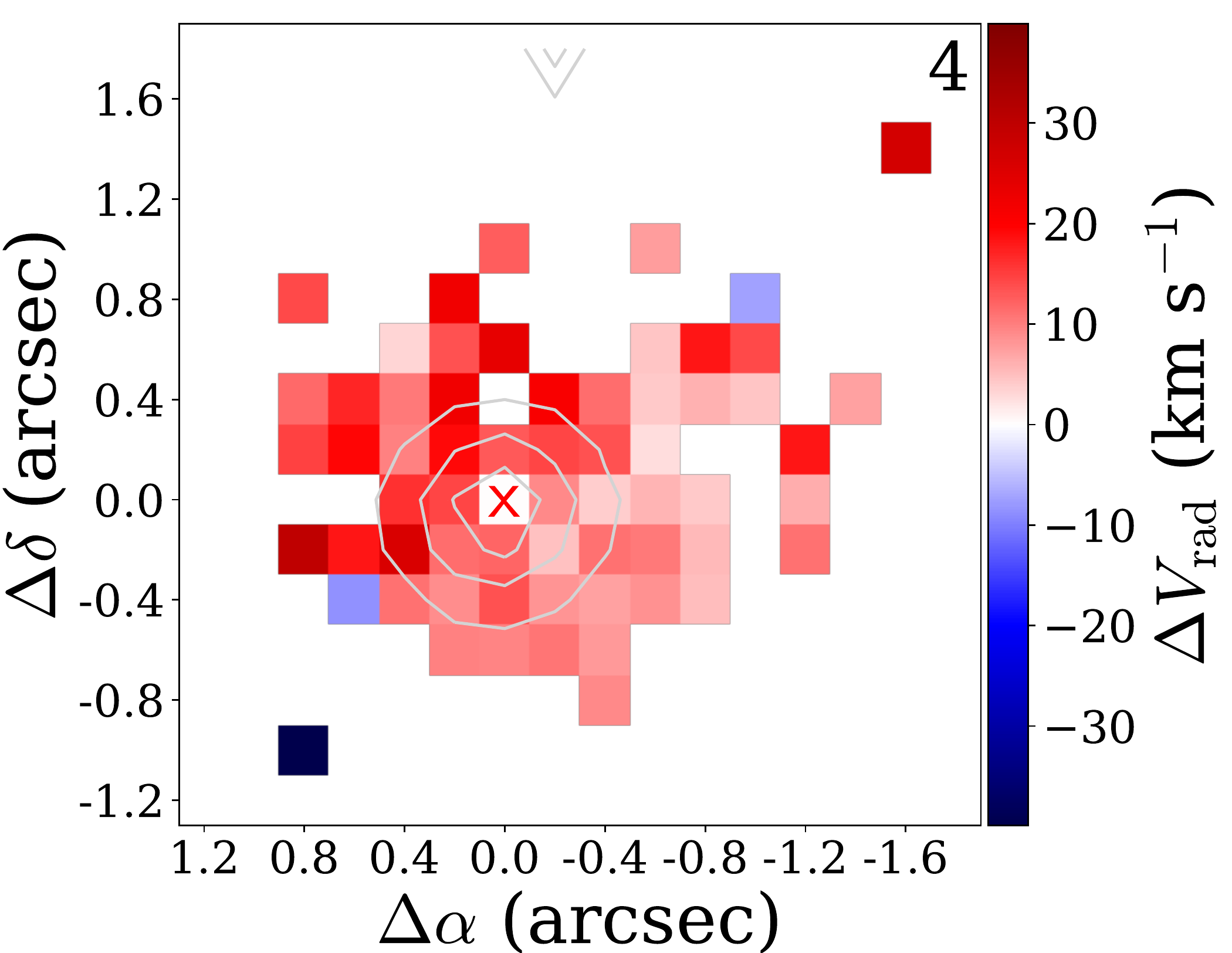}\hspace{-0.1cm}
\includegraphics[width=0.25\textwidth]{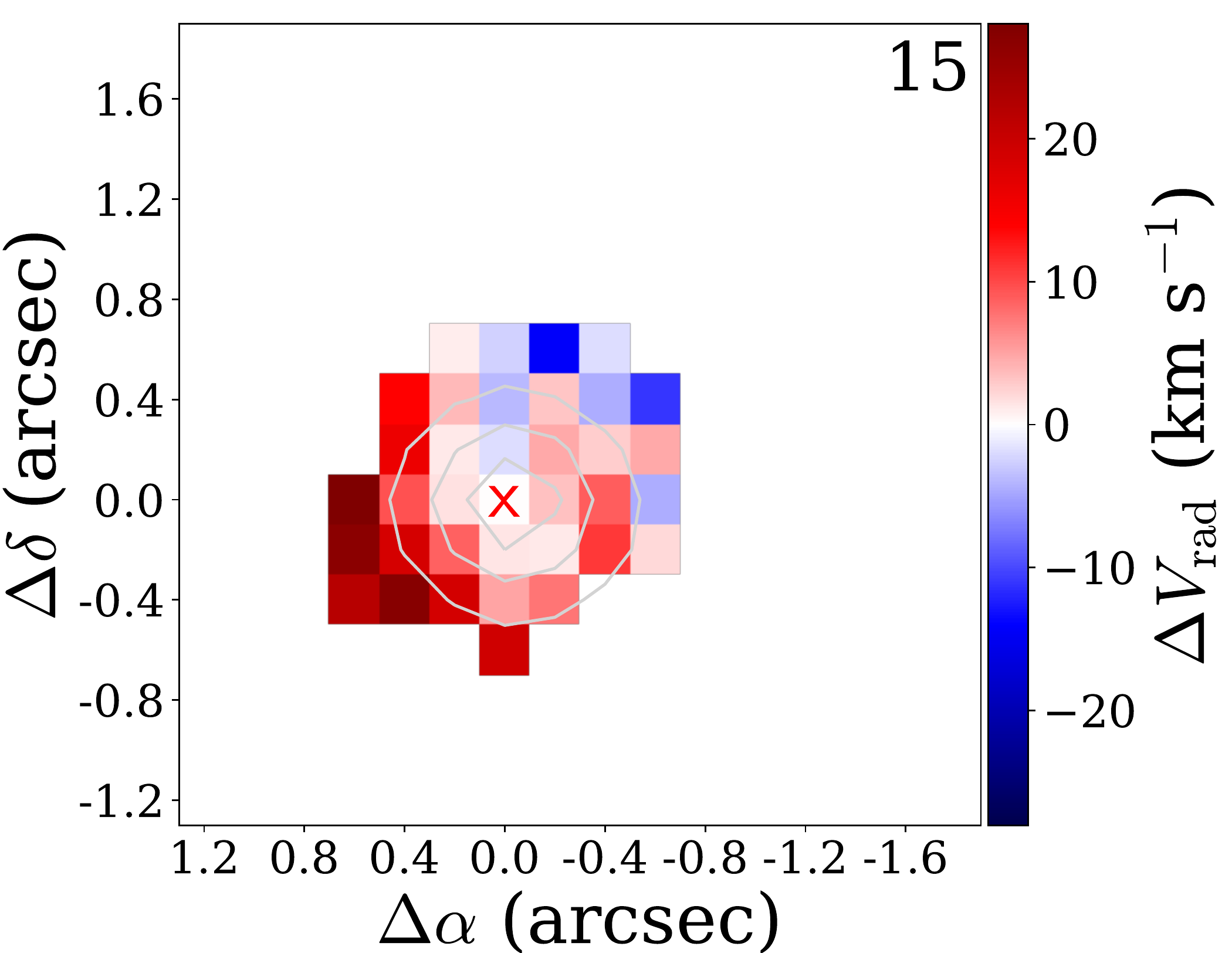}\hspace{-0.1cm}
\includegraphics[width=0.25\textwidth]{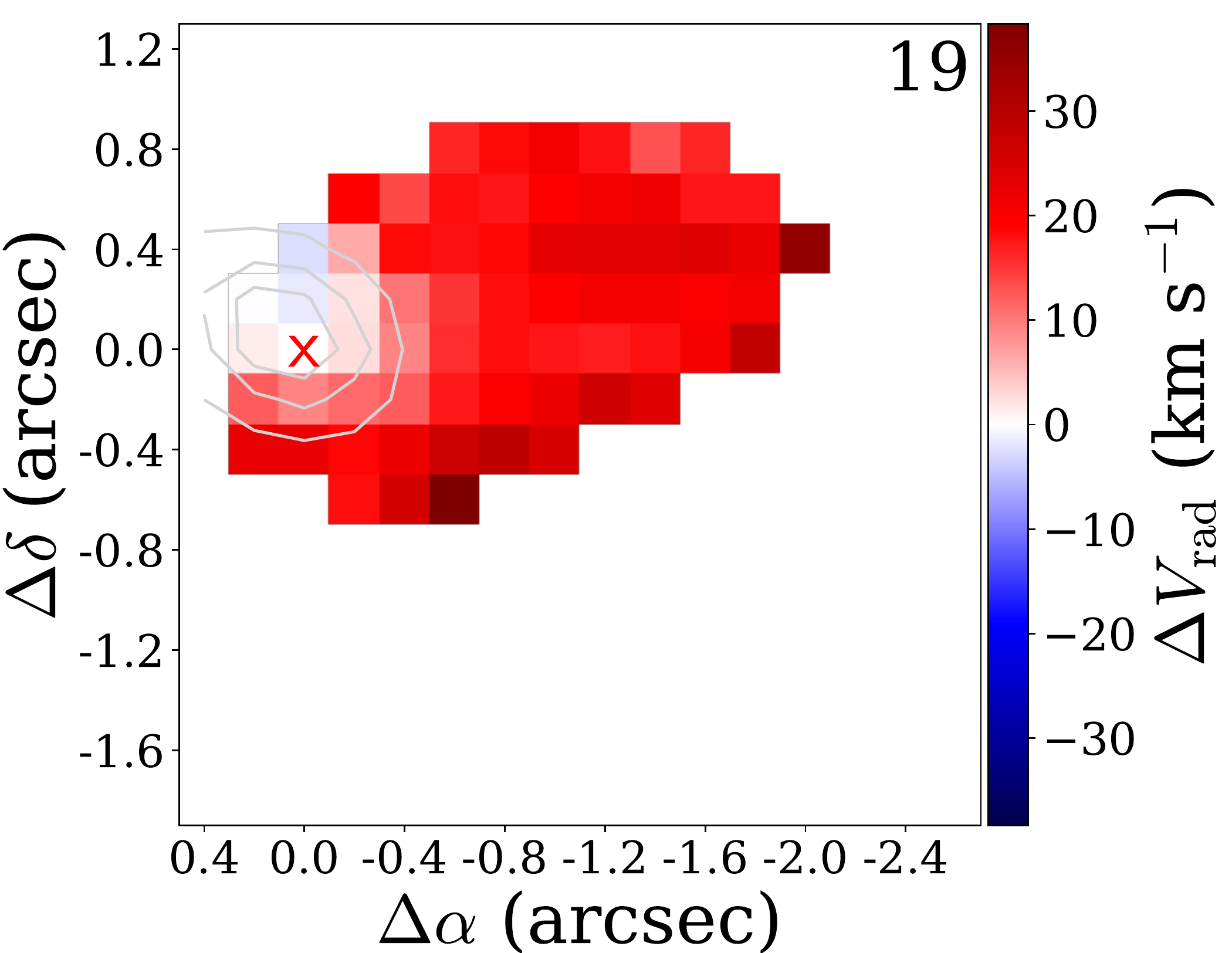}\hspace{-0.1cm}
\includegraphics[width=0.25\textwidth]{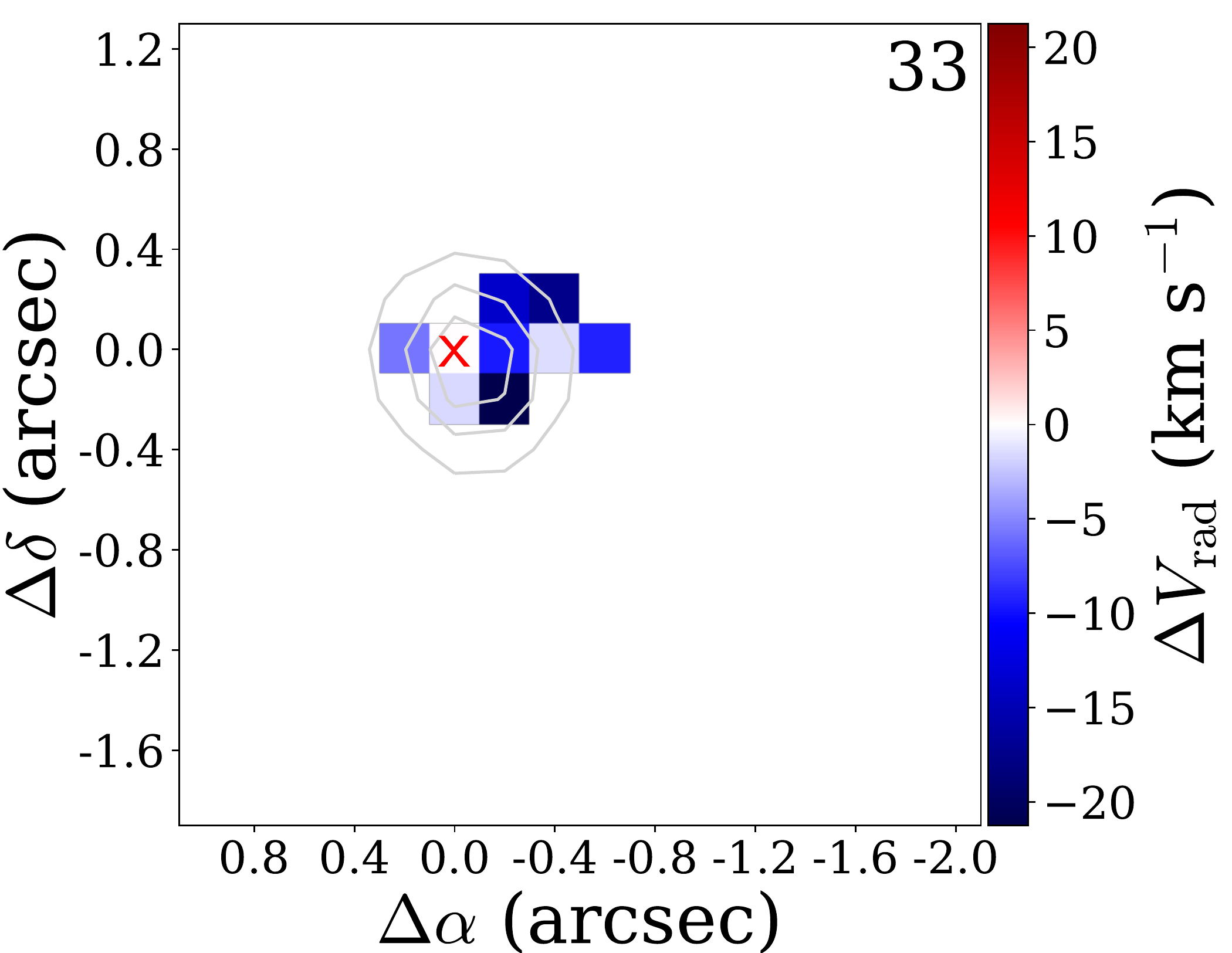}\hspace{-0.1cm} 
\includegraphics[width=0.25\textwidth]{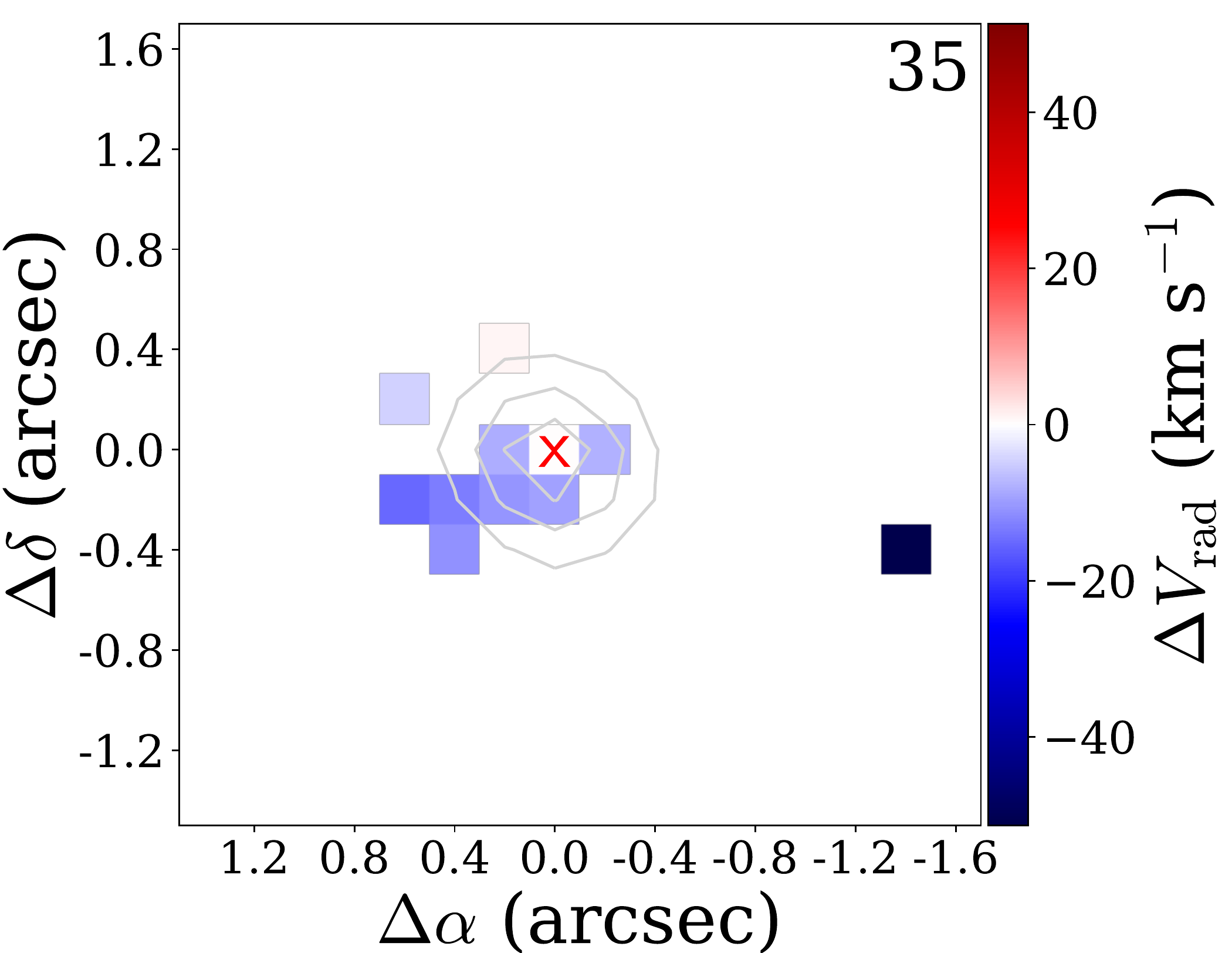}\hspace{-0.1cm} 
\includegraphics[width=0.25\textwidth]{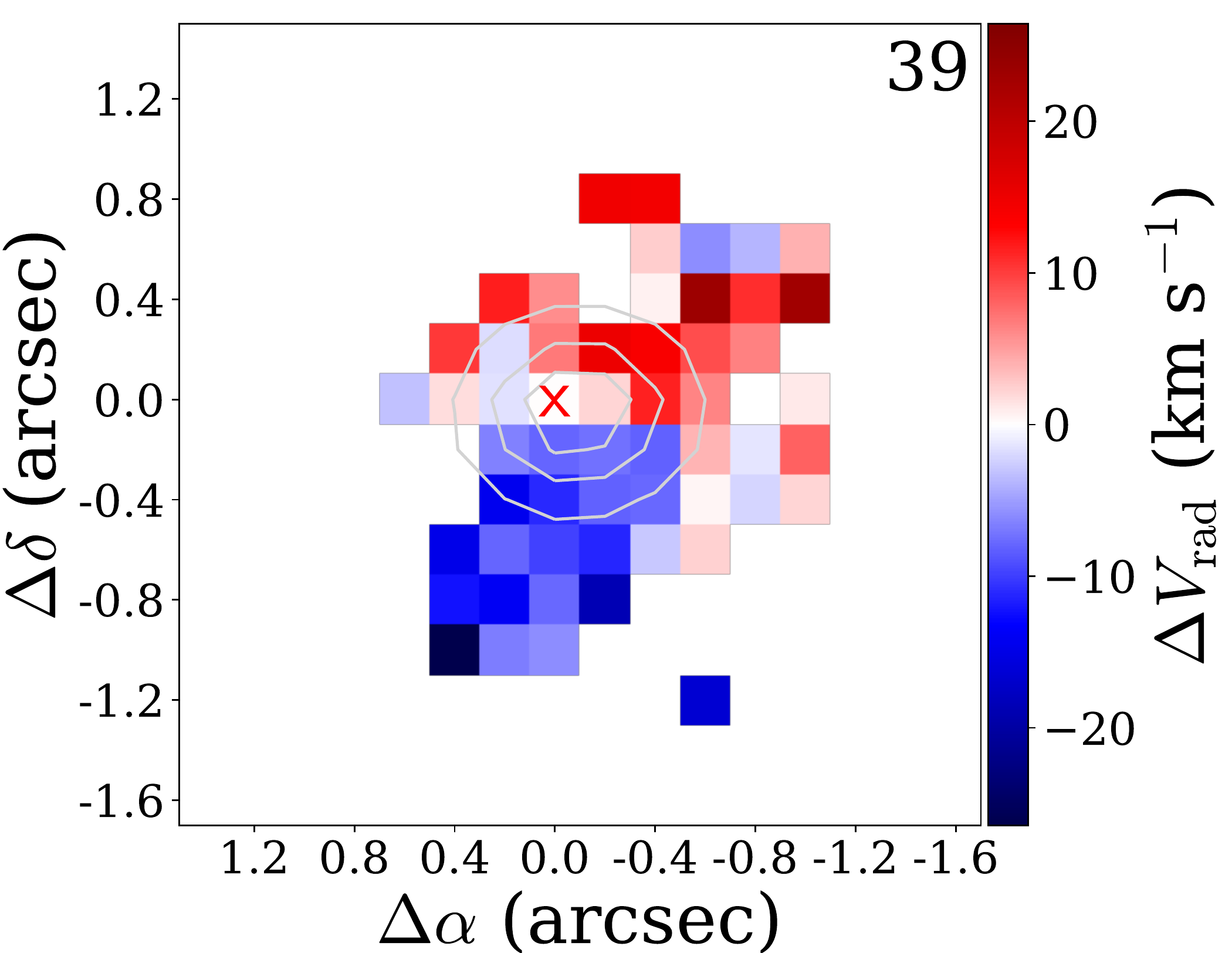}\hspace{-0.1cm}
\includegraphics[width=0.25\textwidth]{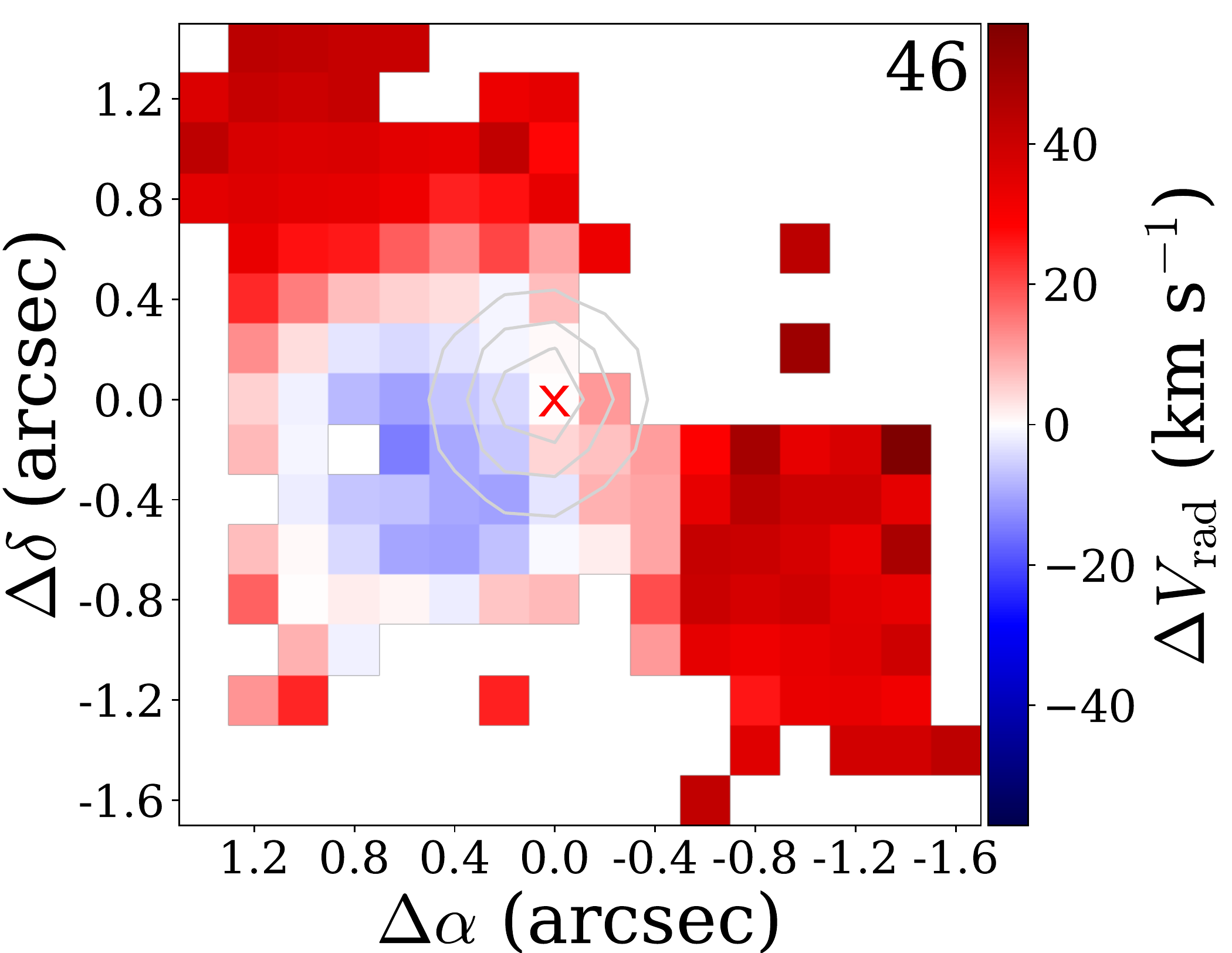}\hspace{-0.1cm}
\includegraphics[width=0.25\textwidth]{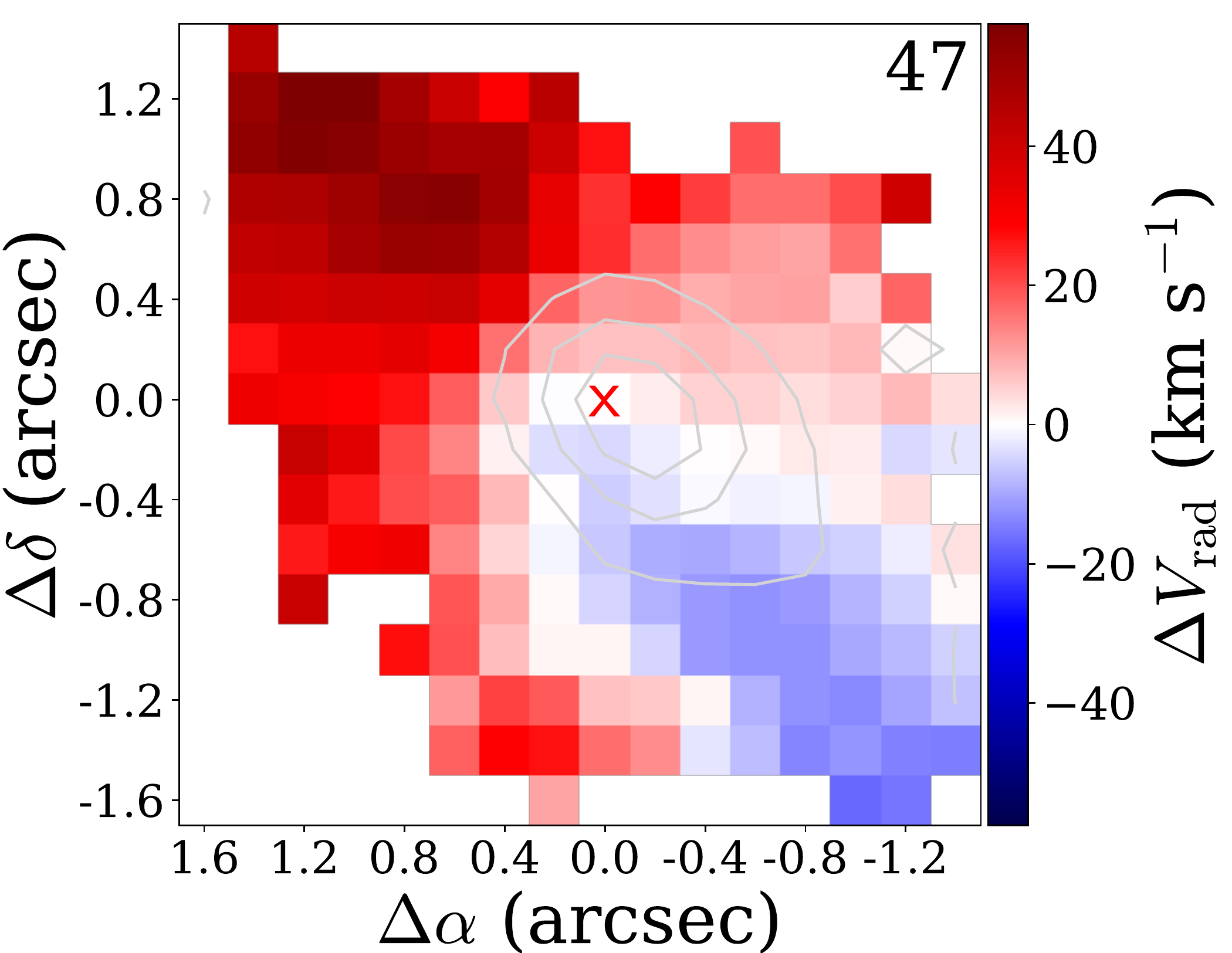}\hspace{-0.1cm}
\includegraphics[width=0.25\textwidth]{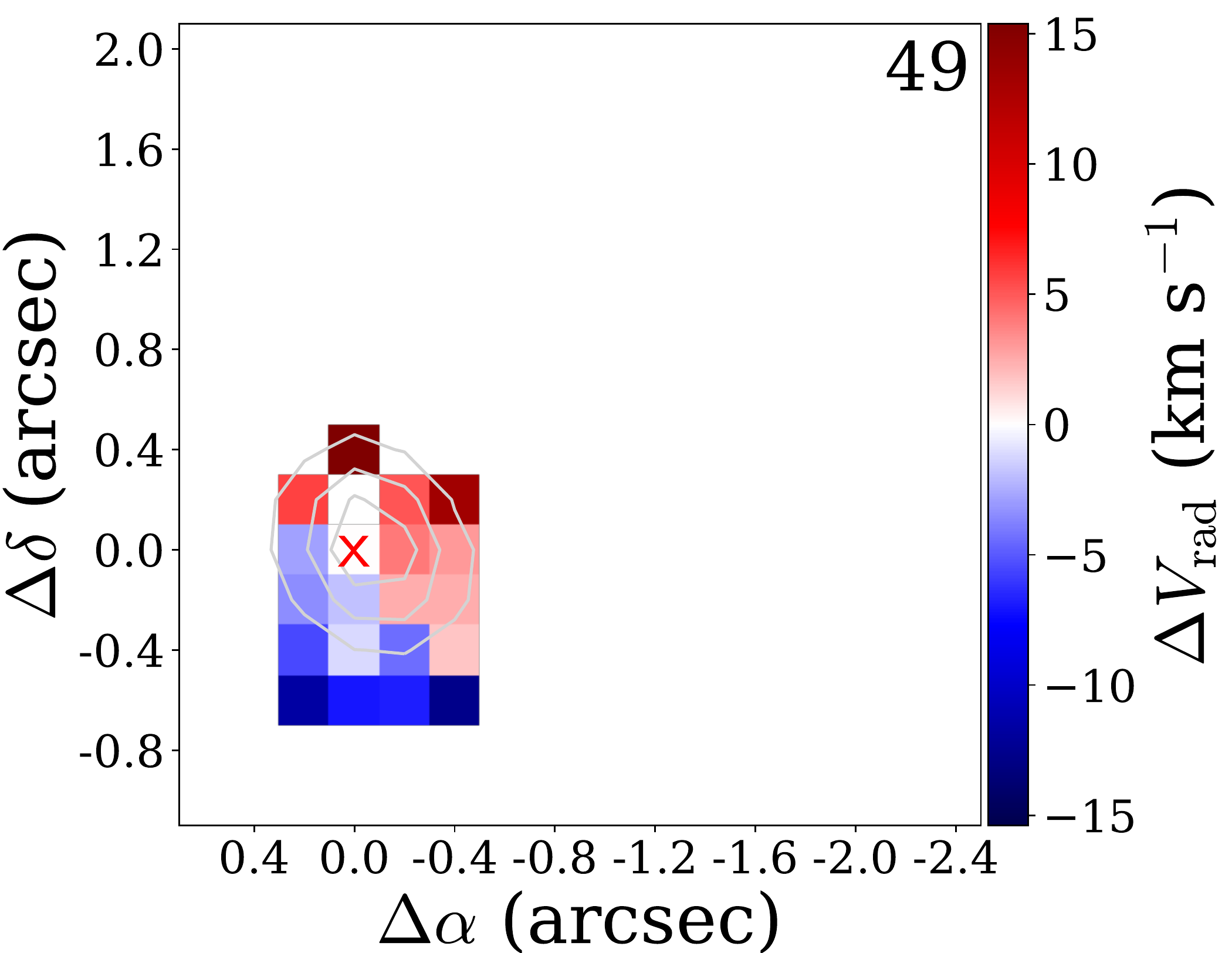}\hspace{-0.1cm} 
\includegraphics[width=0.25\textwidth]{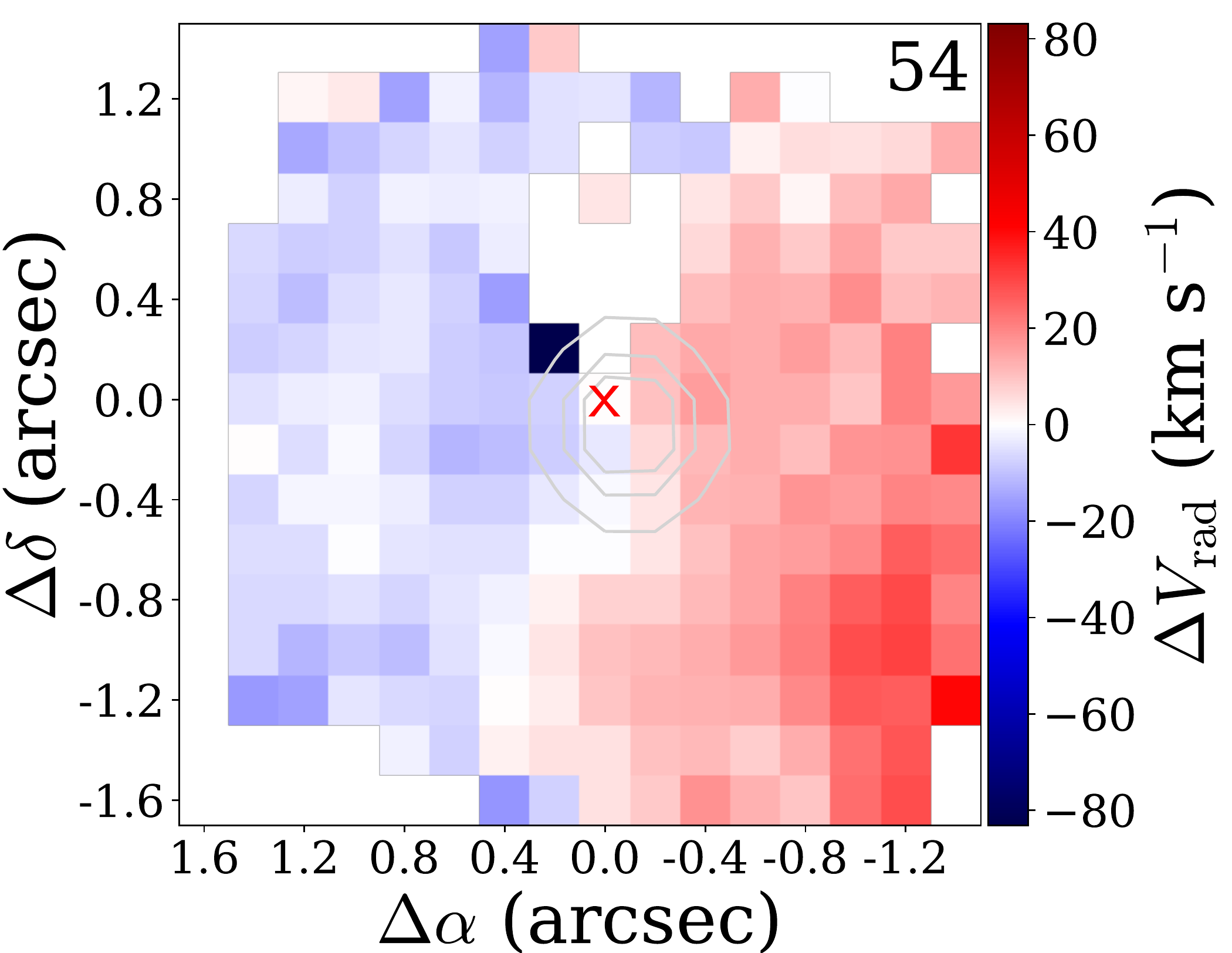}\hspace{-0.1cm}
\includegraphics[width=0.25\textwidth]{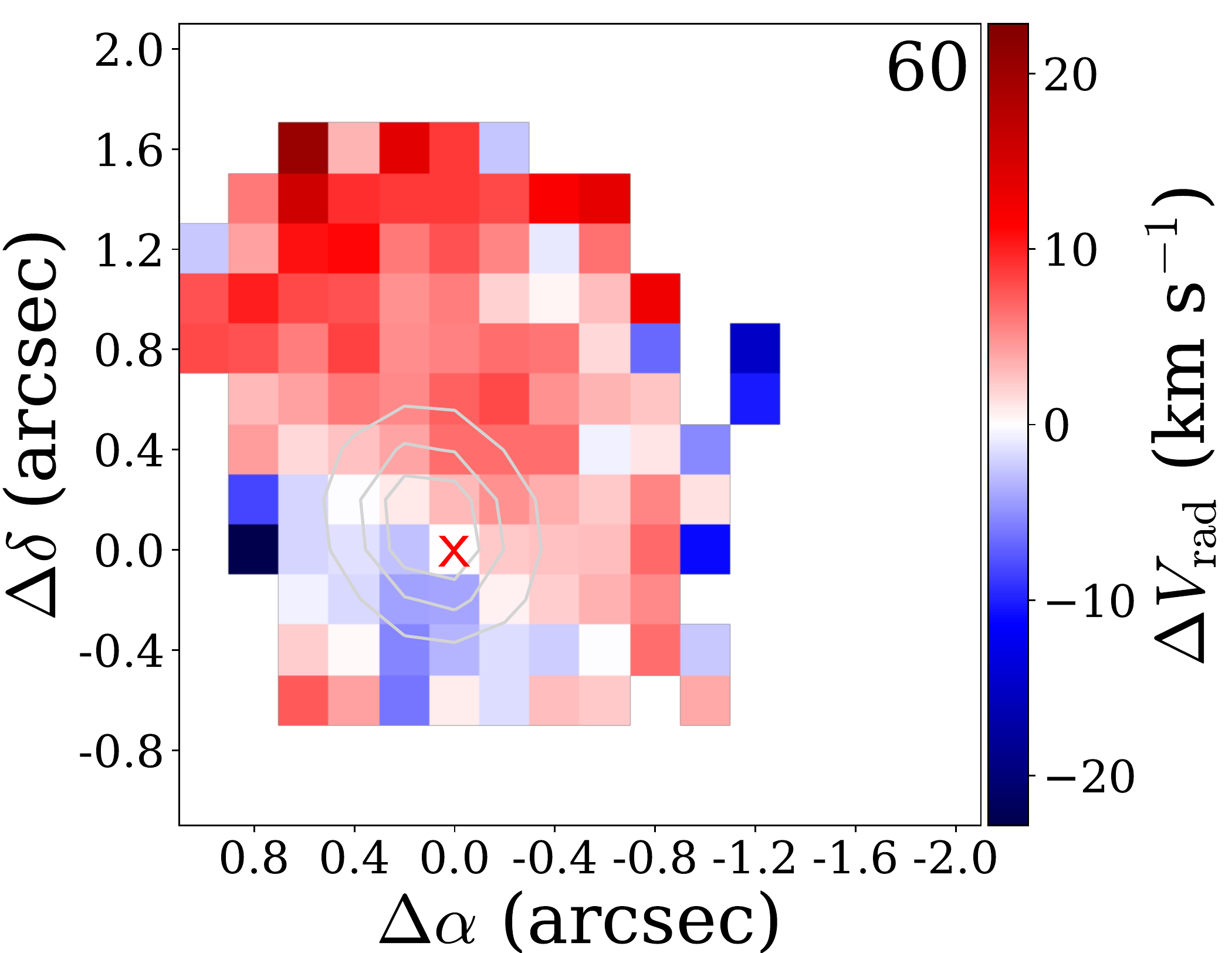}\hspace{-0.1cm}
\includegraphics[width=0.25\textwidth]{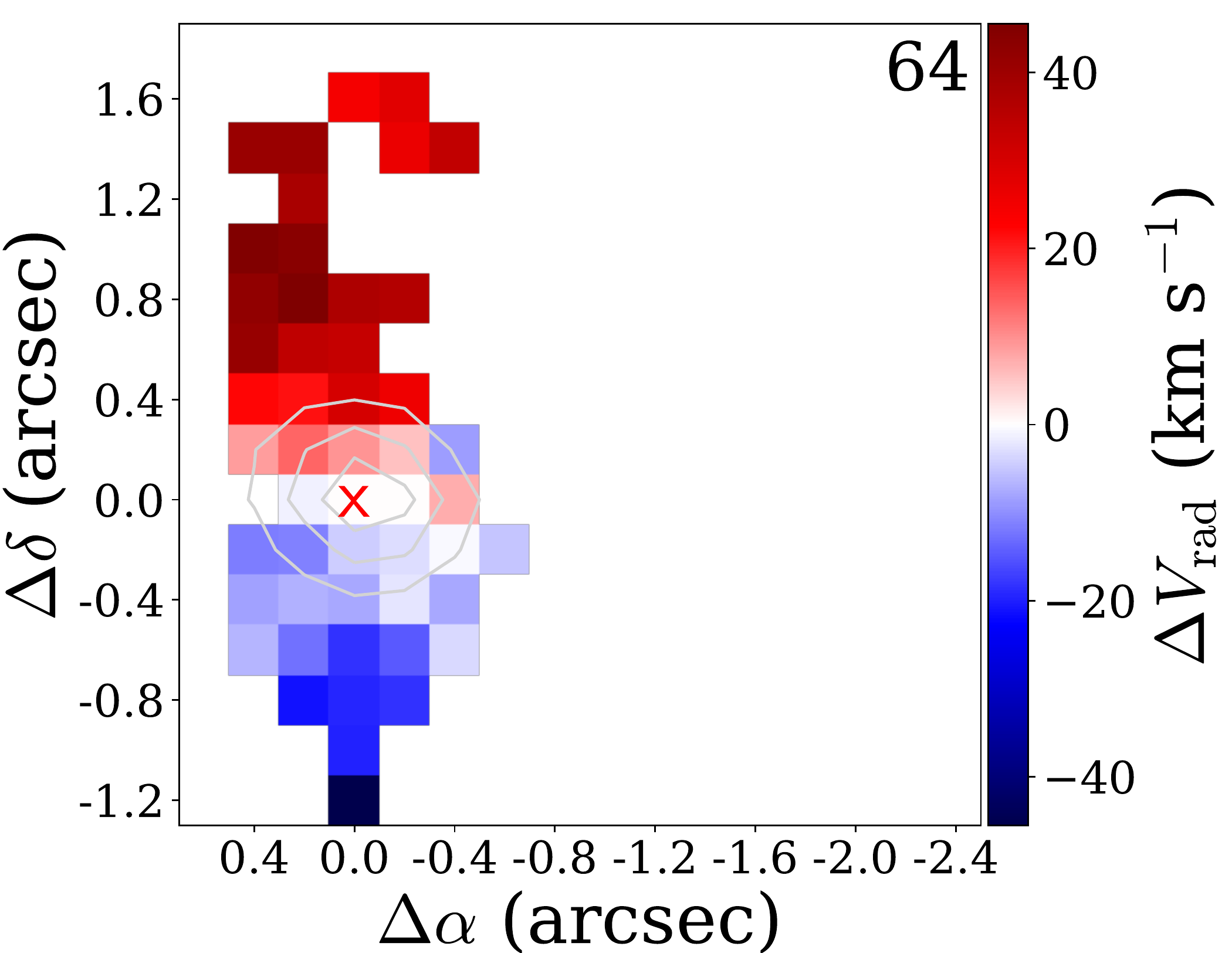}\hspace{-0.1cm}
\includegraphics[width=0.25\textwidth]{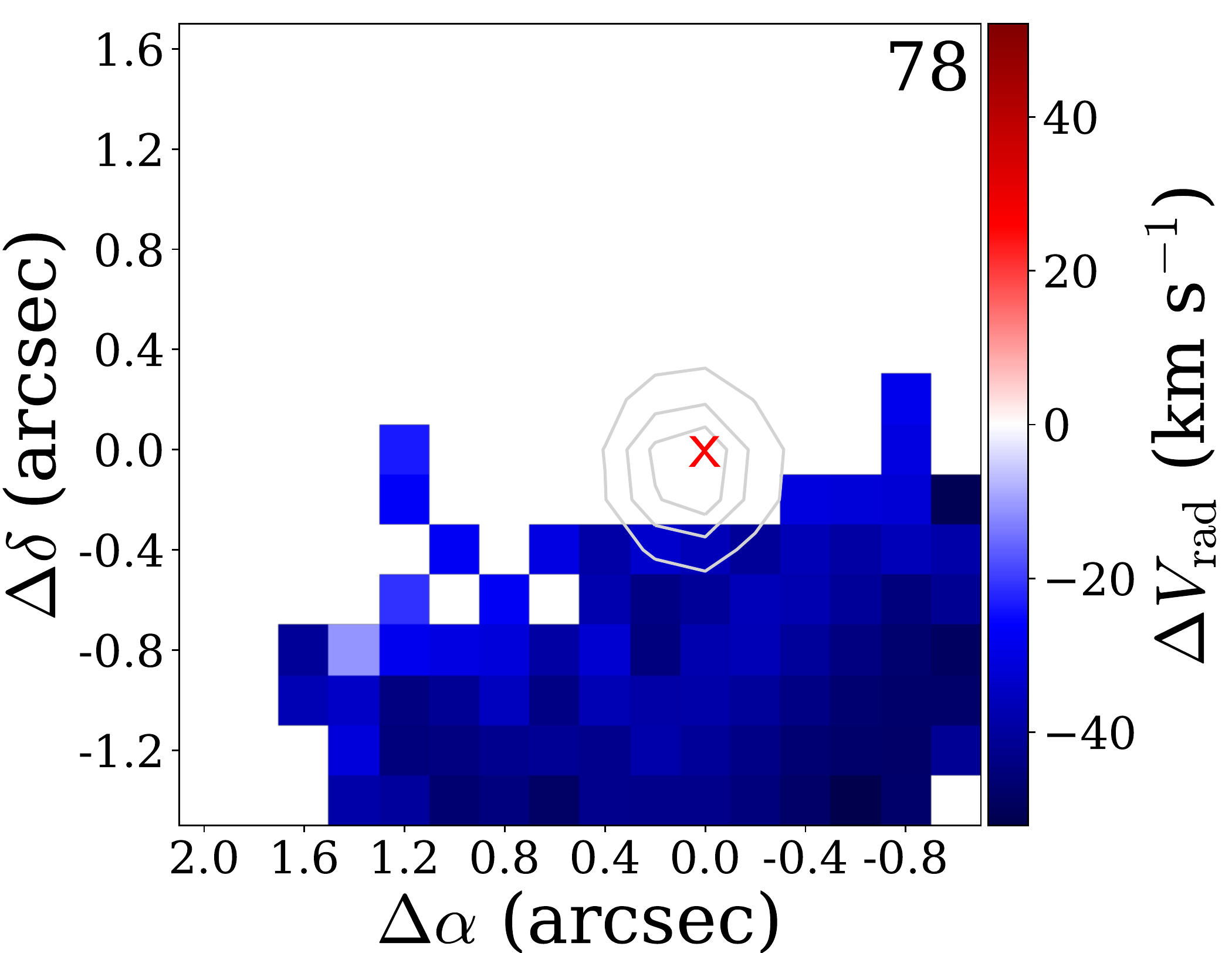}\hspace{-0.1cm}
\includegraphics[width=0.25\textwidth]{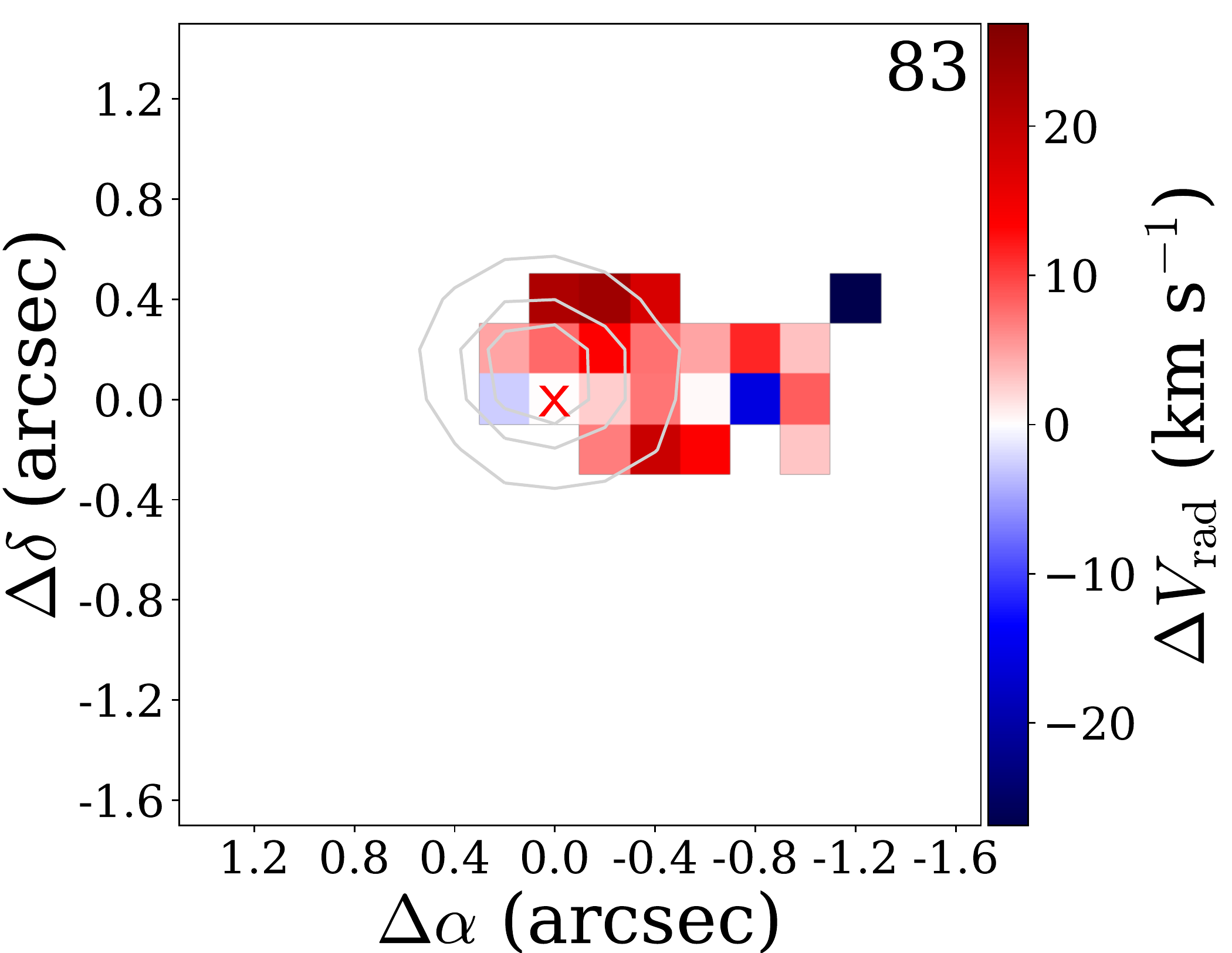}\hspace{-0.1cm}
\includegraphics[width=0.25\textwidth]{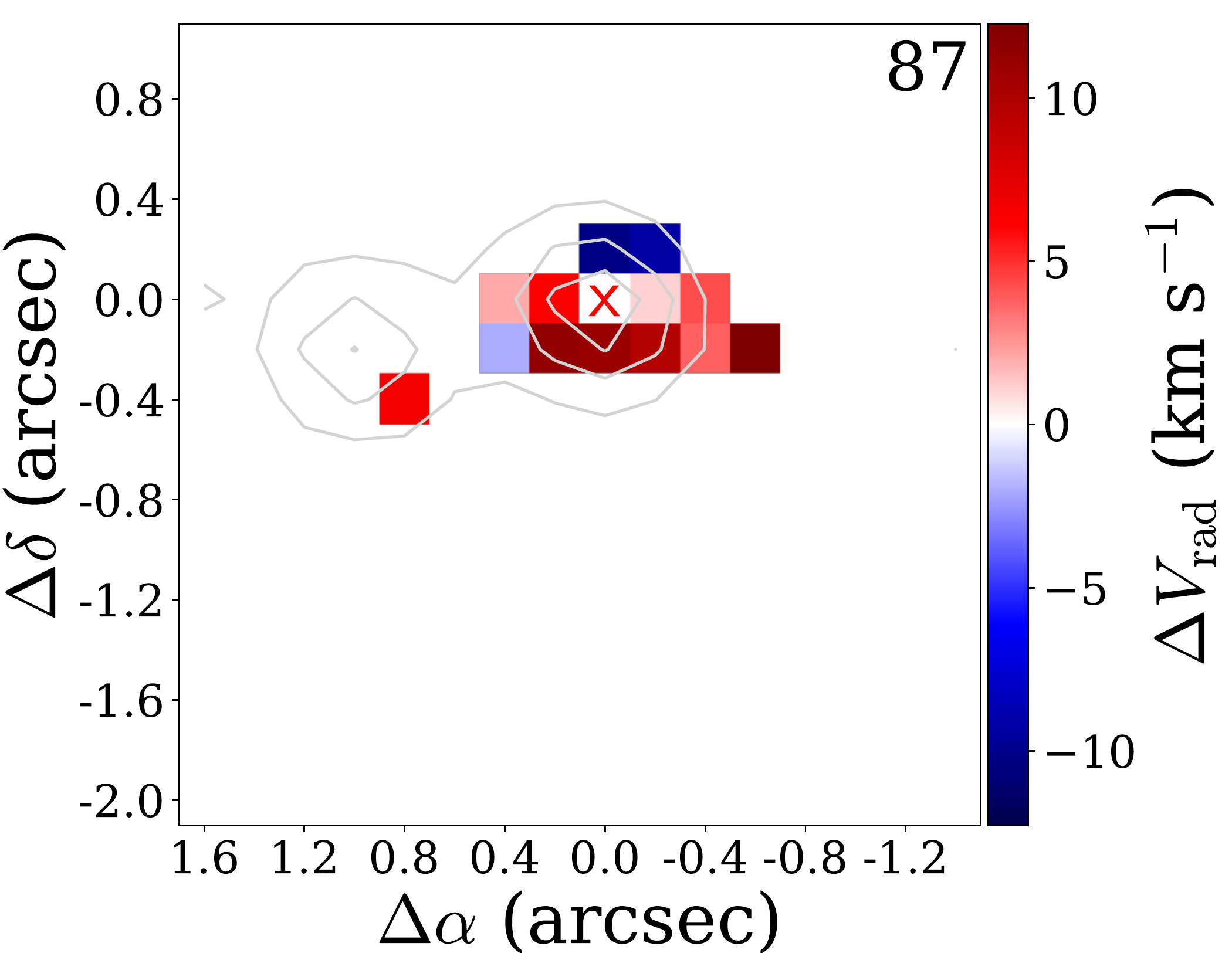}\hspace{-0.1cm}
\includegraphics[width=0.25\textwidth]{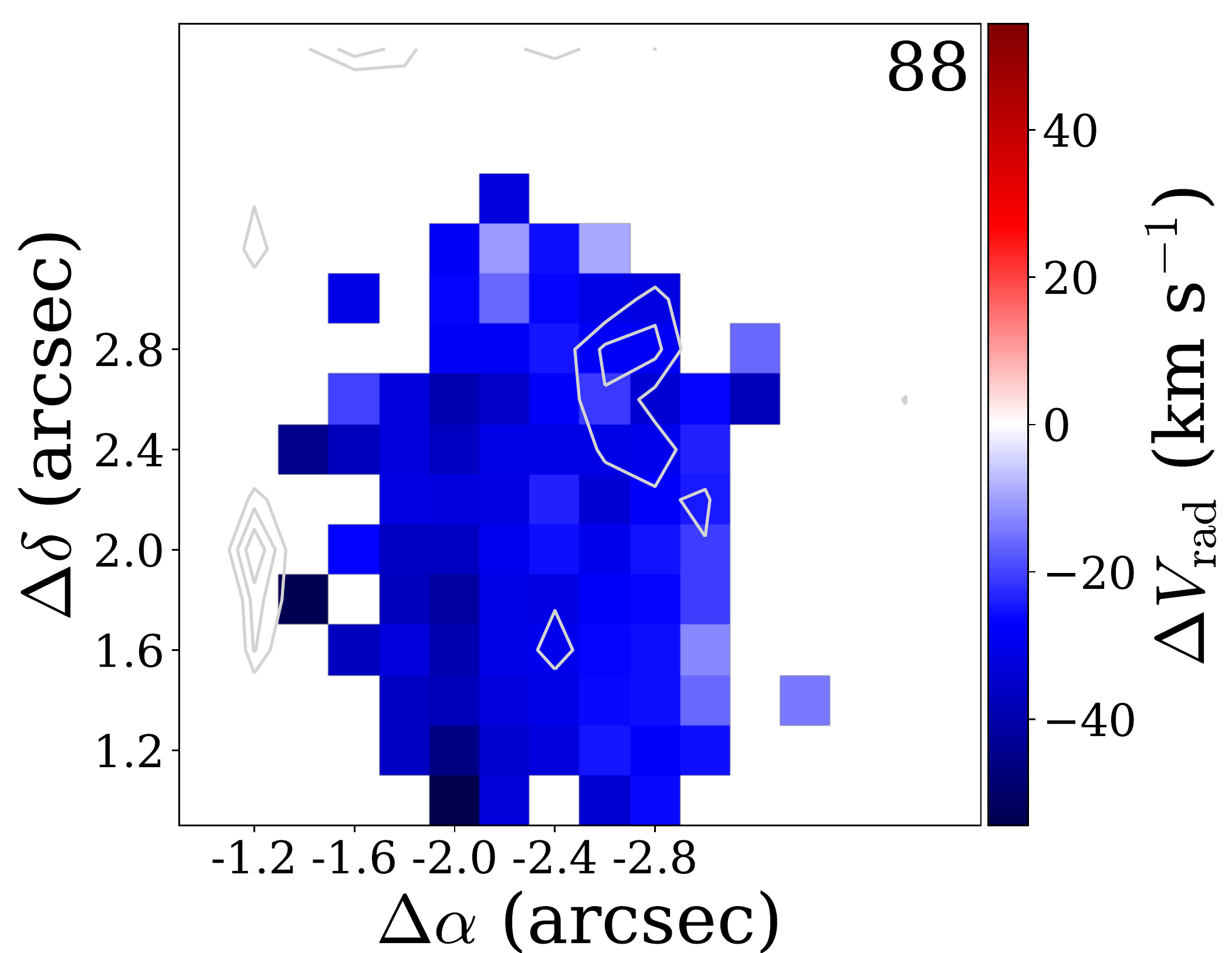}\hspace{-0.1cm}
\includegraphics[width=0.25\textwidth]{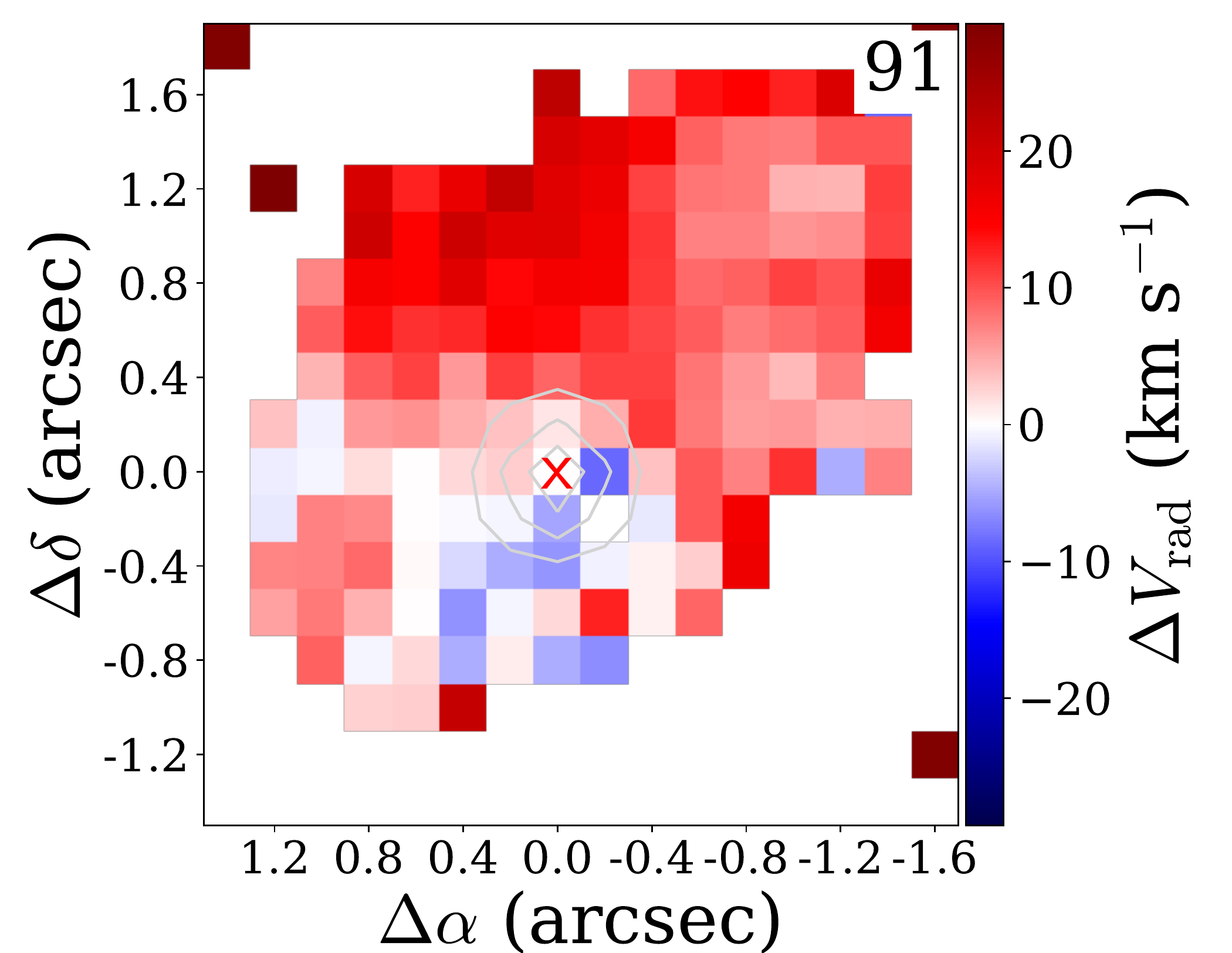}\hspace{-0.1cm}
\includegraphics[width=0.25\textwidth]{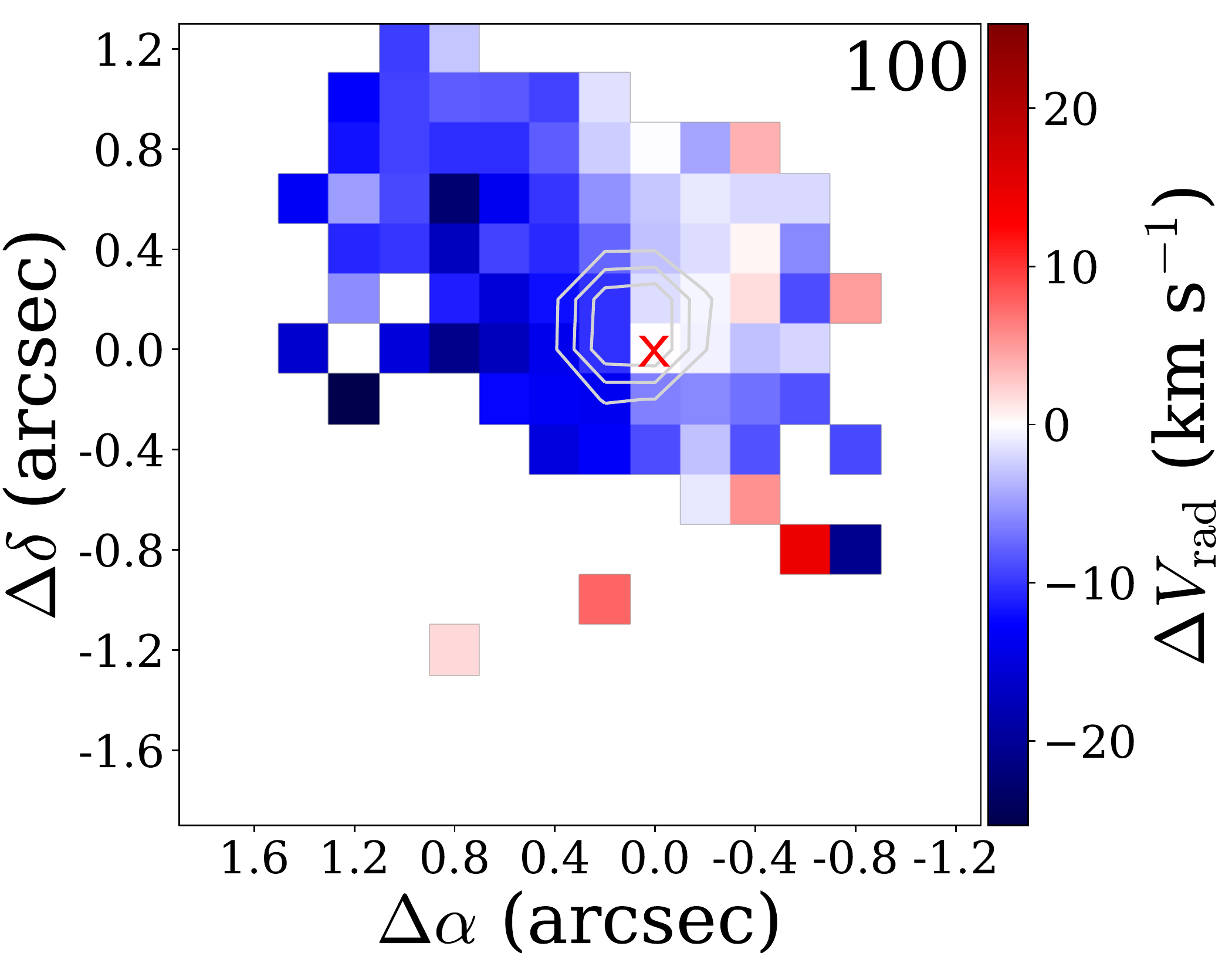}\hspace{-0.1cm}
\includegraphics[width=0.25\textwidth]{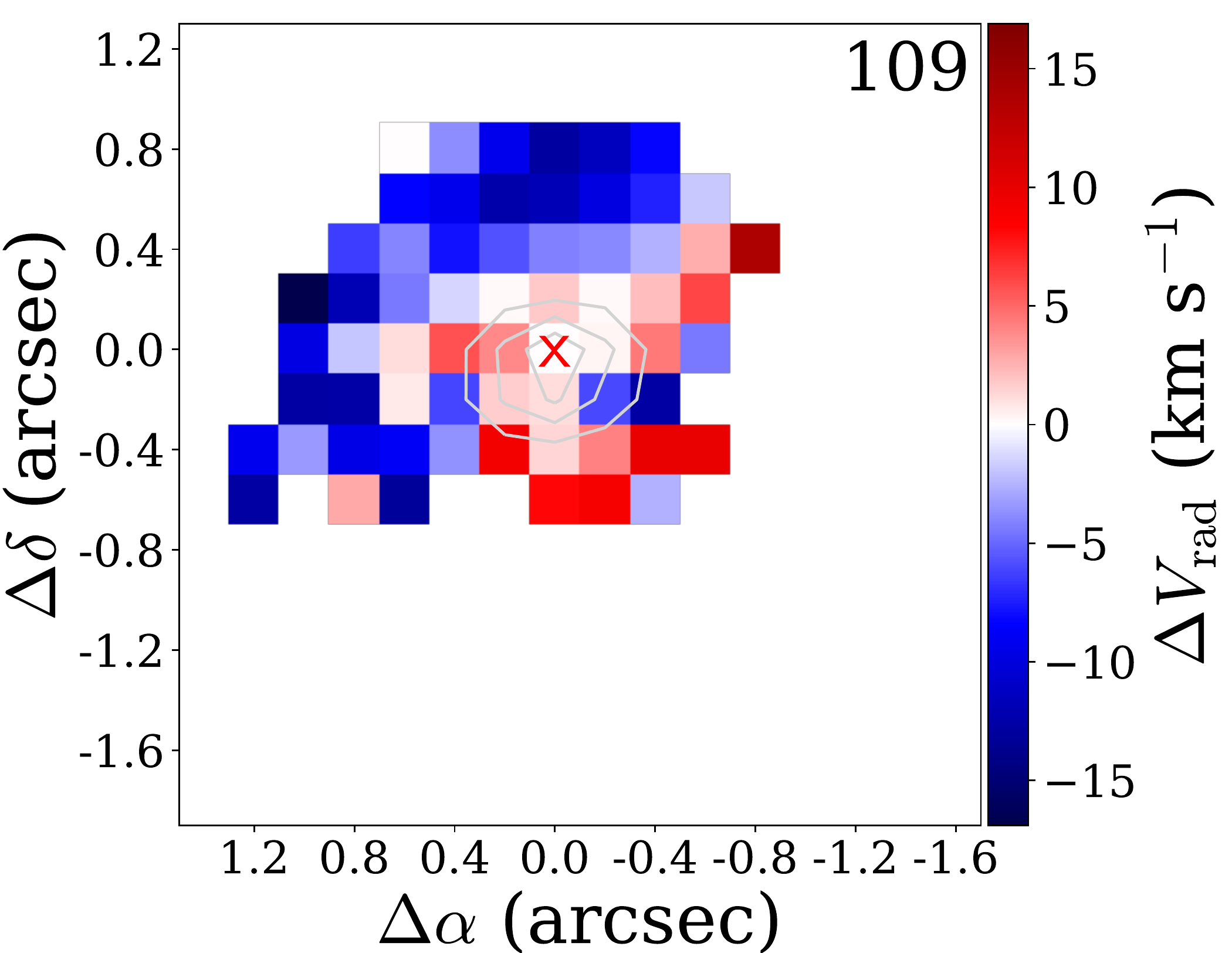}\hspace{-0.1cm}
\includegraphics[width=0.25\textwidth]{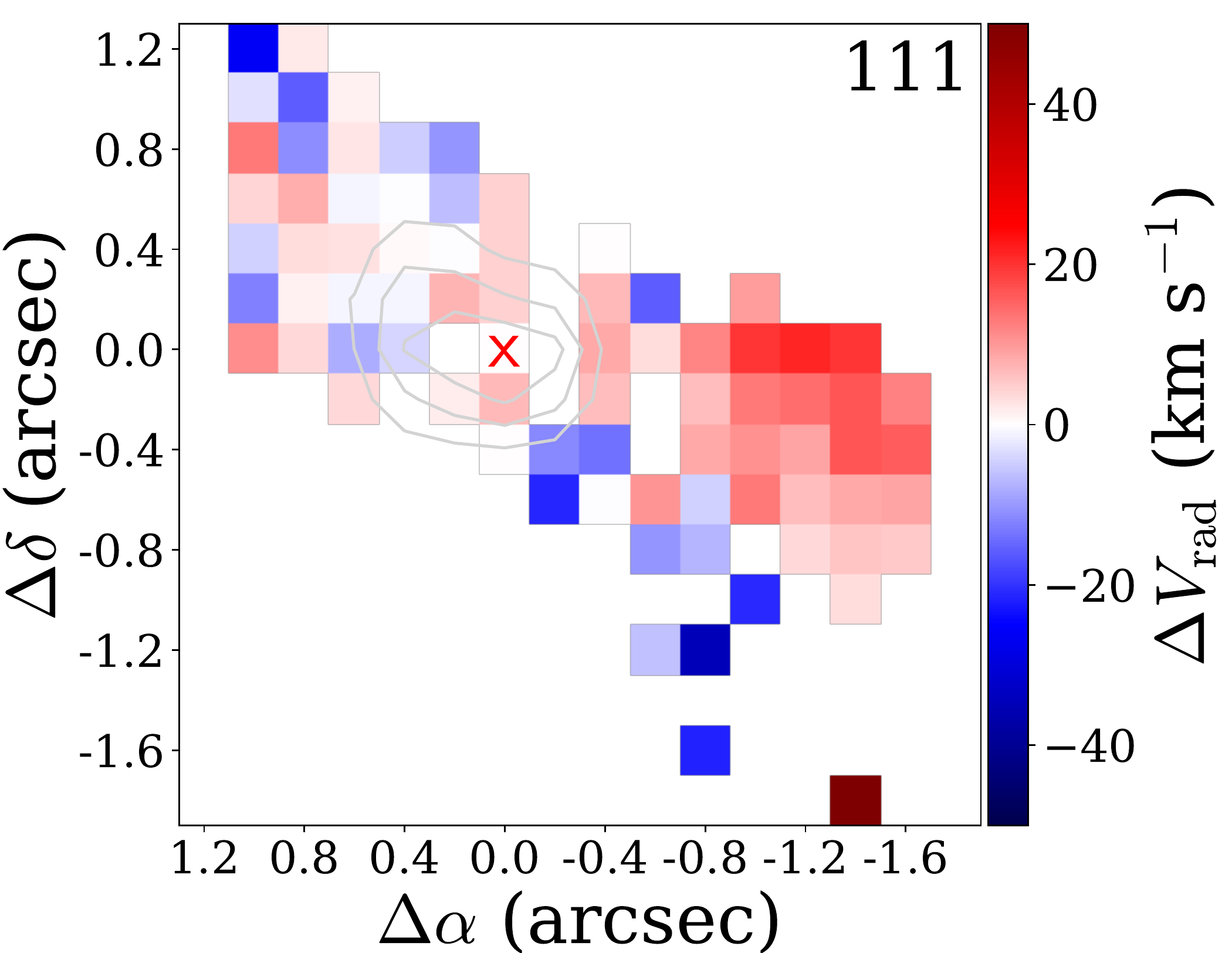}\hspace{-0.1cm}
\includegraphics[width=0.25\textwidth]{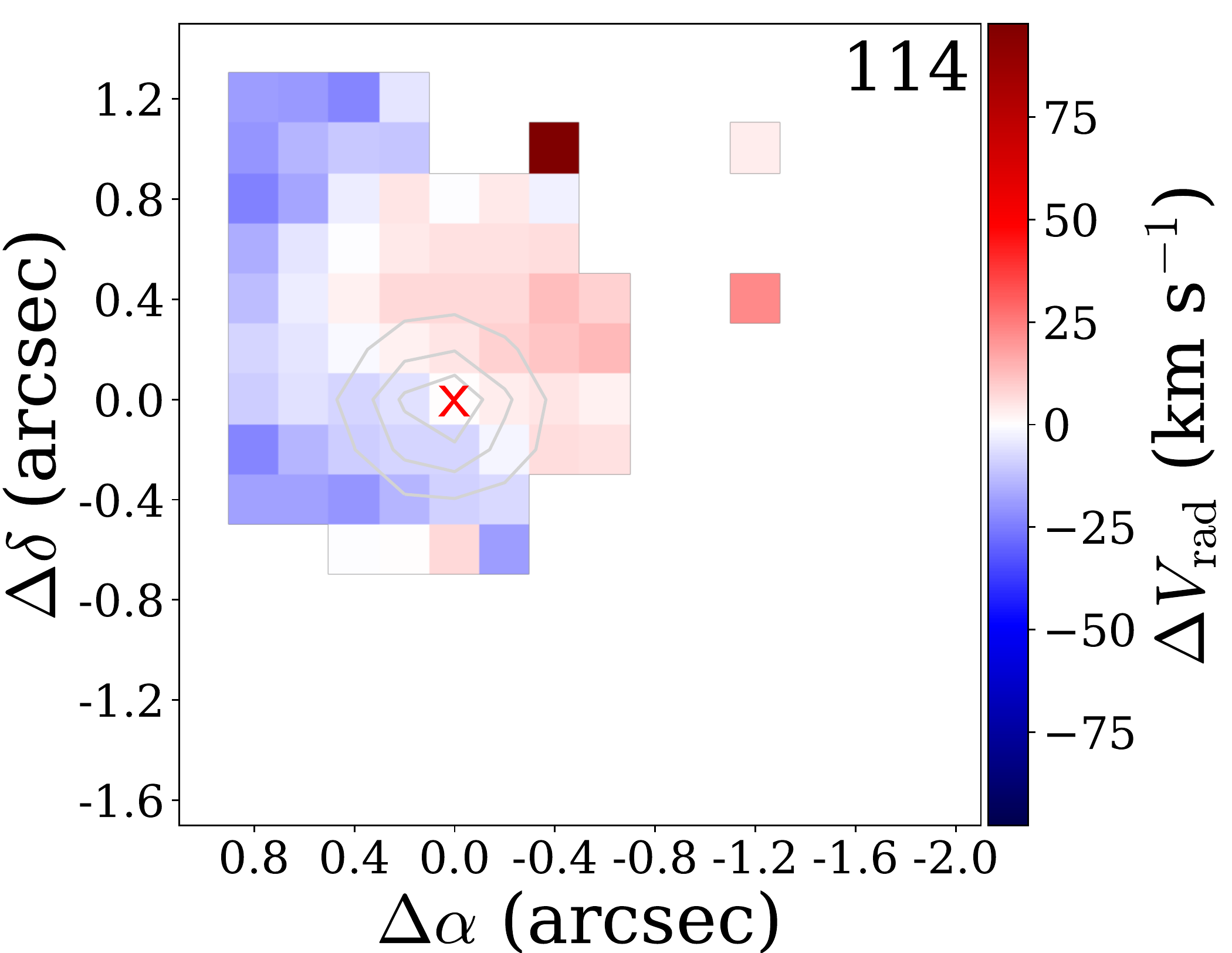}\hspace{-0.1cm}
\caption{The H$_2$ 1-0 S(1) velocity centroid maps for all YSOs associated with the H$_2$ emission; the velocities are calculated with respect to the radial velocity of the corresponding source. Only pixels with the H$_2$ detection above 3$\sigma$ are shown. The contours show the $K$-band continuum emission, and the red crosses the positions of the YSO candidates (the peak of the $K$-band continuum emission). The source names are indicated in the upper right corners. Radial velocities of the gas associated with sources No. 78 and 88 cannot be estimated, and here only the apparent radial velocities are shown. The KMOS spectral sampling at 2.12~$\mu$m corresponds to 39.67~km/s.}
\label{fig:velmap}
\end{figure*}

\subsection{Extinction} 
\label{subsec:ext}

The line fluxes in the near-IR are affected by extinction. We adopt $A_{\rm{V}}$ obtained from the spectral energy distribution (SED) modelling by \cite{sewilo2019} for 85 sources in our sample. Our $K$-band observations cover several H$_2$ lines that are often used to estimate $A_{\rm{V}}$; however, after assessing the quality of our H$_2$ data, we decided against using this method due to the small separation between the two lines ($\sim0.2$ $\mu$m), uncertainty in telluric corrections near the Q-branch H$_2$ emission, and small number of sources for which this method can be used. 

We assess the uncertainty of extinction estimated from the SED fitting using the work by \cite{furlan2016}. There, the {\it R} statistics was used to calculate a mode -- the value with the highest frequency within a certain range from the best-fit {\it R} value (see their Fig. 48). Almost all their $A_{\rm{V}}$ uncertainties are consistent with a 1:1 correlation within a 10~mag error. We adopt this conservative value in the subsequent analysis, in particular in the calculation of the mass accretion 
(Sections \ref{subsec:acc}).

We note that extinction is expected to differ between the position of the infrared continuum source and the outflow \citep{yoon22}. The ratios of the hydrogen Br$\gamma$ and Pa$\alpha$ lines are routinely used to estimate $A_{\rm{V}}$ toward the accretion region \citep{fiorellino2021}, but the latter line is located in the $H$-band and not covered by our observations. 
{Since SED modelling provides the estimate of $A_{\rm{V}}$ toward the central star and H$_2$ emission is often extended, such $A_{\rm{V}}$ estimate is likely not representative for the true reddening of the H$_2$ lines. Therefore, we do not apply it for those lines. Thus, Table \ref{tab:H2flux-ex} shows the H$_2$ line fluxes, which are not corrected for extinction.}

\subsection{Gas Spatial Distribution and Kinematics}
\label{subsec:linemap}

The spatial distribution of the line emission varies significantly between different species detected with KMOS (Fig.~\ref{fig:rgb}). The Br$\gamma$ emission is spatially unresolved for the majority of the sources, in line with the origin in the inner disk. Similarly, the CO bandhead emission is also compact and co-spatial with the continuum emission. In contrast to the Br$\gamma$ and CO bandhead emission, the H$_2$ emission is { usually} extended.

The extended distribution of the H$_2$ emission resembles the structure of the molecular outflows from Class 0/I protostars for 14 out of 33 sources with the H$_2$ line detection (42\% or 11\% of the entire sample; see Fig. \ref{fig:rgb}). In five sources, the extended H$_2$ emission is not associated with the targeted source (see Table \ref{tab:coordinates}) and likely originates from an outflow from another YSO. {Those five cases often show only one lobe within the field of view}.  
{Among other sources with the  detection of H$_2$, the emission pattern does not show two prominent outflow lobes. } 
The lack of the bipolar pattern might be due to too low spatial resolution or the projection effects. 
{In general, a contribution from a photo-dissociation region is unlikely among the low-mass YSOs found in CMa-$\ell$224.} The emission maps for the Br$\gamma$, H$_2$, and CO bandhead lines for the entire KMOS YSO sample in CMa-$\ell$224 are presented in Appendix \ref{app:emiss}.

We use the brightest and {spectroscopically} well-resolved H$_2$ line in $K$-band, the 1-0 S(1) transition at 2.12~$\mu$m, to calculate radial velocities of the emitting gas. Figure \ref{fig:chan47} shows selected channel maps for an example source (No. 47) associated with an extended H$_2$ emission. {Here, the blue-shifted emission transitions into red-shifted one on the other side of the central source, which is a clear signature of a bipolar ejection.} The images reveal a velocity structure consistent with an outflow with velocities up to 50 km s$^{-1}$. A similar kinematic structure has been detected toward most of the other sources with the extended H$_2$ emission (No. 2, 15, 19, 46, 47, 54, 60, 64, 91 and 114; see Figure \ref{fig:velmap}). The large spatial scales of those structures cannot be due to the emission from protoplanetary disks since they are too cold at such large distances from the central star to excite H$_2$. 

The H$_2$ emission structure associated with sources No. 78 and 88 resembles the single outflow lobes; however, the continuum source is not seen in the map of the YSO candidate No. 88, while for No. 78, the H$_2$ emission does not seem to be physically connected to the source. Additional observations with a larger field-of-view are needed to identify sources responsible for the extended H$_2$ emission in these fields.

The estimated relative velocities do not exceed $\sim$50~km/s, with velocities increasing towards lobe ends, typical for outflow-driven shock waves. The moderate spectral resolution of KMOS does not allow for a detailed study on outflow kinematics, including any minor variations in kinematic patterns. {The spatial extent of the H$_2$ gas supports the origin in outflow shocks. The limited wavelength coverage of our spectra, however, does not allow to properly estimate extinction as a function of position along the outflow (see Section \ref{subsec:ext}).  

Finally, we note that the bipolar structure in several H$_2$ lines toward source No. 47 (Appendix \ref{app:emiss}) is consistent with its classification as a Class I source using SED models \citep{sewilo2019}. The prominent H$_2$ outflow suggest that the source might be one of the youngest in our sample. 
{It would be a suitable candidate for a detailed spectral modeling using shock models; however, this is outside of the scope of this paper.}

\subsection{Gas Accretion}
\label{subsec:acc}

Gas accretion onto Class II YSOs is typically measured using UV continuum excess or line emission in e.g., H, He or \ion{Ca}{2} at optical wavelengths \citep[e.g.,][]{alcala2014,ni18}. In more embedded Class I YSOs, the accretion is more often quantified using hydrogen lines in the near-IR which are less affected by dust extinction {\citep{muzerolle1998,natta2004}}.

We estimate the accretion luminosity ($L_\mathrm{acc}$ in units of L$_\odot$) for YSOs in CMa-$\ell$224 by measuring the flux of the hydrogen Br$\gamma$ line. {As illustrated in Figure 3, the Br$\gamma$ emission arises from the same area as the continuum source; we therefore assume it traces primarily the unresolved accretion region (see also Section \ref{subseubsec:brg}). 
We use the relation between $L_{\mathrm{Br}\gamma}$ and $L_{\mathrm{acc}}$ for low- and intermediate-mass YSOs from  \citep{alcala2017} to estimate $L_\mathrm{acc}$}:

\begin{equation}
\label{eq:lacc}
\log{(L_{acc})} = (1.19\pm0.10)\cdot\log{L_{Br\gamma}} + (4.02\pm0.51)
\end{equation}
The mass accretion rates, $\dot{M}_{acc}$, are then calculated using the formula:
\begin{equation}
\label{eq:macc}
\dot{M}_{acc} = L_{acc} \times \left(1-\frac{R_*}{R_\mathrm{in}}\right)^{-1} \frac{R_*}{GM_*} 
\end{equation}
where  $R_*$ and $M_*$ are the stellar radius and mass, respectively, and $R_{\mathrm{in}}$ is the inner-disk radius; typically $R_{\mathrm{in}}$ is assumed to equal 5$R_*$ \citep{gullbring1998}. {Although the relation was found for Class II objects, it is a specific case of the general formula describing efficiency of converting kinetic energy of falling matter onto a stellar photosphere into radiation and can be applied to Class I sources as well (see the review of \citealt{hartmann2016} or a recent paper of \citealt{fiorellino2022}).}
\begin{figure}
\includegraphics[width=\columnwidth]{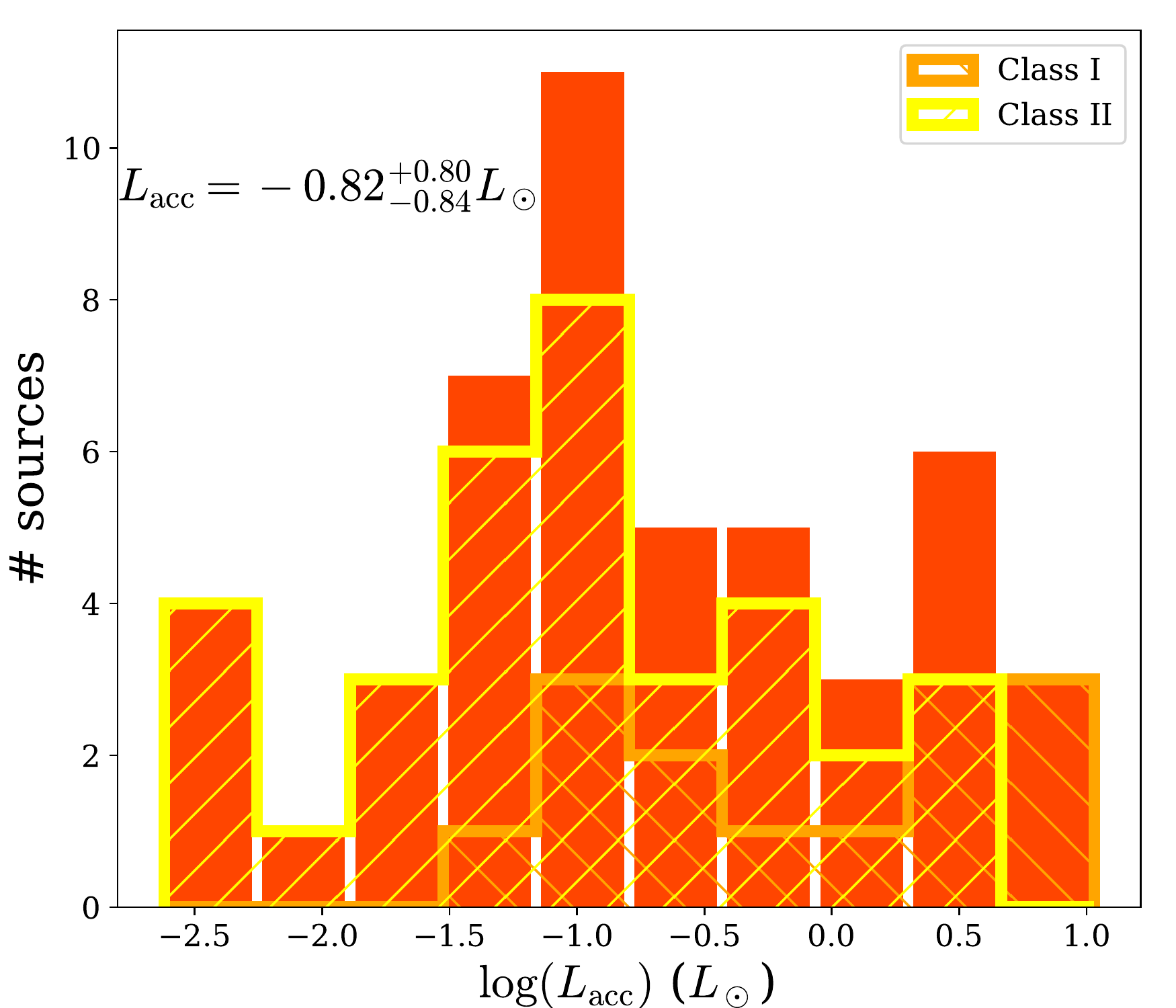}
\caption{Histogram of the accretion luminosities ($L_\mathrm{acc}$) based on Br$\gamma$ line luminosity (Table \ref{tab:acc-eject}) measured toward YSOs in CMa-$\ell$224 (in dark orange). The median value of $L_\mathrm{acc}$ is indicated at the top of the plot. The orange and yellow  hatched histograms show the distribution of $L_\mathrm{acc}$ for sources classified by \cite{sewilo2019} as, respectively, Class I and Class II YSO candidates.}
\label{fig:hist:lacc}
\end{figure}

{To calculate the mass accretion rates, knowledge of stellar masses and radii is needed. Stellar parameters for 17 sources from our sample have been estimated from the SED fitting\footnote{The spatial resolution of SED datapoints does not allow us to resolve visually multiple systems on the KMOS maps.
We therefore adopted the same stellar parameters for the two stars in a binary candidate, source  55 \citep{sewilo2019}. Proper characterization of binaries is beyond the scope of this paper.}
\citep{sewilo2019}; only half of them are associated with the Br$\gamma$ emission. The typical masses are $\sim2~M_\odot$ and radii are $\sim3~R_\odot$.} {The median logarithm of the accretion luminosity is -0.82$_{-0.84}^{+0.80}$
~$L_\odot$} 
and the median {logarithm of the} accretion rate is {-7.92$^{+1.06}_{-1.09}$
~M$_\odot$yr$^{-1}$ for sources with $A_V$ estimates}. 


\startlongtable
\begin{deluxetable}{lccchh}
\tabletypesize{\normalsize}
\tablecaption{Accretion properties of YSOs in CMa-$\ell$224. \label{tab:acc-eject}}
\tablehead{ 
\colhead{No.} & 
\colhead{Br$\gamma$} & 
\colhead{$\log{(L_{\rm{acc}})}$} & \colhead{$\log{(\dot{M}_{\mathrm{acc}})}$} & \nocolhead{$\log{(M_{\rm{out}})}$} & \nocolhead{$\log{(\dot{M}_{\rm{out}})}$} \\
\colhead{~} & 
\colhead{10$^{-16}$ (erg s$^{-1}$ cm$^{-2}$)} & \colhead{(L$_\odot$)} & \colhead{(M$_\odot$yr$^{-1}$)} & \nocolhead{(M$_\odot$)} & \nocolhead{(M$_\odot$yr$^{-1}$)} 
}
\startdata
2 & 13.00 $\pm$ 0.32 & -1.30$_{-0.82}^{+0.76}$ & \nodata & -6.07$_{-0.12}^{+0.07}$ & -8.56$_{-0.15}^{+0.10}$ \\
3 & 13.89 $\pm$ 0.32 & -1.26$_{-0.73}^{+0.72}$ & \nodata & \nodata & \nodata \\
5 & -397.29 $\pm$ 0.55 & \nodata & \nodata & \nodata & \nodata \\
6 & 19.08 $\pm$ 0.25 & -1.09$_{-0.81}^{+0.79}$ & \nodata & \nodata & \nodata \\
9 & 341.80 $\pm$ 0.79 & 0.40$_{-0.86}^{+0.86}$ & \nodata & -7.98$_{-0.17}^{+0.05}$ & -10.15$_{-0.27}^{+0.13}$ \\
10 & 54.74 $\pm$ 0.45 & -0.55$_{-0.86}^{+0.84}$ & \nodata & \nodata & \nodata \\
12 & 352.30 $\pm$ 1.35 & 0.41$_{-0.90}^{+0.89}$ & \nodata & \nodata & \nodata \\
13 & 28.33 $\pm$ 0.77 & -0.89$_{-0.80}^{+0.79}$ & \nodata & -7.45$_{-0.10}^{+0.03}$ & -10.17$_{-0.32}^{+0.18}$ \\
15 & 521.01 $\pm$ 0.35 & 0.62$_{-0.88}^{+0.87}$ & \nodata & -6.37$_{-0.15}^{+0.03}$ & -8.42$_{-0.30}^{+0.14}$ \\
16 & 108.77 $\pm$ 1.57 & -0.19$_{-0.87}^{+0.85}$ & -7.50$_{-1.07}^{+1.05}$ & \nodata & \nodata \\
18* & 1.52 $\pm$ 0.34 & -2.40$_{-0.10}^{+0.08}$ & \nodata & \nodata & \nodata \\
19 & 68.62 $\pm$ 0.21 & -0.43$_{-0.84}^{+0.83}$ & \nodata & -5.90$_{-0.14}^{+0.02}$ & -8.32$_{-0.25}^{+0.11}$ \\
20 & 12.93 $\pm$ 0.85 & -1.29$_{-0.84}^{+0.80}$ & \nodata & \nodata & \nodata \\
22* & 1.63 $\pm$ 0.56 & -2.36$_{-0.19}^{+0.13}$ & \nodata & \nodata & \nodata \\
23* & 4.91 $\pm$ 1.36 & -1.79$_{-0.19}^{+0.15}$ & \nodata & \nodata & \nodata \\
24 & 100.01 $\pm$ 0.29 & -0.24$_{-0.80}^{+0.80}$ & \nodata & \nodata & \nodata \\
26 & 18.52 $\pm$ 0.33 & -1.11$_{-0.93}^{+0.85}$ & \nodata & \nodata & \nodata \\
28 & 33.00 $\pm$ 0.60 & -0.81$_{-0.75}^{+0.74}$ & \nodata & \nodata & \nodata \\
30 & 1.00 $\pm$ 0.20 & -2.62$_{-0.74}^{+0.69}$ & \nodata & \nodata & \nodata \\
31* & 1.16 $\pm$ 0.16 & -2.54$_{-0.03}^{+0.02}$ & \nodata & \nodata & \nodata \\
32 & 27.83 $\pm$ 3.14 & -0.90$_{-0.86}^{+0.82}$ & \nodata & -9.32$_{-0.11}^{+0.07}$ & -11.27$_{-0.19}^{+0.14}$ \\
33 & 17.17 $\pm$ 0.31 & -1.15$_{-0.82}^{+0.80}$ & \nodata & -8.57$_{-0.15}^{+0.06}$ & -10.99$_{-0.38}^{+0.21}$ \\
34* & 4.82 $\pm$ 0.34 & -1.80$_{-0.06}^{+0.06}$ & \nodata & \nodata & \nodata \\
35 & -20.72 $\pm$ 0.54 & \nodata & \nodata & -8.73$_{-0.14}^{+0.05}$ & -10.99$_{-0.59}^{+0.26}$ \\
36A* & 65.58 $\pm$ 2.41 & -0.46$_{-0.15}^{+0.15}$ & -7.69$_{-0.35}^{+0.35}$ & \nodata & \nodata \\
38A* & 2.36 $\pm$ 0.35 & -2.17$_{-0.07}^{+0.06}$ & \nodata & \nodata & \nodata \\
38B* & 0.36 $\pm$ 0.09 & -3.15$_{-0.06}^{+0.03}$ & \nodata & \nodata & \nodata \\
39 & 33.48 $\pm$ 3.62 & -0.80$_{-0.95}^{+0.88}$ & -8.10$_{-1.15}^{+1.08}$ & -7.39$_{-0.11}^{+0.06}$ & -9.66$_{-0.32}^{+0.20}$ \\
40 & -2119.07 $\pm$ 12.85 & \nodata & \nodata & \nodata & \nodata \\
41* & 10.11 $\pm$ 0.59 & -1.42$_{-0.08}^{+0.08}$ & \nodata & \nodata & \nodata \\
43 & -1.29 $\pm$ 0.27 & \nodata & \nodata & \nodata & \nodata \\
45 & 4.54 $\pm$ 0.27 & -1.83$_{-1.09}^{+0.87}$ & \nodata & \nodata & \nodata \\
46 & 307.57 $\pm$ 5.92 & 0.34$_{-0.87}^{+0.86}$ & -6.97$_{-1.07}^{+1.06}$ & -6.39$_{-0.12}^{+0.05}$ & -8.60$_{-0.20}^{+0.12}$ \\
47 & 373.10 $\pm$ 0.37 & 0.44$_{-0.89}^{+0.86}$ & \nodata & -4.52$_{-0.13}^{+0.03}$ & -6.70$_{-0.17}^{+0.06}$ \\
48 & 1.37 $\pm$ 0.33 & -2.45$_{-0.80}^{+0.73}$ & \nodata & \nodata & \nodata \\
49 & 119.13 $\pm$ 0.60 & -0.15$_{-0.90}^{+0.87}$ & \nodata & -7.09$_{-0.21}^{+0.09}$ & -9.53$_{-0.55}^{+0.27}$ \\
50 & 94.51 $\pm$ 0.52 & -0.27$_{-0.78}^{+0.78}$ & \nodata & \nodata & \nodata \\
52A* & 2.38 $\pm$ 0.45 & -2.17$_{-0.10}^{+0.08}$ & \nodata & \nodata & \nodata \\
52B* & 1.56 $\pm$ 0.50 & -2.39$_{-0.17}^{+0.12}$ & \nodata & \nodata & \nodata \\
53 & -107.15 $\pm$ 0.76 & \nodata & \nodata & \nodata & \nodata \\
54* & 493.43 $\pm$ 19.10 & 0.59$_{-0.24}^{+0.24}$ & \nodata & -6.32$_{-0.09}^{+0.08}$ & -8.34$_{-0.18}^{+0.15}$ \\
55A & 9.30 $\pm$ 1.35 & -1.46$_{-0.79}^{+0.76}$ & -8.41$_{-0.99}^{+0.96}$ & \nodata & \nodata \\
56 & 19.69 $\pm$ 0.40 & -1.08$_{-0.86}^{+0.82}$ & \nodata & \nodata & \nodata \\
57* & 3.87 $\pm$ 0.93 & -1.92$_{-0.15}^{+0.12}$ & \nodata & \nodata & \nodata \\
58 & 7.93 $\pm$ 1.03 & -1.55$_{-0.79}^{+0.76}$ & \nodata & \nodata & \nodata \\
60 & 1163.21 $\pm$ 1.90 & 1.03$_{-0.87}^{+0.85}$ & \nodata & -5.55$_{-0.10}^{+0.03}$ & -8.06$_{-0.21}^{+0.11}$ \\
61 & -3.22 $\pm$ 0.45 & \nodata & \nodata & \nodata & \nodata \\
63 & -34.89 $\pm$ 0.53 & \nodata & \nodata & \nodata & \nodata \\
64 & 413.58 $\pm$ 0.40 & 0.50$_{-0.71}^{+0.71}$ & \nodata & -7.40$_{-0.12}^{+0.08}$ & -9.74$_{-0.24}^{+0.17}$ \\
68 & 2.63 $\pm$ 0.39 & -2.12$_{-0.83}^{+0.76}$ & \nodata & \nodata & \nodata \\
71 & -69.23 $\pm$ 0.25 & \nodata & \nodata & -6.28$_{-0.15}^{+0.03}$ & -8.09$_{-0.25}^{+0.11}$ \\
72* & 1.66 $\pm$ 0.58 & -2.36$_{-0.20}^{+0.13}$ & \nodata & -8.53$_{-0.09}^{+0.08}$ & -11.22$_{-0.22}^{+0.18}$ \\
73 & 44.53 $\pm$ 0.30 & -0.66$_{-0.81}^{+0.80}$ & \nodata & -8.64$_{-0.15}^{+0.04}$ & -11.33$_{-0.54}^{+0.24}$ \\
75 & -58.02 $\pm$ 0.27 & \nodata & \nodata & \nodata & \nodata \\
76 & 30.09 $\pm$ 0.38 & -0.86$_{-0.84}^{+0.81}$ & \nodata & \nodata & \nodata \\
77 & 276.96 $\pm$ 0.21 & 0.29$_{-0.85}^{+0.85}$ & \nodata & -8.43$_{-0.14}^{+0.02}$ & -11.16$_{-0.56}^{+0.56}$ \\ 
79 & 9.33 $\pm$ 0.31 & -1.46$_{-0.77}^{+0.75}$ & -8.48$_{-0.97}^{+0.95}$ & \nodata & \nodata \\
80* & 2.65 $\pm$ 0.73 & -2.11$_{-0.16}^{+0.12}$ & \nodata & -8.68$_{-0.10}^{+0.08}$ & -10.95$_{-0.21}^{+0.17}$ \\
81 & 65.21 $\pm$ 0.82 & -0.46$_{-0.90}^{+0.87}$ & -7.74$_{-1.10}^{+1.07}$ & \nodata & \nodata \\
83 & 625.84 $\pm$ 0.35 & 0.71$_{-0.84}^{+0.78}$ & \nodata & -8.38$_{-0.11}^{+0.05}$ & -10.60$_{-0.25}^{+0.15}$ \\
84 & -109.85 $\pm$ 0.87 & \nodata & \nodata & \nodata & \nodata \\
85* & 8.34 $\pm$ 0.50 & -1.52$_{-0.08}^{+0.07}$ & \nodata & \nodata & \nodata \\
86B & 1.10 $\pm$ 0.14 & -2.57$_{-0.72}^{+0.68}$ & \nodata & \nodata & \nodata \\
89A* & 0.81 $\pm$ 0.23 & -2.72$_{-0.12}^{+0.07}$ & \nodata & \nodata & \nodata \\
89B* & 0.24 $\pm$ 0.08 & -3.36$_{-0.10}^{+0.04}$ & \nodata & \nodata & \nodata \\
91A* & -6.12 $\pm$ 1.11 & \nodata & \nodata & \nodata & \nodata \\
91B* & 6.90 $\pm$ 0.50 & -1.62$_{-0.08}^{+0.07}$ & \nodata & -7.25$_{-0.09}^{+0.08}$ & -9.87$_{-0.53}^{+0.29}$ \\
92 & -14.75 $\pm$ 0.51 & \nodata & \nodata & \nodata & \nodata \\
93 & 115.79 $\pm$ 0.52 & -0.16$_{-0.84}^{+0.83}$ & \nodata & \nodata & \nodata \\
94 & -3.35 $\pm$ 0.25 & \nodata & \nodata & \nodata & \nodata \\
96 & -56.89 $\pm$ 0.71 & \nodata & \nodata & \nodata & \nodata \\
97 & 29.29 $\pm$ 1.65 & -0.87$_{-0.89}^{+0.84}$ & \nodata & \nodata & \nodata \\
98 & 4.91 $\pm$ 0.42 & -1.80$_{-0.75}^{+0.73}$ & \nodata & \nodata & \nodata \\
99* & 1.42 $\pm$ 0.25 & -2.44$_{-0.07}^{+0.05}$ & \nodata & \nodata & \nodata \\
102 & 273.94 $\pm$ 3.28 & 0.28$_{-0.85}^{+0.84}$ & \nodata & \nodata & \nodata \\
104 & -6.90 $\pm$ 0.81 & \nodata & \nodata & \nodata & \nodata \\
109 & 716.77 $\pm$ 0.24 & 0.78$_{-0.73}^{+0.73}$ & \nodata & -6.59$_{-0.13}^{+0.06}$ & -9.20$_{-0.73}^{+0.30}$ \\
111 & 51.99 $\pm$ 0.37 & -0.58$_{-0.76}^{+0.76}$ & \nodata & -6.91$_{-0.33}^{+0.32}$ & -9.39$_{-0.66}^{+0.50}$ \\
112A & 9.55 $\pm$ 0.16 & -1.45$_{-0.72}^{+0.71}$ & \nodata & \nodata & \nodata \\
112B & 30.86 $\pm$ 0.14 & -0.84$_{-0.76}^{+0.76}$ & \nodata & \nodata & \nodata \\
114 & 187.72 $\pm$ 0.76 & 0.09$_{-1.08}^{+0.97}$ & -7.06$_{-1.28}^{+1.17}$ & -5.68$_{-0.20}^{+0.07}$ & -7.69$_{-8.25}^{+8.25}$ \\ 
116 & -3.91 $\pm$ 0.28 & \nodata & \nodata & \nodata & \nodata \\
117 & -4.64 $\pm$ 0.51 & \nodata & \nodata & \nodata & \nodata \\
118 & 9.01 $\pm$ 0.43 & -1.48$_{-0.78}^{+0.76}$ & \nodata & \nodata & \nodata \\
\enddata
\tablecomments{Negative flux values indicate absorption. Measurements for targets with asterisk have not been corrected for extinction. Flux uncertainties do not cover the extinction uncertainty of 10~mag, which is included in uncertainties of accretion luminosity and mass accretion rates. Uncertainties of $\log{(L_{\mathrm{acc}})}$ and $\log{(\dot{M}_{\mathrm{acc}})}$ do not include the systematic distance uncertainty.} 
\end{deluxetable}


All accretion luminosities and, where available, mass accretion rates {and associated uncertainties} are listed in Table \ref{tab:acc-eject}. {The uncertainties of mass accretion rates 
account for the uncertainties in 
line fluxes, $L_{\rm Br\gamma}$--$L_{\rm acc}$ relation, extinction ($A_\mathrm{V}$ of 10~mag, corresponding to $\sim$1~mag in $K$-band), and 0.1~dex for both M$_*$ and R$_*$.} 
We do not include, however, the uncertainty of the distance to the region, which is systematic.
The adopted value of 0.92 kpc from \cite{sewilo2019} is consistent with those from stellar studies 
\citep[1.05$\pm$0.15~kpc, 0.99$\pm$0.05~kpc, 1.15$\pm$0.06~kpc,][respectively]{shevchenko1999,kaltcheva2000,lombardi2011}, and recent estimates using ALMA observations \citep[a median of 0.92~kpc for cores within our pointings,][]{olmi2023}.

We also cross-checked the distance to CMa-l224 with the results from the 3D dust maps using photometry from Pan-STARRS 1 and 2MASS, as well as the Gaia parallaxes\footnote{The interactive web interface: \url{http://argonaut.skymaps.info/}} \citep{green2019}. The median distance  estimated this way for several targets in our study yields 1.26~kpc. The value  exceeds our adopted distance of 0.92 kpc, with the difference that is larger than uncertainties of previous studies and the dispersion of individual distance values. Thus, we consider 0.34 kpc as the uncertainty of the distance; it corresponds to the uncertainty of 0.27 dex in accretion luminosity and mass accretion rate.

\begin{figure*}
\includegraphics[width=\textwidth, height=0.45\textheight]{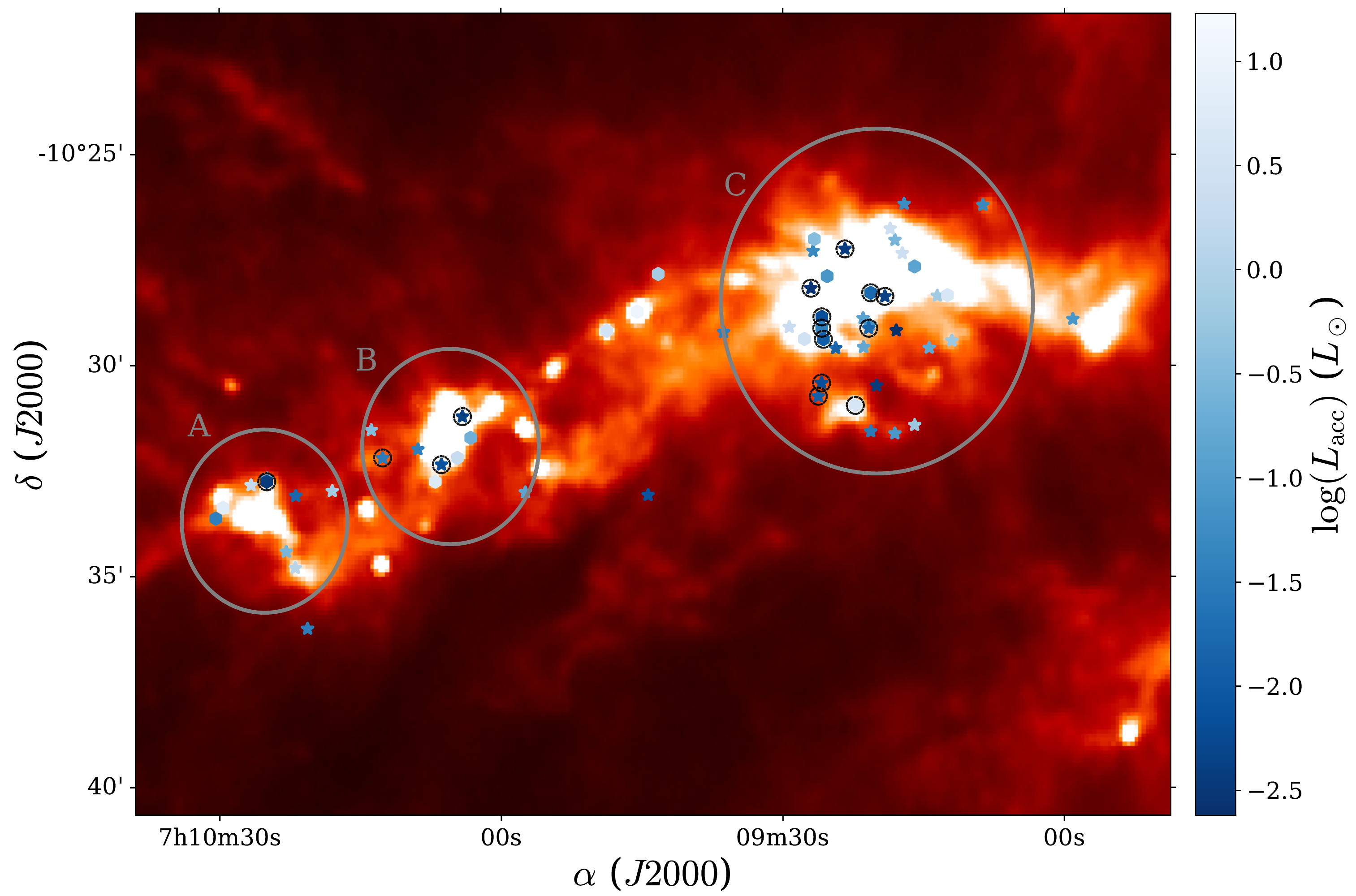} 
\caption{Distribution of accretion luminosity in the main filament in CMa-$\ell$224 (blue color scale). Class I sources are marked with hexagon and Class II sources with star symbols. { Black circles mark targets without $A_{\rm{V}}$ estimation.} Marked with grey circles are three regions identified by \cite{sewilo2019}.}
\label{fig:Lacc-sky}
\end{figure*}

Figure \ref{fig:hist:lacc} shows the distribution of accretion luminosities among Class I and II YSOs, which range from { -2.5 to 0.6 $L_{\odot}$ in logarithm.} 
The distributions of $L_{\mathrm{acc}}$ are similar for both Class I and II YSOs, 
but the number of Class II YSOs in this sample is two times higher than Class I YSOs. We do not find any correlation between accretion luminosity and stellar mass, but the number of mass measurements is too small { and the mass measurements are too uncertain} 
to draw reliable conclusions. Accretion rate estimations are feasible {only} for {8} Class~II YSOs (including one binary candidate), with available mass and radius estimates; their {logarithmic} values range from {-8.4 to -7.0 
~M$_\odot$yr$^{-1}$}.

\section{Discussion} \label{sec:dis}

\subsection{Star Formation in CMa-$\ell$224}
\label{subsec:SF}

Star formation in Canis Majoris is closely linked with a supernova explosion $\sim$1.5 Myr ago \citep{herbst77,comeron98}. The expanding shell, coincident with CMa-$\ell$224, might have compressed the interstellar medium in a turbulent flow and triggered the filament formation \citep{fischer16,sewilo2019}. The main filament in the region is highly supercritical \citep{olmi16}, which leads to gravitational fragmentation and the formation of clumps and cores \citep{andre17}. Overall, the filament is associated with $\sim$700 starless cores and $\sim$250 prestellar cores \citep{elia2013}, and the largest { concentration} 
of Class I YSOs in the entire Canis Major star-forming region \citep{fischer16,sewilo2019}. 

Integral Field Spectroscopy in $K$-band at high-angular resolution provides unique insights into star formation processes on scales of individual YSOs in the main filament of CMa-$\ell$224. As shown in Section \ref{sec:res}, the compact emission in the Br$\gamma$ line traces the on-going accretion onto the young stars, whereas the spatial distribution and kinematics of H$_2$ pin-points molecular outflows and mass ejection. The question remains, to what extent the mass accretion and ejections rates in YSOs are linked with the properties of the filament and its possible evolution. 

Figure \ref{fig:Lacc-sky} shows the distribution of accretion luminosity in the main filament in CMa-$\ell$224 (Section \ref{subsec:acc}). YSOs are preferentially located within three \lq\lq clusters'' associated with bright far-IR continuum emission. Following the nomenclature introduced in \cite{sewilo2019}, cluster C is known to coincide also with C$^{18}$O 2-1 emission and the peak of $^{13}$CO 1-0, with the latter encompassing the entire filament \citep{olmi16}. The {$^{13}$CO and C$^{18}$O} line observations revealed a velocity gradient along the main filament, which extends from cluster C to A \citep{olmi16} and is likely responsible for the formation of YSOs at the ridge connecting clusters B and C. Another velocity gradient was identified from a secondary filament which is roughly perpendicular to the main one at the position of cluster B \citep{olmi16}. \cite{sewilo2019} suggest that this smaller filament has provided molecular gas reservoir necessary to initiate the star formation for cluster B, as evidenced by the most evolved YSO population in this cluster.  

To investigate this scenario, we calculate cumulative distributions of accretion luminosities, $L_{\rm{acc}}$, 
for the three clusters (Figure \ref{fig:cum-distr}). {The distributions of accretion luminosity are similar for clusters A and B, whereas cluster C seems to be shifted towards lower values. 
Observed similarity between the accretion distributions of clusters A and B is confirmed by the two-sample Kolmogorov–Smirnov (KS) test that yields statistic of  0.2 and $p$-value of 0.99}. 
{The KS test for the cluster A and C, and cluster B and C, yield 
statistic of 0.30--0.35 and $p$-value of 0.55-0.47, not allowing to statistically distinguish the two distributions. 
The median accretion luminosities also gradually increase from Cluster C (-0.89~$\log{(L_\odot)}$), to B (-0.56~$\log{(L_\odot)}$) and A (-0.37~$\log{(L_\odot)}$). Higher accretion rate is expected from less evolved stars, assuming similar masses. Thus, the evolutionary stage of Clusters A and B could be earlier than those of Cluster C, contrary to earlier works \citep{sewilo2019}.}

However, the distributions can be affected by the number of targets within the Clusters, i.e. the small number of statistics. 
Adopting the YSO classification from \cite{sewilo2019}, Cluster C contains 5 Class~I YSOs and 20 Class~II YSOs for which we calculated accretion luminosities. 
In Cluster A, we measured $L_{\rm{acc}}$ in 2 Class I and 6 Class II YSOs. 
Similarly, in Cluster B -- 3 Class I and 3 Class II YSOs have accretion estimates. 
Thus, Cluster C contains significantly more YSOs with measured accretion luminosities (25) than the two other clusters combined (14). As a result, it has the most representative sample of sources, characterized by a broad range of accretion  
properties.

\begin{figure}
\includegraphics[width=\columnwidth, trim={0 0.5cm 14.5cm 0}, clip]{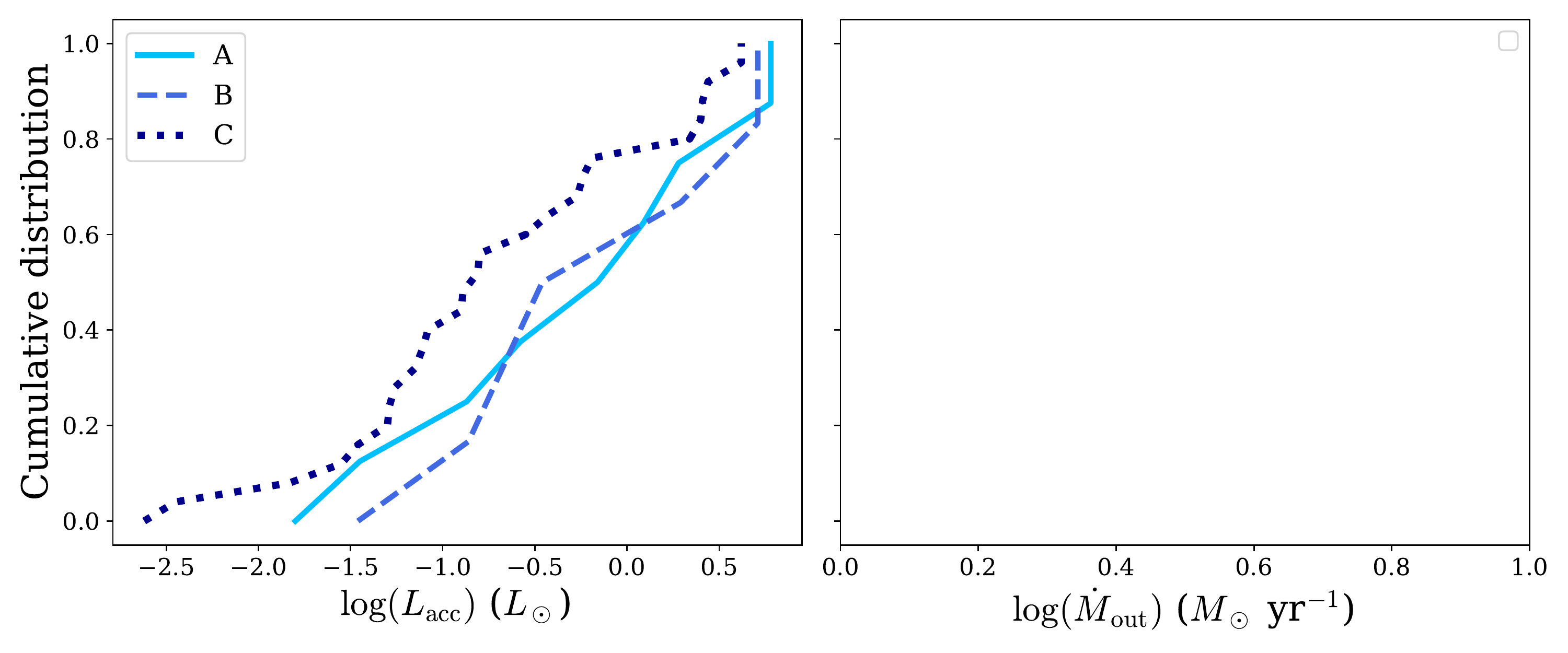}
\caption{Cumulative distributions of accretion luminosities  
for clusters A (solid), B (dashed) and C (dotted line), as defined in \cite{sewilo2019}. Sources without $A_{\rm{V}}$ estimates are excluded.}
\label{fig:cum-distr}
\end{figure}

\subsection{A Possible Impact of Low Metallicity}

Observations of star formation in external galaxies suggest that subsolar metallicity has a clear impact on the physical and chemical conditions in molecular clouds \citep[e.g.,][]{madden2013,roman2014}. In particular, a lower dust content in the low-metallicity environments reduces the shielding from ultraviolet radiation and translates to a decrease of molecular abundances due to photodissociation \citep[\lq\lq CO-dark gas'', e.g., ][]{wolfire2010}. The formation rate of molecules on the dust grains also decreases, which is the main reason for the reduction of H$_2$ abundances in the low-metallicity clouds \citep{glover2012}. 

The Large and Small Magellanic Clouds, with Z of $\sim$0.3-0.5~Z$_{\odot}$ \citep[LMC;][]{russell1992,westerlund1997,rolleston2002} and 0.2 Z$_{\odot}$ \citep[SMC;][]{russell1992}, are routinely used as laboratories for studying star formation in low-metallicity environments. Observations show a decrease of molecular gas cooling in YSO envelopes, {UV radiation penetrating deeper into the clouds}, and an increase in average dust temperatures with respect to those in local star forming regions \citep[e.g.,][]{vanLoon2010a,vanLoon2010b,oliveira2019}. However, 
the distances of 50.0$\pm$1.1 kpc (LMC, \citealt{pietrzyn2013}) and 62.1$\pm$2.0 kpc (SMC, \citealt{graczyk2014}) {hinder characterisation of} individual low-mass YSOs. 
\begin{figure}
\includegraphics[width=\columnwidth]{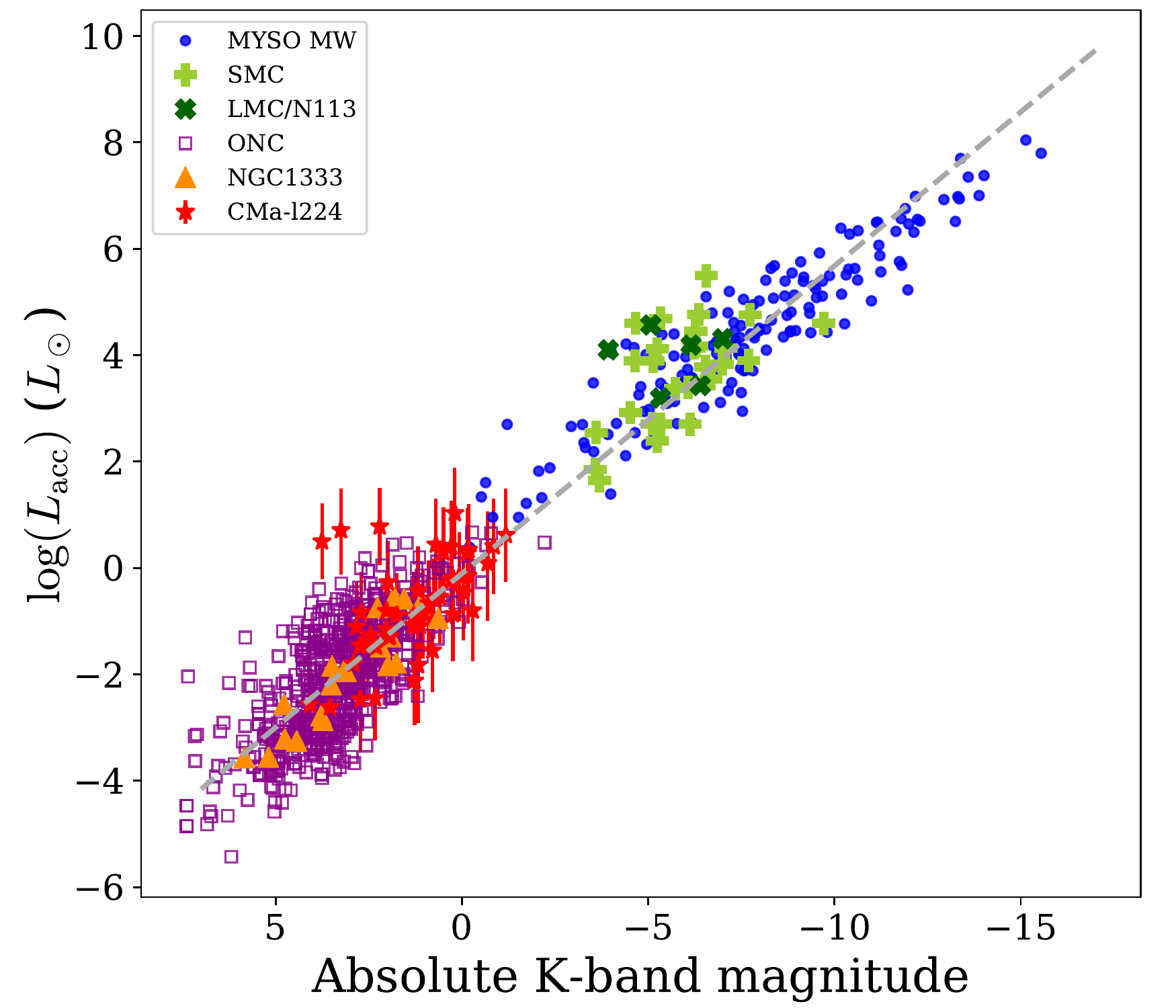}\hspace{0.0cm}
\caption{Accretion luminosity versus absolute $K$-band magnitude. Red stars show YSOs in CMa-$\ell$224, blue circles show MYSOs \citep{cooper2013}, light-green \lq\lq +'' symbols show MYSOs from SMC \citep{ward2017}, green crosses show MYSOs from LMC \citep{ward2016}, violet open squares show YSOs in Orion \citep{manara2012}, and orange triangles show low-mass YSOs in Perseus \citep{fiorellino2021}. The dashed lines show linear fits separately to each of the datasets with the same colors. The grey line show a linear fit to all datapoints. All luminosities have been corrected for extinction.}
\label{fig:Lacc-K}
\end{figure}

\begin{figure}
\includegraphics[width=\columnwidth]{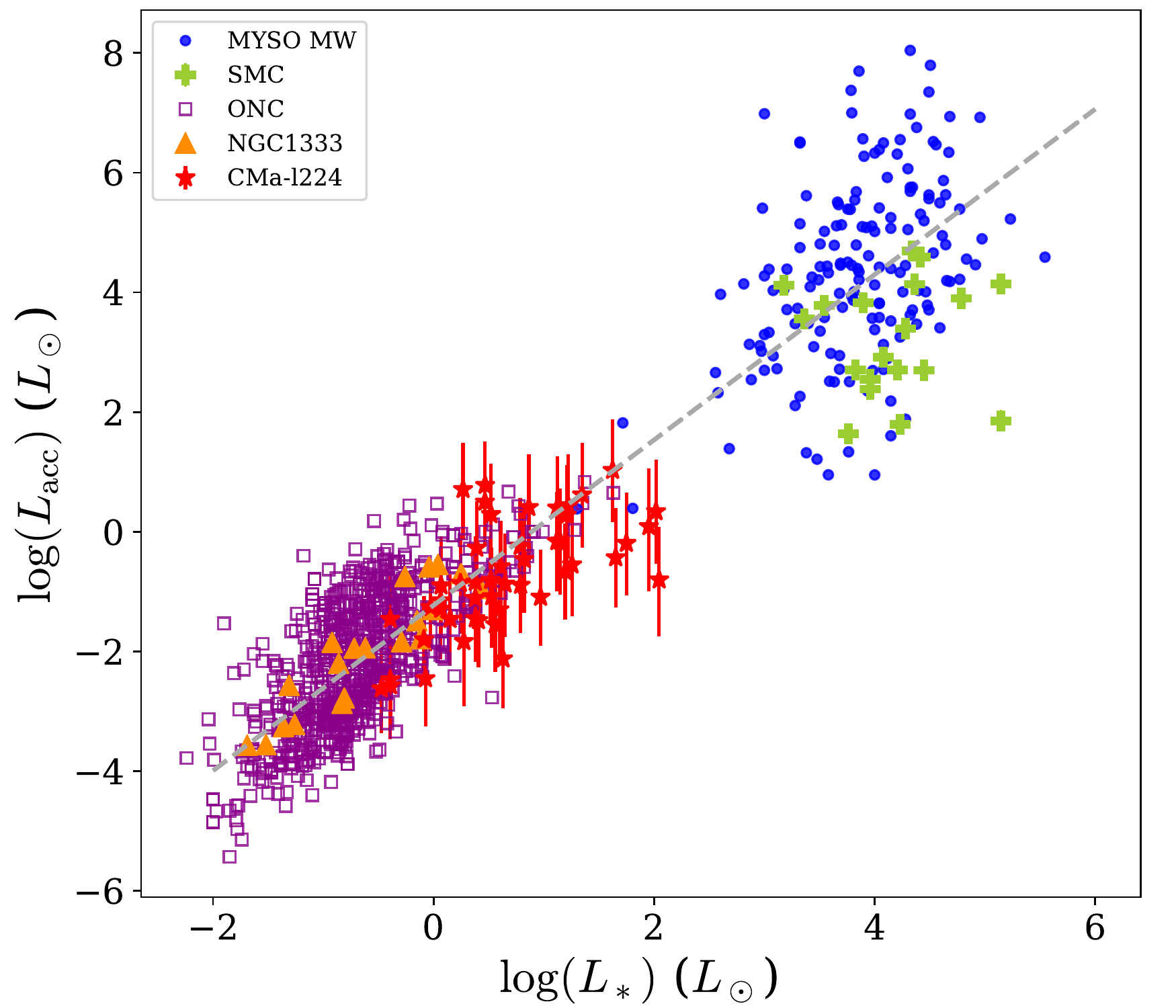}\hspace{0.cm}
\caption{Accretion luminosity versus star luminosity. The symbols are the same as in Fig. \ref{fig:Lacc-K}. The dashed line show a linear fit to the data. All luminosities have been corrected for extinction.}
\label{fig:Lacc-Lstar}
\end{figure}
The outer Galaxy also consists of star forming regions with the subsolar metallicities, and so offers the opportunity to study the impact of low metal content on star formation \citep{sodroski1997}. The metallicity of CMa-$\ell$224 has not been measured, {so} here we adopt the O/H Galactocentric radial gradients obtained from the observations of HII regions { accross the Milky Way}. Depending on the adopted survey, the metallicity of CMa-$\ell$224 would be 0.55-0.58~Z$_{\odot}$ \citep{esteban18}, 0.68 Z$_{\odot}$ \citep{fernandez2017}, or 0.67-0.73 Z$_{\odot}$ \citep{balser11}. The O/H and Fe/H gradients obtained using the observations of Cepheids suggest much higher metallicities of 1.11-1.17~Z$_{\odot}$ \citep{maciel2019} or 1.13-1.27 Z$_{\odot}$ \citep{luck2011}, yet they are not representative for star forming regions. Therefore, we assume that the metallicity of CMa-$\ell$224 is intermediate between the metallicity of the LMC and the metallicity in the Solar neighborhood.

A possible effect of metallicity on YSO accretion has been investigated in the LMC and SMC based on the near-IR observations of massive YSOs \citep{ward2016,ward2017}, {and optical observations with H$\alpha$ in YSOs with a wide range of masses \citep{deMarchi2011,deMarchi2017,biazzo2019}}. The Br$\gamma$ luminosities for Magellanic YSOs seem higher than those found in the Galactic high-mass YSOs from the RMS survey {for the same absolute magnitudes} \citep{cooper2013}. Here, we extend the measurements toward lower-mass YSOs, which are located in CMa-$\ell$224 and more nearby star-forming regions (Figs. \ref{fig:Lacc-K} and \ref{fig:Lacc-Lstar}). We adopt the same empirical relations between the Br$\gamma$ luminosity and accretion luminosities for various samples \citep{alcala2017}; however, $L_{\mathrm{acc}}$ from Orion has been calculated using H$\alpha$ and two-color diagrams \citep{manara2012}. The $K$-band magnitudes of YSOs from Orion are adopted from \cite{robberto2010}. 

Figures \ref{fig:Lacc-K} and \ref{fig:Lacc-Lstar} show the increase of $L_\mathrm{acc}$ with both the absolute $K$-band magnitude and $L_*$, with the YSOs from CMa-$\ell$224 located between the least massive YSOs in the NGC~1333 cluster in Perseus \citep{fiorellino2021} and in Orion \citep{manara2012}, and the high-mass YSOs both in the Milky Way \citep{cooper2013} and the Magellanic Clouds \citep{ward2016,ward2017}. {The Pearson coefficient of $|r|=0.96$ (30$\sigma$)} is obtained for the absolute $K$-band magnitude vs. $L_\mathrm{acc}$ relation,  reflecting a strong correlation (Fig. \ref{fig:Lacc-K}). Similarly, the Pearson coefficient of $|r|=0.93$ (30$\sigma$) characterises the $L_*$ vs $L_\mathrm{acc}$ relation.

Accretion luminosities in CMa-$\ell$224 are consistent with those in the Solar neighbourhood (e.g., Perseus, Orion) and do not show a clear enhancement of accretion luminosity expected in the low-metallicity environments \citep{ward2016,ward2017}. Most of the Br$\gamma$ emitters are however Class II sources, characterized by lower $L_{\mathrm{acc}}$ with respect to less evolved, Class I sources. 
{The metallicity of CMa-$\ell$224 \citep[$\sim$0.6-0.7 Z$_\odot$,][]{esteban18,fernandez2017,balser11} might be too similar to the Solar neighbourhood to exhibit a clear, measurable difference in accretion properties. } 
Follow-up observations of additional clouds in the outer Galaxies, spanning a broad range of metallicities {and sufficient number of sources}, would be critical to {draw robust conclusions about the impact of the lower metallicity on accretion rates in the Outer Galaxy.}

\section{Conclusions} \label{sec:sum}

We presented VLT/KMOS near-IR spectroscopy of YSOs in CMa-$\ell$224, illustrating the instrument capability to study low-mass star formation, and where applicable, to detect molecular outflows in the outer Galaxy. We summarise the findings as follows:

\begin{itemize}

    \item {The Br$\gamma$ line is detected in emission} toward 47\% and in absorption toward 13\% of the sources.
    The H$_2$ 1-0 S(1) line is detected {on-source} in 33 sources, including 27 sources where the emission is extended. The CO bandhead at 2.3~$\mu$m is detected in emission in 5\% of YSOs and in absorption in 60\% of YSOs.

    \item {The extent and velocity structures of H$_2$ emission suggest an origin of the emission in outflows and shocks. Follow-up observations are necessary, however, to estimate the extinction along the outflows.}

    \item {We find a gradual increase in accretion luminosities from Cluster C to A and B with median values of -0.89, -0.56, and -0.37 $\log{L_\odot}$, respectively. Cluster C might be the most evolved part of CMa-$\ell$224.} 
    
    \item Accretion luminosities  
    {do not show the impact of sub-Solar metallicity in the CMa-$\ell$224. 
    It is likely due to insufficient difference in metallicity between the region and
 the Solar neighbourhood.}

\end{itemize}

Large-scale galactic surveys, such as the Outer Galaxy High Resolution Survey \citep{colombo21}, are starting to provide a complete census of molecular clouds and star-forming filaments spanning a range of metallicities. IFU observations of YSOs in those regions, either with ground-based facilities or the James Webb Space Telescope, will be critical to confirm the impact of metallicity on low-mass star formation within our Galaxy.

\newpage
\acknowledgments
{The authors thank the referee for a careful reading of the manuscript.} 
DI, AK, MK, and NL acknowledge support from the First TEAM grant of the Foundation for Polish Science No. POIR.04.04.00-00-5D21/18-00. This work was supported by the Polish National Science Center grants 2014/15/B/ST9/02111 and 2016/21/D/ST9/01098. The material is based upon work supported by NASA under award number 80GSFC21M0002 (MS). This article has been supported by the Polish National Agency for Academic Exchange under Grant No. PPI/APM/2018/1/00036/U/001. DI was partly funded by the European Research Council (ERC) via the ERC Synergy Grant {\em ECOGAL} (grant 855130). 
The research of LEK is supported by a research grant (19127) from VILLUM FONDEN. GJH is supported by general grant 12173003 awarded by the National Natural Science Foundation of China. 
WRMR thanks the financial support from the Leiden Observatory. 
Based on observations made with ESO Telescopes at the La Silla Paranal Observatory under programme ID 0102.C-0914(A).

\vspace{5mm}
\facility{VLT(KMOS)}

\vspace{5mm}
\software{astropy\footnote{\url{http://www.astropy.org}} \citep{astropy:2013, astropy:2018},
~matplotlib\footnote{\url{https://matplotlib.org/}} \citep{matplotlib2007},
~numpy\footnote{\url{https://numpy.org/}} \citep{numpy2006, numpy2011}, 
~esorex\footnote{\url{https://www.eso.org/sci/software/cpl/esorex.html}},
~KARMA\footnote{\url{https://www.eso.org/sci/observing/phase2/SMGuidelines/KARMA.html}}
}

\bibliography{kmos2020-bib}{}
\bibliographystyle{aasjournal}

\appendix
\counterwithin{figure}{section}

\section{KMOS $K$-Band Continuum Maps} \label{app:cont}

Figure~\ref{fig:cont} shows the $K$-band continuum flux density maps for YSO candidates in CMa-$\ell$224 without stellar companions. Similar maps for the fields with the detection of two sources (the binary star candidates) are shown in Figure~\ref{fig:cont-double}. { Figure~\ref{fig:cont-int} shows the integrated continuum maps for YSO candidates with the very weak $K$-band continuum.}


\begin{figure*}[h!]
\centering
\includegraphics[width=0.2\textwidth]{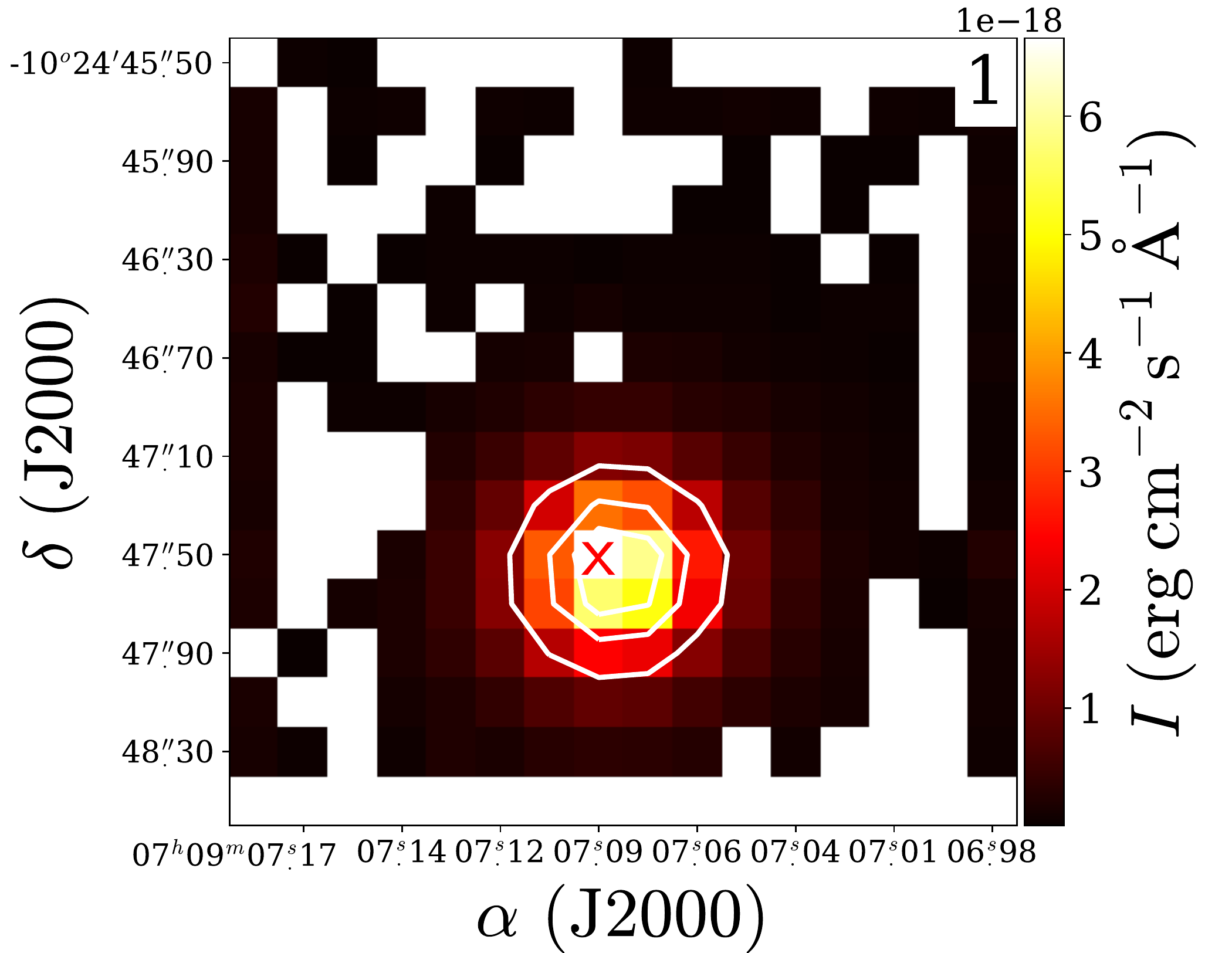}\hspace{-0.1cm}
\includegraphics[width=0.2\textwidth]{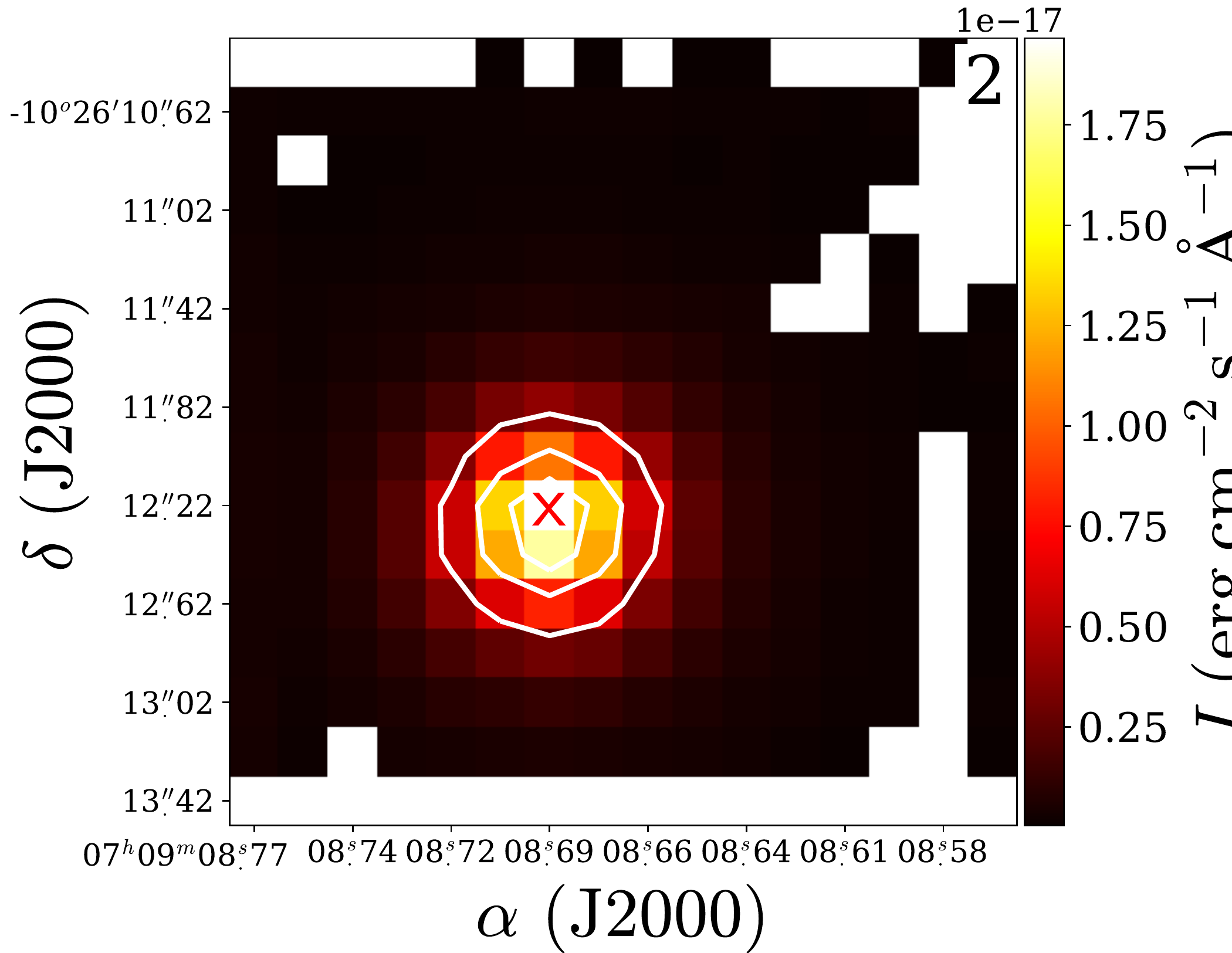}\hspace{-0.1cm}
\includegraphics[width=0.2\textwidth]{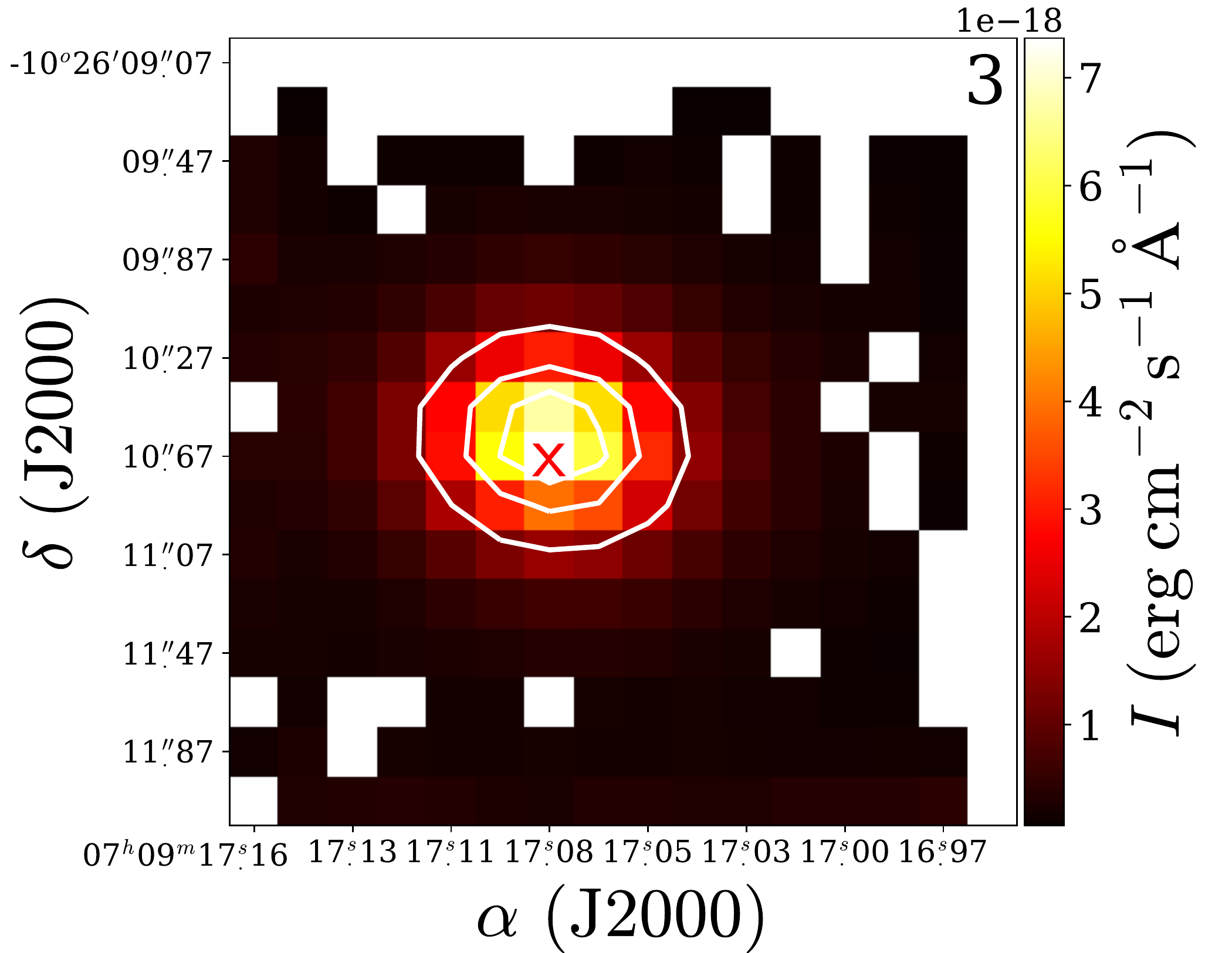}\hspace{-0.1cm}
\includegraphics[width=0.2\textwidth]{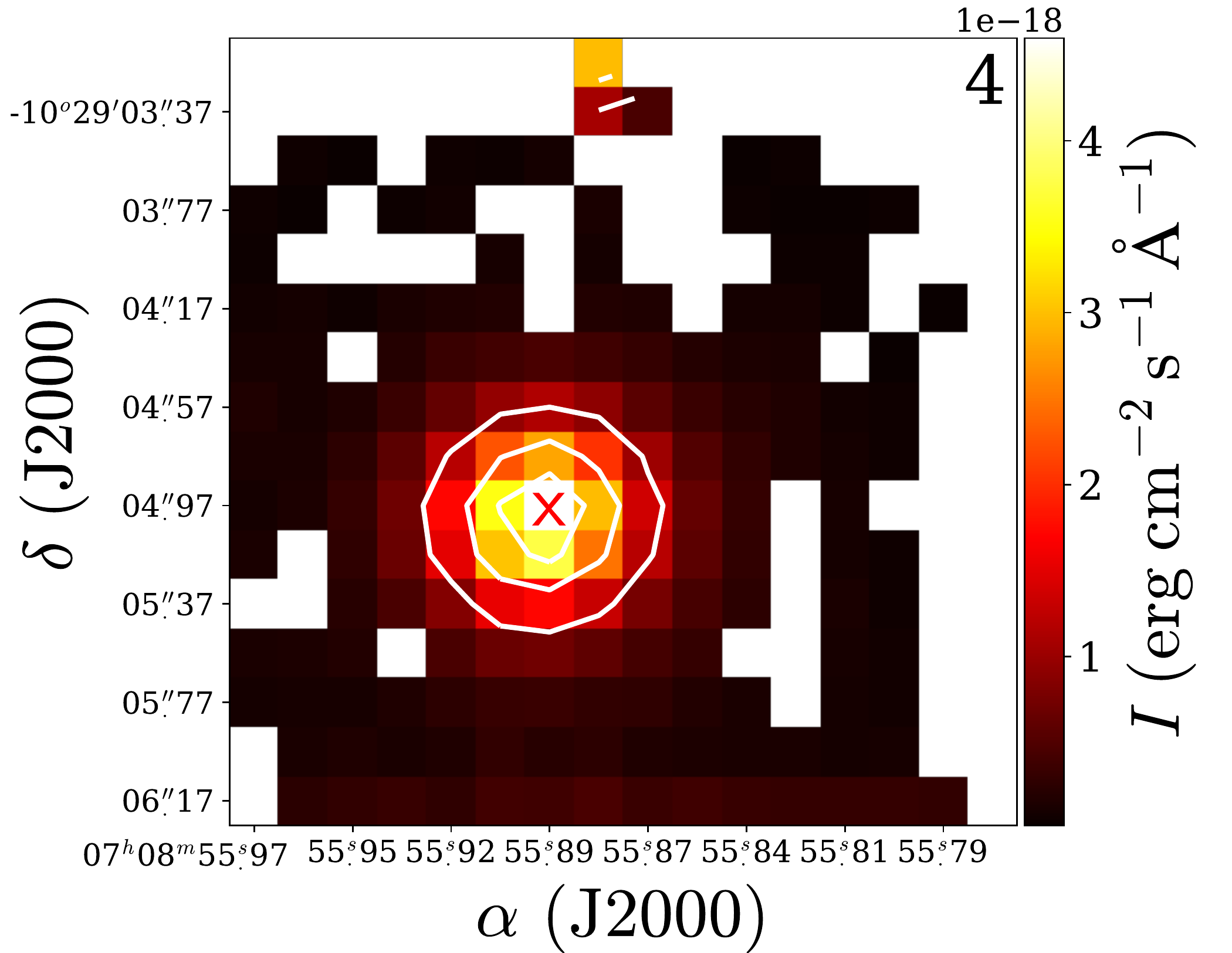}\hspace{-0.1cm}
\includegraphics[width=0.2\textwidth]{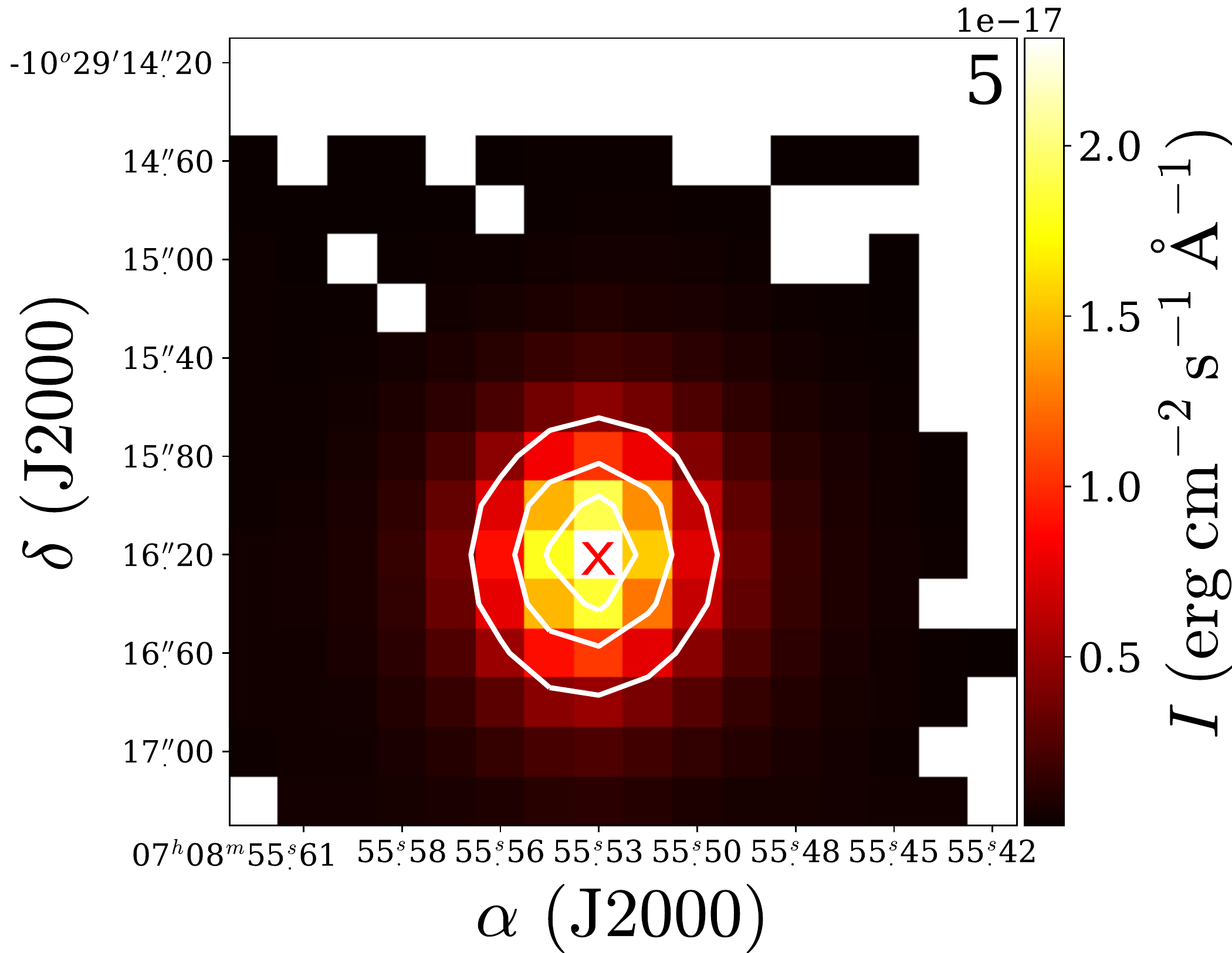}\hspace{-0.1cm}
\includegraphics[width=0.2\textwidth]{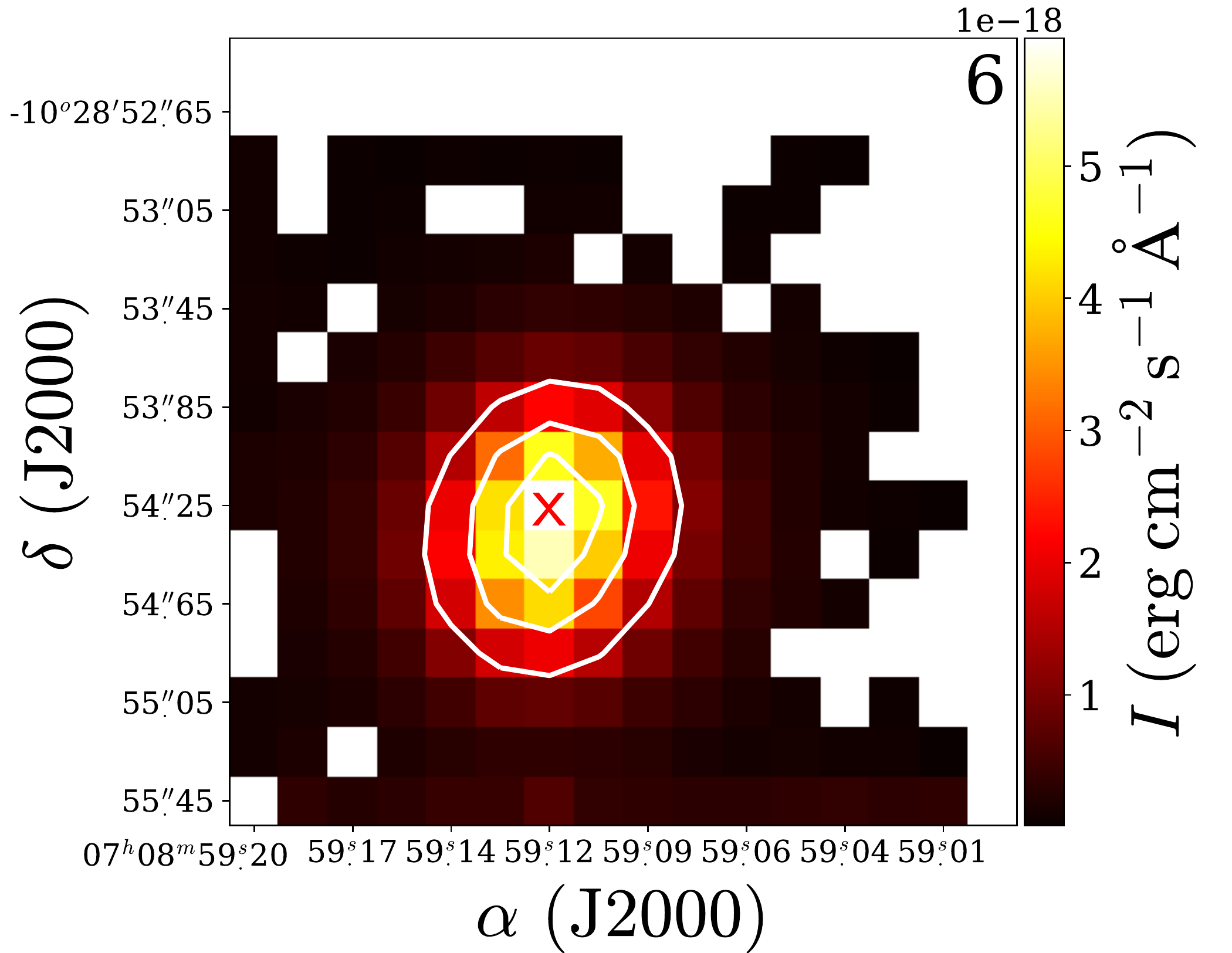}\hspace{-0.1cm}
\includegraphics[width=0.2\textwidth]{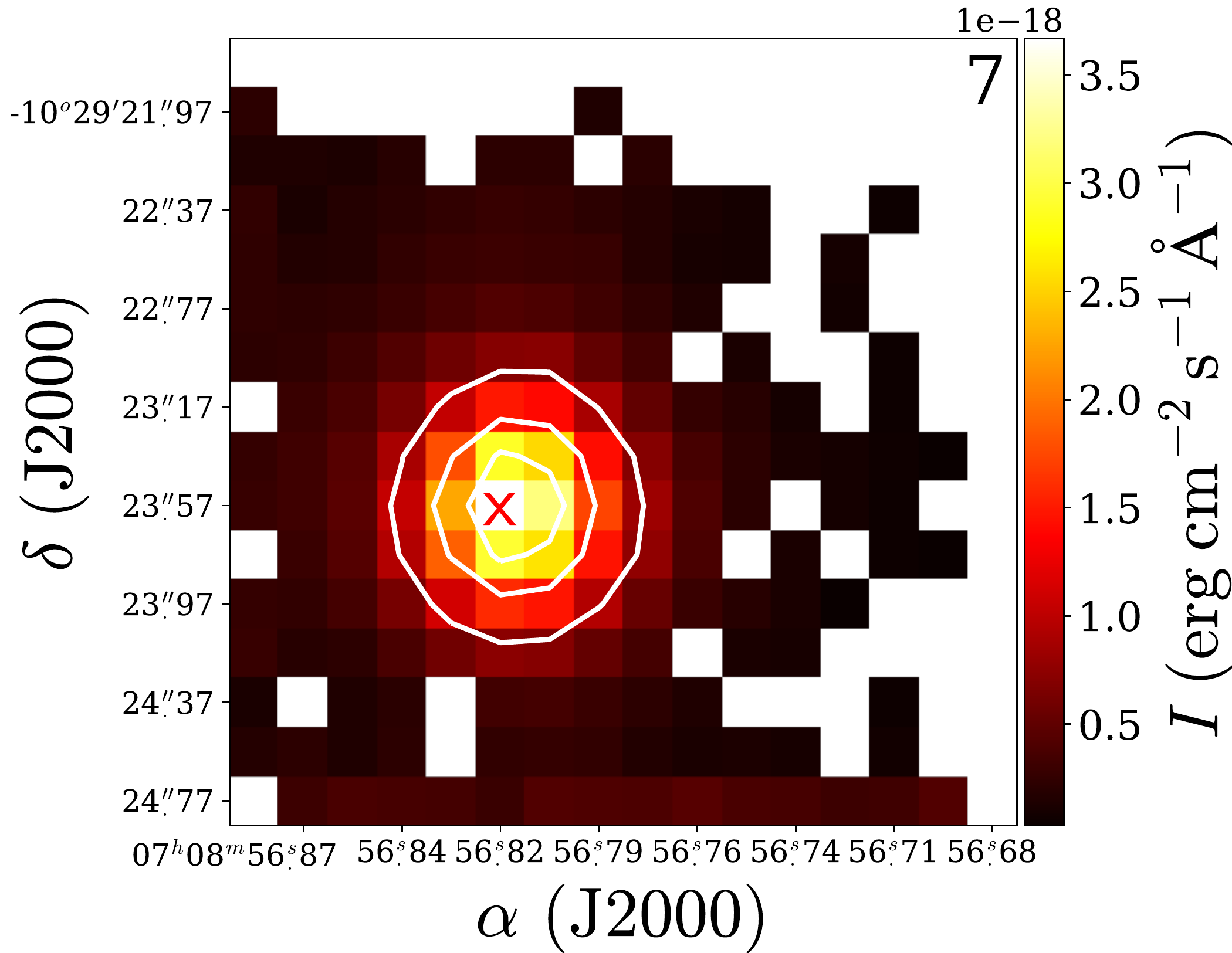}\hspace{-0.1cm}
\includegraphics[width=0.2\textwidth]{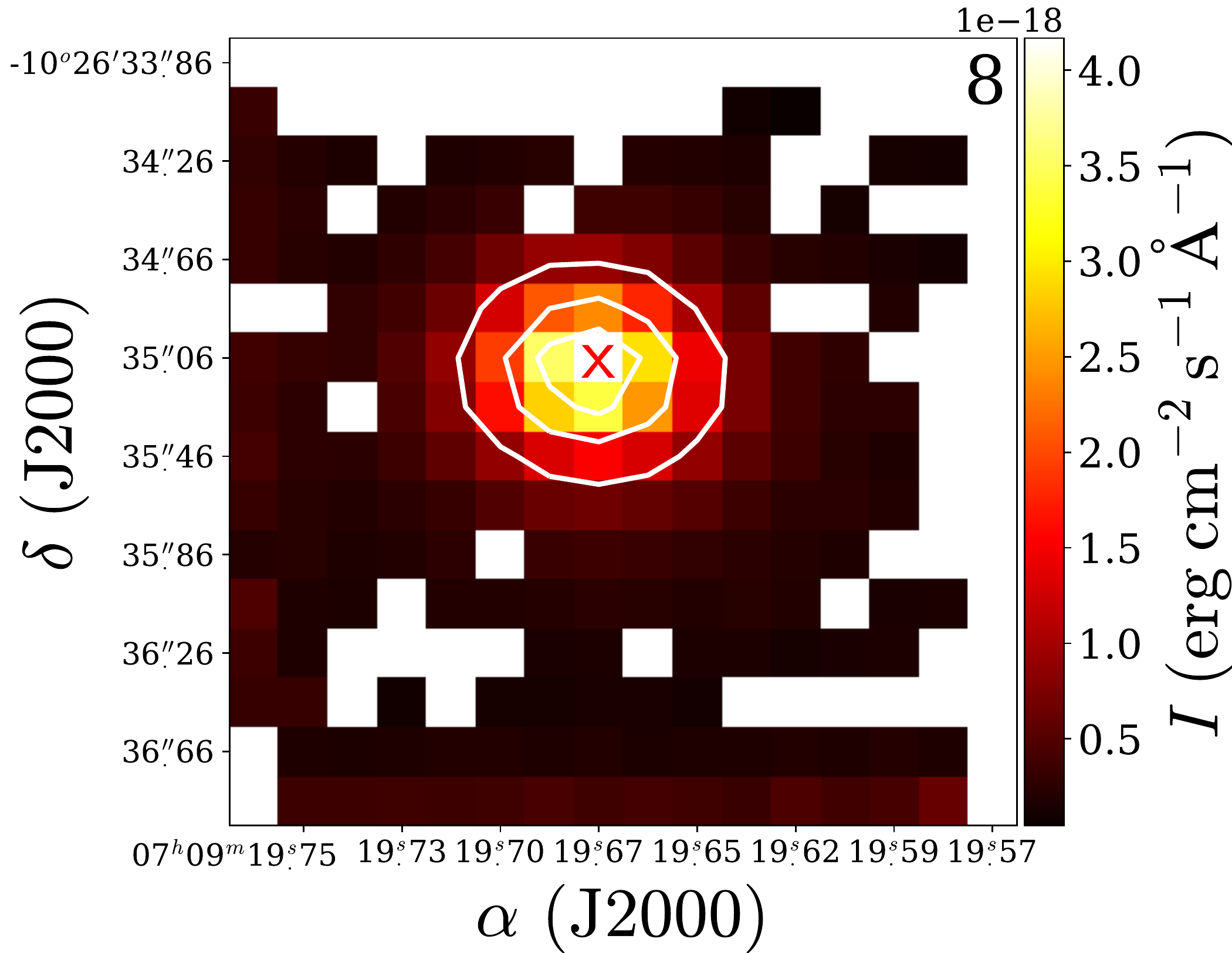}\hspace{-0.1cm}
\includegraphics[width=0.2\textwidth]{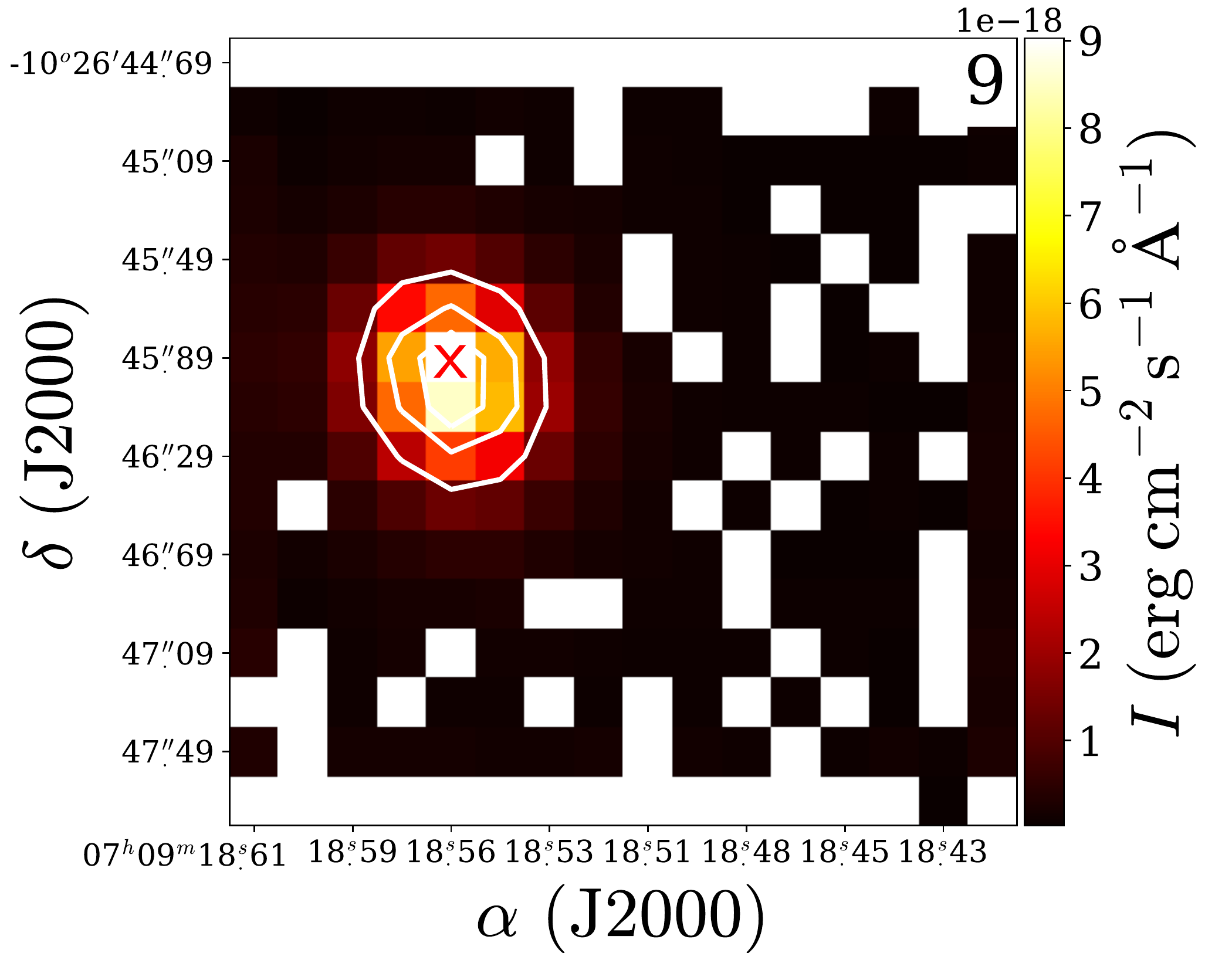}\hspace{-0.1cm}
\includegraphics[width=0.2\textwidth]{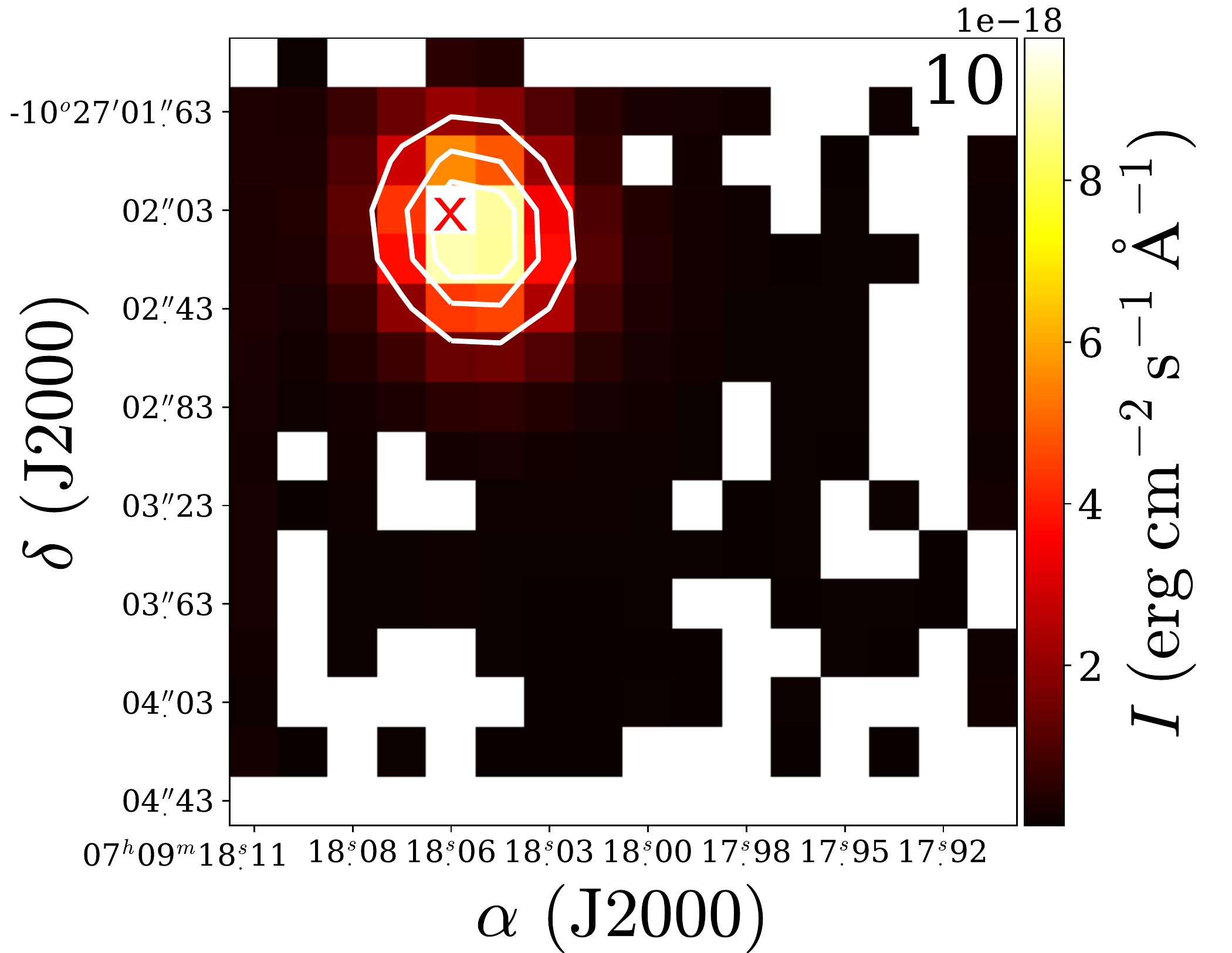}\hspace{-0.1cm}
\includegraphics[width=0.2\textwidth]{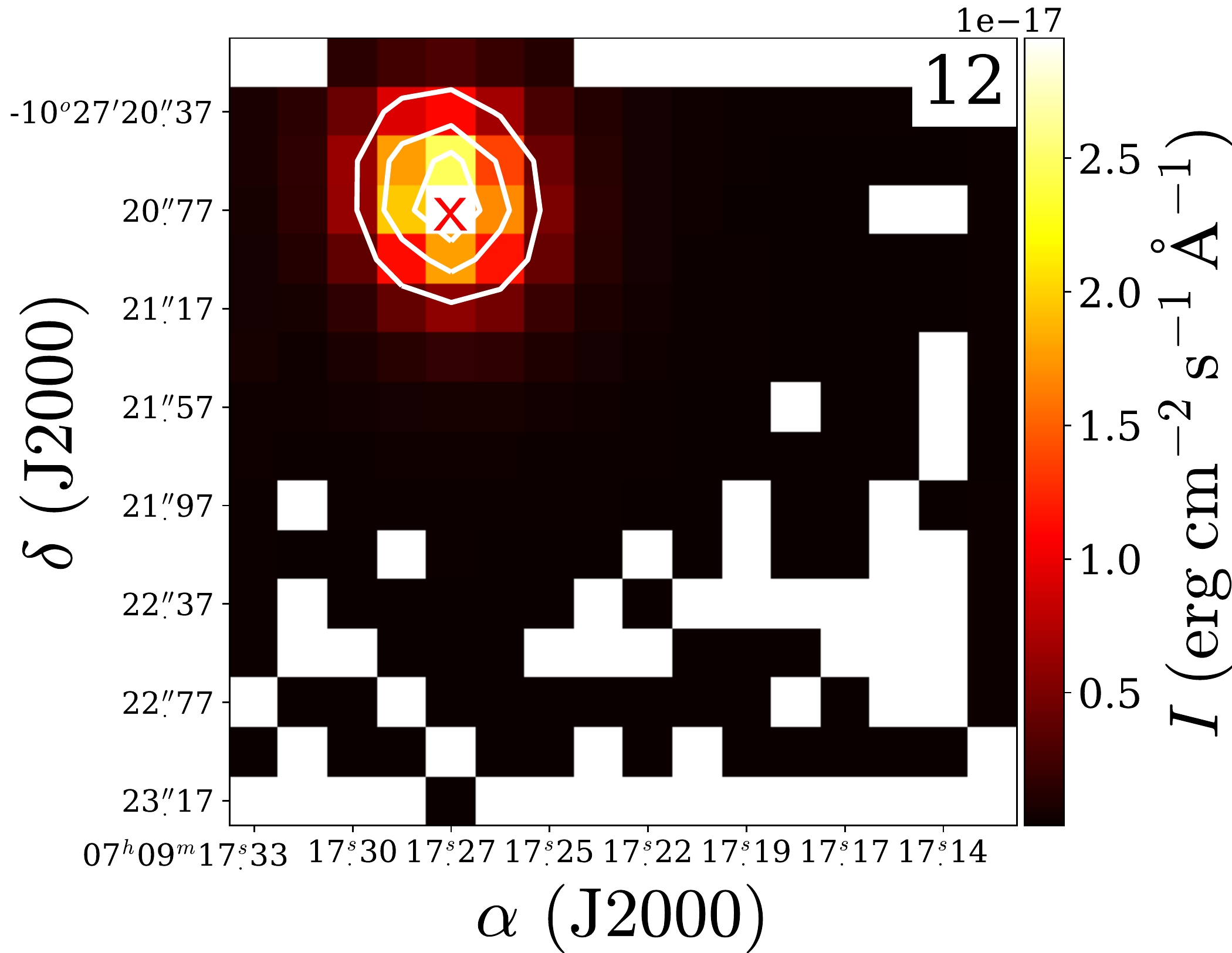}\hspace{-0.1cm}
\includegraphics[width=0.2\textwidth]{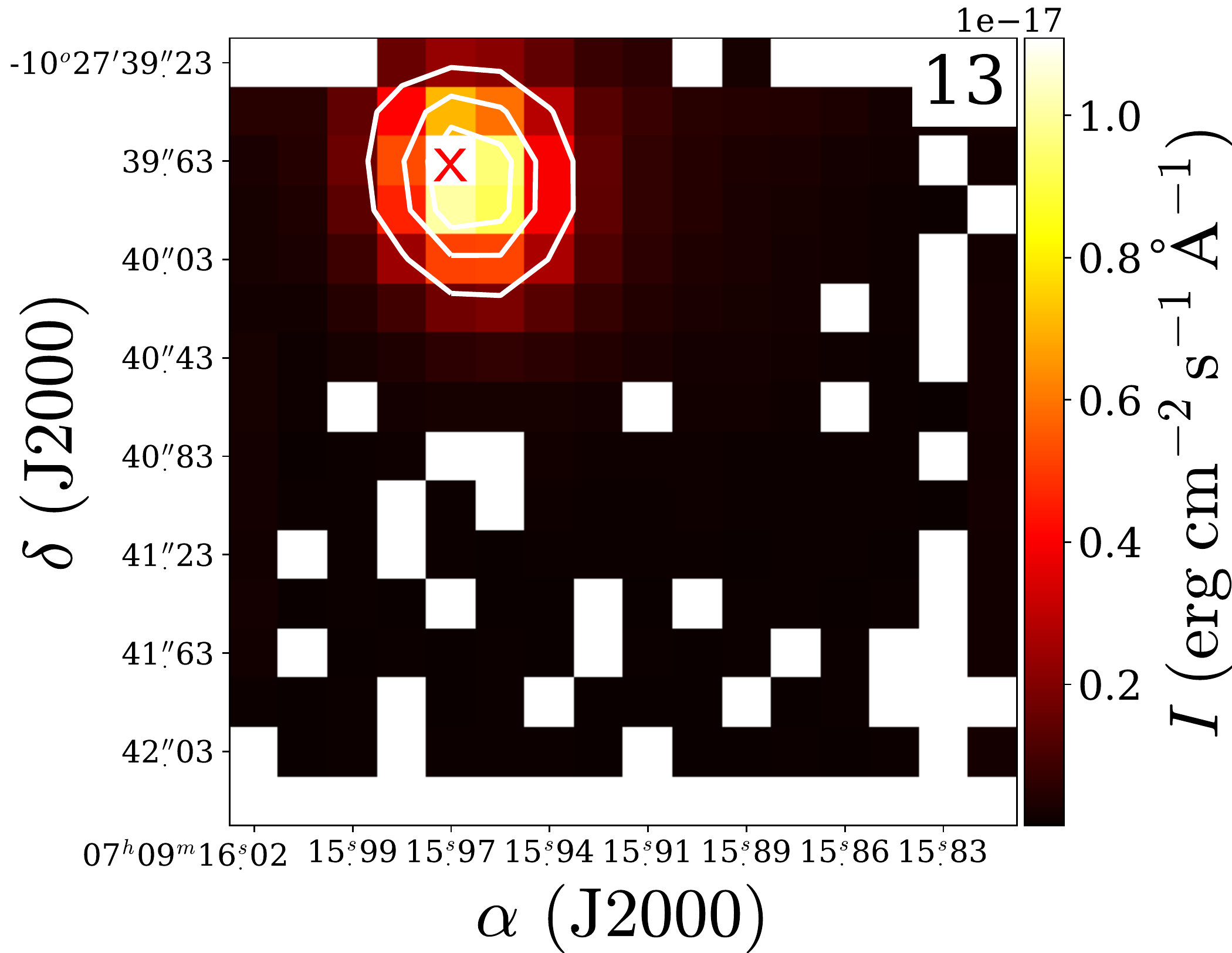}\hspace{-0.1cm}
\includegraphics[width=0.2\textwidth]{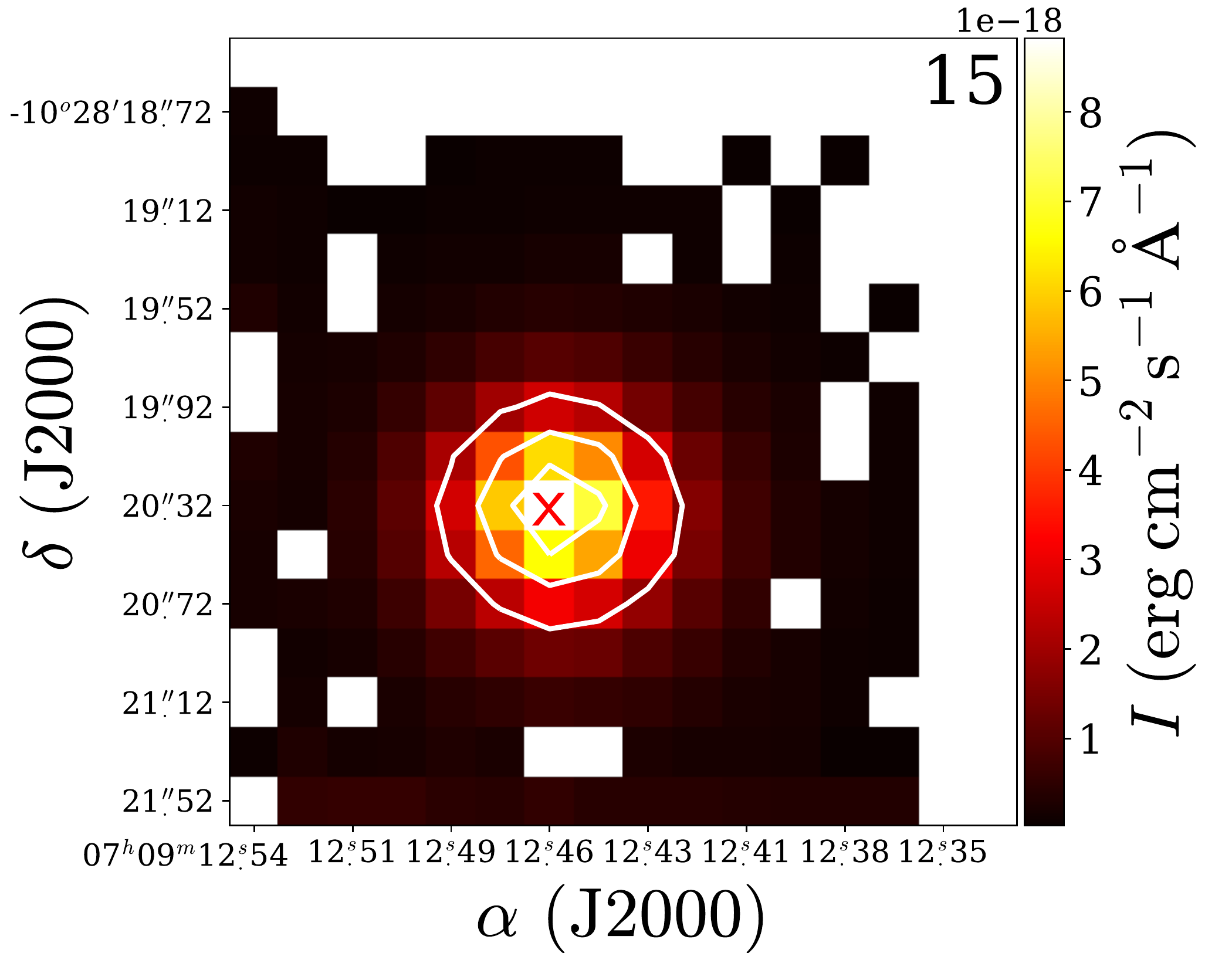}\hspace{-0.1cm}
\includegraphics[width=0.2\textwidth]{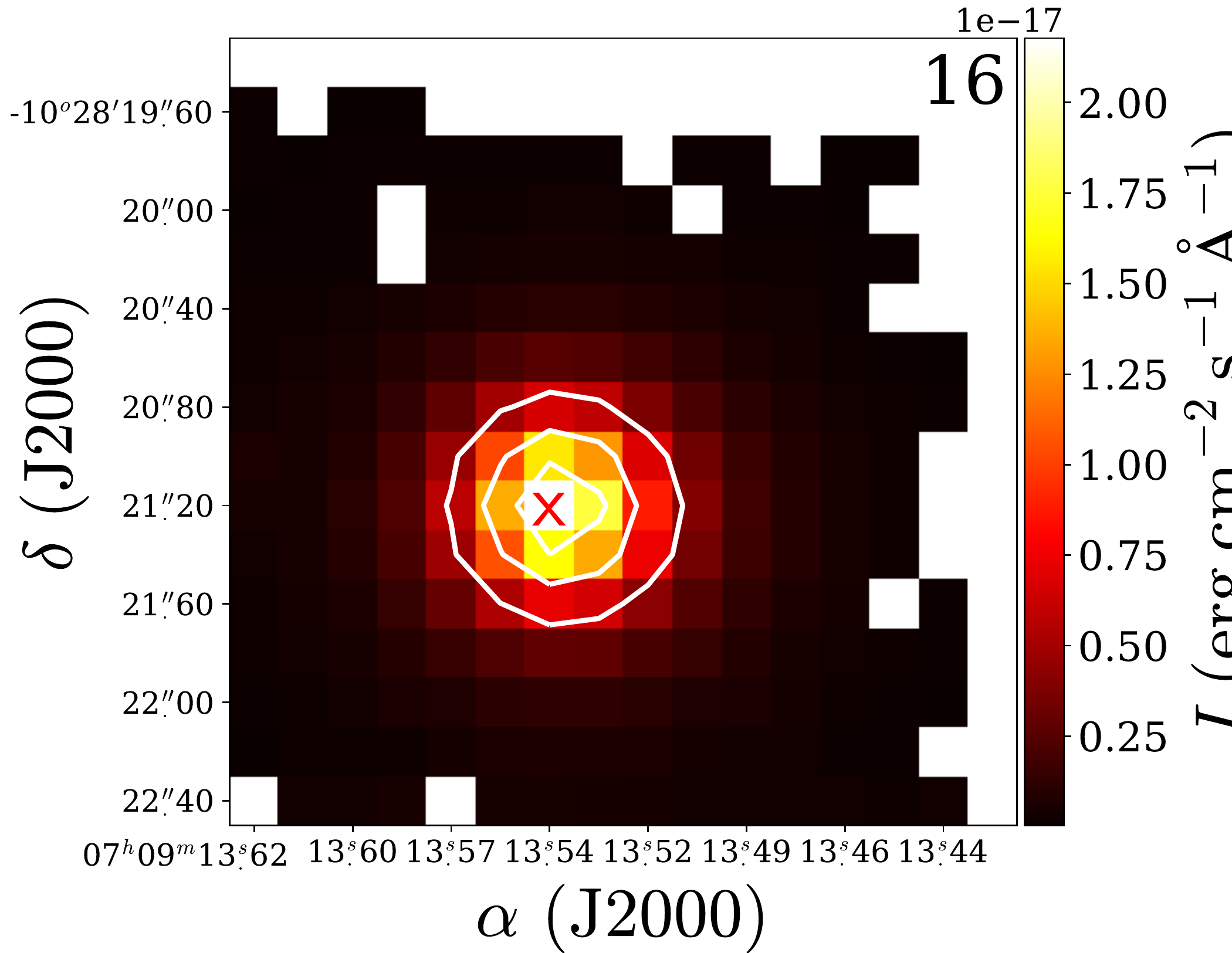}\hspace{-0.1cm}
\includegraphics[width=0.2\textwidth]{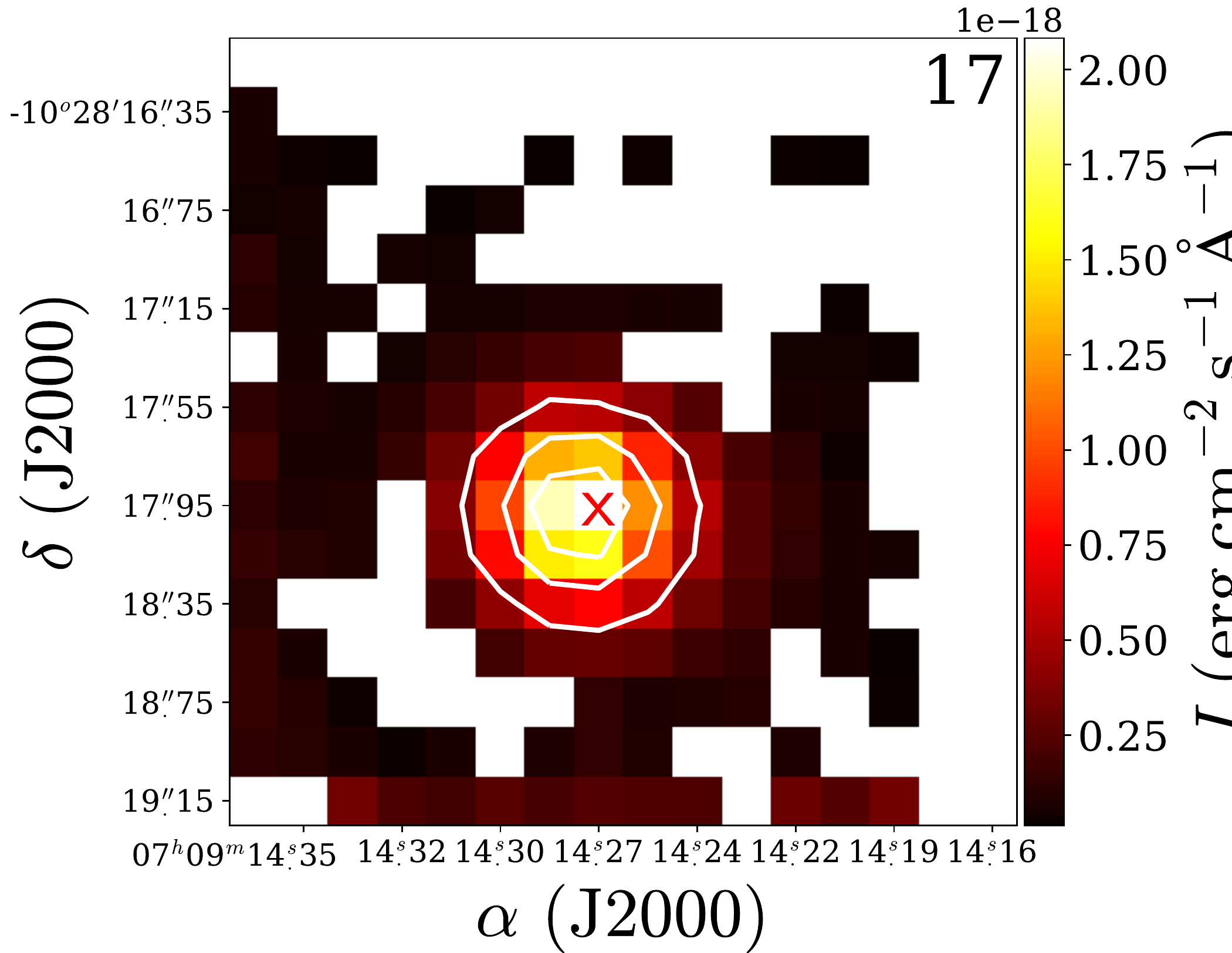}\hspace{-0.1cm}
\includegraphics[width=0.2\textwidth]{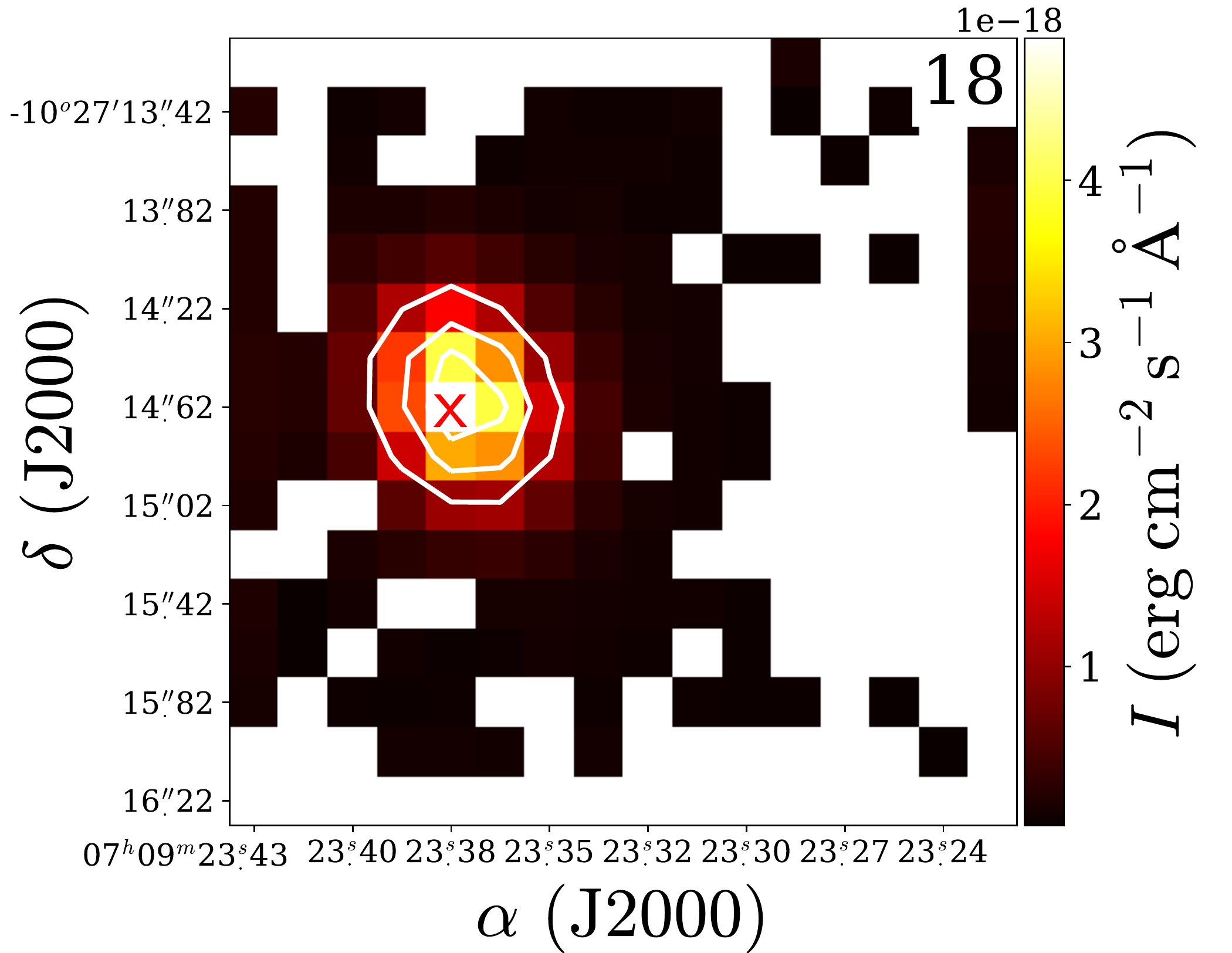}\hspace{-0.1cm}
\includegraphics[width=0.2\textwidth]{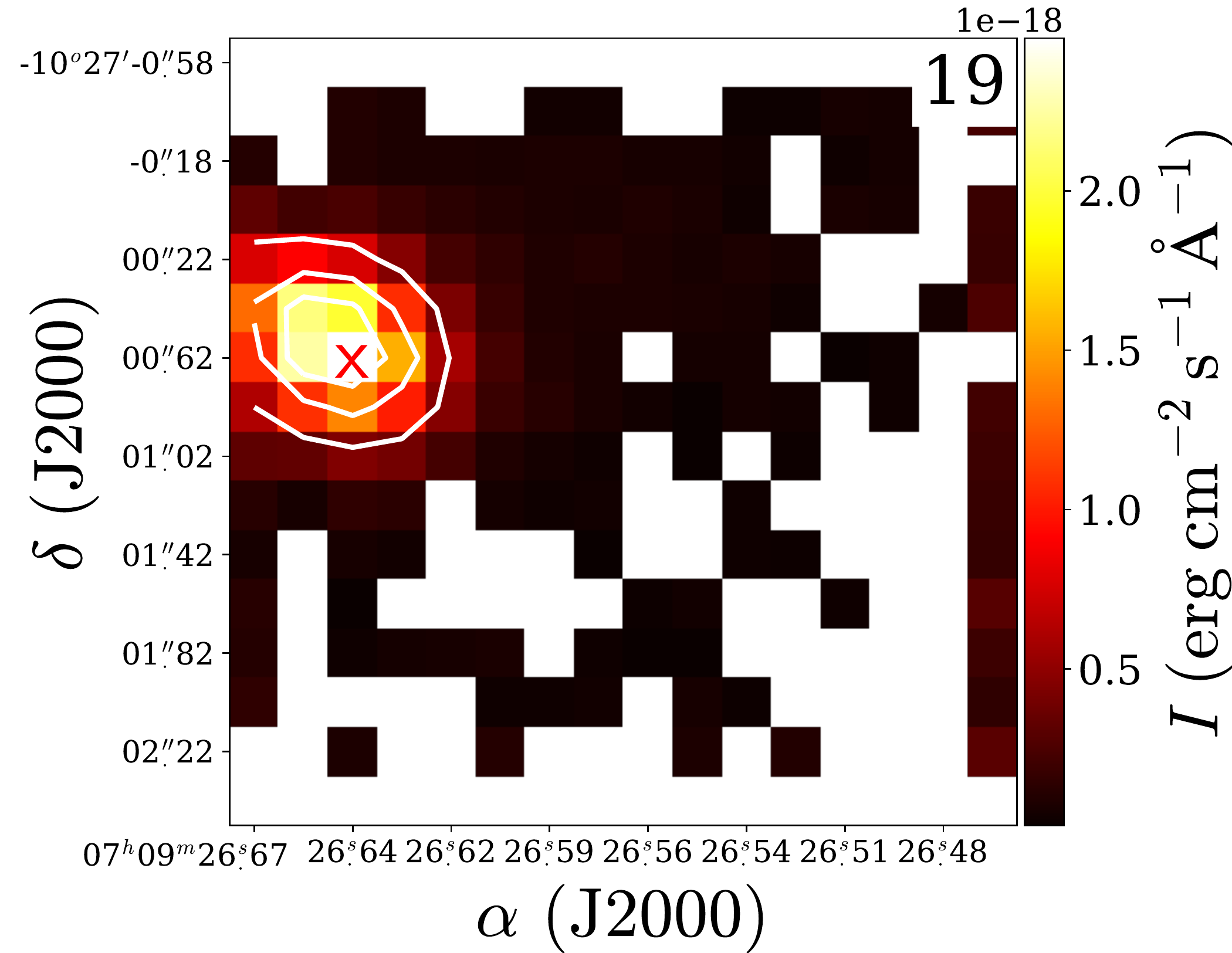}\hspace{-0.1cm}
\includegraphics[width=0.2\textwidth]{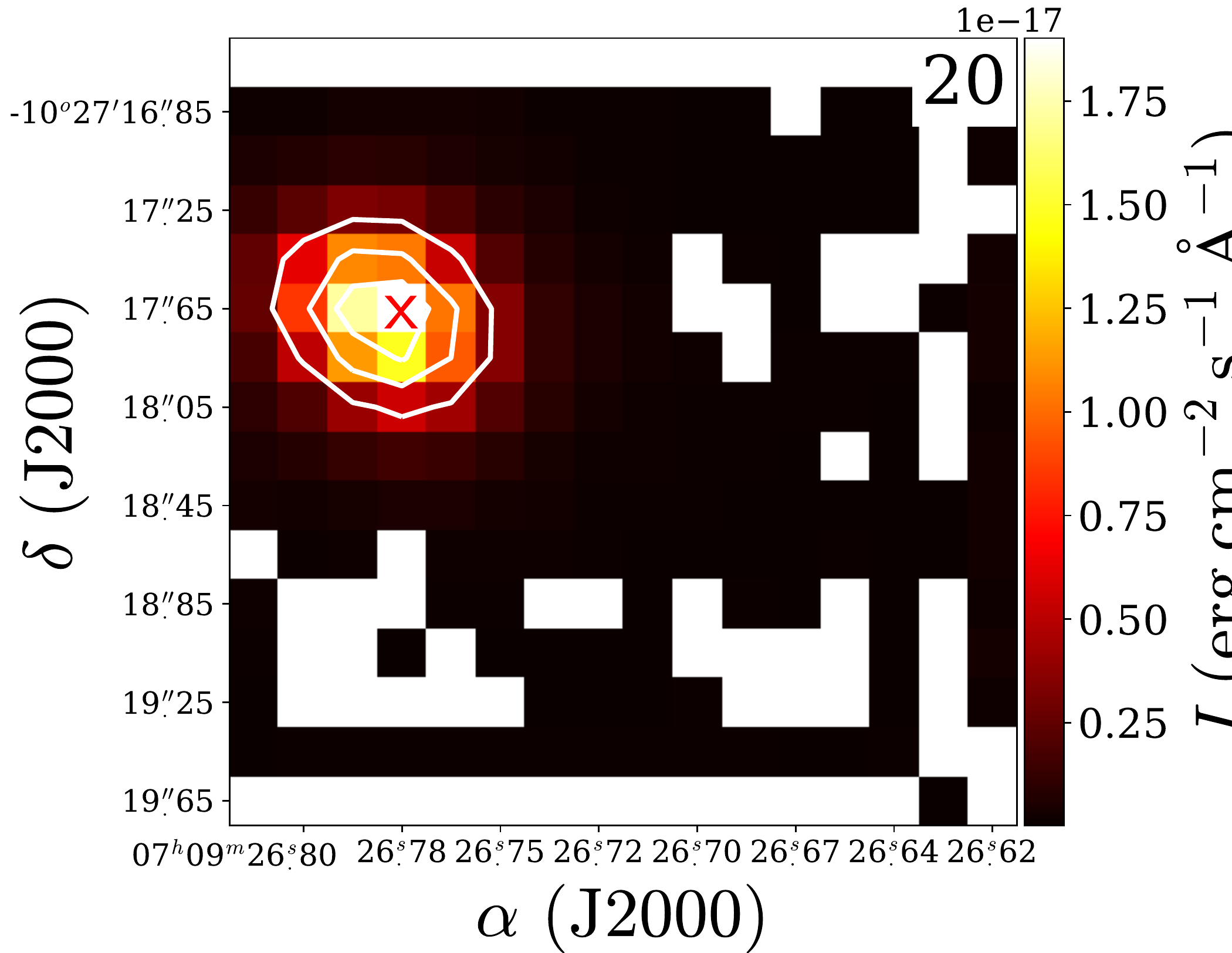}\hspace{-0.1cm}
\includegraphics[width=0.2\textwidth]{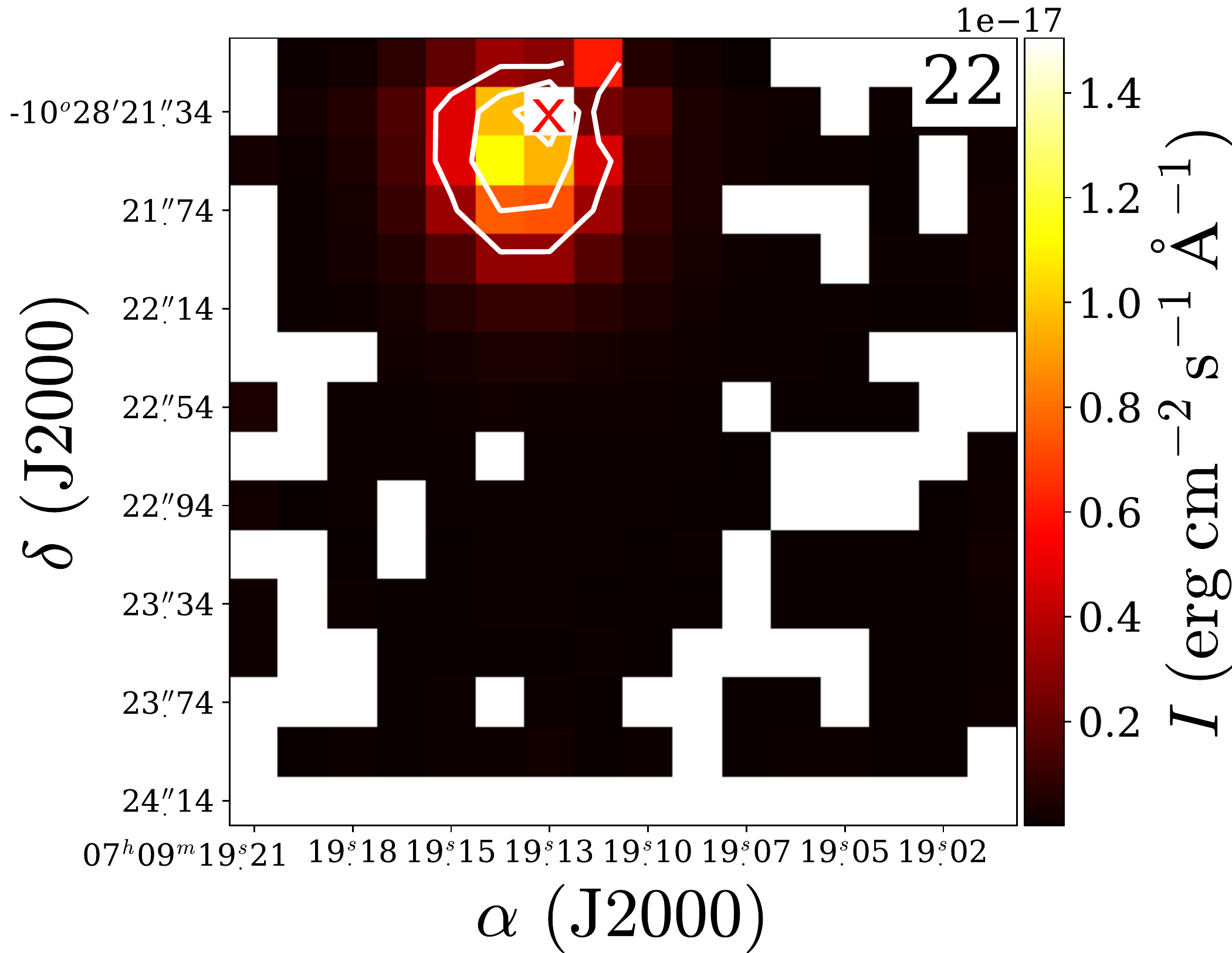}\hspace{-0.1cm}
\includegraphics[width=0.2\textwidth]{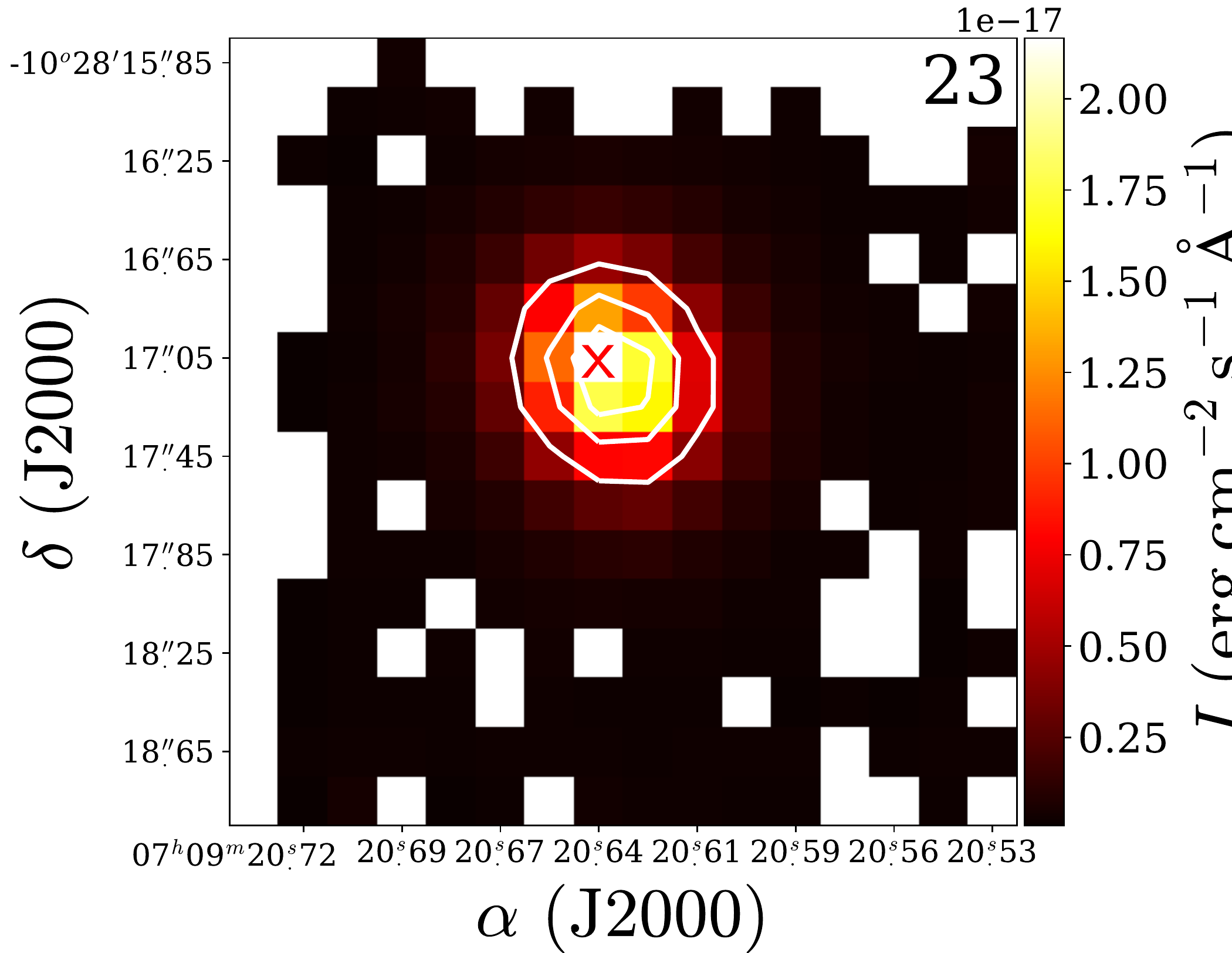}\hspace{-0.1cm}
\includegraphics[width=0.2\textwidth]{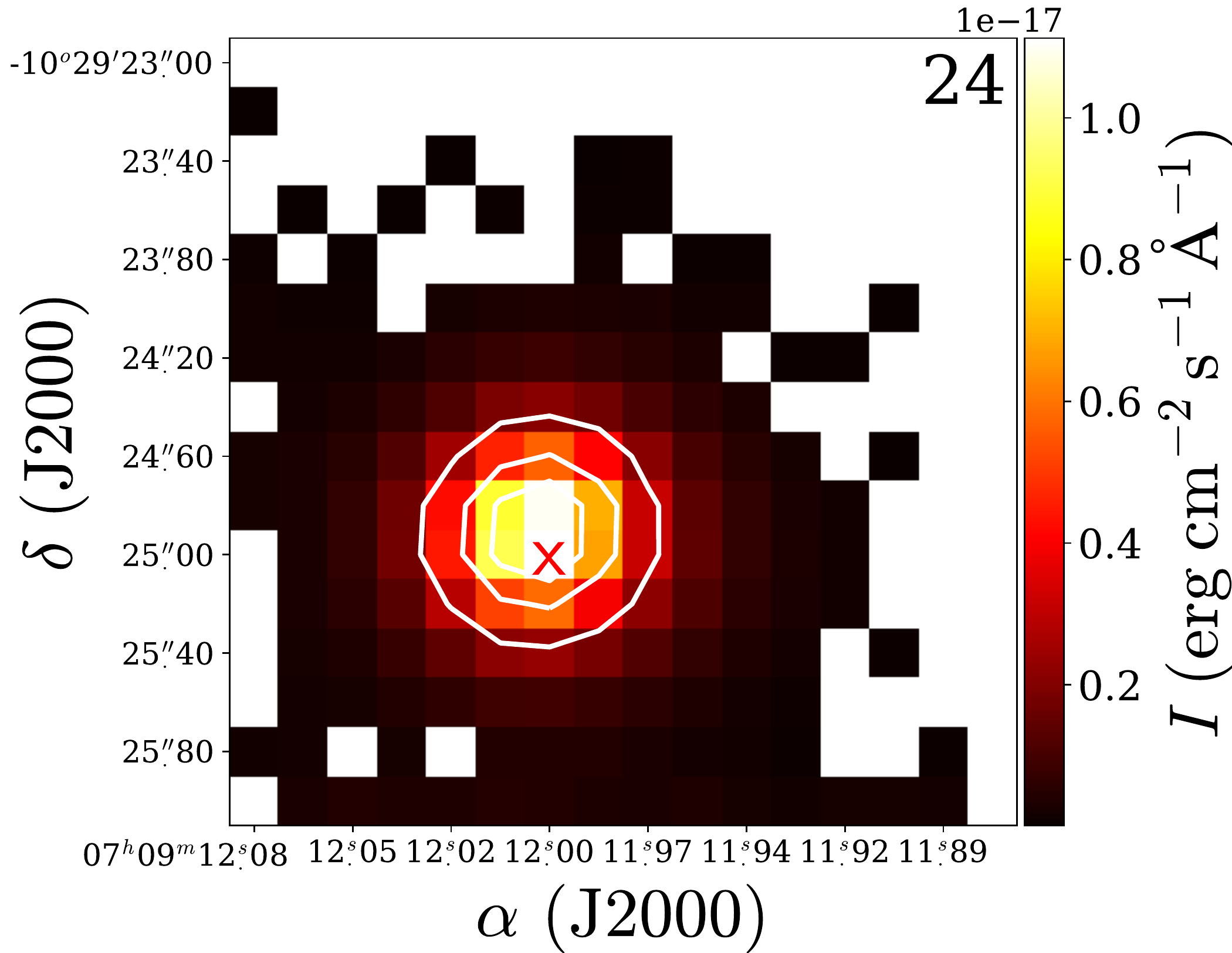}\hspace{-0.1cm}
\includegraphics[width=0.2\textwidth]{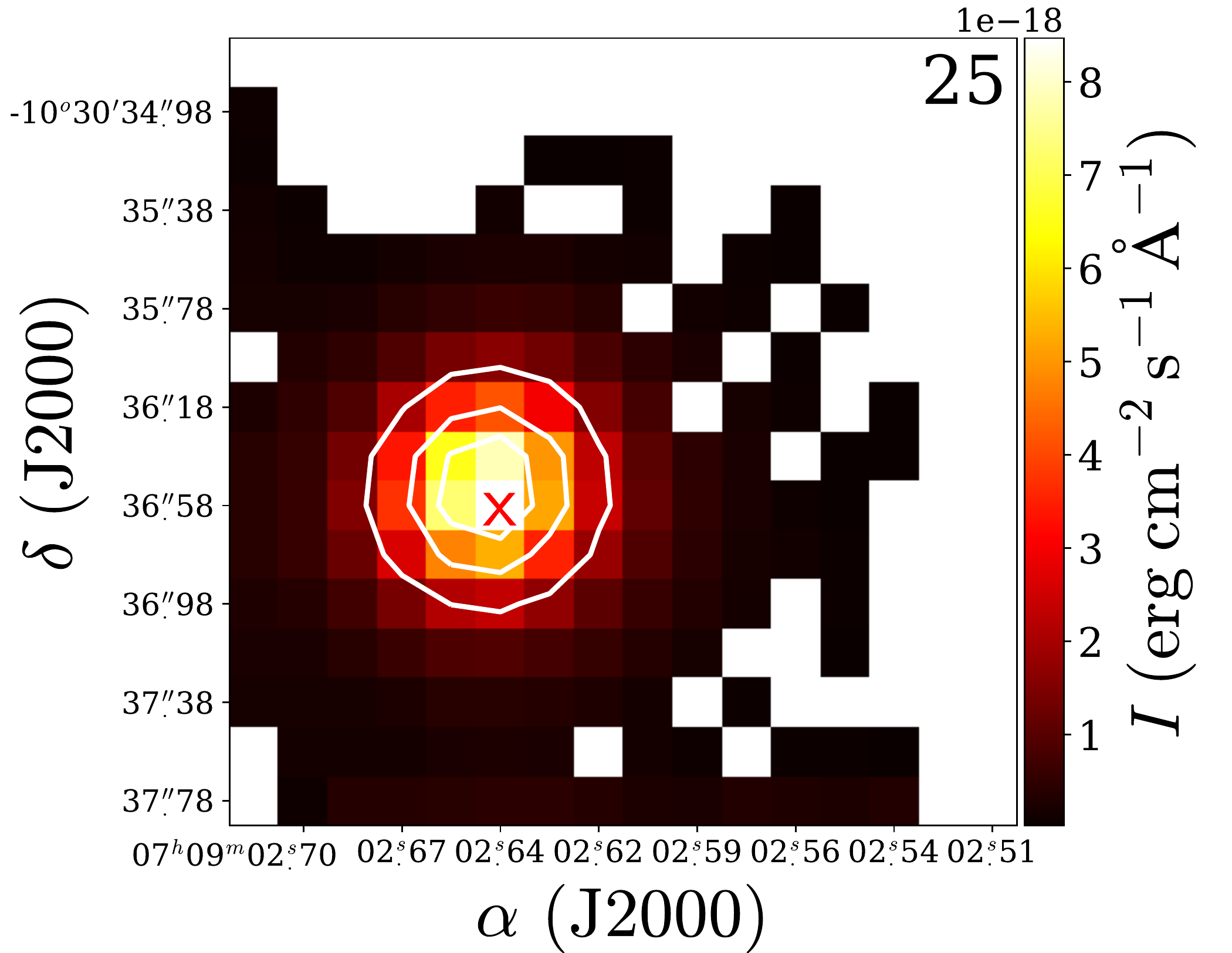}\hspace{-0.1cm}
\includegraphics[width=0.2\textwidth]{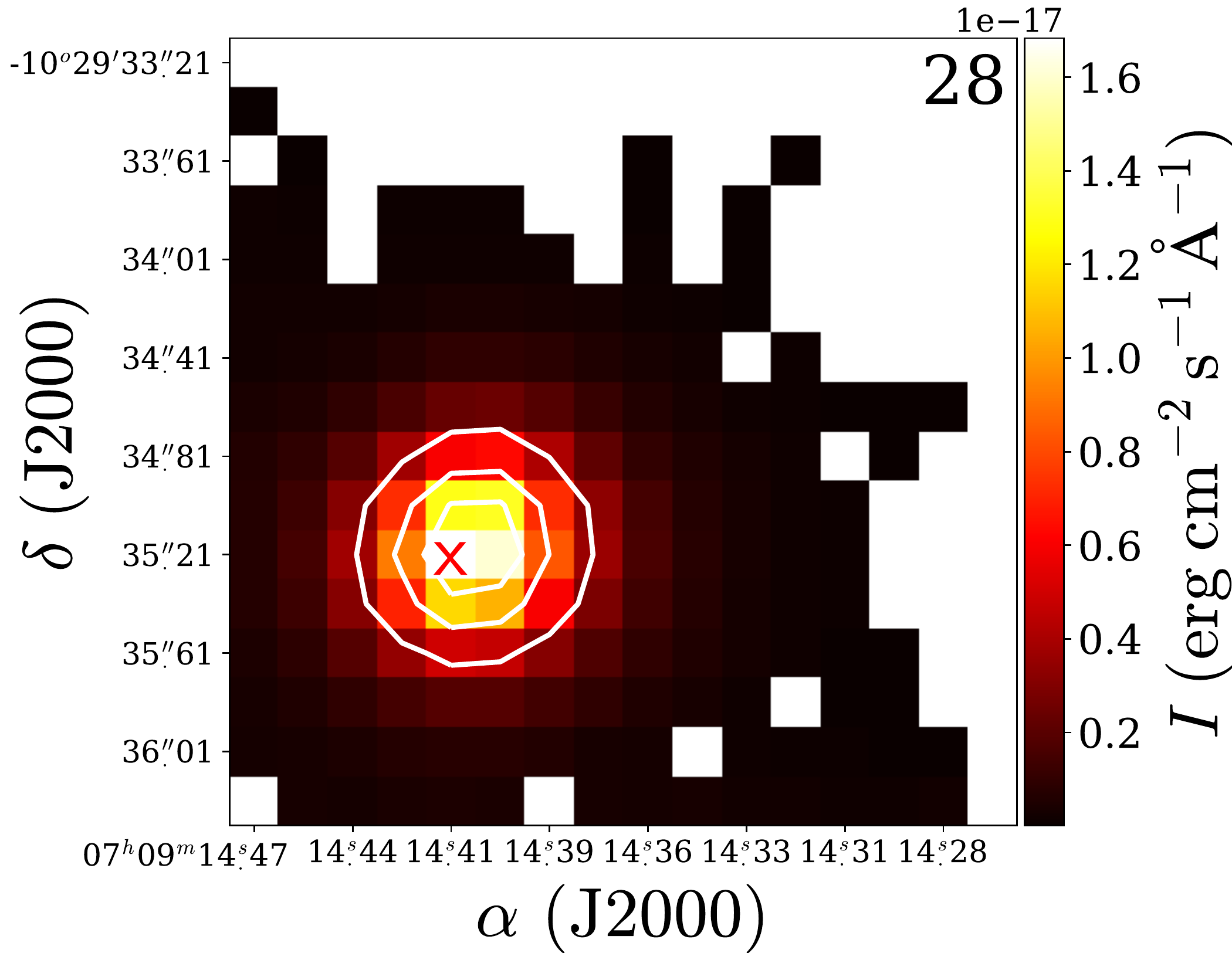}\hspace{-0.1cm}
\includegraphics[width=0.2\textwidth]{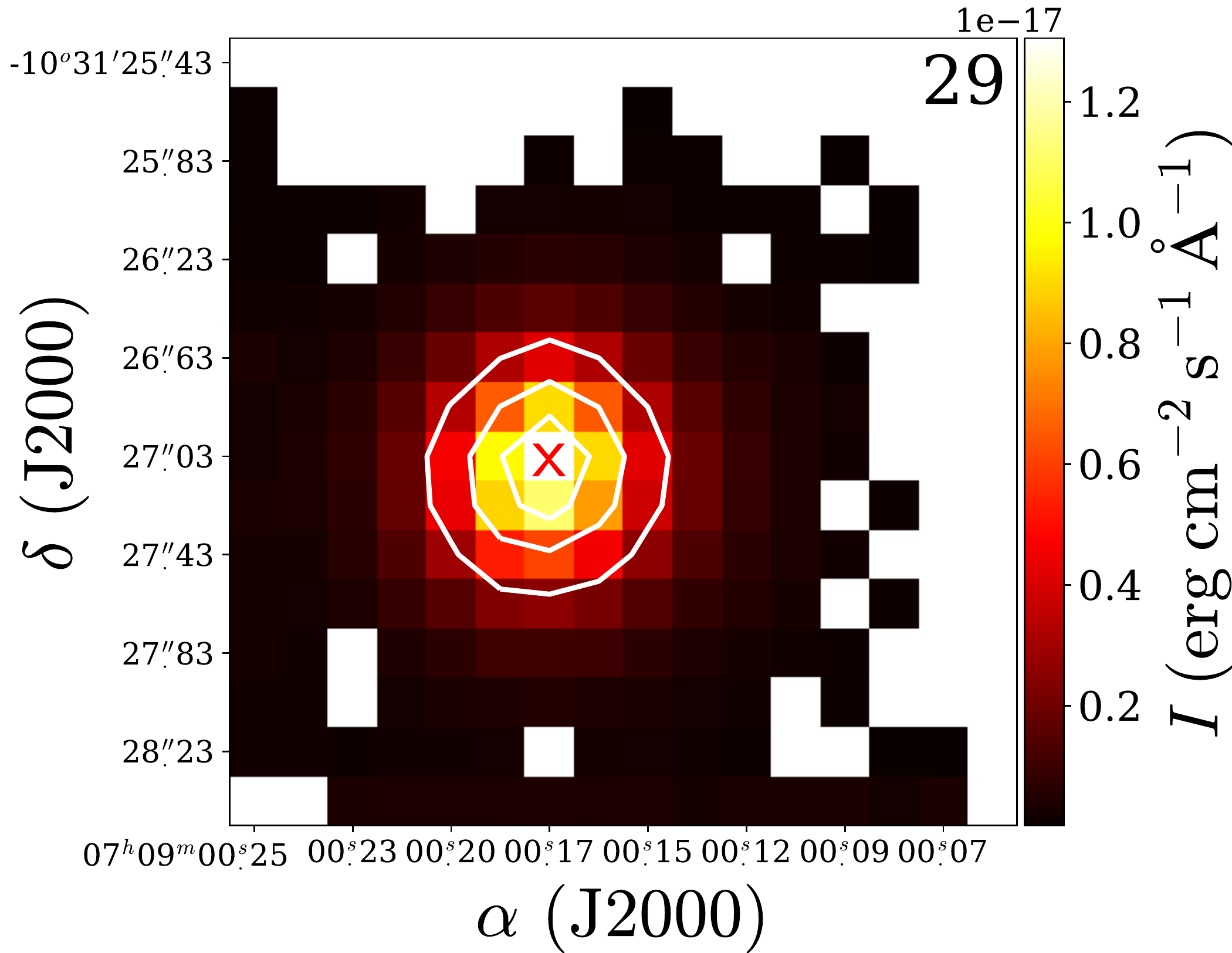}\hspace{-0.1cm}
\includegraphics[width=0.2\textwidth]{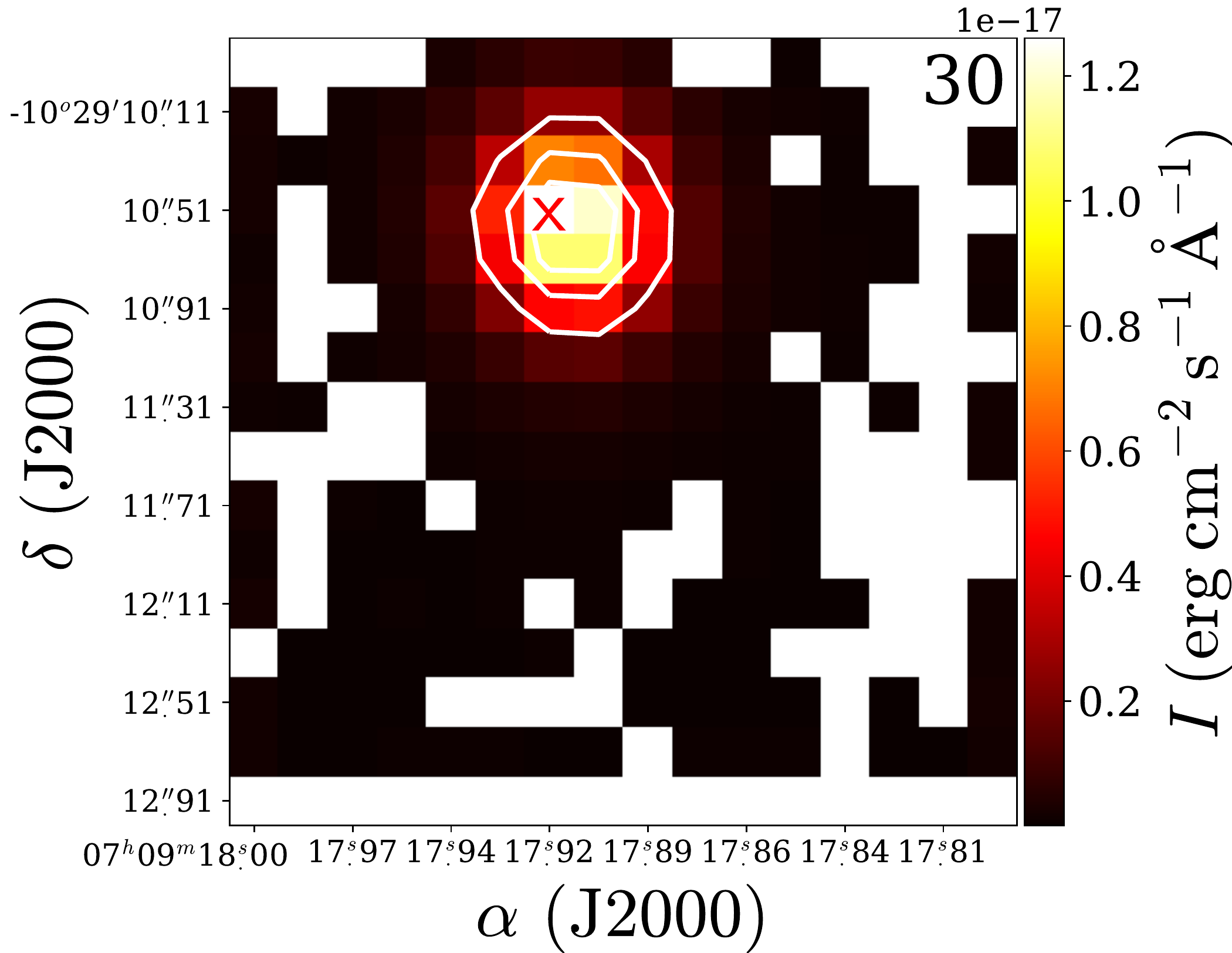}\hspace{-0.1cm}
\includegraphics[width=0.2\textwidth]{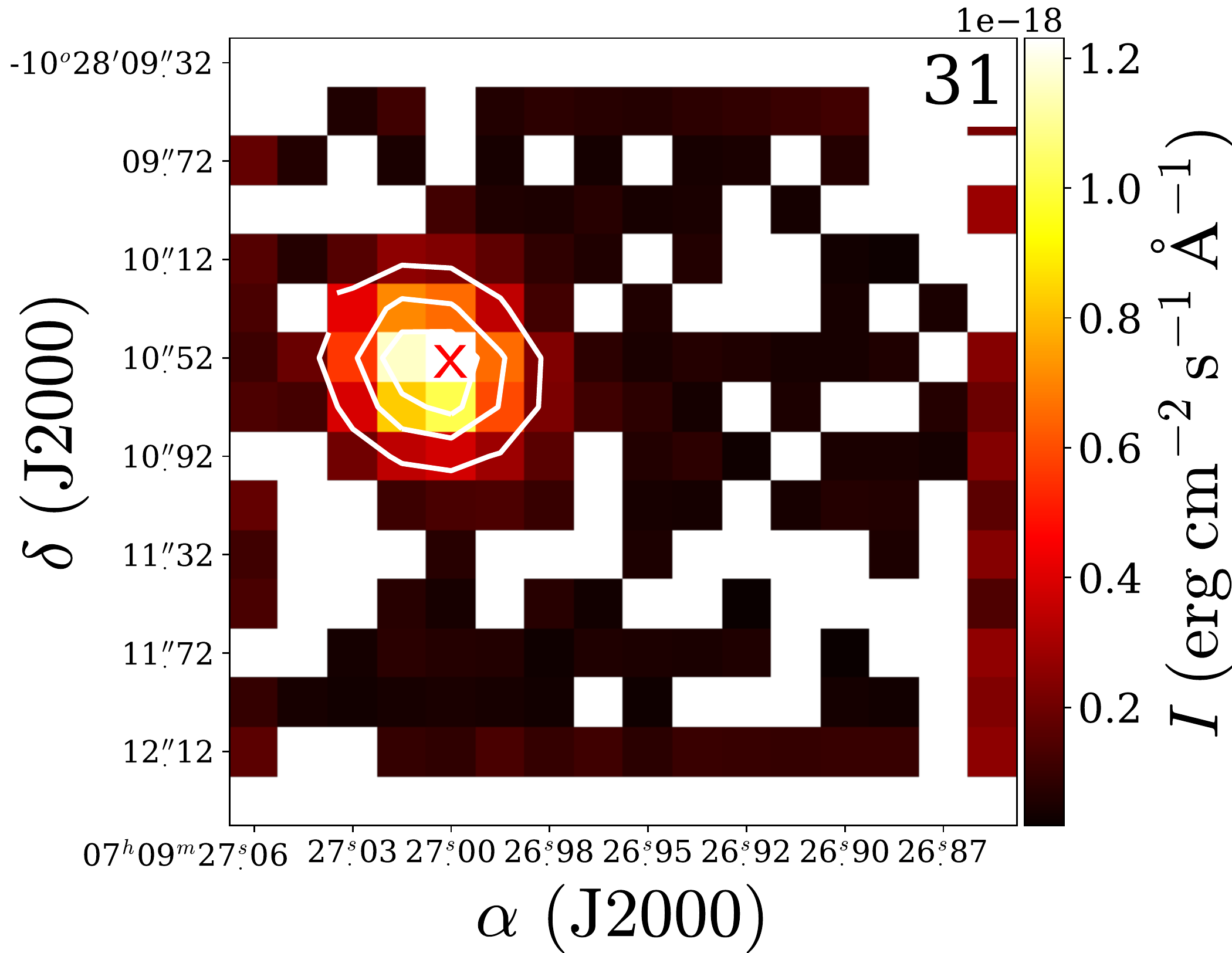}\hspace{-0.1cm}
\includegraphics[width=0.2\textwidth]{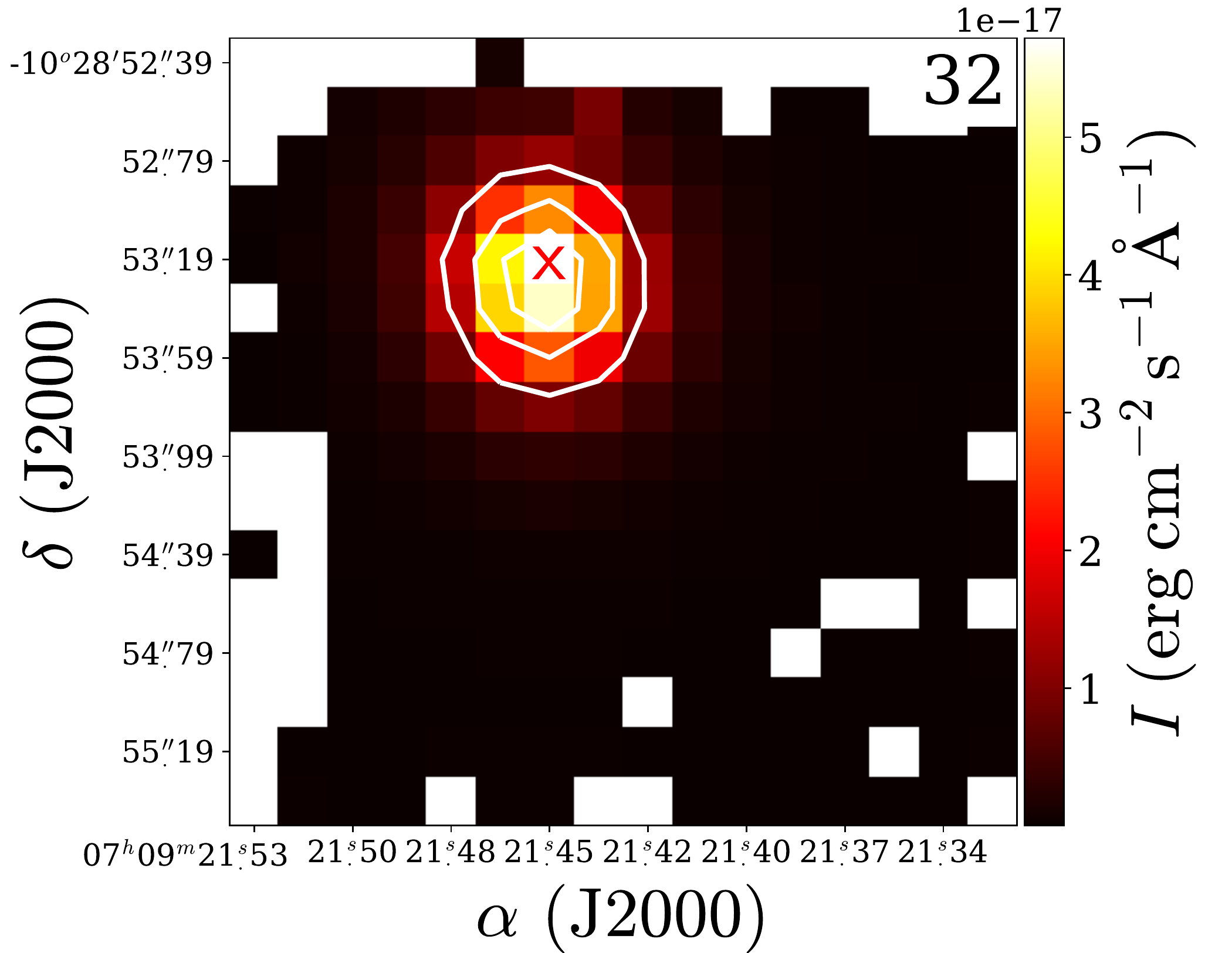}\hspace{-0.1cm}
\includegraphics[width=0.2\textwidth]{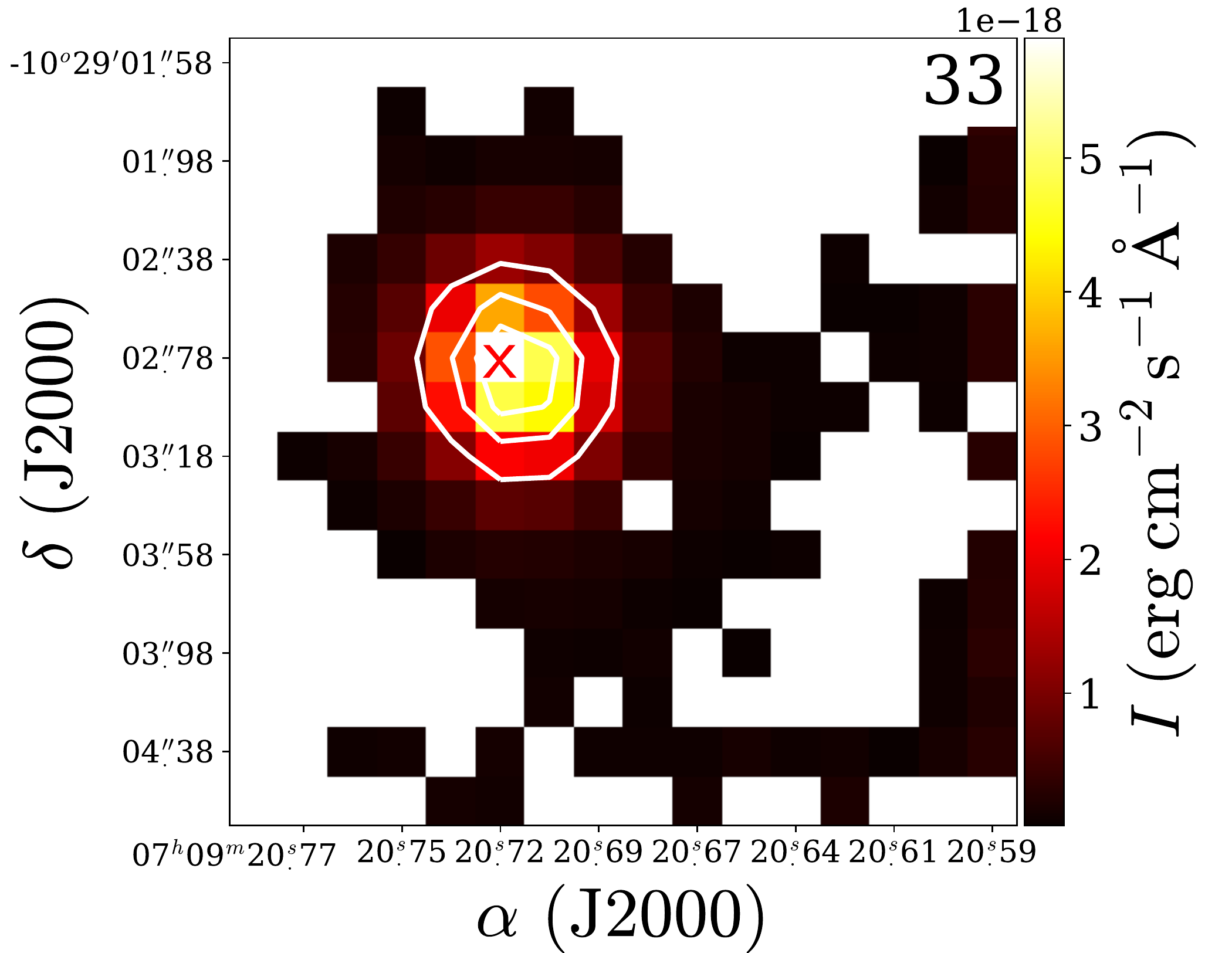}\hspace{-0.1cm}
\includegraphics[width=0.2\textwidth]{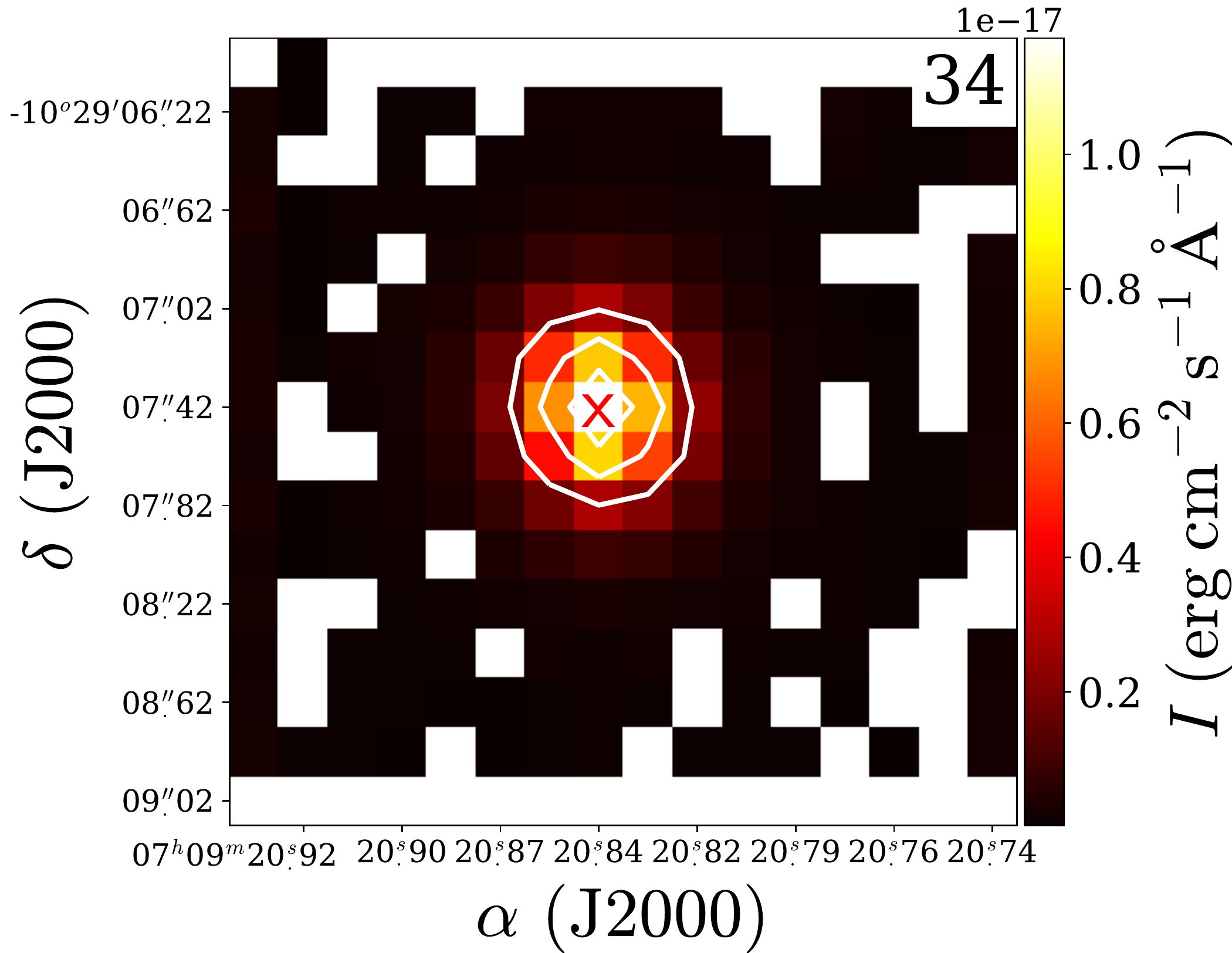}\hspace{-0.1cm}
\includegraphics[width=0.2\textwidth]{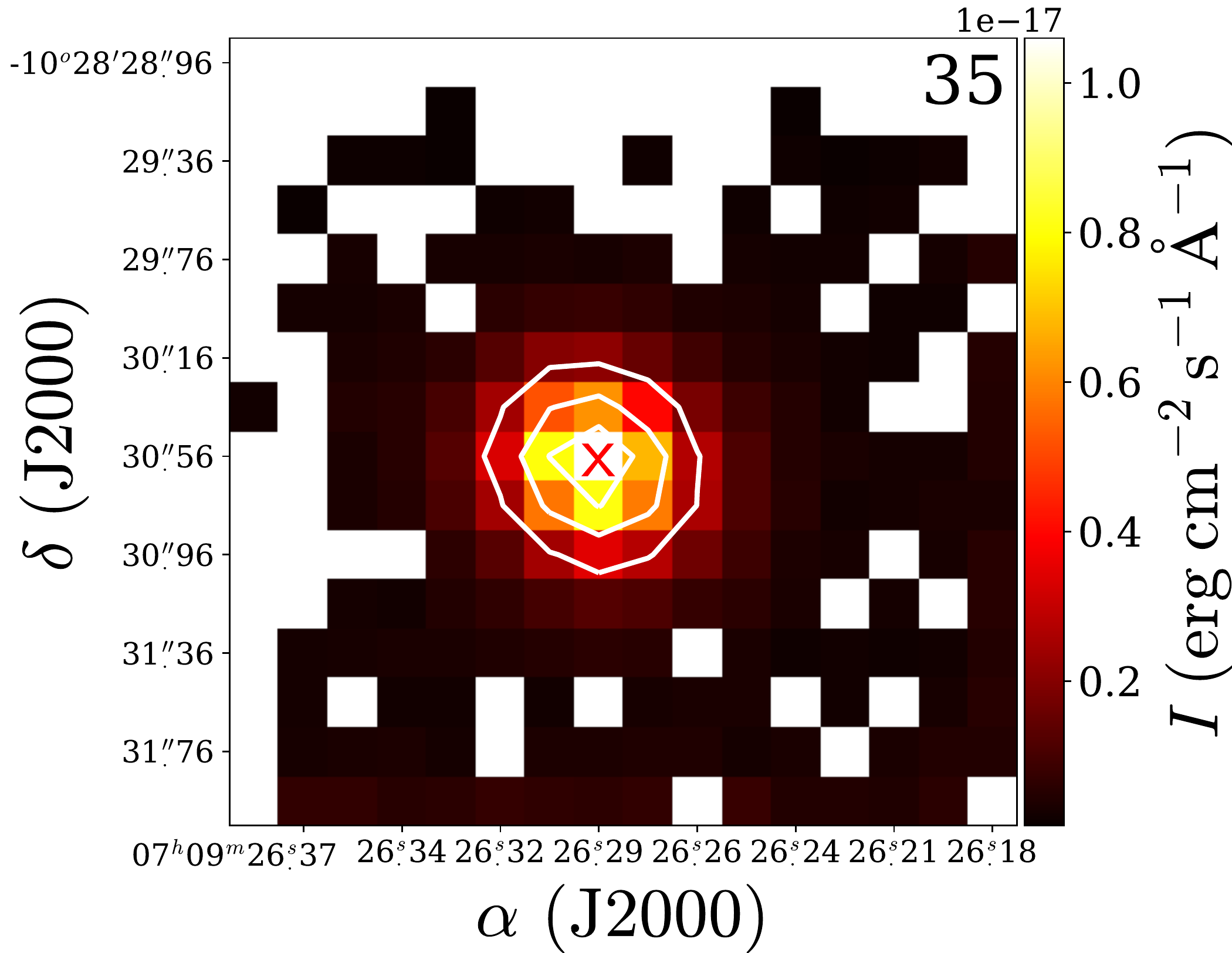}\hspace{-0.1cm}
\includegraphics[width=0.2\textwidth]{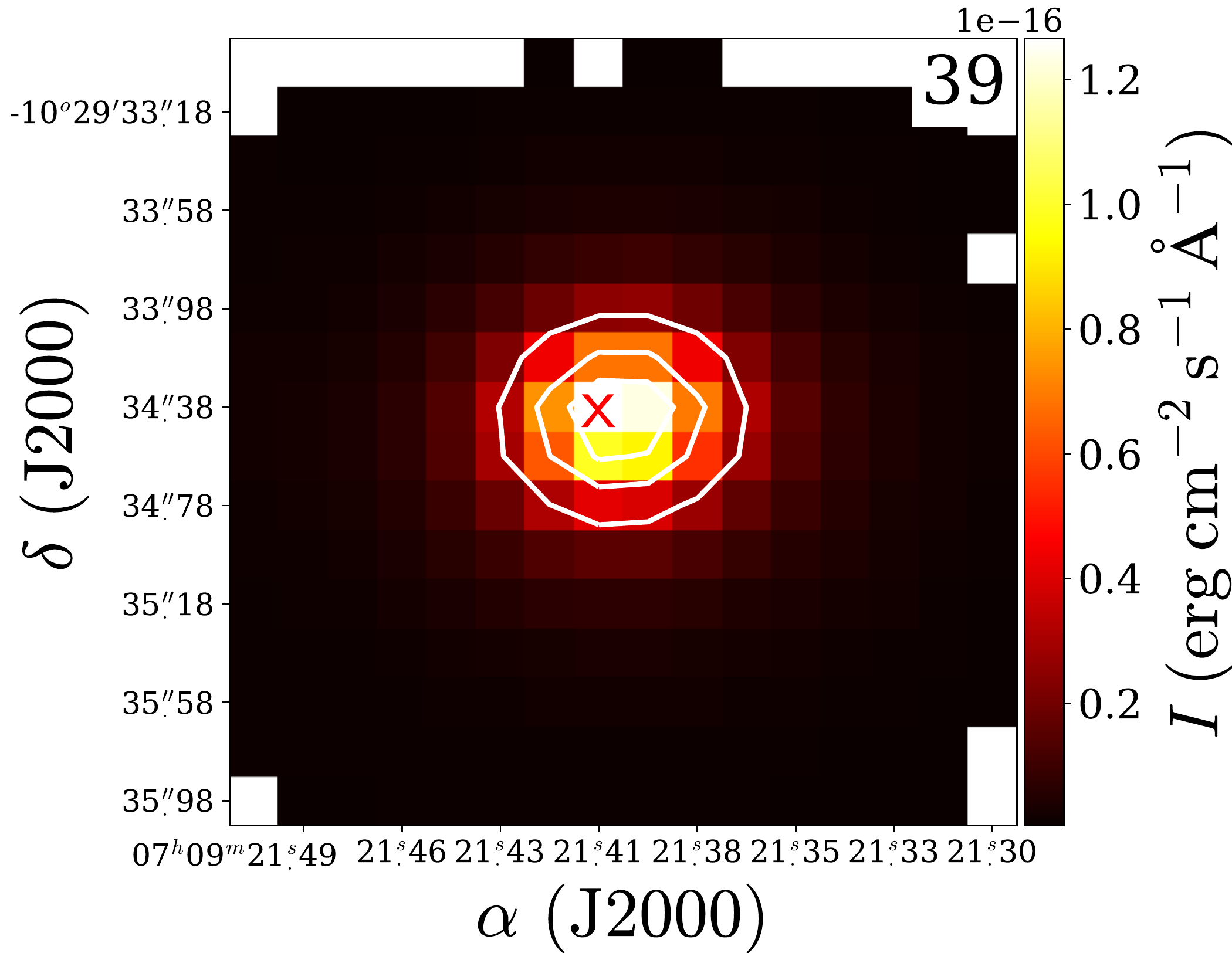}\hspace{-0.1cm}
\includegraphics[width=0.2\textwidth]{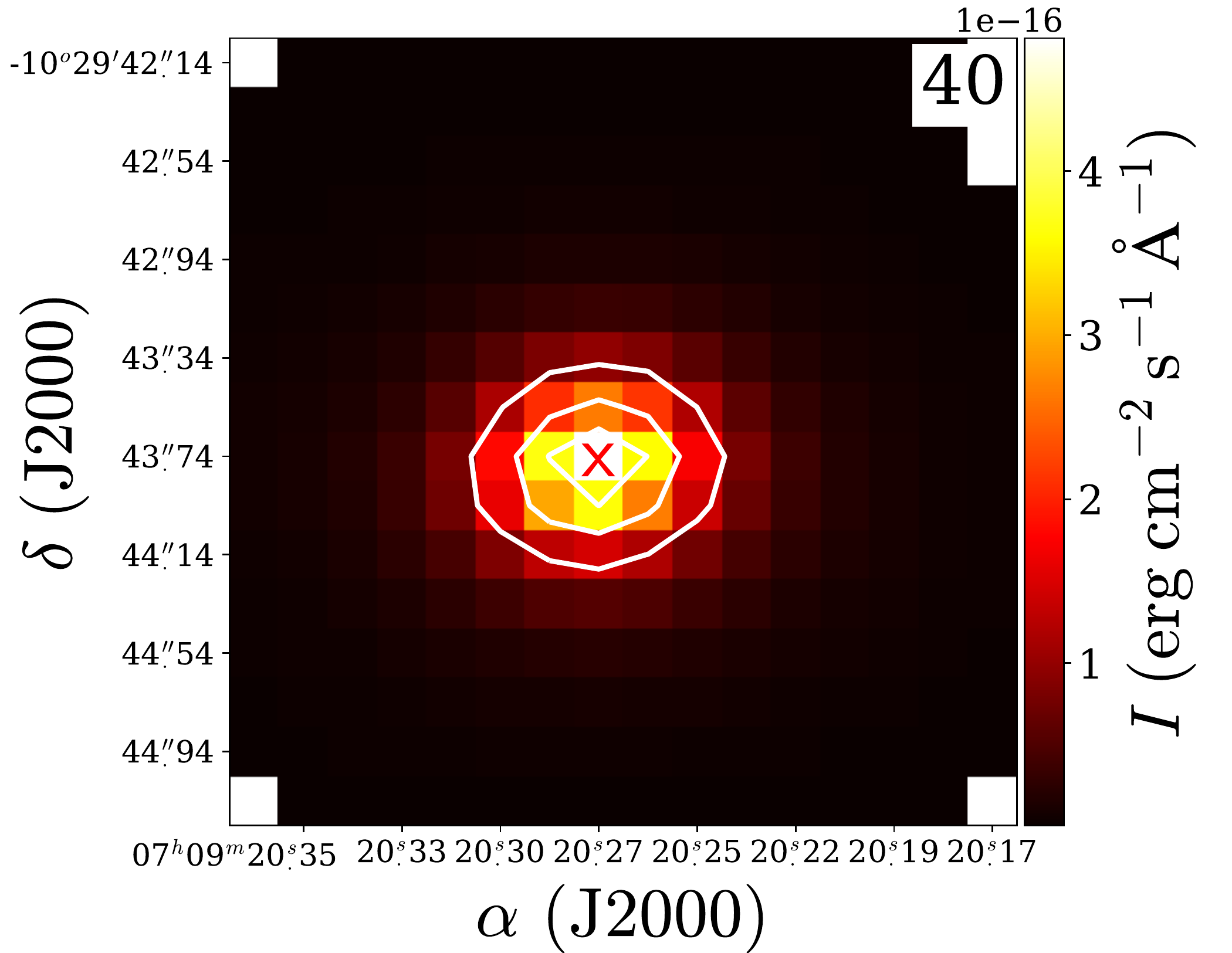}\hspace{-0.1cm}
\includegraphics[width=0.2\textwidth]{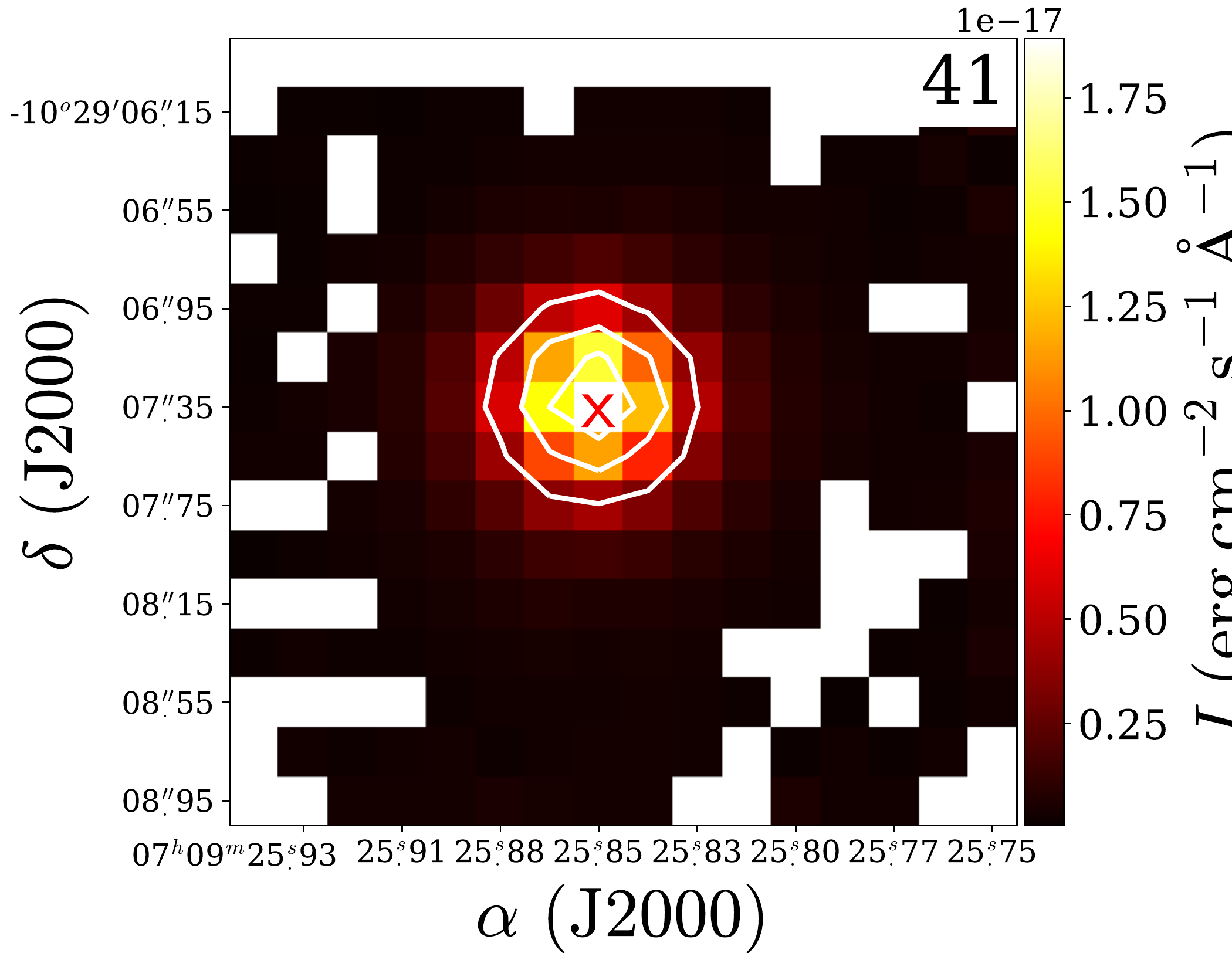}\hspace{-0.1cm}
\includegraphics[width=0.2\textwidth]{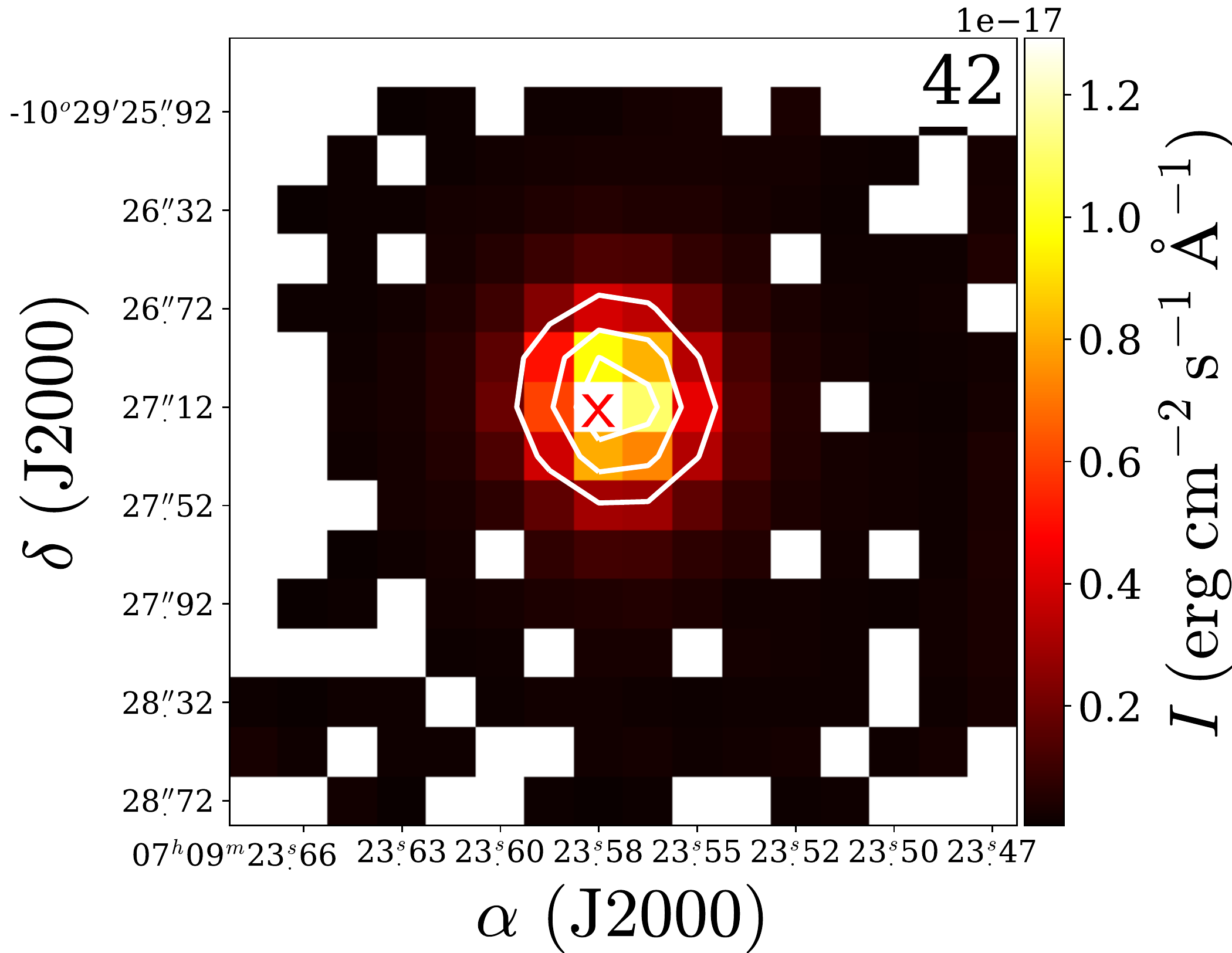}\hspace{-0.1cm}
\includegraphics[width=0.2\textwidth]{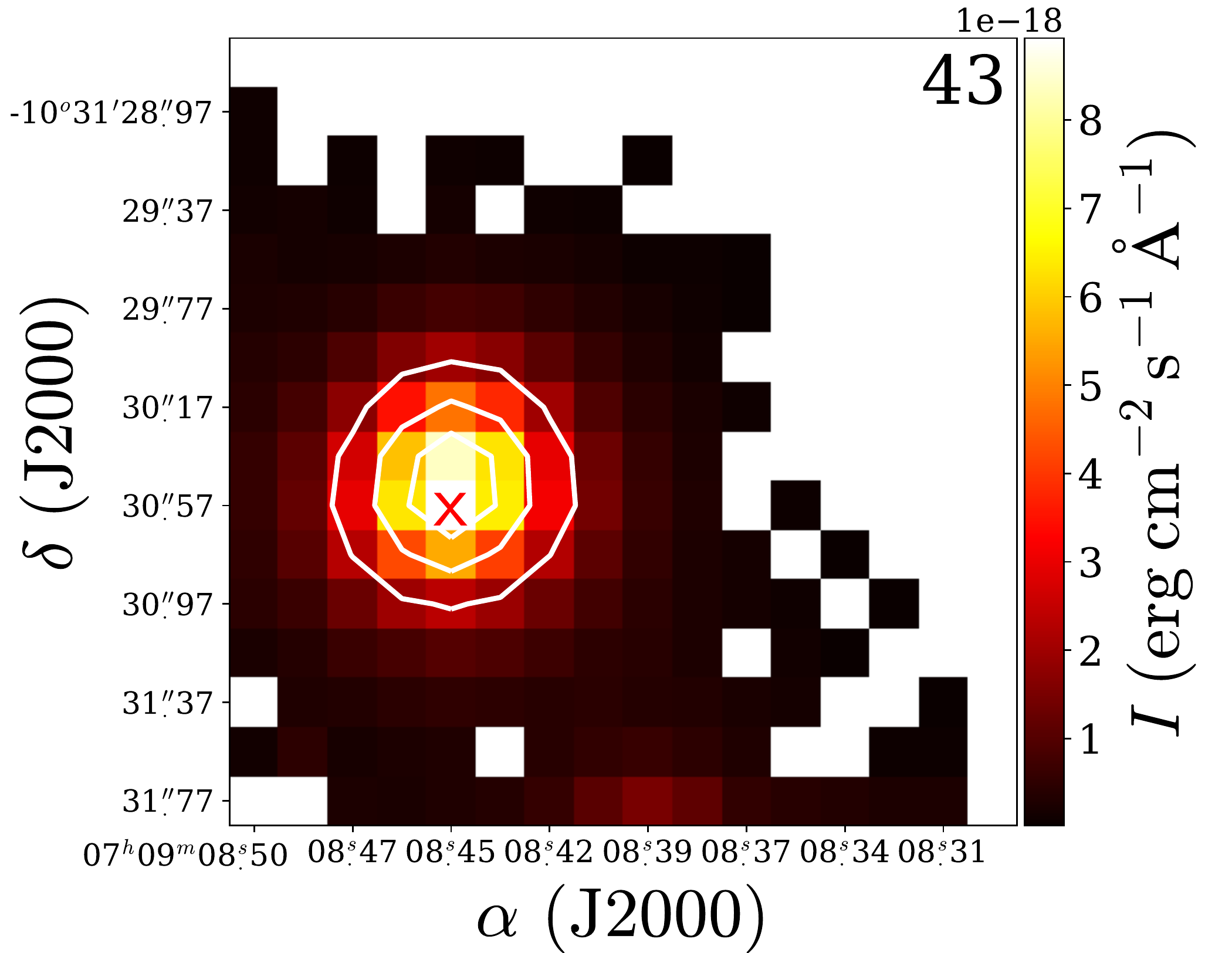}\hspace{-0.1cm}
\includegraphics[width=0.2\textwidth]{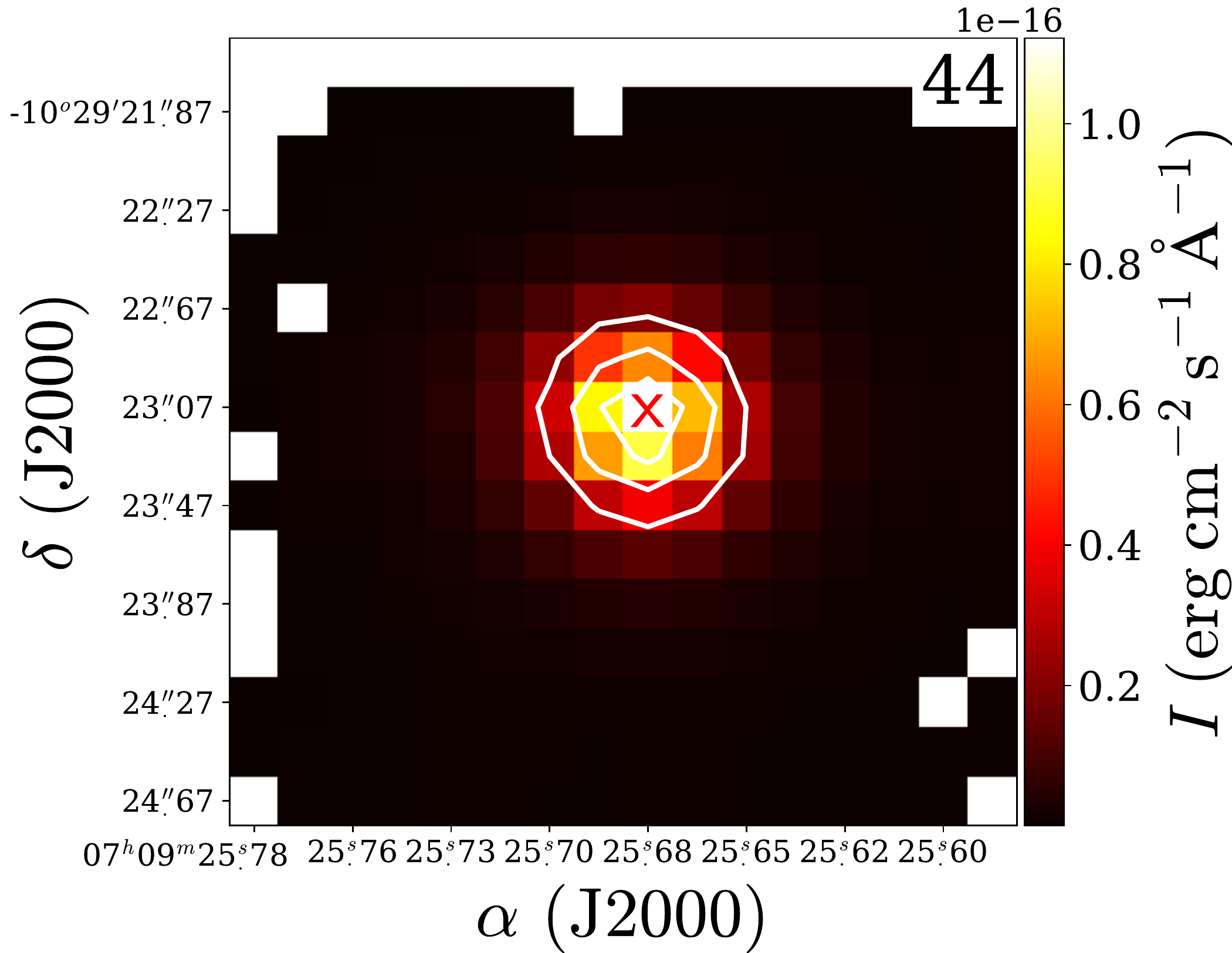}\hspace{-0.1cm}
\includegraphics[width=0.2\textwidth]{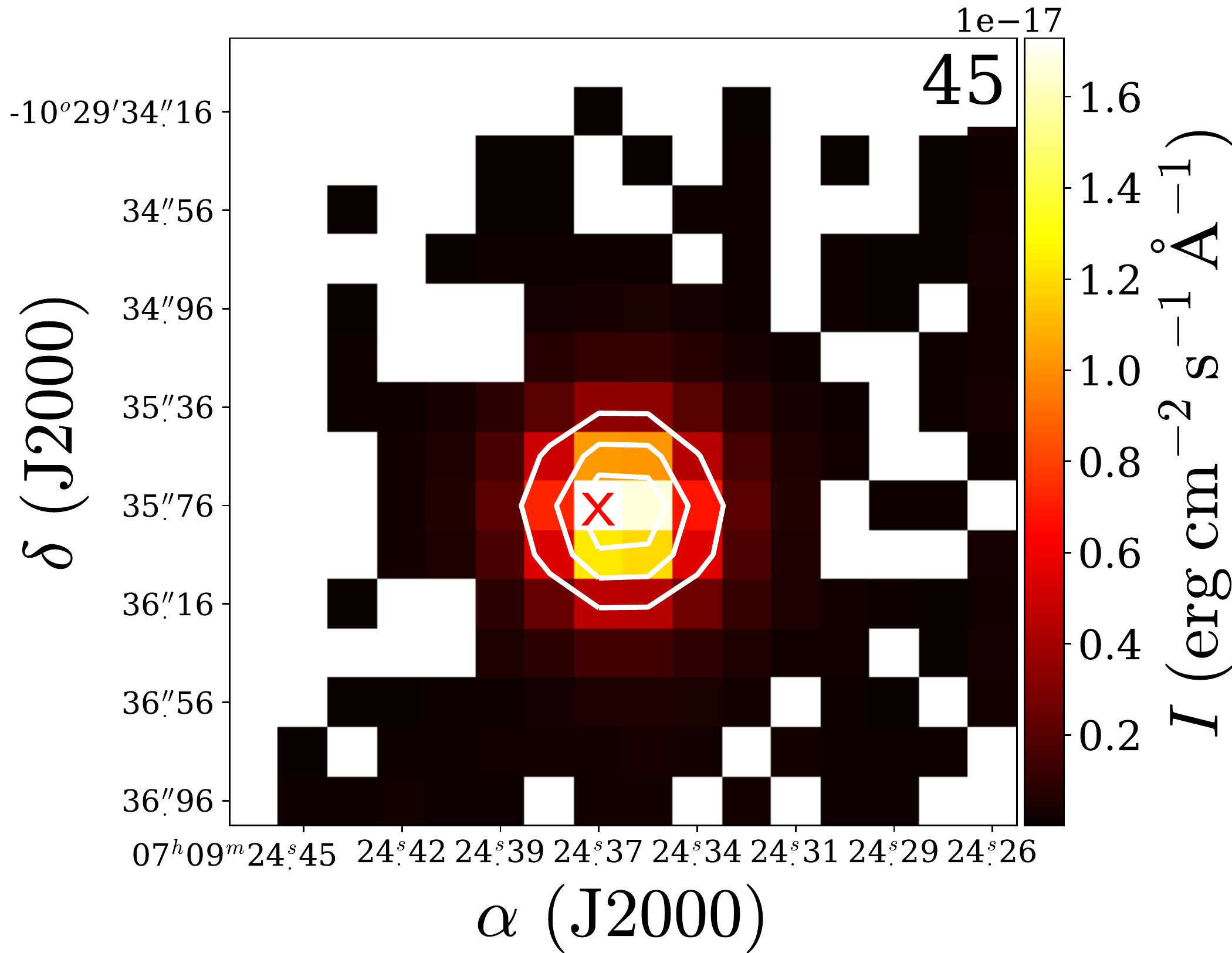}\hspace{-0.1cm}
\includegraphics[width=0.2\textwidth]{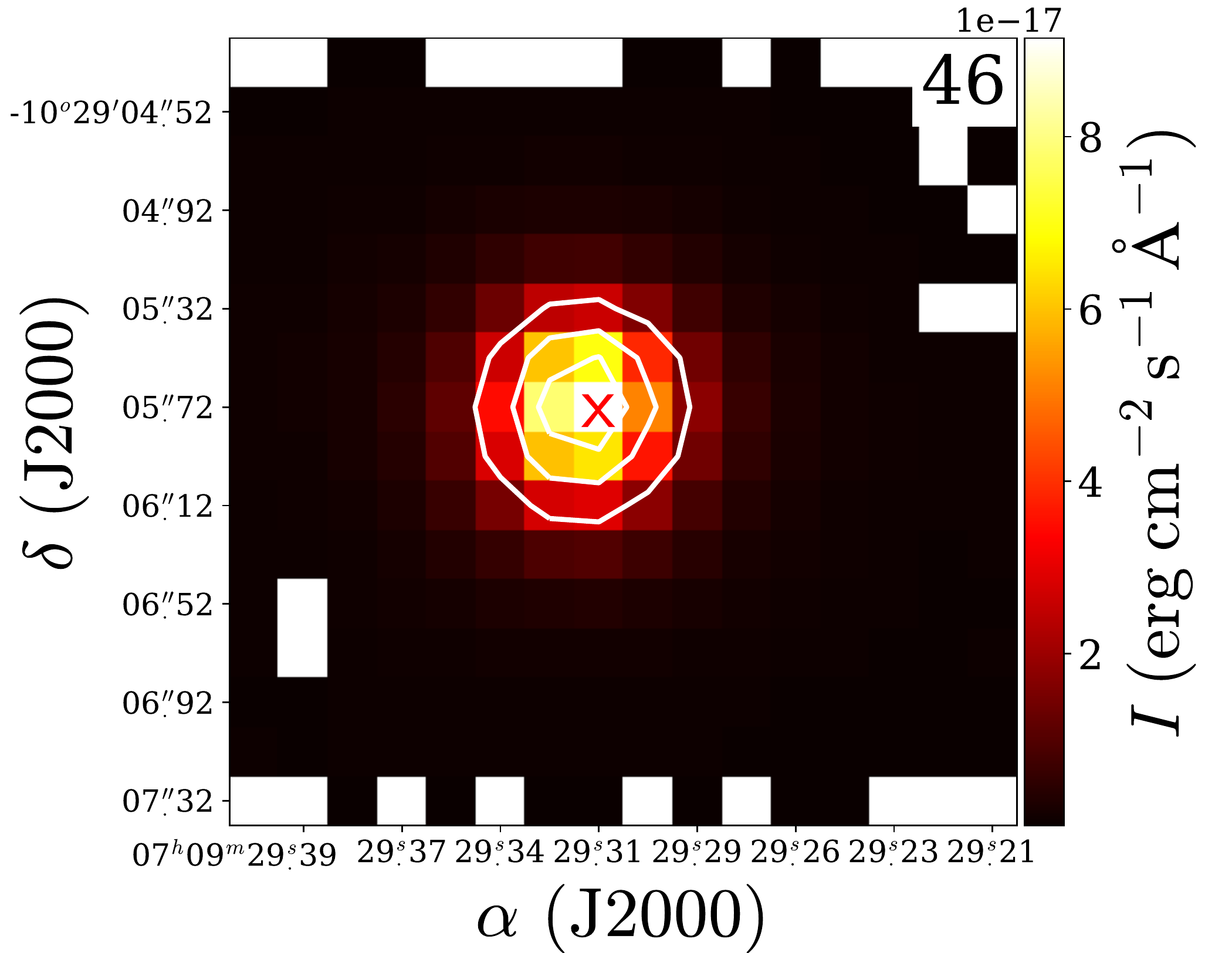}\hspace{-0.1cm}
\includegraphics[width=0.2\textwidth]{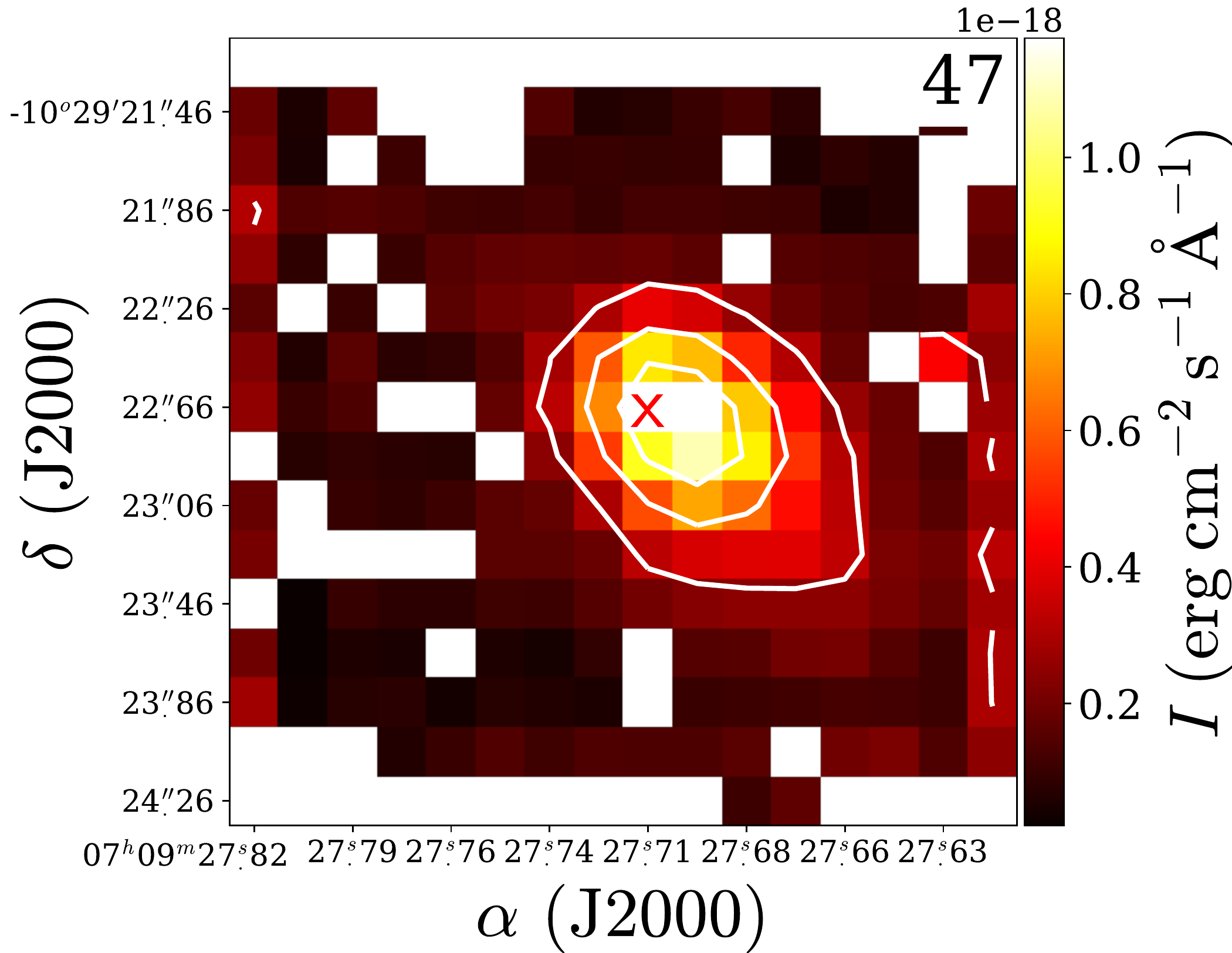}\hspace{-0.1cm}
\includegraphics[width=0.2\textwidth]{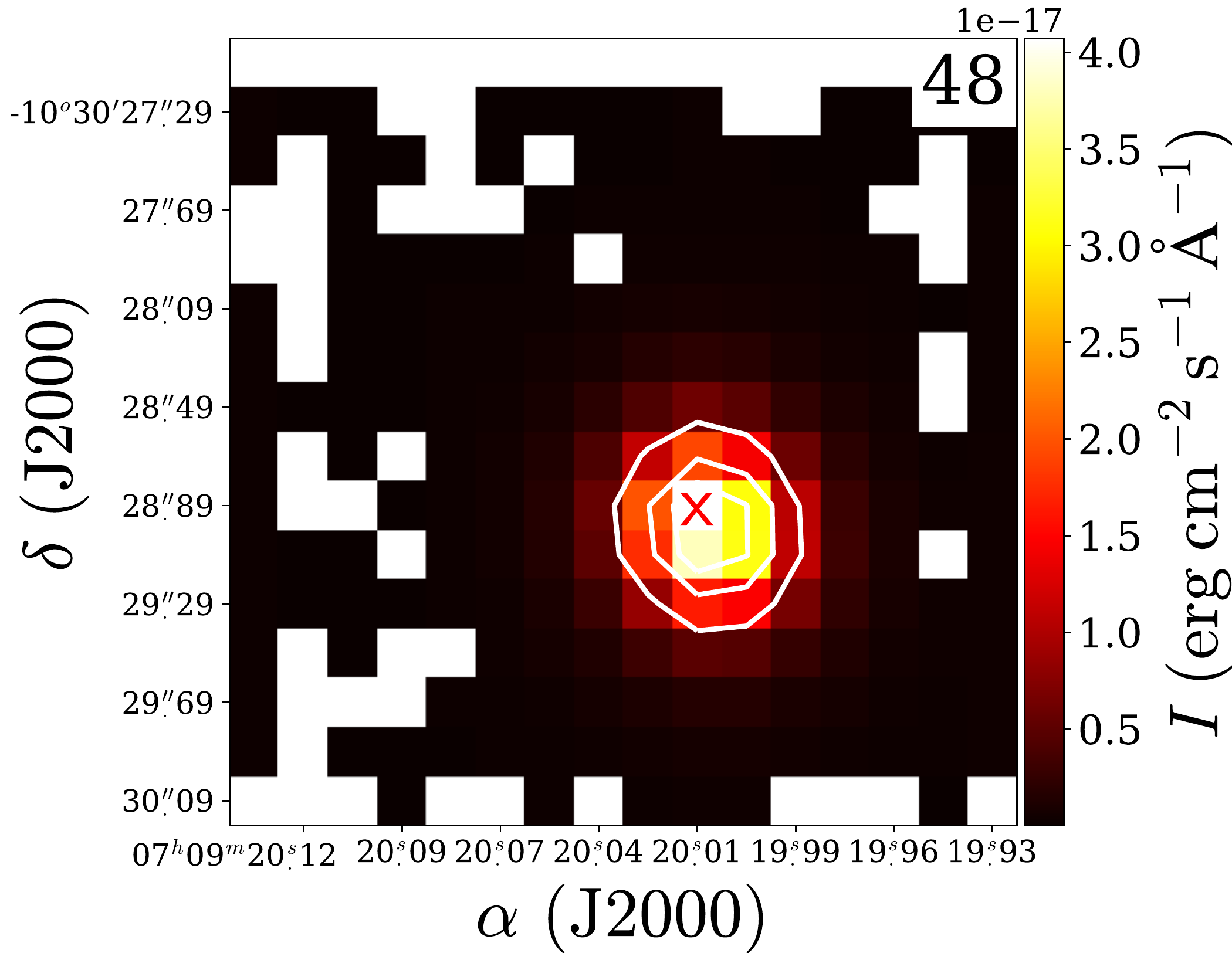}\hspace{-0.1cm}
\caption{KMOS $K$-band continuum maps for single YSO candidates. Source names are provided in the upper right corners (No. in Table \ref{tab:coordinates}). The contour levels are 0.25, 0.5, and 0.75$\times$ the  $K$-band continuum peak. Only pixels with the continuum flux density above 3$\sigma$ are shown.}
\label{fig:cont}
\end{figure*}

\addtocounter{figure}{-1}
\begin{figure*}
\centering
\includegraphics[width=0.2\textwidth]{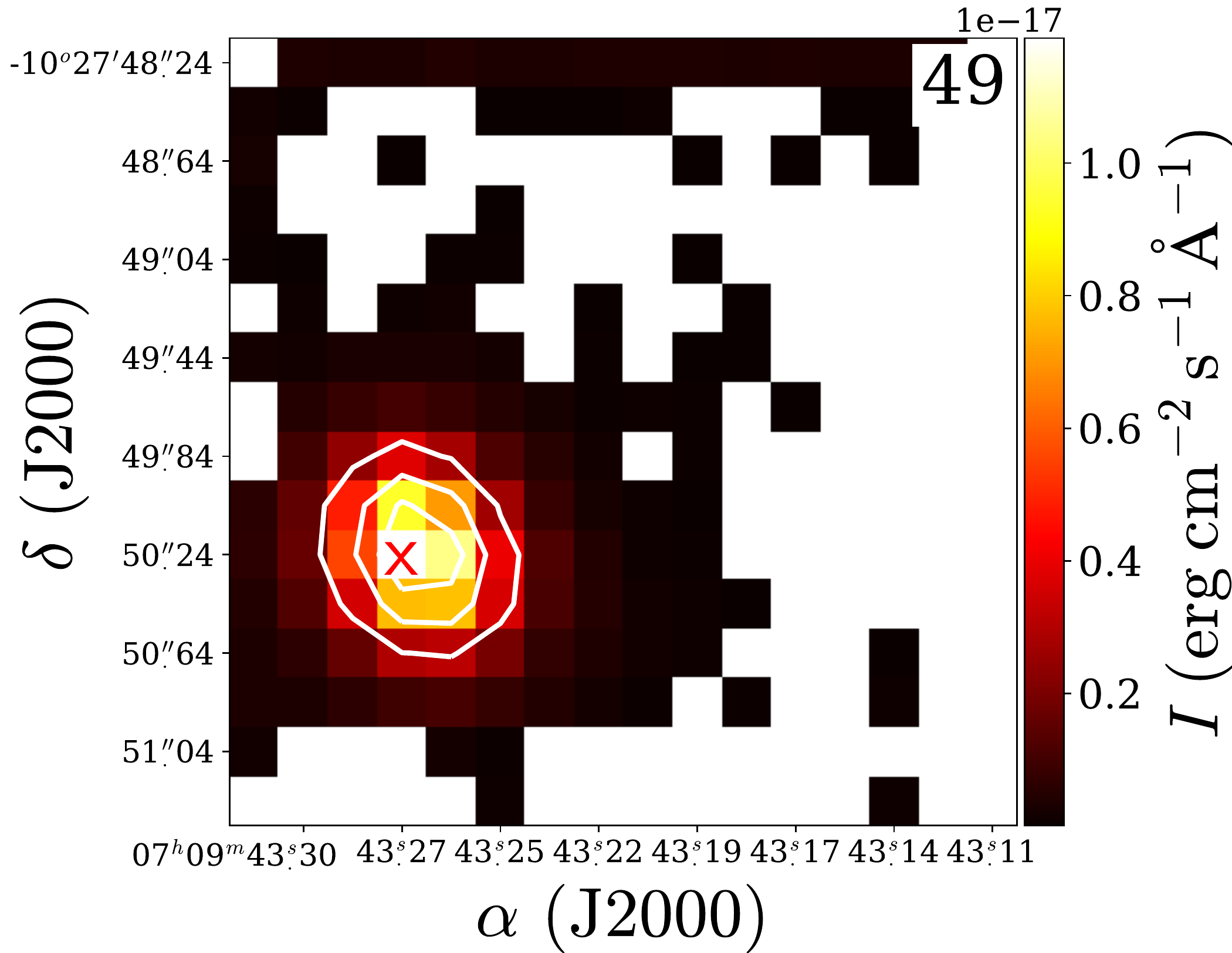}\hspace{-0.1cm}
\includegraphics[width=0.2\textwidth]{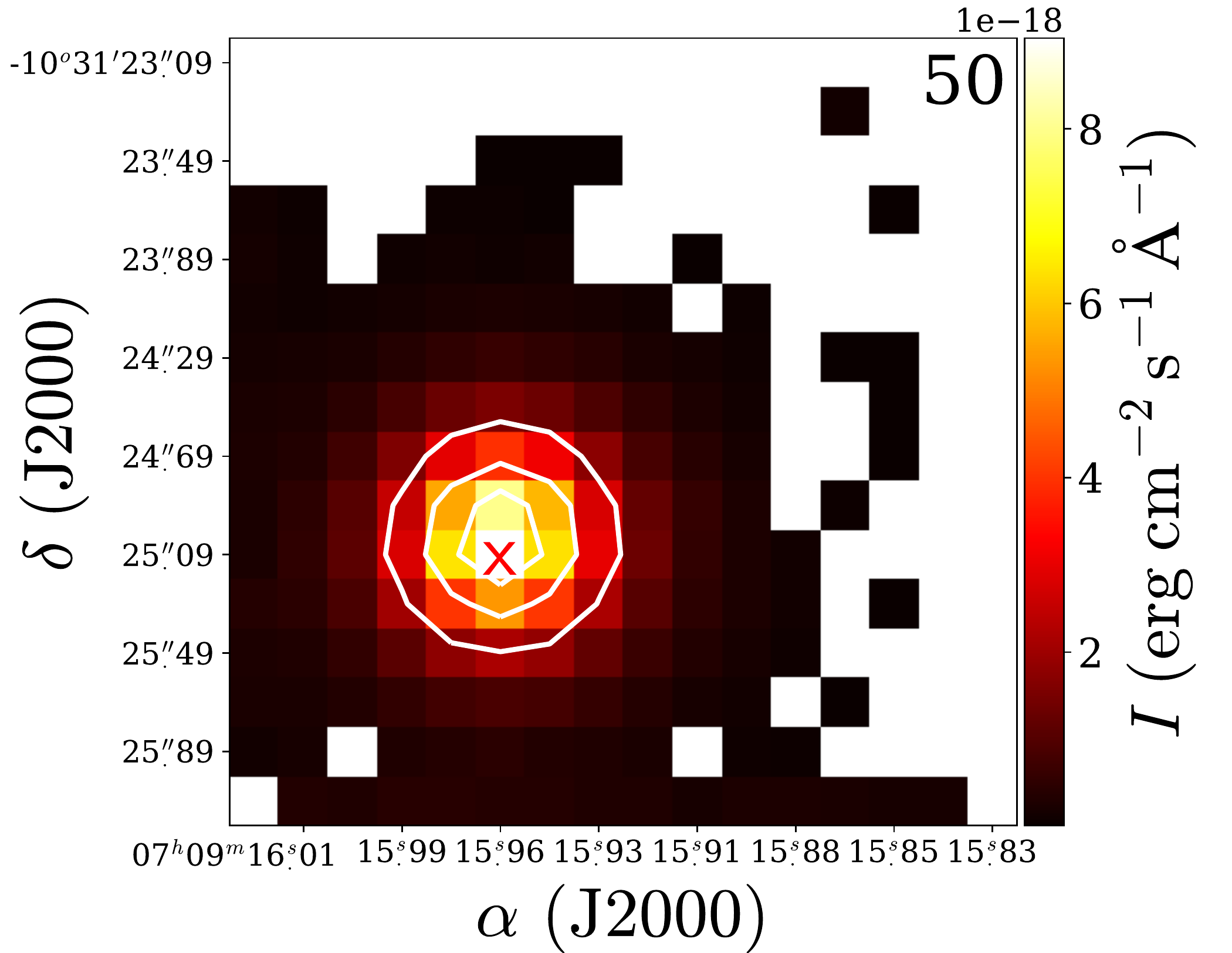}\hspace{-0.1cm}
\includegraphics[width=0.2\textwidth]{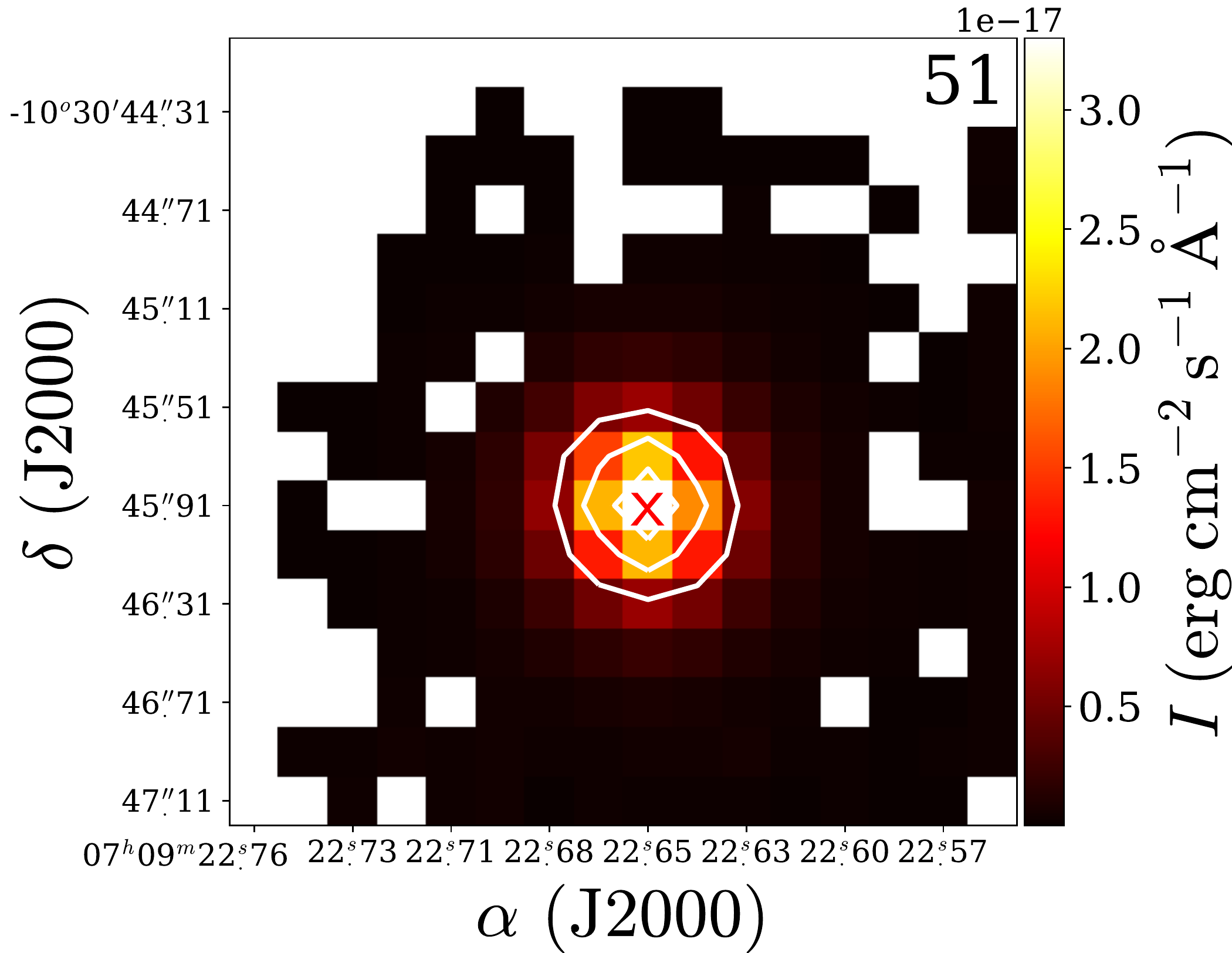}\hspace{-0.1cm}
\includegraphics[width=0.2\textwidth]{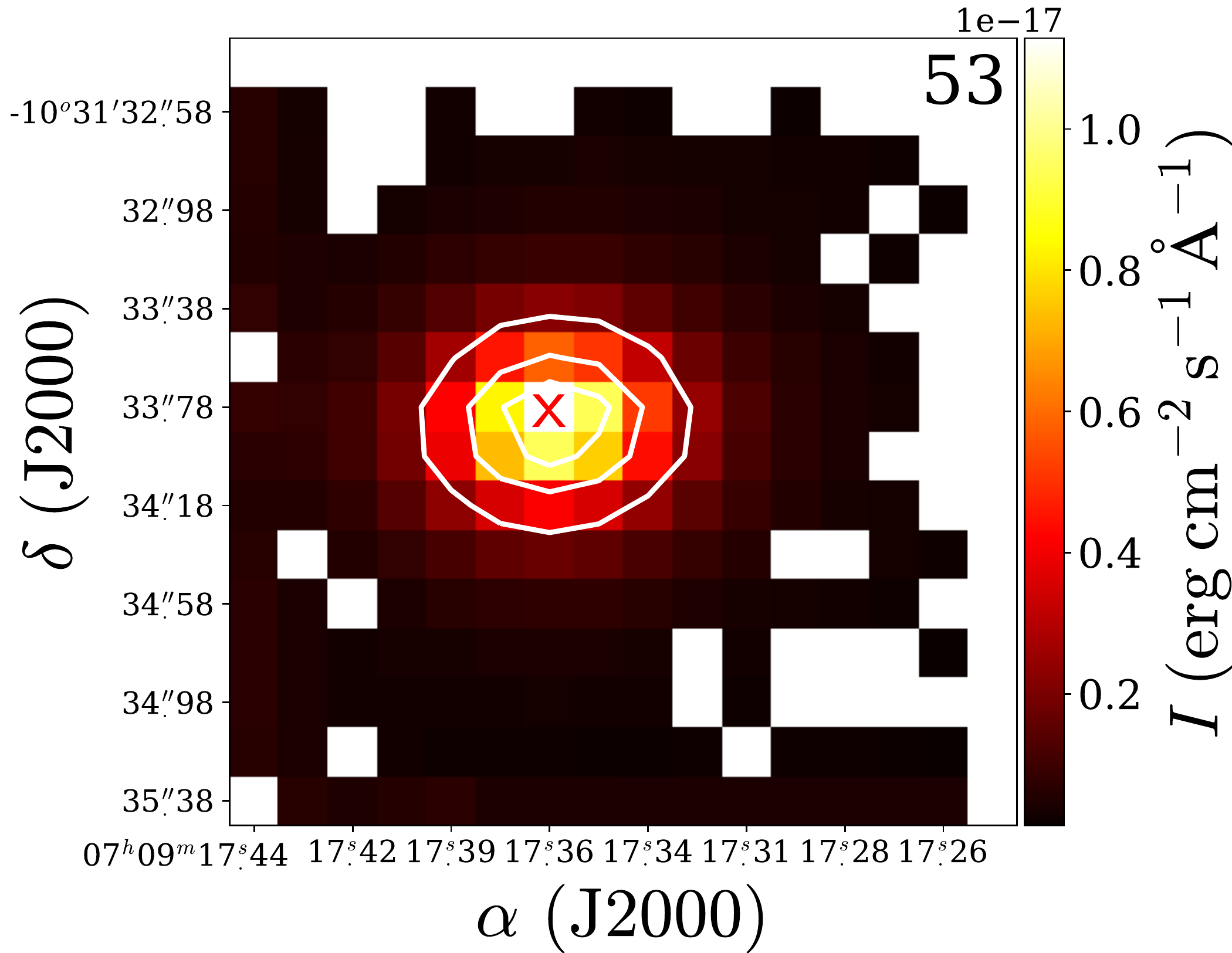}\hspace{-0.1cm}
\includegraphics[width=0.2\textwidth]{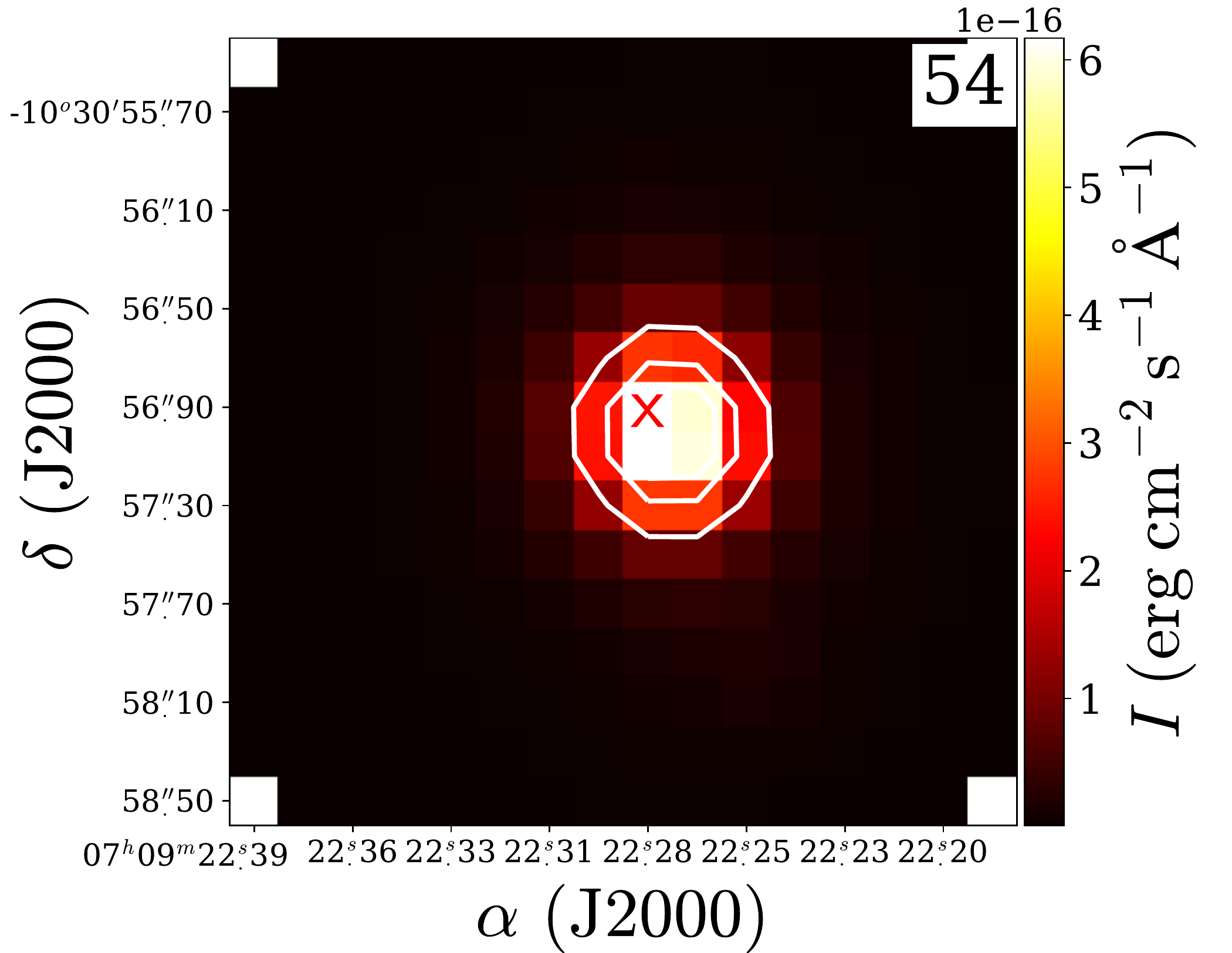}\hspace{-0.1cm}
\includegraphics[width=0.2\textwidth]{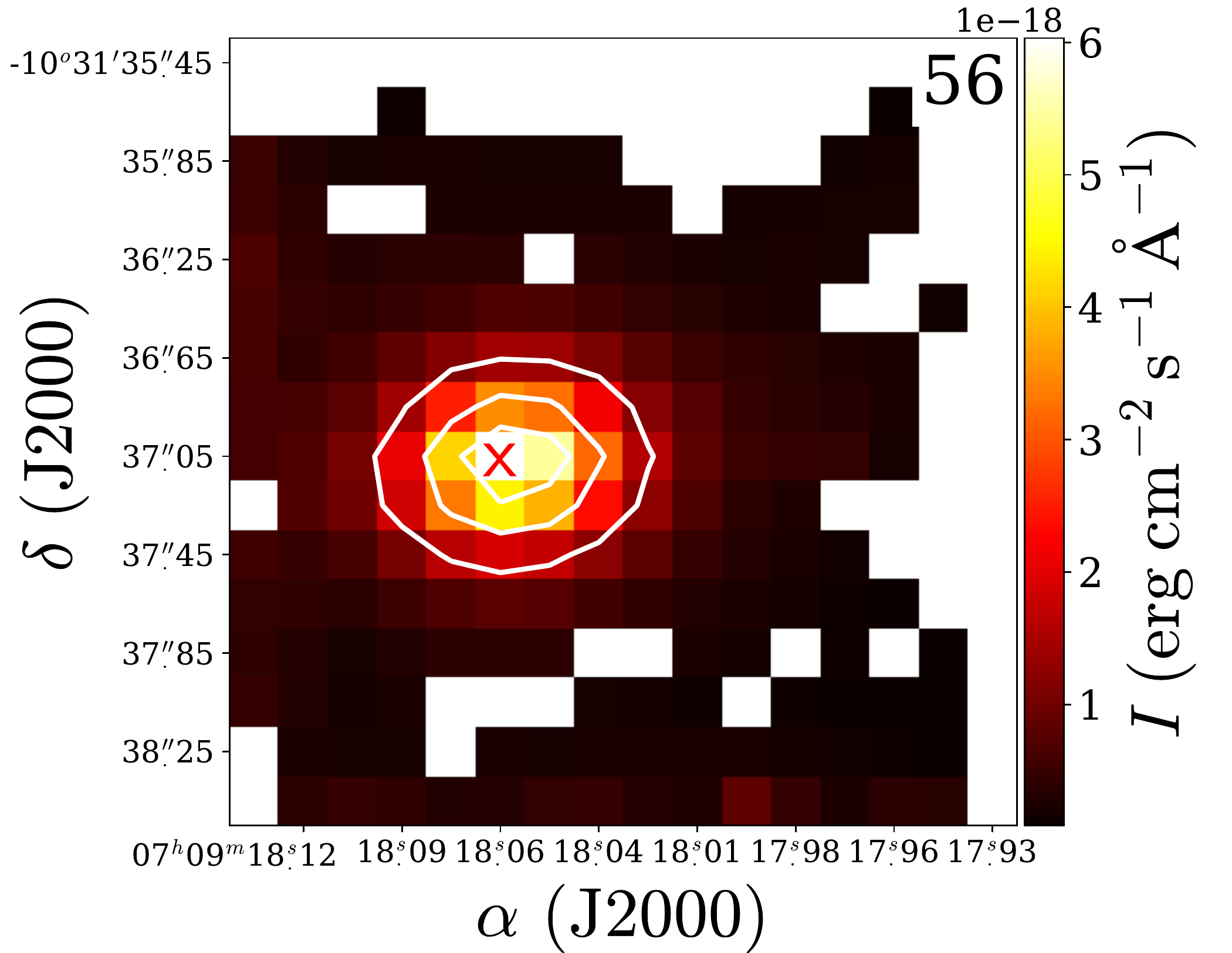}\hspace{-0.1cm}
\includegraphics[width=0.2\textwidth]{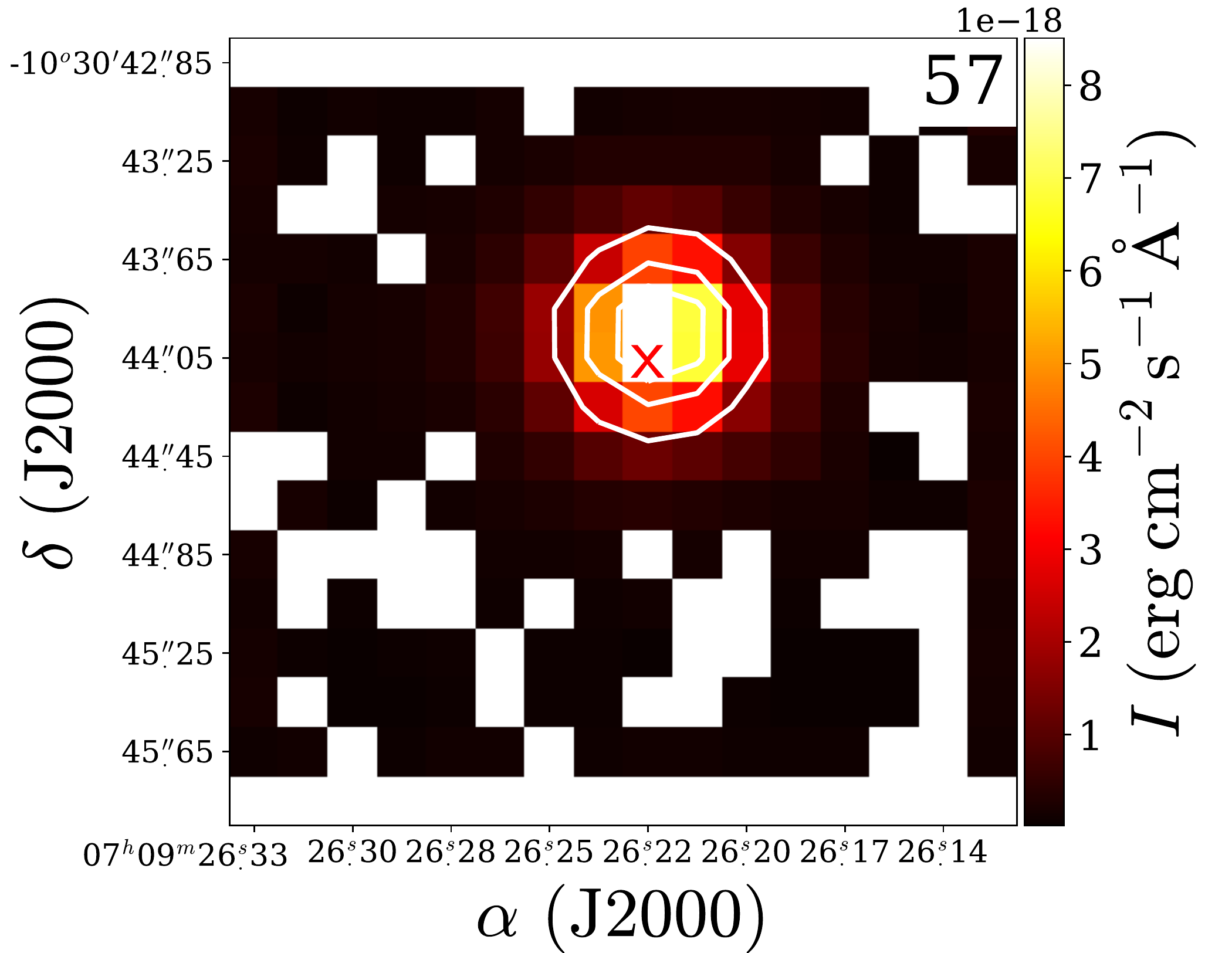}\hspace{-0.1cm}
\includegraphics[width=0.2\textwidth]{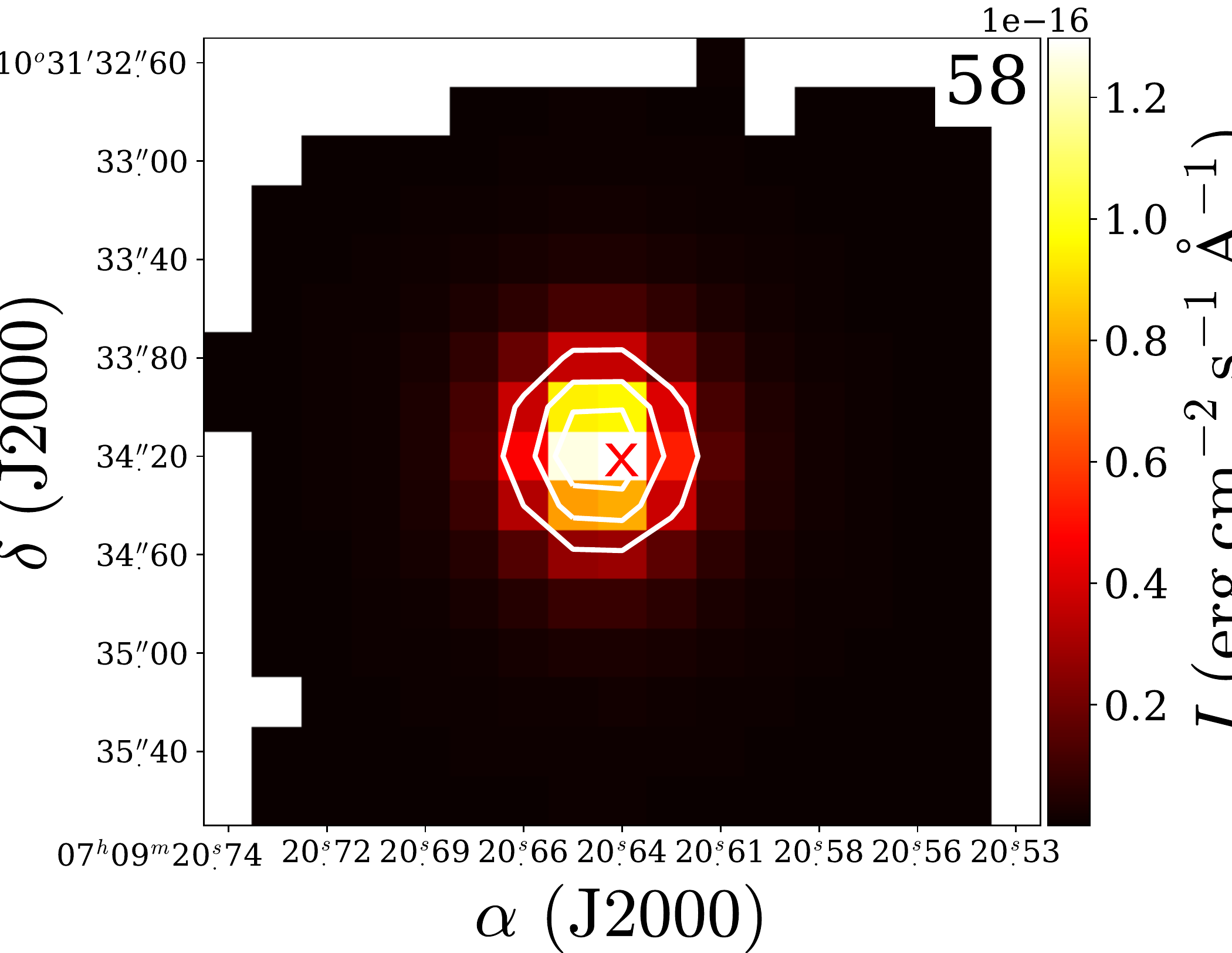}\hspace{-0.1cm}
\includegraphics[width=0.2\textwidth]{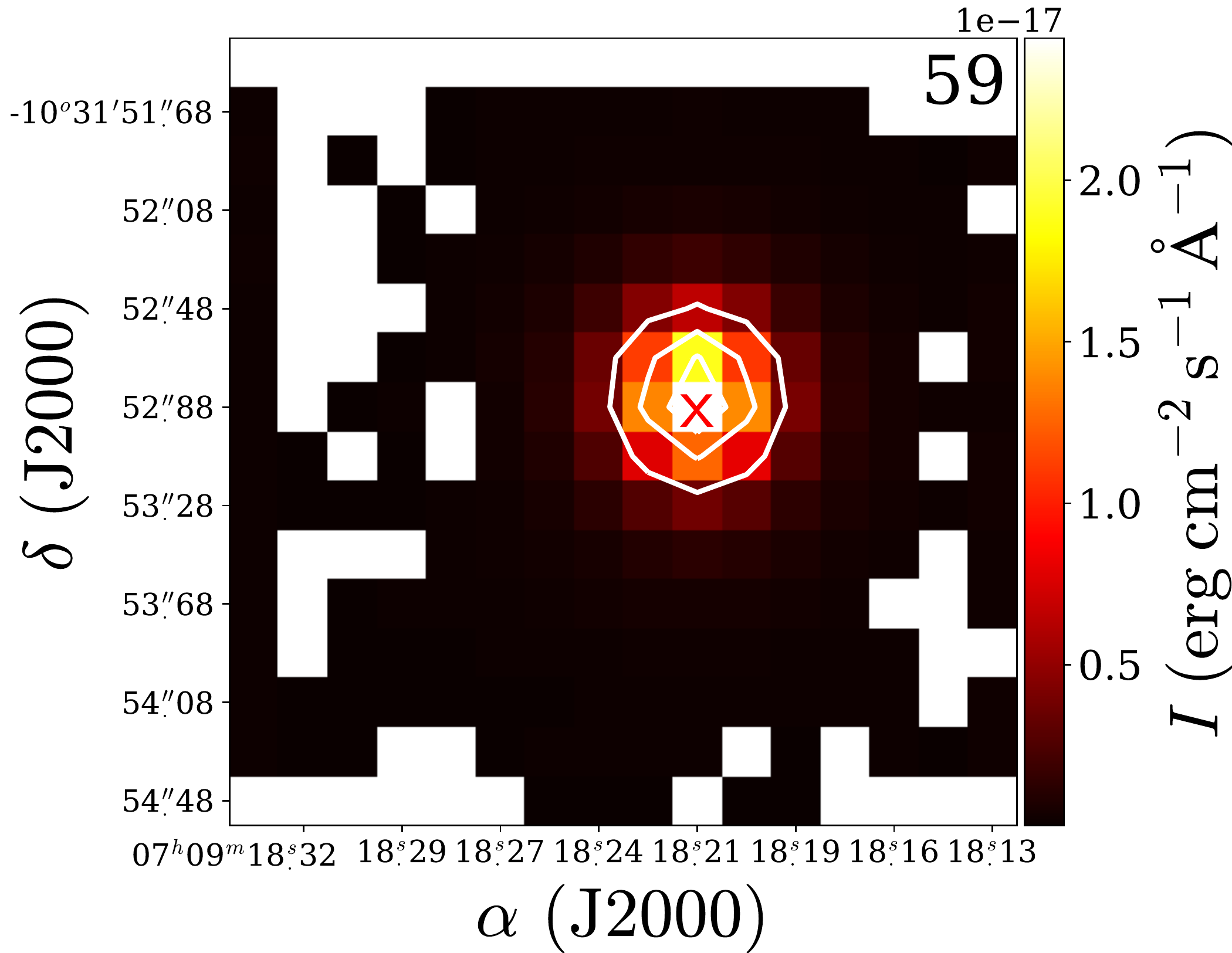}\hspace{-0.1cm}
\includegraphics[width=0.2\textwidth]{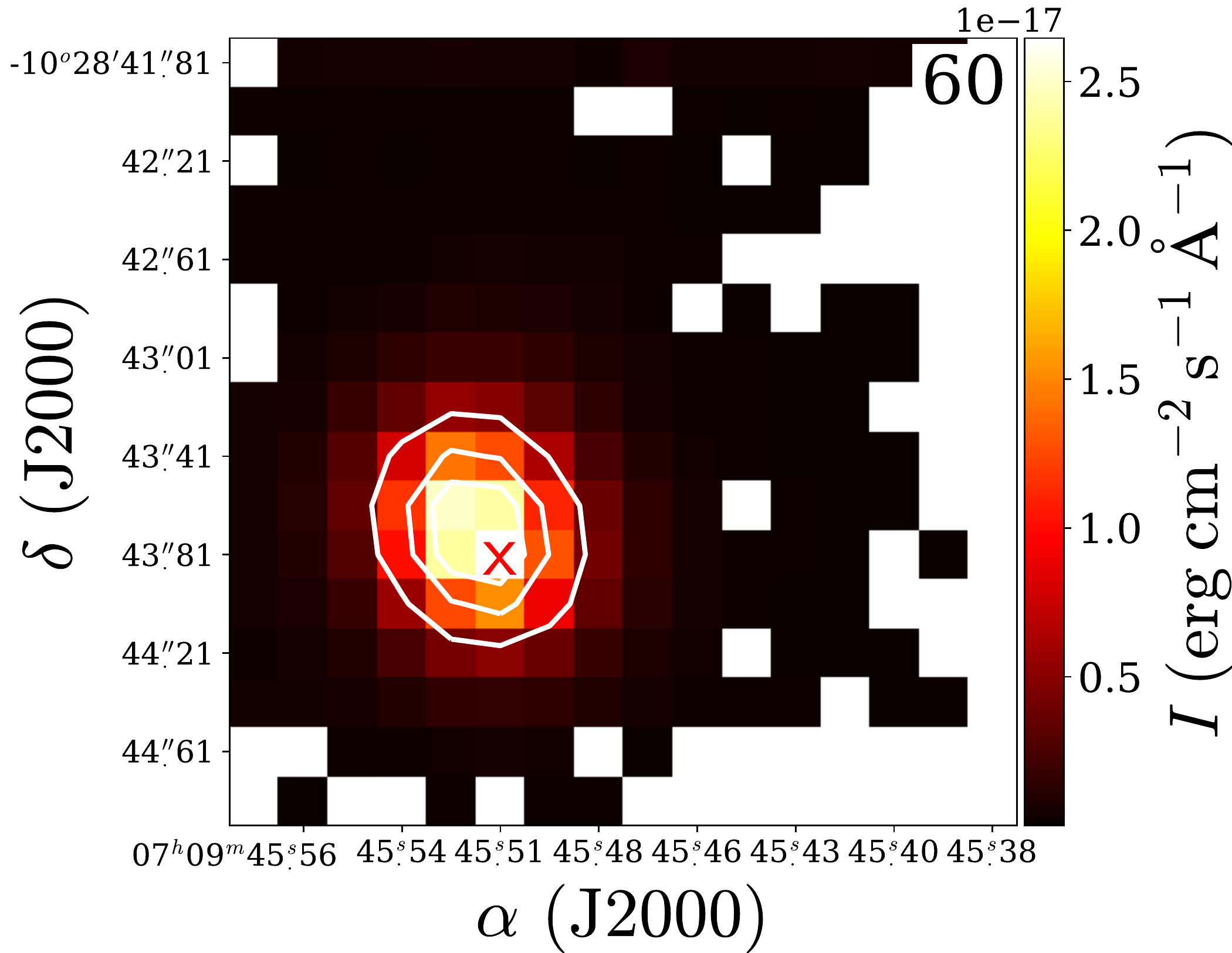}\hspace{-0.1cm}
\includegraphics[width=0.2\textwidth]{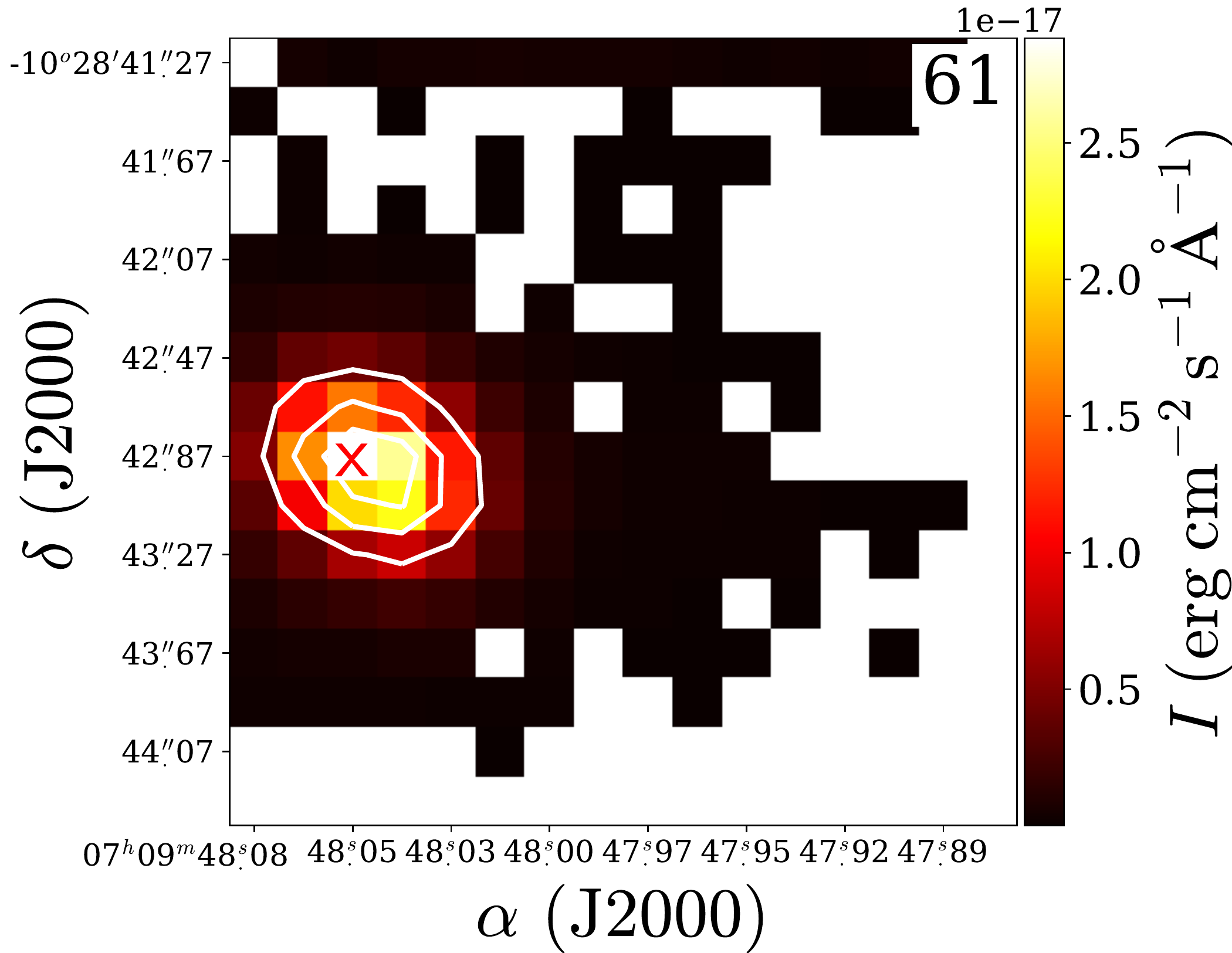}\hspace{-0.1cm}
\includegraphics[width=0.2\textwidth]{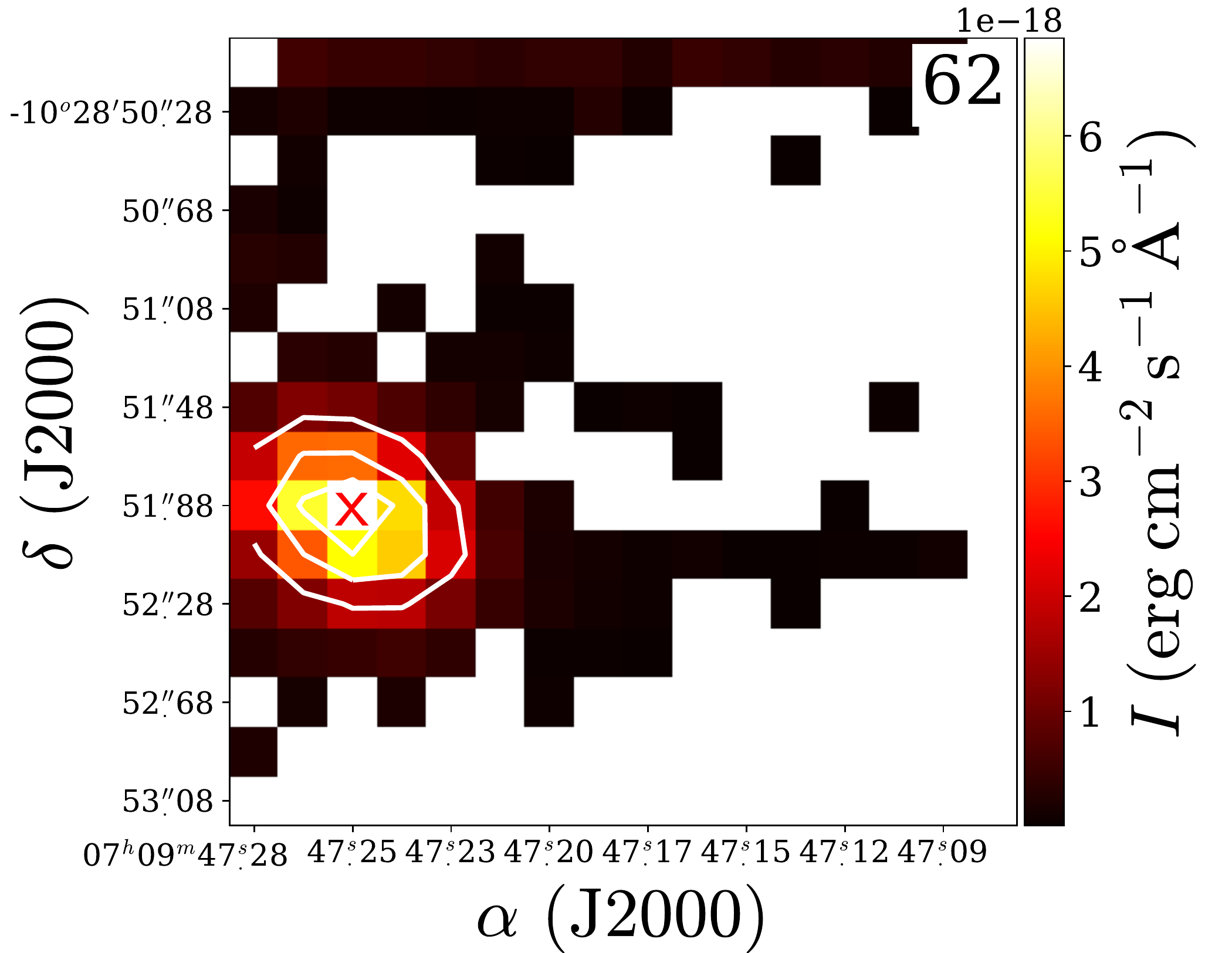}\hspace{-0.1cm}
\includegraphics[width=0.2\textwidth]{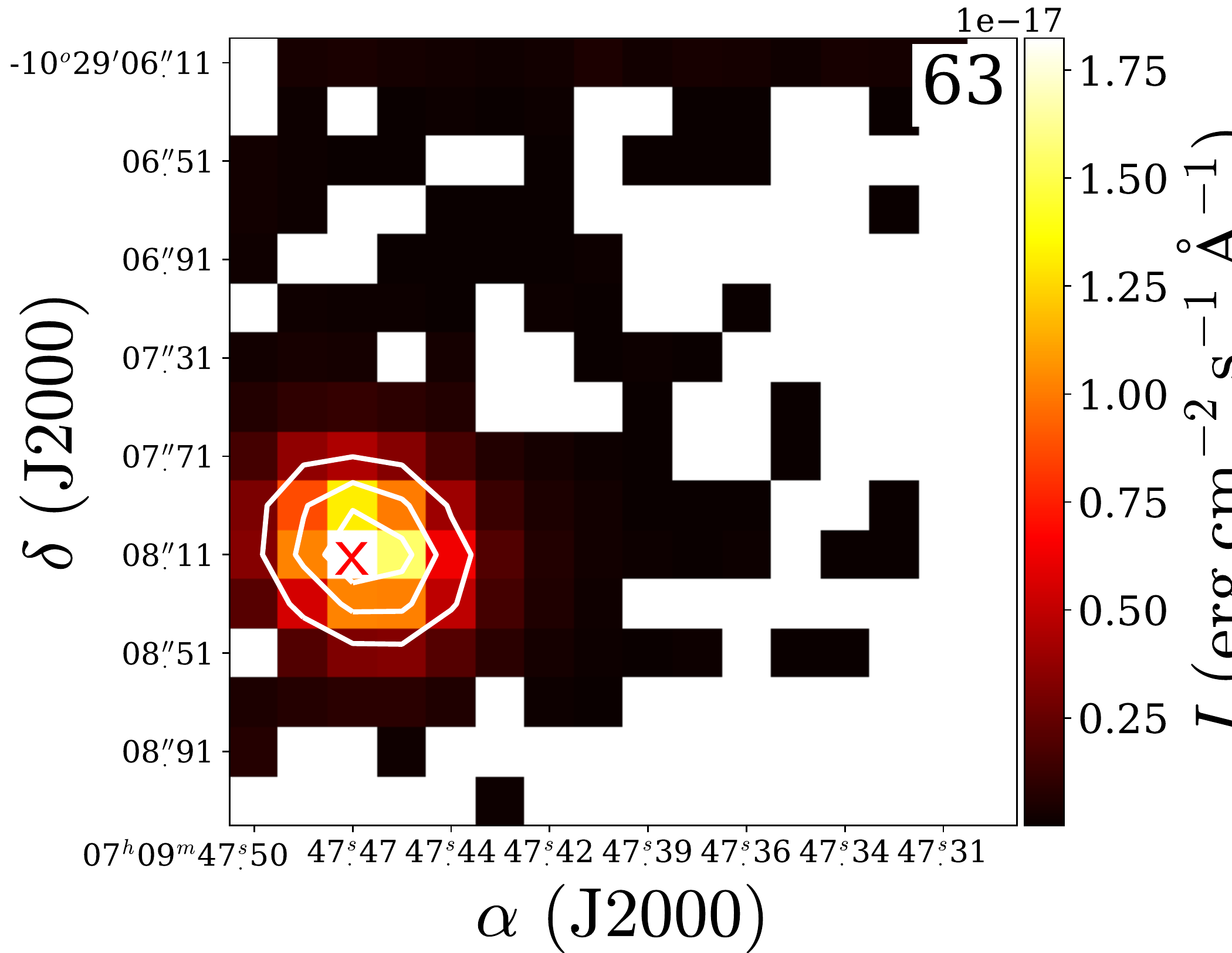}\hspace{-0.1cm}
\includegraphics[width=0.2\textwidth]{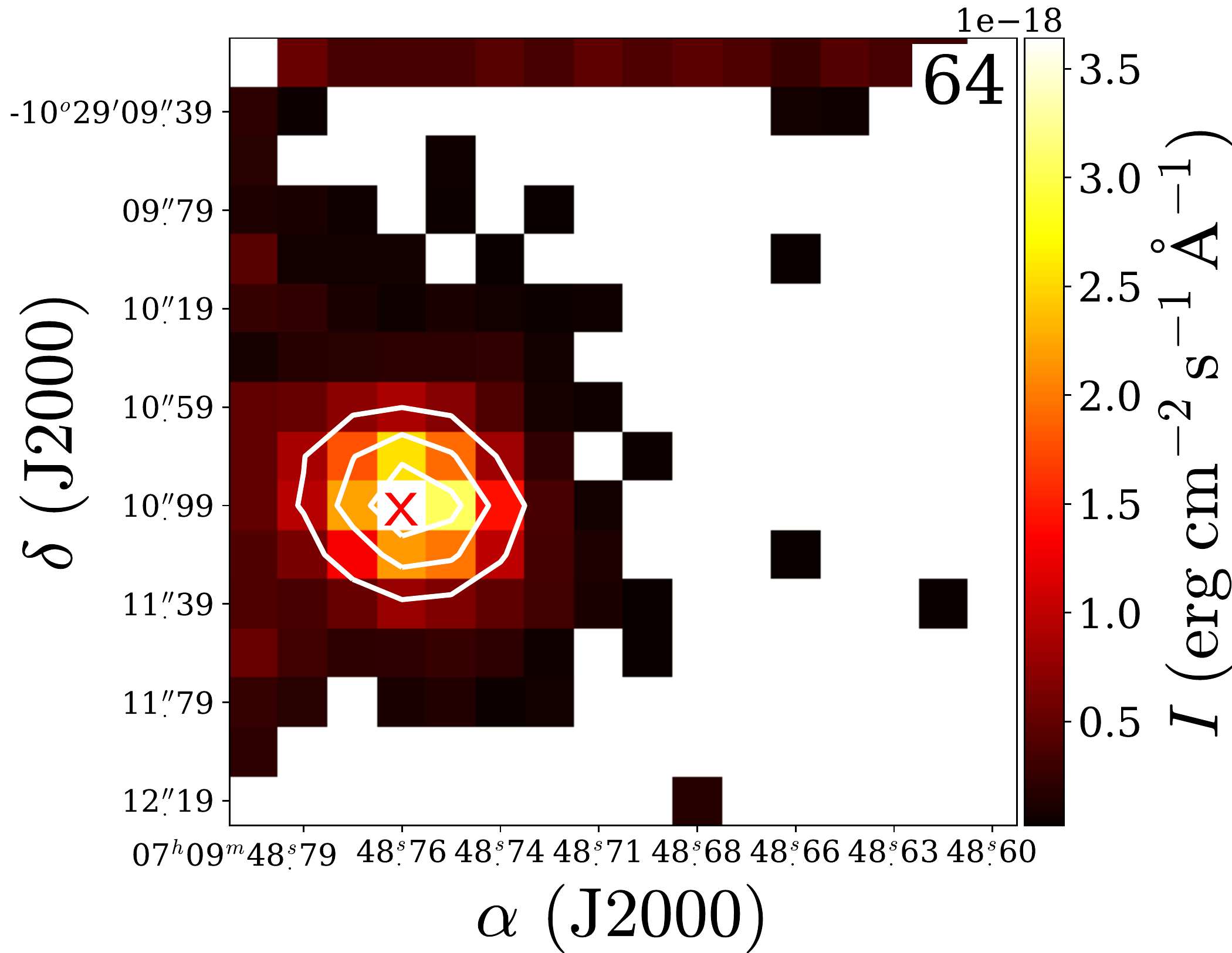}\hspace{-0.1cm}
\includegraphics[width=0.2\textwidth]{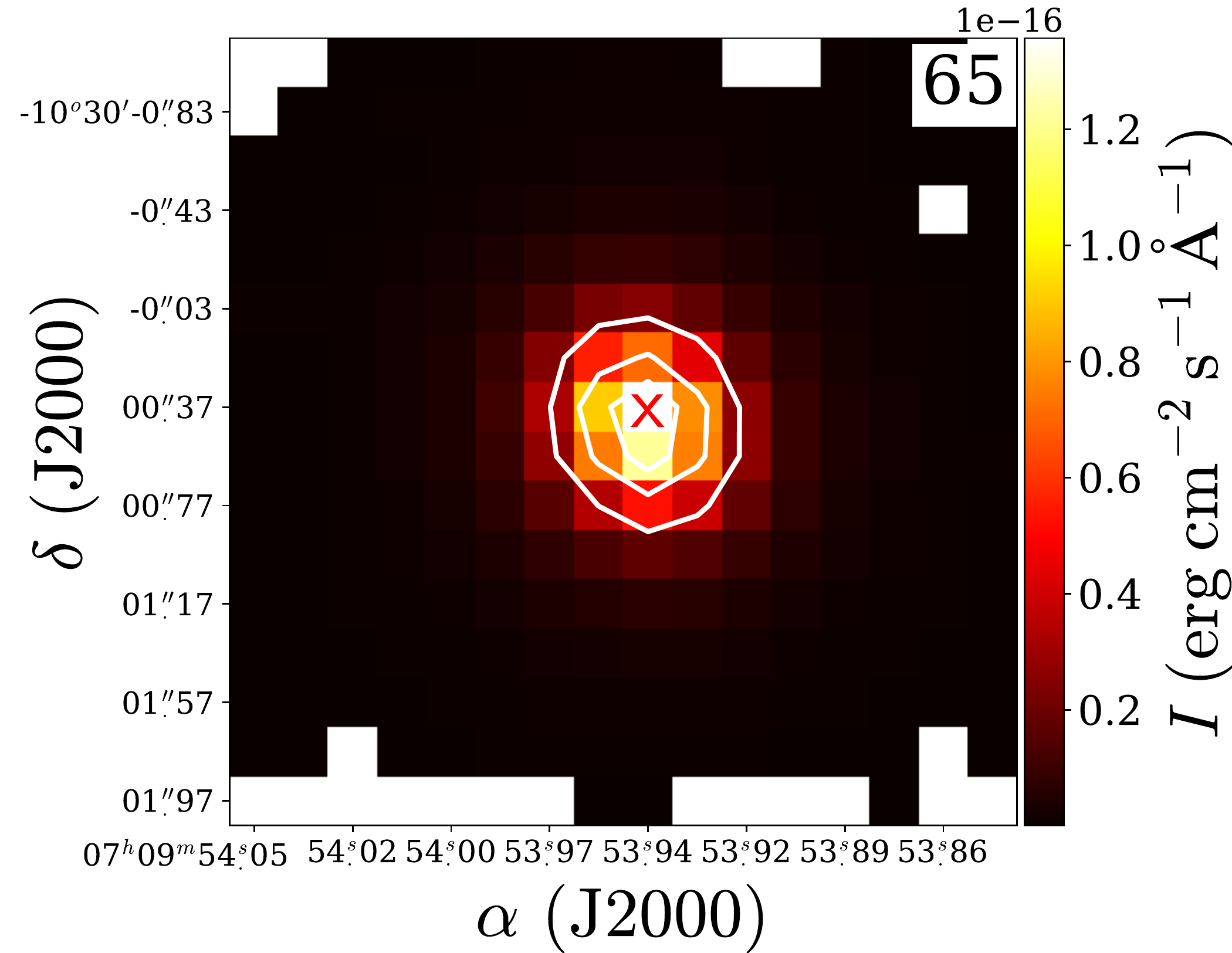}\hspace{-0.1cm}
\includegraphics[width=0.2\textwidth]{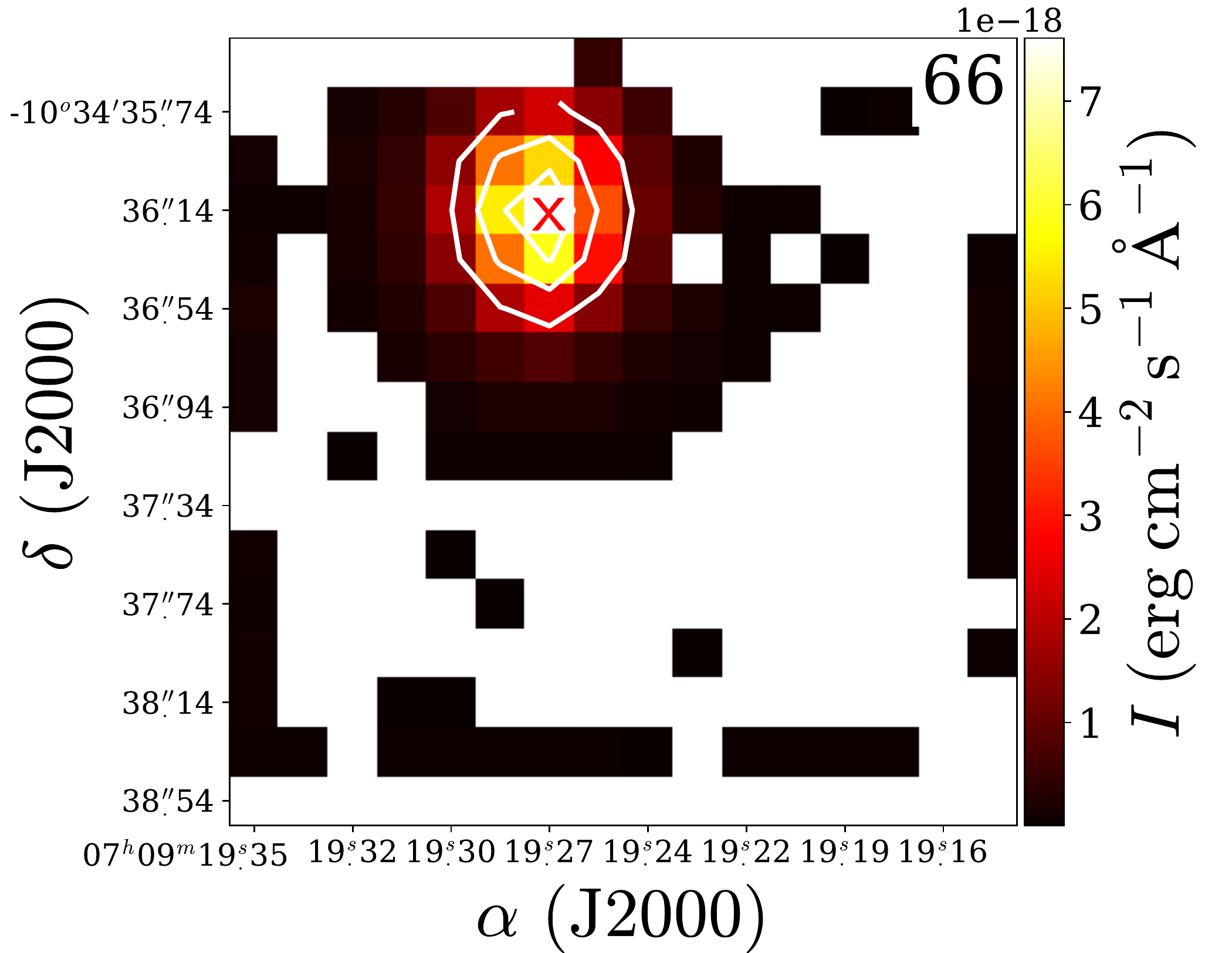}\hspace{-0.1cm}
\includegraphics[width=0.2\textwidth]{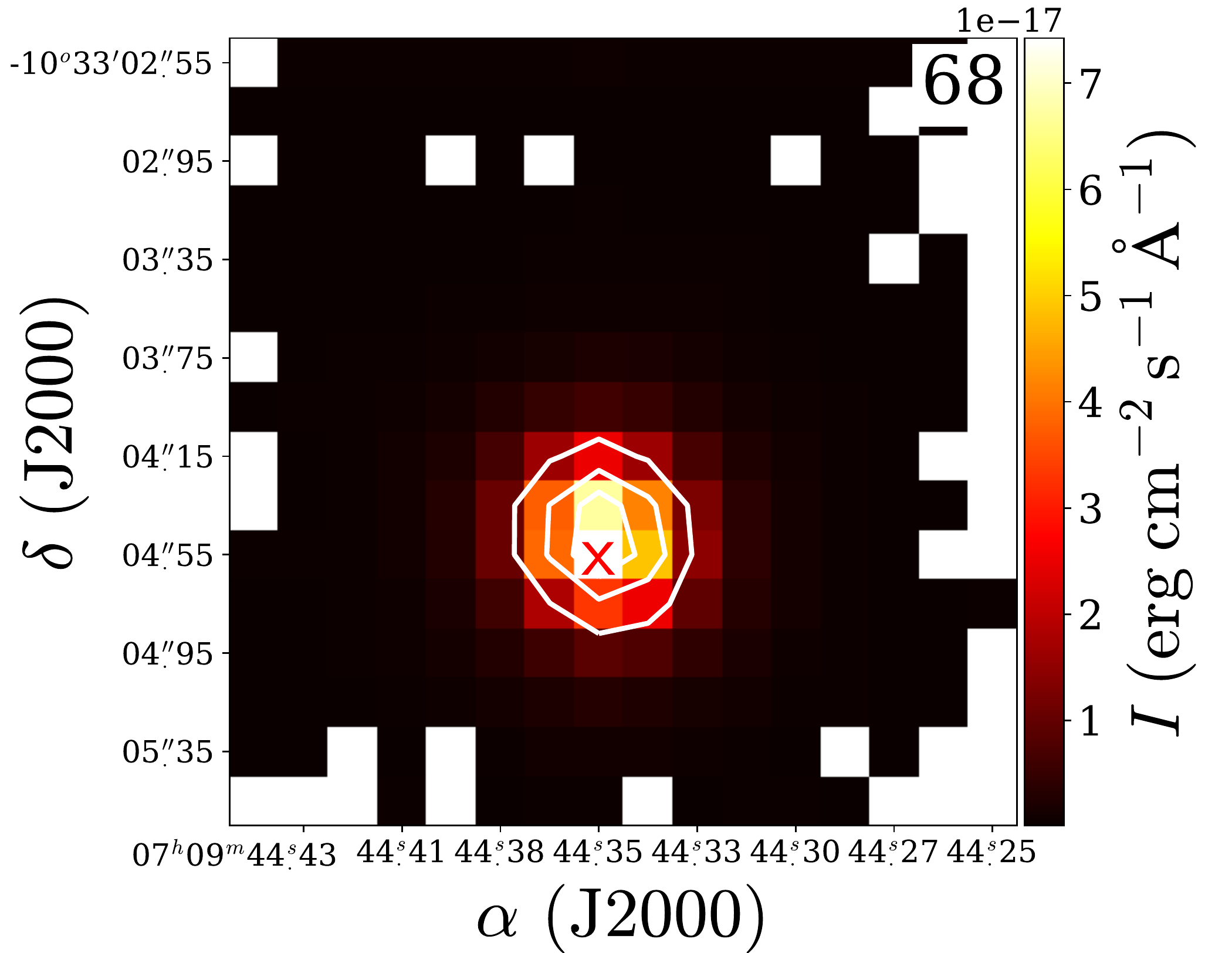}\hspace{-0.1cm}
\includegraphics[width=0.2\textwidth]{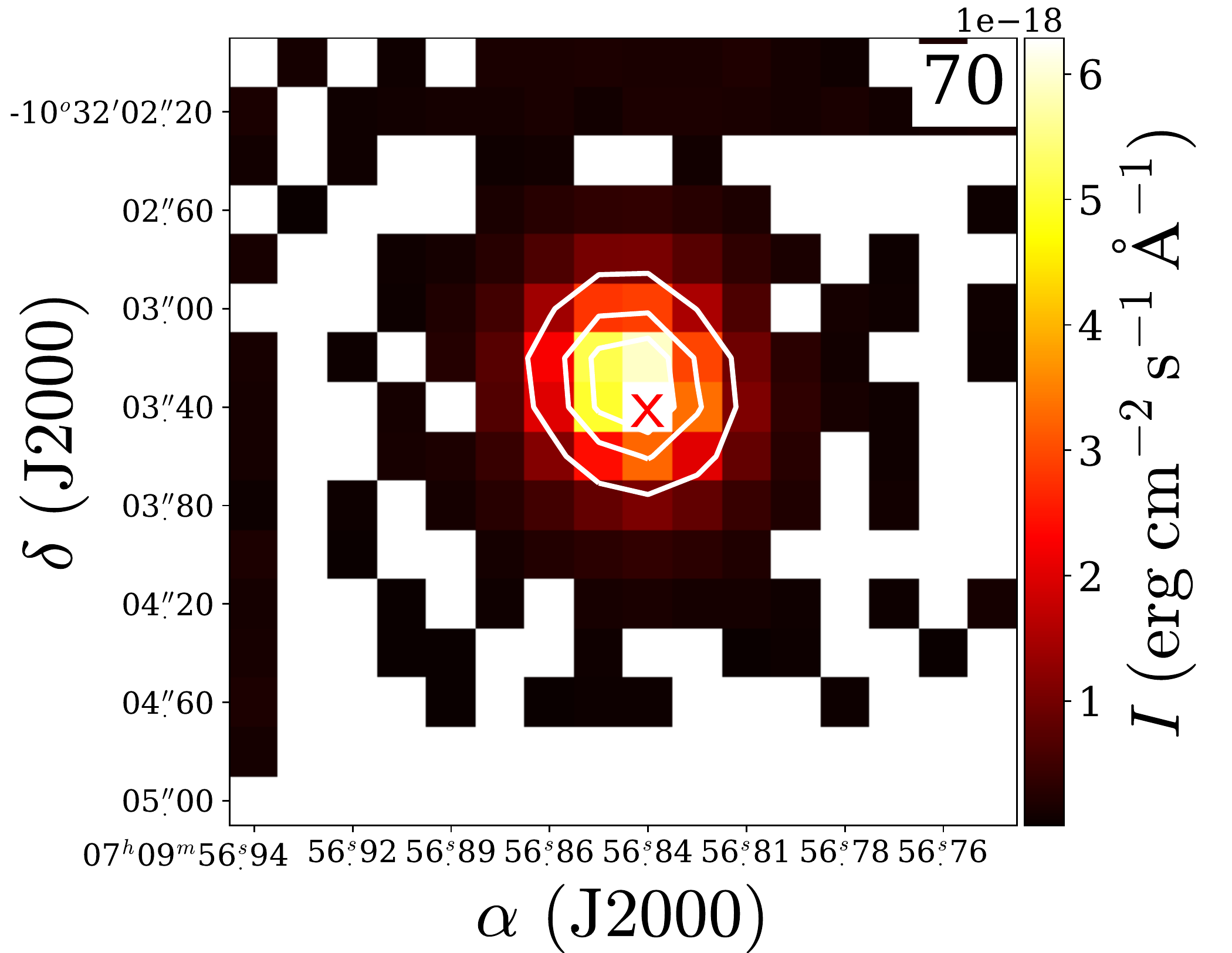}\hspace{-0.1cm}
\includegraphics[width=0.2\textwidth]{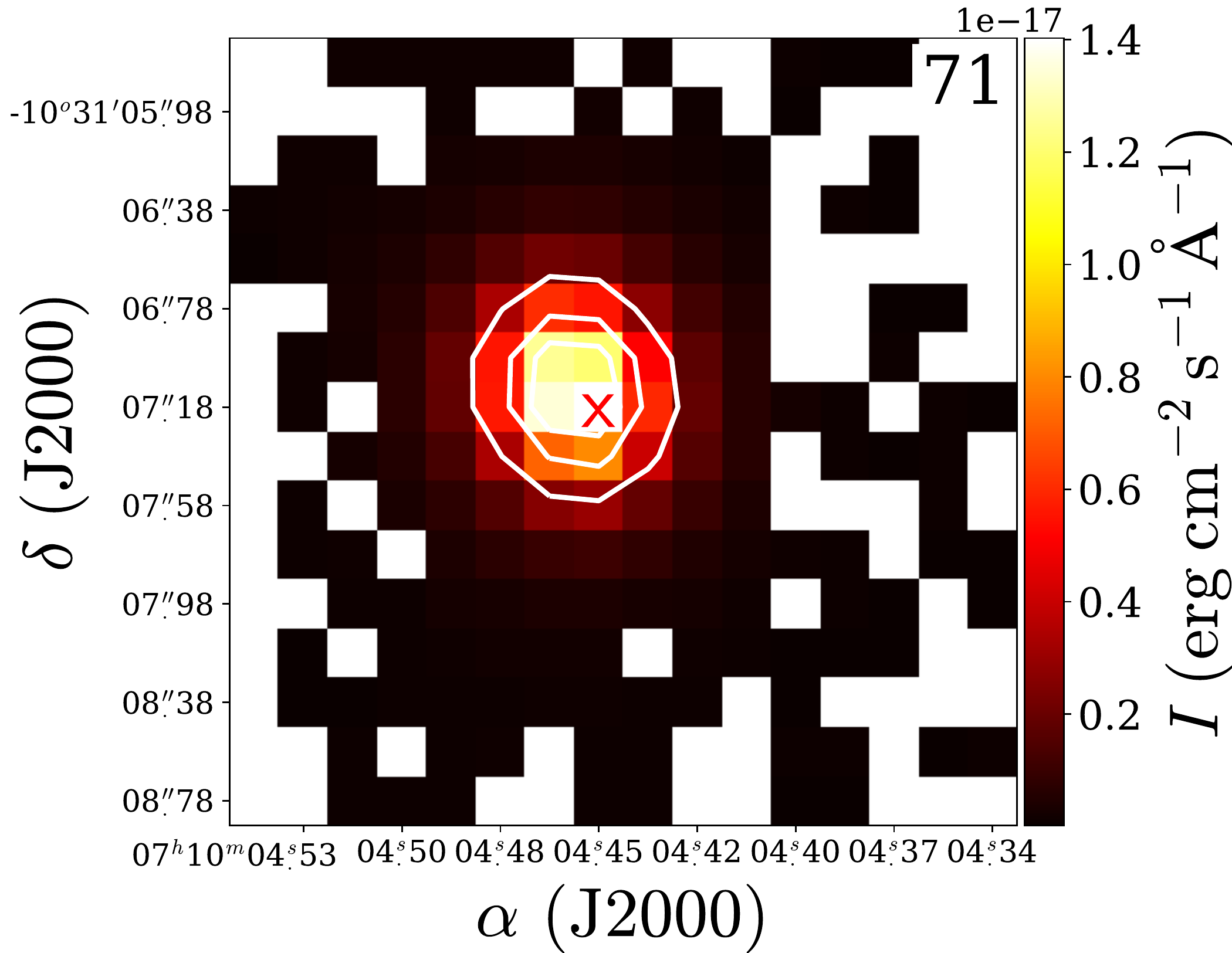}\hspace{-0.1cm}
\includegraphics[width=0.2\textwidth]{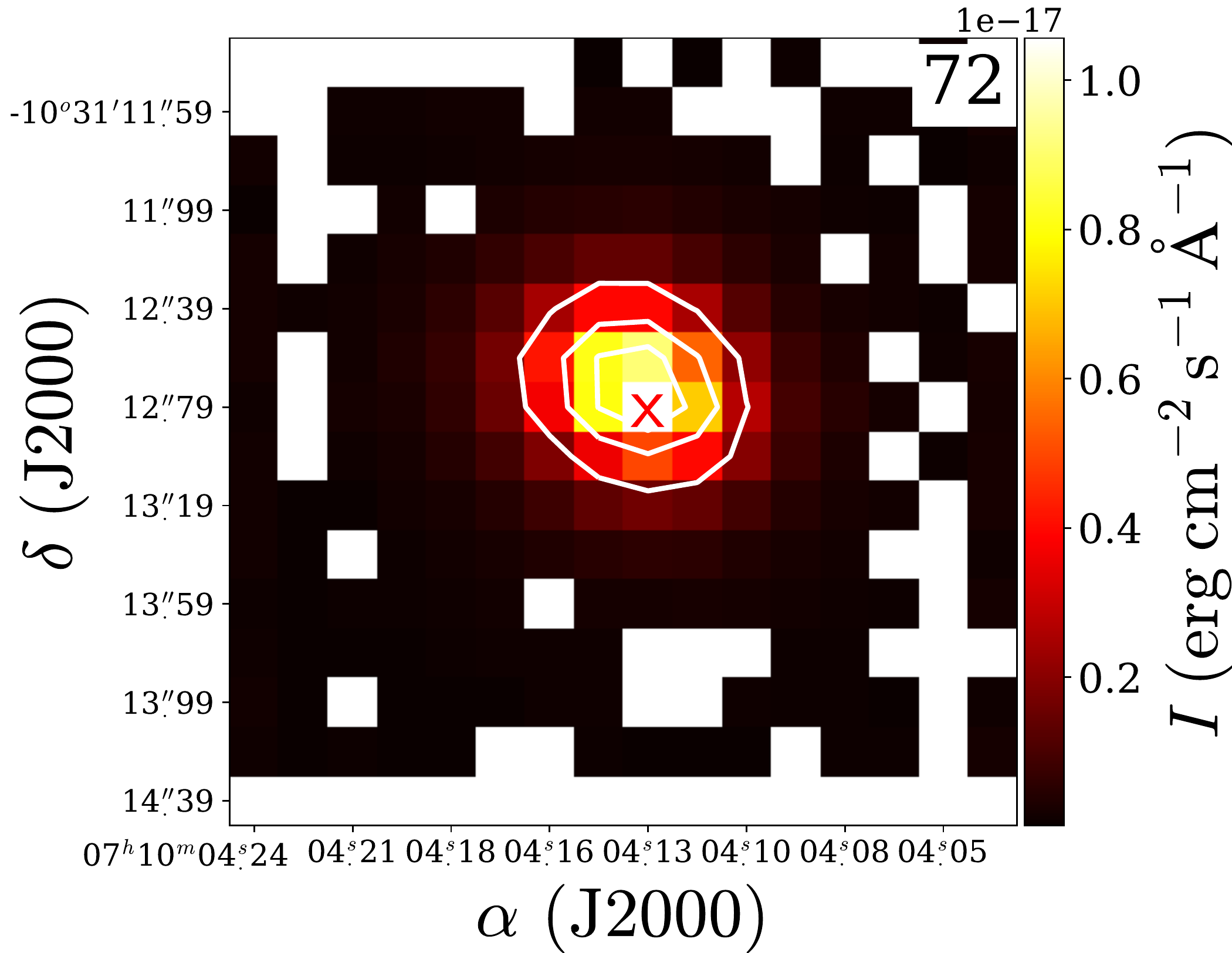}\hspace{-0.1cm}
\includegraphics[width=0.2\textwidth]{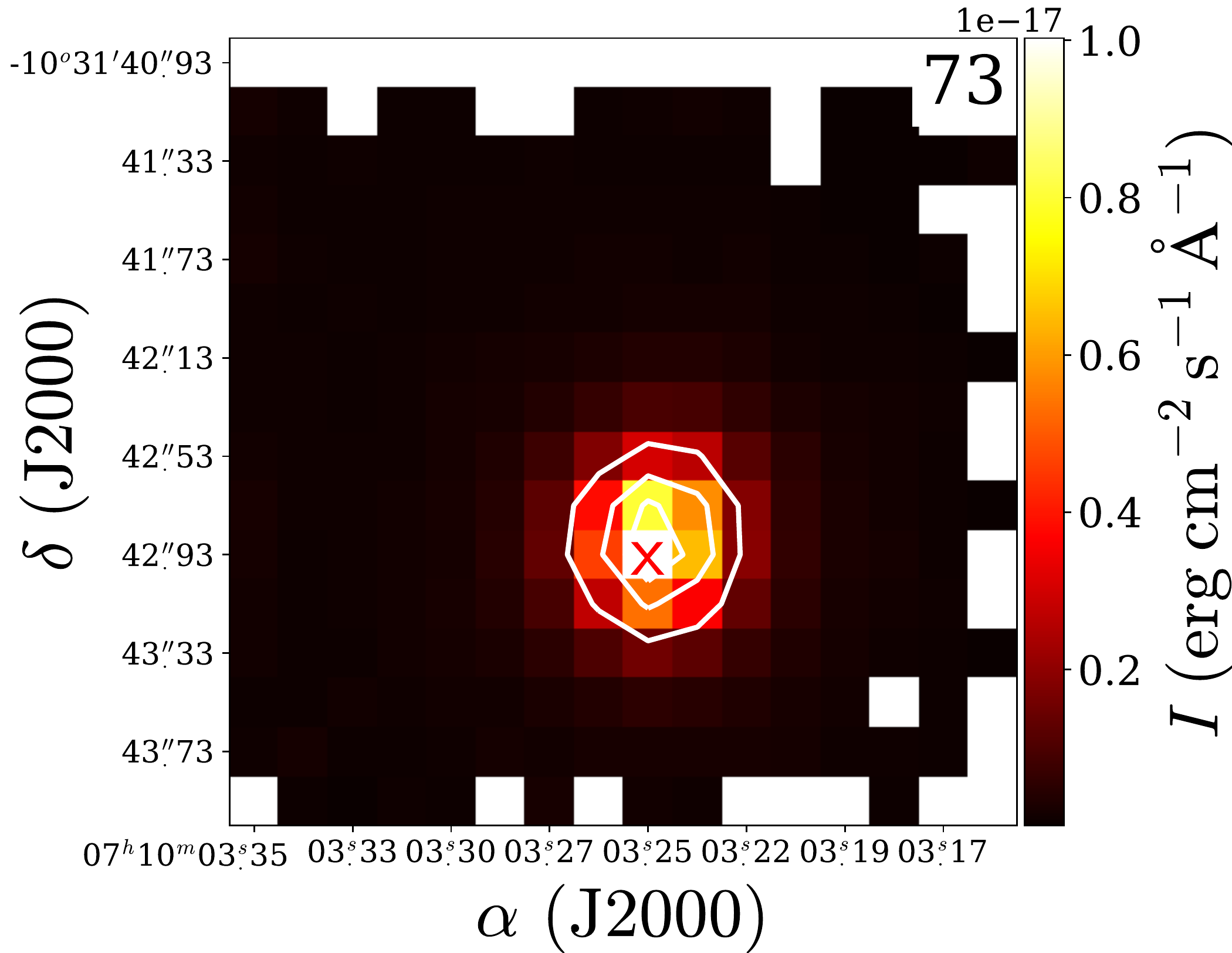}\hspace{-0.1cm}
\includegraphics[width=0.2\textwidth]{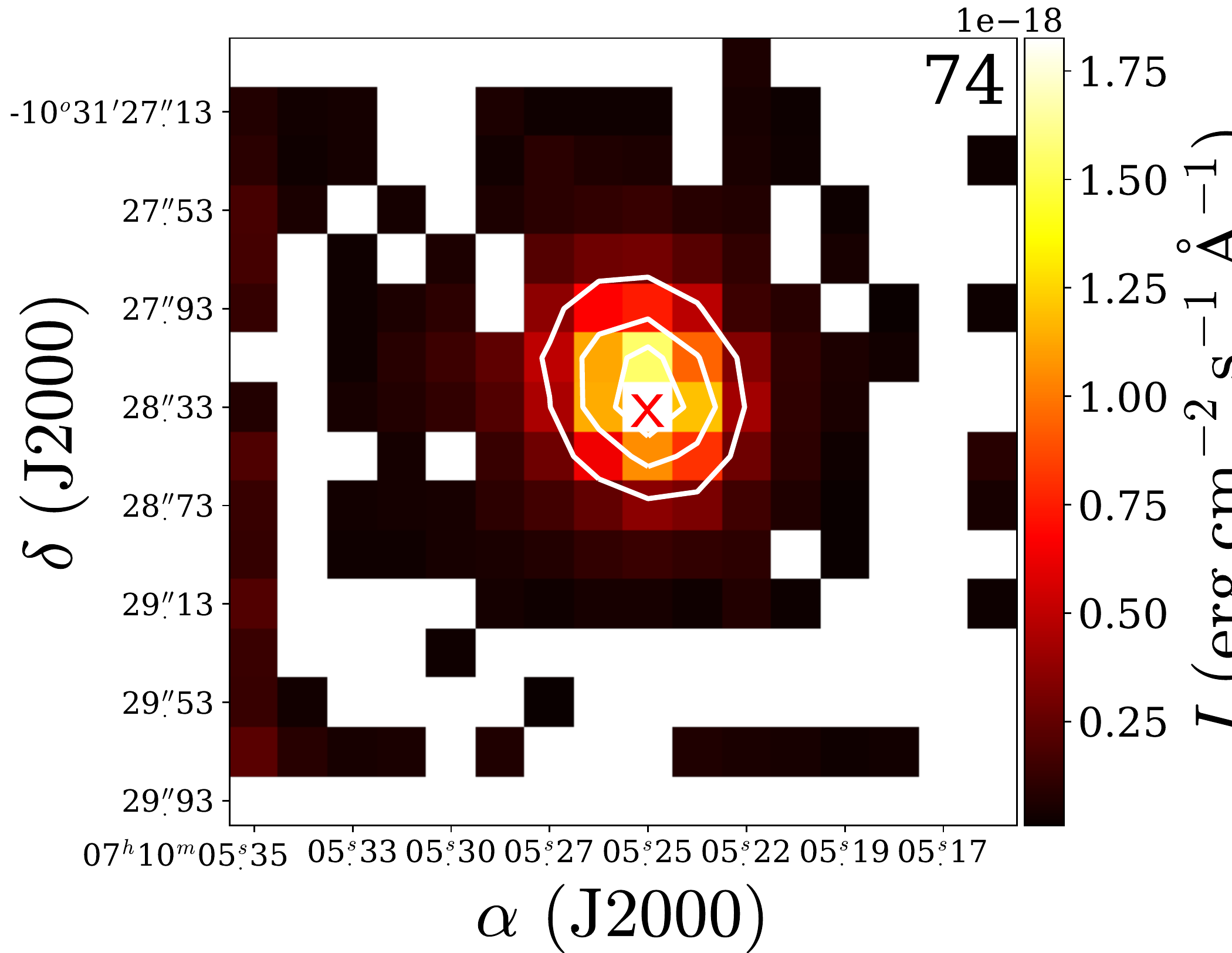}\hspace{-0.1cm}
\includegraphics[width=0.2\textwidth]{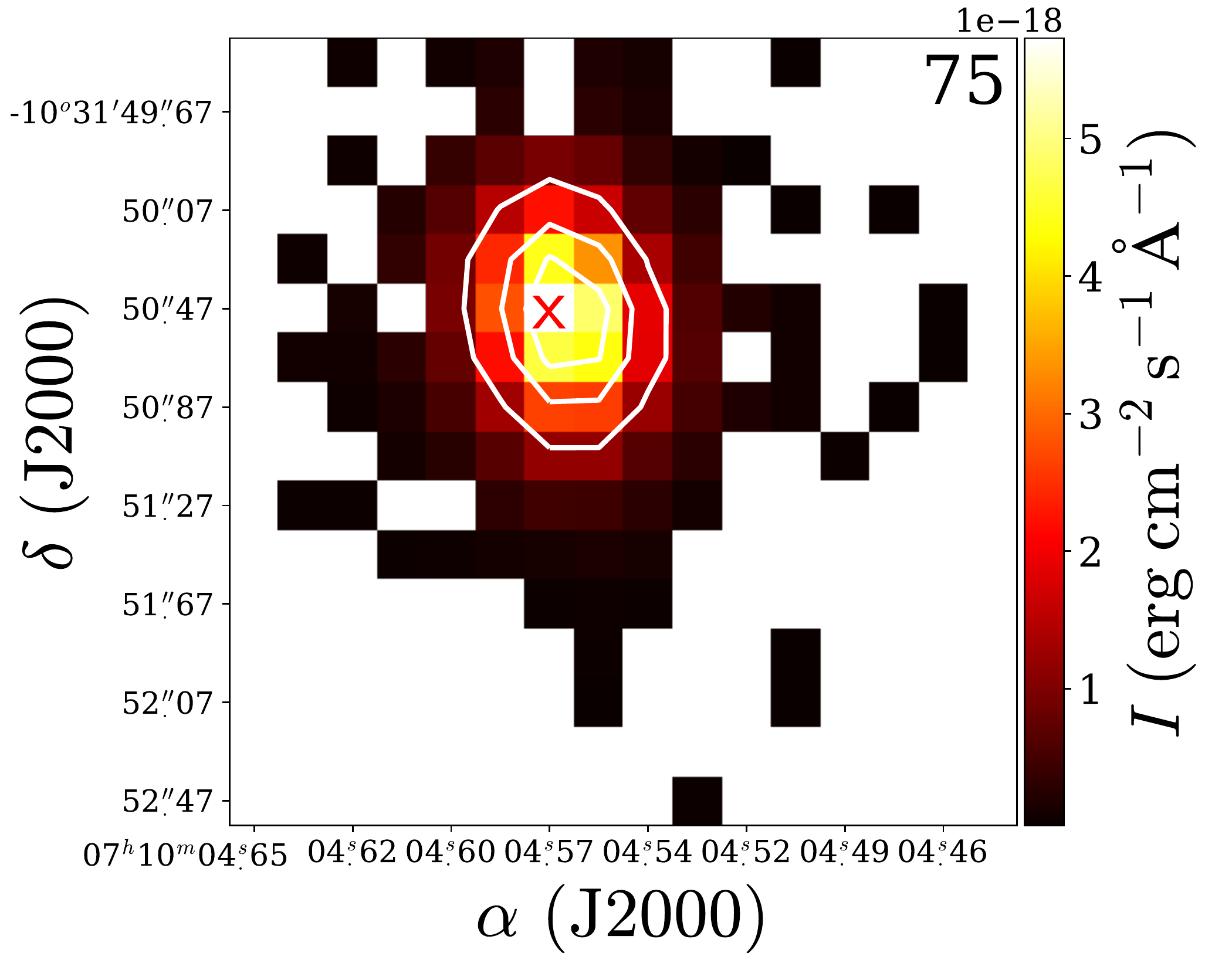}\hspace{-0.1cm}
\includegraphics[width=0.2\textwidth]{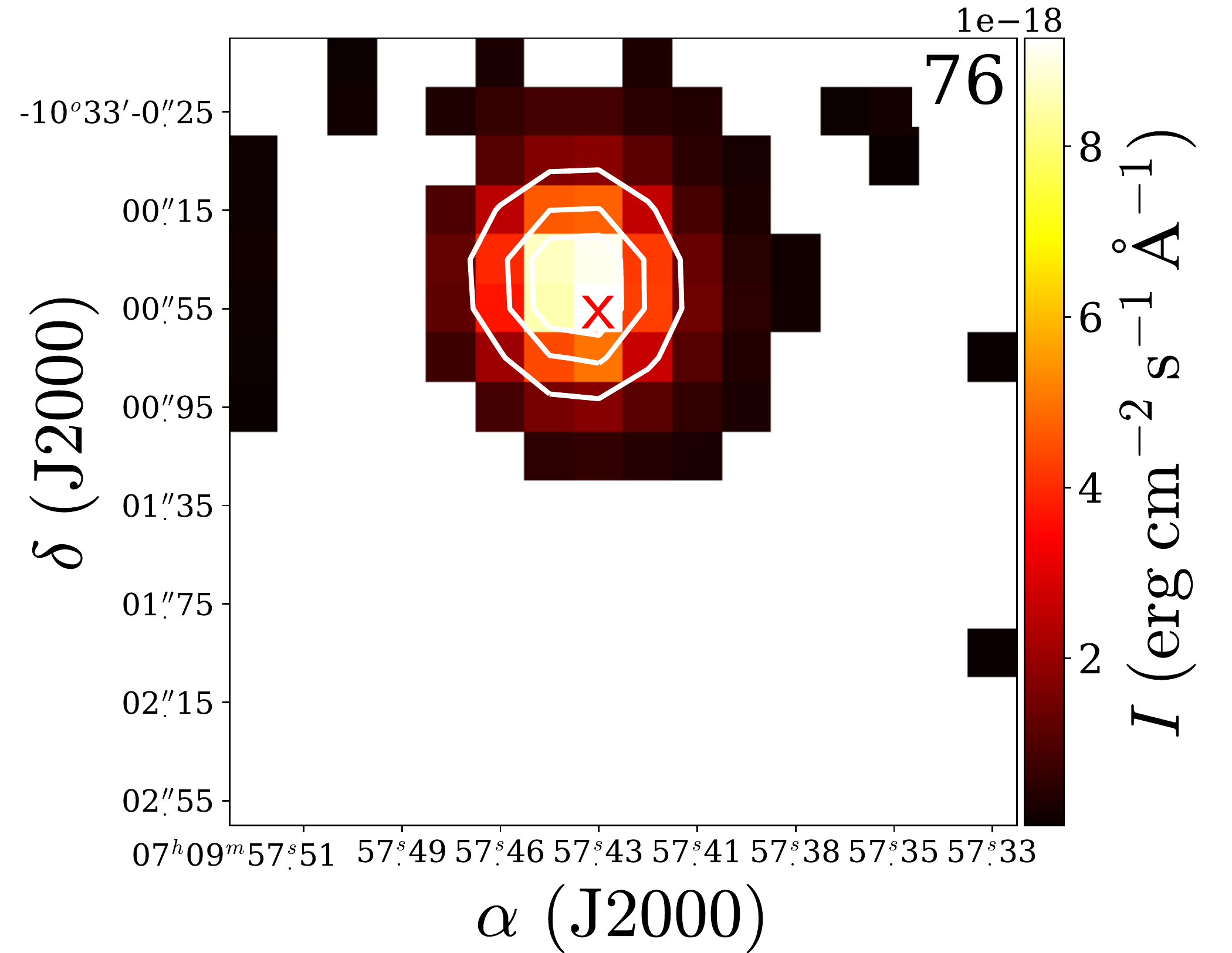}\hspace{-0.1cm}
\includegraphics[width=0.2\textwidth]{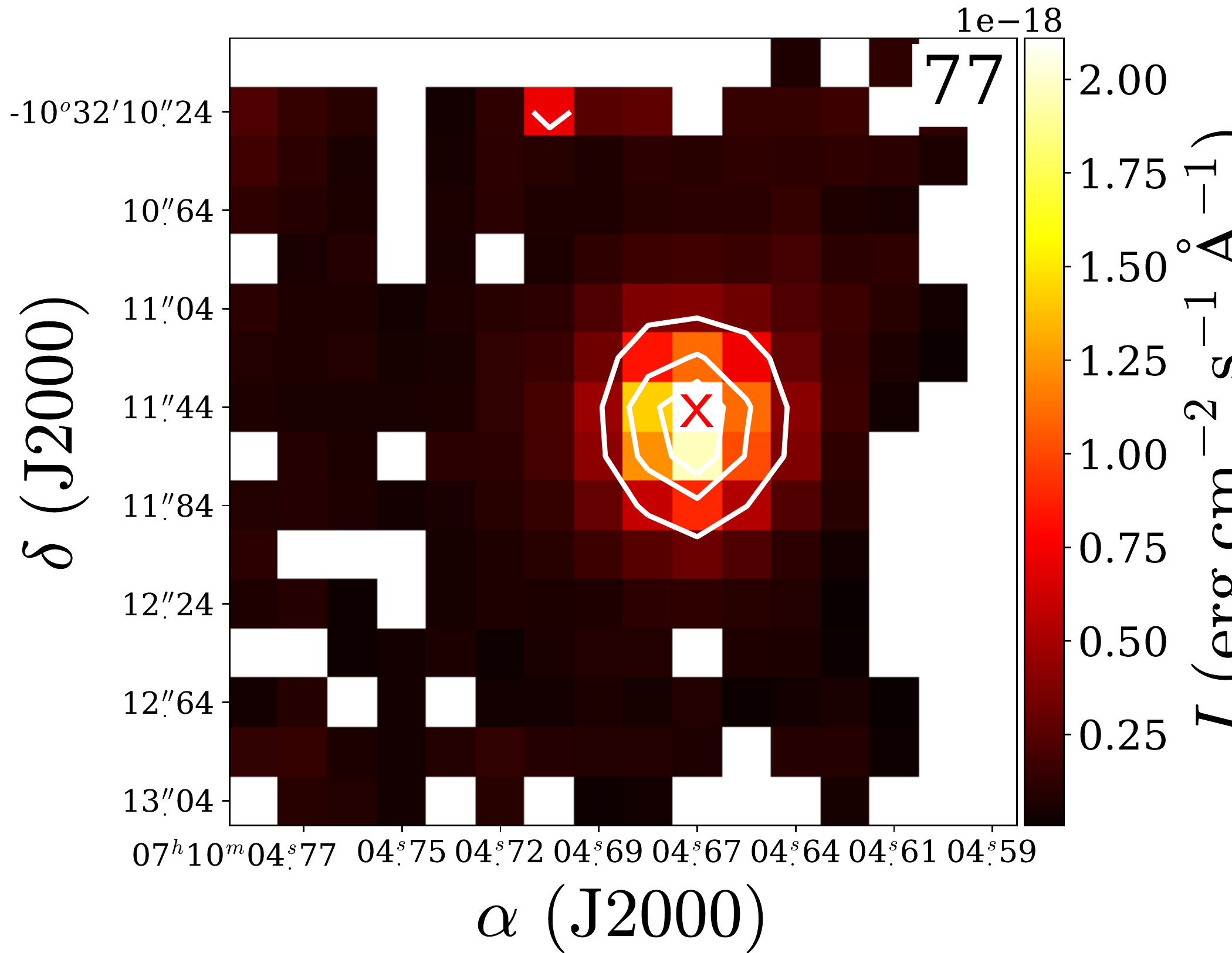}\hspace{-0.1cm}
\includegraphics[width=0.2\textwidth]{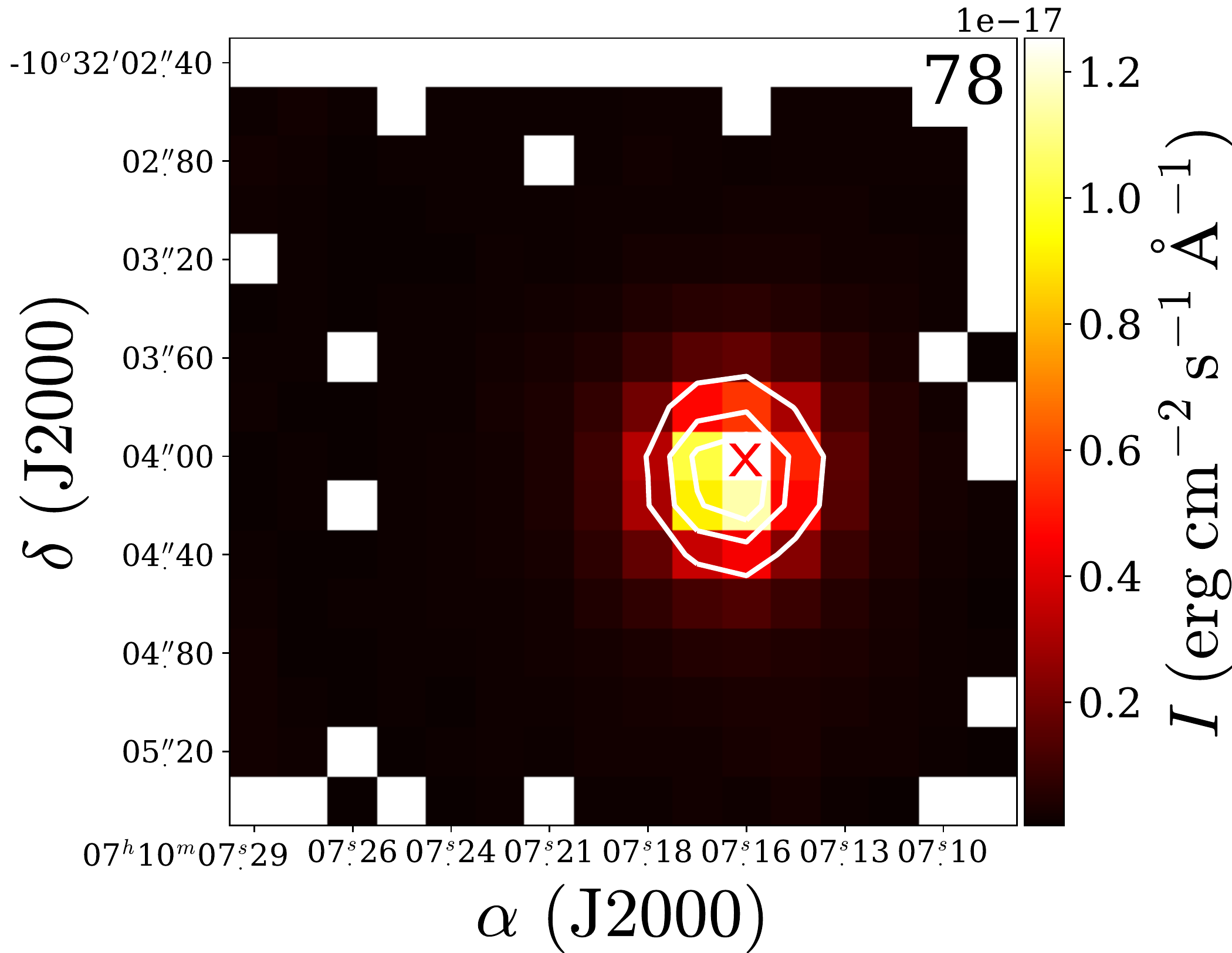}\hspace{-0.1cm}
\includegraphics[width=0.2\textwidth]{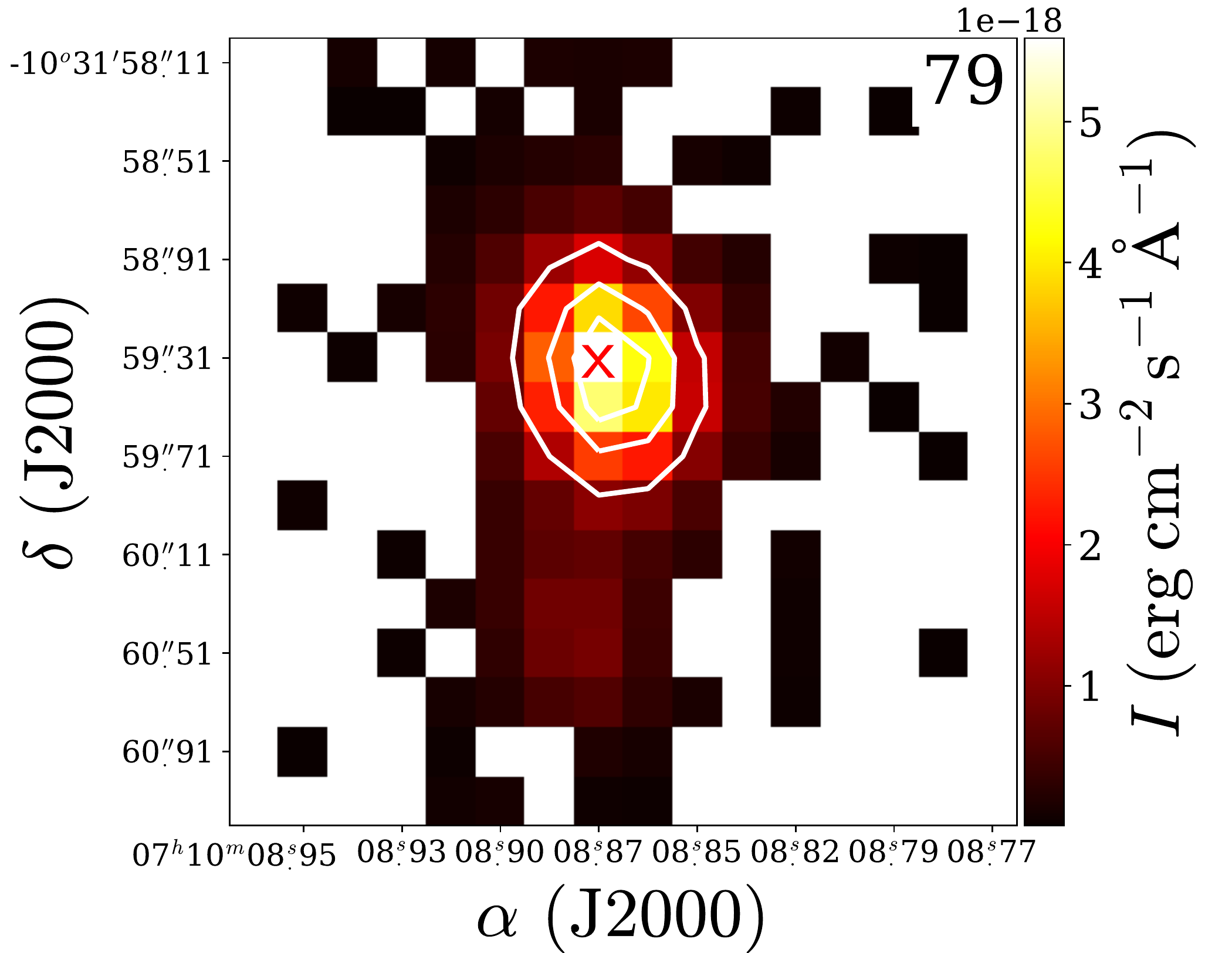}\hspace{-0.1cm}
\includegraphics[width=0.2\textwidth]{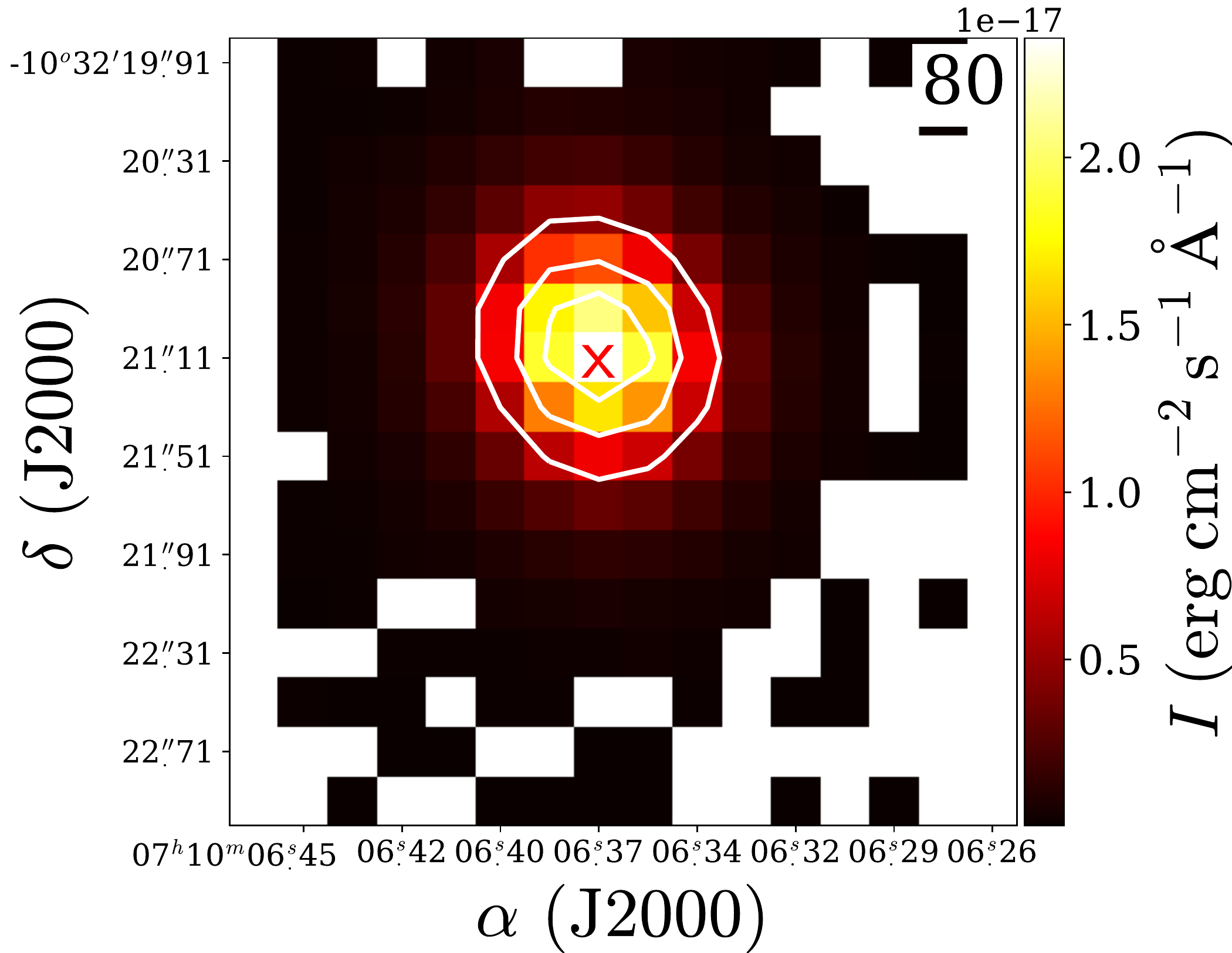}\hspace{-0.1cm}
\includegraphics[width=0.2\textwidth]{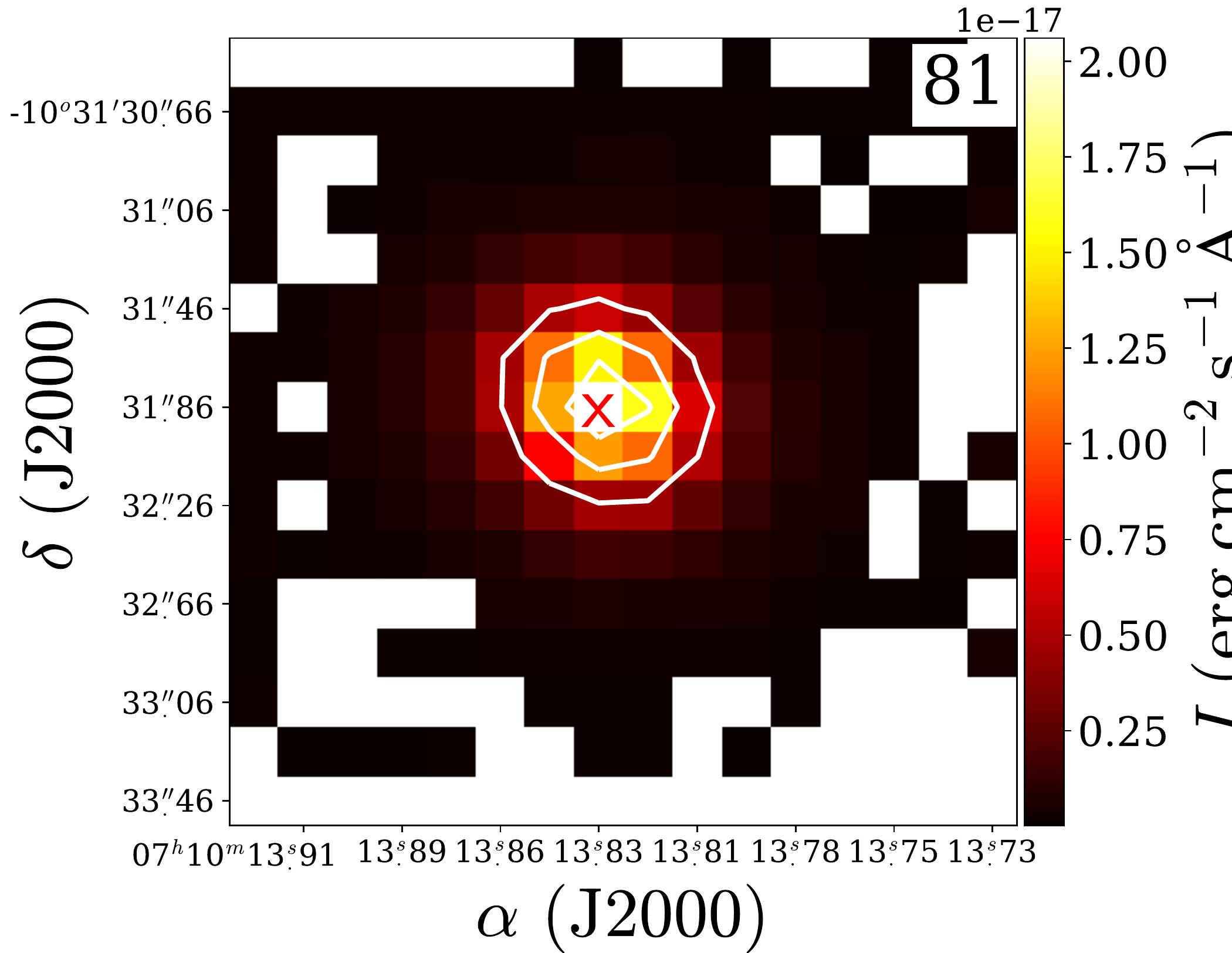}\hspace{-0.1cm}
\includegraphics[width=0.2\textwidth]{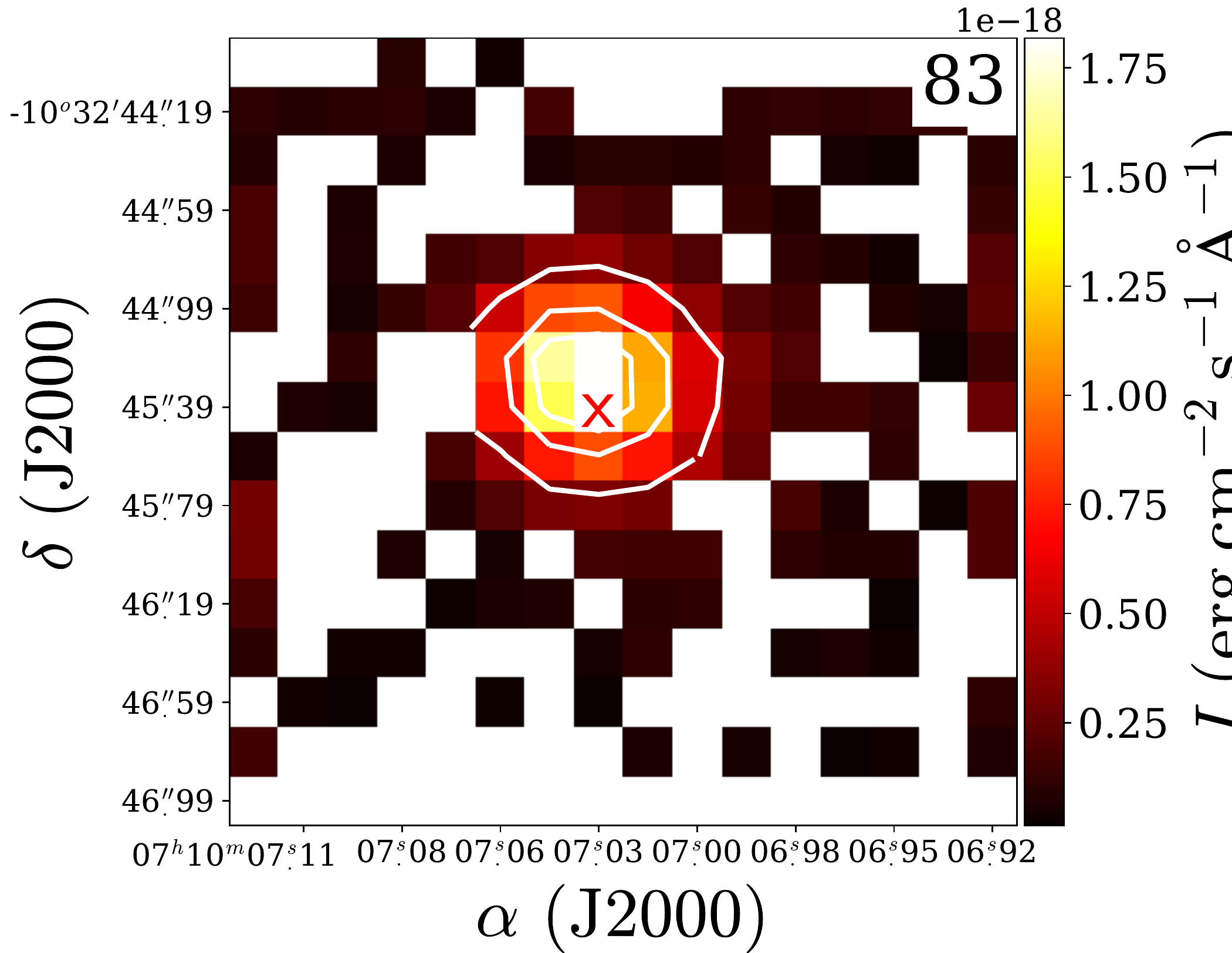}\hspace{-0.1cm}
\includegraphics[width=0.2\textwidth]{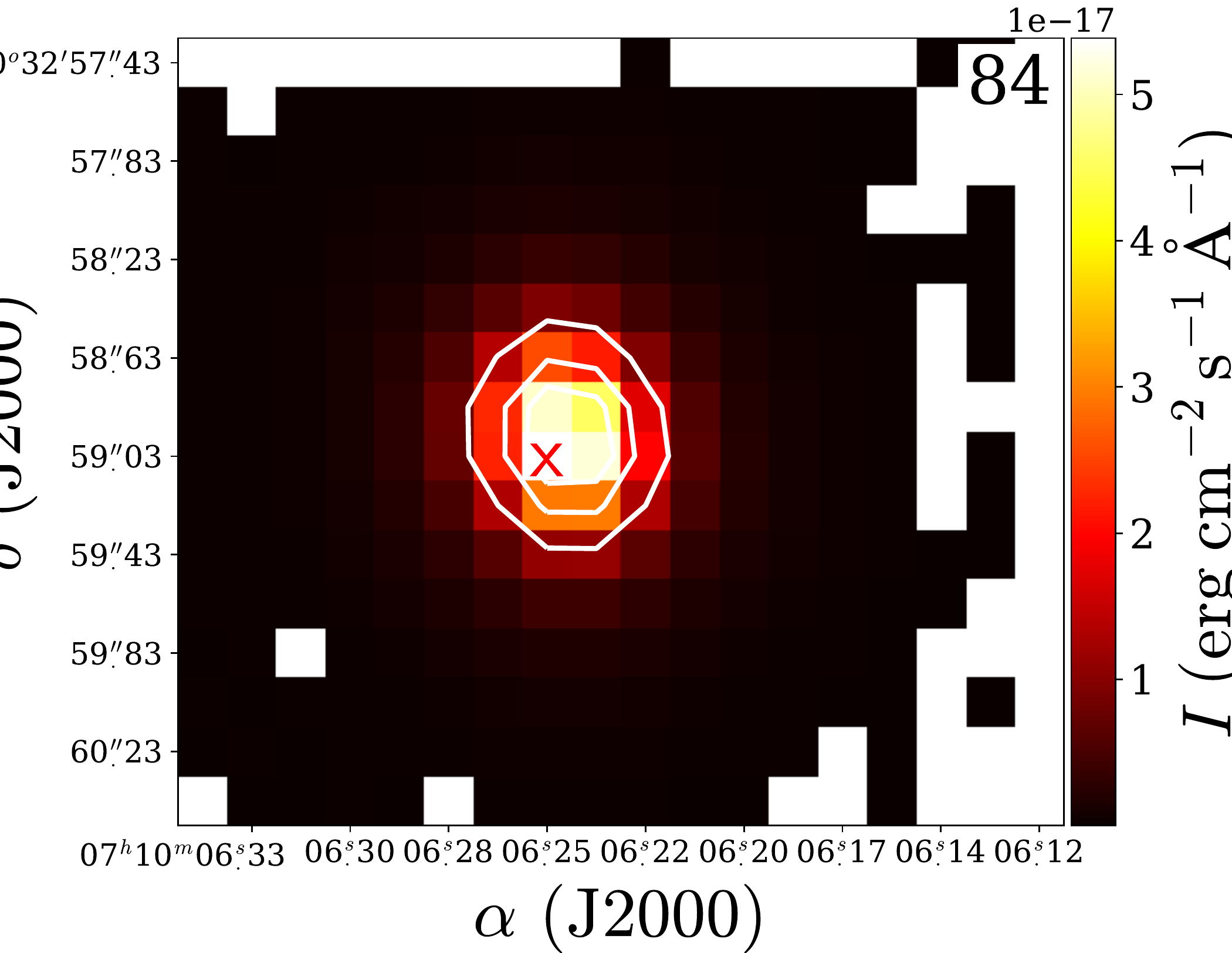}\hspace{-0.1cm}
\includegraphics[width=0.2\textwidth]{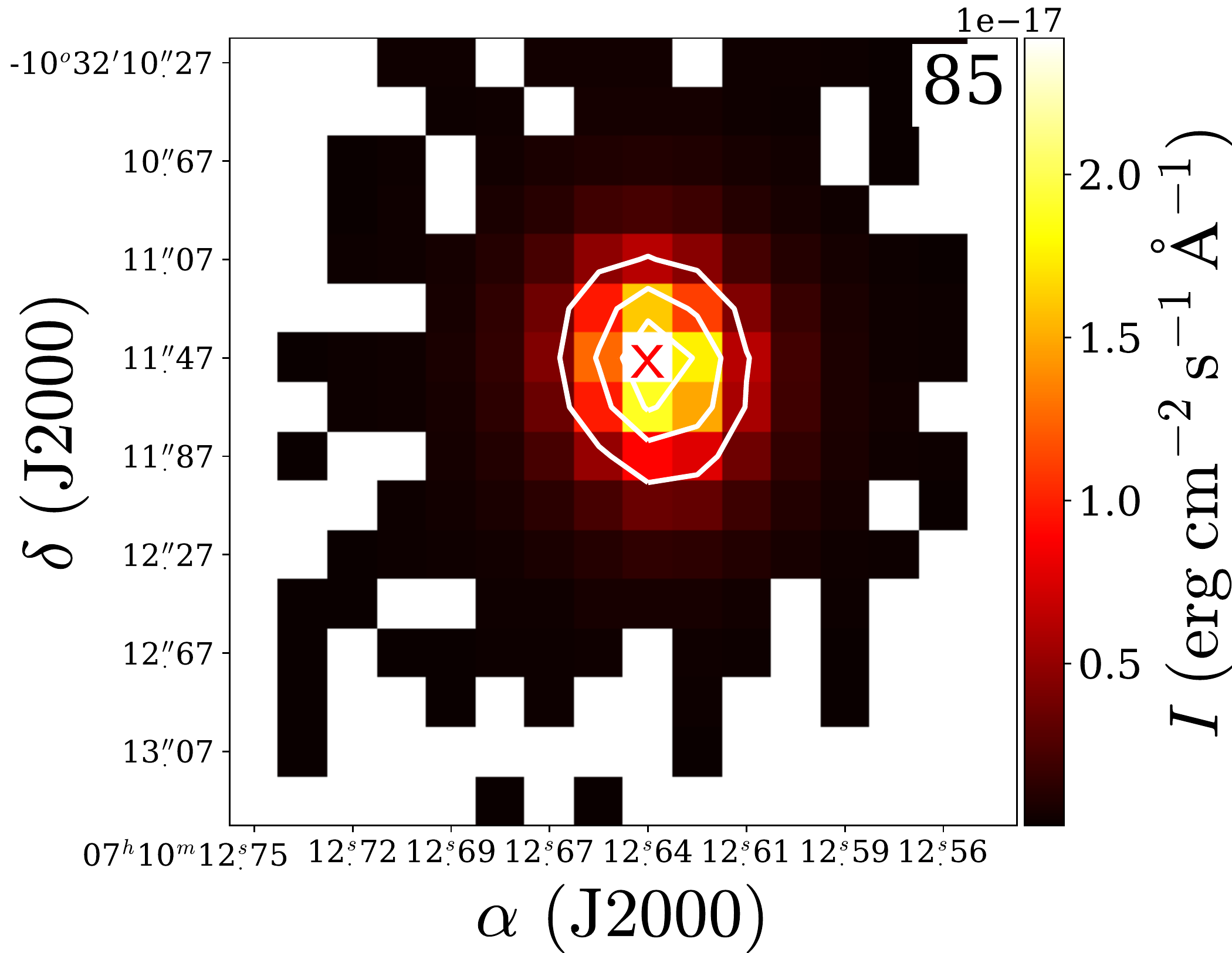}\hspace{-0.1cm}
\includegraphics[width=0.2\textwidth]{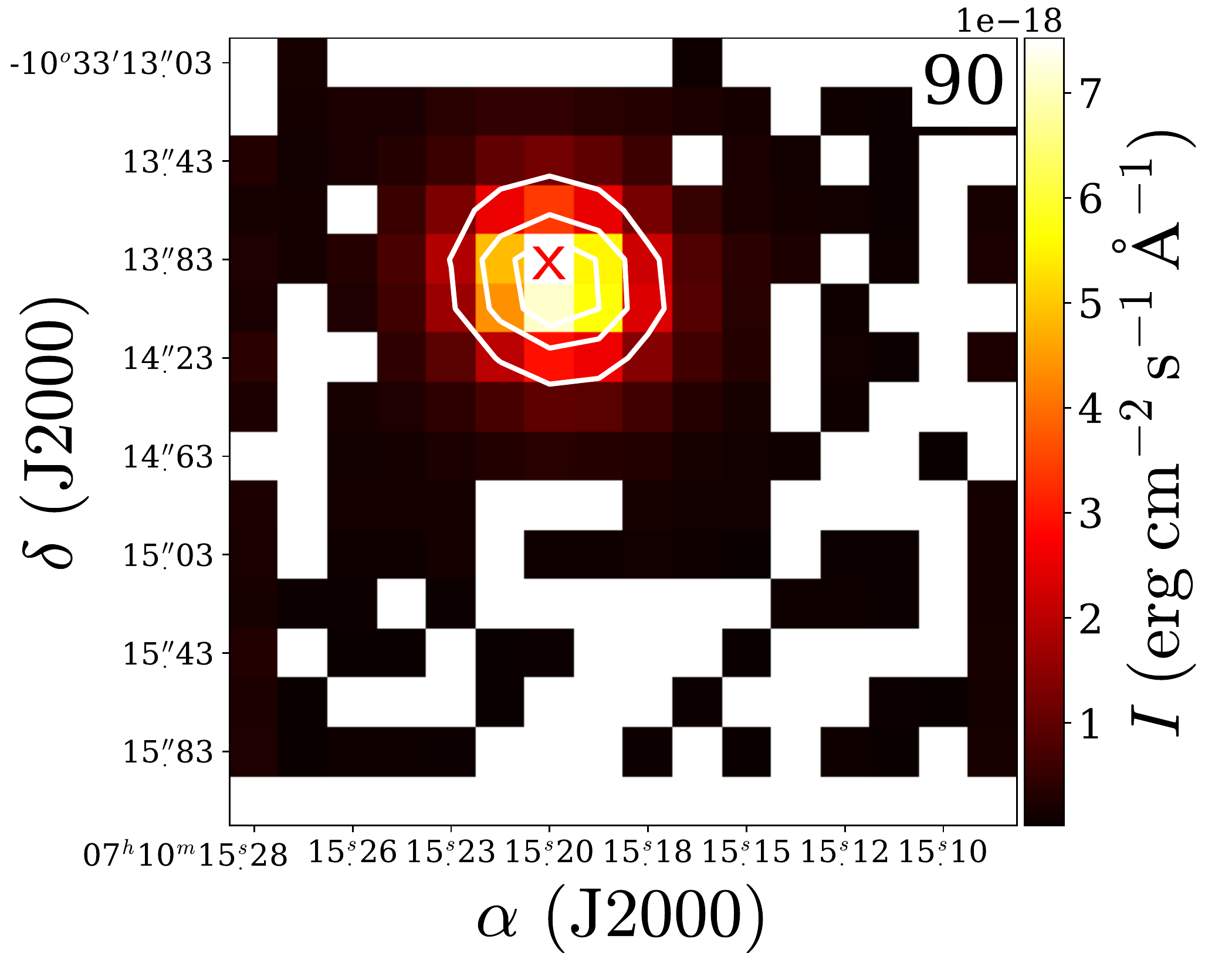}\hspace{-0.1cm}
\includegraphics[width=0.2\textwidth]{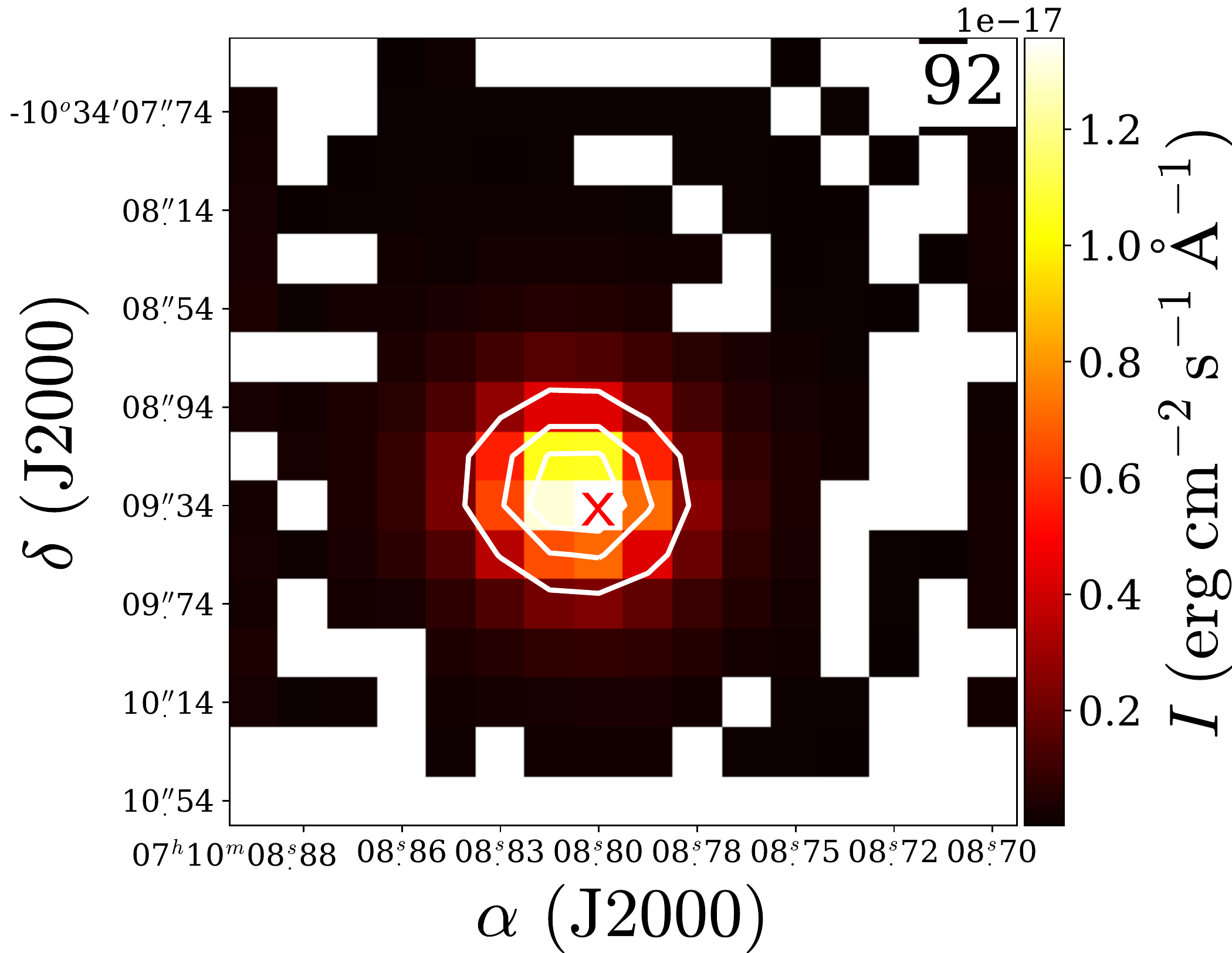}\hspace{-0.1cm}
\includegraphics[width=0.2\textwidth]{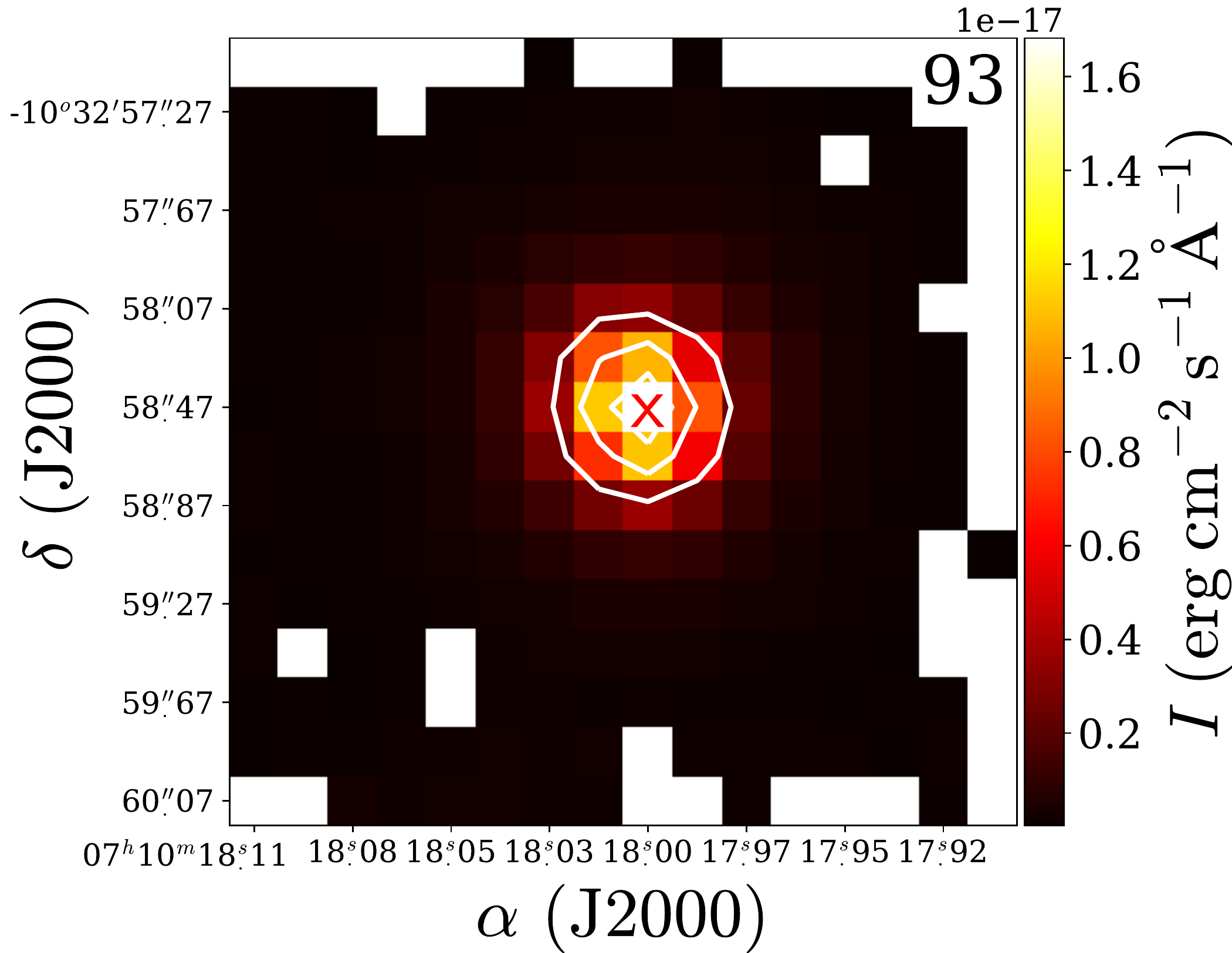}\hspace{-0.1cm}
\includegraphics[width=0.2\textwidth]{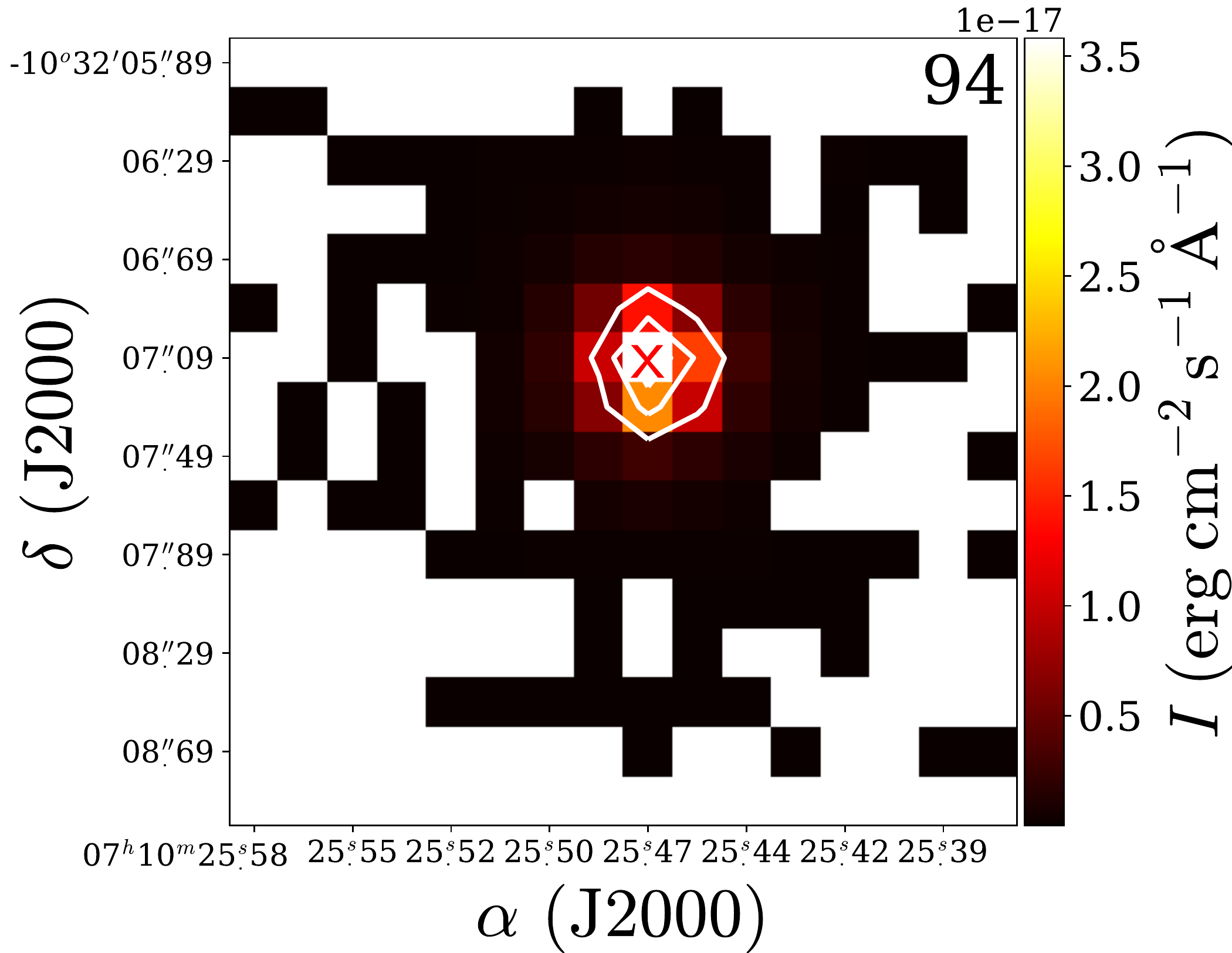}\hspace{-0.1cm}
\includegraphics[width=0.2\textwidth]{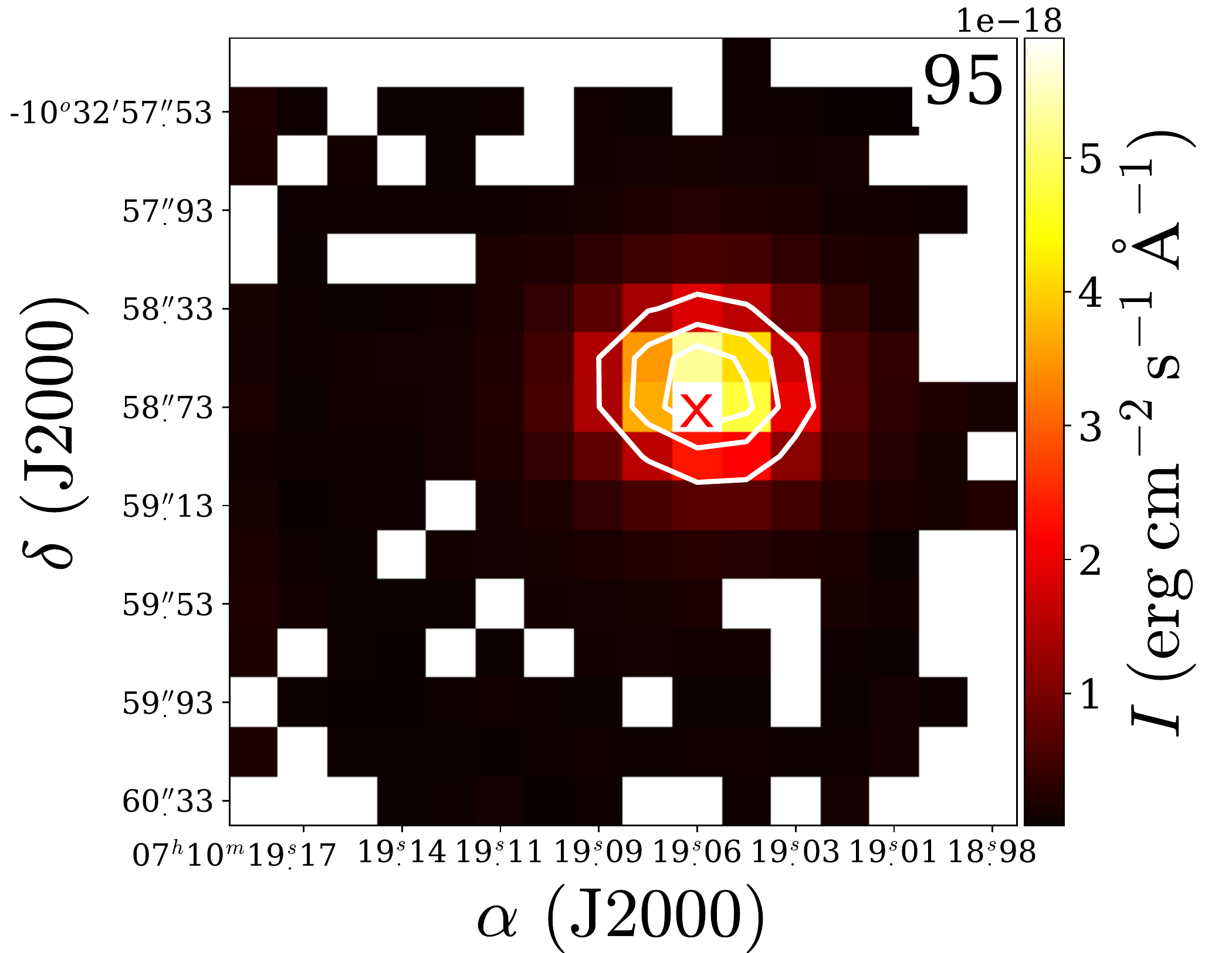}\hspace{-0.1cm}
\includegraphics[width=0.2\textwidth]{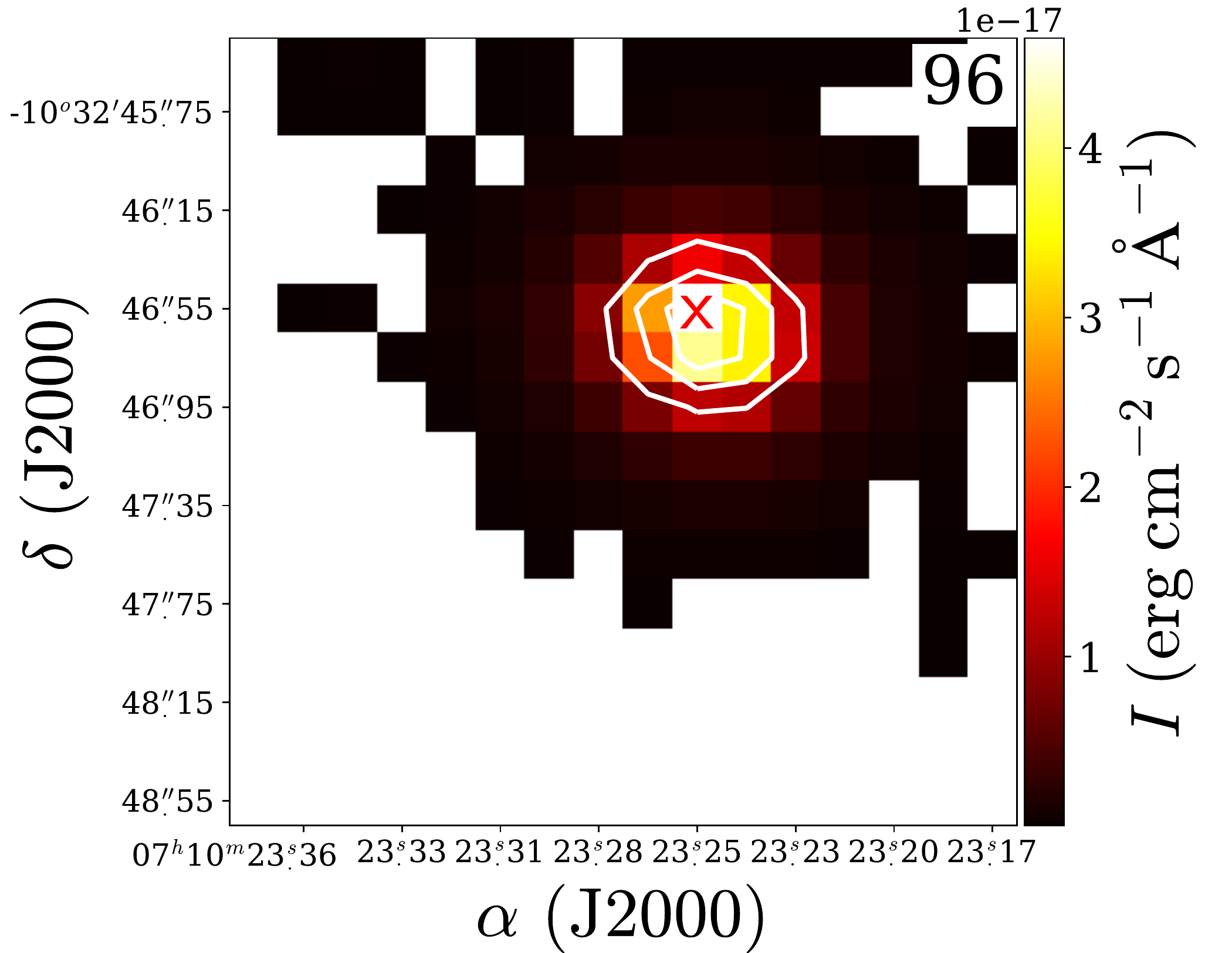}\hspace{-0.1cm}
\includegraphics[width=0.2\textwidth]{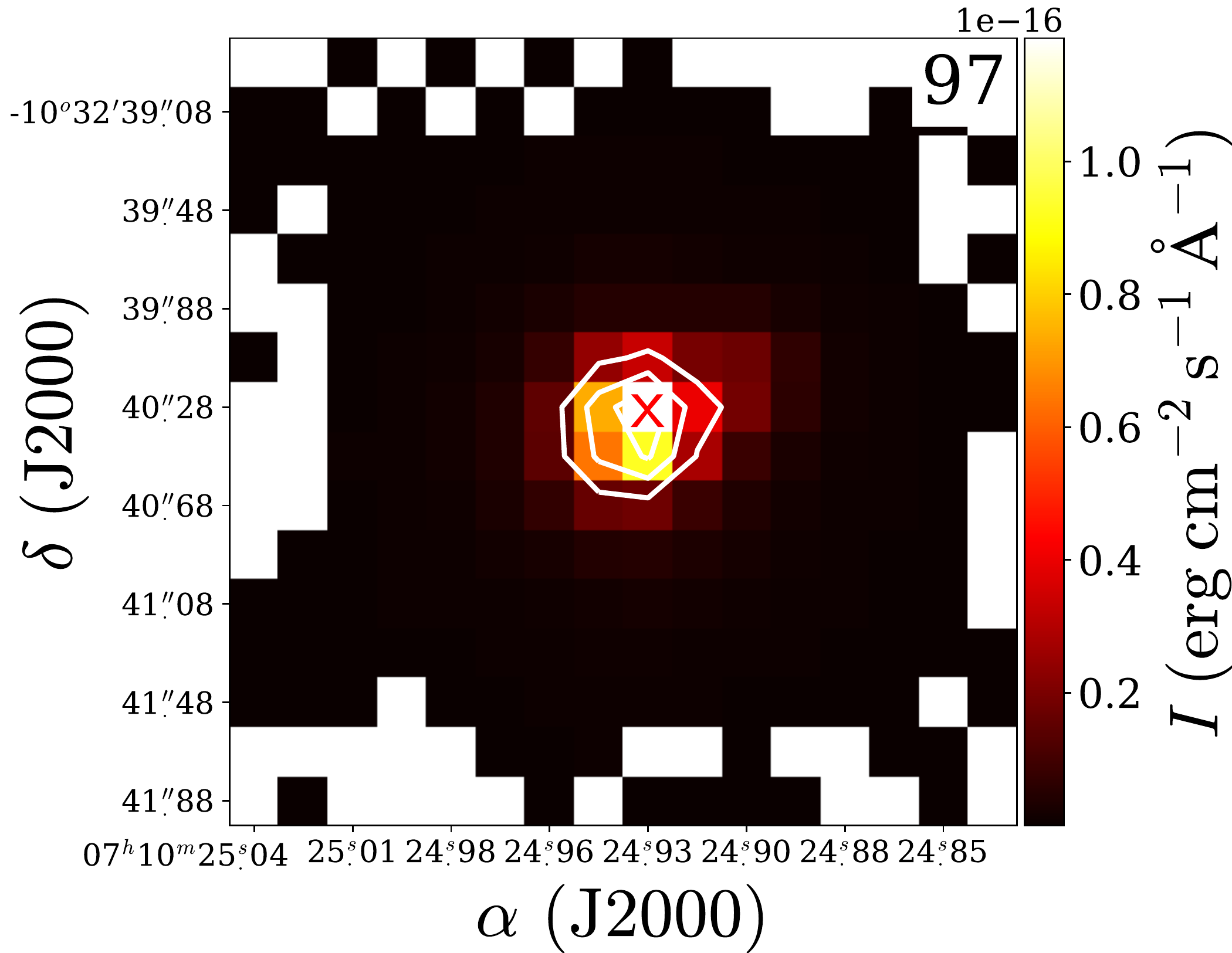}\hspace{-0.1cm}
\includegraphics[width=0.2\textwidth]{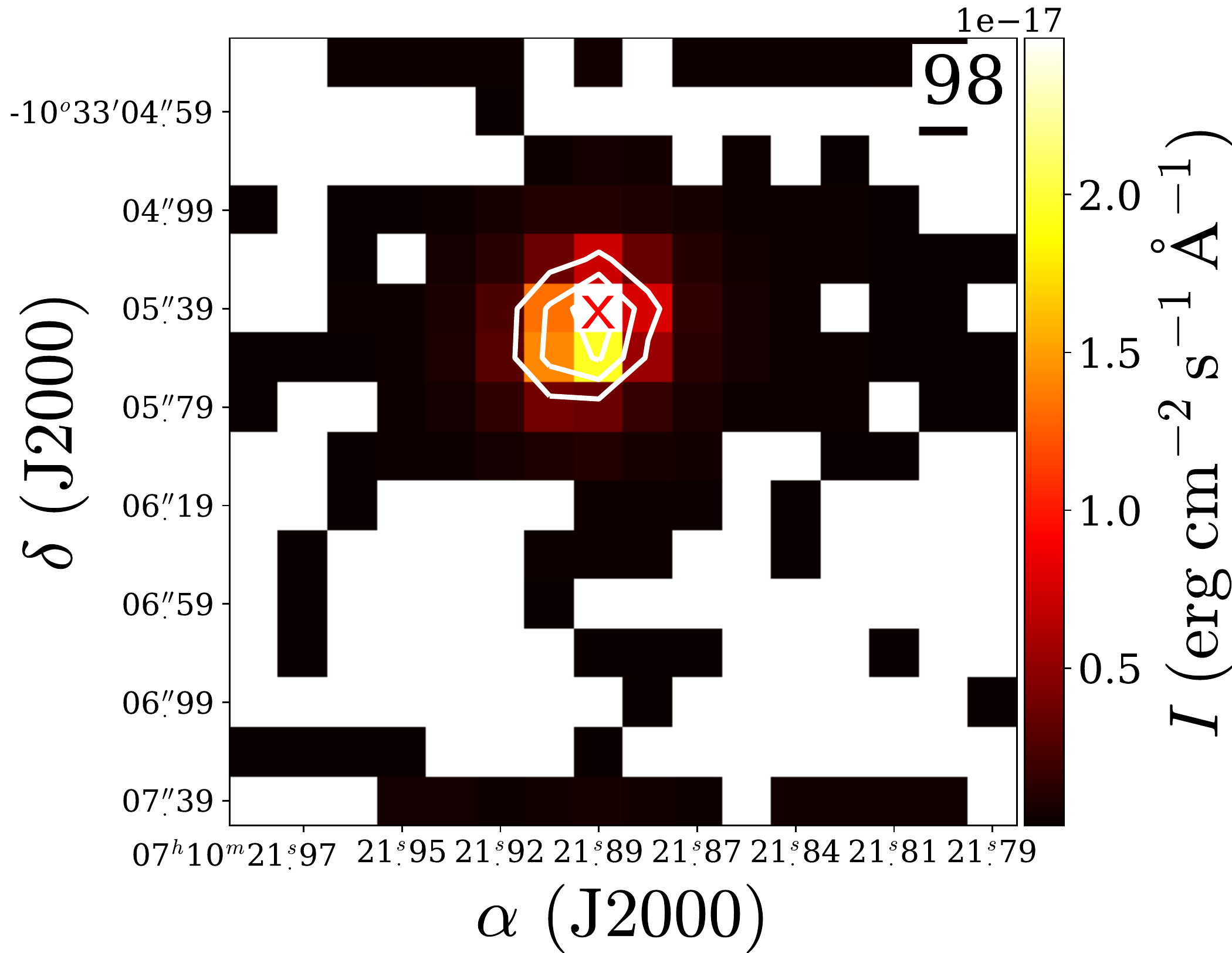}\hspace{-0.1cm}
\caption{Continued}
\end{figure*}
\addtocounter{figure}{-1}
\begin{figure*}
\includegraphics[width=0.2\textwidth]{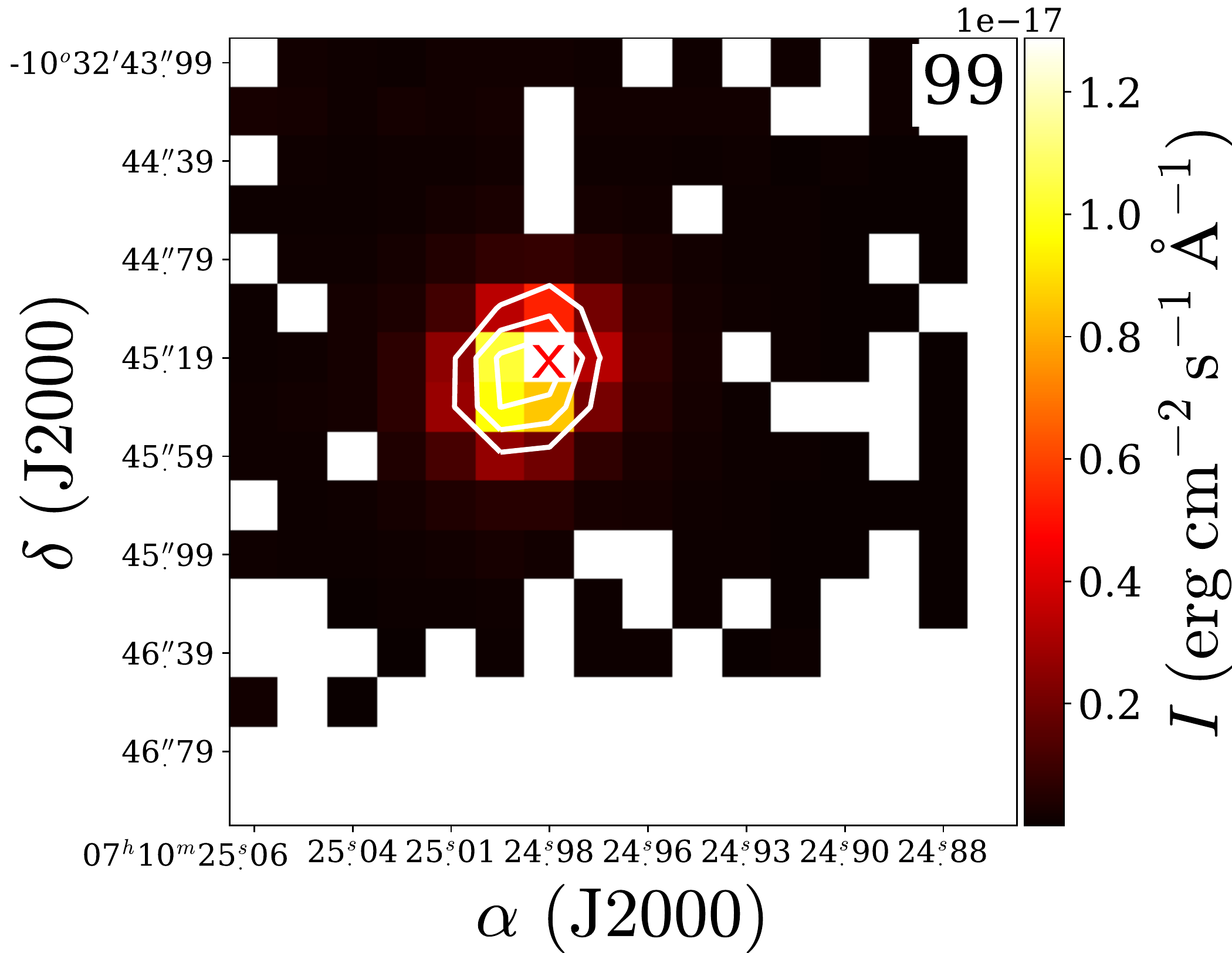}\hspace{-0.1cm}
\includegraphics[width=0.2\textwidth]{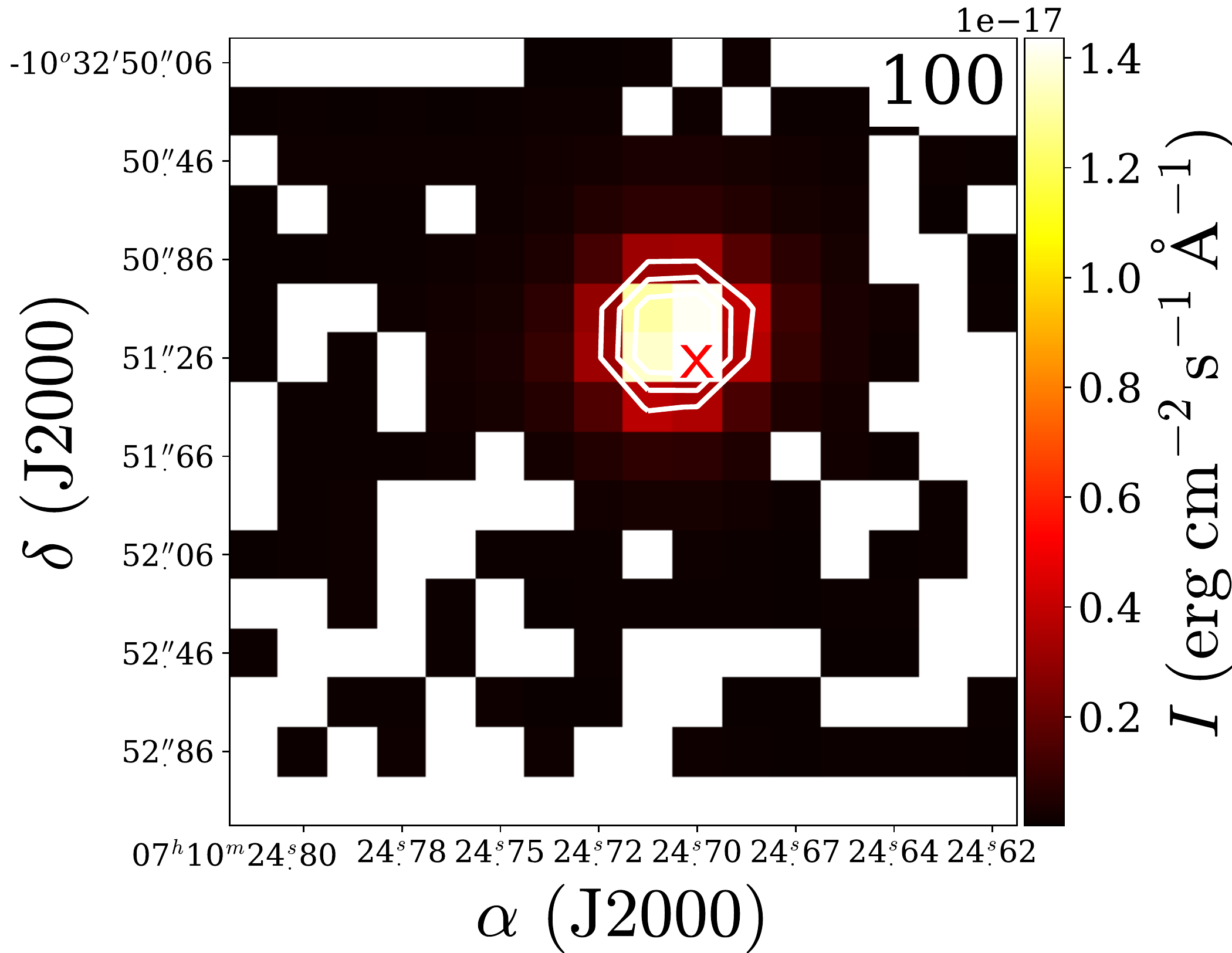}\hspace{-0.1cm}
\includegraphics[width=0.2\textwidth]{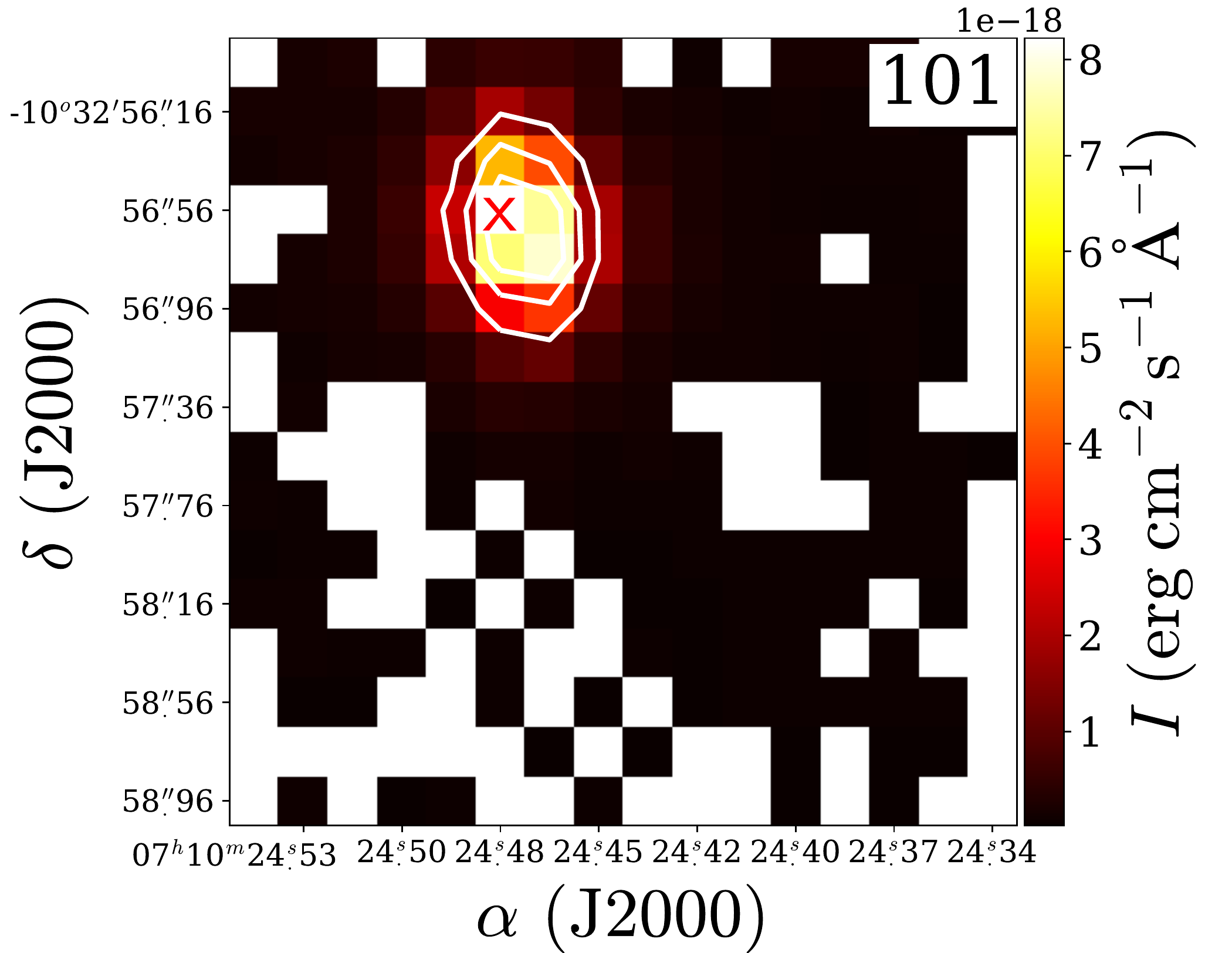}\hspace{-0.1cm}
\includegraphics[width=0.2\textwidth]{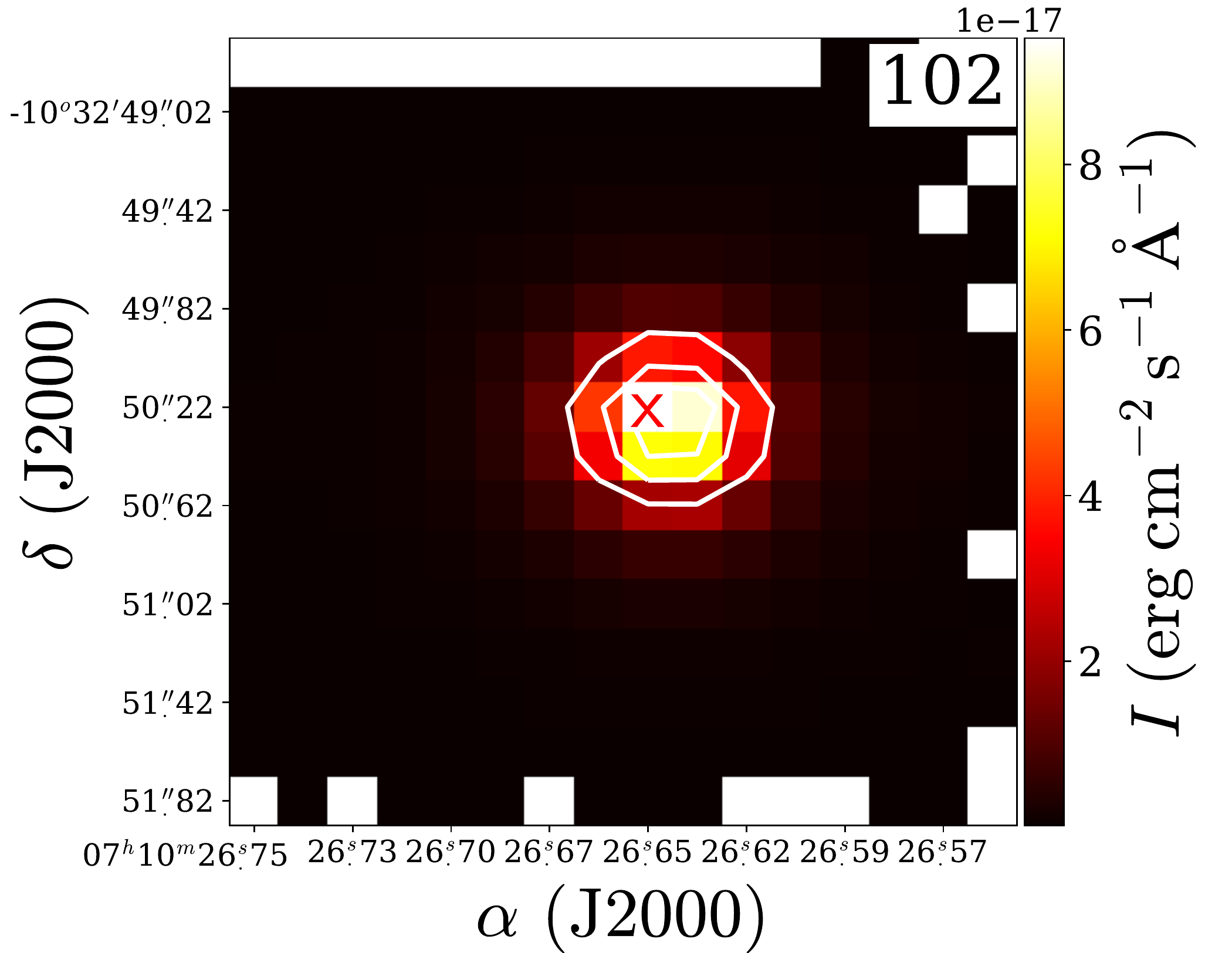}\hspace{-0.1cm}
\includegraphics[width=0.2\textwidth]{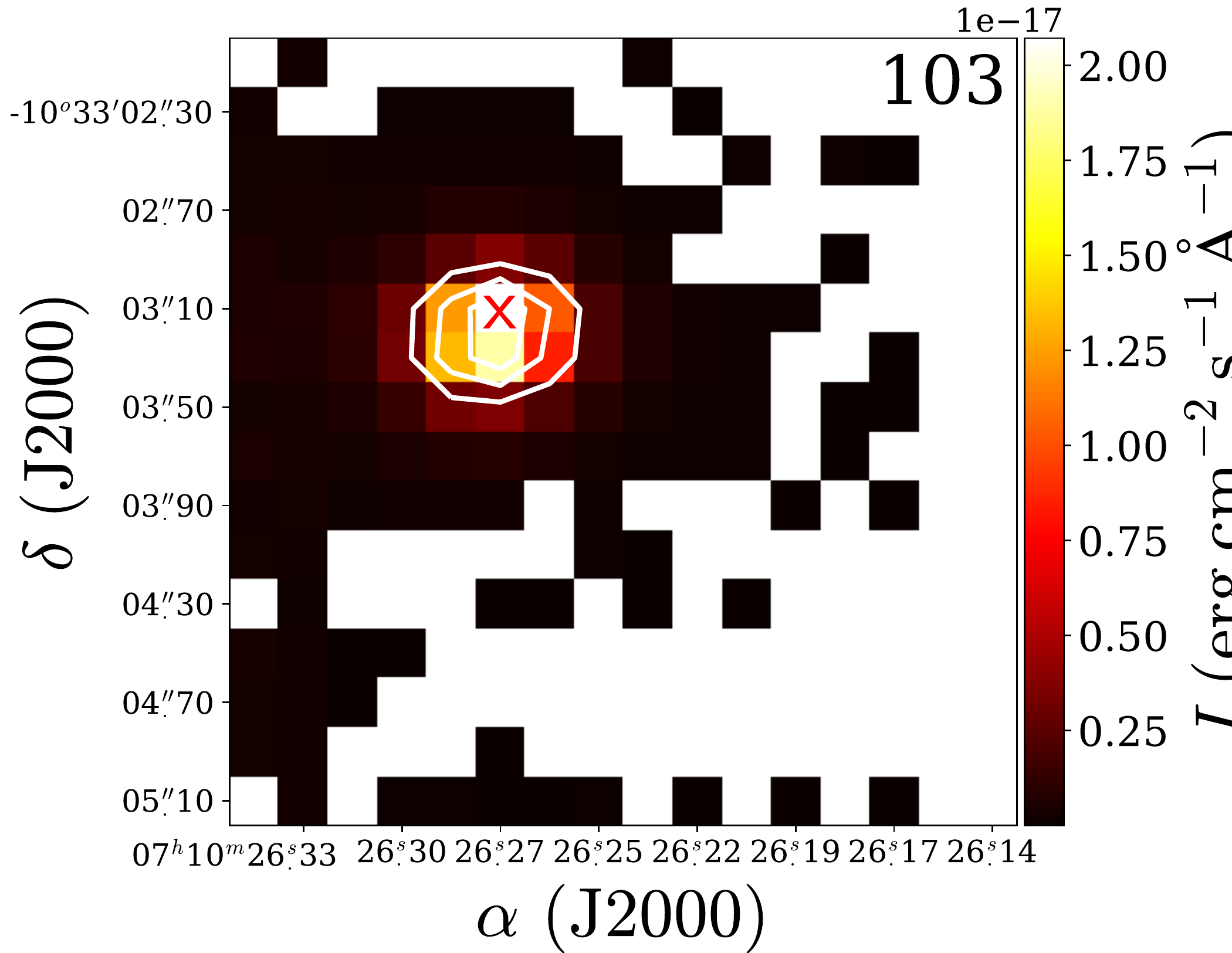}\hspace{-0.1cm}
\includegraphics[width=0.2\textwidth]{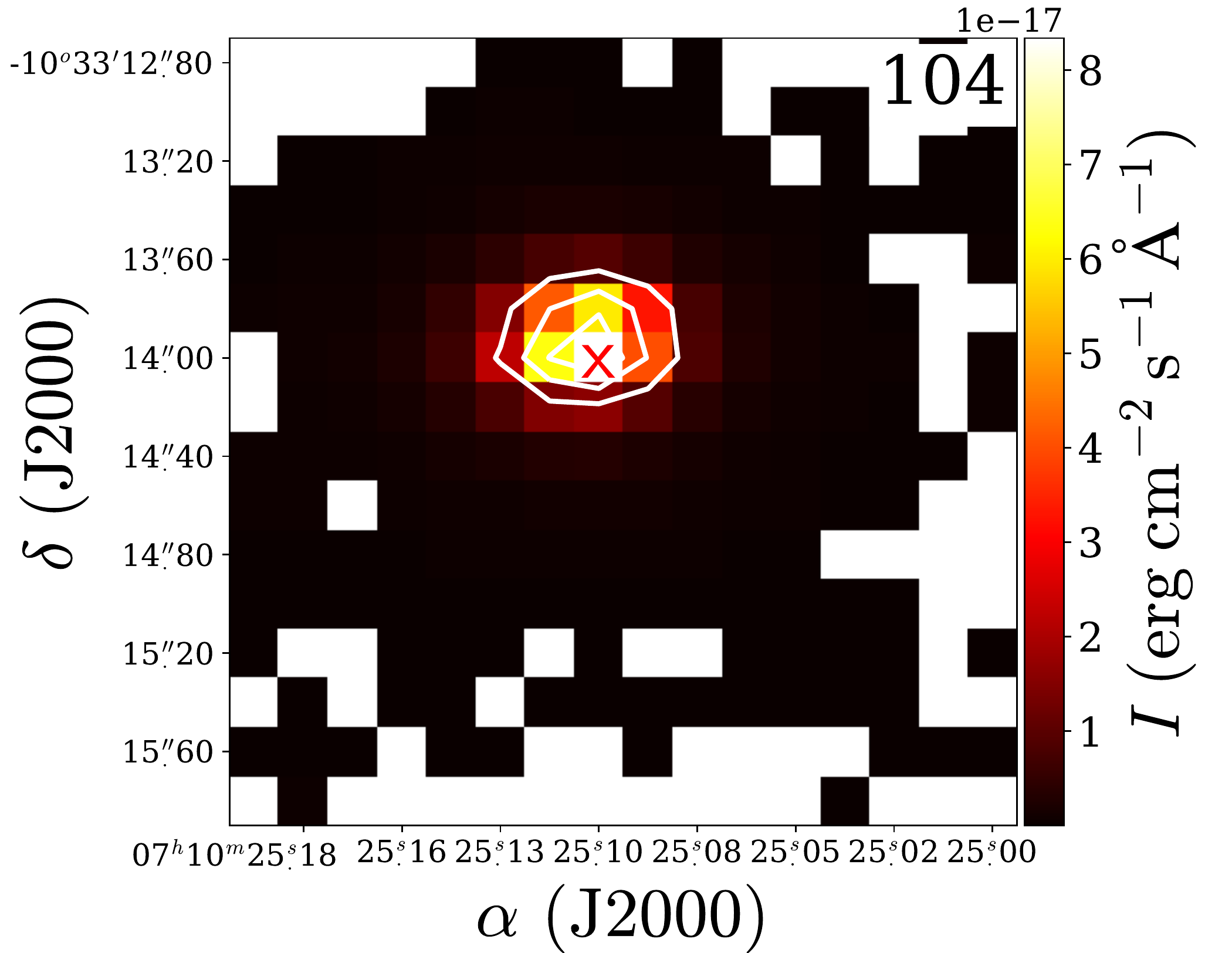}\hspace{-0.1cm}
\includegraphics[width=0.2\textwidth]{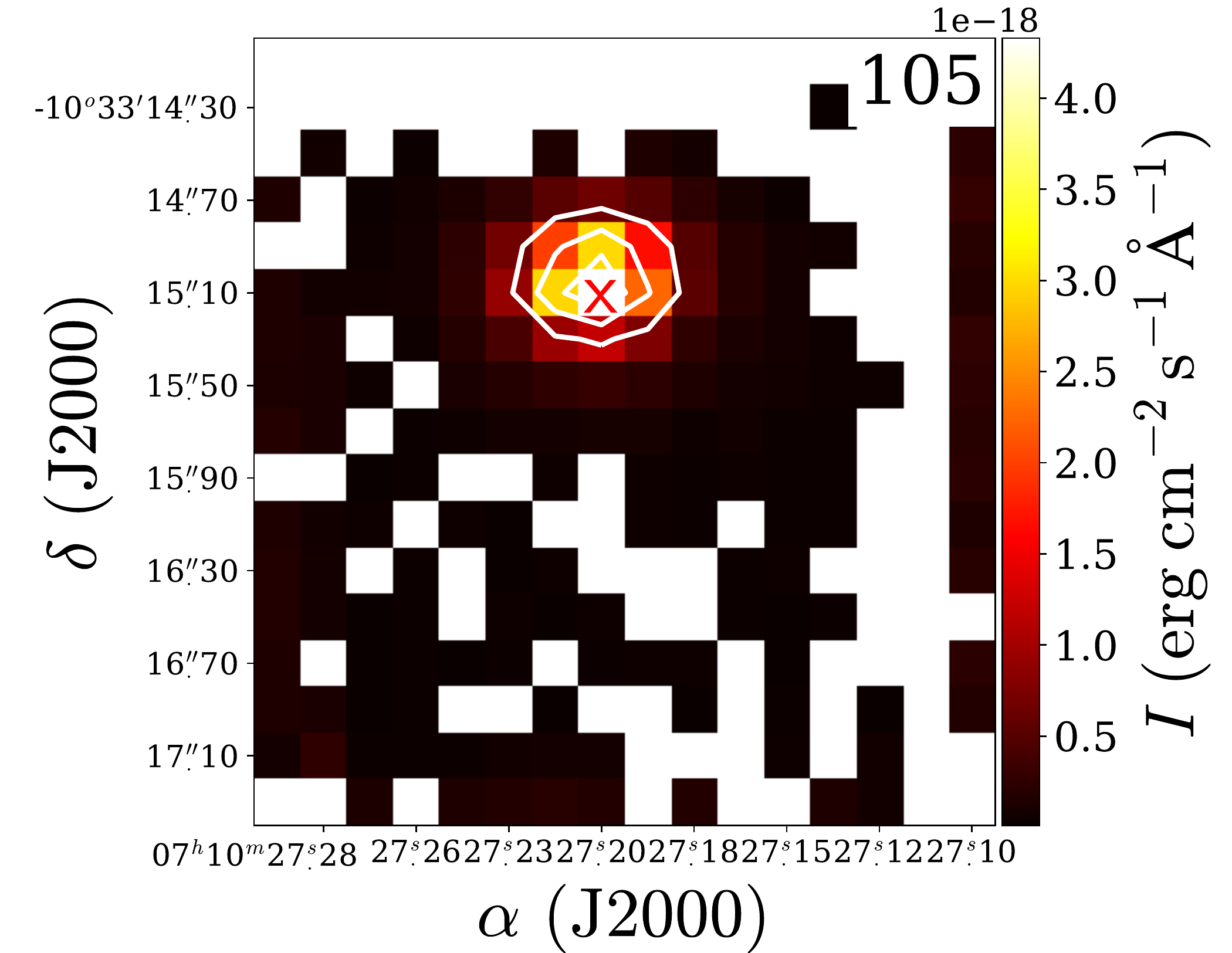}\hspace{-0.1cm}
\includegraphics[width=0.2\textwidth]{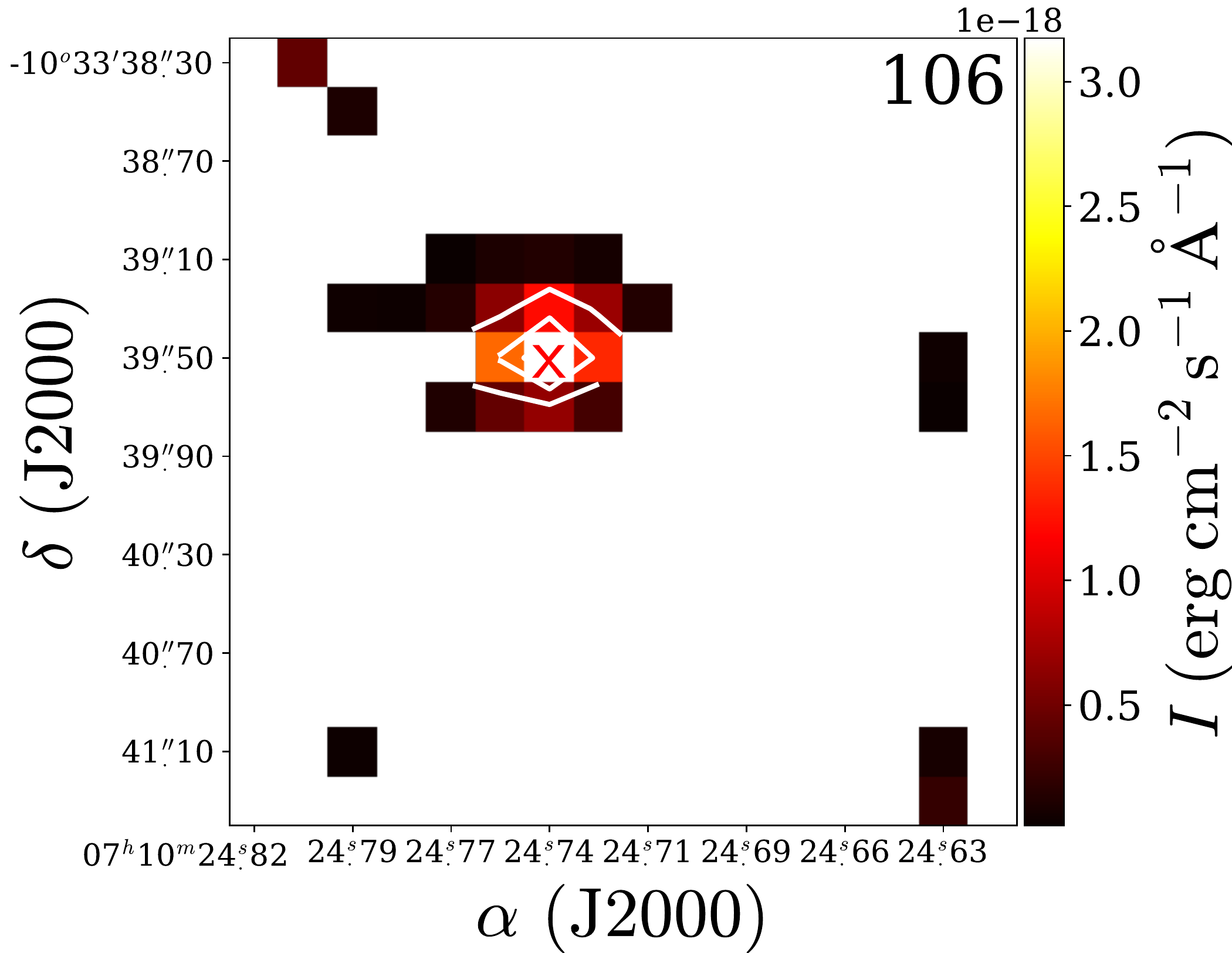}\hspace{-0.1cm}
\includegraphics[width=0.2\textwidth]{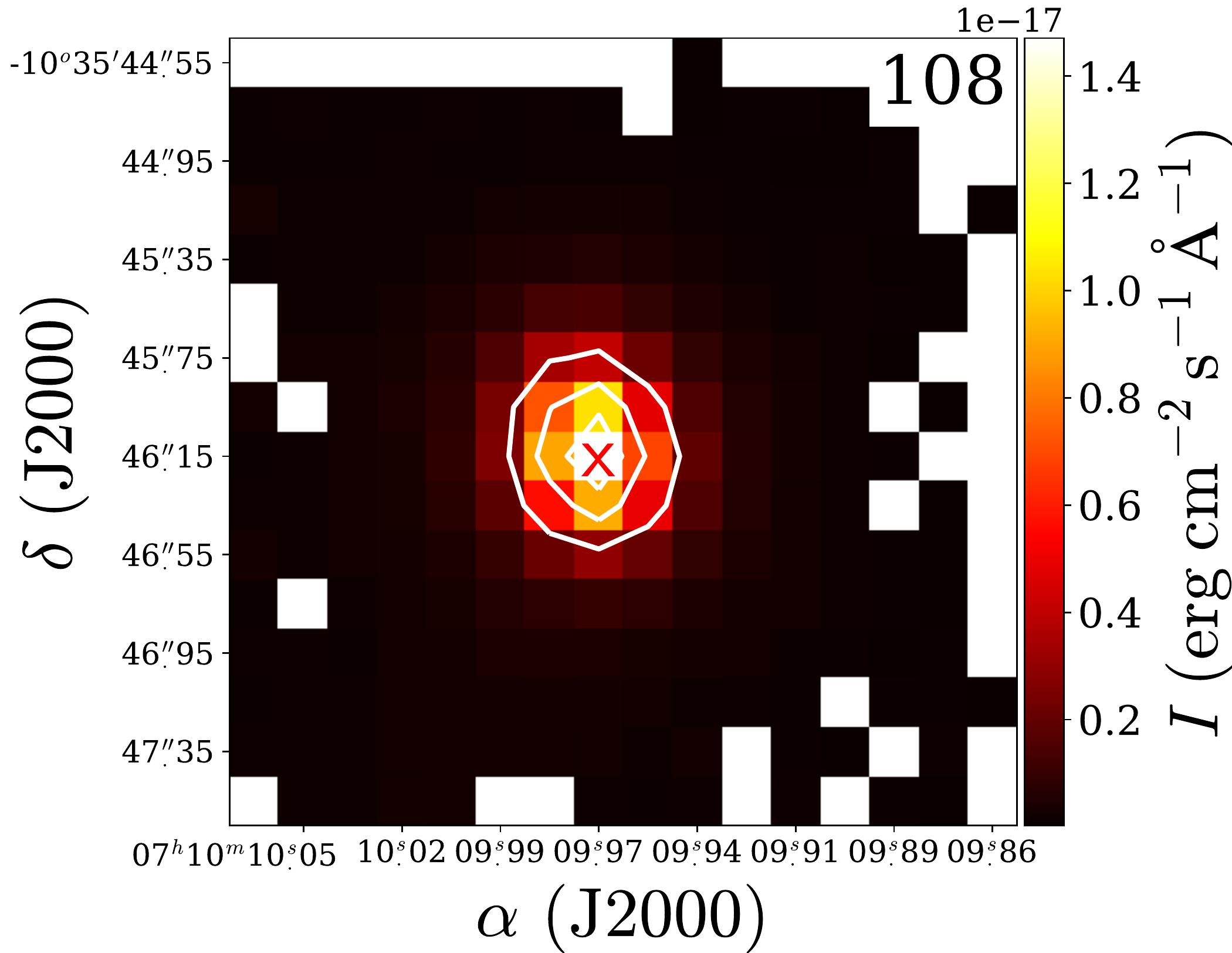}\hspace{-0.1cm}
\includegraphics[width=0.2\textwidth]{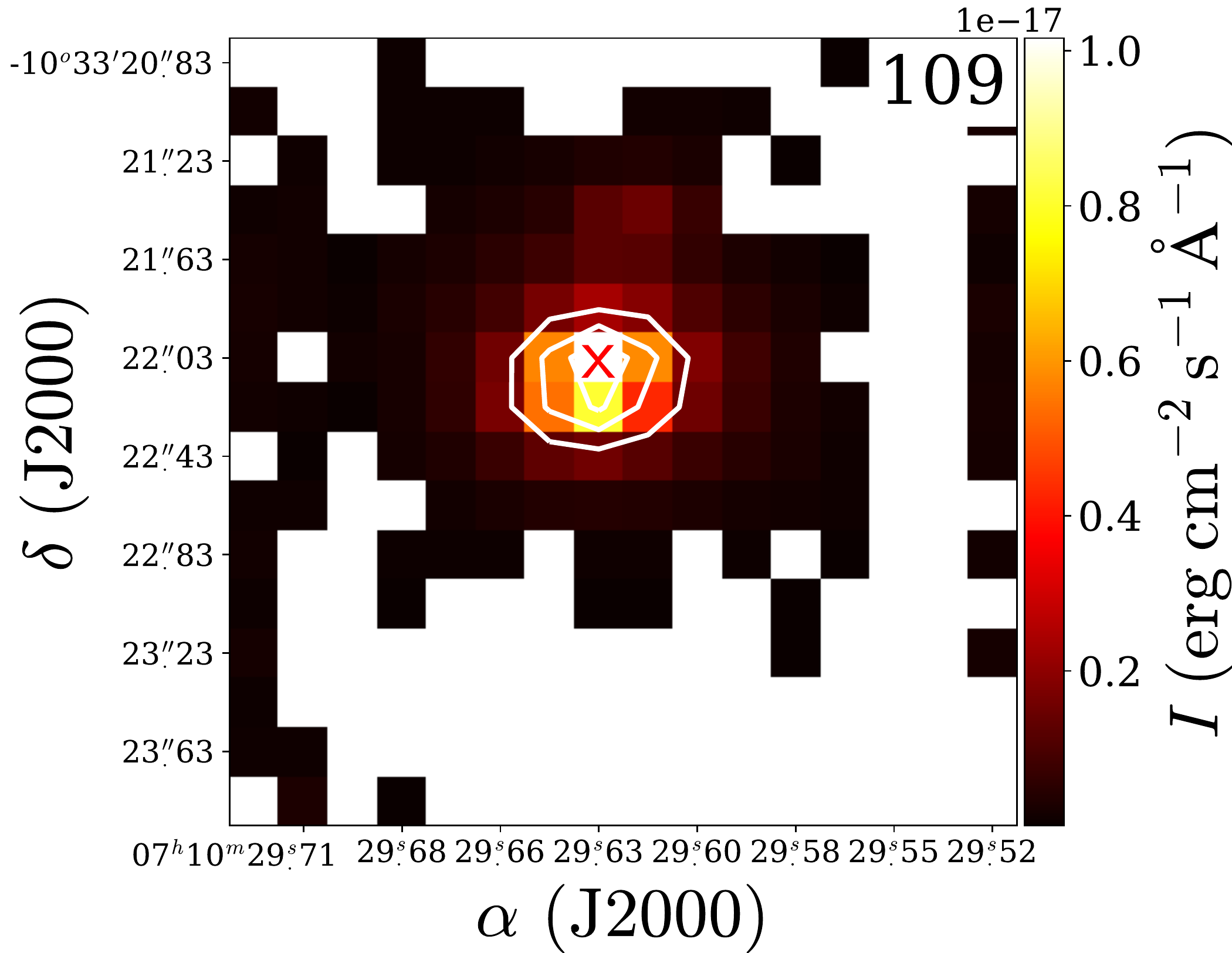}\hspace{-0.1cm}
\includegraphics[width=0.2\textwidth]{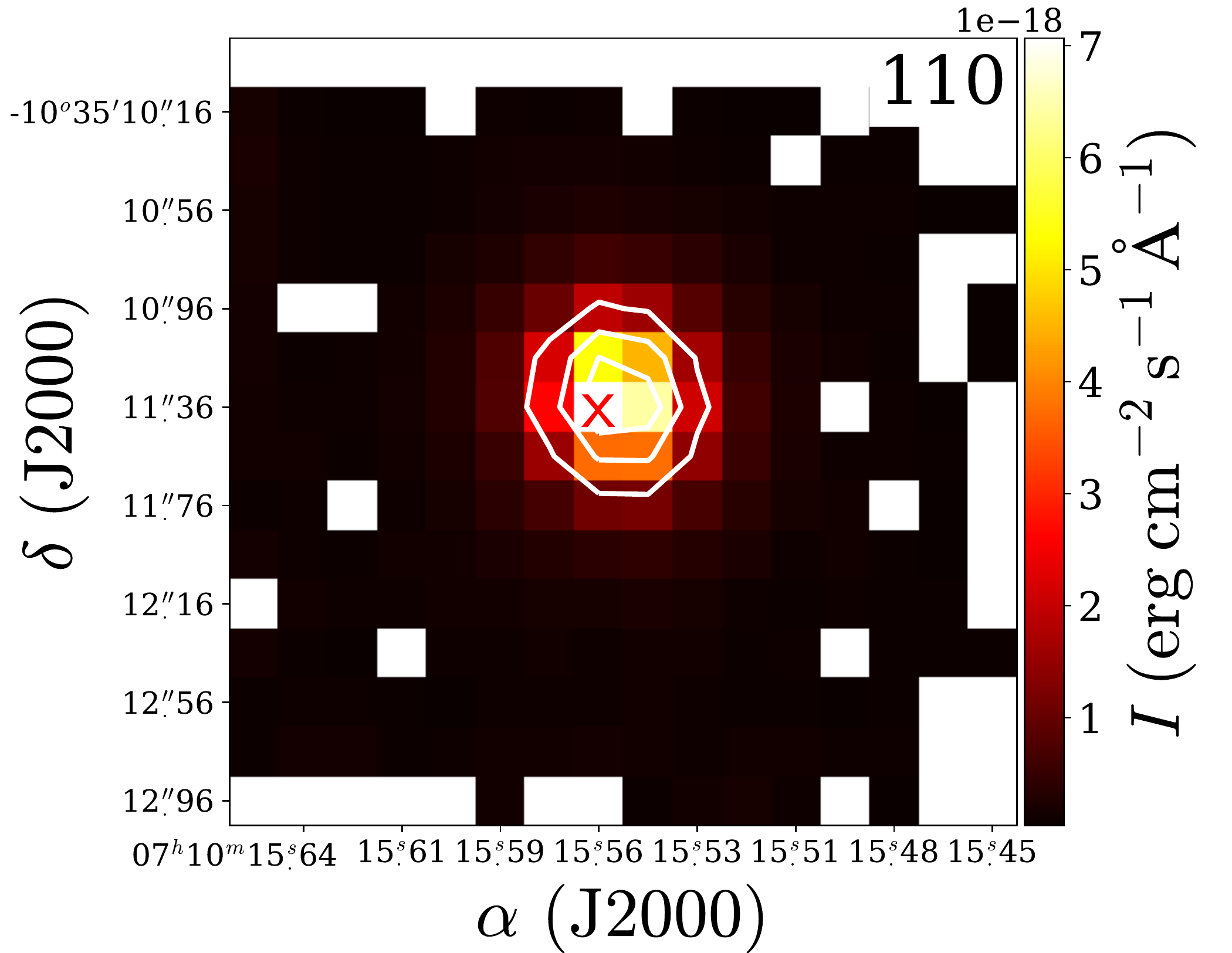}\hspace{-0.1cm}
\includegraphics[width=0.2\textwidth]{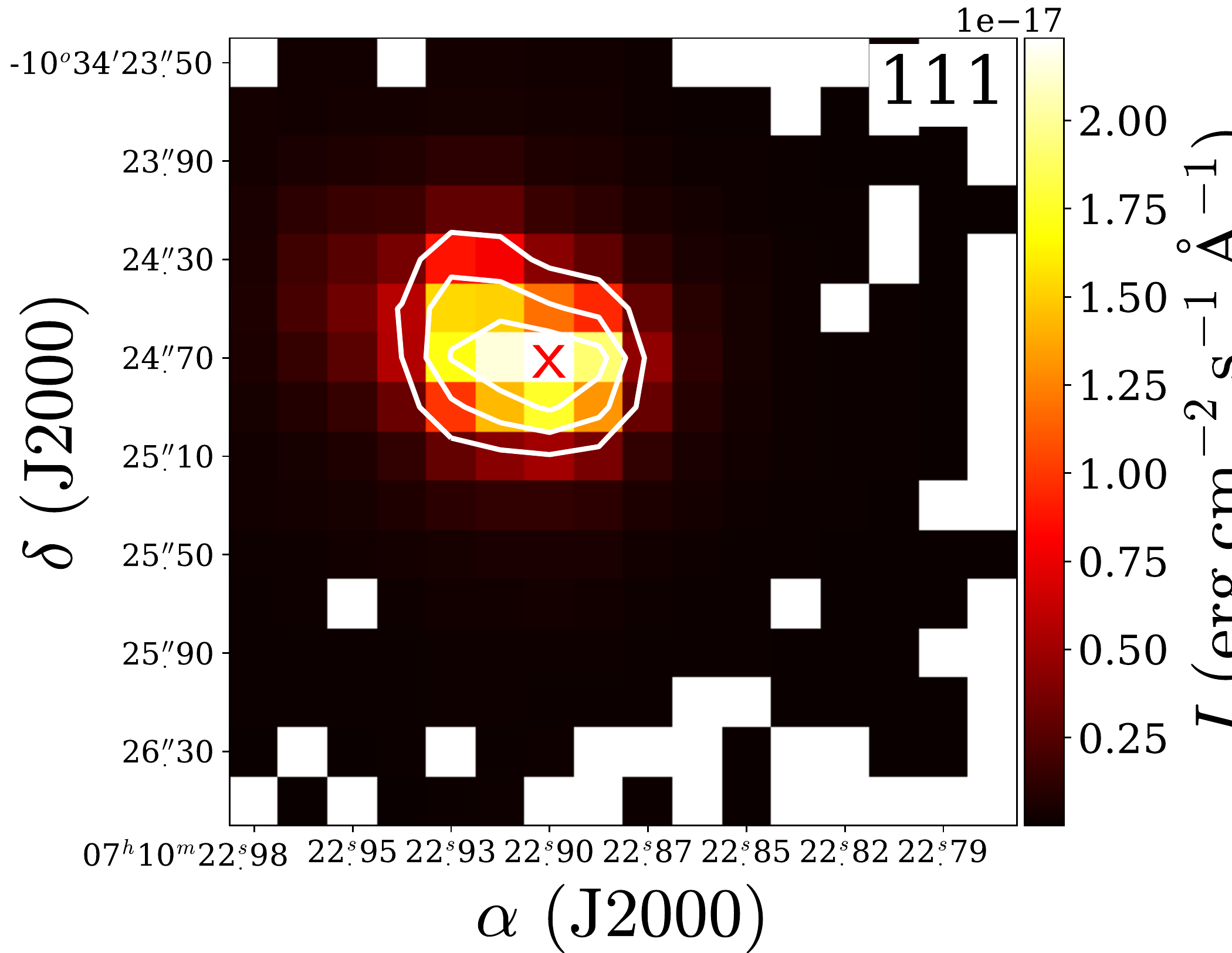}\hspace{-0.1cm}
\includegraphics[width=0.2\textwidth]{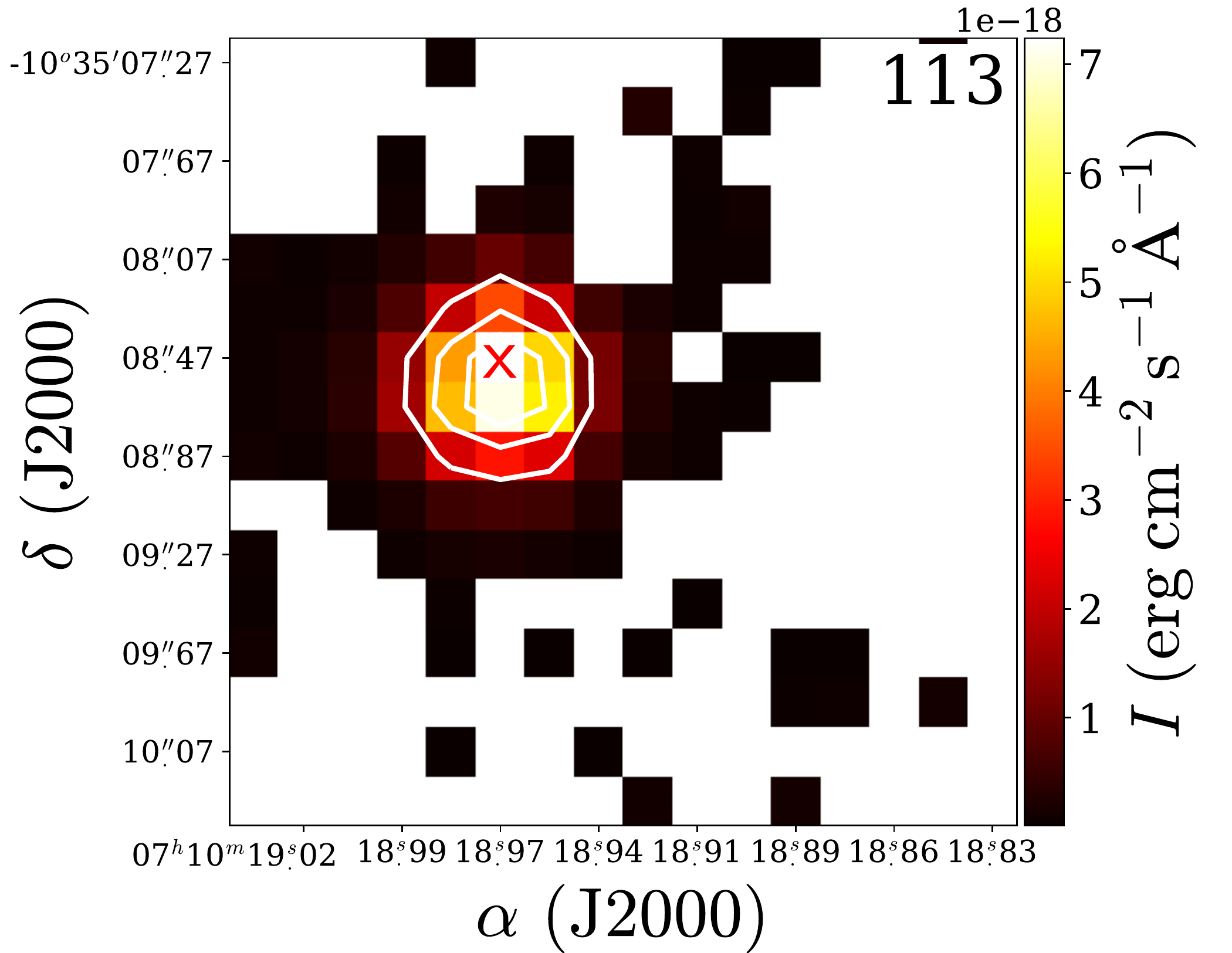}\hspace{-0.1cm}
\includegraphics[width=0.2\textwidth]{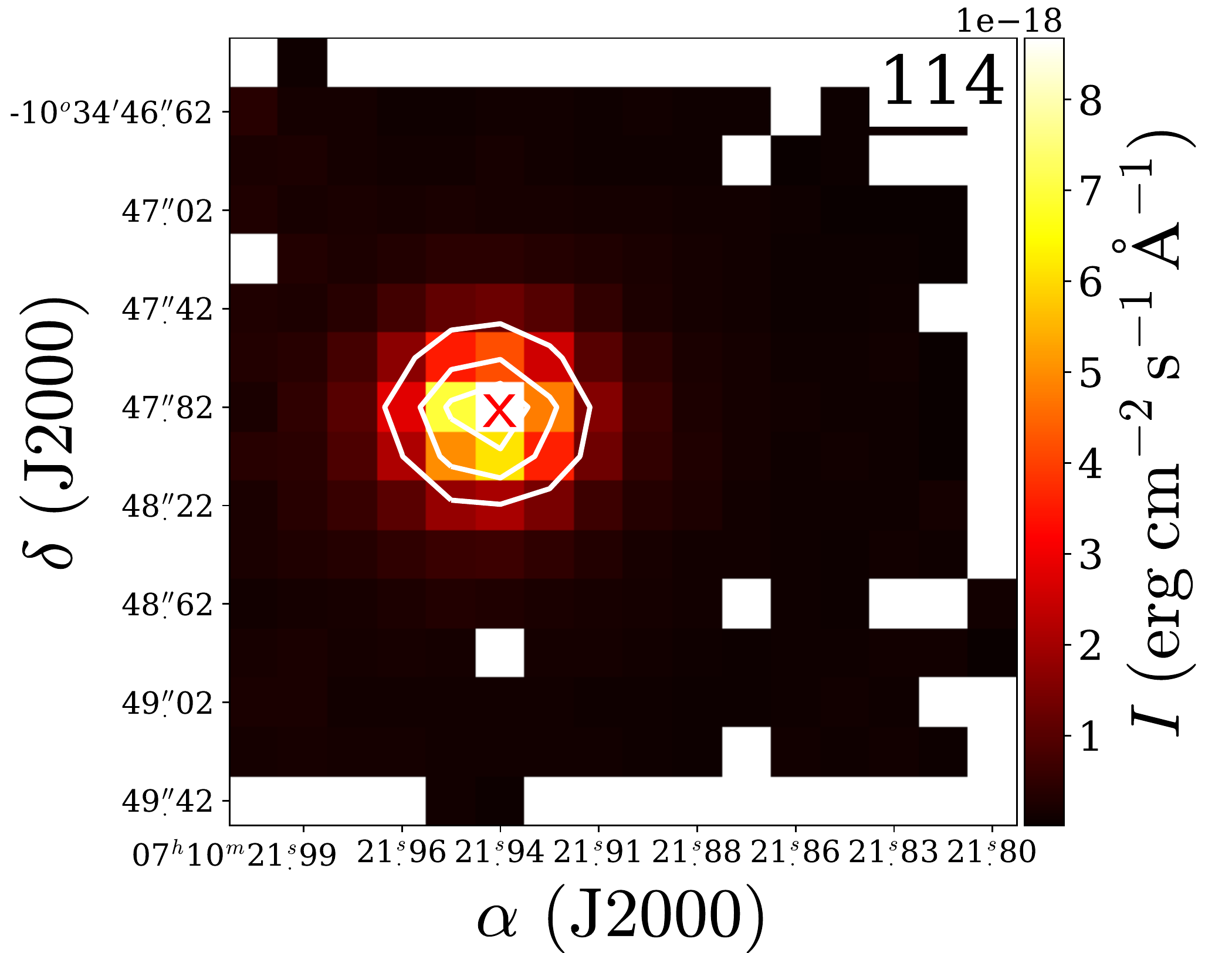}\hspace{-0.1cm}
\includegraphics[width=0.2\textwidth]{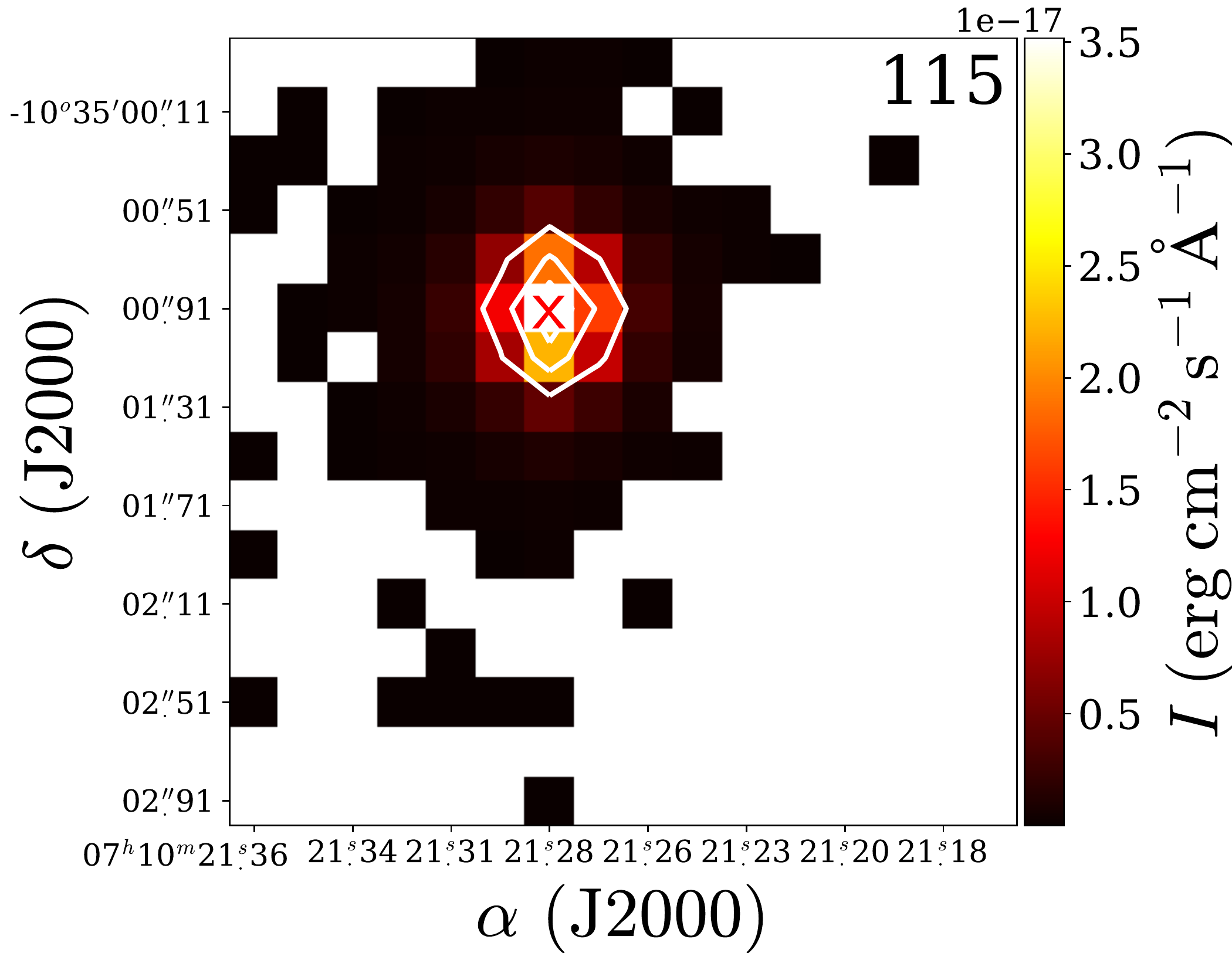}\hspace{-0.1cm}
\includegraphics[width=0.2\textwidth]{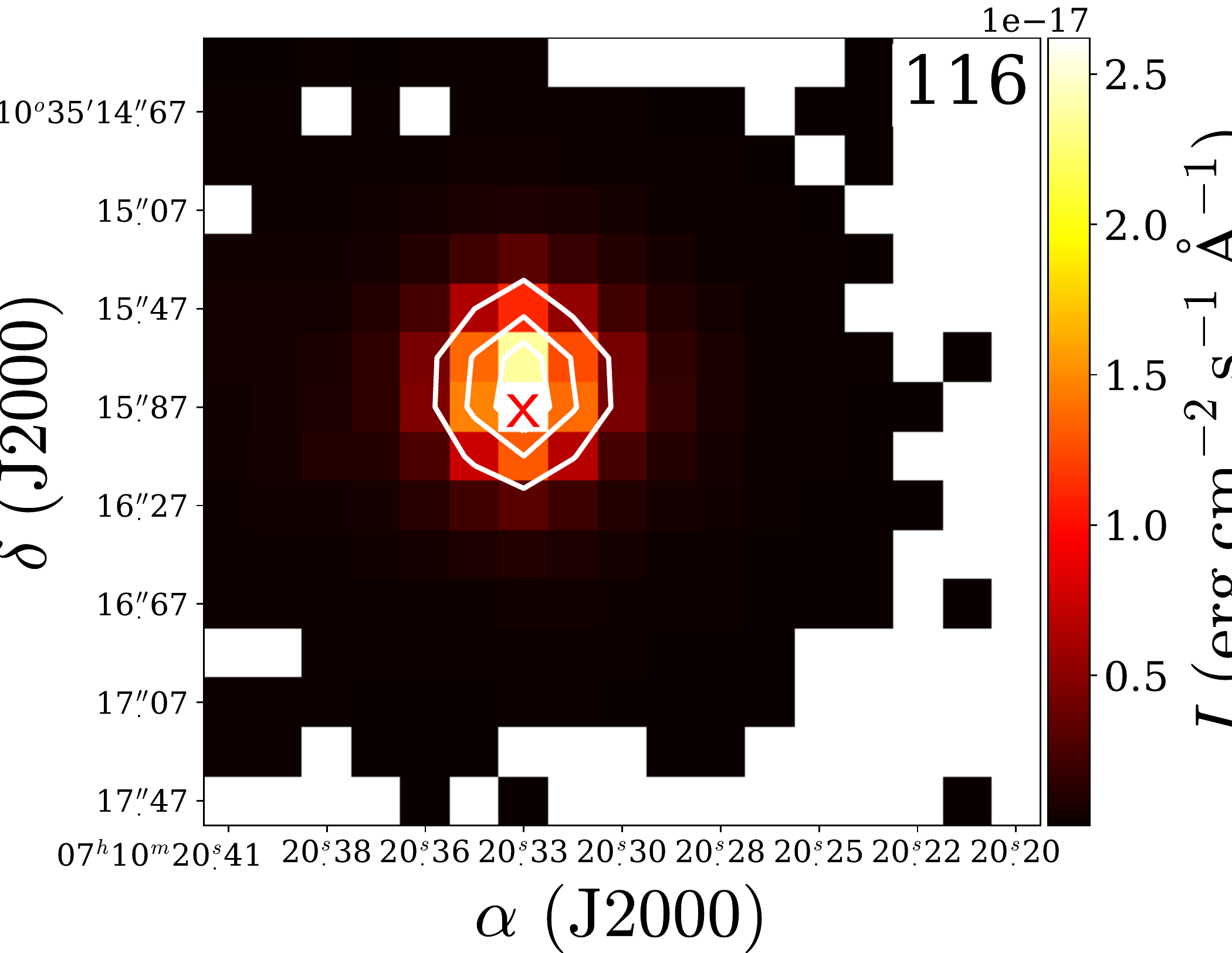}\hspace{-0.1cm}
\includegraphics[width=0.2\textwidth]{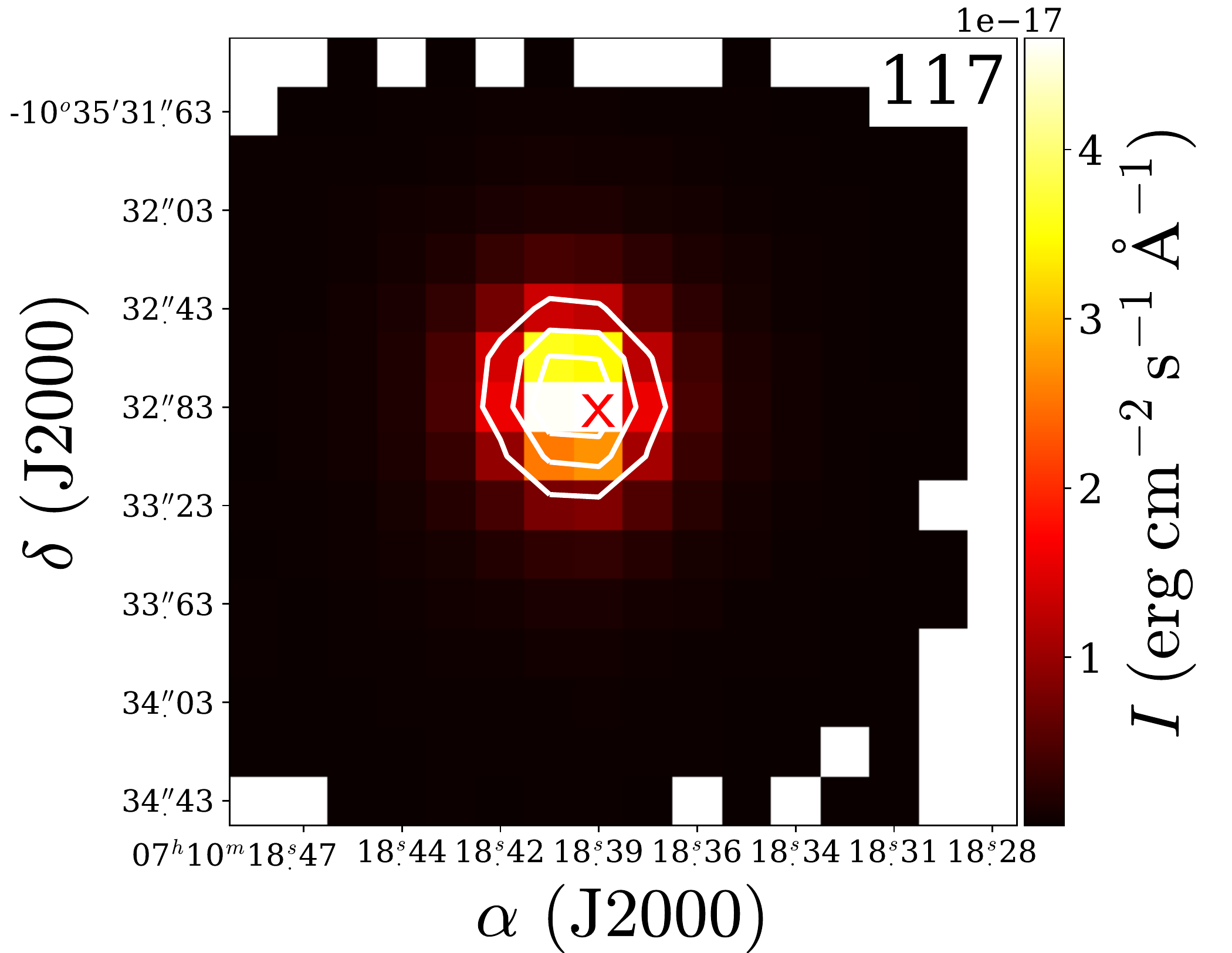}\hspace{-0.1cm}
\includegraphics[width=0.2\textwidth]{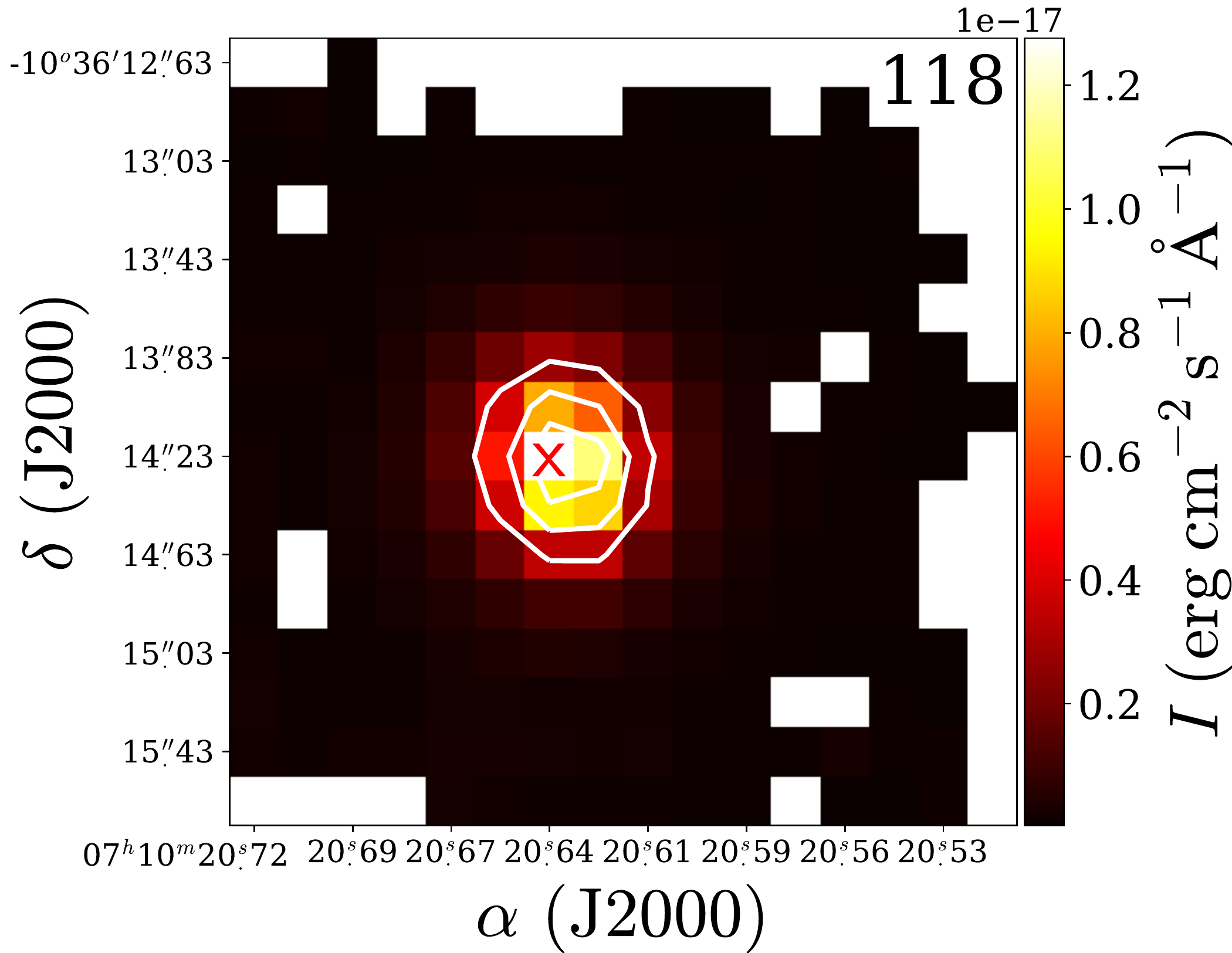}\hspace{-0.1cm}
\caption{Continued}
\end{figure*}
\begin{figure*}
\centering
\includegraphics[width=0.2\textwidth]{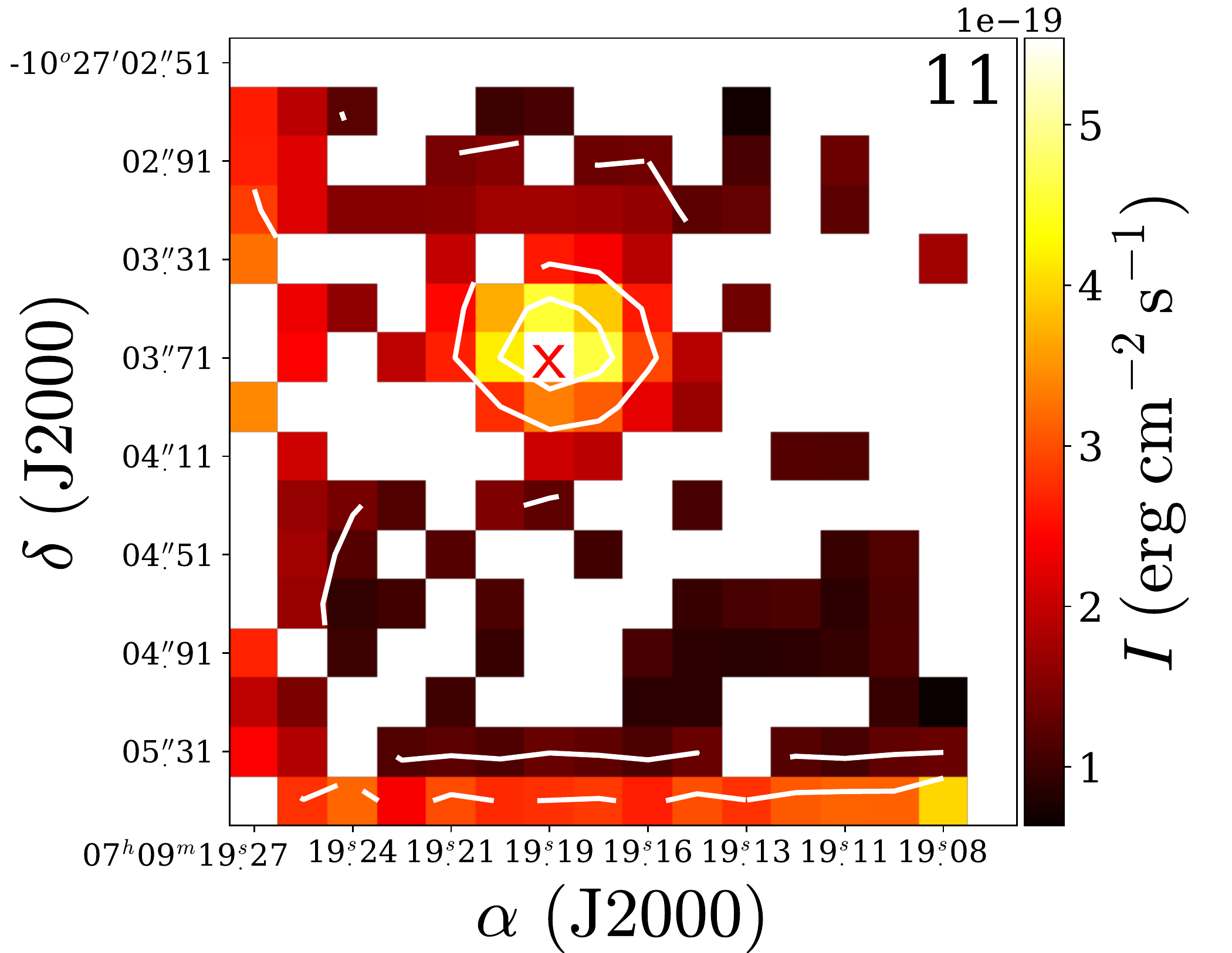}\hspace{-0.1cm} 
\includegraphics[width=0.2\textwidth]{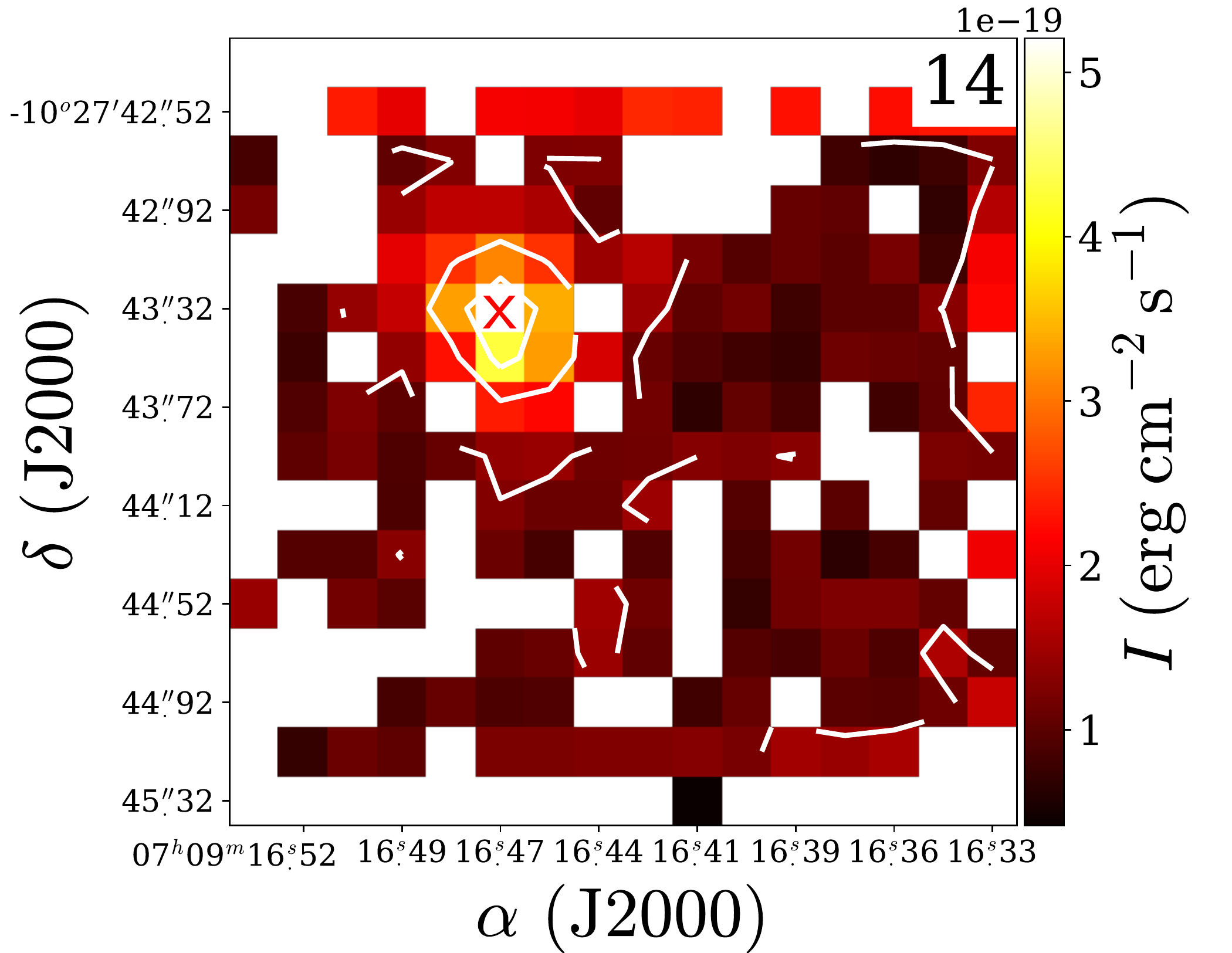}\hspace{-0.1cm} 
\includegraphics[width=0.2\textwidth]{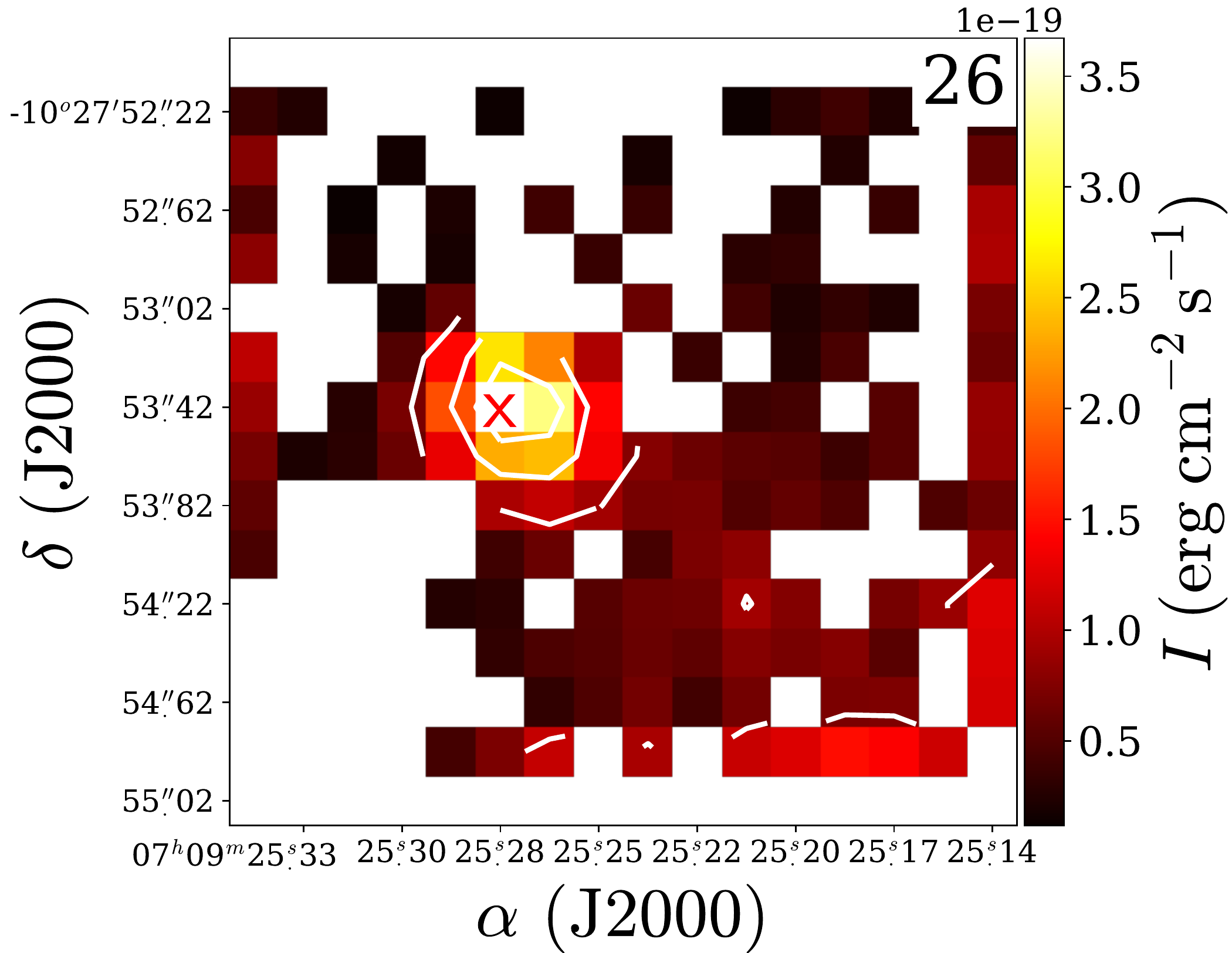}\hspace{-0.1cm}
\includegraphics[width=0.2\textwidth]{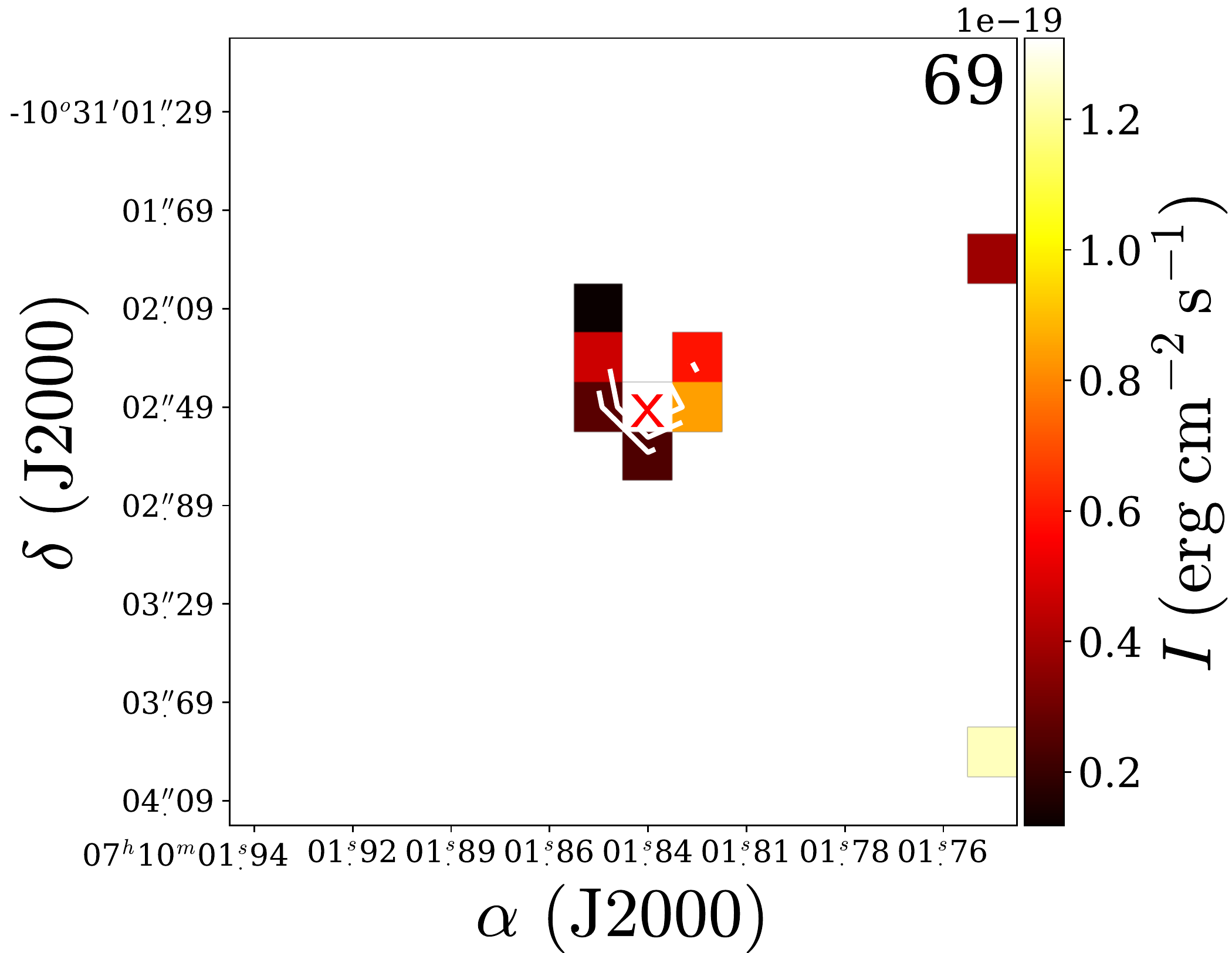}\hspace{-0.1cm}	
\caption{Similar to Figure \ref{fig:cont}, but for the faintest sources in the sample: No. 11, 14, 26, and 69 (see Table \ref{tab:coordinates}). The continuum was integrated over the entire $K$-band spectrum in every pixel to increase the signal-to-noise ratio.}
\label{fig:cont-int}
\end{figure*}


\section{Spectra}
\label{app:spec}

Figures~\ref{spec:1}-\ref{spec:10} show the $K$-band spectra of all YSO candidates in CMa-$\ell$224 observed with KMOS. The spectra are presented in subgroups having similar characteristics -- line detections and their profiles. In particular, 
Figure~\ref{spec:1} shows spectra of sources with the Br$\gamma$ line in emission, as represented by source No. 12 in Figure~\ref{fig:spec:ex}. 
Figure \ref{spec:2} presents spectra of sources with the detections of both the Br$\gamma$ and H$_2$ at 2.12 $\mu$m lines in emission, but a lack of the CO bandhead emission. 
Figure \ref{spec:3} shows spectra of two sources with the strong H$_2$ emission from outflows, but a lack of the Br$\gamma$ or CO bandhead lines. 
Figure \ref{spec:4} shows spectra of three sources with the Br$\gamma$ and H$_2$ lines in emission, and the CO bandhead lines in absorption. 
Figure \ref{spec:5} features spectra of presumably more evolved objects without the H$_2$ emission lines, but with the Br$\gamma$ line in emission and the CO bandhead lines in absorption. 
Figure \ref{spec:6} shows spectra of sources with the H$_2$ emission and the CO bandhead absorption lines, but a lack of the Br$\gamma$ emission. 
Figure \ref{spec:7} shows spectra of sources with the detection of Br$\gamma$ in absorption, and Figure \ref{spec:8}, the only five single sources with spectra characterised by the CO bandhead in emission. Figure \ref{spec:9} shows spectra of sources with CO in absorption, but non-detections of the H$_2$ or Br$\gamma$ lines. Finally, Figure \ref{spec:10} shows spectra of YSO candidates which do not show any signatures of accretion or ejection, and require follow-up observations to confirm their status.

Spectra of the individual components of the binary YSO candidates are presented in Figures~\ref{spec:13}-\ref{spec:15}. While all spectra of single objects were extracted from the spectral cubes within the aperture of radius of 3 pixels, due to their proximity, for double stars different approach has been followed. For the most distant pairs (sources No. 21, 38, 38, 86, and 87) the same aperture radius was used. For sources No. 82, 91, and 112 we used the radius of 2 pixels. To maximise the collected signal, we used the radius of 1.5~pixels for pairs No. 52 and 55. Spectra of the barely resolved binary, source No. 89, were extracted using the 1~pixel radius. Their signal can be contaminated. 

Table~\ref{tab:detect} provides an inventory of the key line detections in all sources, while Table \ref{tab:H2flux-ex} lists the measured H$_2$ line fluxes not corrected for extinction. We include only detections above 3$\sigma$.

\clearpage

\startlongtable
\begin{deluxetable}{llllllllll}
\tablecaption{Detections. \label{tab:detect}} 
\tablehead{
  \colhead{No.} &
  \colhead{H$_2$}&
  \colhead{Br$_\gamma$}&
  \colhead{He I} &
  \colhead{Ca I} &
  \colhead{Mg I} &
  \colhead{Na I} &
  \colhead{Si I} &
  \colhead{K I} &
  \colhead{CO} 
 }
\startdata 
2	&	E	&	N	&	N	&	A	&	A	&	A	&	N	&	N	&	N\\
47	&	E	&	N	&	N	&	N	&	N	&	N	&	N	&	N	&	N\\
87A	&	E	&	N	&	A	&	A	&	N	&	N	&	N	&	A	&	N\\
87B	&	E	&	N	&	A	&	A	&	A	&	A?	&	N	&	A	&	N\\
\hline 
9	&	E	&	E	&	E	&	A	&	A	&	A	&	N	&	N	&	N\\
13	&	E	&	E	&	A	&	A?	&	N	&	N	&	N	&	A	&	N\\
15	&	E	&	E	&	A	&	A	&	A	&	A	&	N	&	A	&	N\\
26	&	E	&	E	&	?	&	A?	&	N	&	A	&	N	&	E	&	N\\
33	&	E	&	E	&	E	&	A	&	N	&	N	&	N	&	E	&	N\\
50	&	E	&	E	&	A	&	A	&	A	&	A	&	A?	&	A	&	N\\
54	&	E	&	E	&	N	&	A	&	?	&	N	&	N	&	A	&	N\\
60	&	E	&	E	&	E?	&	A	&	N	&	N	&	N	&	E?	&	N\\
83	&	E	&	E	&	A	&	A?	&	A?	&	N	&	?	&	A	&	N\\
111	&	E	&	E	&	E?	&	A	&	A	&	A	&	A?	&	E	&	N\\
112B	&	E?	&	E	&	A	&	?	&	N	&	N	&	N	&	A	&	N\\
114	&	E	&	E	&	E	&	?	&	N	&	A?	&	N	&	E	&	N\\
\hline 
4	&	E	&	N	&	A?	&	A	&	A	&	A	&	N	&	A	&	A\\
7	&	E	&	N	&	A	&	A	&	A	&	A	&	A?	&	A	&	A\\
35	&	E	&	N	&	?	&	A	&	A	&	A	&	N	&	A	&	A\\
72	&	E	&	?	&	A?	&	A	&	A	&	A	&	A?	&	A	&	A\\
78	&	E	&	N	&	E	&	A	&	A	&	A	&	A	&	E	&	A\\
100	&	E	&	N	&	A	&	A	&	A	&	A	&	?	&	A	&	A\\
\hline 
49	&	E	&	E	&	E	&	A	&	A	&	A	&	A?	&	E	&	A\\
73	&	E	&	E	&	E	&	A	&	A	&	A	&	E	&	E	&	A?\\
91B	&	E	&	E	&	N	&	A	&	A?	&	A	&	A	&	?	&	A\\
109	&	E	&	E	&	A	&	A	&	A	&	A	&	N	&	A	&	A\\
\hline 
16	&	N	&	E	&	N	&	A	&	A	&	A	&	A?	&	N	&	N\\
24	&	N	&	E	&	N	&	A	&	A	&	A	&	N	&	A	&	N\\
28	&	N	&	E	&	N	&	A	&	A	&	A	&	A	&	A	&	N\\
31	&	N	&	E	&	?	&	A	&	N	&	N	&	N	&	A?	&	N\\
36A	&	N	&	E	&	N	&	A	&	A	&	A	&	A	&	N	&	N\\
41	&	N	&	E	&	N	&	A	&	A	&	N	&	N	&	A	&	N\\
56	&	N	&	E	&	A	&	A	&	A	&	A	&	N	&	A	&	N\\
79	&	N	&	E	&	A	&	A	&	N	&	A	&	N	&	A	&	N\\
81	&	N	&	E	&	A	&	A	&	A	&	A?	&	N	&	A?	&	N\\
85	&	N	&	E	&	A	&	A	&	A	&	A	&	N	&	A	&	N\\
93	&	N	&	E	&	E	&	A	&	N	&	A	&	E?	&	E	&	N\\
102	&	N	&	E	&	N	&	A	&	A	&	A	&	N	&	N	&	N\\
118	&	N	&	E	&	E	&	A	&	A	&	A	&	A?	&	E	&	N\\
\hline 
3	&	N	&	E	&	N	&	A	&	A	&	A	&	A?	&	A	&	A\\
6	&	N	&	E	&	N	&	A	&	A	&	A	&	A?	&	A	&	A\\
10	&	N	&	E	&	E	&	A	&	A	&	A	&	N	&	E?	&	A?\\
12	&	N	&	E	&	E?	&	A	&	A	&	A	&	A?	&	N	&	A\\
18	&	N	&	E	&	E	&	A	&	A	&	A	&	A	&	E	&	A\\
20	&	N	&	E	&	A?	&	A	&	A	&	A	&	A	&	N	&	A\\
22	&	N	&	E	&	E	&	A	&	A	&	A	&	A?	&	E	&	A\\
23	&	N	&	E	&	E	&	A	&	A	&	A	&	A?	&	E?	&	A\\
30	&	N	&	E?	&	E?	&	A	&	A	&	A	&	A?	&	N	&	A\\
32	&	N	&	E	&	N	&	A	&	A	&	A	&	A	&	N	&	A\\
34	&	N	&	E	&	E	&	A	&	A	&	A	&	A	&	E?	&	A\\
38A	&	N	&	E	&	A	&	A	&	A	&	A	&	A?	&	E	&	A\\
38B	&	N	&	E	&	A	&	A	&	A	&	A?	&	A?	&	E	&	A\\
48	&	N	&	E	&	E?	&	A	&	A	&	A	&	A	&	N	&	A\\
52A	&	N	&	E	&	A	&	A	&	A	&	A	&	A	&	A	&	A\\
52B	&	N	&	E	&	A	&	A	&	A	&	A	&	A	&	A	&	A\\
55A	&	N	&	E	&	E?	&	A	&	A	&	A	&	A	&	E?	&	A\\
57	&	N	&	E	&	A?	&	A	&	A	&	A	&	N	&	A?	&	A\\
80	&	N	&	E	&	A	&	A	&	A	&	A	&	A	&	A	&	A\\
89A	&	N	&	E?	&	A	&	A	&	A	&	A	&	A?	&	A	&	A\\
97	&	N	&	E	&	A	&	A	&	A	&	A	&	A?	&	A?	&	A\\
98	&	N	&	E	&	A	&	A	&	A	&	A	&	A?	&	A	&	A\\
99	&	N	&	E	&	E	&	A	&	?	&	A	&	E?	&	E?	&	A\\
\hline 
5	&	N	&	A	&	N	&	A	&	A	&	A	&	A	&	N	&	A\\
40	&	N	&	A	&	N	&	A	&	A	&	A	&	A	&	N	&	A\\
43	&	N	&	A	&	N	&	A	&	A	&	A	&	A	&	A?	&	A\\
53	&	N	&	A	&	N	&	A	&	A	&	N	&	N	&	A	&	N\\
61	&	N	&	A	&	E?	&	A	&	A	&	A	&	A	&	E	&	A\\
63	&	N	&	A	&	E?	&	A	&	A	&	A	&	A	&	E	&	A\\
65	&	N	&	A	&	N	&	A	&	A	&	A	&	A	&	N	&	A\\
71	&	N	&	A	&	A	&	A	&	A	&	A?	&	?	&	A	&	A\\
75	&	N	&	A	&	A	&	A?	&	A	&	N	&	A?	&	A	&	N\\
84	&	N	&	A	&	N	&	A	&	A	&	A	&	A	&	E?	&	A\\
91A	&	E	&	A	&	A?	&	A	&	A	&	A	&	A	&	E?	&	A\\
92	&	N	&	A	&	A	&	A	&	A	&	N	&	N	&	A	&	N\\
94	&	N	&	A	&	A	&	A	&	A	&	A	&	A?	&	A?	&	A\\
96	&	N	&	A	&	A	&	A	&	A	&	A	&	A?	&	A	&	A\\
104	&	N	&	A	&	A?	&	A	&	A	&	A	&	A	&	E?	&	A\\
116	&	N	&	A	&	A	&	A	&	A	&	A	&	A	&	E	&	A?\\
\hline 
19	&	E	&	E	&	E	&	N	&	N	&	N	&	N	&	E	&	E\\
39	&	E?	&	E	&	N	&	A	&	A	&	A	&	N	&	N	&	E\\
46	&	E	&	E	&	N	&	A	&	A	&	N	&	N	&	N	&	E\\
64	&	E	&	E	&	E	&	A	&	A?	&	N	&	N	&	E	&	E\\
77	&	E	&	E	&	E	&	A?	&	A?	&	A?	&	E?	&	E	&	E\\
112A	&	E?	&	E	&	A	&	?	&	N	&	N	&	N	&	A	&	E\\
\hline 
25	&	N	&	N	&	N	&	A	&	A	&	A	&	A	&	A	&	A\\
29	&	N	&	N	&	A?	&	A	&	A	&	A	&	A	&	A	&	A\\
42	&	N	&	N	&	E?	&	A	&	A	&	A	&	N	&	E?	&	A\\
44	&	N	&	N	&	N	&	A	&	A	&	A	&	A?	&	A?	&	A\\
45	&	N	&	N	&	N	&	A	&	A	&	A	&	N	&	N	&	A\\
51	&	N	&	N	&	N	&	A	&	A	&	A	&	A	&	N	&	A\\
55B	&	N	&	N	&	E	&	A	&	A	&	A	&	A?	&	E?	&	A\\
58	&	N	&	N	&	E?	&	A	&	A	&	A	&	A?	&	N	&	A\\
59	&	N	&	N	&	?	&	A	&	A	&	A	&	A?	&	N	&	A\\
66	&	N	&	N	&	A	&	A	&	A	&	A	&	A	&	N	&	A\\
68	&	N	&	N	&	N	&	A	&	A	&	A	&	A	&	E?	&	A\\
69	&	N	&	N	&	E?	&	A	&	A?	&	A?	&	N	&	A	&	A?\\
70	&	N	&	N	&	A	&	A	&	A	&	A	&	A?	&	A	&	A\\
74	&	N	&	N	&	E?	&	A	&	A	&	A	&	?	&	A	&	A\\
82A	&	N	&	N	&	A	&	A	&	A	&	A	&	A?	&	A	&	A\\
82B	&	N	&	N	&	A	&	A	&	A	&	A	&	A?	&	A	&	A\\
86A	&	N	&	N	&	N	&	A	&	A	&	A	&	A?	&	A	&	A\\
86B	&	N	&	N	&	?	&	A	&	A?	&	A	&	A?	&	A?	&	A\\
89B	&	N	&	N	&	?	&	A	&	A	&	A	&	A?	&	A	&	A\\
95	&	N	&	N	&	E	&	A	&	A	&	A	&	N	&	E	&	A?\\
101	&	N	&	N	&	E?	&	A	&	N	&	A	&	E?	&	E	&	A\\
103	&	N	&	N	&	A?	&	A	&	A	&	A	&	A?	&	E	&	A\\
105	&	N	&	N	&	E	&	A	&	A	&	A	&	A	&	E	&	A\\
108	&	N	&	N	&	E	&	A	&	A	&	A	&	A	&	E	&	A\\
110	&	N	&	N	&	E	&	A	&	A	&	A	&	A?	&	E	&	A\\
113	&	N	&	N	&	A	&	A	&	A	&	A	&	A	&	A	&	A\\
115	&	N	&	N	&	A	&	A	&	A	&	A	&	A	&	A	&	A\\
117	&	N	&	N	&	E	&	A	&	A	&	A	&	A	&	E	&	A\\
\hline 
1	&	N	&	N	&	?	&	A	&	A	&	A	&	A	&	E?	&	N\\
8	&	N	&	N	&	N	&	A	&	A?	&	A?	&	N	&	A	&	N\\
11	&	N	&	N	&	E	&	A	&	A	&	N	&	N	&	A	&	N\\
14	&	N	&	N	&	E	&	A?	&	A	&	N	&	E?	&	E	&	N\\
17	&	N	&	N	&	A	&	A	&	A	&	A	&	N	&	A	&	N\\
21A	&	N	&	N	&	A?	&	?	&	N	&	N	&	N	&	A	&	N\\
21B	&	N	&	N	&	A?	&	?	&	A	&	N	&	N	&	A	&	N\\
62	&	N	&	N	&	E	&	A	&	A	&	A	&	N	&	E	&	N\\
76	&	N	&	N	&	?	&	N	&	A?	&	N	&	?	&	A?	&	N\\
90	&	N	&	N	&	A	&	A?	&	A?	&	N	&	N	&	A	&	N\\
106	&	N	&	N	&	A	&	A	&	N	&	A?	&	N	&	A	&	N\\
\enddata
\tablecomments{E indicates emission line, A - absorption, N - no detection, ? - possible detection.}
\end{deluxetable}


\movetabledown=-3cm
\begin{splitdeluxetable*}{lcccccccBlccccccc} \tabletypesize{\fontsize{8}{6.8}\selectfont}
\tablecaption{H$_2$ Emission Line Fluxes. \label{tab:H2flux-ex}}
\tablehead{
\colhead{No.} & \colhead{1-0 S(0)} & \colhead{1-0 S(1)}  & 
\colhead{1-0 S(2)} & \colhead{1-0 S(3)} & 
\colhead{2-1 S(1)} & \colhead{2-1 S(2)} & 
\colhead{2-1 S(3)} & 
\colhead{No.} & 
\colhead{1-0 Q(1)} & \colhead{1-0 Q(2)} & 
\colhead{1-0 Q(3)} & \colhead{1-0 Q(4)} & 
\colhead{1-0 Q(5)} & \colhead{1-0 Q(6)} \\ 
\cline{2-8} \cline{10-15}
\colhead{} & \multicolumn{7}{c}{($\rm 10^{-16}~erg~s^{-1}cm^{-2}$)} & \colhead{} & \multicolumn{6}{c}{($\rm 10^{-16}~erg~s^{-1}cm^{-2}$)}
}
\startdata
2 & 5.35 $\pm$ 0.33 & 25.61 $\pm$ 0.58 & 10.18 $\pm$ 0.38 & 36.90 $\pm$ 3.39 & \nodata & \nodata & 2.24 $\pm$ 0.87 & 2 & 25.77 $\pm$ 1.54 & 5.97 $\pm$ 1.86 & 22.60 $\pm$ 1.10 & \nodata & 16.72 $\pm$ 1.14 & 5.40 $\pm$ 1.37 \\ 
4* & 1.55 $\pm$ 0.16 & 3.36 $\pm$ 0.20 & 0.85 $\pm$ 0.11 & 6.27 $\pm$ 1.19 & 0.97 $\pm$ 0.26 & \nodata & \nodata & 4* & 5.38 $\pm$ 0.84 & \nodata & 4.99 $\pm$ 0.65 & \nodata & 2.98 $\pm$ 0.56 & \nodata \\ 
7* & \nodata & 0.87 $\pm$ 0.15 & \nodata & \nodata & \nodata & \nodata & \nodata & 7* & \nodata & \nodata & \nodata & \nodata & 2.57 $\pm$ 0.76 & \nodata \\ 
9 & \nodata & 1.62 $\pm$ 0.12 & \nodata & 1.83 $\pm$ 0.51 & \nodata & \nodata & \nodata & 9 & \nodata & \nodata & \nodata & \nodata & \nodata & \nodata \\ 
13 & \nodata & 2.33 $\pm$ 0.56 & \nodata & 3.26 $\pm$ 0.93 & \nodata & \nodata & \nodata & 13 & \nodata & \nodata & \nodata & \nodata & \nodata & \nodata \\ 
15 & \nodata & 4.42 $\pm$ 0.27 & \nodata & 7.84 $\pm$ 2.26 & 1.91 $\pm$ 0.31 & \nodata & \nodata & 15 & 5.70 $\pm$ 1.22 & \nodata & 6.59 $\pm$ 1.45 & \nodata & 4.21 $\pm$ 0.92 & \nodata \\ 
19 & 1.52 $\pm$ 0.25 & 5.30 $\pm$ 0.22 & 2.62 $\pm$ 0.24 & 5.14 $\pm$ 0.84 & 1.21 $\pm$ 0.31 & \nodata & \nodata & 19 & 6.21 $\pm$ 0.83 & \nodata & 6.71 $\pm$ 0.90 & \nodata & 3.62 $\pm$ 1.02 & 2.24 $\pm$ 0.67 \\ 
26 & \nodata & 1.51 $\pm$ 0.20 & 0.86 $\pm$ 0.20 & \nodata & \nodata & \nodata & \nodata & 26 & \nodata & \nodata & \nodata & \nodata & \nodata & \nodata \\ 
33 & 0.94 $\pm$ 0.30 & 2.17 $\pm$ 0.16 & 0.83 $\pm$ 0.15 & 5.43 $\pm$ 1.39 & \nodata & \nodata & \nodata & 33 & 3.53 $\pm$ 0.75 & \nodata & 3.30 $\pm$ 1.00 & \nodata & \nodata & \nodata\\ 
35 & \nodata & 3.24 $\pm$ 0.31 & \nodata & \nodata & \nodata & \nodata & \nodata & 35 & \nodata & \nodata & \nodata & \nodata & \nodata & \nodata \\ 
39 & \nodata & 12.53 $\pm$ 1.57 & \nodata & \nodata & \nodata & \nodata & \nodata & 39 & \nodata & \nodata & 22.78 $\pm$ 2.78 & \nodata & \nodata & \nodata\\  
46 & \nodata & 12.92 $\pm$ 0.60 & 6.16 $\pm$ 0.81 & \nodata & \nodata & \nodata & \nodata & 46 & \nodata & \nodata & \nodata & \nodata & \nodata & \nodata \\ 
47 & 17.48 $\pm$ 0.52 & 69.91 $\pm$ 1.05 & 22.15 $\pm$ 0.51 & 76.05 $\pm$ 2.38 & 7.43 $\pm$ 0.40 & 3.09 $\pm$ 0.36 & 6.19 $\pm$ 0.34 & 47 & 85.63 $\pm$ 2.80 & 24.76 $\pm$ 0.95 & 76.93 $\pm$ 2.74 & 20.75 $\pm$ 3.33 & 26.15 $\pm$ 1.24 & 10.96 $\pm$ 2.36 \\
49 & \nodata & 4.65 $\pm$ 0.34 & 3.14 $\pm$ 0.47 & \nodata & \nodata & \nodata & 3.30 $\pm$ 0.59 & 49 & 6.04 $\pm$ 0.53 & \nodata & 4.87 $\pm$ 0.53 & \nodata & \nodata & \nodata \\ 
50 & \nodata & 1.47 $\pm$ 0.18 & 0.49 $\pm$ 0.17 & 4.37 $\pm$ 1.22 & \nodata & \nodata & \nodata & 50 & \nodata & \nodata & 2.09 $\pm$ 0.56 & \nodata & 4.97 $\pm$ 1.02 & \nodata \\ 
54* & 23.05 $\pm$ 6.16 & 75.32 $\pm$ 14.15 & \nodata & 123.66 $\pm$ 27.71 & \nodata & \nodata & \nodata & 54* & \nodata & \nodata & \nodata & \nodata & \nodata & \nodata \\ 
60 & 14.58 $\pm$ 0.69 & 60.13 $\pm$ 0.47 & 23.57 $\pm$ 0.71 & 99.90 $\pm$ 13.36 & 7.30 $\pm$ 0.79 & 2.74 $\pm$ 0.36 & 5.84 $\pm$ 1.29 & 60 & 65.22 $\pm$ 3.30 & 12.80 $\pm$ 1.80 & 58.16 $\pm$ 2.53 & 39.66 $\pm$ 9.77 & 24.78 $\pm$ 5.70 & 13.01 $\pm$ 4.73 \\
64 & 3.11 $\pm$ 0.42 & 9.67 $\pm$ 0.45 & 4.86 $\pm$ 0.23 & \nodata & \nodata & 0.79 $\pm$ 0.25 & 2.45 $\pm$ 0.77 & 64 & 10.93 $\pm$ 1.08 & 4.87 $\pm$ 1.14 & 11.93 $\pm$ 0.62 & \nodata & \nodata & 3.51 $\pm$ 1.05 \\ 
73 & \nodata & 1.64 $\pm$ 0.32 & 2.45 $\pm$ 0.26 & \nodata & \nodata & 0.69 $\pm$ 0.13 & \nodata & 73 & 2.78 $\pm$ 0.58 & \nodata & 2.49 $\pm$ 0.33 & \nodata & 2.37 $\pm$ 0.53 & \nodata \\ 
77 & \nodata & 1.37 $\pm$ 0.14 & 2.43 $\pm$ 0.26 & \nodata & 0.40 $\pm$ 0.14 & \nodata & \nodata & 77 & 1.64 $\pm$ 0.46 & \nodata & \nodata & \nodata & \nodata & \nodata \\ 
83 & \nodata & 2.88 $\pm$ 0.26 & \nodata & 3.25 $\pm$ 1.01 & \nodata & \nodata & \nodata & 83 & \nodata & \nodata & \nodata & \nodata & \nodata & \nodata \\
87A & 0.49 $\pm$ 0.17 & 2.11 $\pm$ 0.19 & \nodata & \nodata & \nodata & \nodata & \nodata & 87A & 1.52 $\pm$ 0.52 & \nodata & 1.17 $\pm$ 0.33 & \nodata & \nodata & \nodata \\ 
87B & \nodata & 0.66 $\pm$ 0.08 & \nodata & \nodata & \nodata & \nodata & \nodata & 87B & 0.40 $\pm$ 0.17 & \nodata & \nodata & \nodata & \nodata & \nodata \\ 
91A* & \nodata & 11.73 $\pm$ 1.70 & \nodata & 17.49 $\pm$ 4.68 & \nodata & 5.43 $\pm$ 1.05 & \nodata & 91A* & 20.27 $\pm$ 6.68 & \nodata & 17.11 $\pm$ 4.79 & \nodata & \nodata & \nodata \\ 
91B* & 1.21 $\pm$ 0.32 & 4.84 $\pm$ 0.26 & 1.93 $\pm$ 0.32 & 4.14 $\pm$ 0.56 & \nodata & \nodata & 0.79 $\pm$ 0.23 & 91B* & 5.87 $\pm$ 0.50 & \nodata & 4.77 $\pm$ 0.25 & \nodata & \nodata & \nodata \\ 
100* & 2.56 $\pm$ 0.28 & 10.07 $\pm$ 0.25 & 3.77 $\pm$ 0.54 & 6.33 $\pm$ 0.92 & 1.40 $\pm$ 0.25 & \nodata & \nodata & 100* & 8.61 $\pm$ 0.48 & 2.63 $\pm$ 0.67 & 6.22 $\pm$ 0.38 & \nodata & \nodata & \nodata \\ 
109 & 2.12 $\pm$ 0.18 & 9.18 $\pm$ 0.46 & 2.61 $\pm$ 0.27 & 7.89 $\pm$ 1.44 & 0.80 $\pm$ 0.14 & \nodata & \nodata & 109 & 9.26 $\pm$ 0.42 & 2.91 $\pm$ 0.56 & 6.85 $\pm$ 0.66 & \nodata & \nodata & \nodata \\
111 & \nodata & 3.93 $\pm$ 0.90 & 2.14 $\pm$ 0.49 & \nodata & \nodata & \nodata & \nodata & 111 & \nodata & \nodata & \nodata & \nodata & \nodata & \nodata \\ 
112A & 0.27 $\pm$ 0.03 & 0.79 $\pm$ 0.16 & \nodata & \nodata & \nodata & \nodata & \nodata & 112A & \nodata & \nodata & 1.73 $\pm$ 0.34 & \nodata & \nodata & \nodata \\ 
112B & \nodata & 0.58 $\pm$ 0.15 & 0.25 $\pm$ 0.09 & \nodata & \nodata & \nodata & \nodata & 112B & \nodata & \nodata & \nodata & \nodata & \nodata & \nodata \\ 
114 & 2.64 $\pm$ 0.31 & 10.46 $\pm$ 0.65 & 4.93 $\pm$ 0.19 & 6.72 $\pm$ 1.08 & 0.83 $\pm$ 0.25 & 1.69 $\pm$ 0.40 & 2.97 $\pm$ 0.69 & 114 & 12.57 $\pm$ 1.00 & \nodata & 9.87 $\pm$ 0.79 & \nodata & \nodata & \nodata \\
\enddata
\tablecomments{H$_2$ line fluxes have not been corrected for extinction. Reported are only fluxes with detection significance above 3$\sigma$. Targets with asterisk do not have estimated extinction from SED models.} 
\end{splitdeluxetable*}


\begin{figure*}
\includegraphics[width=\textwidth]{{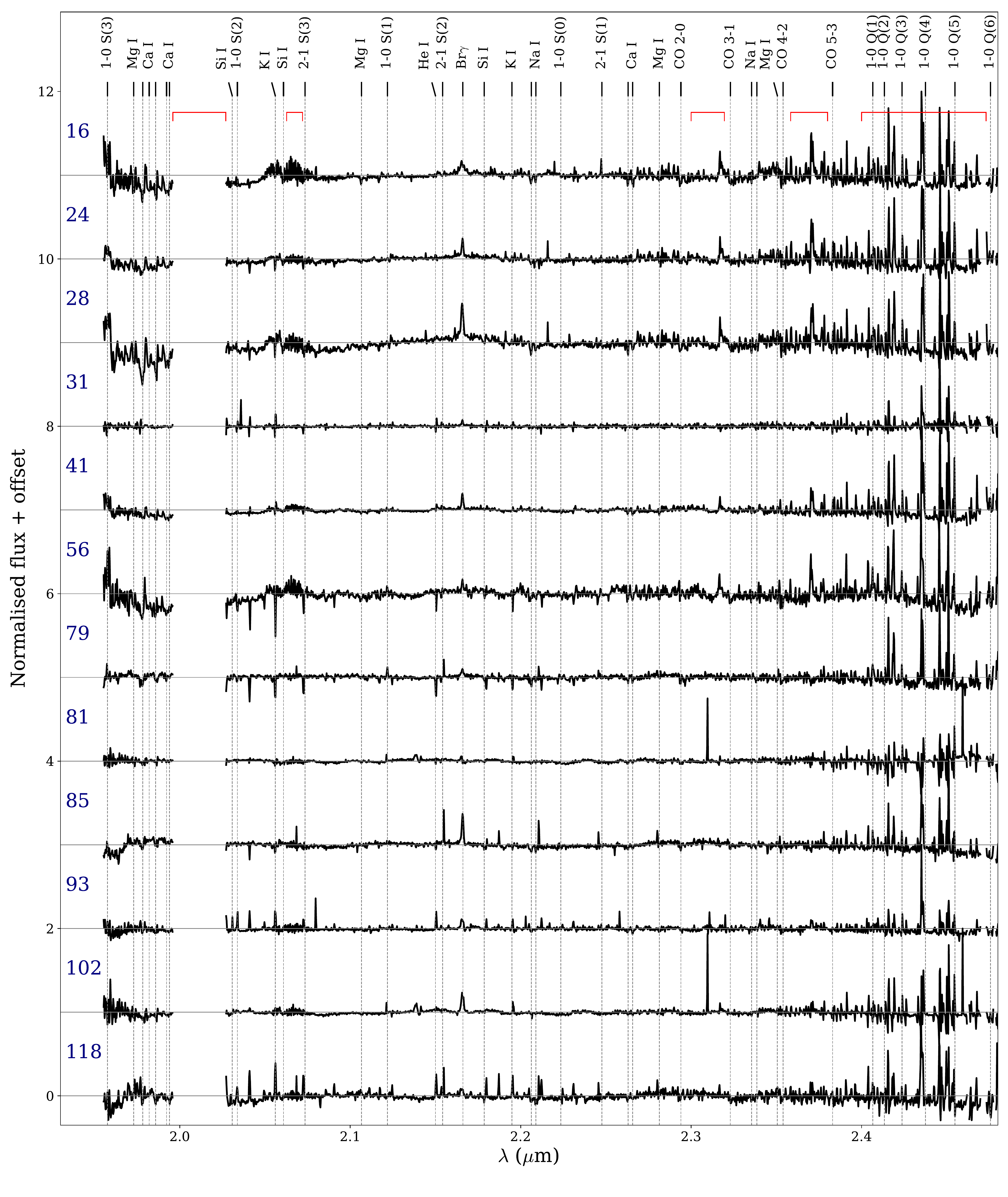}}
\caption{\label{spec:1}$K$-band spectra of YSO candidates in the CMa-$\ell$224 star forming region, which show the H Br$\gamma$ line in emission. All spectra are continuum subtracted and normalised to the peak flux density in the range from 1.97 to 2.47~$\mu$m. Red horizontal lines show spectral ranges most affected by telluric lines; the range at 2.0 $\mu$m has been removed for clarity.}
\end{figure*}
\begin{figure*}
\includegraphics[width=\textwidth]{{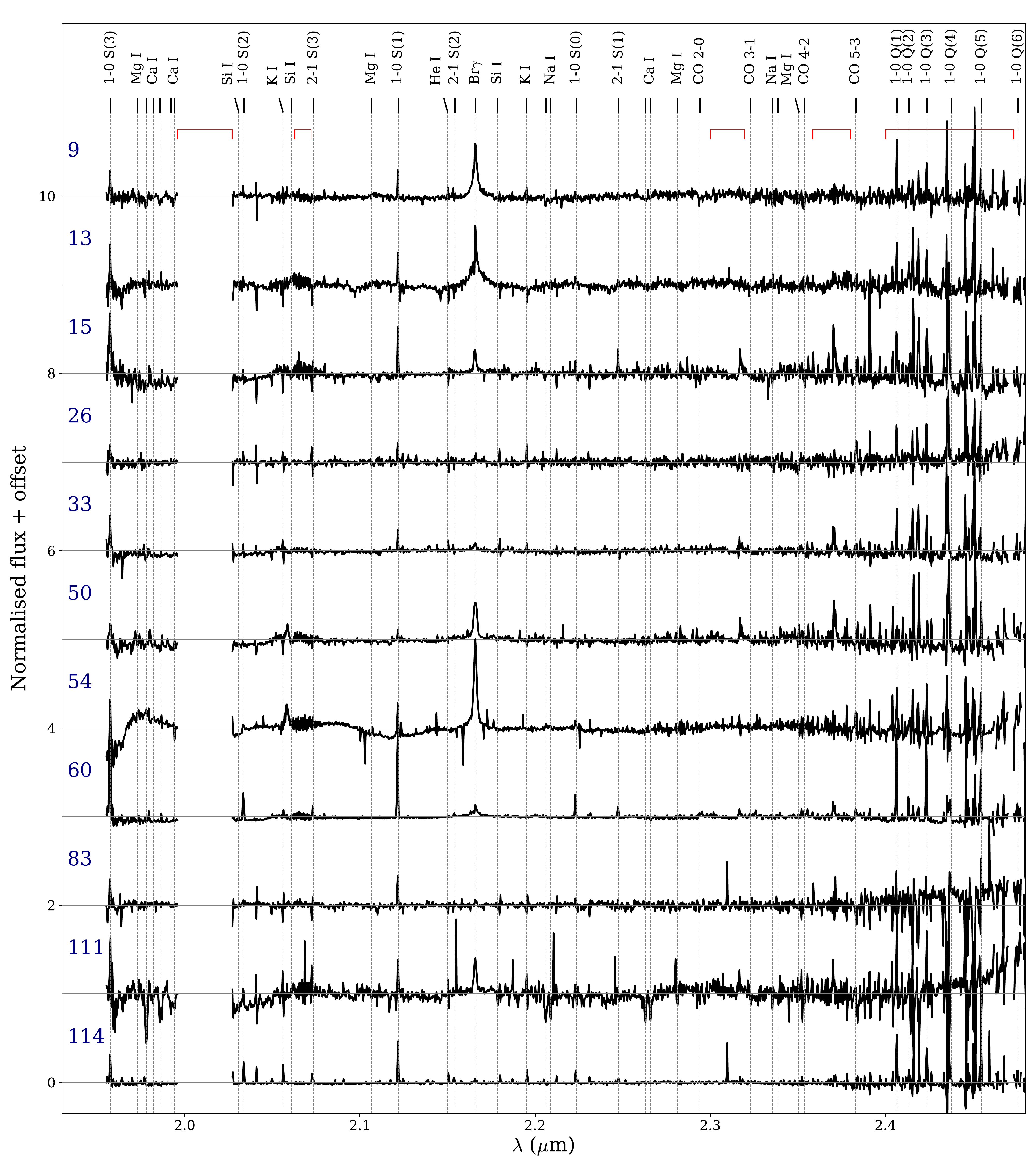}}
\caption{\label{spec:2}Spectra with Br$\gamma$ and H$_2$ emission lines.}
\end{figure*}

\begin{figure*}
\includegraphics[width=\textwidth]{{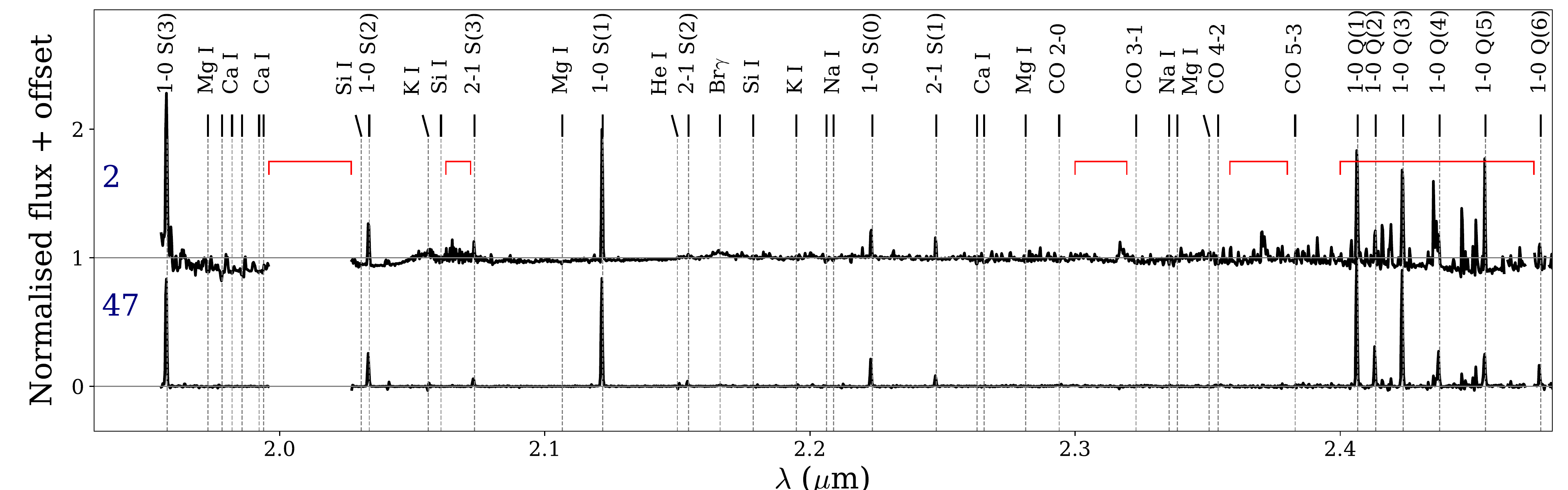}}
\caption{\label{spec:3}Spectra with only H$_2$ emission lines.}
\end{figure*}
\begin{figure*}
\includegraphics[width=\textwidth]{{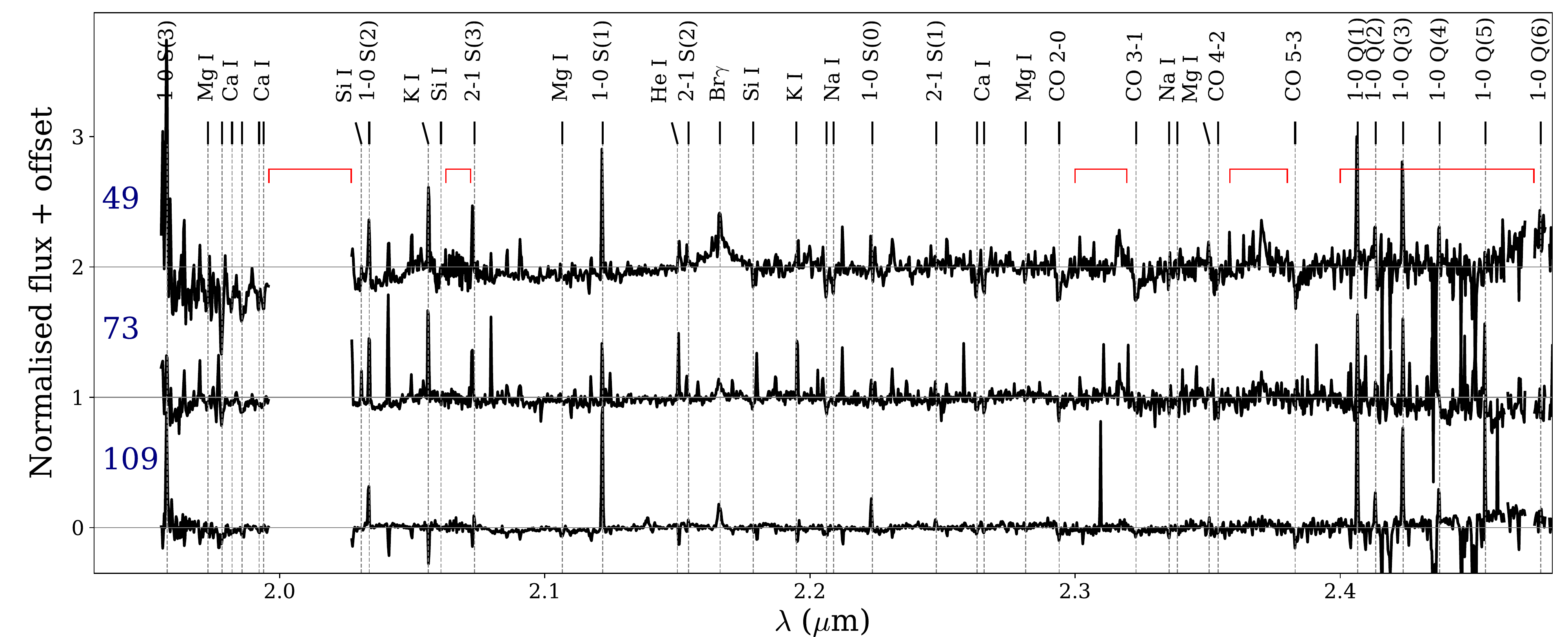}}
\caption{\label{spec:4}Spectra with Br$\gamma$ and H$_2$  emission lines, and CO bandhead in absorption.}
\end{figure*}

\begin{figure*}
\includegraphics[width=\textwidth]{{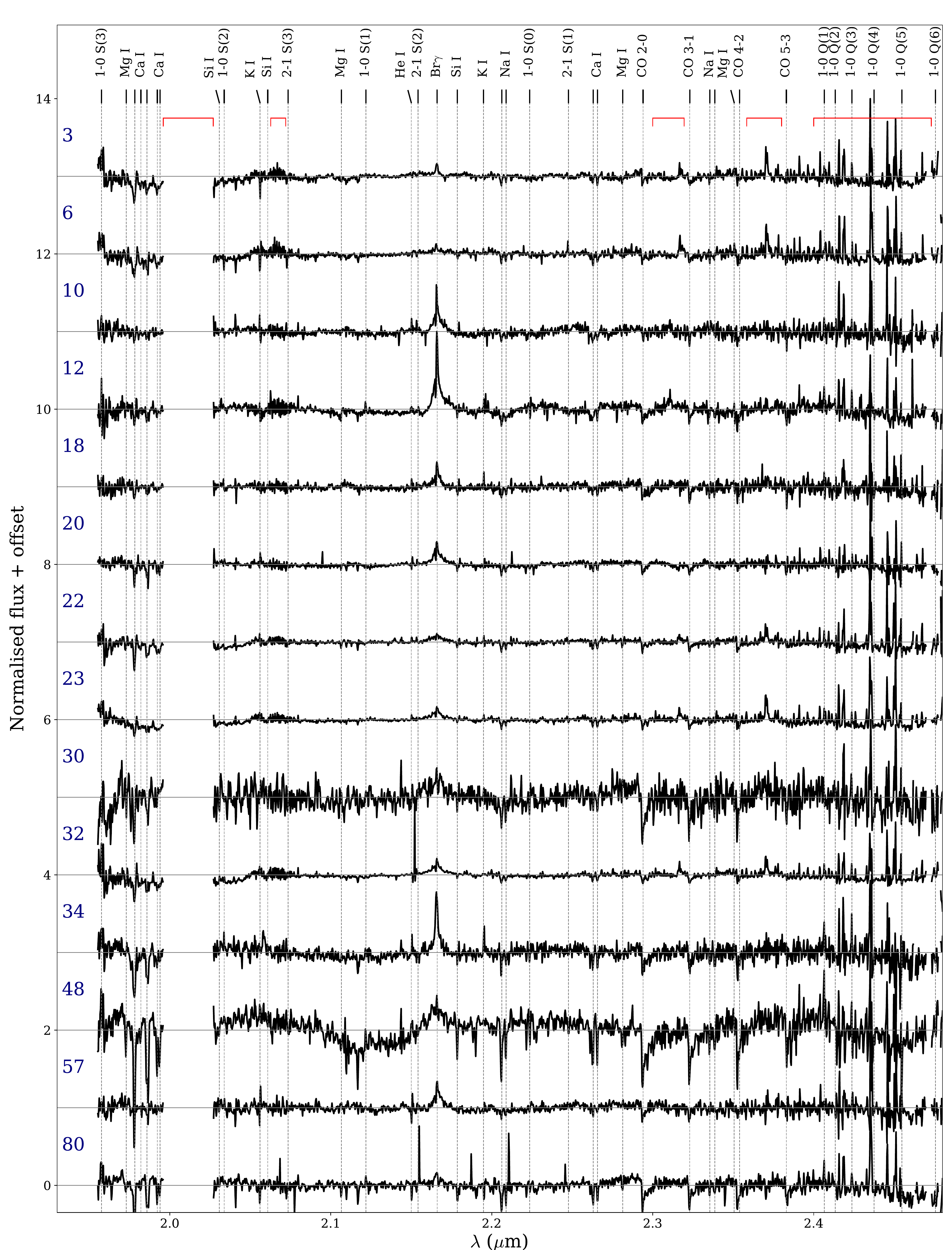}}
\caption{\label{spec:5}Spectra with Br$\gamma$ emission line and CO bandhead in absorption.}
\end{figure*}
\addtocounter{figure}{-1}
\begin{figure*}
\includegraphics[width=\textwidth]{{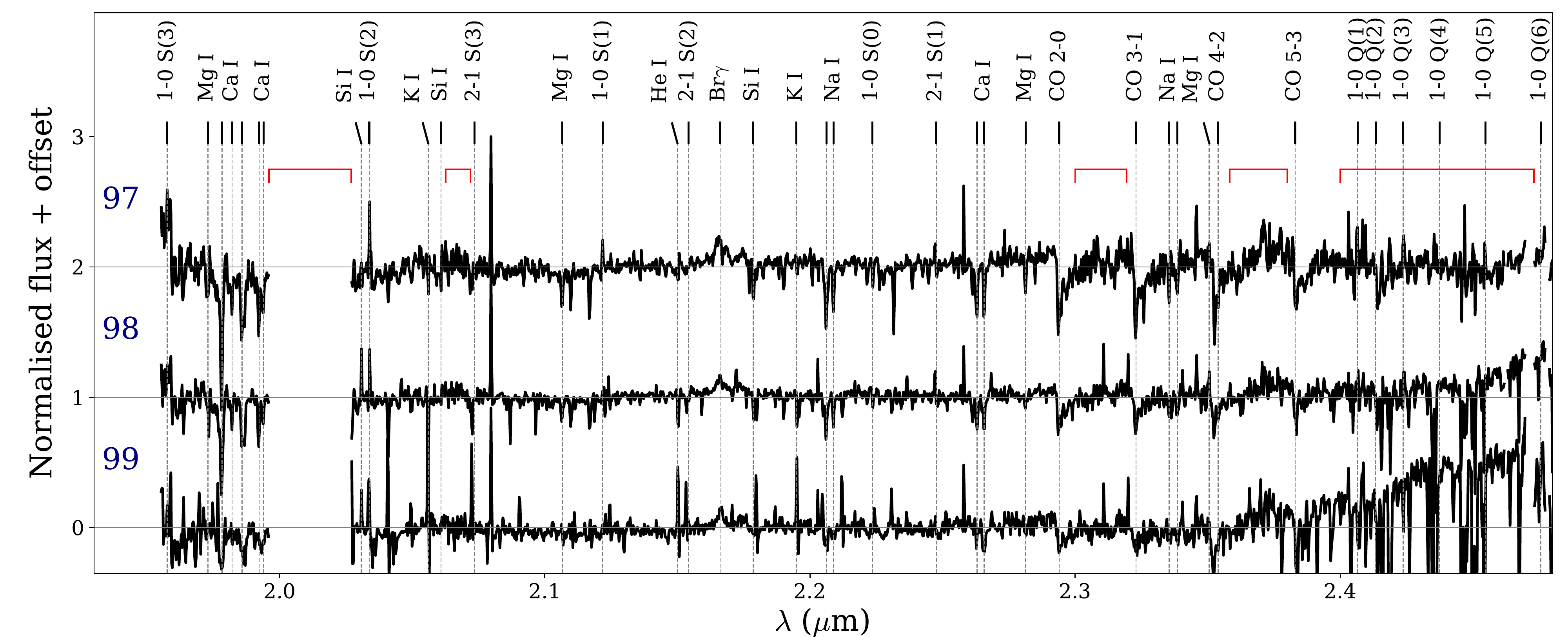}}
\caption{Continuation.}
\end{figure*}

\begin{figure*}
\includegraphics[width=\textwidth]{{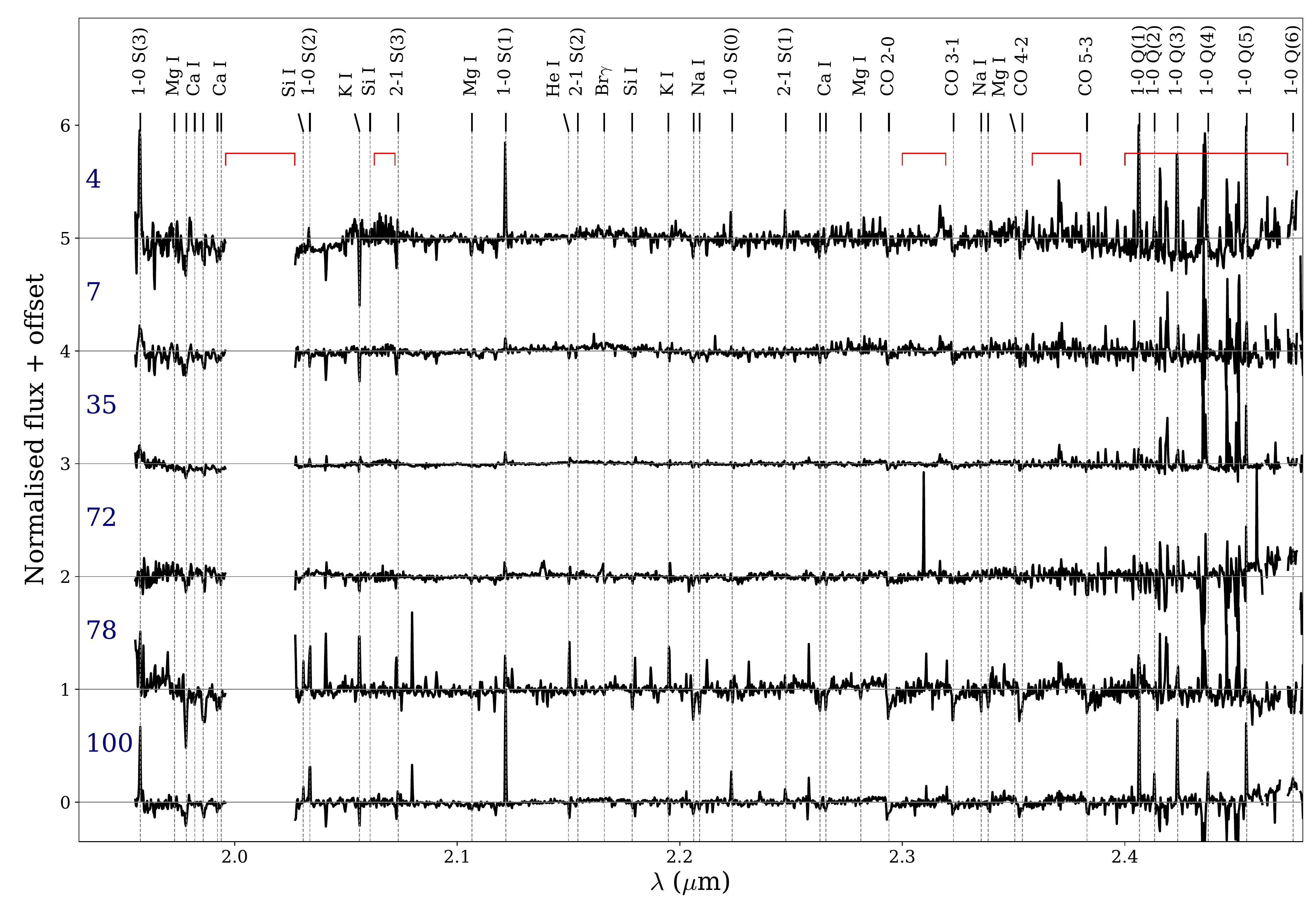}}
\caption{\label{spec:6}Spectra with H$_2$ emission lines and CO bandhead in absorption.}
\end{figure*}

\begin{figure*}
\includegraphics[width=\textwidth]{{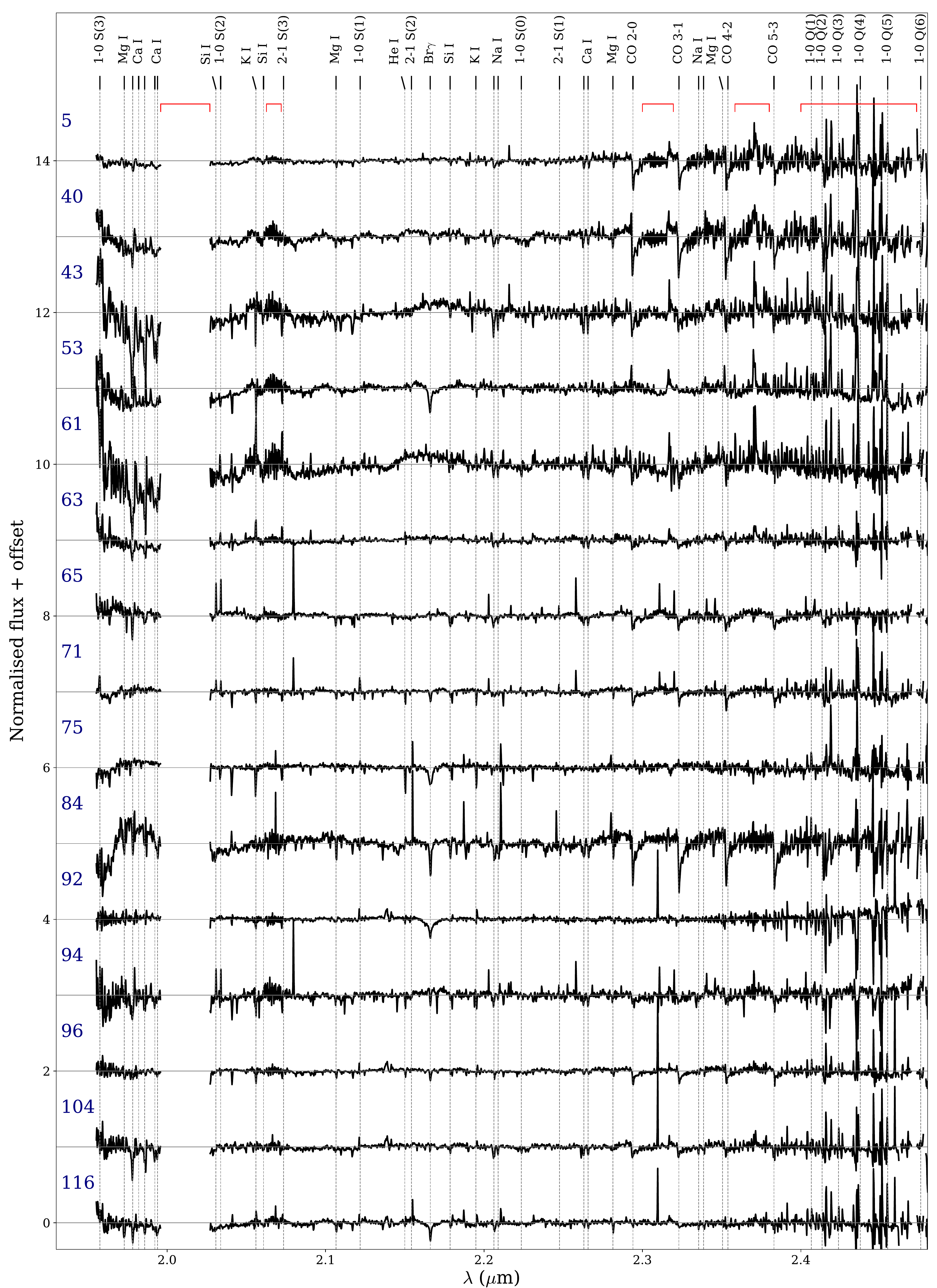}}
\caption{\label{spec:7}Spectra with Br$\gamma$ absorption line.}
\end{figure*}
\begin{figure*}\vspace{-0.3cm}
\includegraphics[width=\textwidth]{{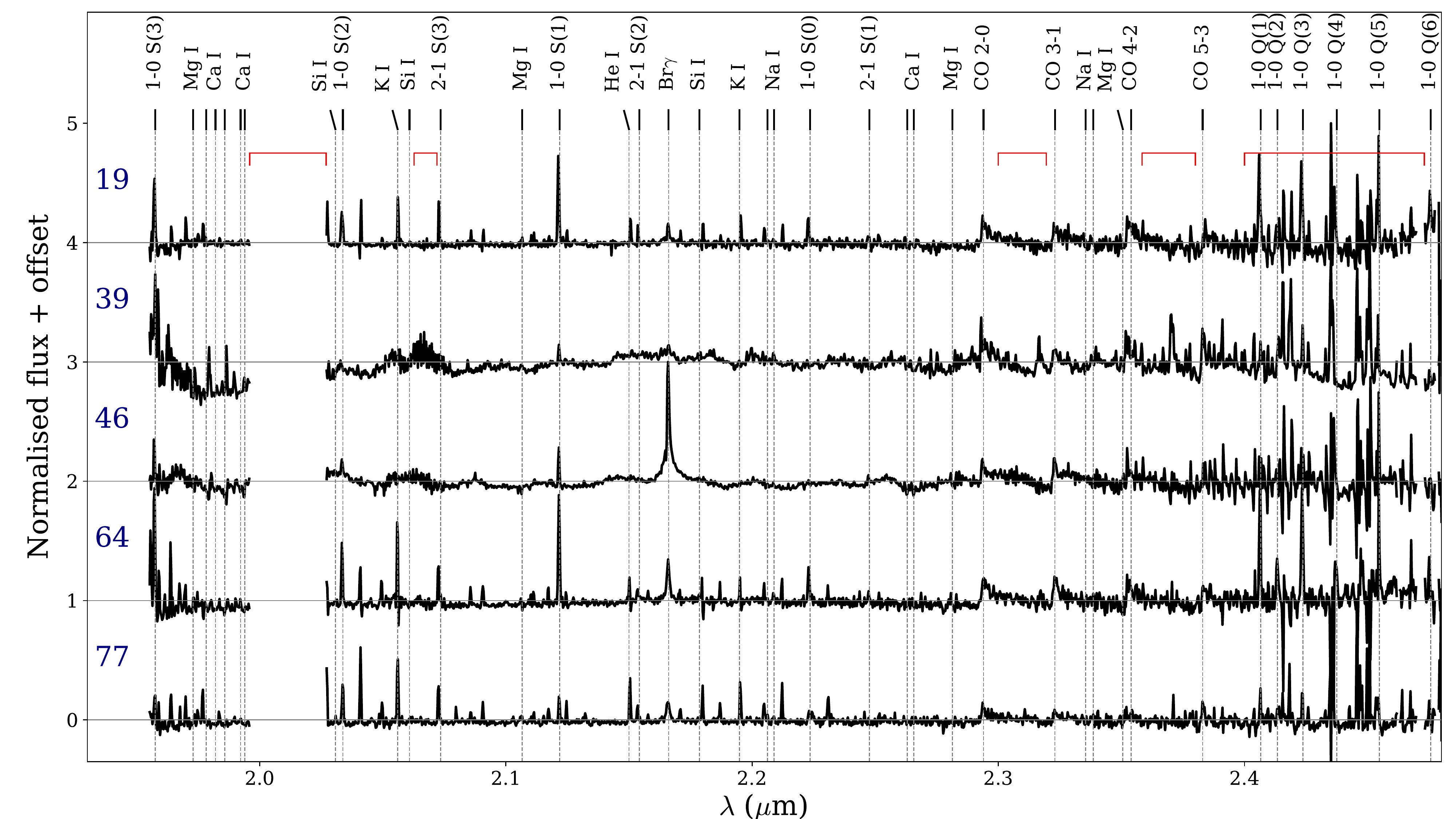}}
\caption{\label{spec:8}Spectra with CO bandhead in emission.}
\end{figure*}

\begin{figure*}\vspace{-.5cm}
\includegraphics[width=\textwidth]{{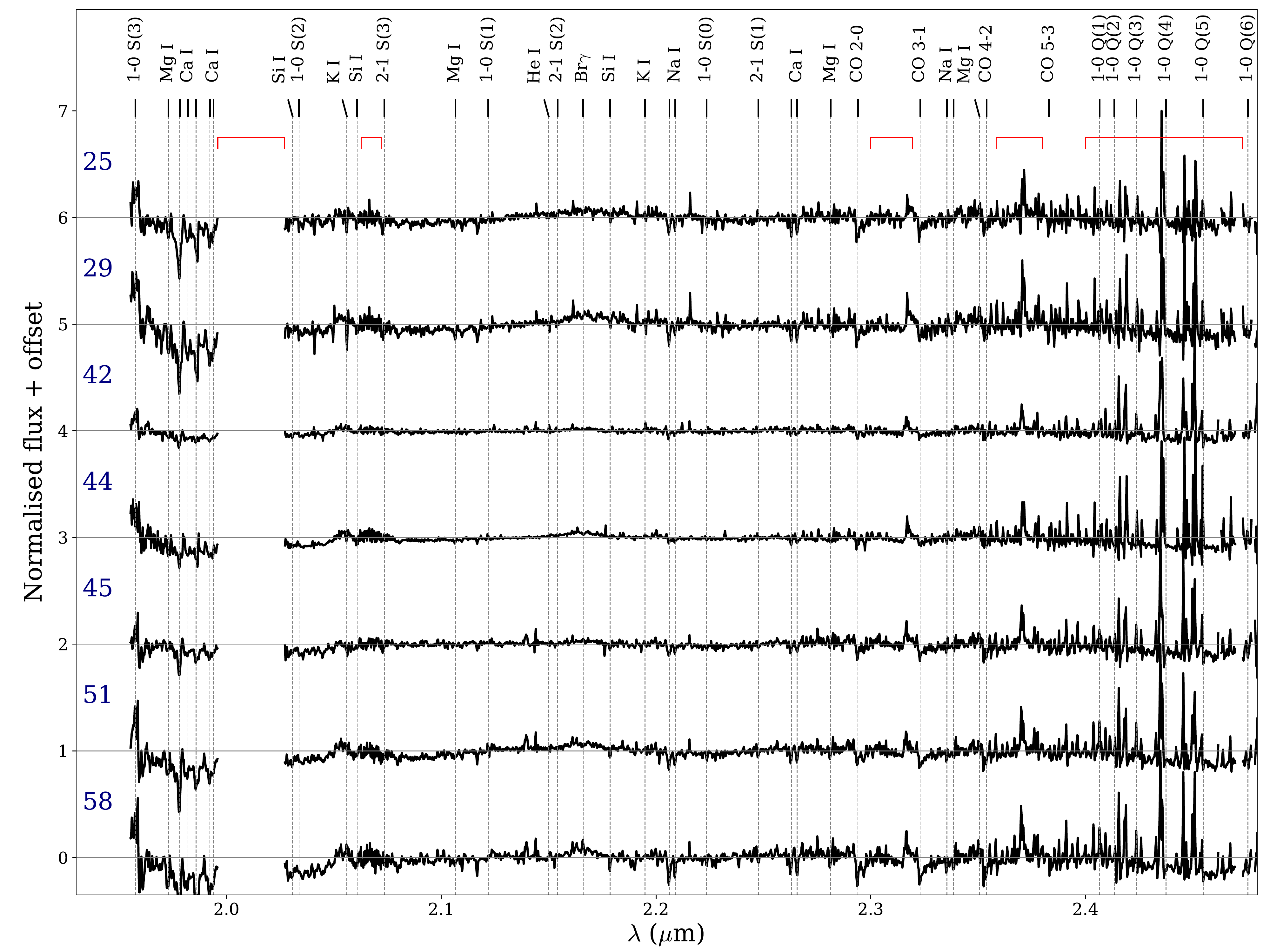}}
\caption{\label{spec:9}Spectra with CO bandhead in absorption.}\vspace{-2.3cm}
\end{figure*}
\addtocounter{figure}{-1}
\begin{figure*}\vspace{-.3cm}
\includegraphics[width=\textwidth]{{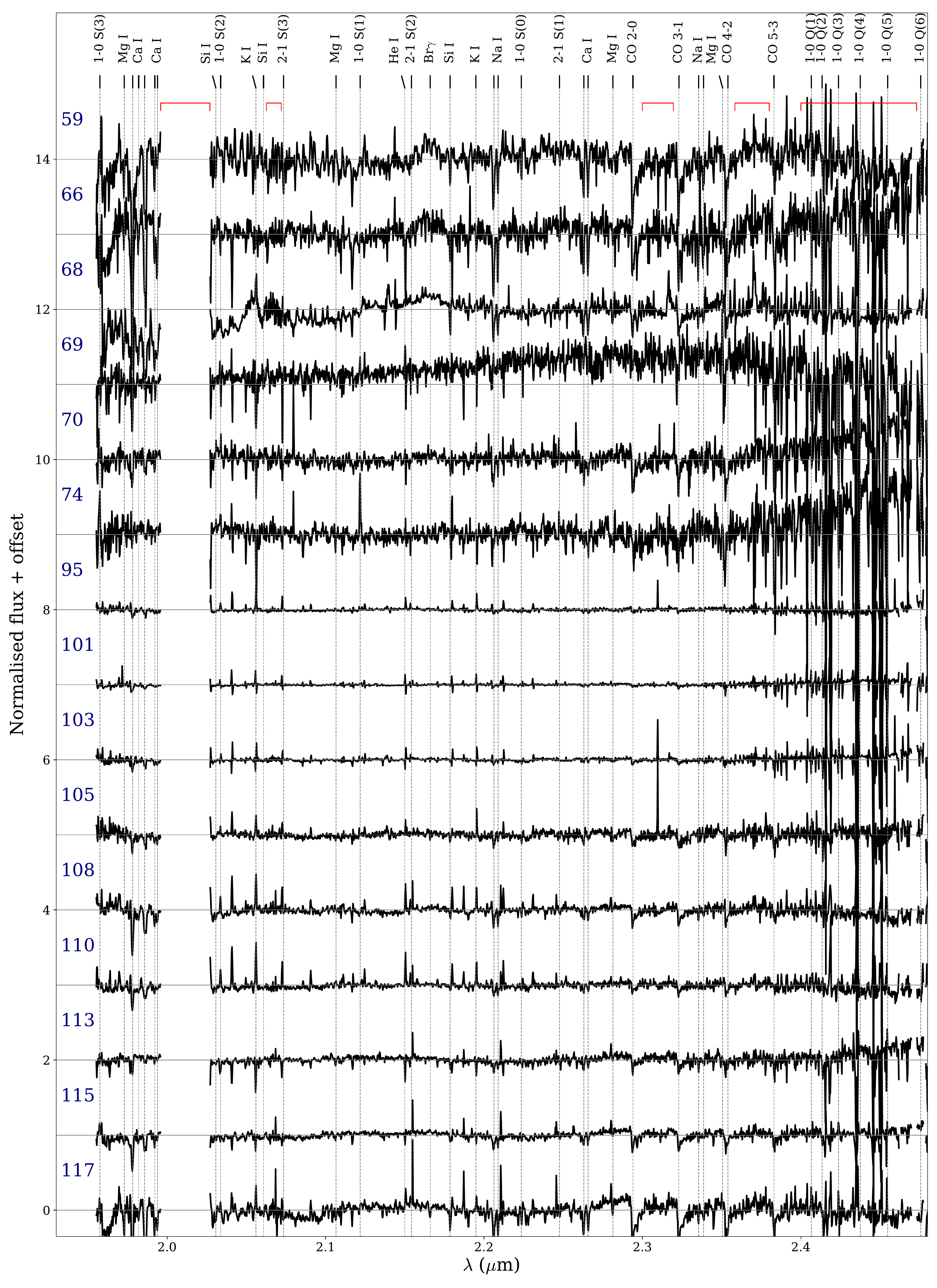}}
\caption{Continuation.}
\end{figure*}

\begin{figure*}
\includegraphics[width=\textwidth]{{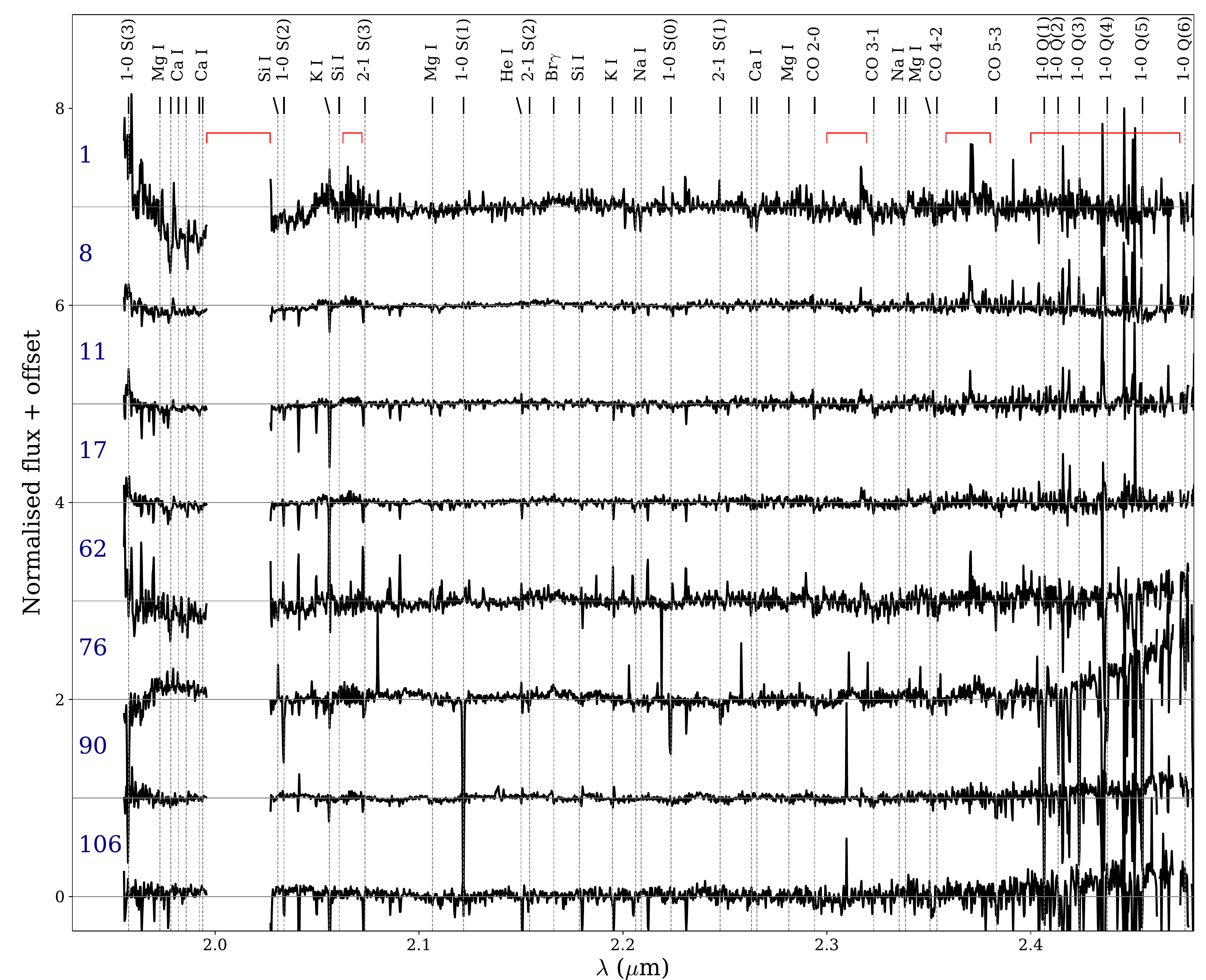}}
\caption{\label{spec:10}Spectra with no emission lines and no CO or Br$\gamma$ absorption lines.}
\end{figure*}

\begin{figure*}
\includegraphics[width=\textwidth]{{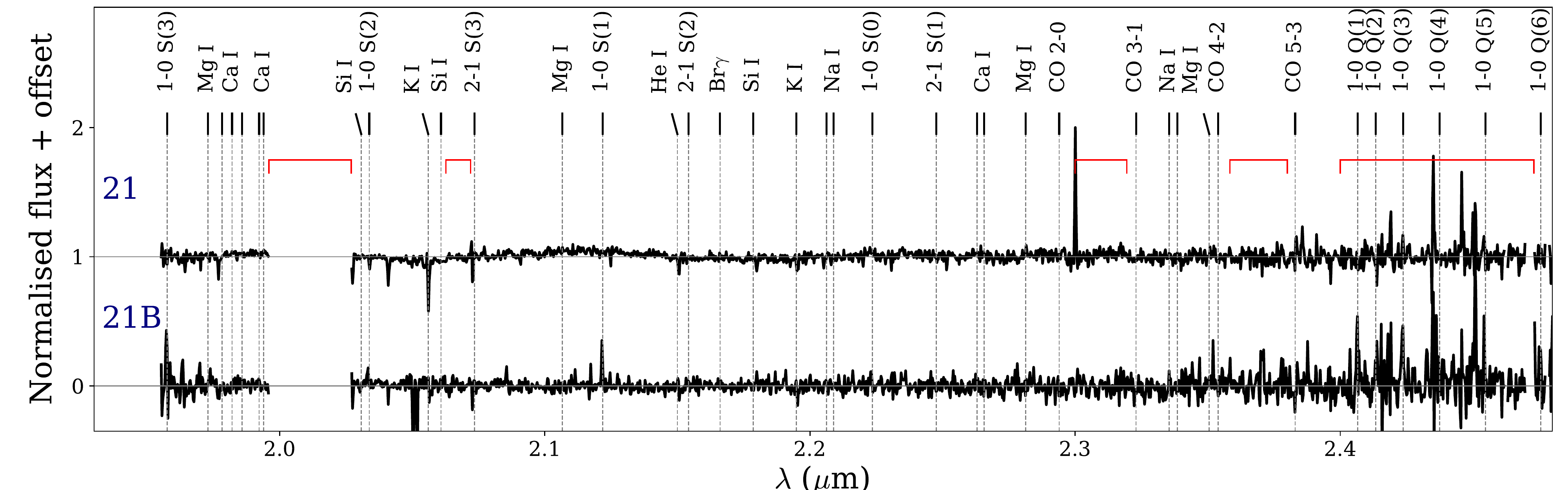}}
\quad
\includegraphics[width=\textwidth]{{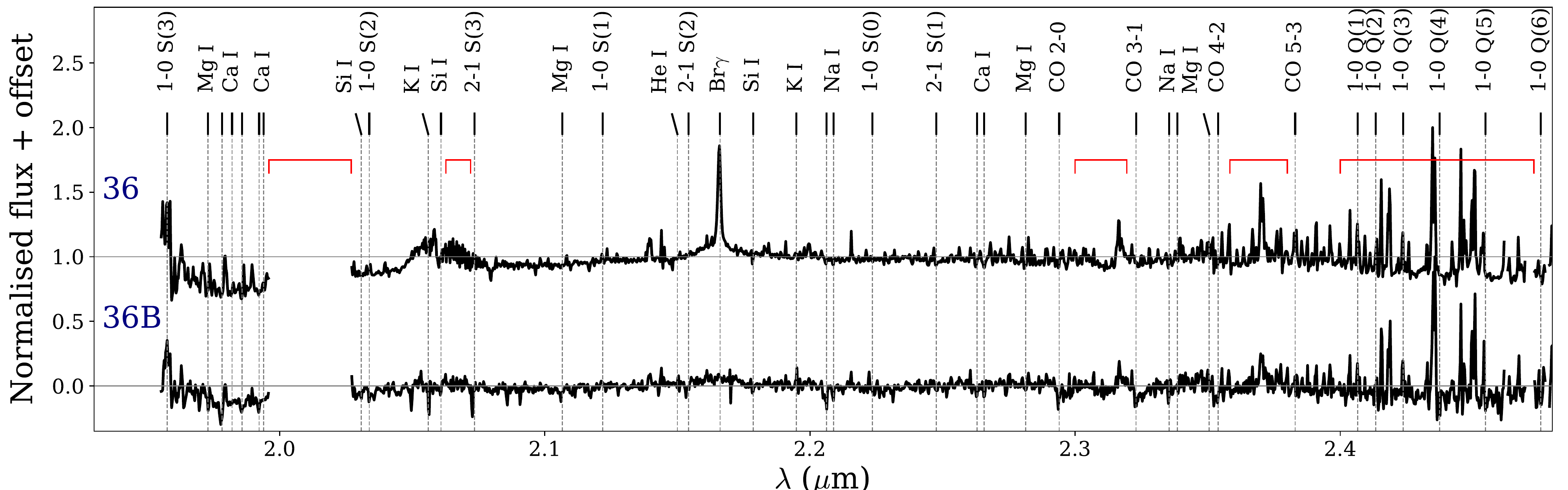}}
\quad
\includegraphics[width=\textwidth]{{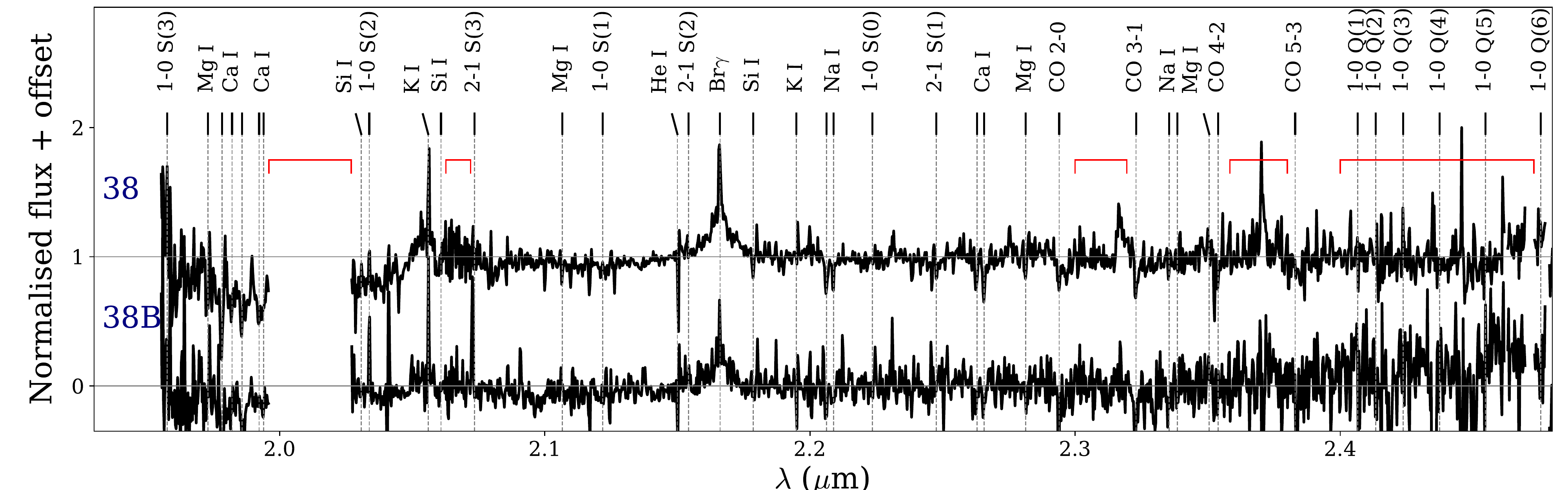}}
\caption{\label{spec:13}Spectra from three pairs of double sources.}
\end{figure*}

\begin{figure*}
\includegraphics[width=\textwidth]{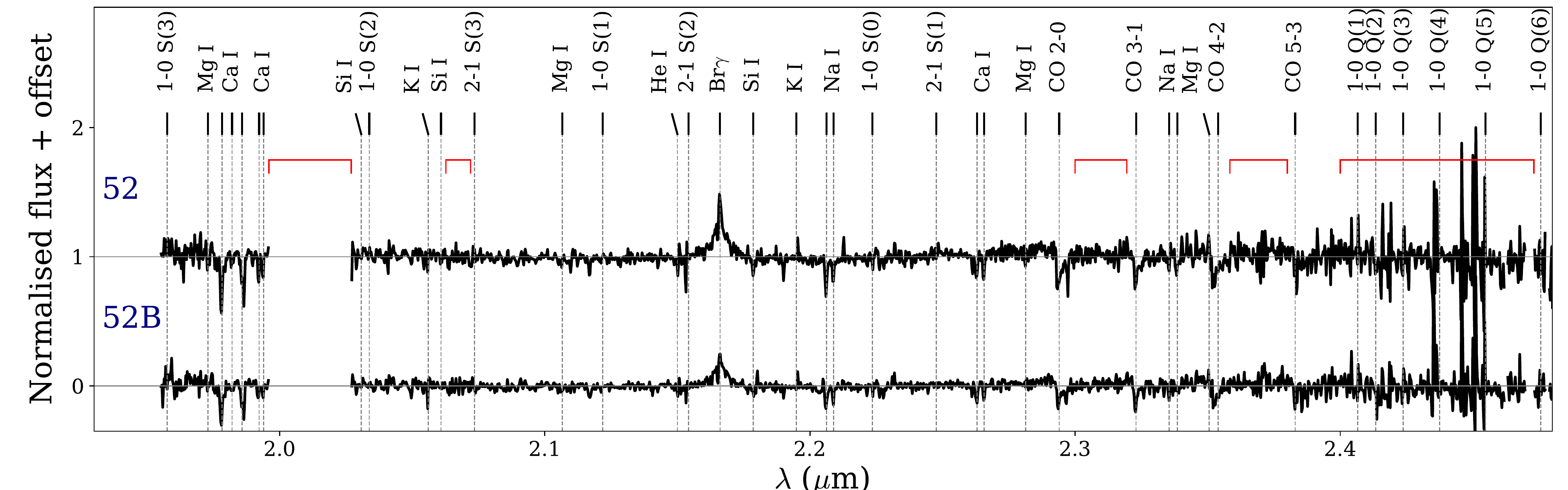}
\quad
\includegraphics[width=\textwidth]{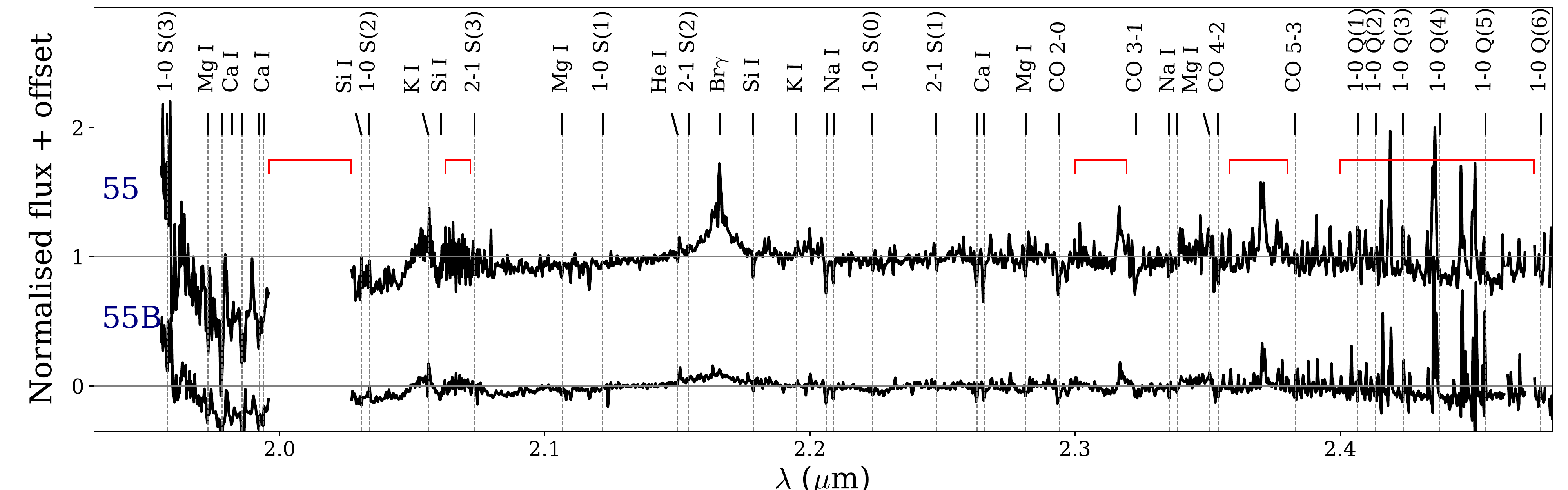}
\quad
\includegraphics[width=\textwidth]{{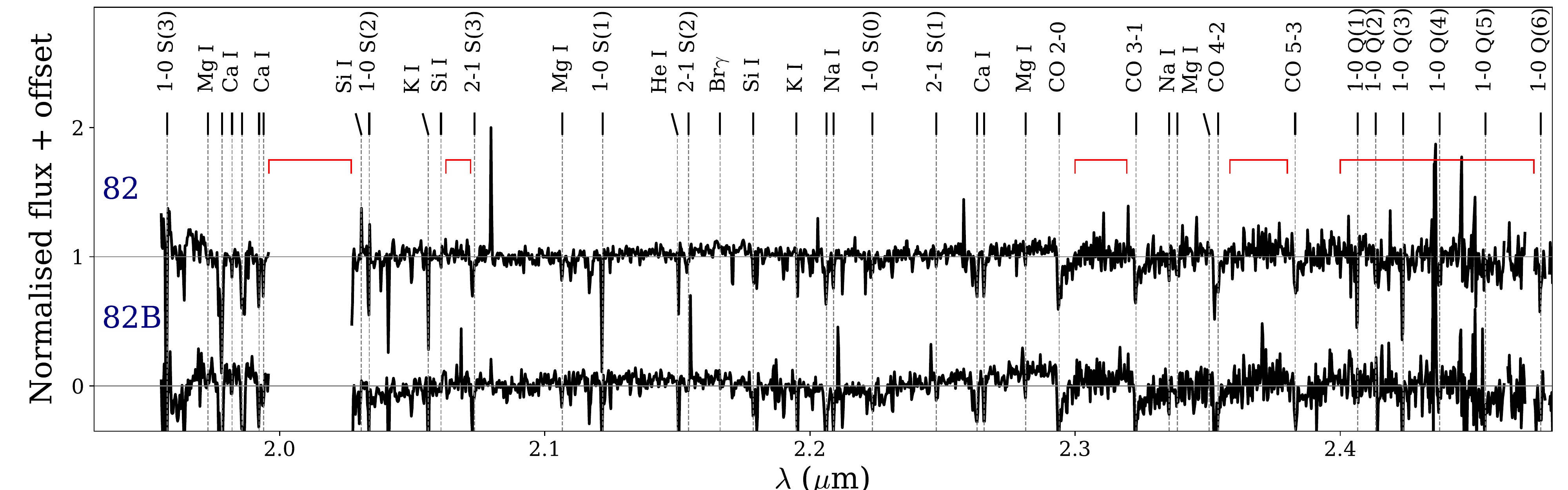}}
\quad
\includegraphics[width=\textwidth]{{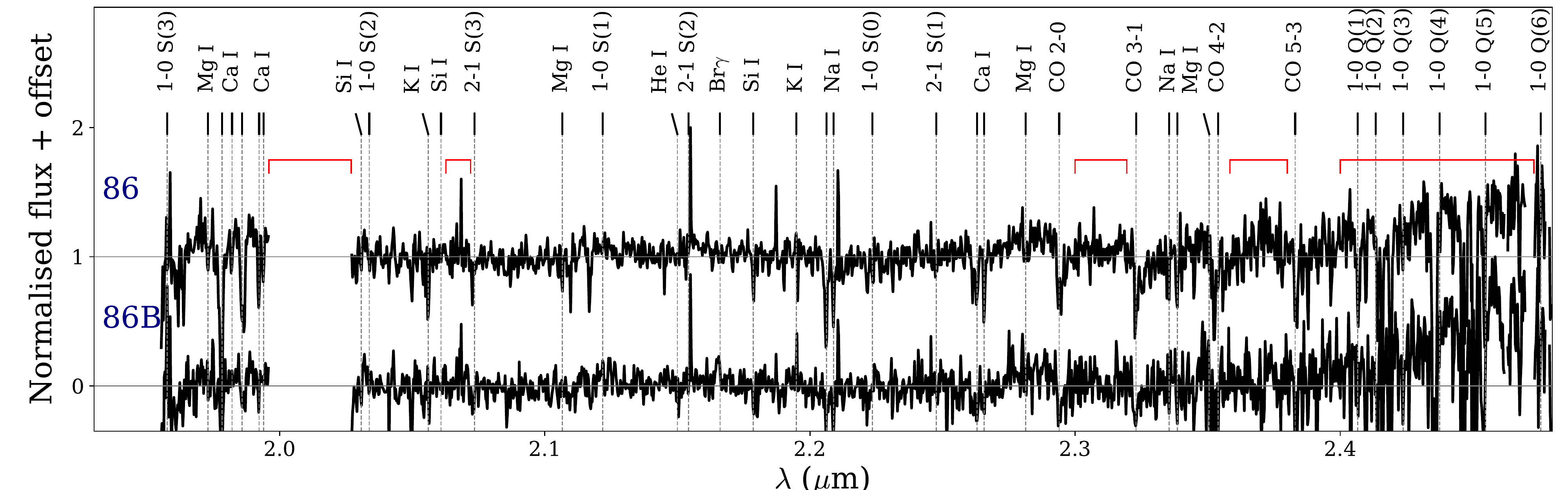}}
\caption{\label{spec:14}Spectra from double sources - continuation.}
\end{figure*}

\addtocounter{figure}{-1}
\begin{figure*}
\includegraphics[width=\textwidth]{{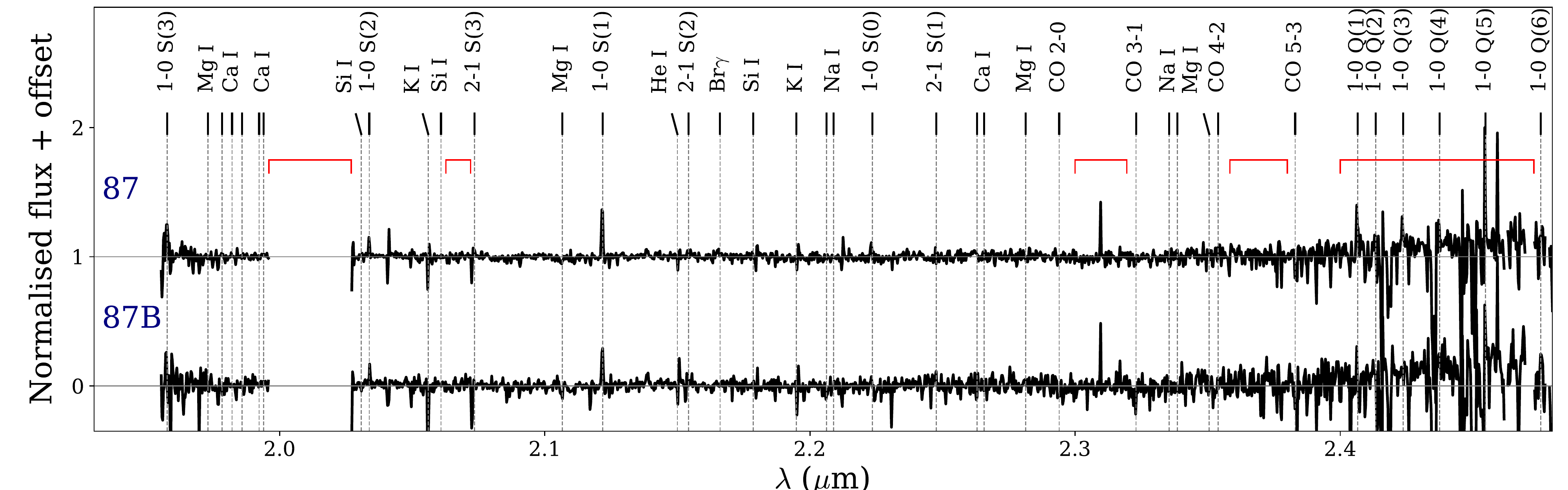}}
\quad
\includegraphics[width=\textwidth]{{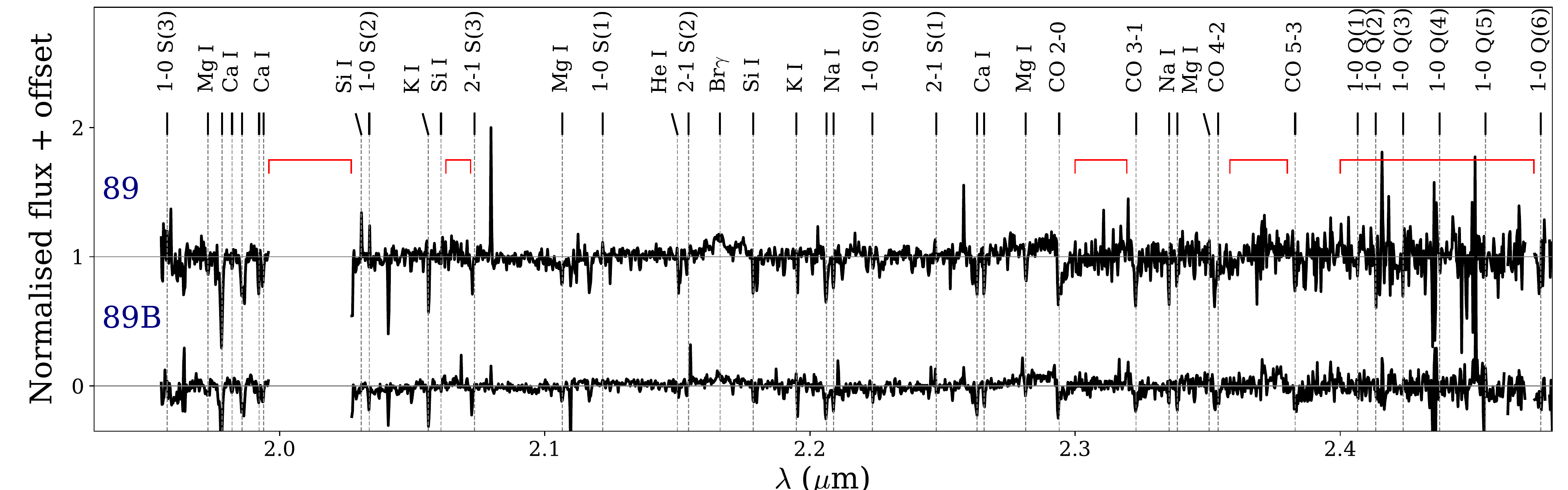}}
\quad
\includegraphics[width=\textwidth]{{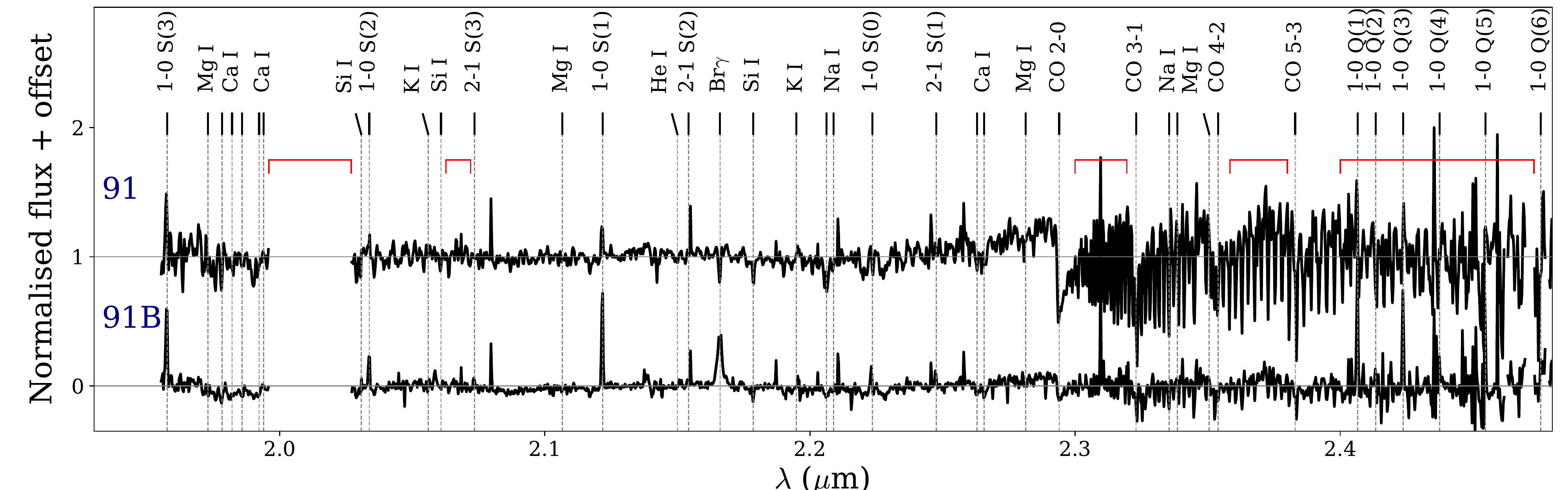}}
\quad
\includegraphics[width=\textwidth]{{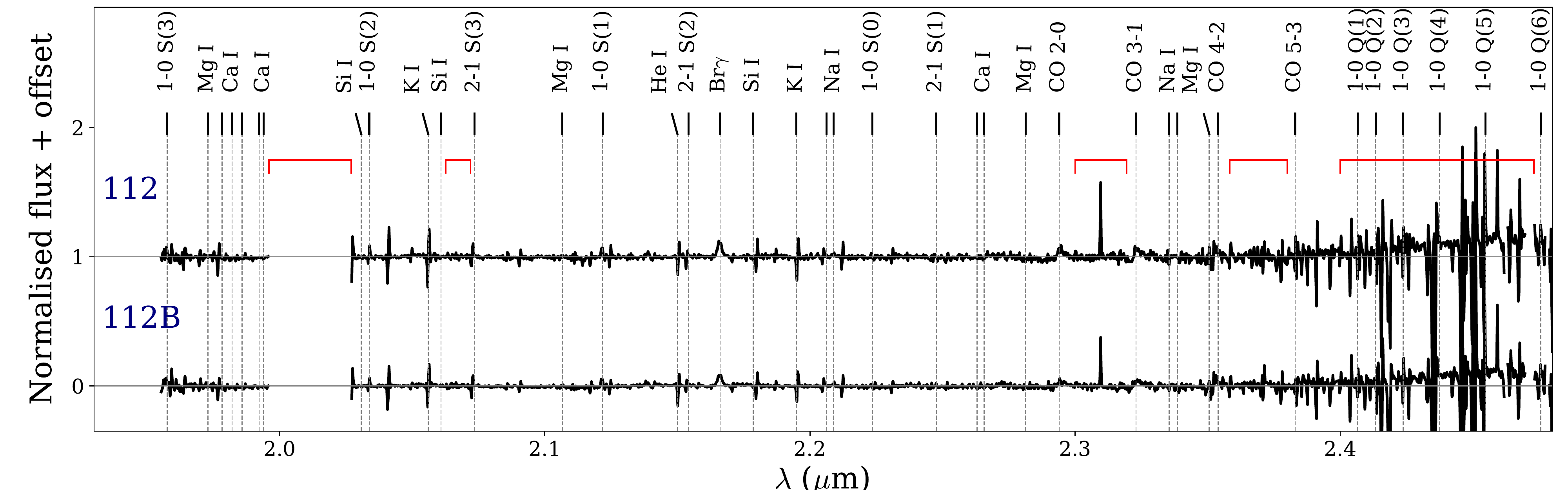}}
\caption{\label{spec:15}Continuation.}
\end{figure*}

\clearpage
\section{Line Emission Maps} \label{app:emiss}

Figures \ref{fig:emiss-2.1218}--\ref{fig:emiss-2.2477} show the emission maps of the most commonly detected H$_2$ lines in the $K$-band range. Figure \ref{fig:emiss-1512183} shows the spatial extent of several transitions of 
H$_2$ toward source No. 47, which has the most molecular-rich spectrum at $K$-band. {Source No. 47 has the {\it K}-band magnitude of 15.2 based on our KMOS observations, thus it is one of the faintest sources in the sample. \cite{sewilo2019} classified it as a Class I YSO. The young evolutionary stage of the source is confirmed by strong bipolar outflows. This source is an interesting case for detailed follow-up studies.}

Figure \ref{fig:emiss-2.1661} shows the hydrogen Br$\gamma$ 2.17 $\mu$m emission maps. Figures \ref{fig:emiss-2.2940}, \ref{fig:emiss-2.3230}, \ref{fig:emiss-2.3540}, and \ref{fig:emiss-2.3830} show the emission maps for several CO bandhead transitions -- CO 2-0, 3-1, 4-2, and CO 5-3, respectively. 


\begin{figure*}
\includegraphics[width=0.2\textwidth]{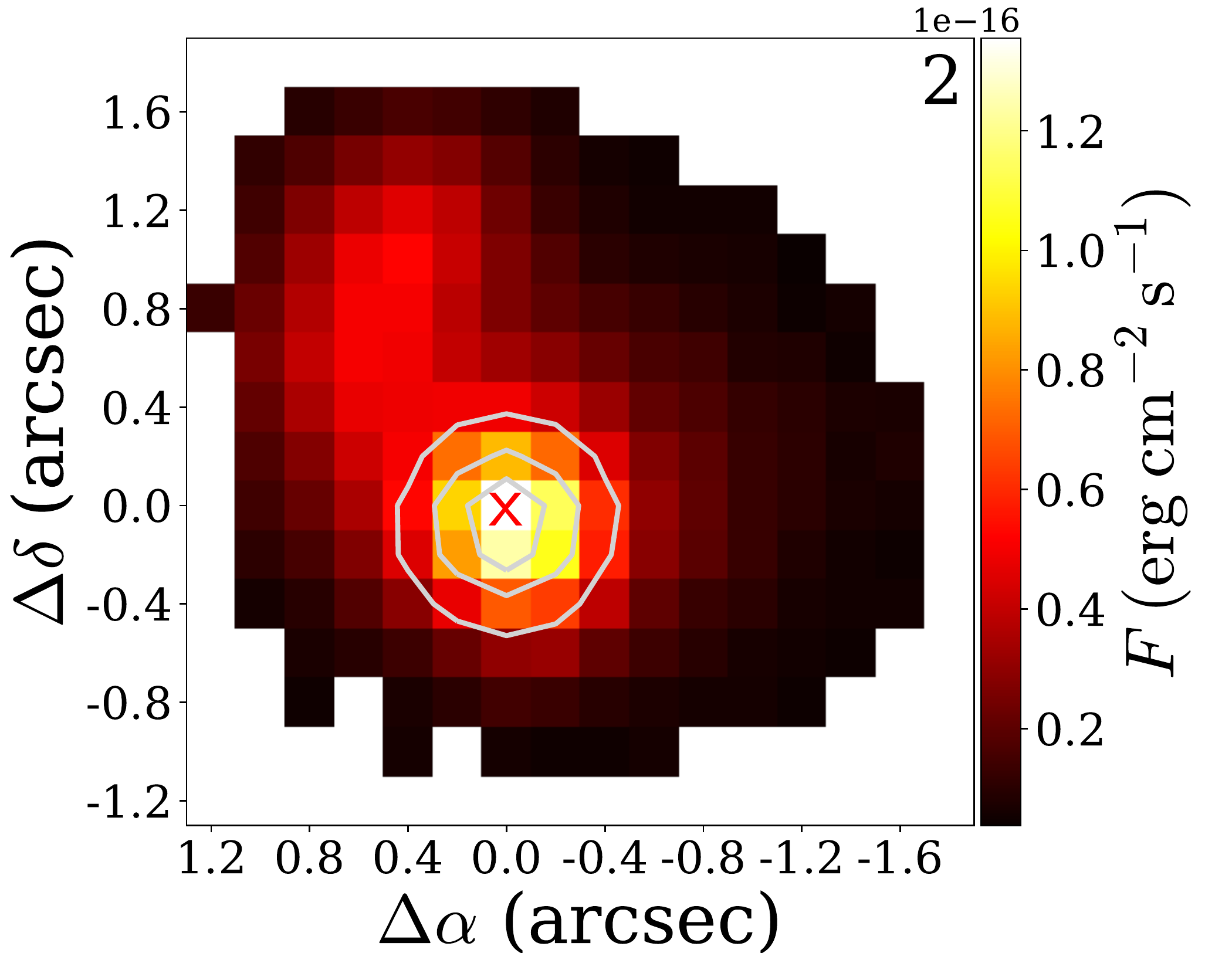}\hspace{-0.1cm}
\includegraphics[width=0.2\textwidth]{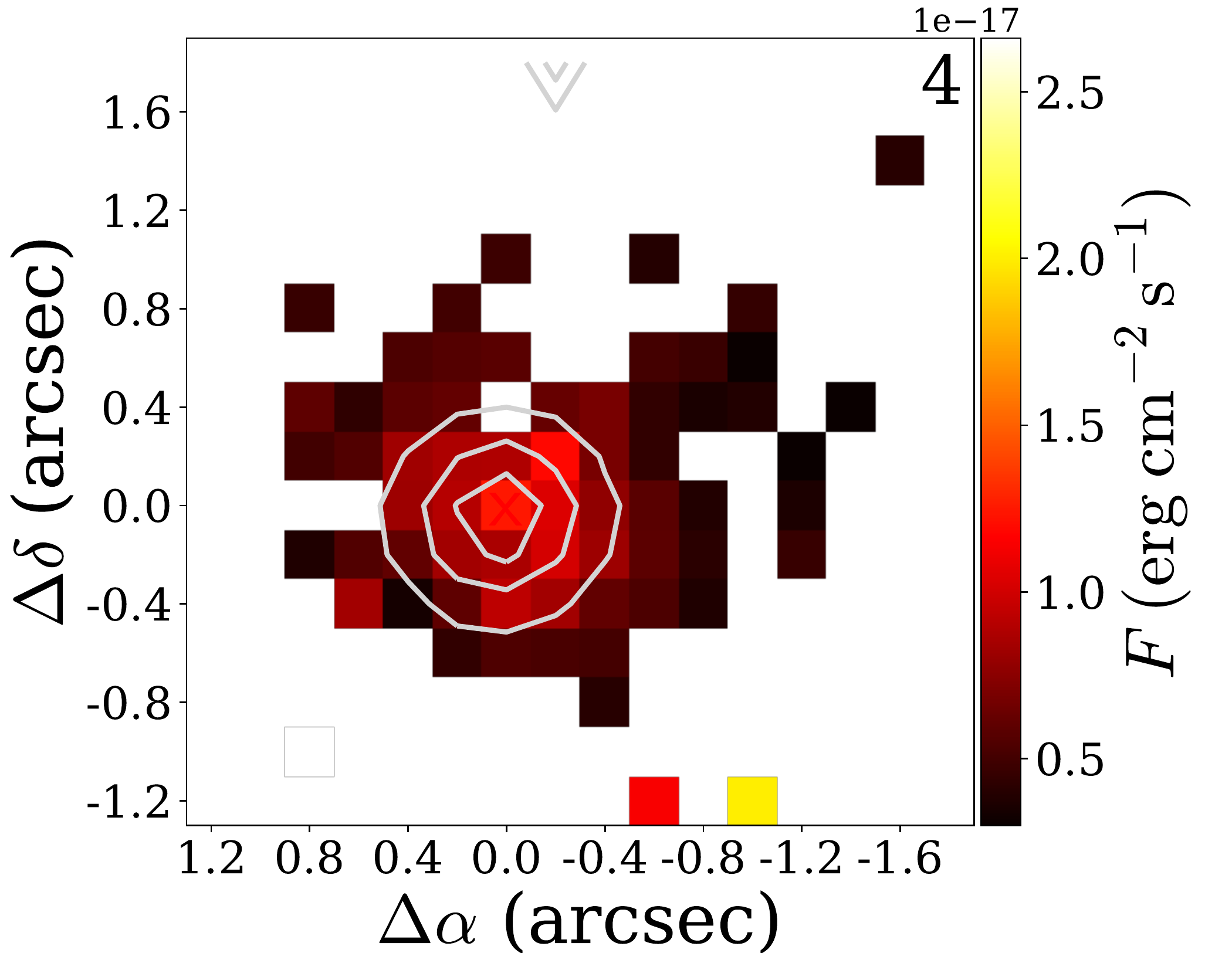}\hspace{-0.1cm}
\includegraphics[width=0.2\textwidth]{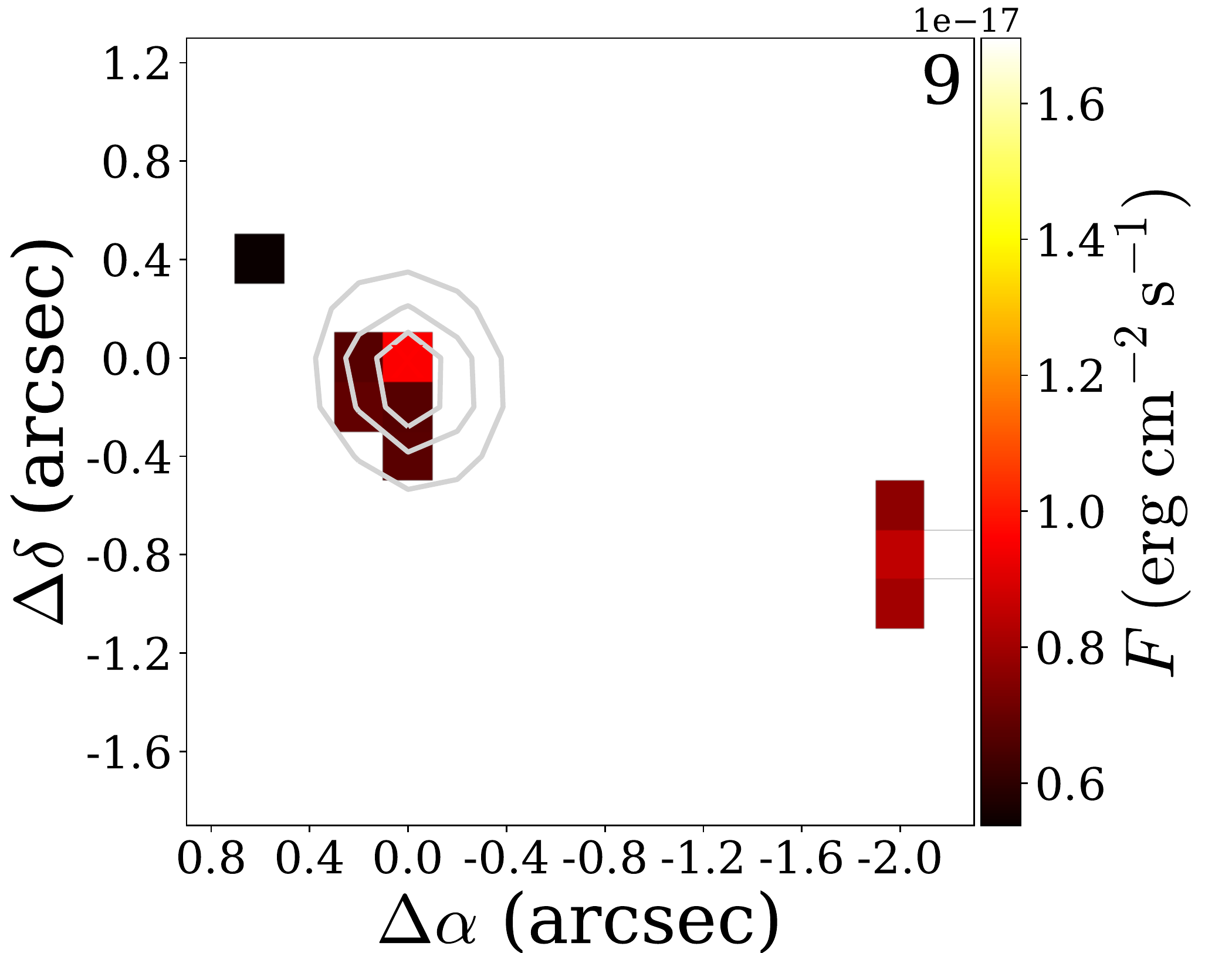}\hspace{-0.1cm}
\includegraphics[width=0.2\textwidth]{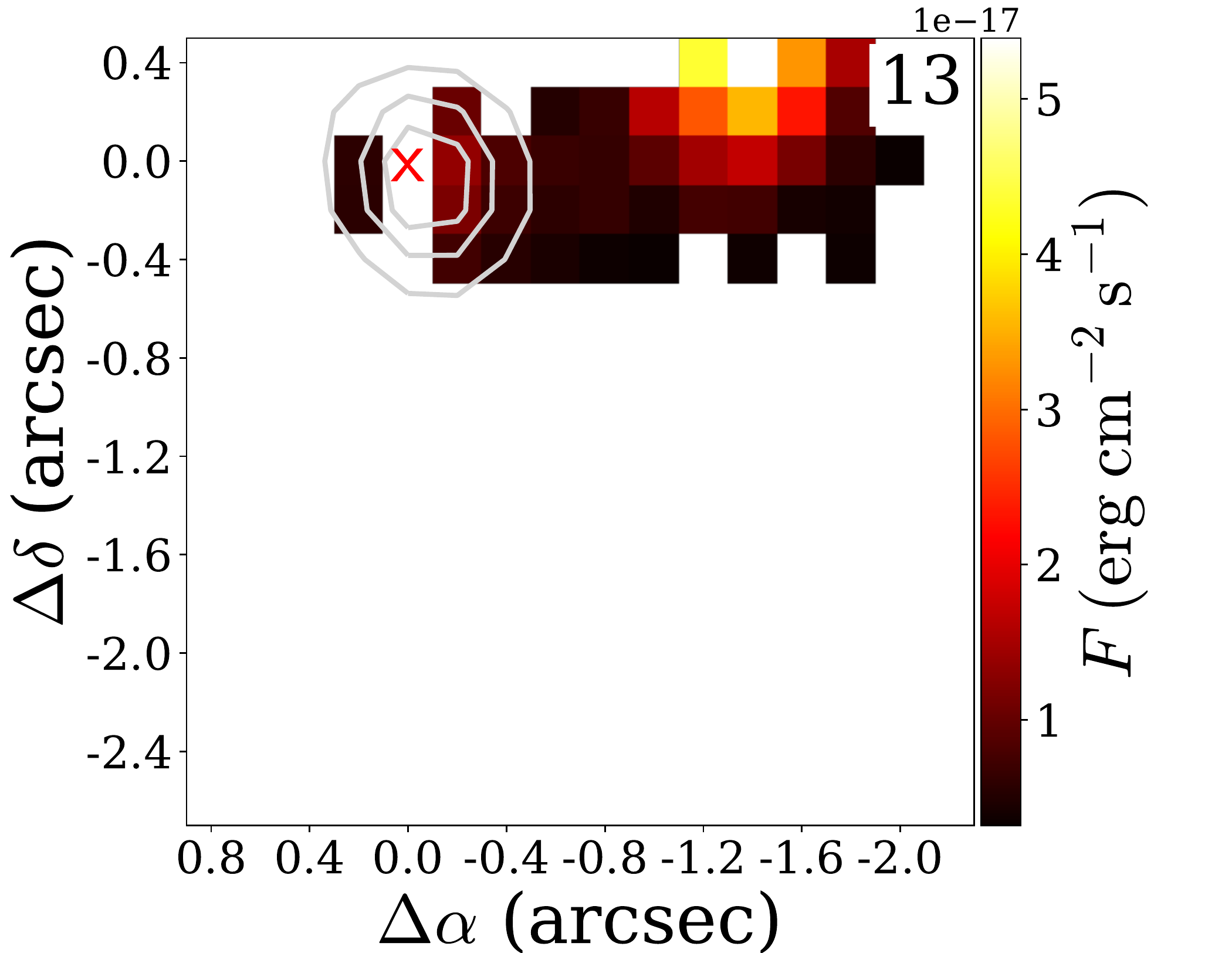}\hspace{-0.1cm}
\includegraphics[width=0.2\textwidth]{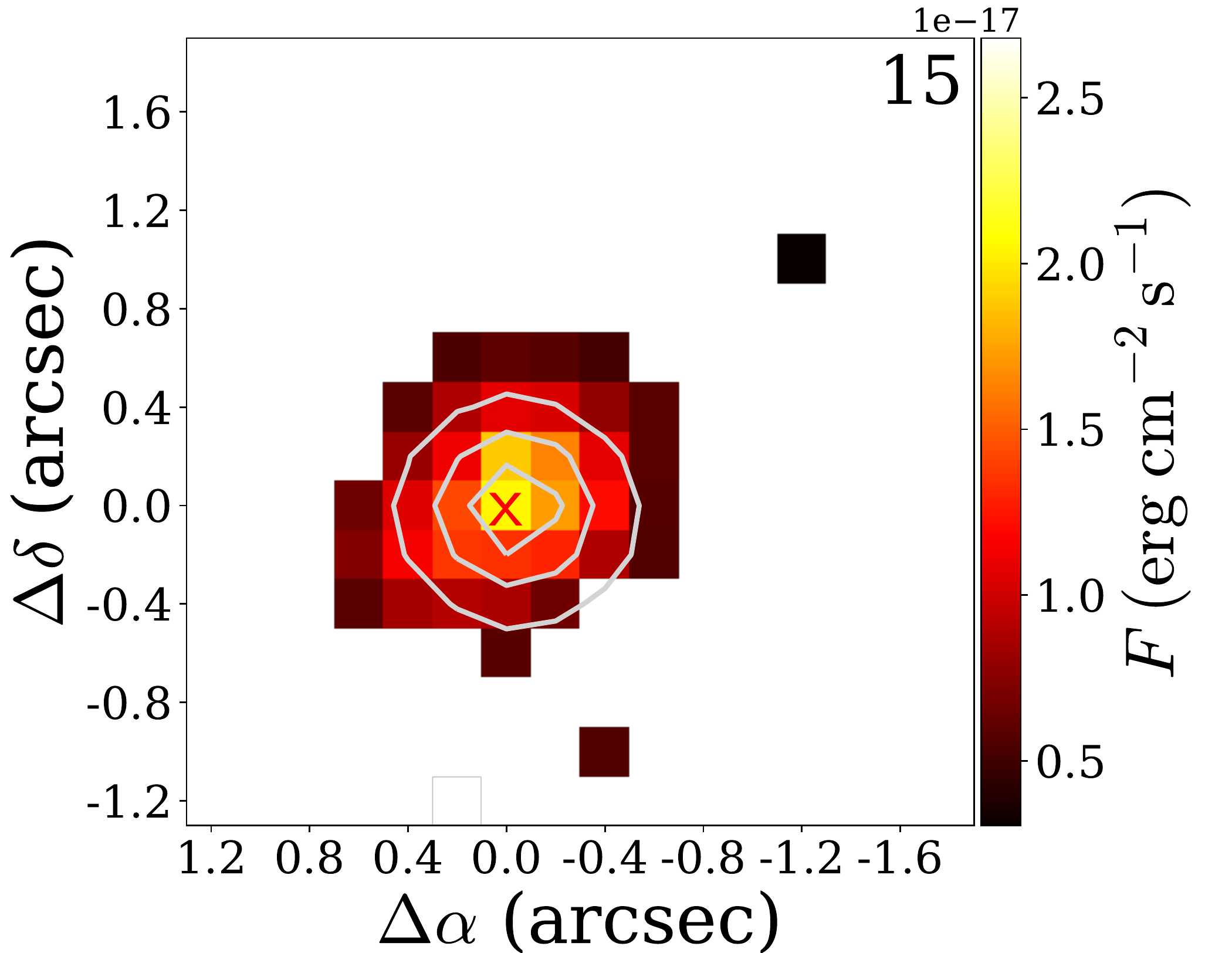}\hspace{-0.1cm}
\includegraphics[width=0.2\textwidth]{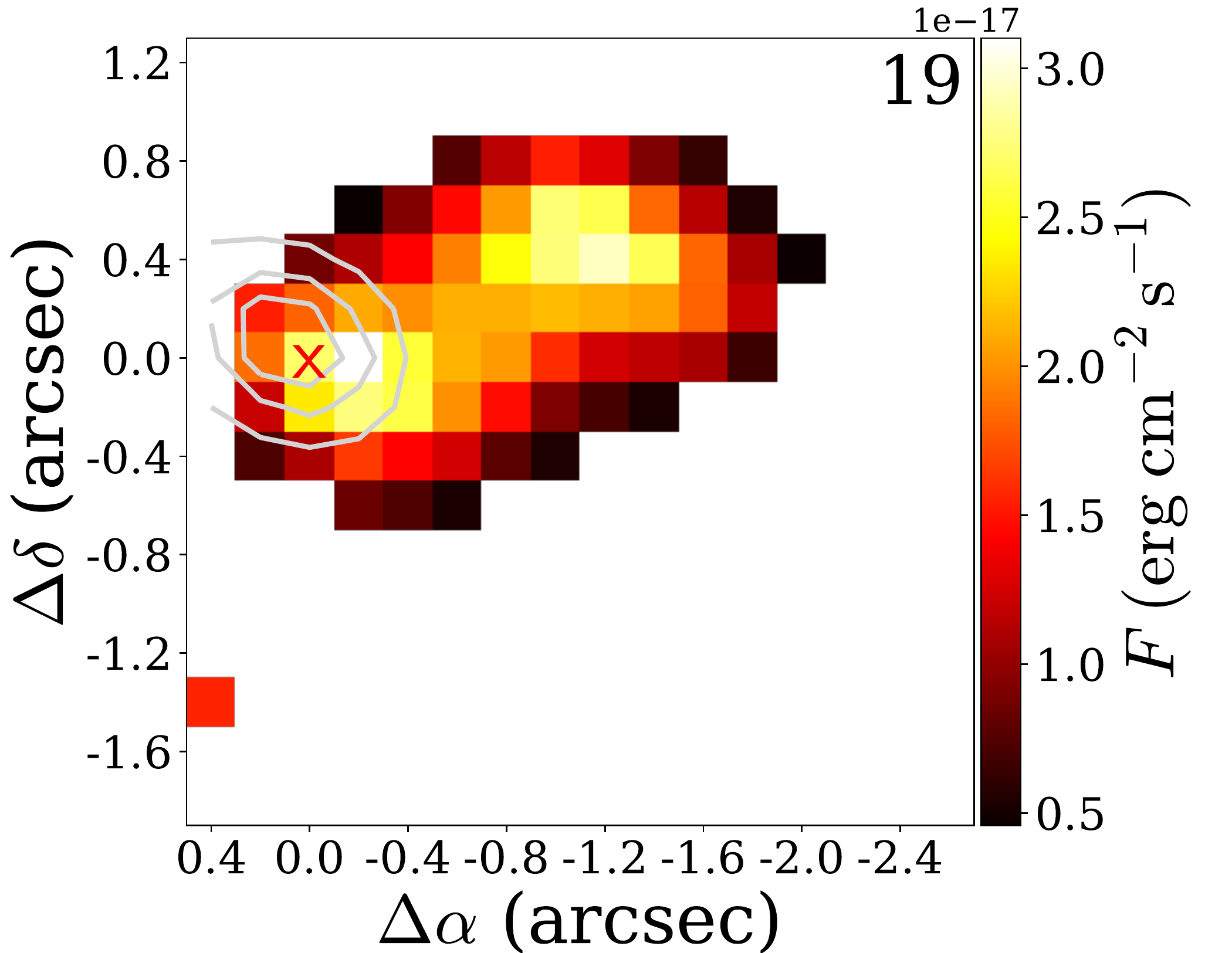}\hspace{-0.3cm}
\includegraphics[width=0.2\textwidth]{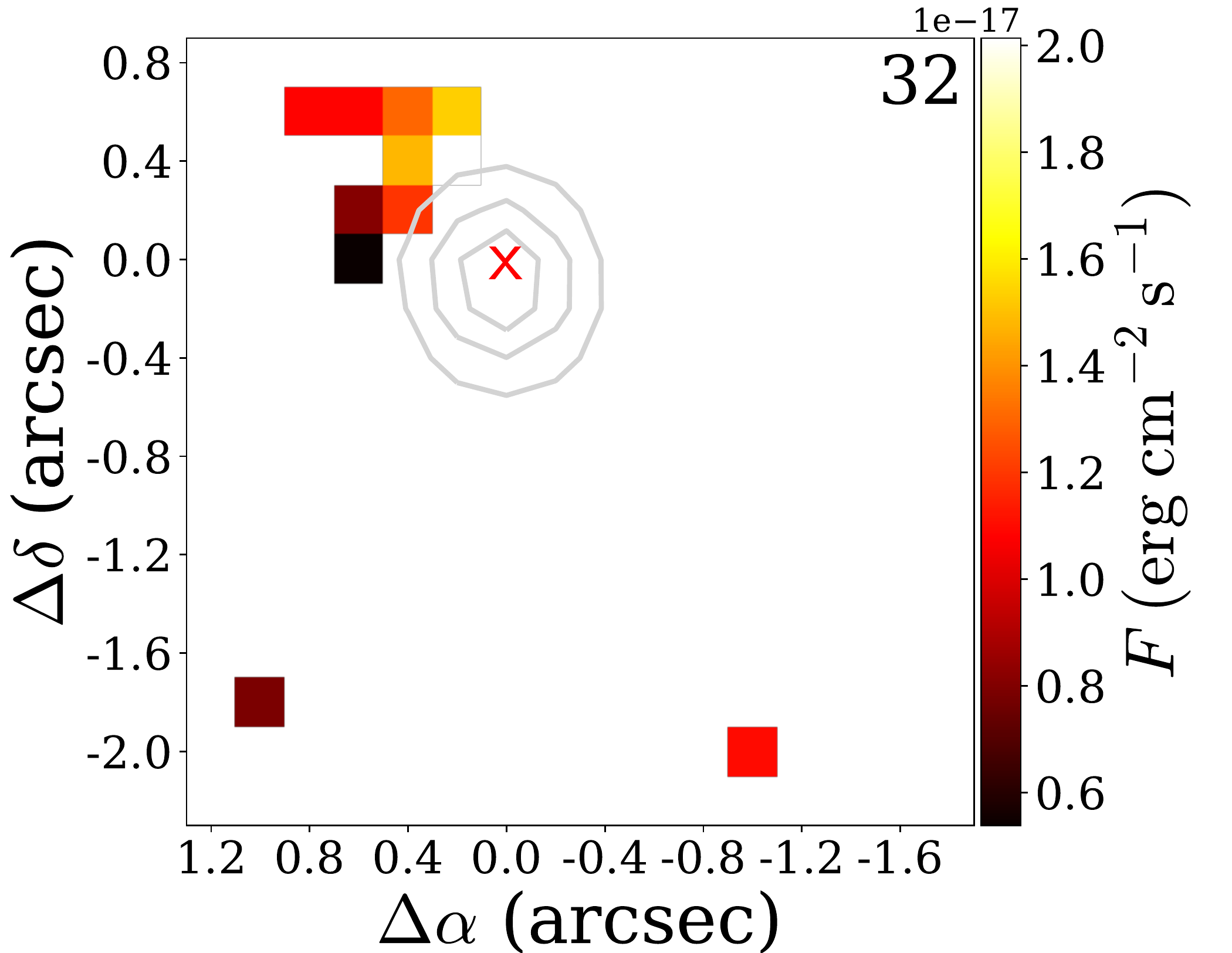}\hspace{-0.cm}
\includegraphics[width=0.2\textwidth]{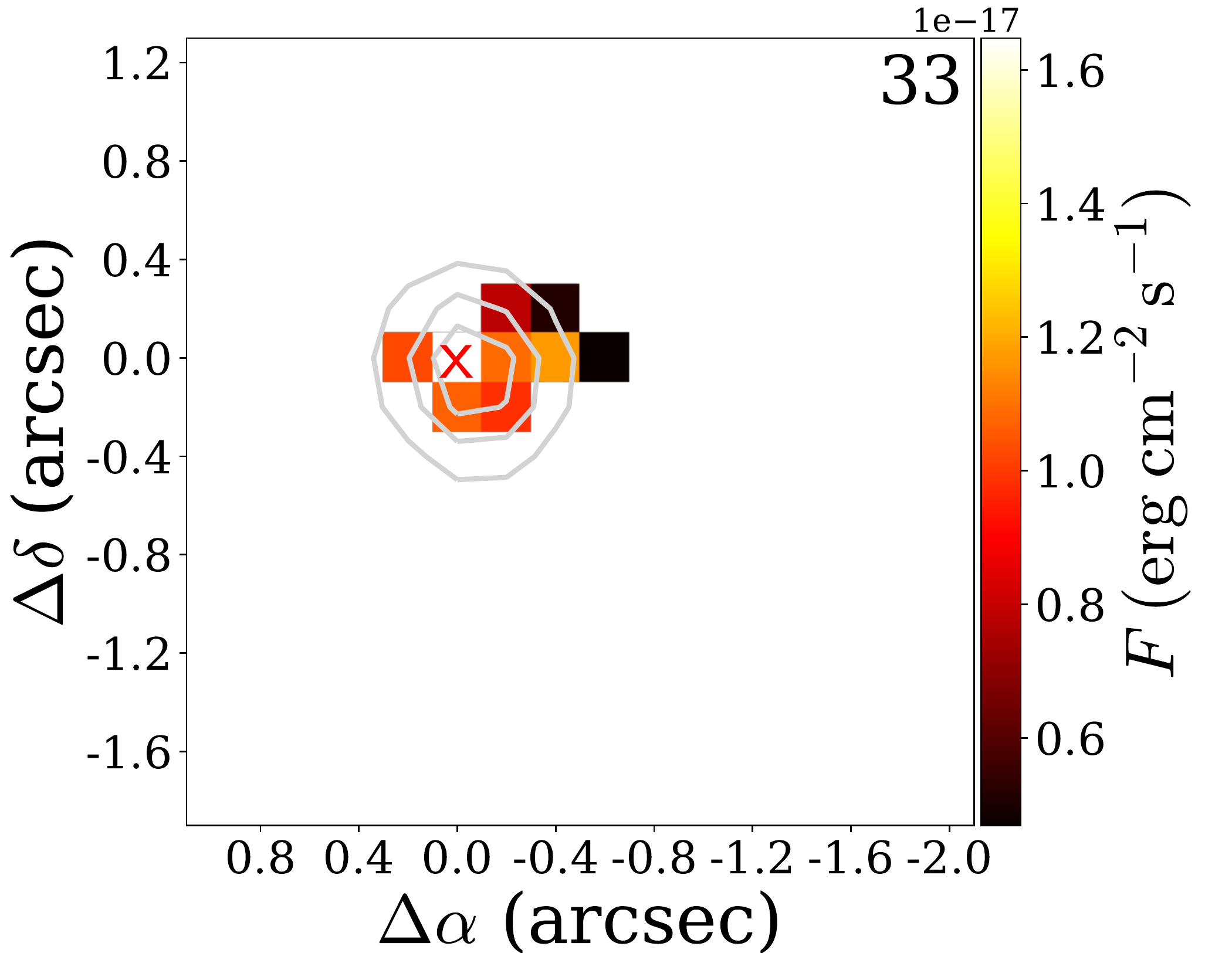}\hspace{-0.1cm}
\includegraphics[width=0.2\textwidth]{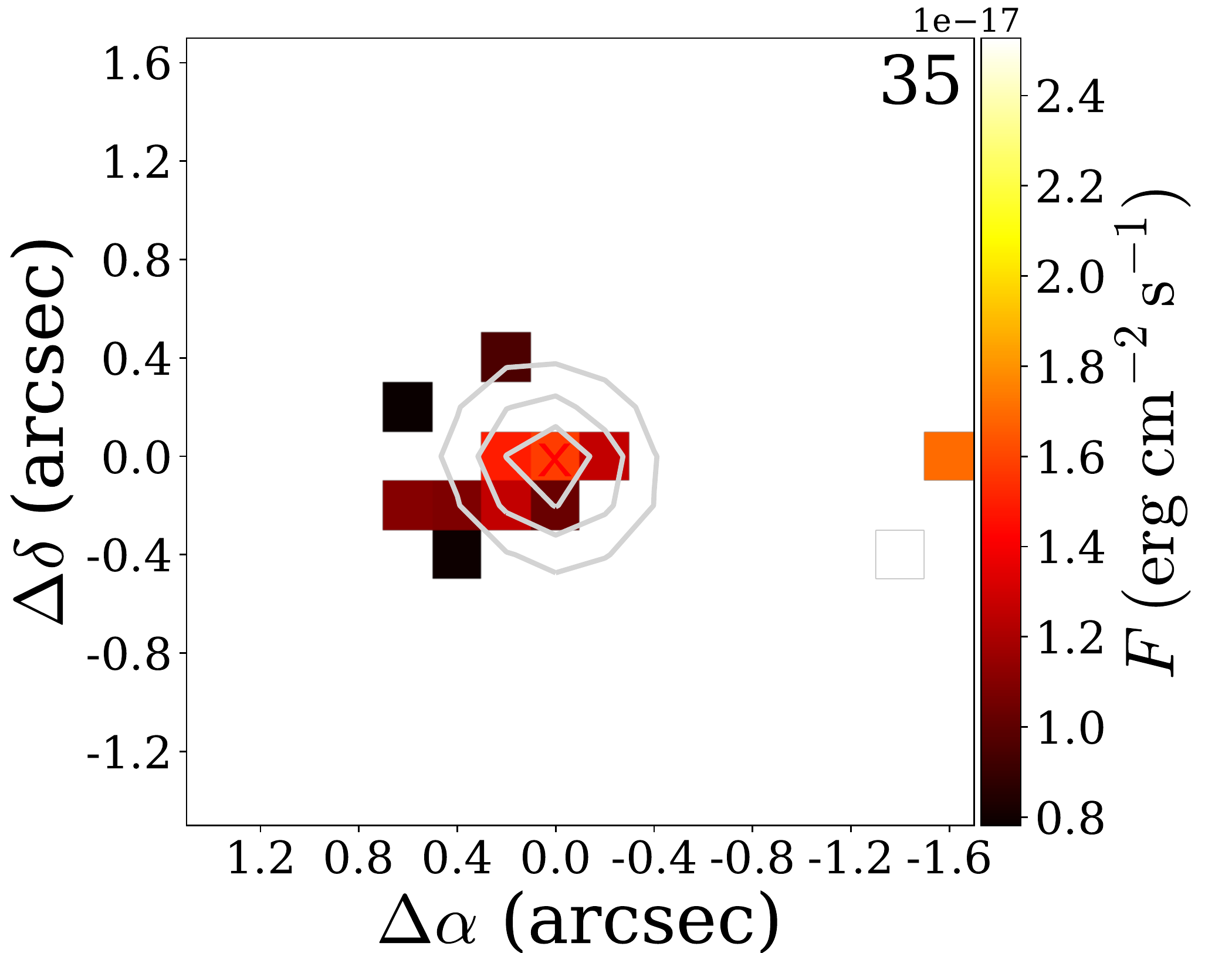}\hspace{-0.1cm}
\includegraphics[width=0.2\textwidth]{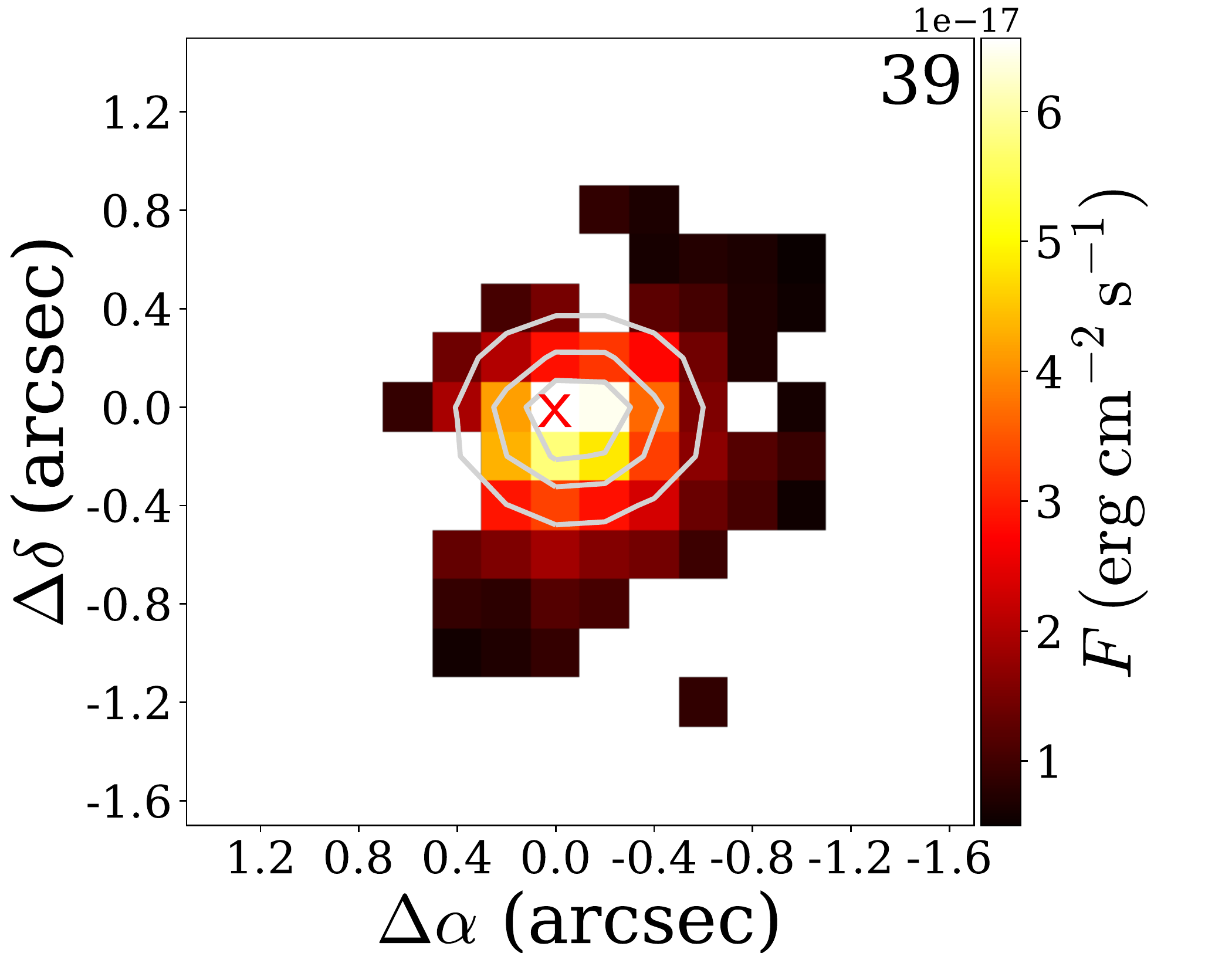}\hspace{-0.1cm}
\includegraphics[width=0.2\textwidth]{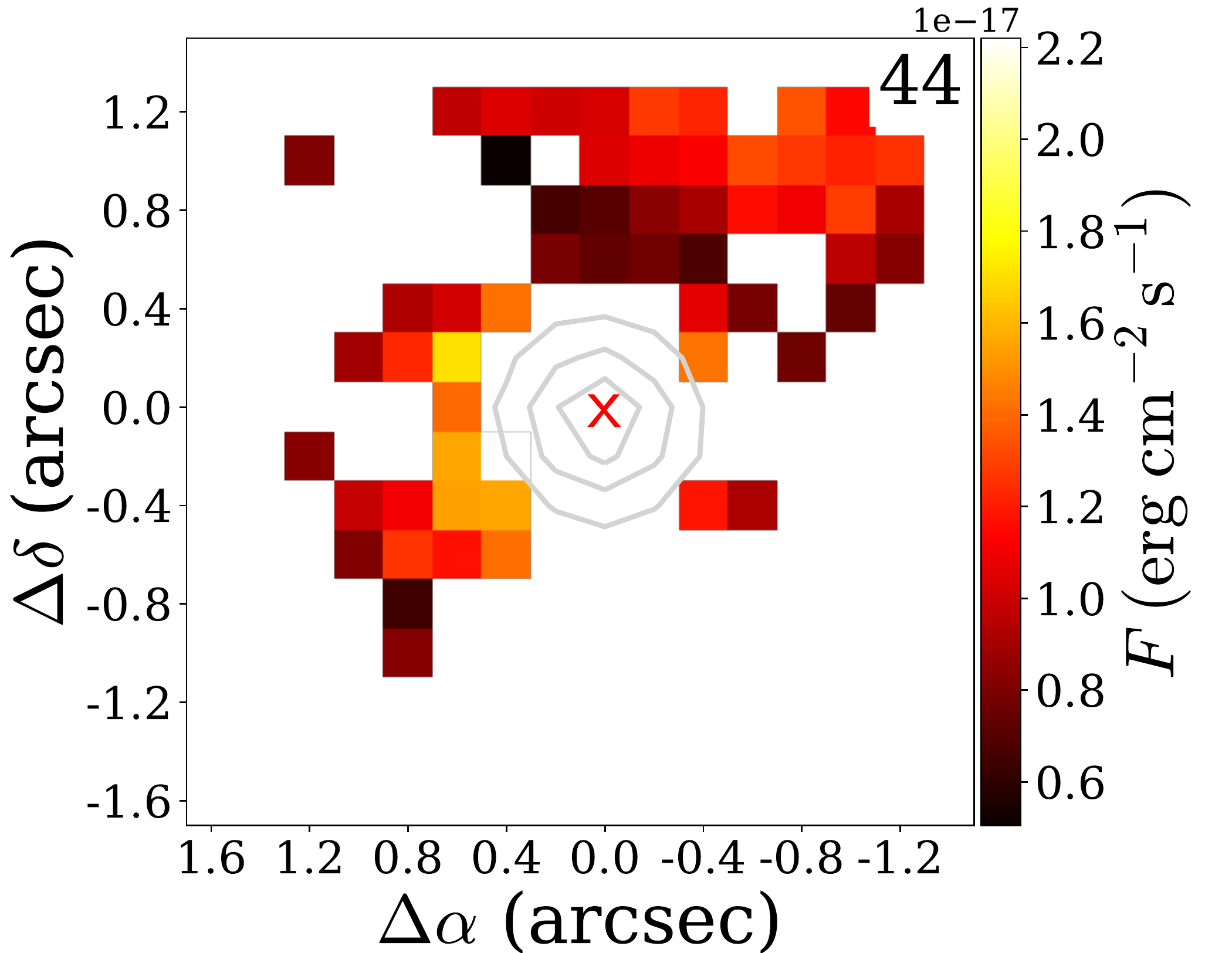}\hspace{-0.1cm}
\includegraphics[width=0.2\textwidth]{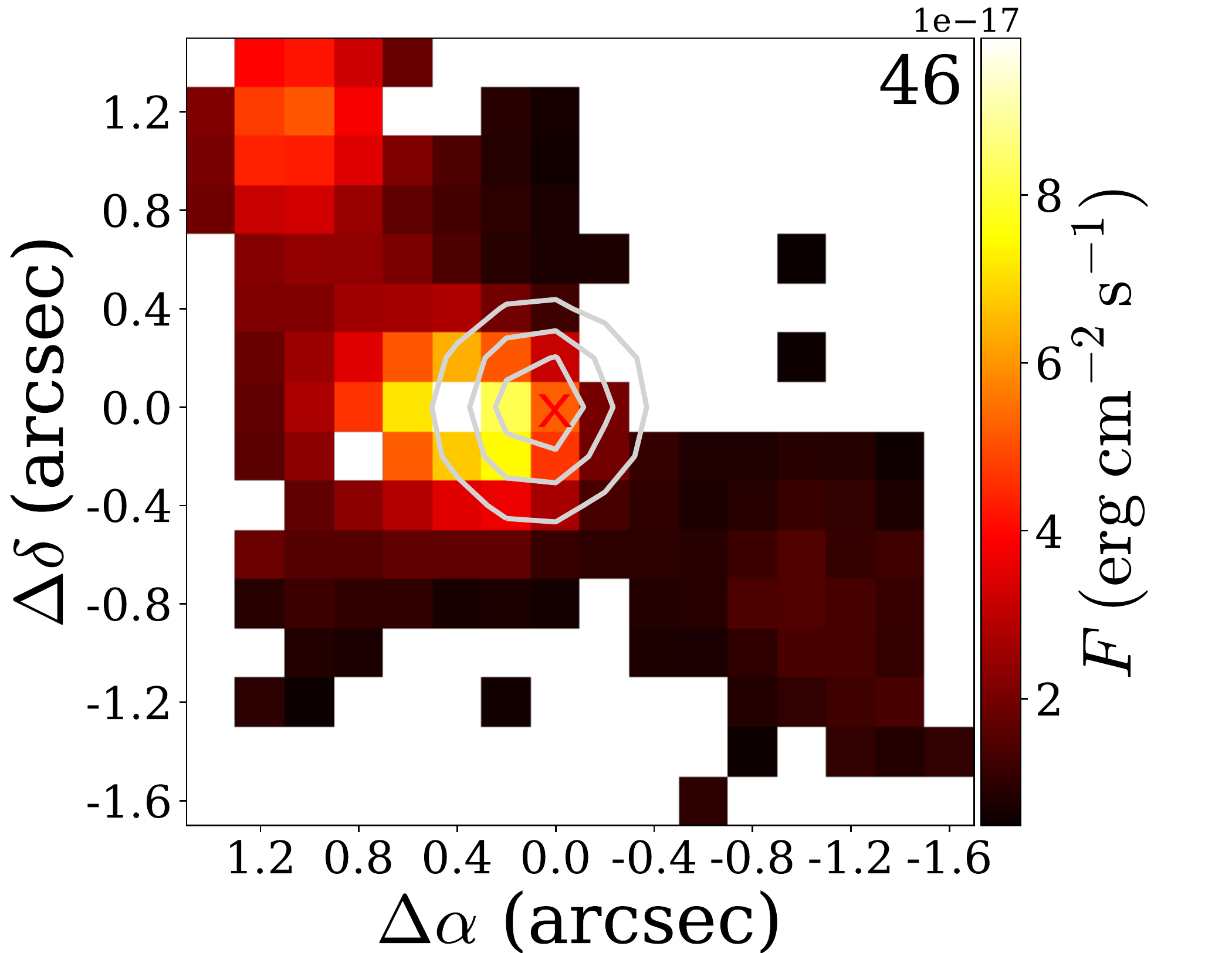}\hspace{-0.1cm}
\includegraphics[width=0.2\textwidth]{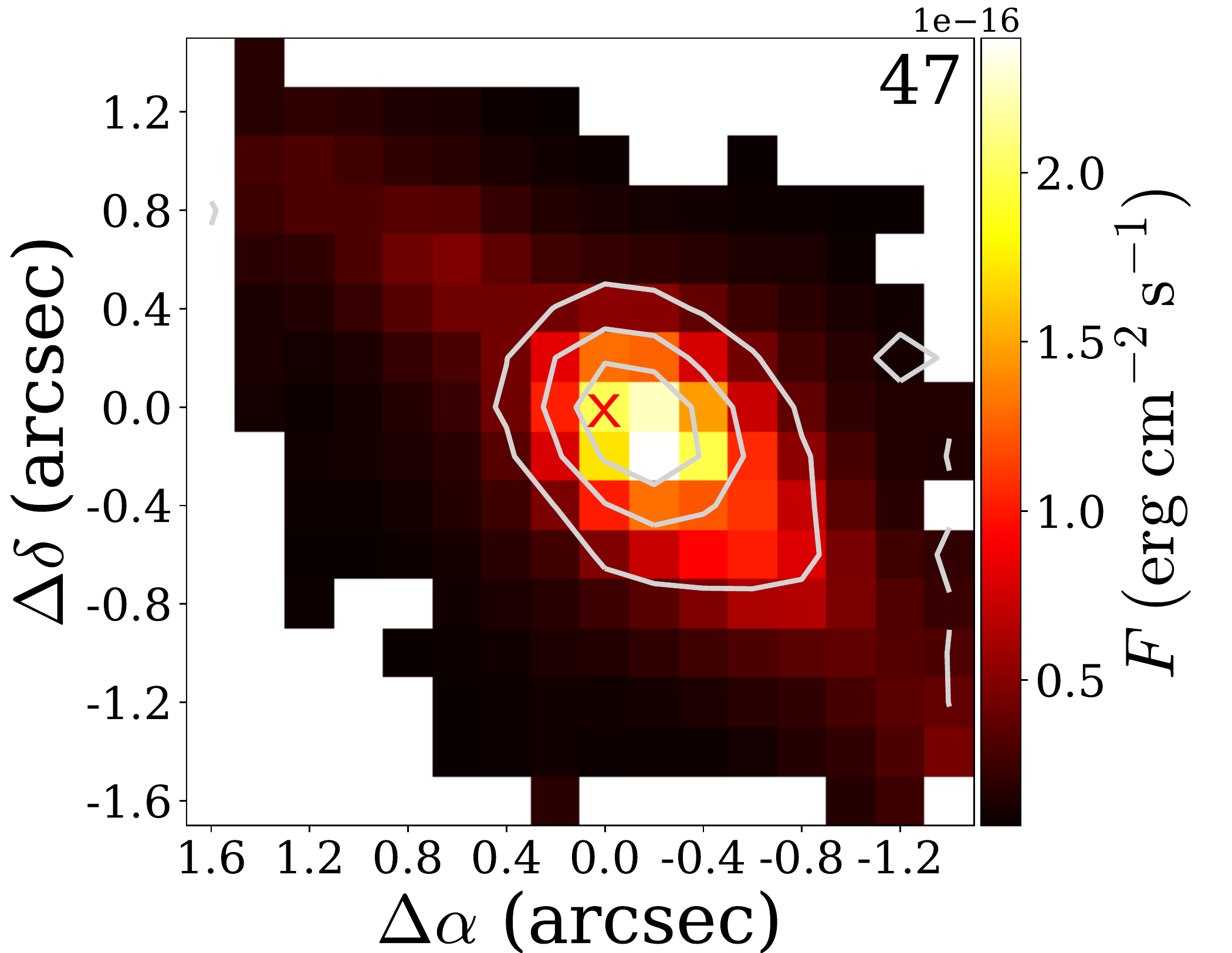}\hspace{-0.1cm}
\includegraphics[width=0.2\textwidth]{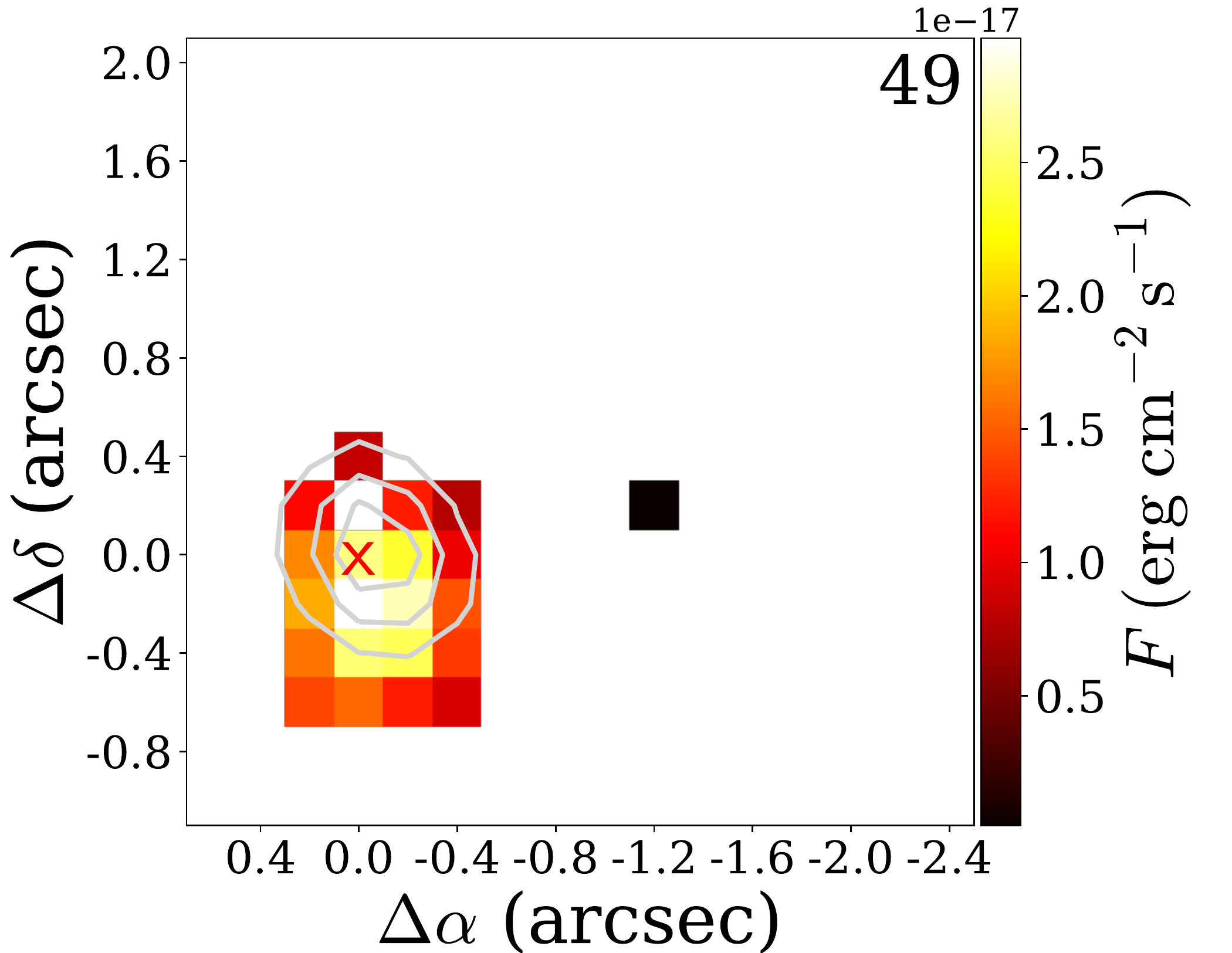}\hspace{-0.1cm}
\includegraphics[width=0.2\textwidth]{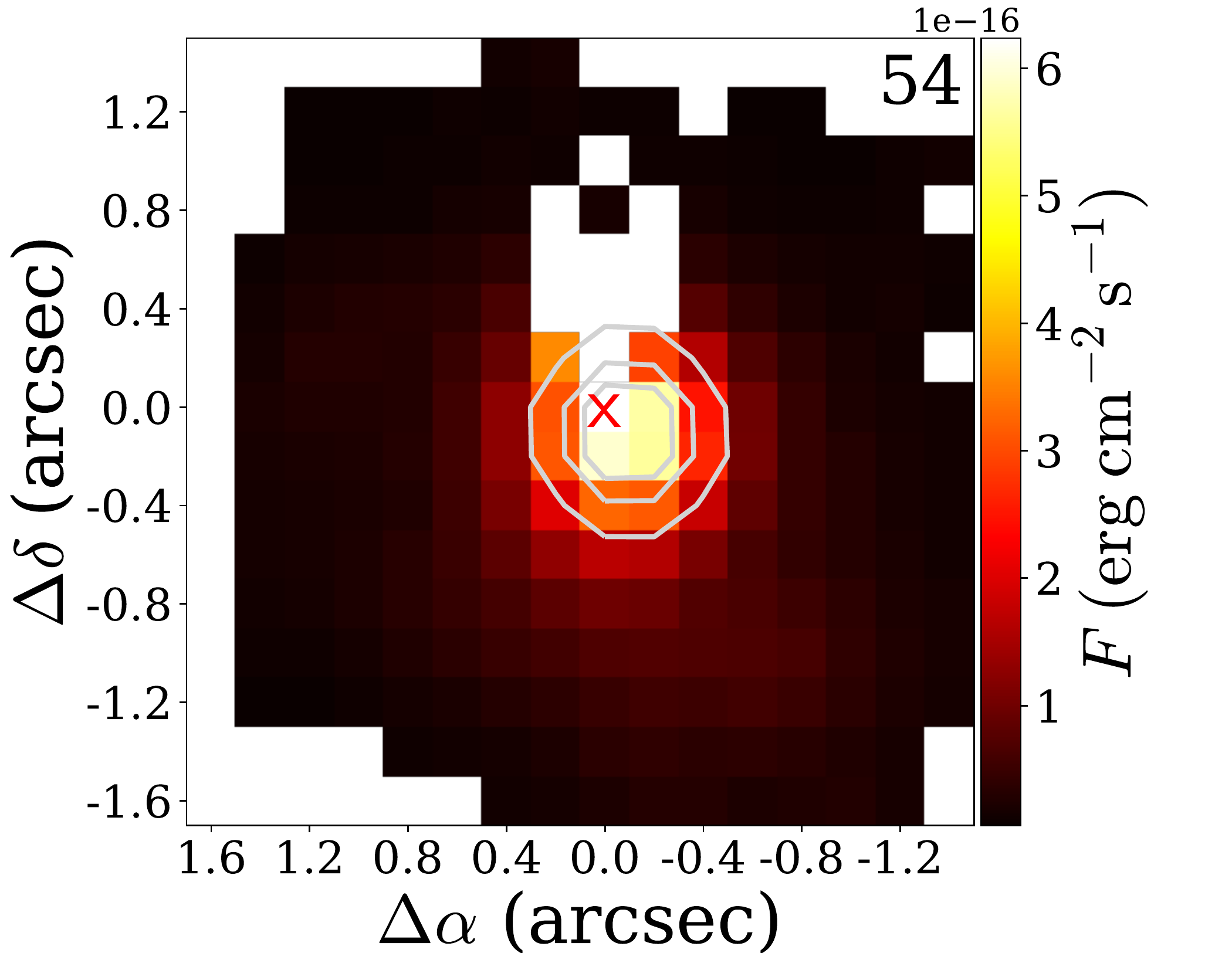}\hspace{-0.1cm}
\includegraphics[width=0.2\textwidth]{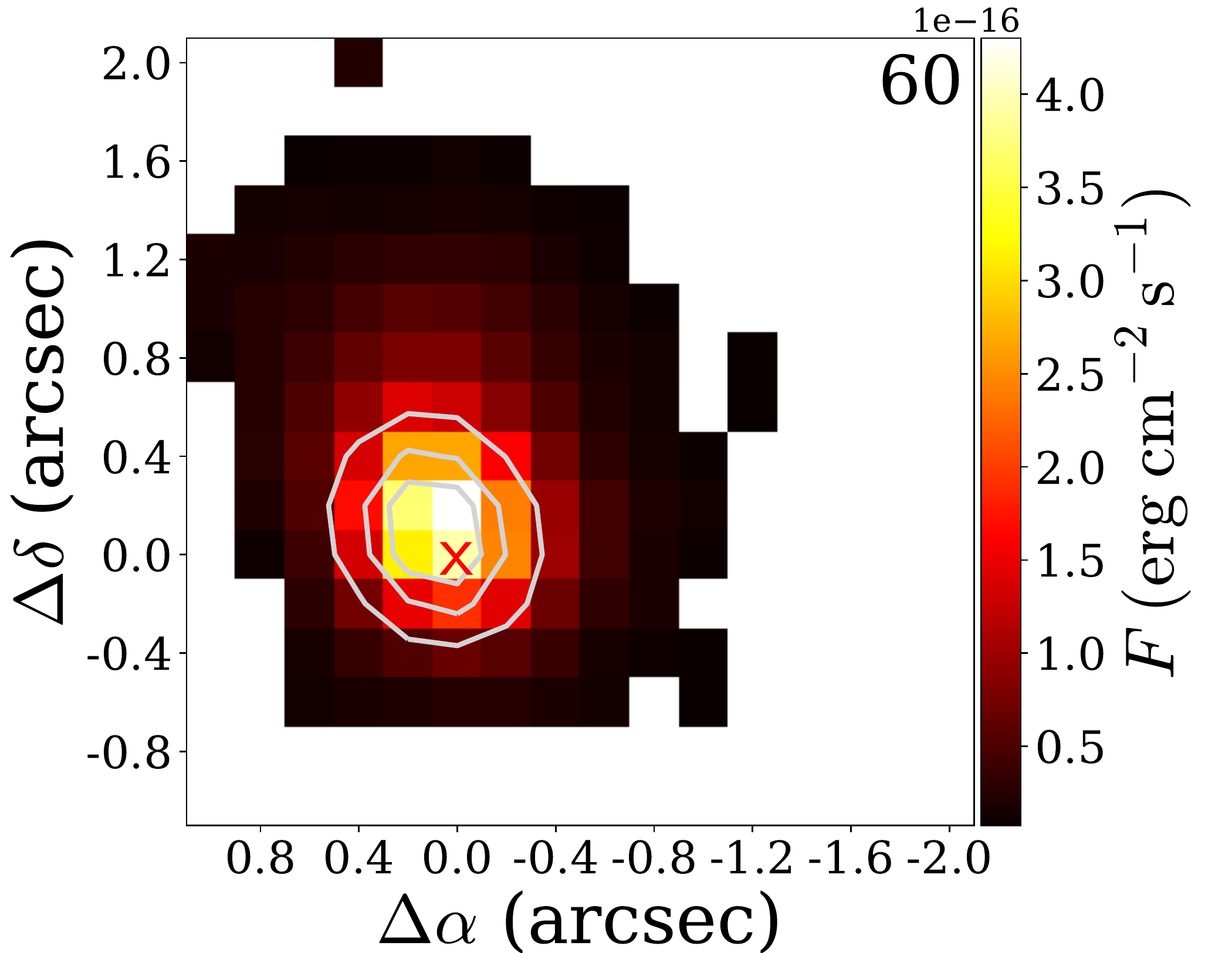}\hspace{-0.1cm}
\includegraphics[width=0.2\textwidth]{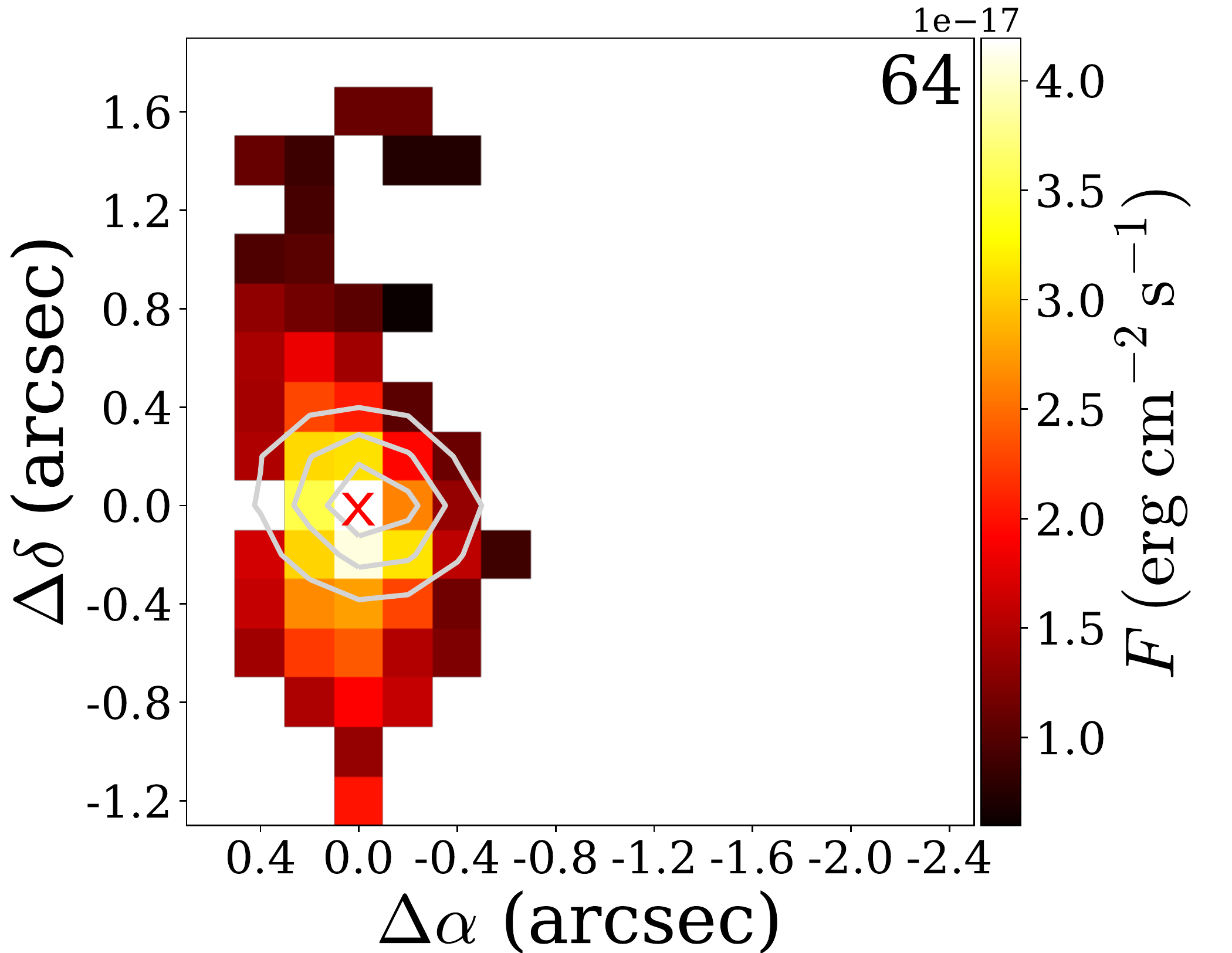}\hspace{-0.1cm}
\includegraphics[width=0.2\textwidth]{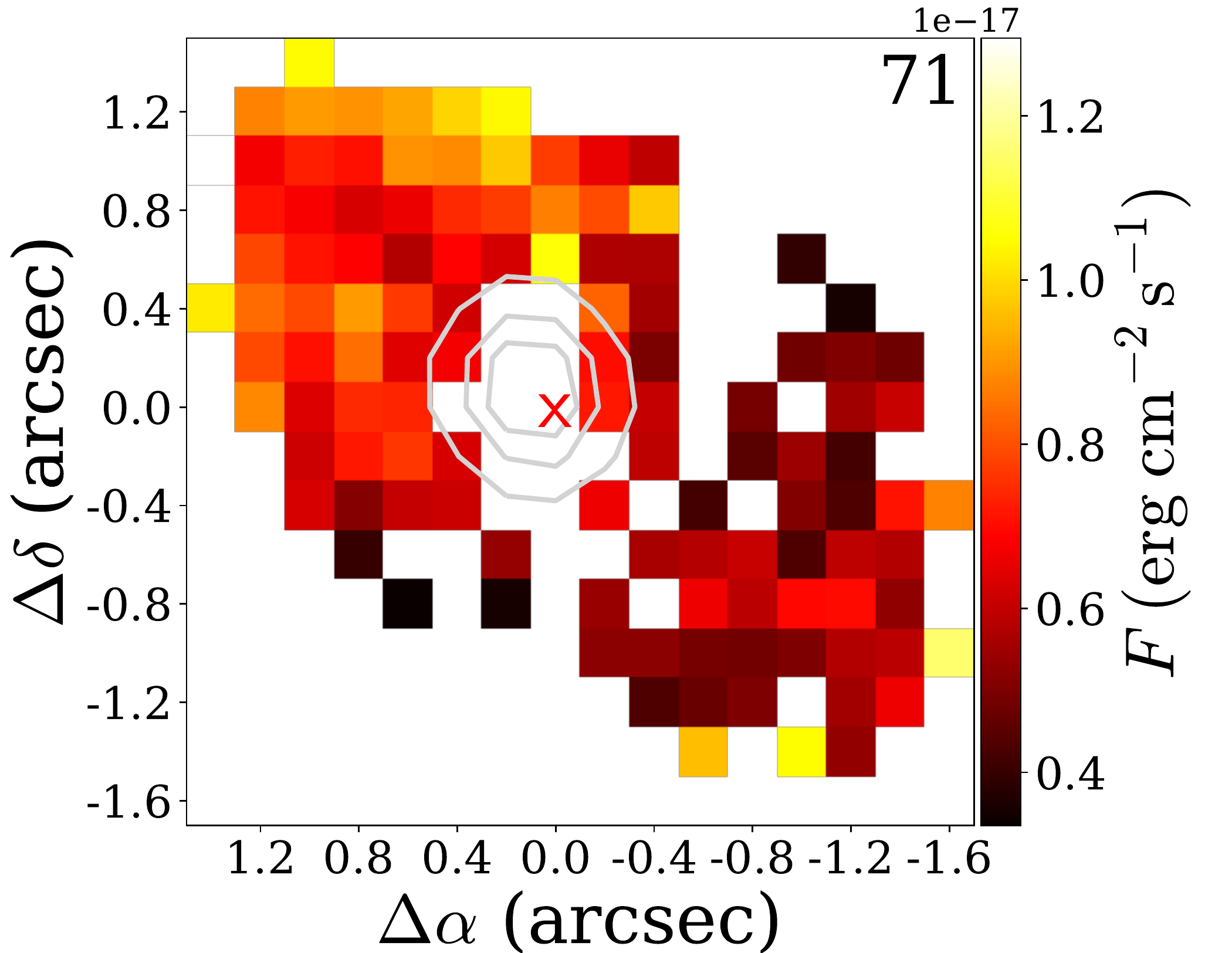}\hspace{-0.1cm}
\includegraphics[width=0.2\textwidth]{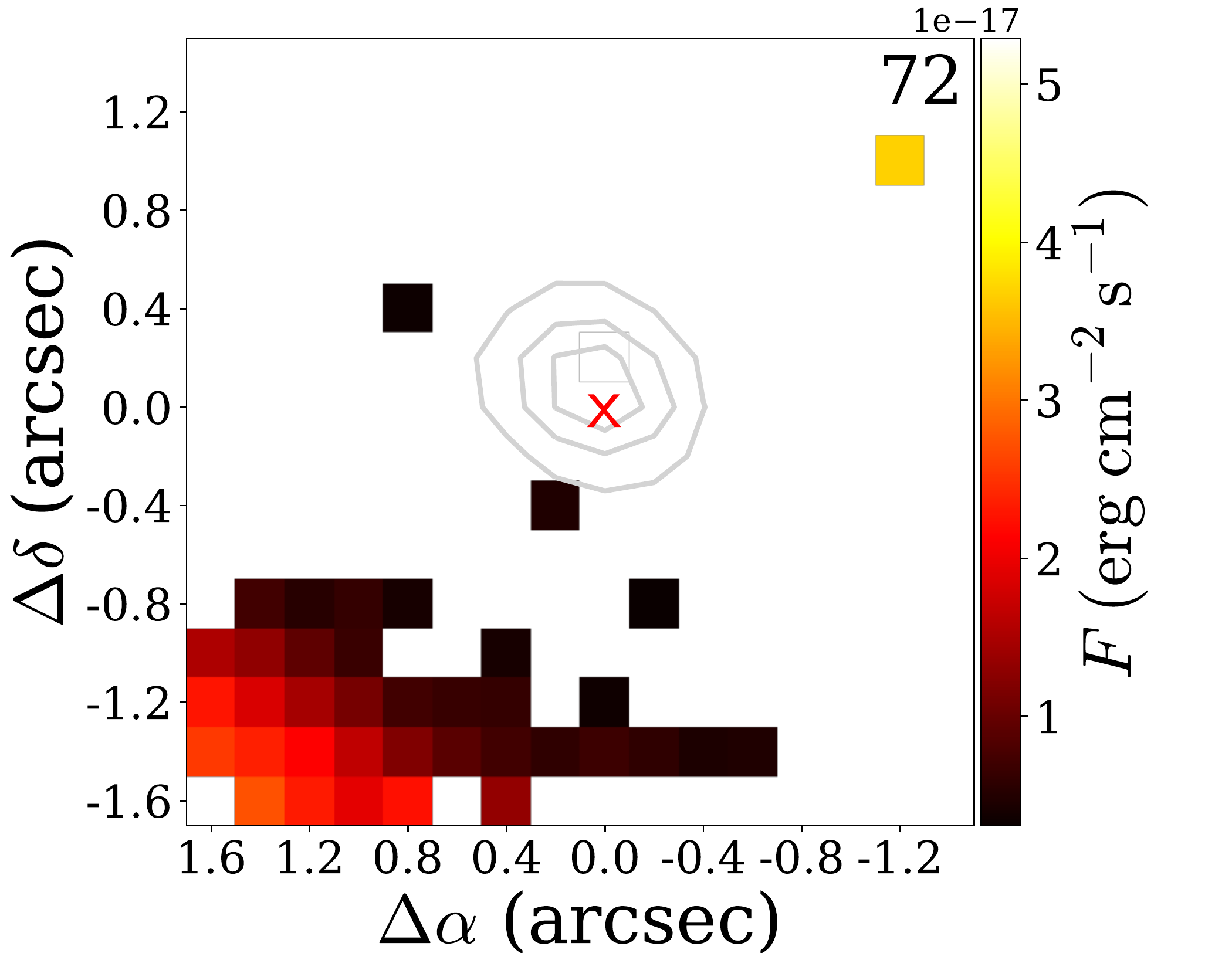}\hspace{-0.1cm}
\includegraphics[width=0.2\textwidth]{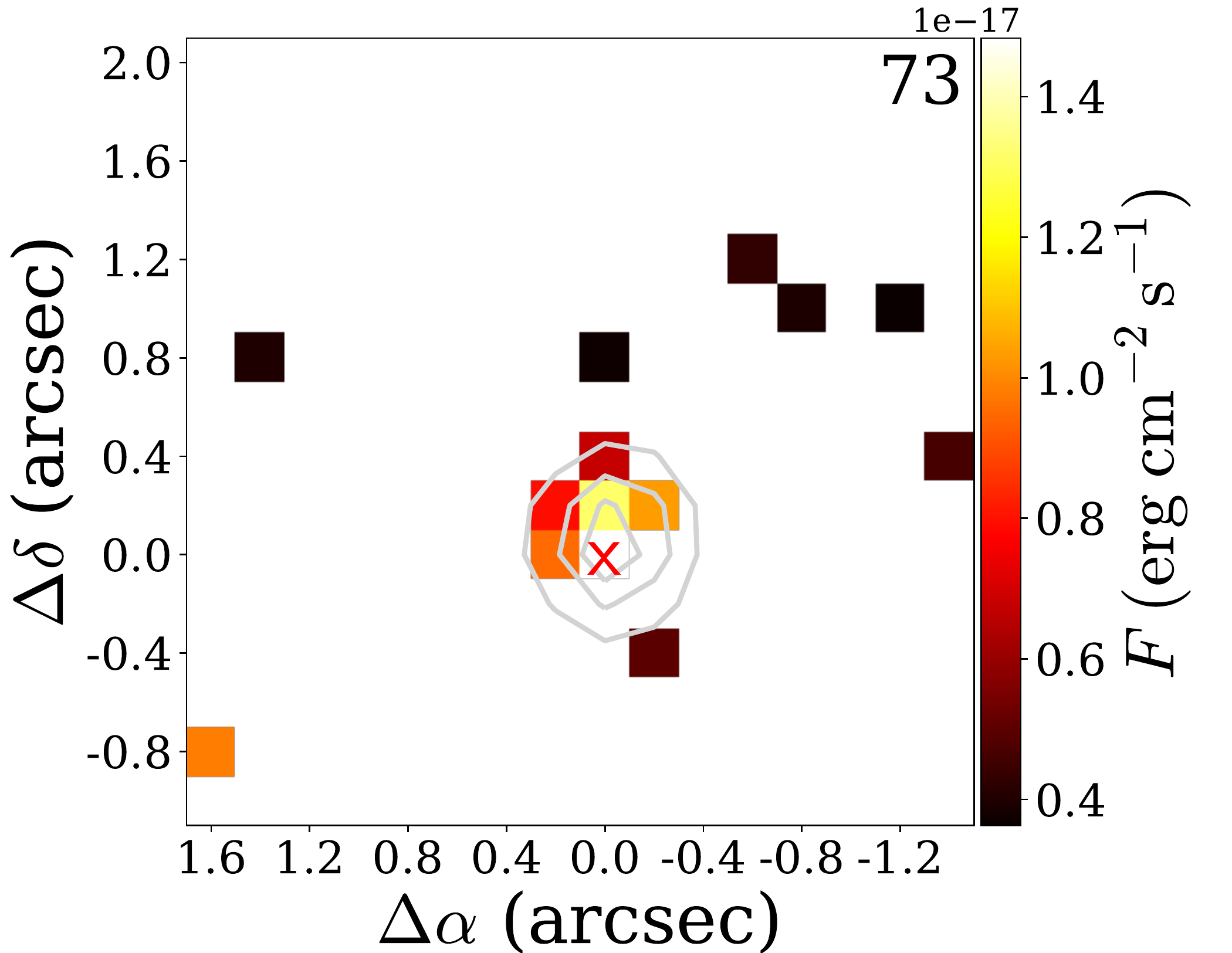}\hspace{-0.1cm}
\includegraphics[width=0.2\textwidth]{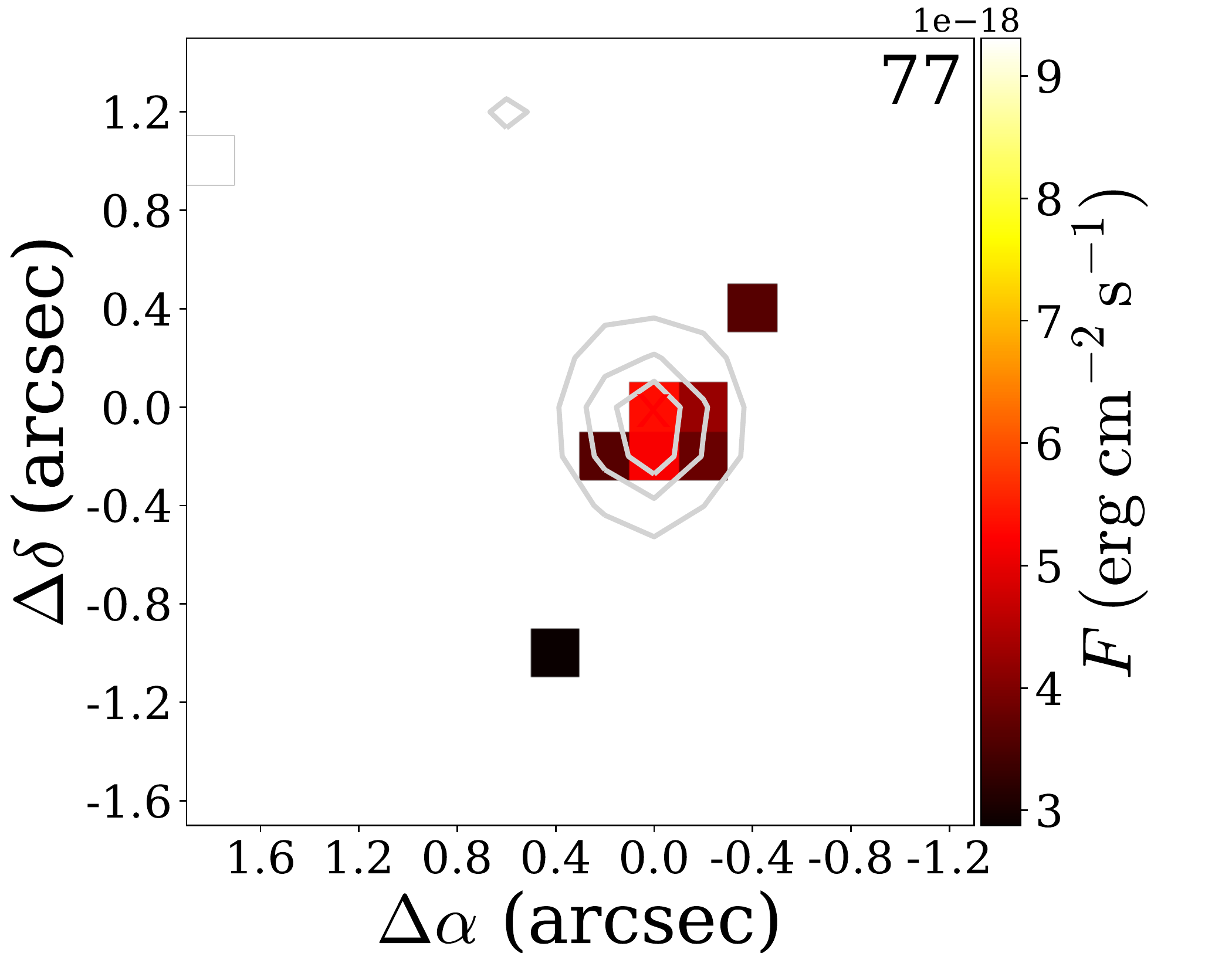}\hspace{-0.1cm}
\includegraphics[width=0.2\textwidth]{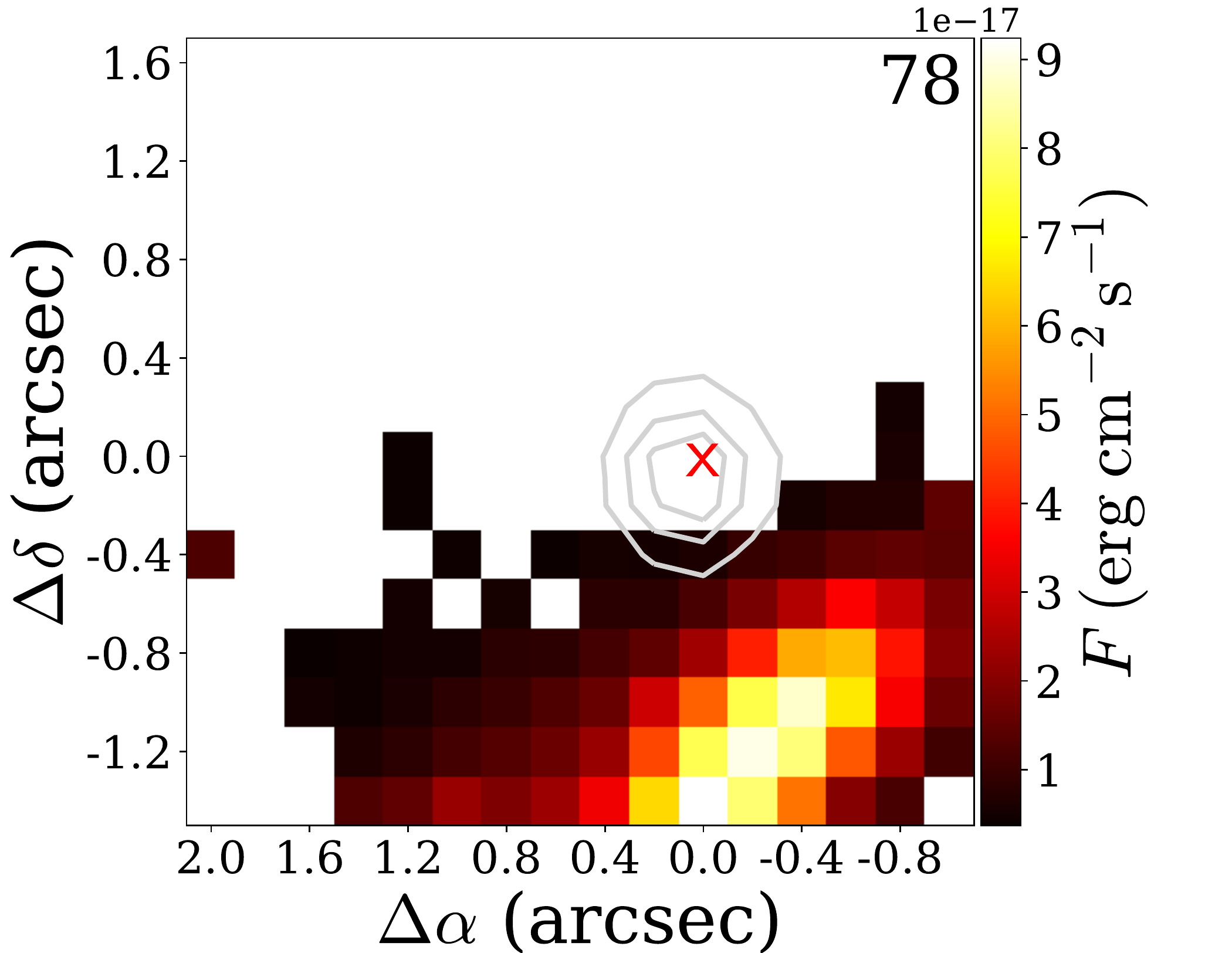}\hspace{-0.1cm}
\includegraphics[width=0.2\textwidth]{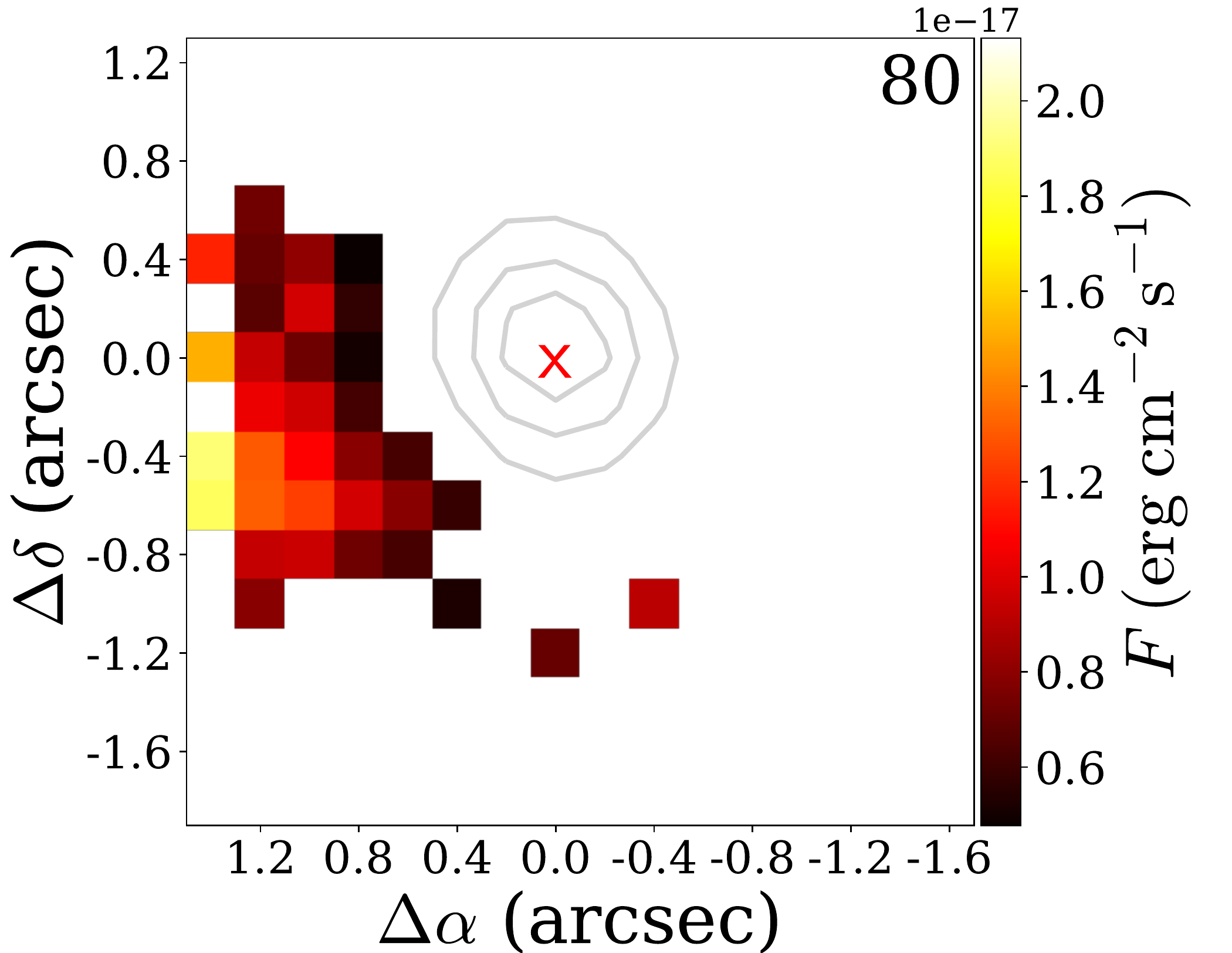}\hspace{-0.1cm}
\includegraphics[width=0.2\textwidth]{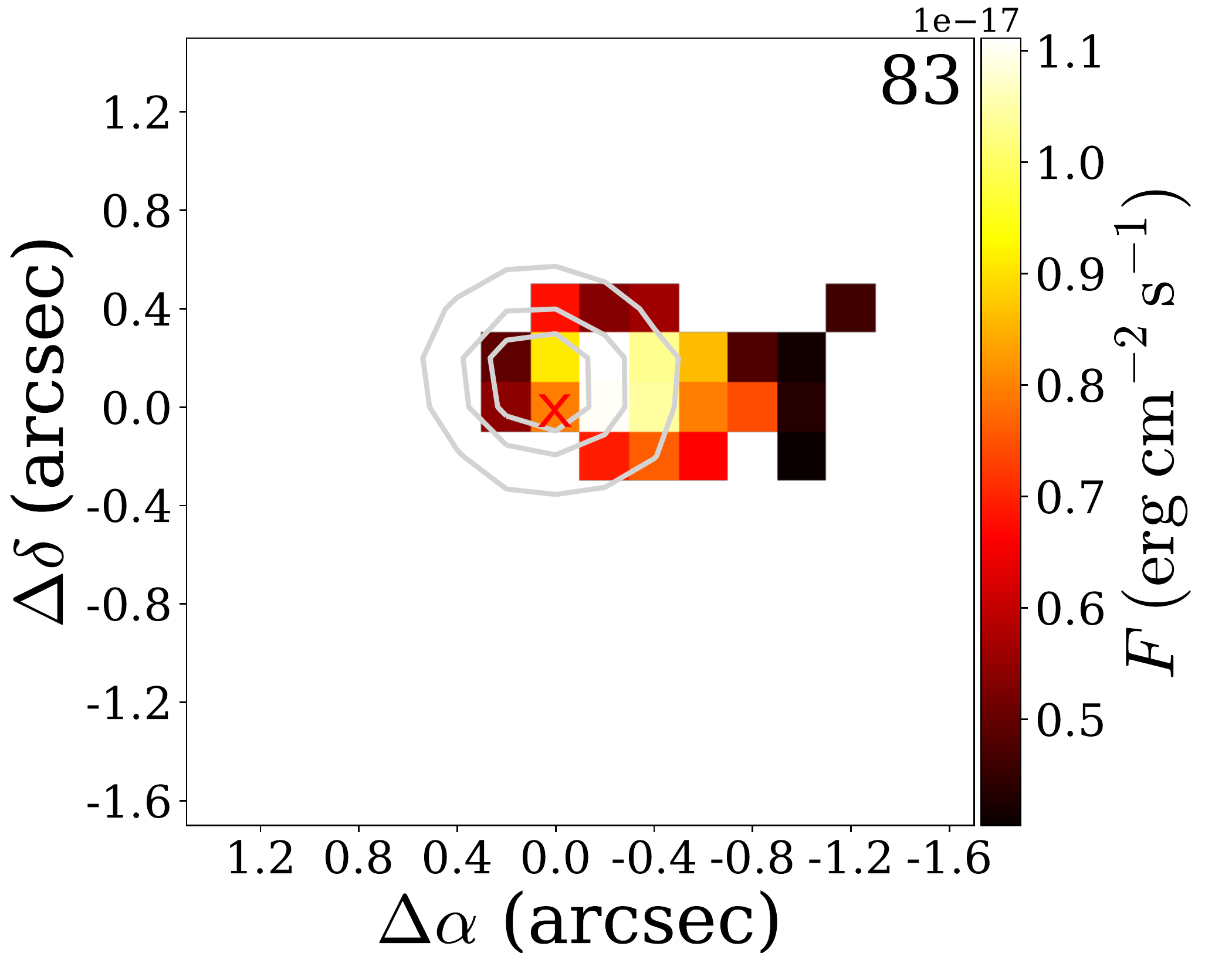}\hspace{-0.1cm}
\includegraphics[width=0.2\textwidth]{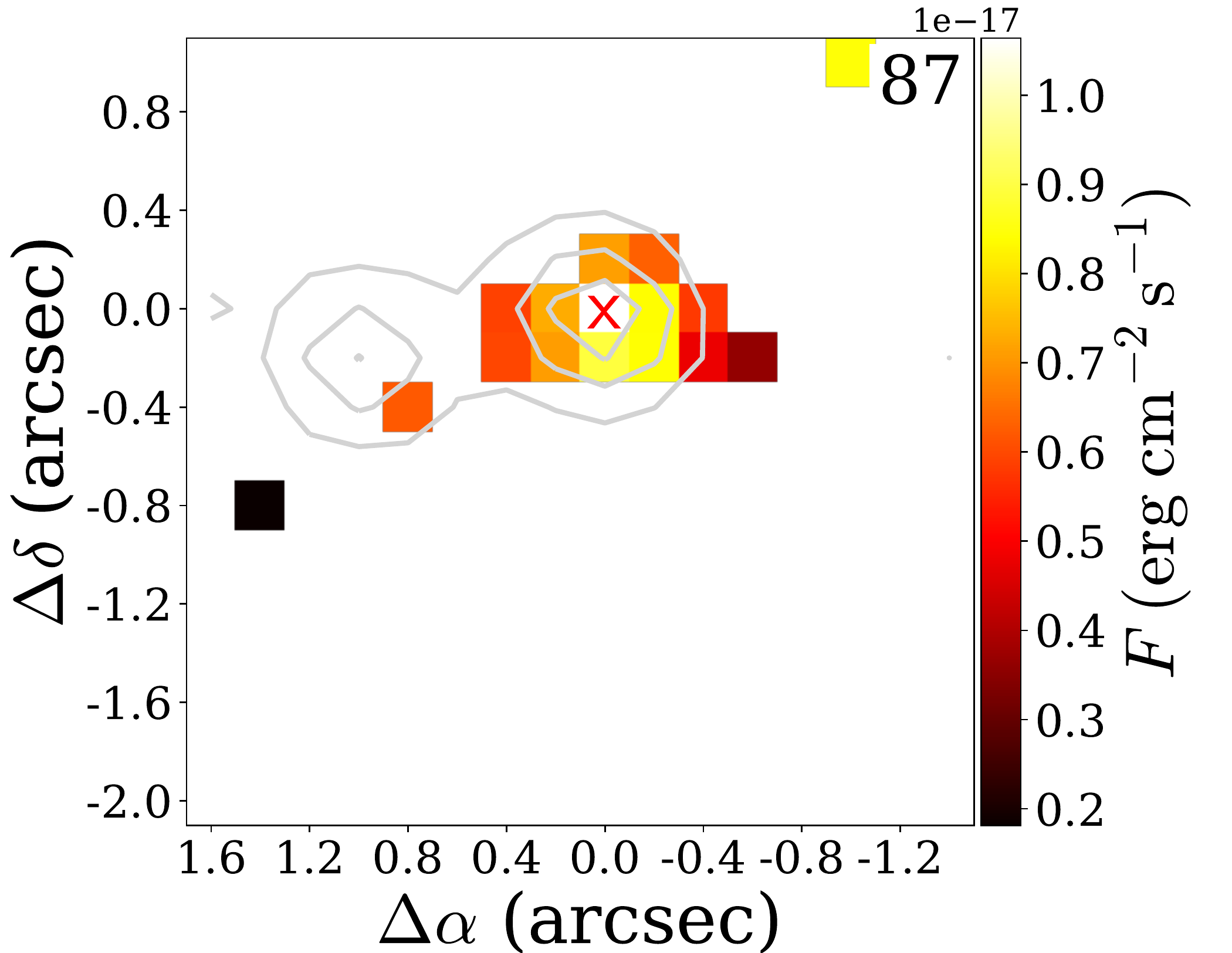}\hspace{-0.1cm}
\includegraphics[width=0.2\textwidth]{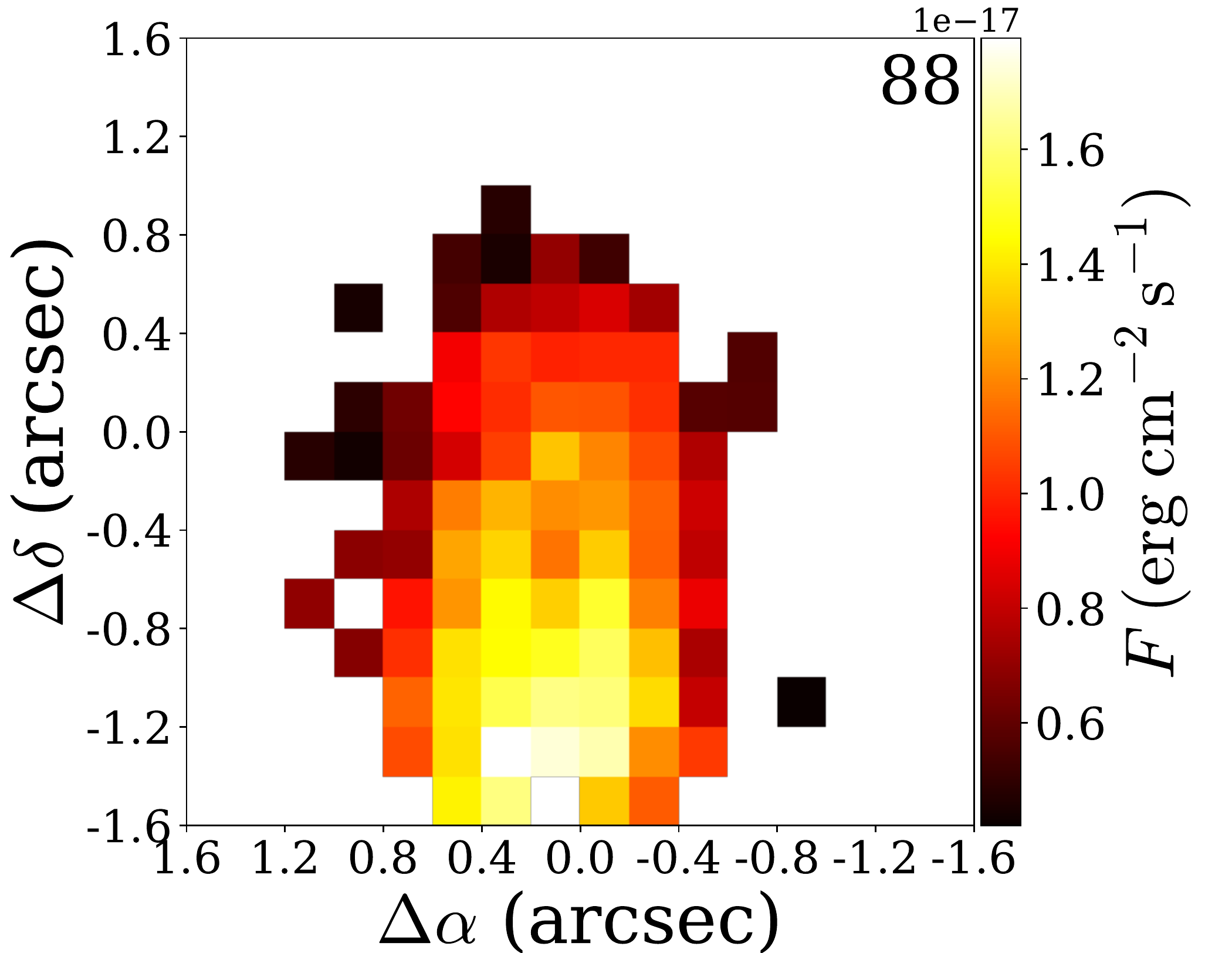}\hspace{-0.1cm}
\includegraphics[width=0.2\textwidth]{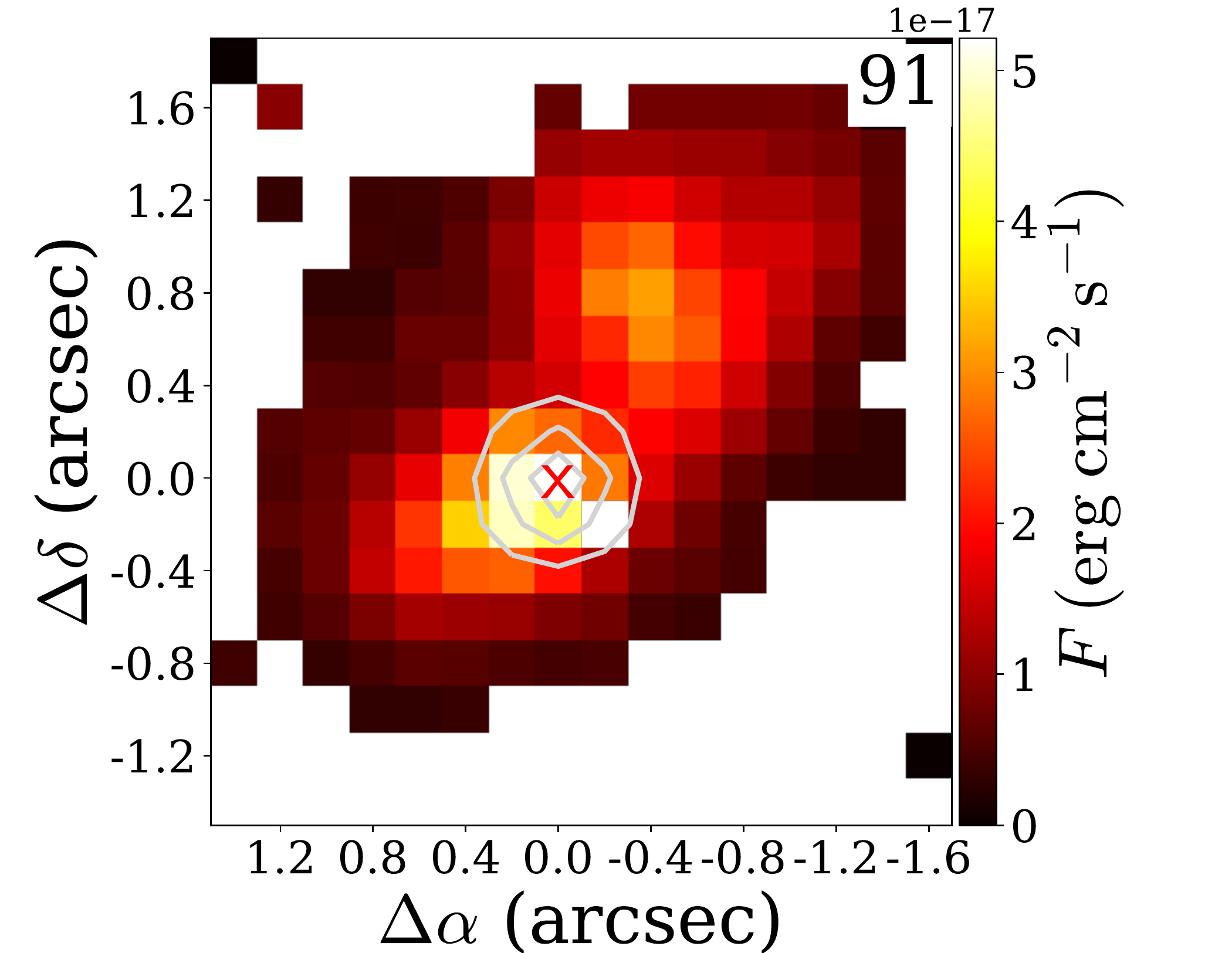}\hspace{-0.1cm}
\includegraphics[width=0.2\textwidth]{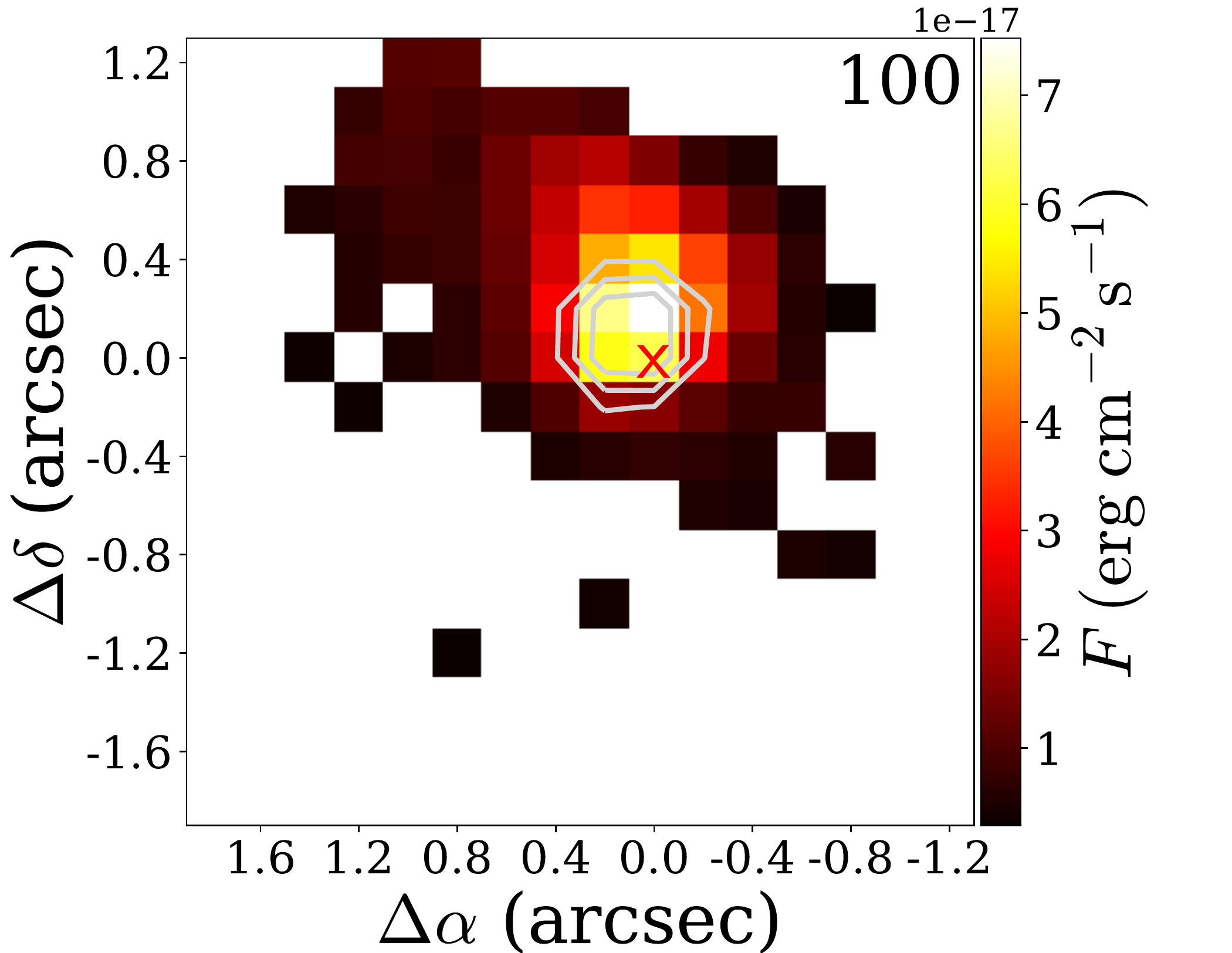}\hspace{-0.1cm}
\includegraphics[width=0.2\textwidth]{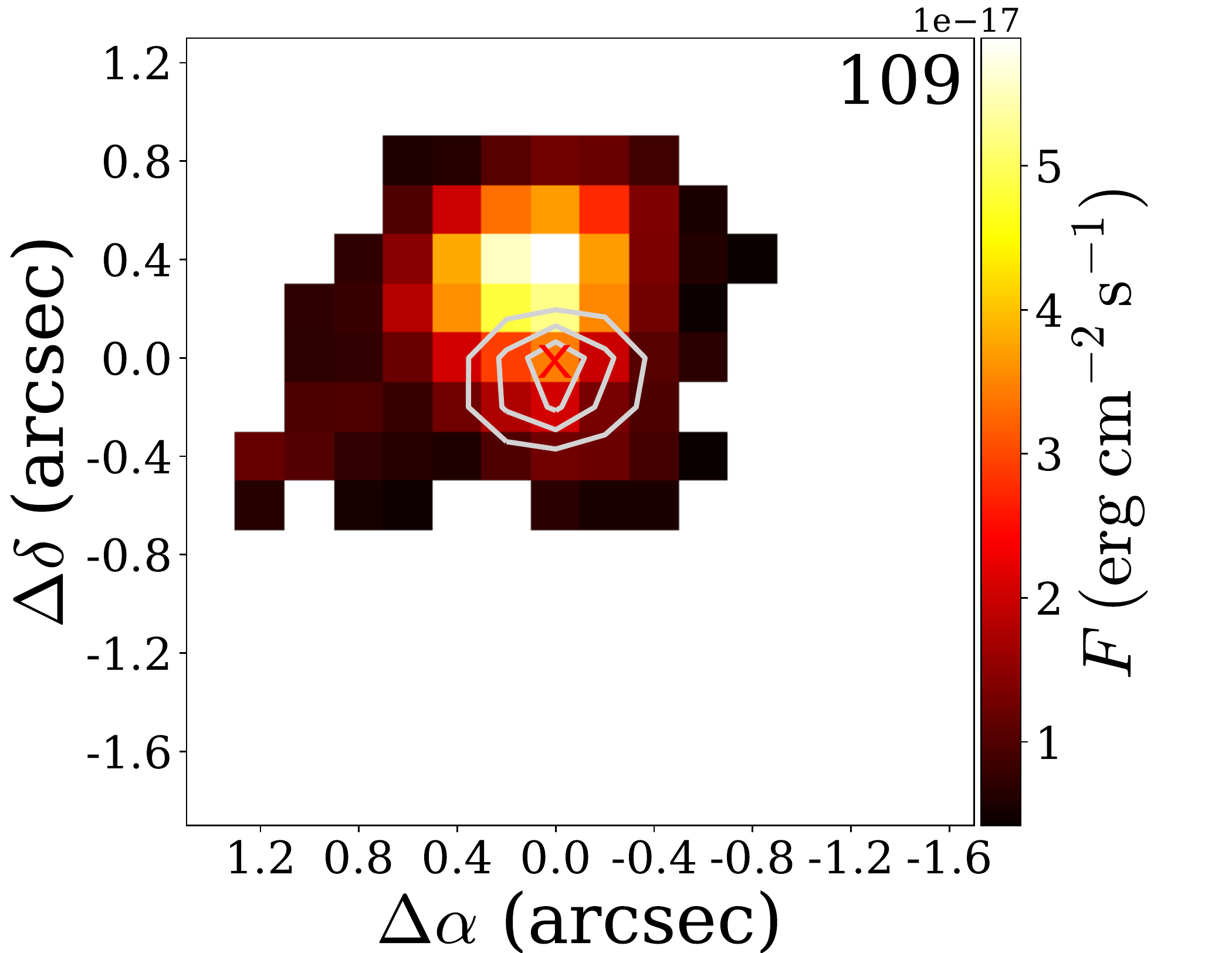}\hspace{-0.1cm}
\includegraphics[width=0.2\textwidth]{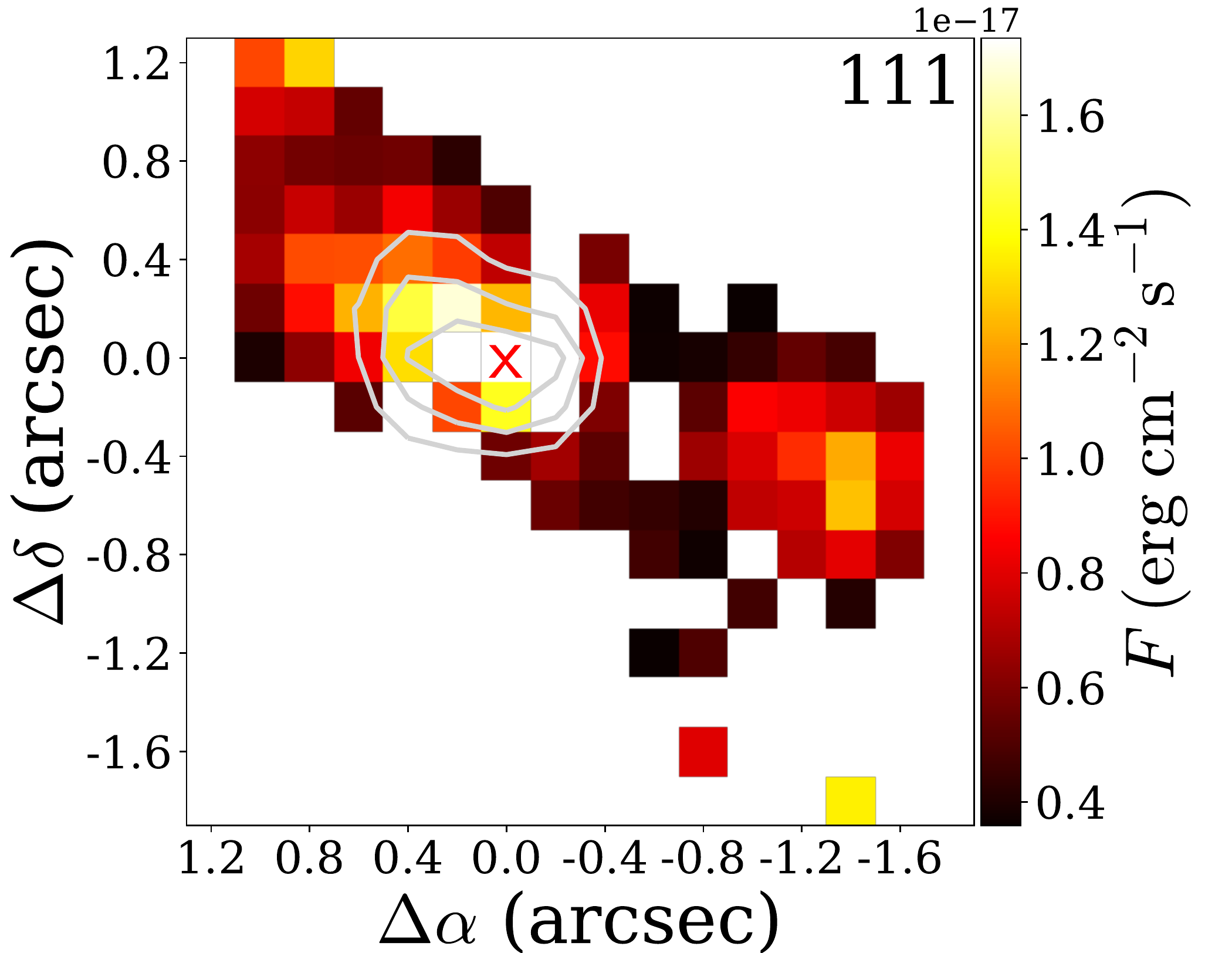}\hspace{-0.1cm}
\includegraphics[width=0.2\textwidth]{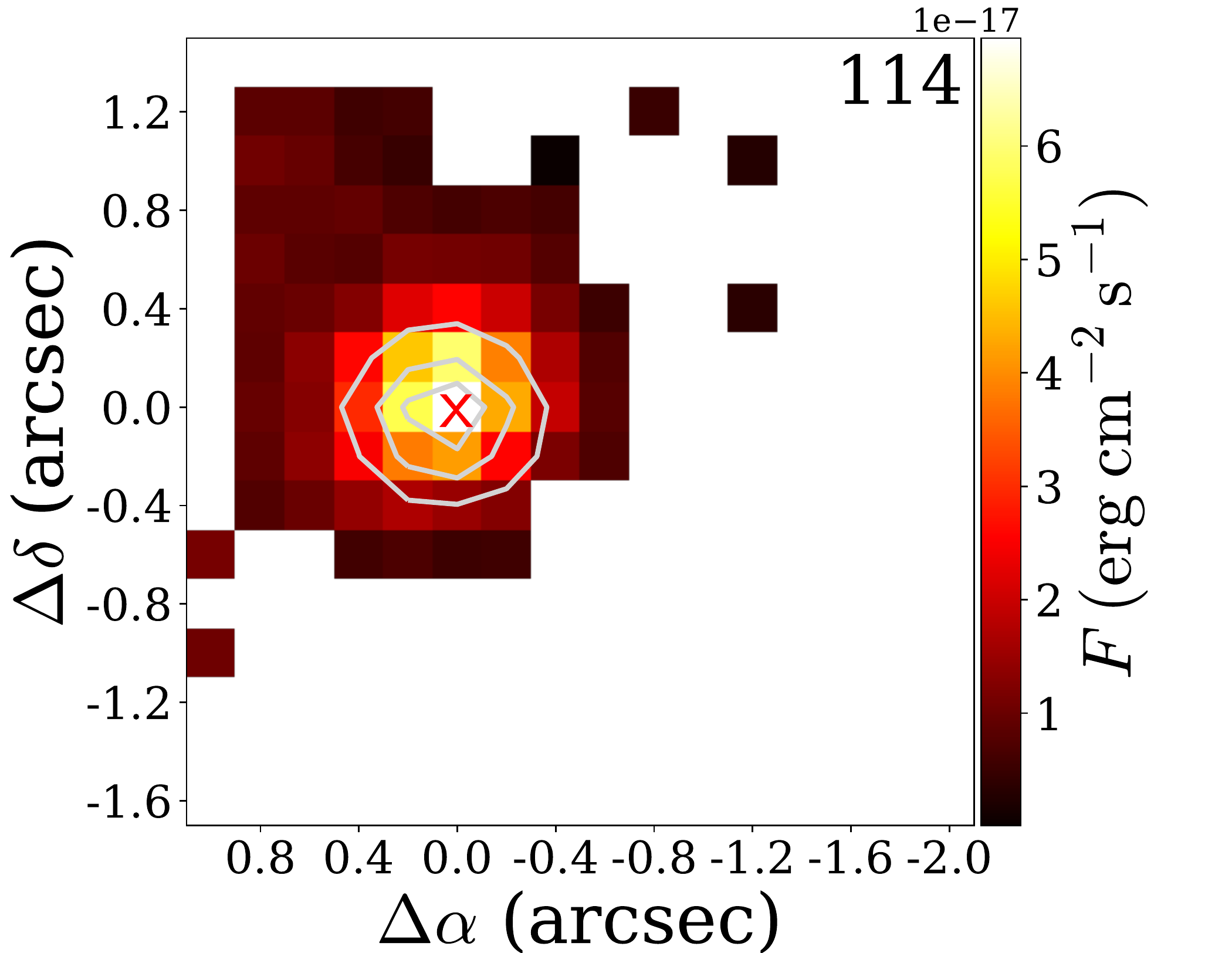}\hspace{-0.1cm}
\includegraphics[width=0.2\textwidth]{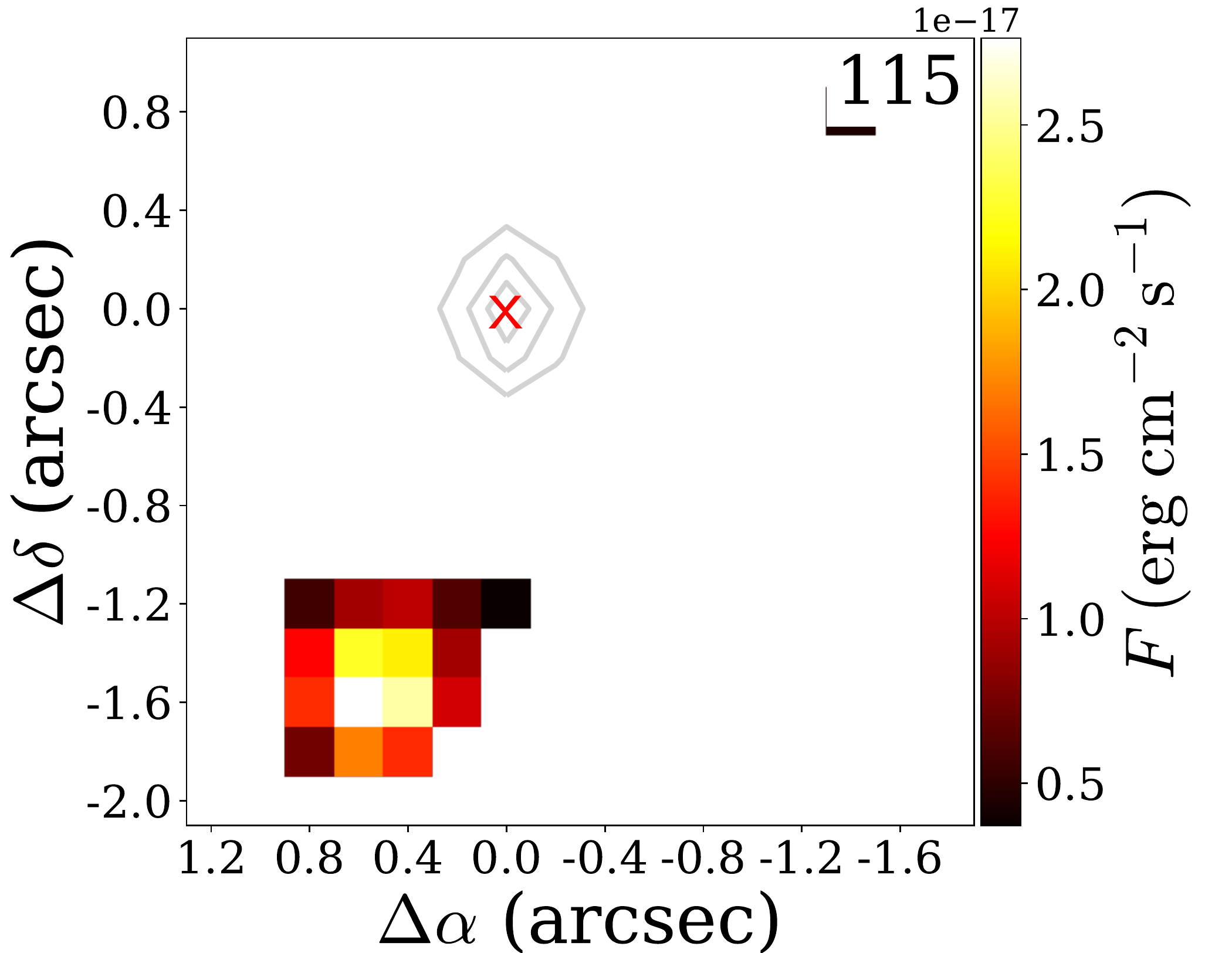}\hspace{-0.1cm}
\caption{Integrated maps of the H$_2$ 1-0 S(1) line at 2.1218 $\mu$m for a subsample of sources where the extended H$_2$ emission is detected (see Table \ref{tab:coordinates}). Colors correspond to the line fluxes at each pixel (detections above 3$\sigma$) and white contours correspond to the continuum emission in $K$-band (see also Appendix A). Source names are provided in the upper right corners (No. in Table \ref{tab:coordinates}). Note that source No. 88 (G224.4514$-$00.7054) has a non-detection of the continuum emission at 2.1218 $\mu$m.
}
\label{fig:emiss-2.1218}
\end{figure*}

\begin{figure*}[h!]
\includegraphics[width=0.2\textwidth]{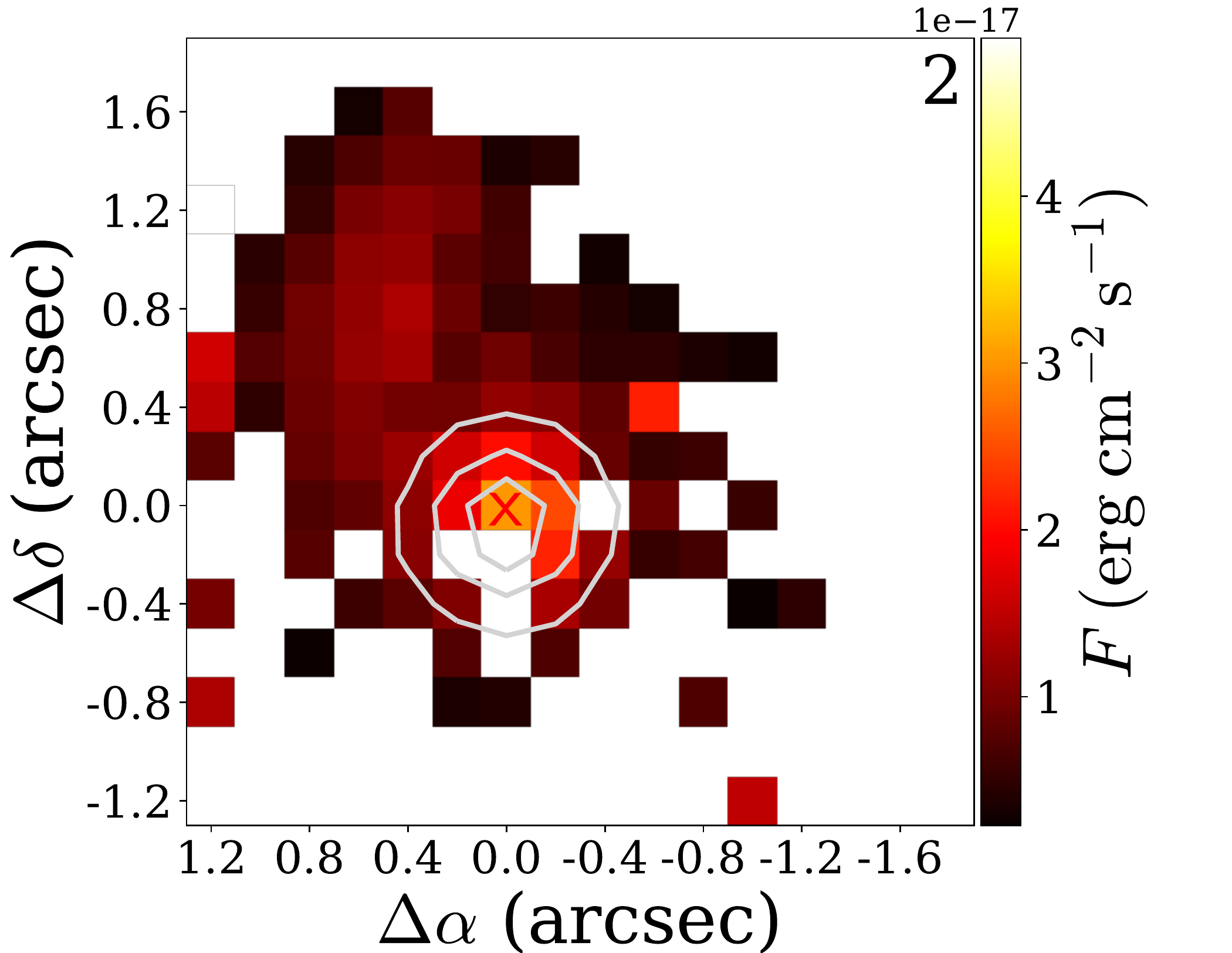}\hspace{-0.2cm}
\includegraphics[width=0.2\textwidth]{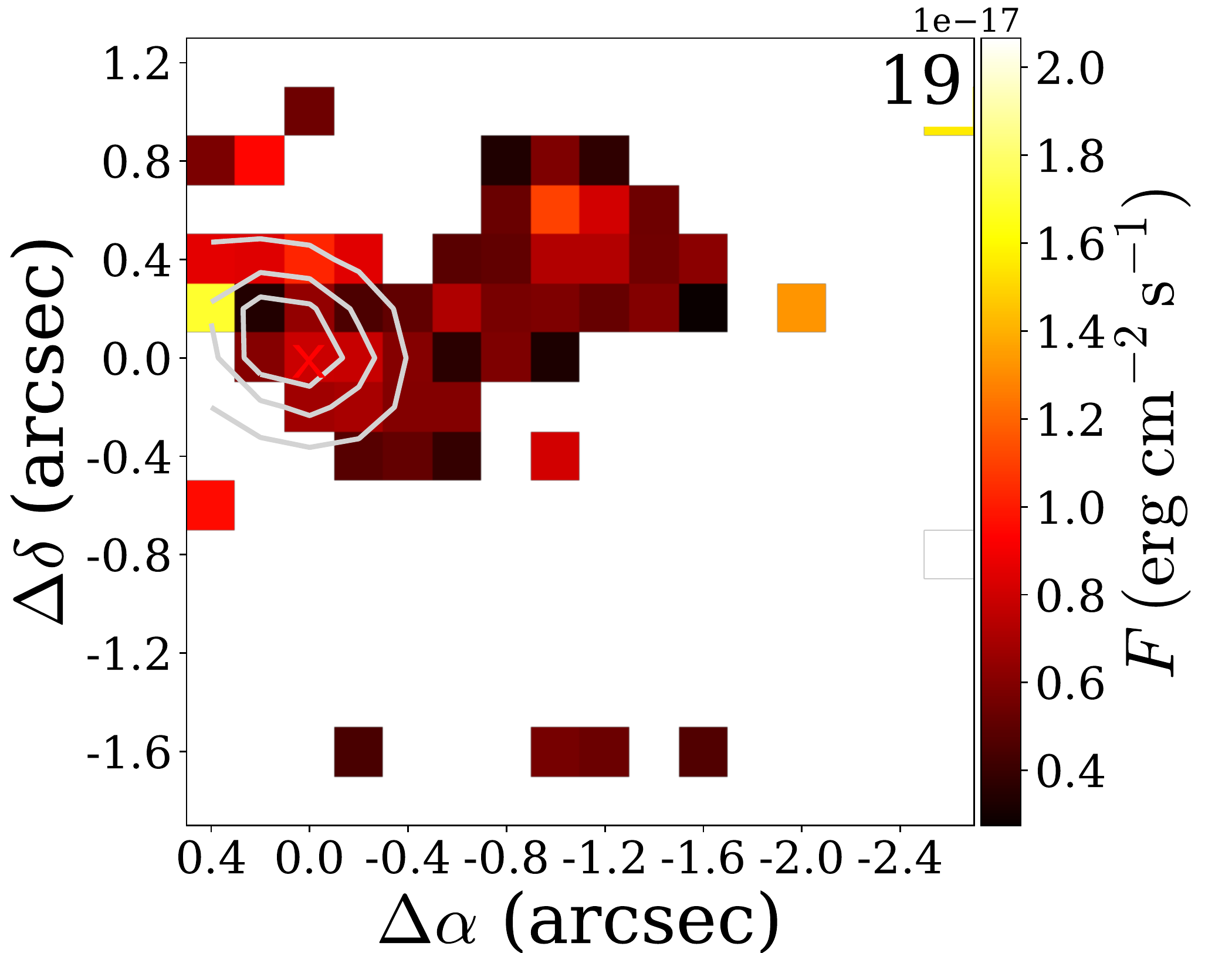}\hspace{-0.2cm}
\includegraphics[width=0.2\textwidth]{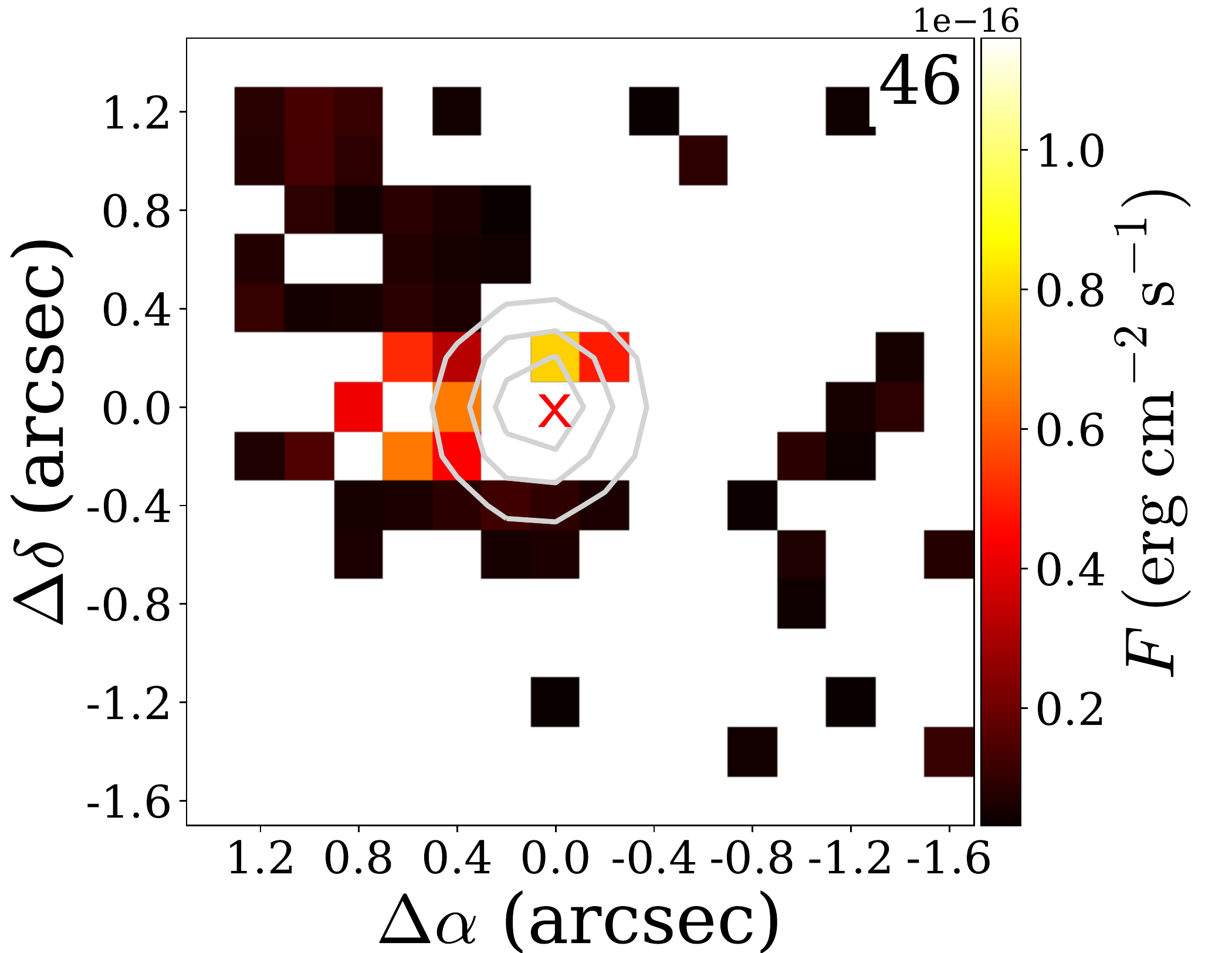}\hspace{-0.2cm}
\includegraphics[width=0.2\textwidth]{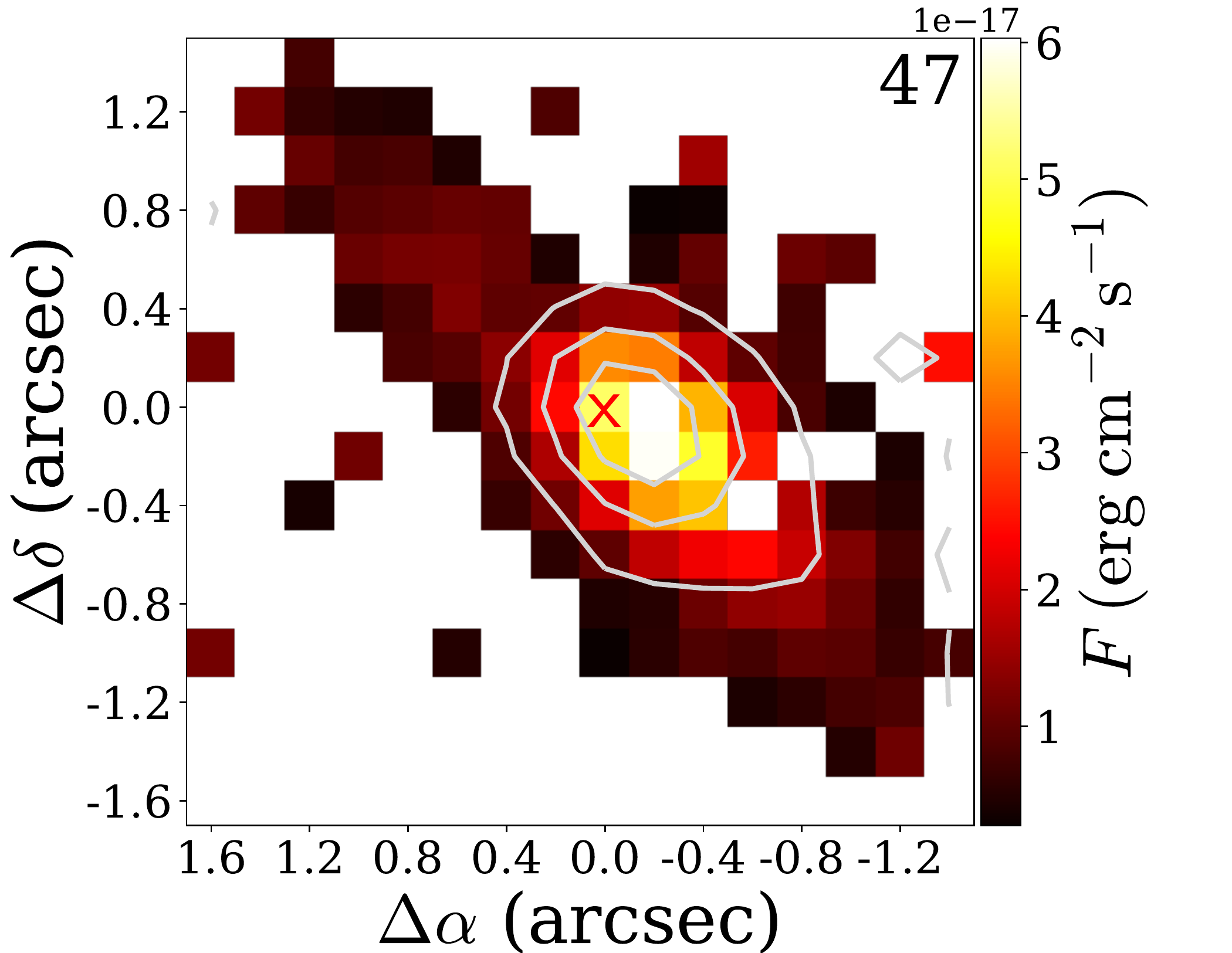}\hspace{-0.2cm}
\includegraphics[width=0.2\textwidth]{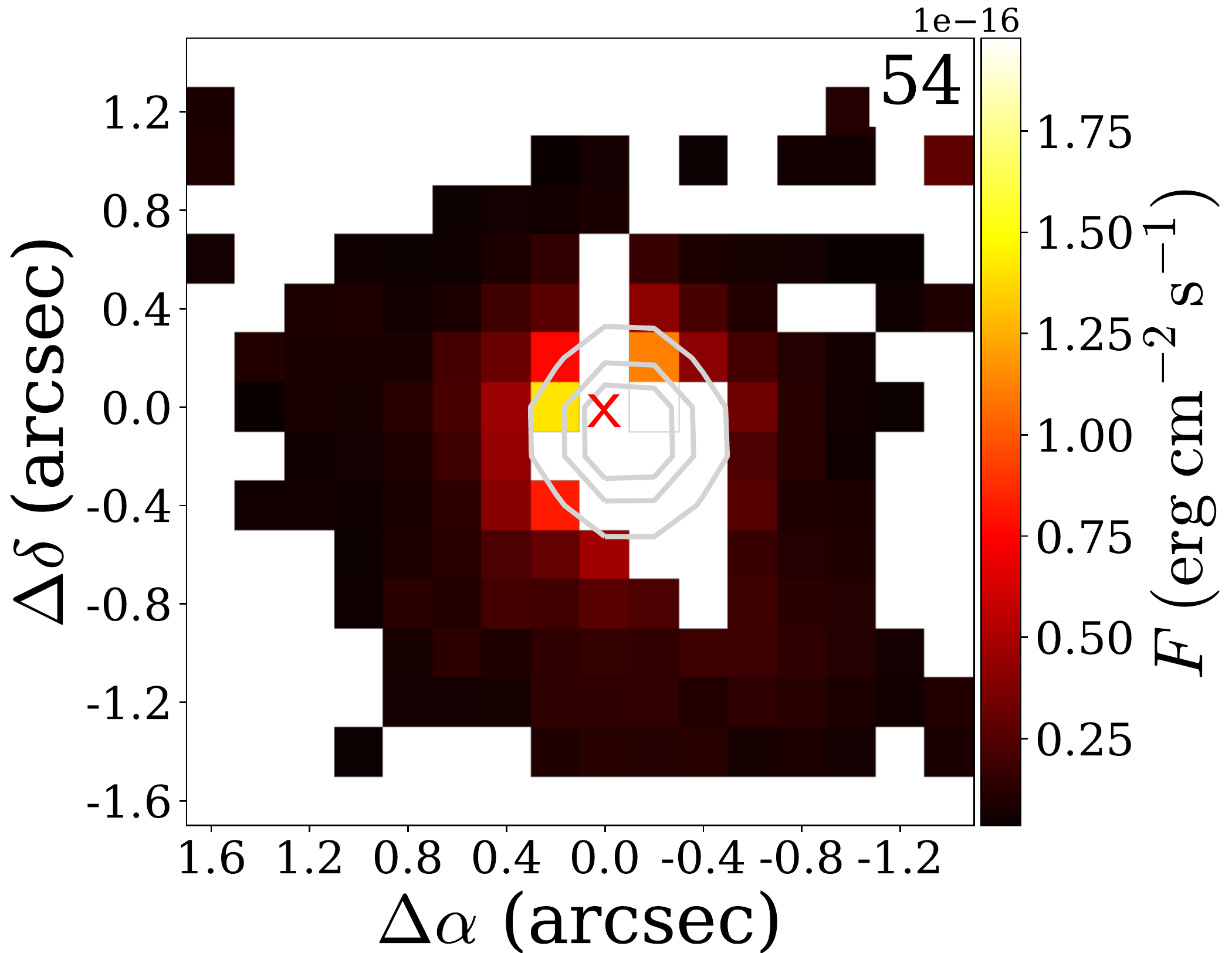}\hspace{-0.2cm}
\includegraphics[width=0.2\textwidth]{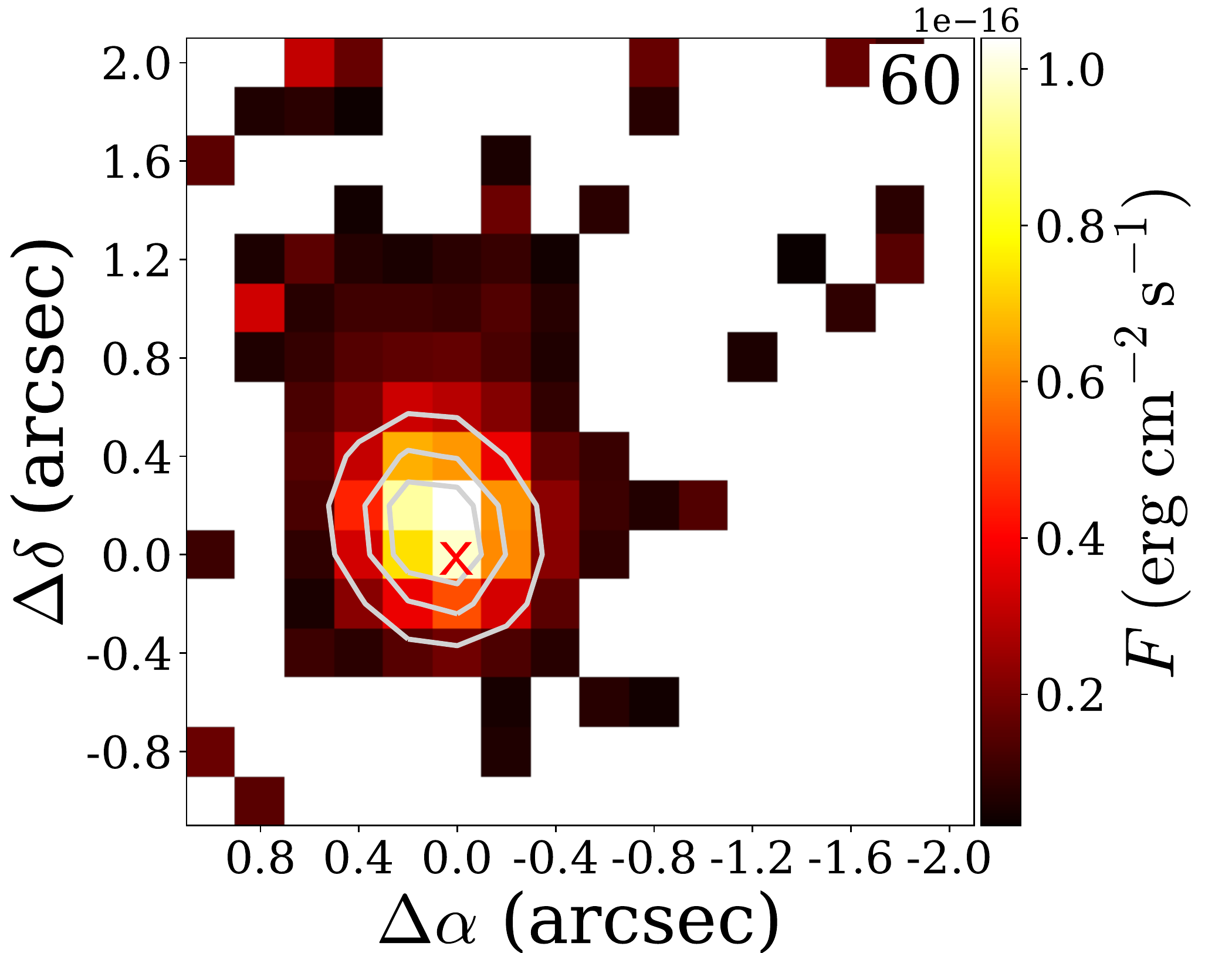}\hspace{-0.2cm}
\includegraphics[width=0.2\textwidth]{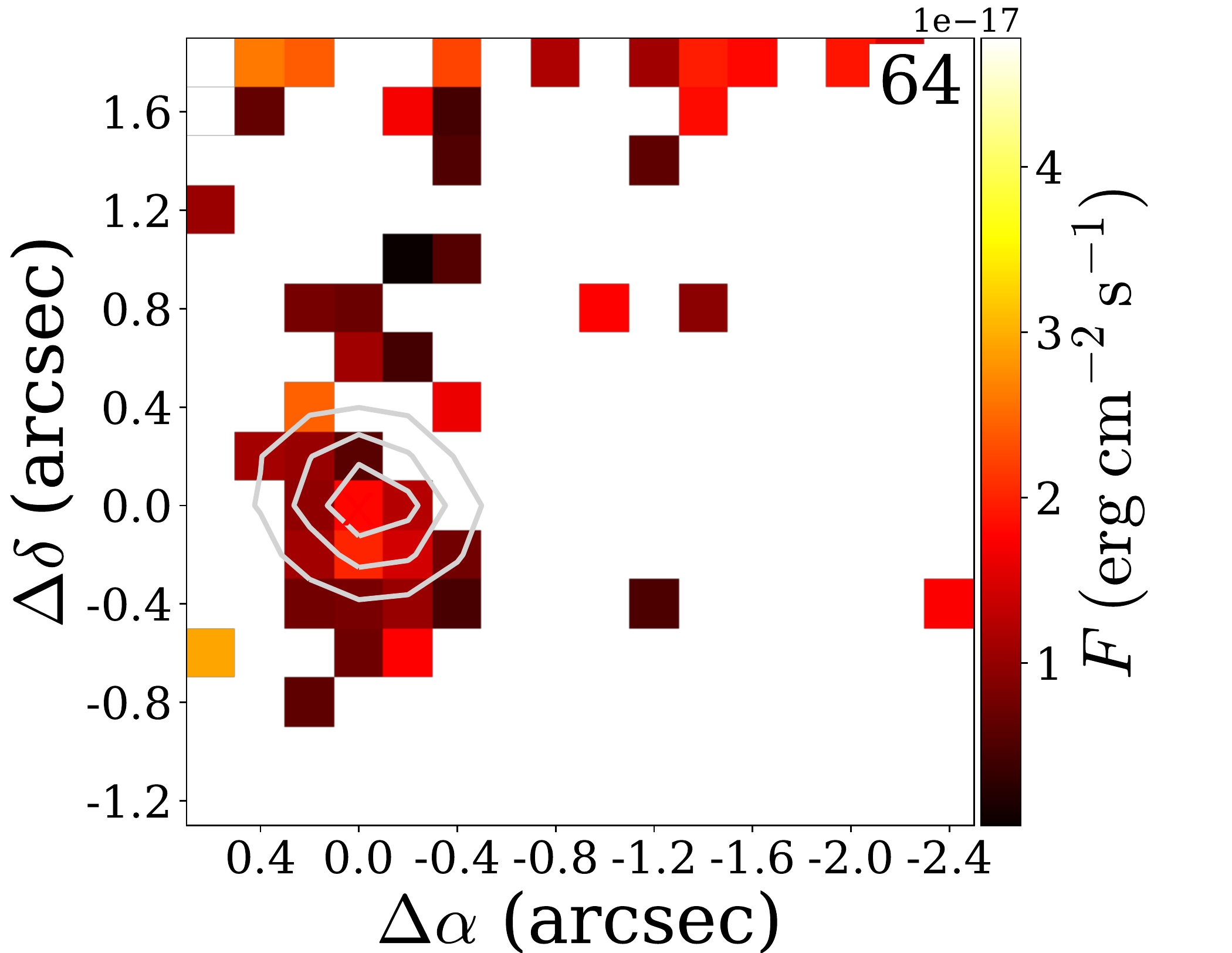}\hspace{-0.1cm}
\includegraphics[width=0.2\textwidth]{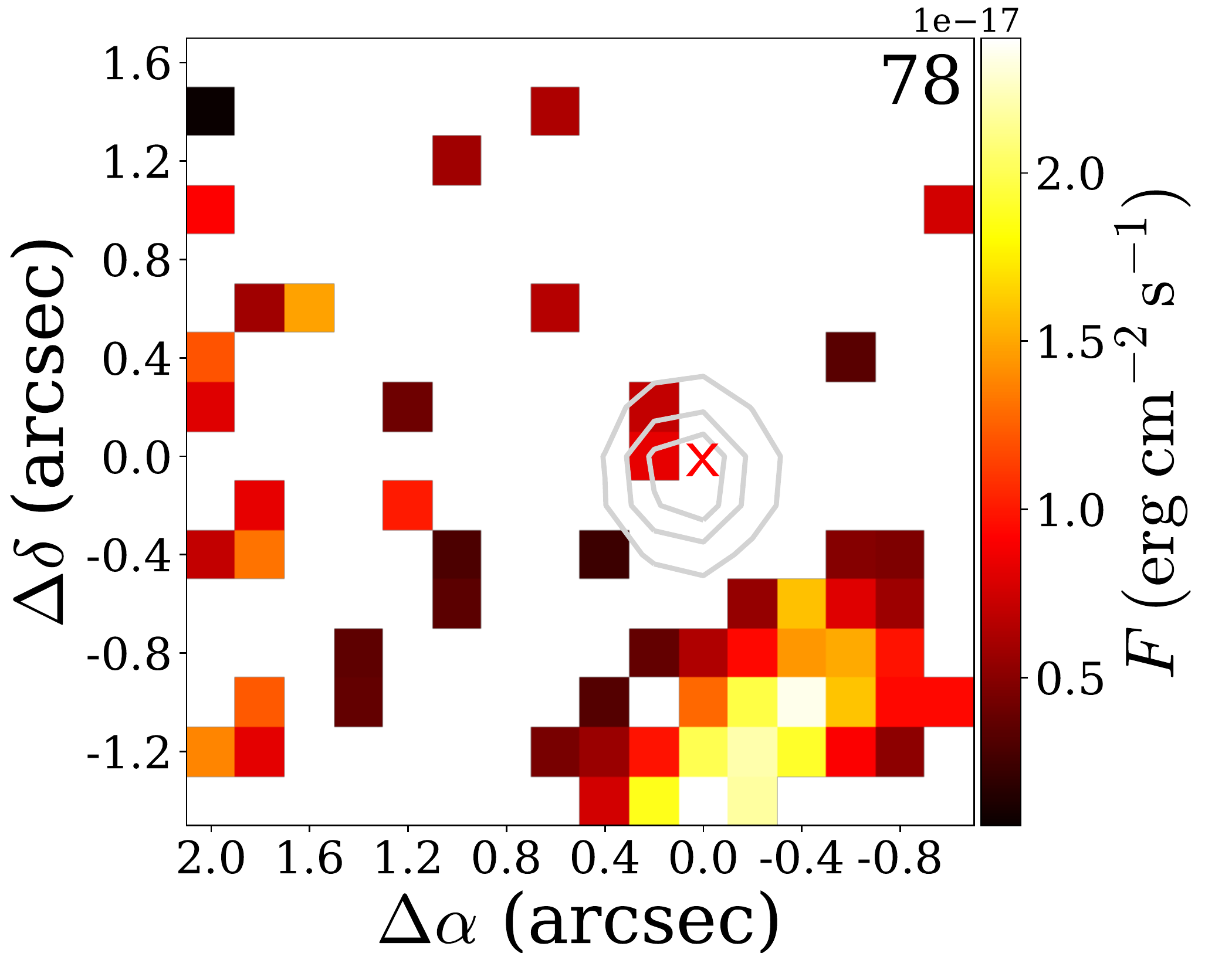}\hspace{-0.2cm}
\includegraphics[width=0.2\textwidth]{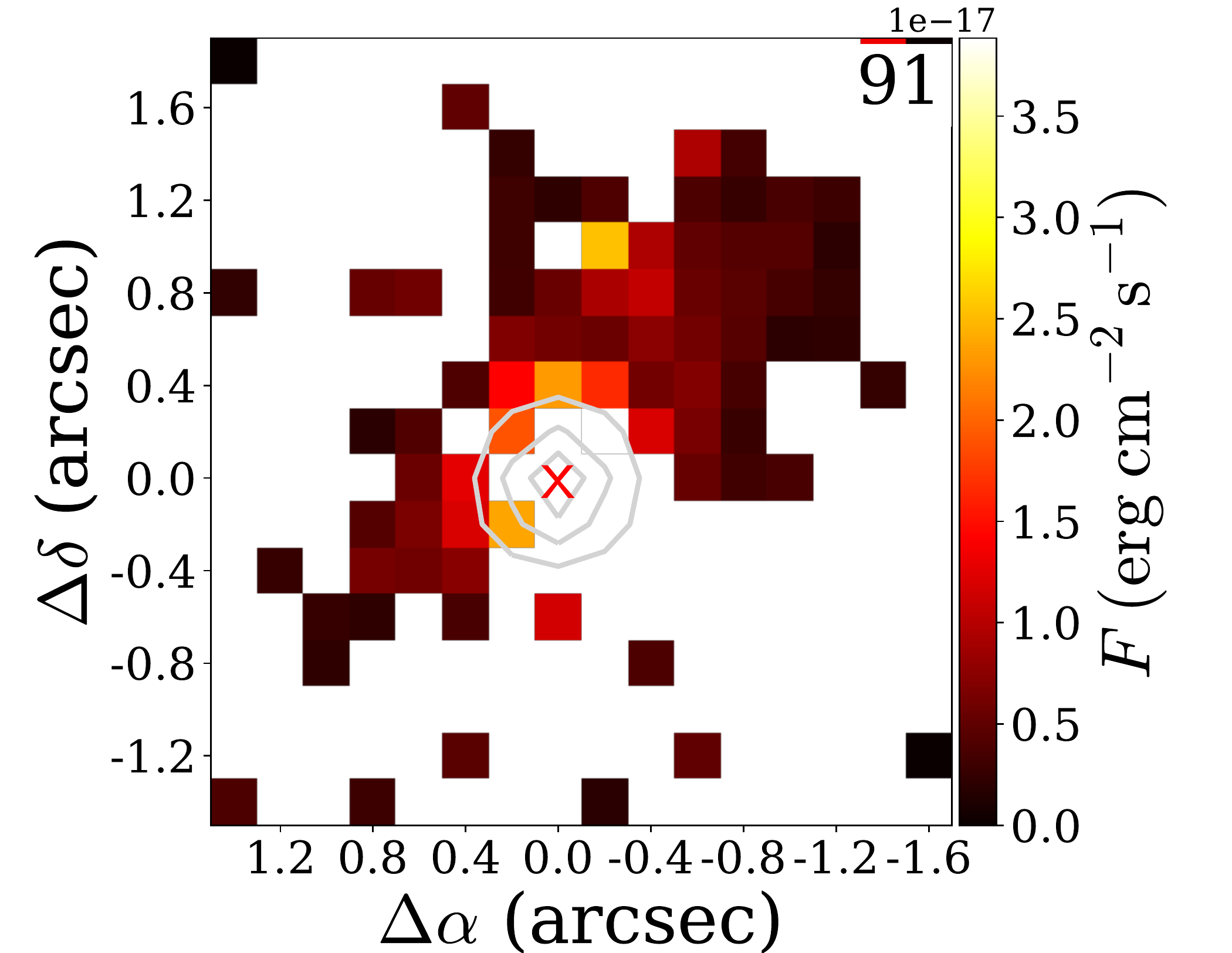}\hspace{-0.2cm}
\includegraphics[width=0.2\textwidth]{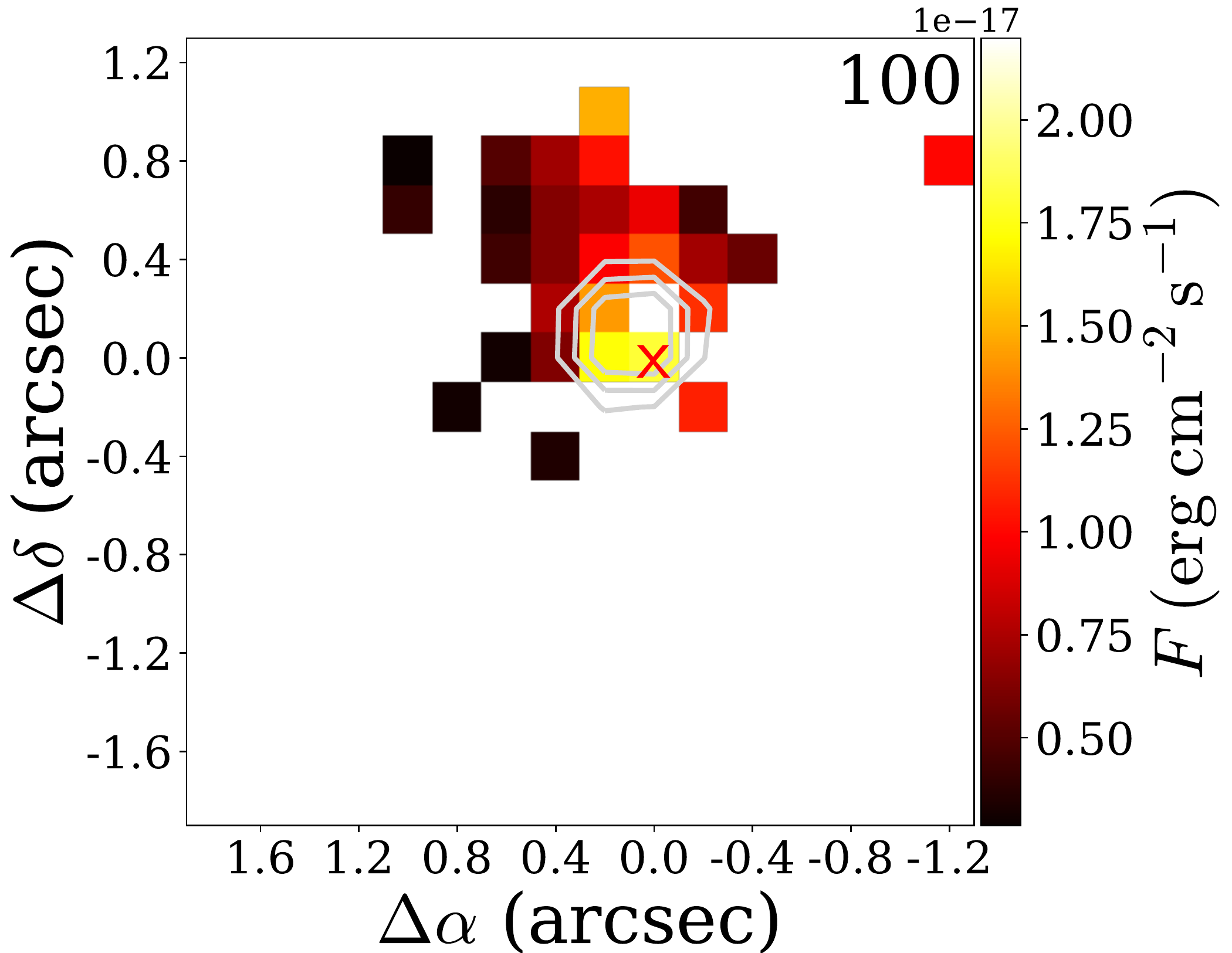}\hspace{-0.1cm}
\includegraphics[width=0.2\textwidth]{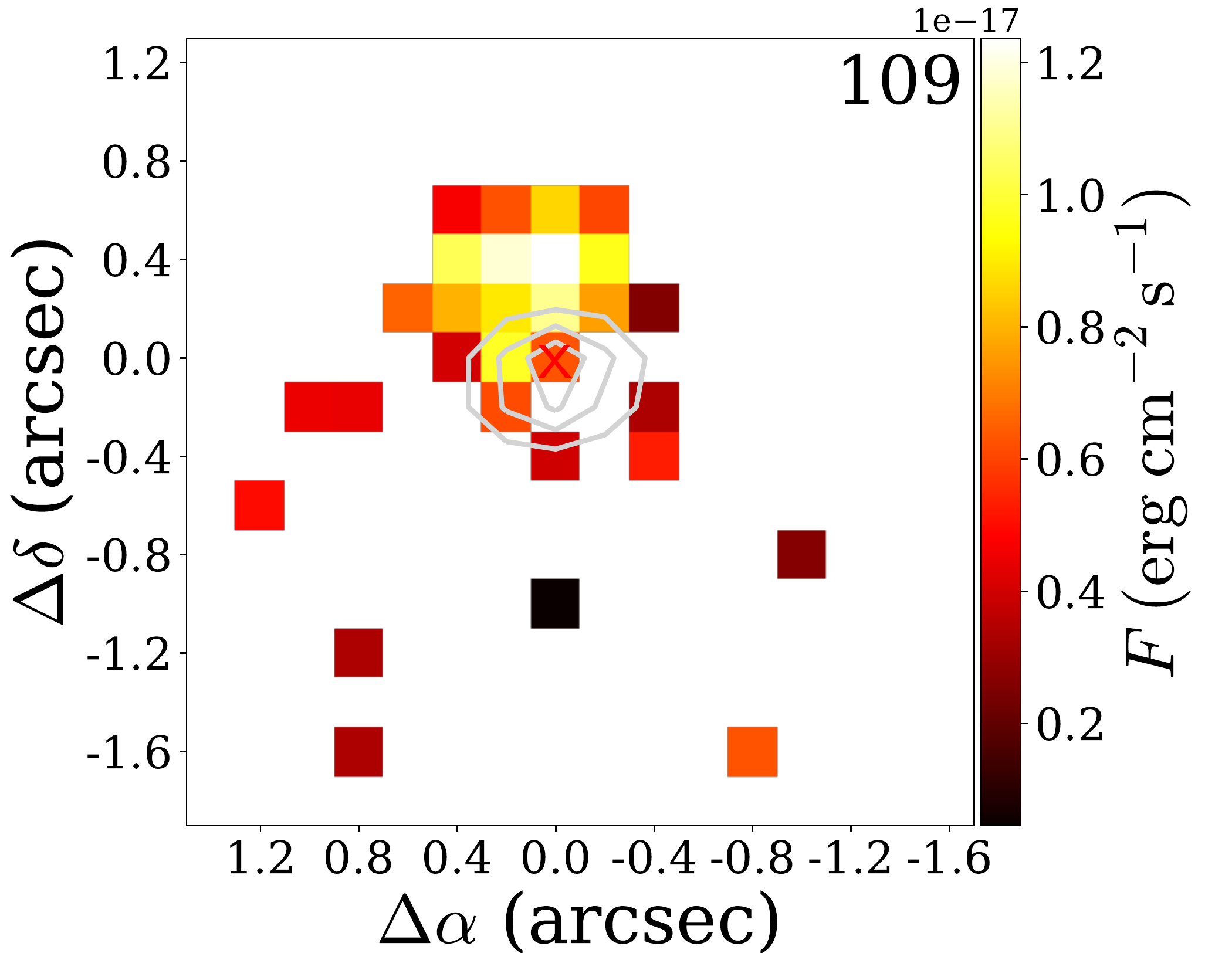}\hspace{-0.1cm}
\includegraphics[width=0.2\textwidth]{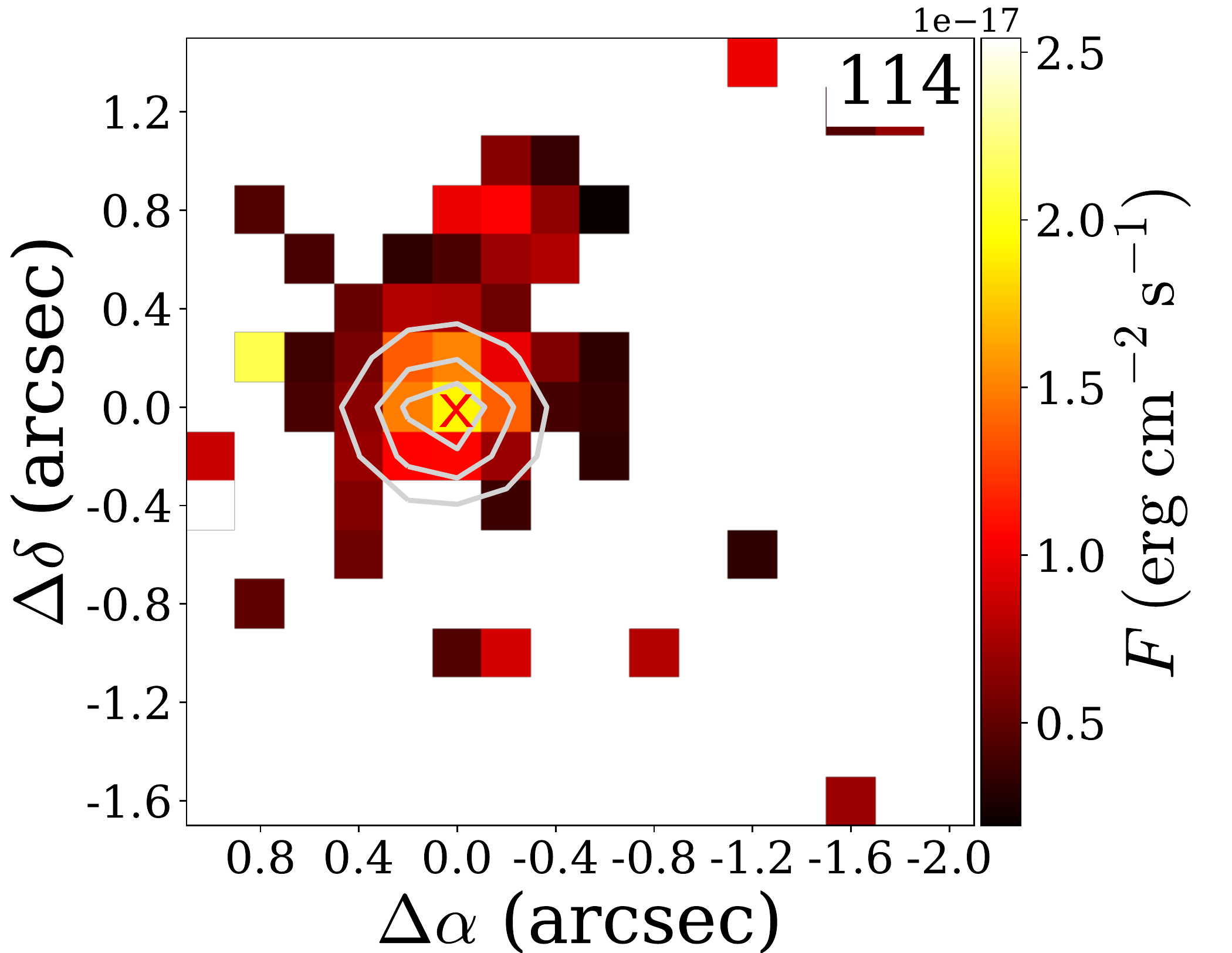}\hspace{-0.1cm}
\includegraphics[width=0.2\textwidth]{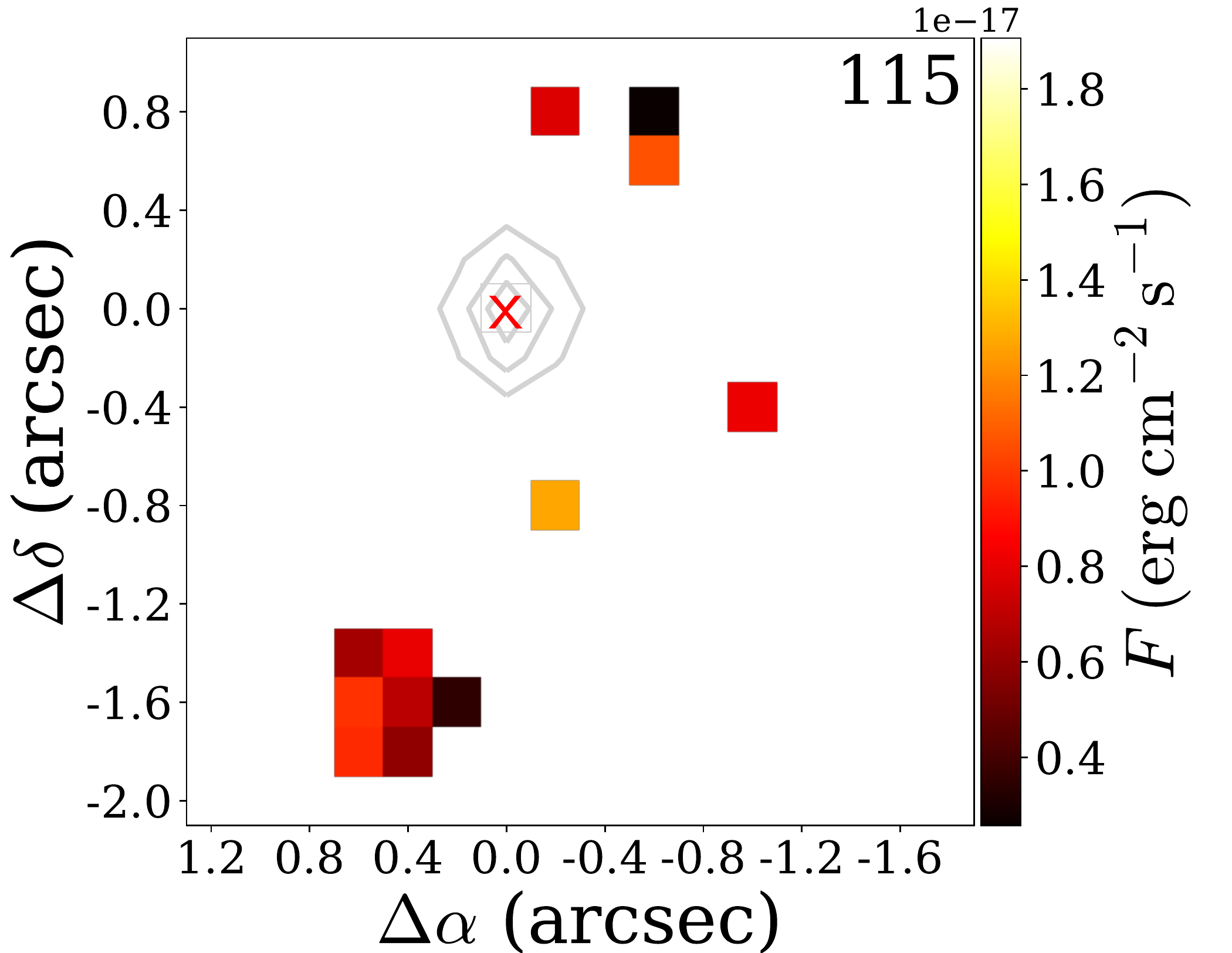}\hspace{-0.1cm}
\caption{Similar to Figure \ref{fig:emiss-2.1218}, but for the H$_2$ 1-0 S(0) line at 2.2235 $\mu$m. {Only emission above 2$\sigma$ is shown.}}
\label{fig:emiss-2.2235}
\end{figure*}

\begin{figure*}[h!]
\includegraphics[width=0.2\textwidth]{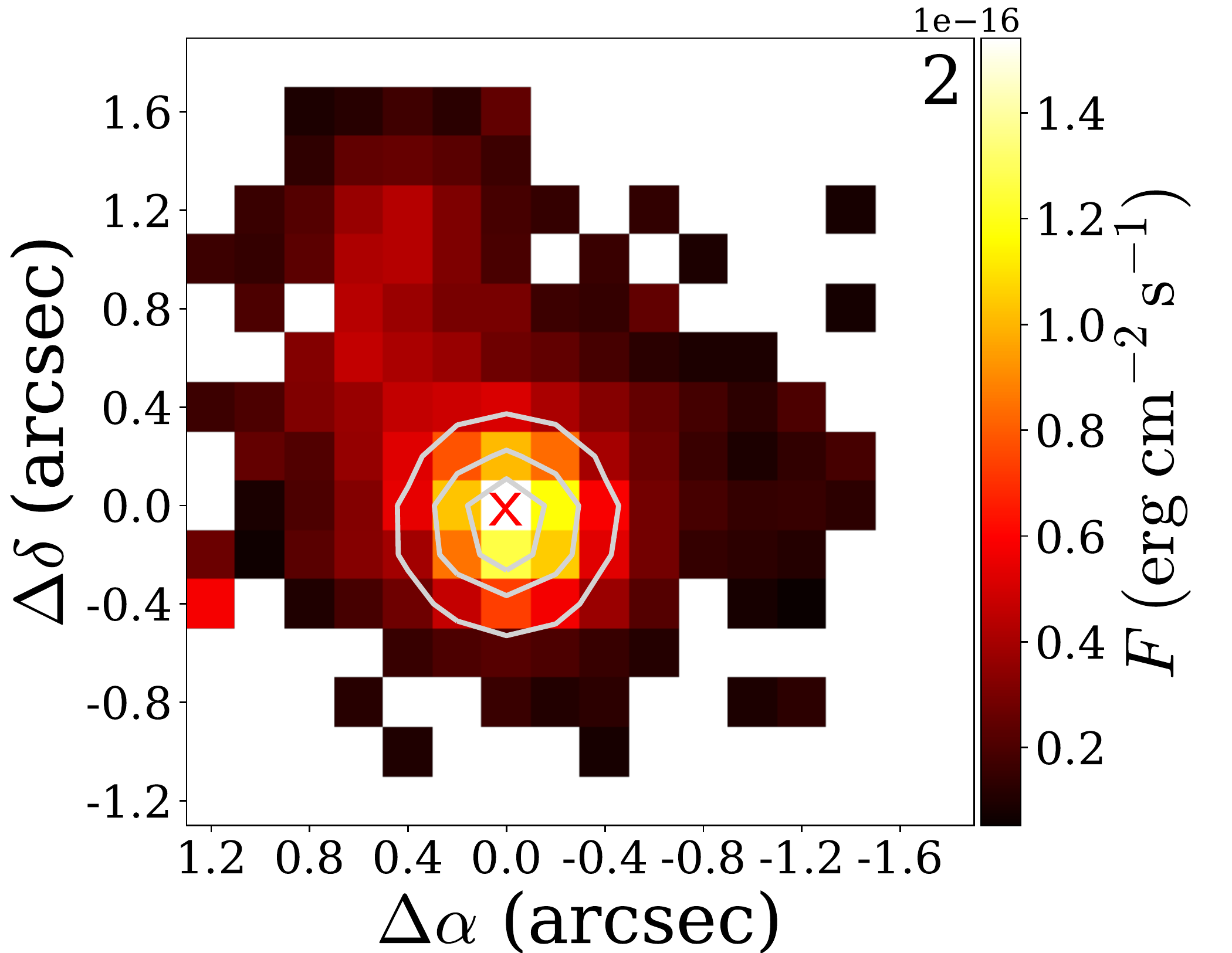}\hspace{-0.2cm}
\includegraphics[width=0.2\textwidth]{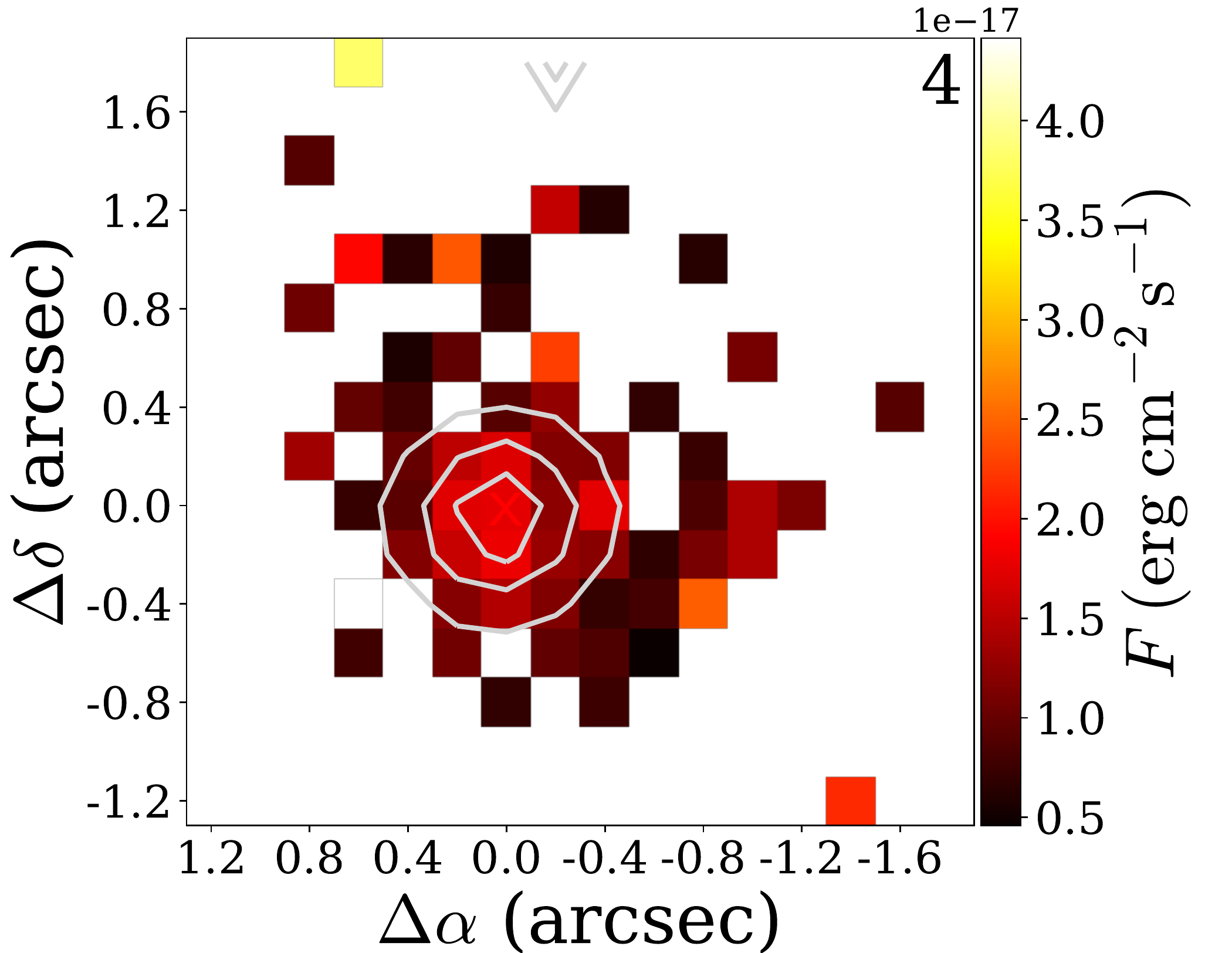}\hspace{-0.2cm}
\includegraphics[width=0.2\textwidth]{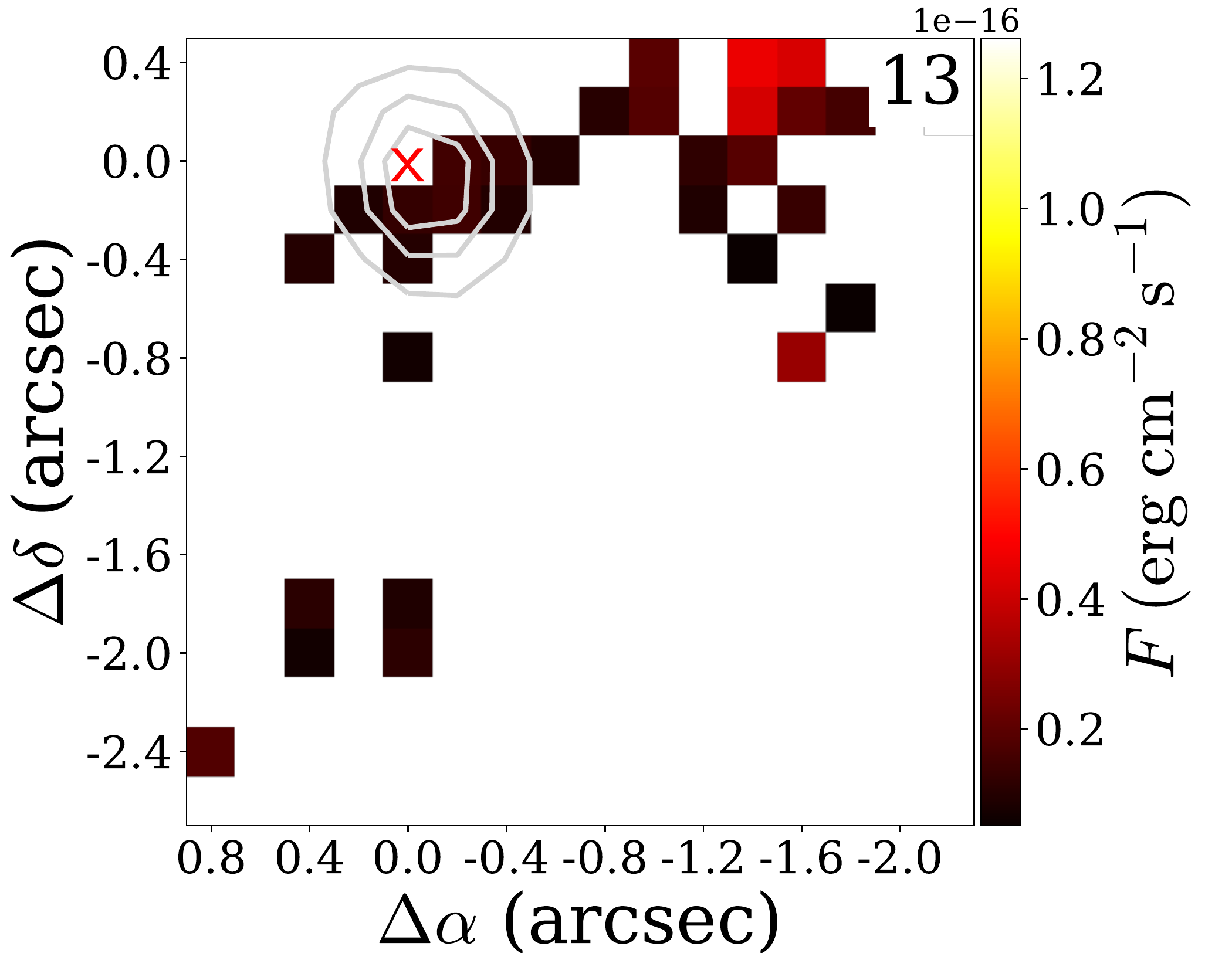}\hspace{-0.2cm}
\includegraphics[width=0.2\textwidth]{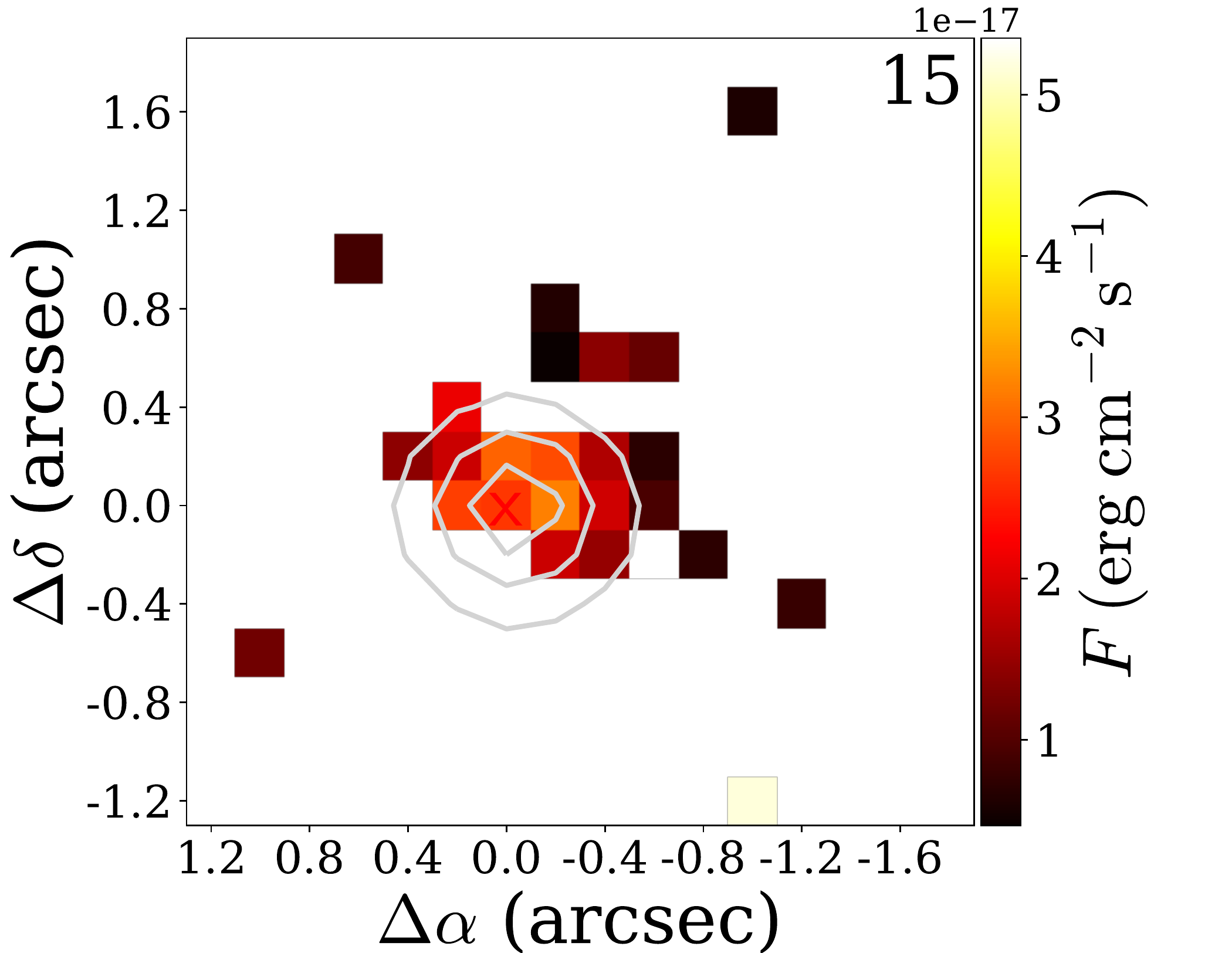}\hspace{-0.2cm}
\includegraphics[width=0.2\textwidth]{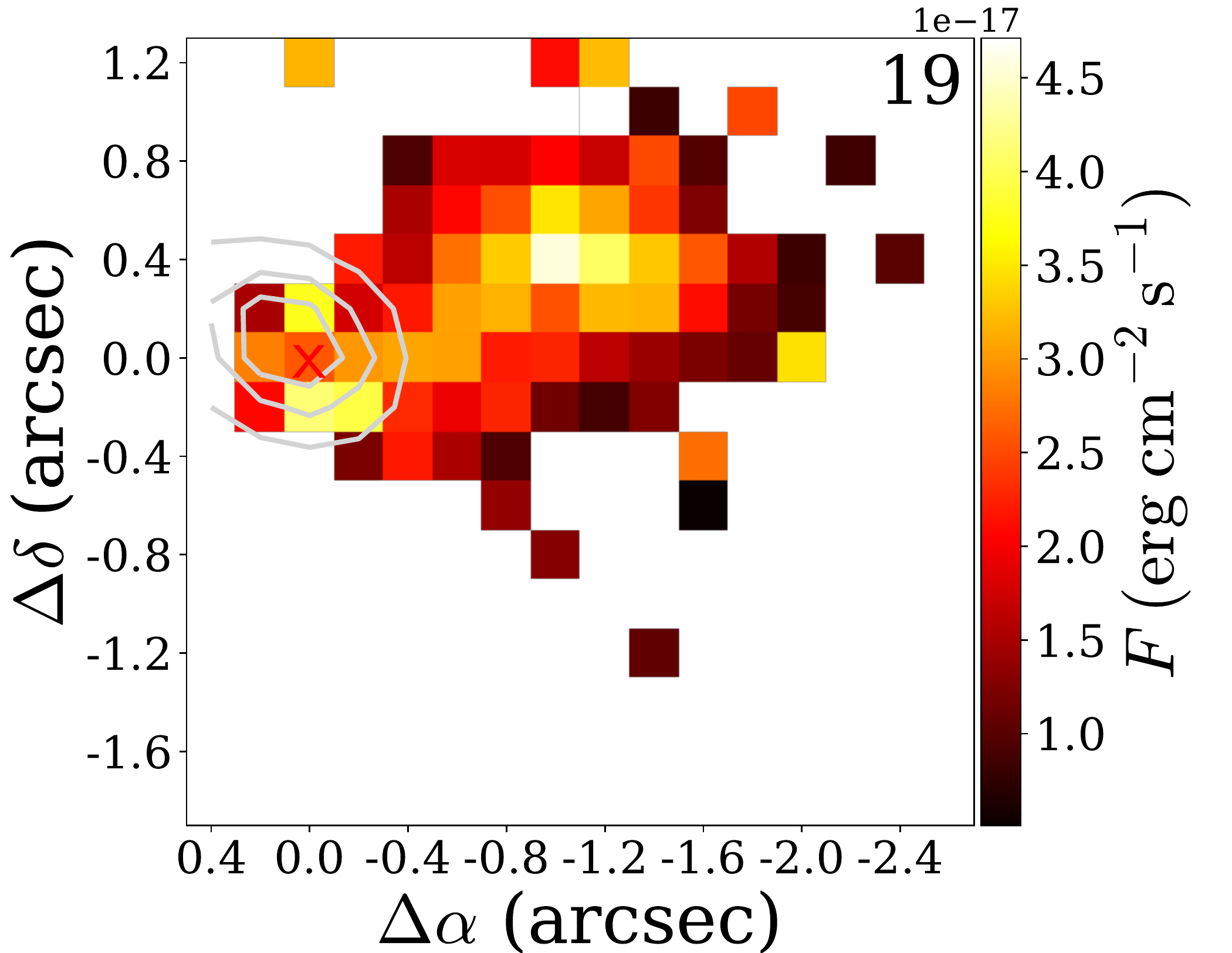}\hspace{-0.2cm}
\includegraphics[width=0.2\textwidth]{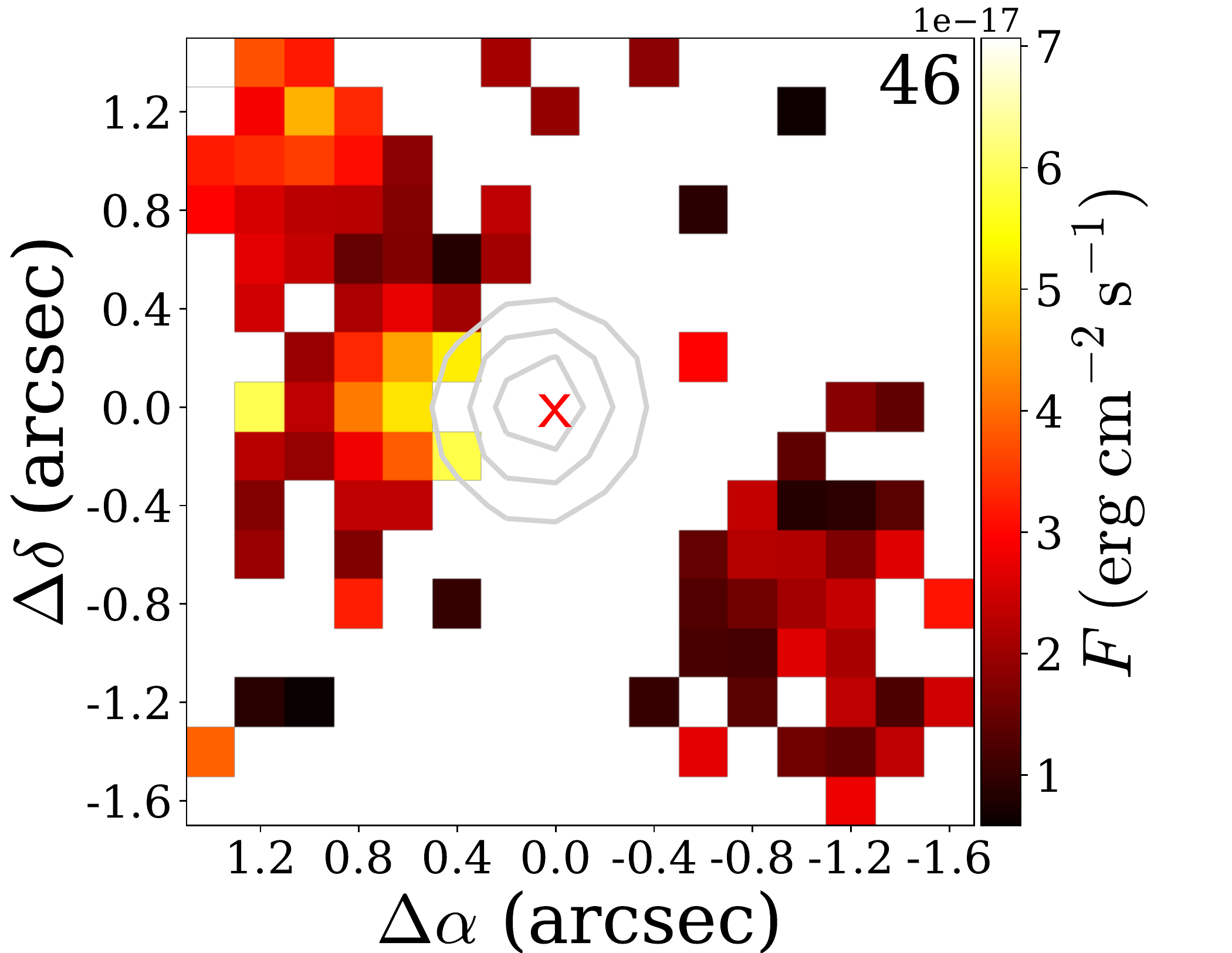}\hspace{-0.2cm}
\includegraphics[width=0.2\textwidth]{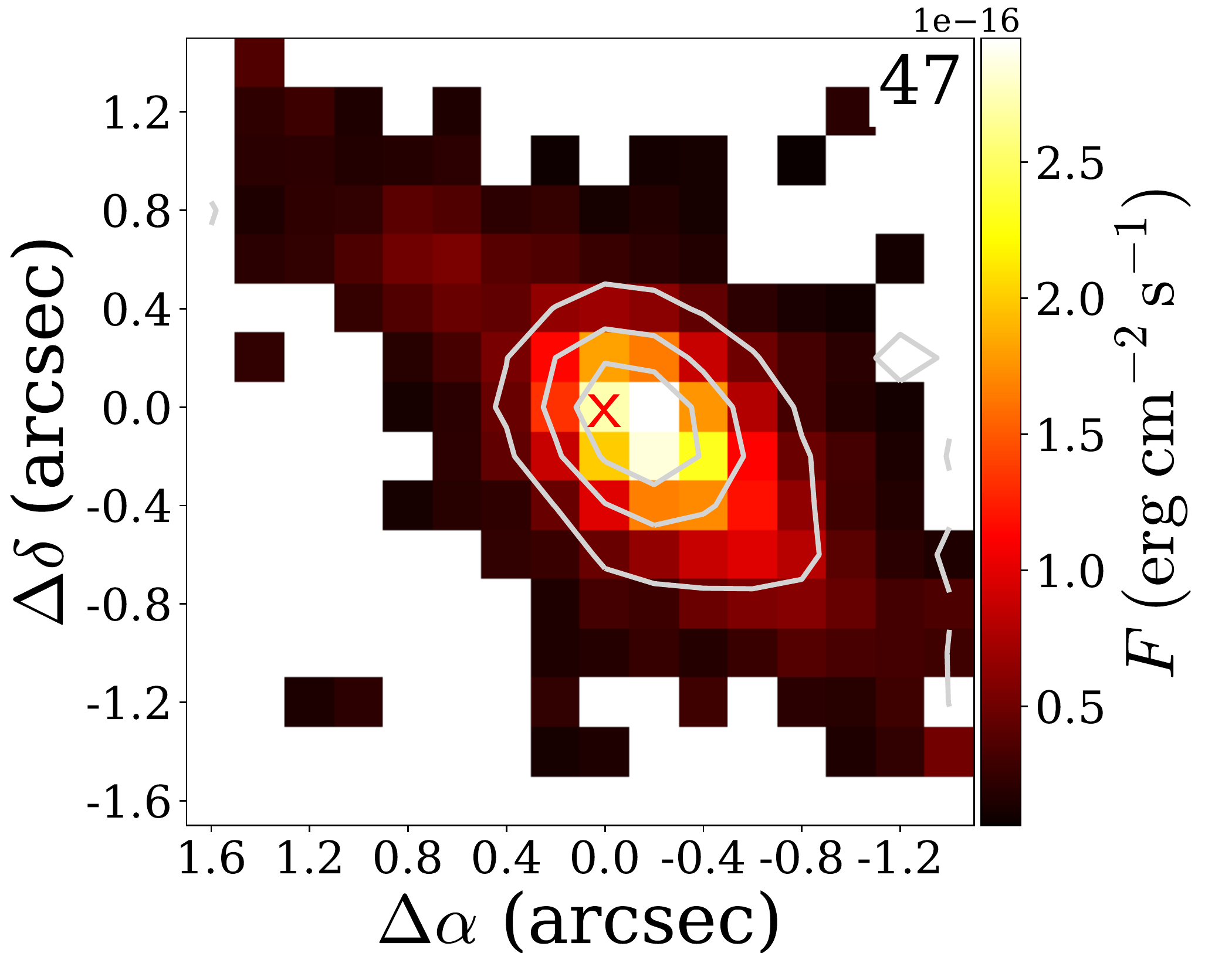}\hspace{-0.2cm}
\includegraphics[width=0.2\textwidth]{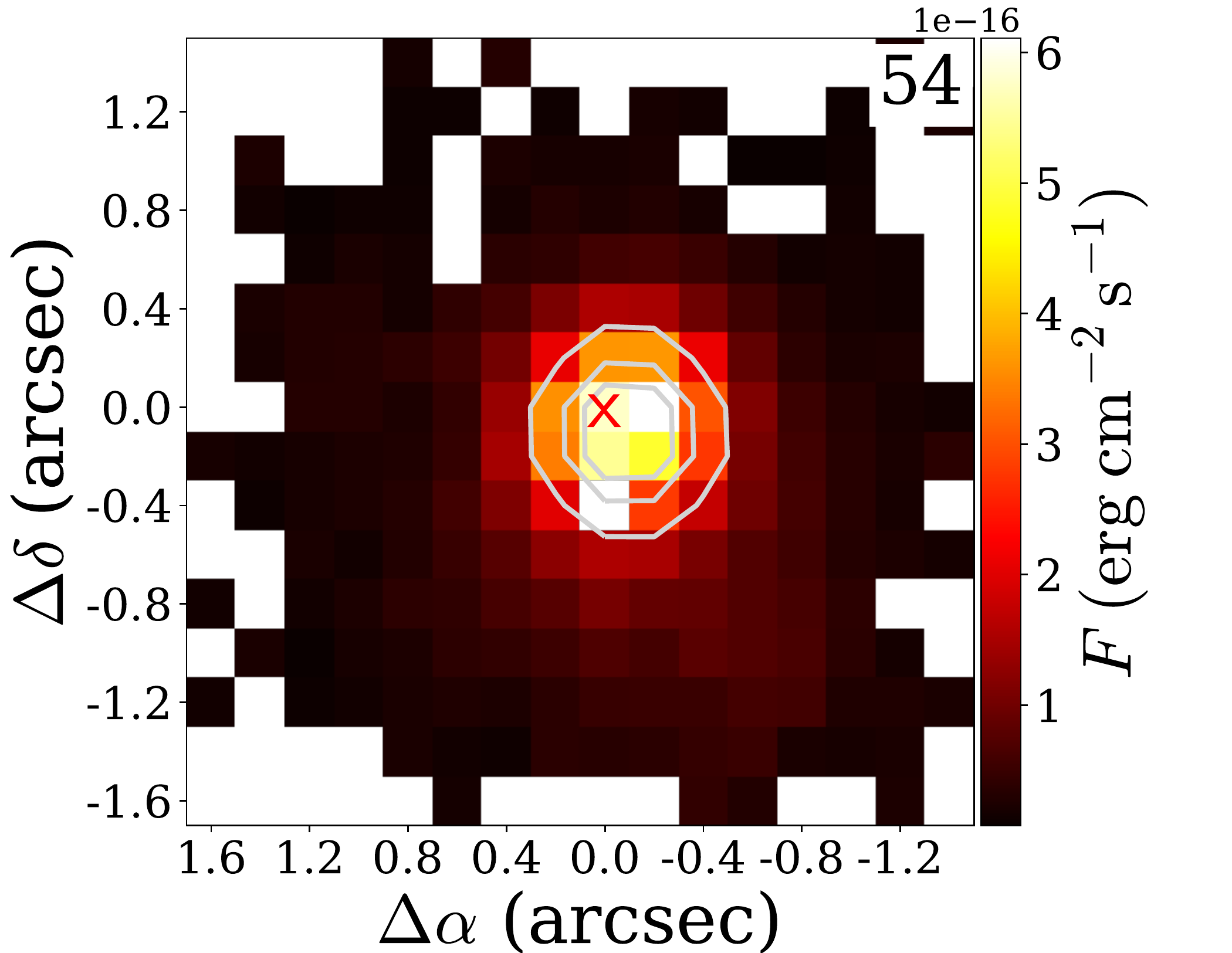}\hspace{-0.2cm}
\includegraphics[width=0.2\textwidth]{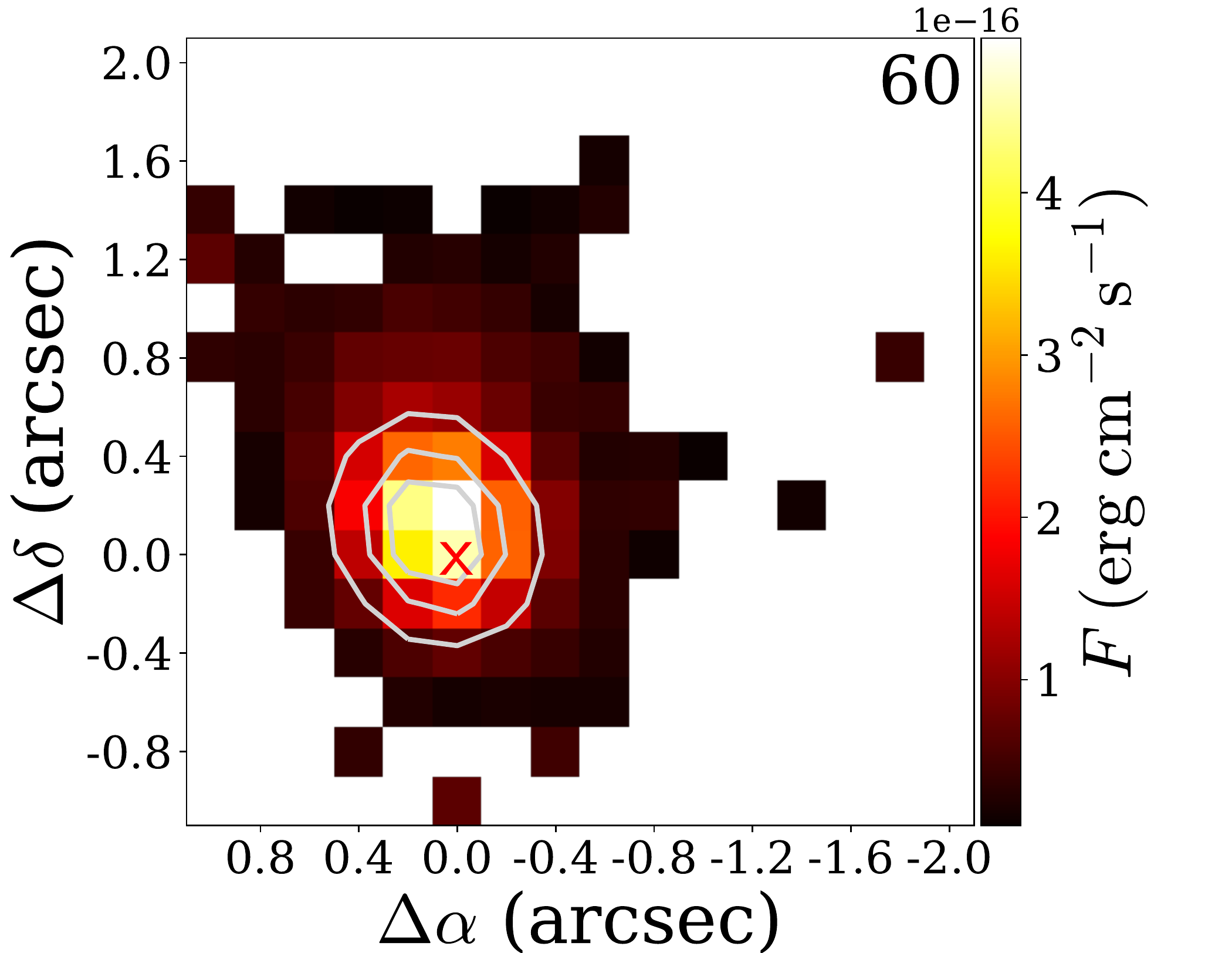}\hspace{-0.2cm}
\includegraphics[width=0.2\textwidth]{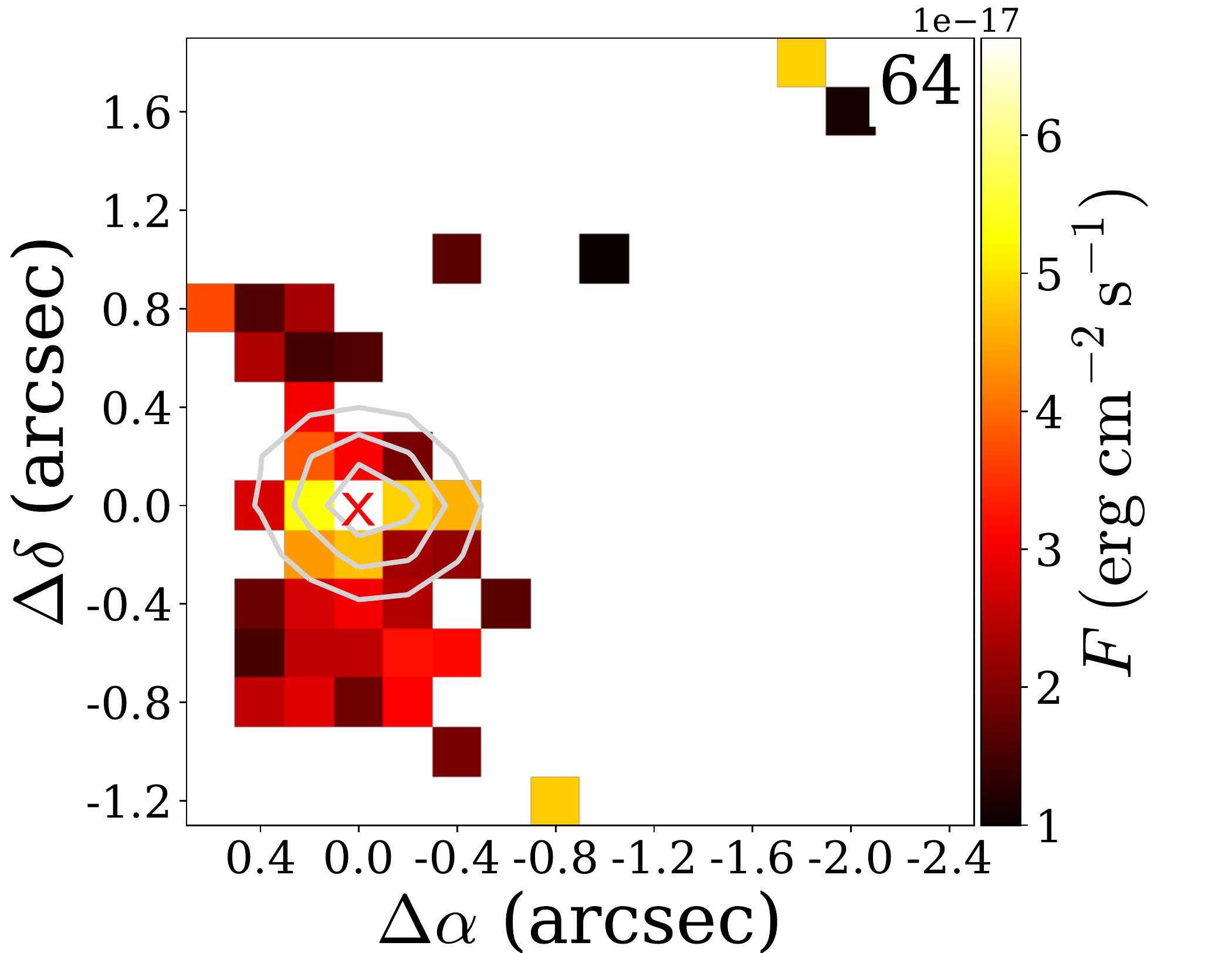}\hspace{-0.2cm}
\includegraphics[width=0.2\textwidth]{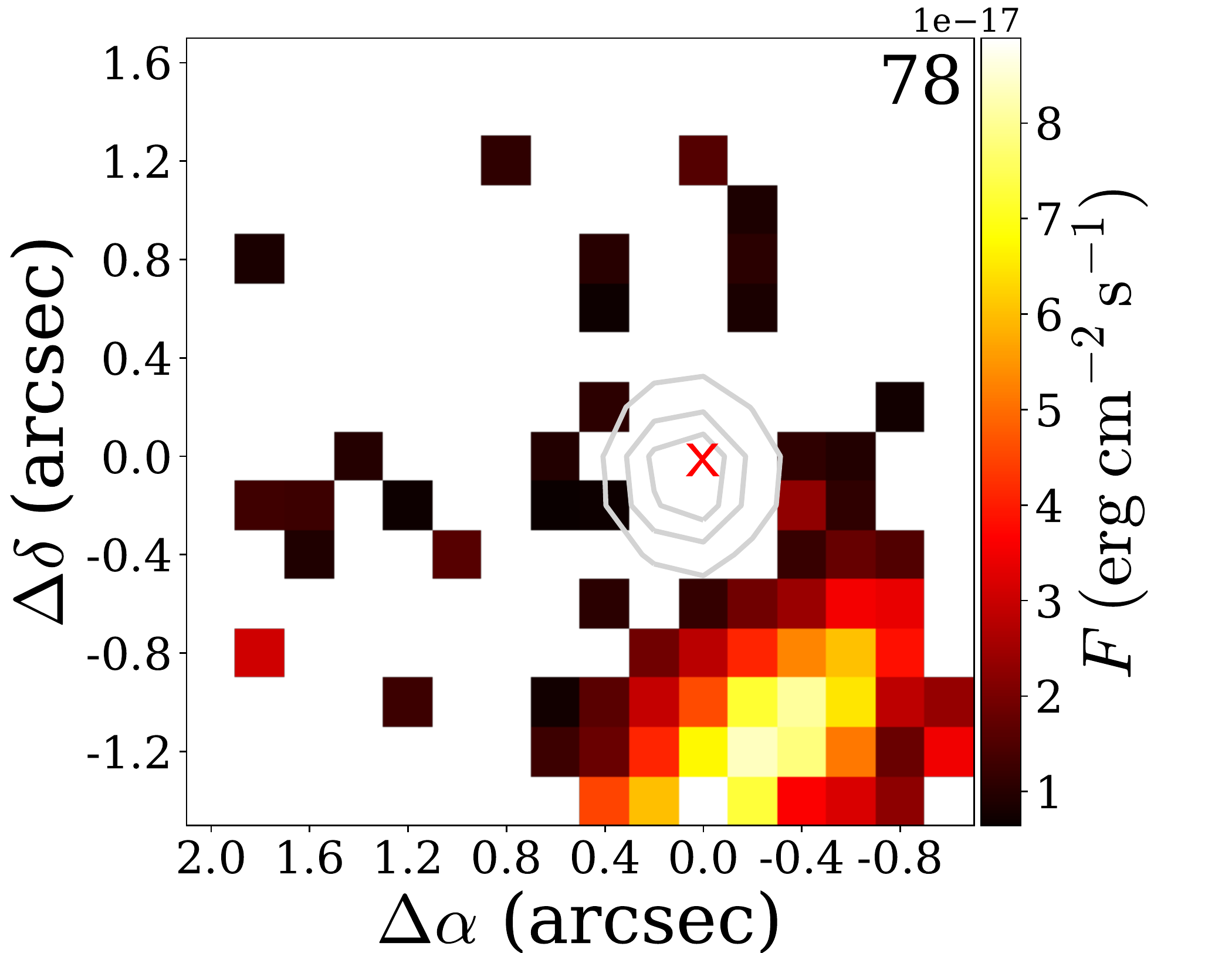}\hspace{-0.1cm}
\includegraphics[width=0.2\textwidth]{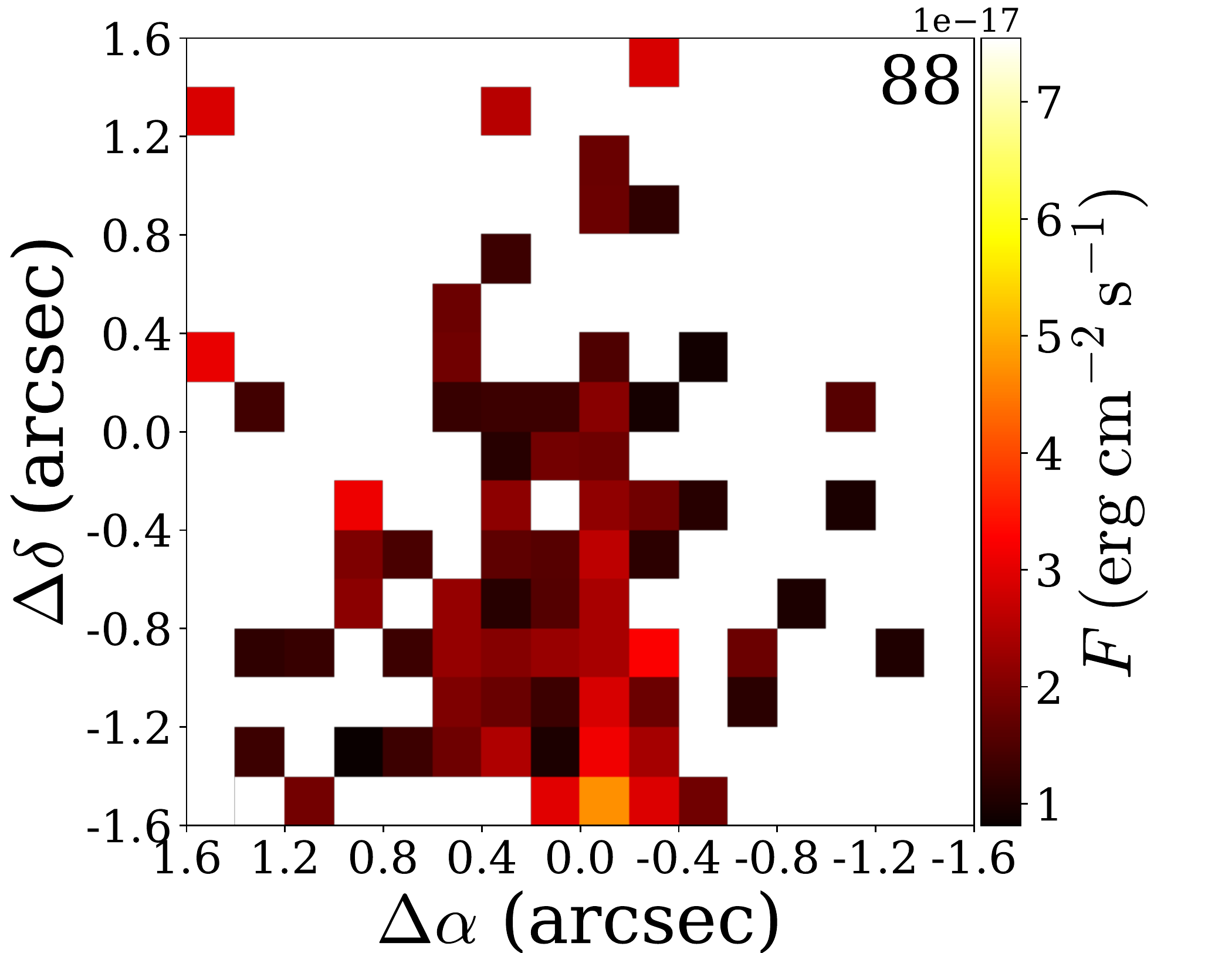}\hspace{-0.2cm}
\includegraphics[width=0.2\textwidth]{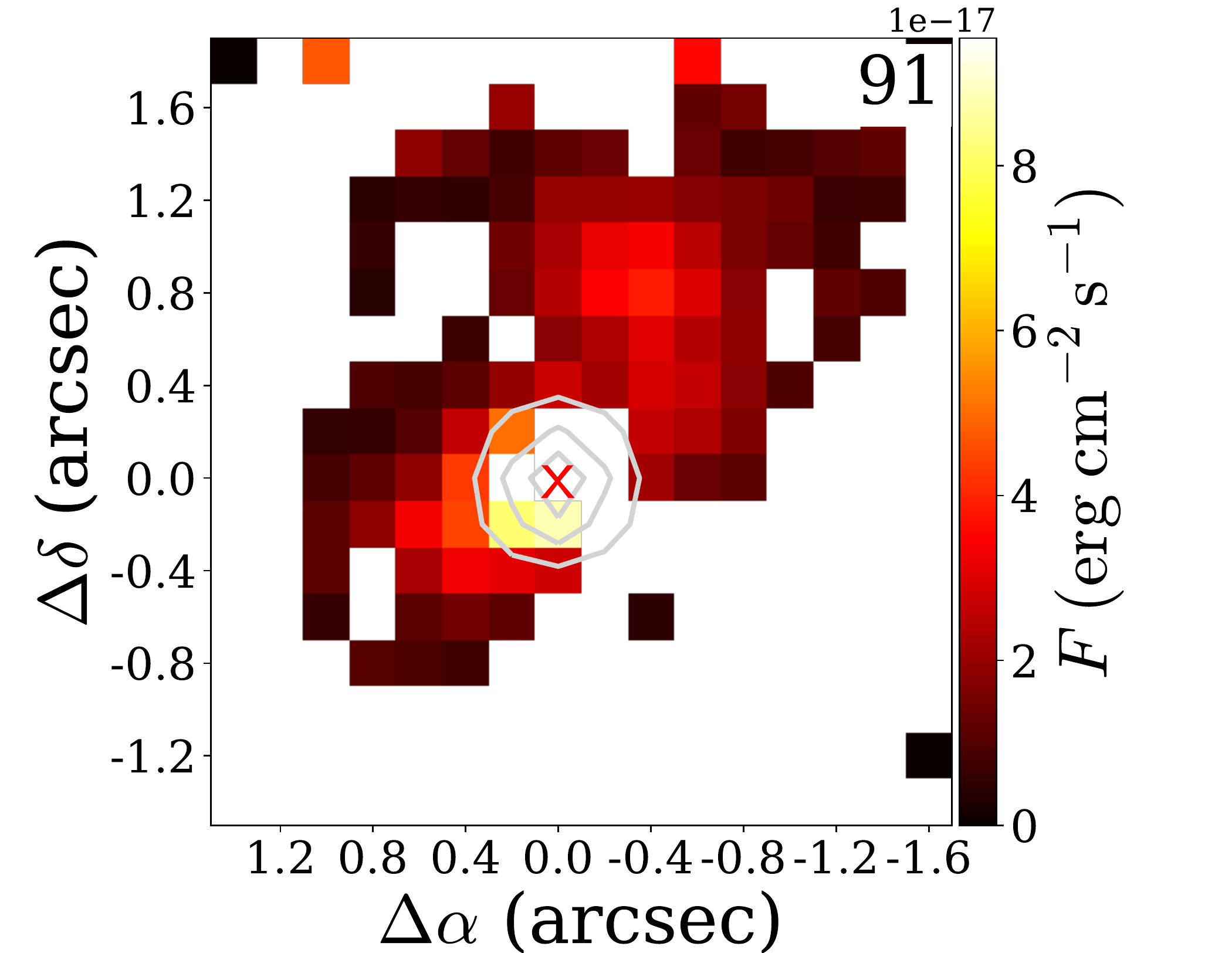}\hspace{-0.2cm}
\includegraphics[width=0.2\textwidth]{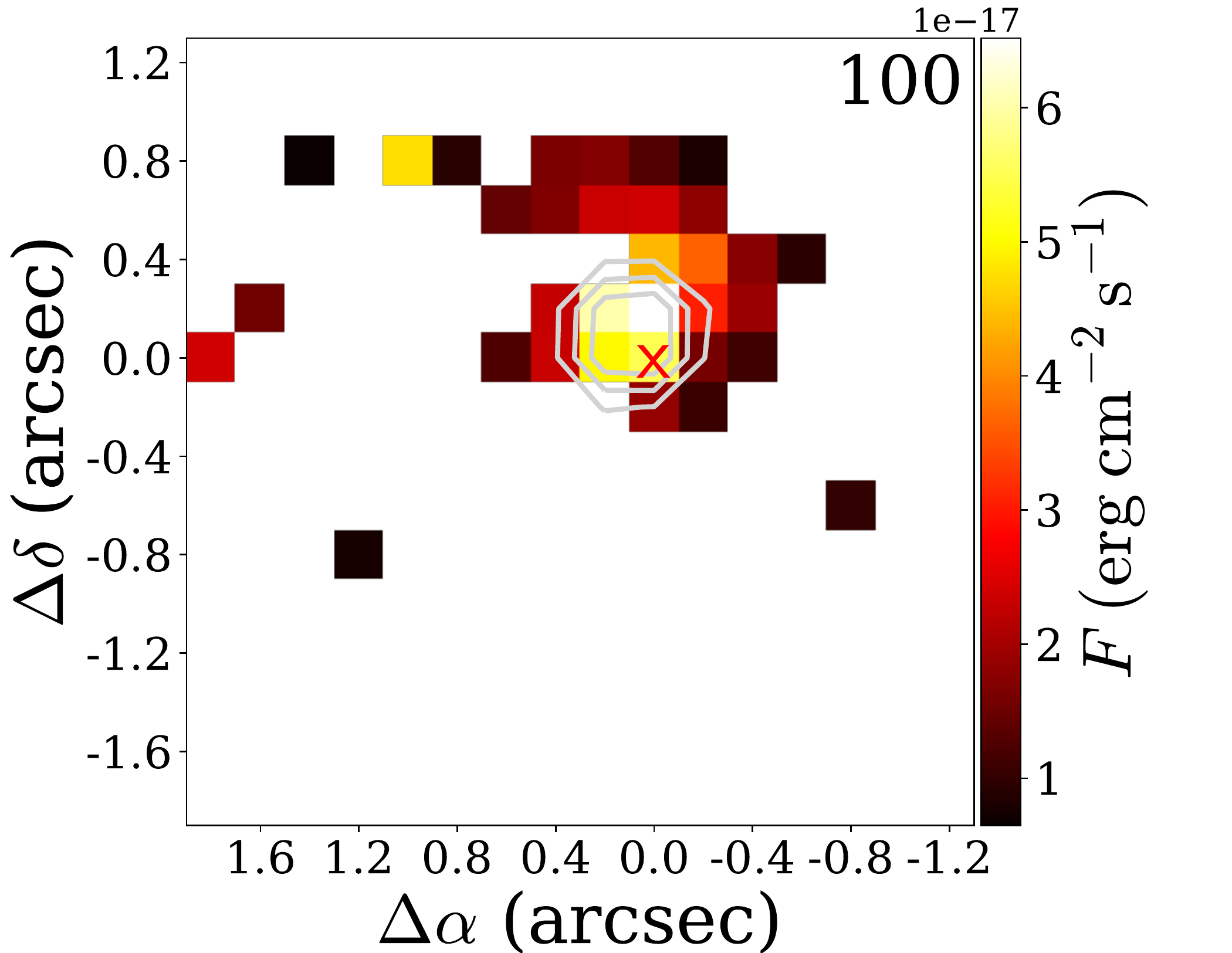}\hspace{-0.2cm}
\includegraphics[width=0.2\textwidth]{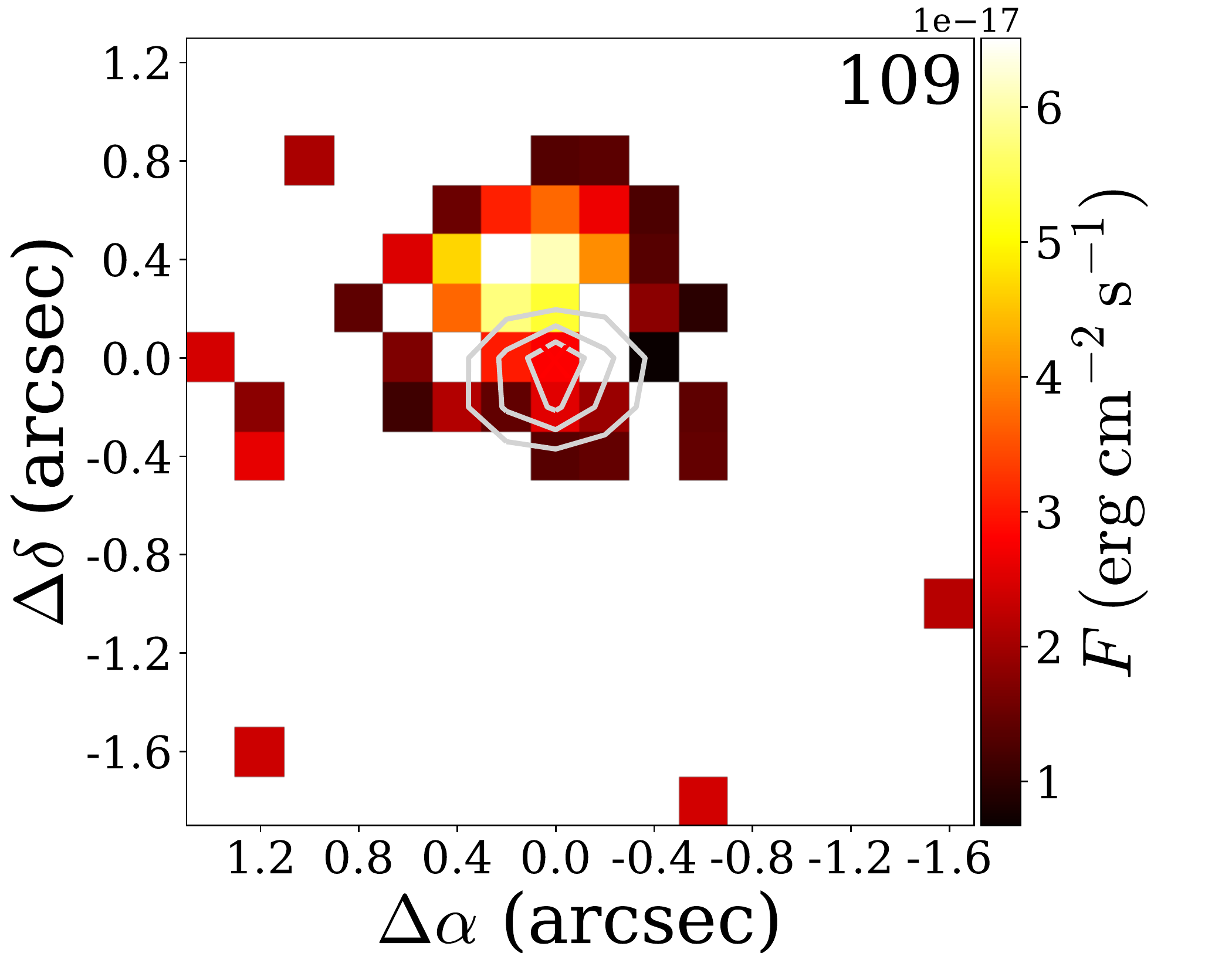}\hspace{-0.2cm}
\includegraphics[width=0.2\textwidth]{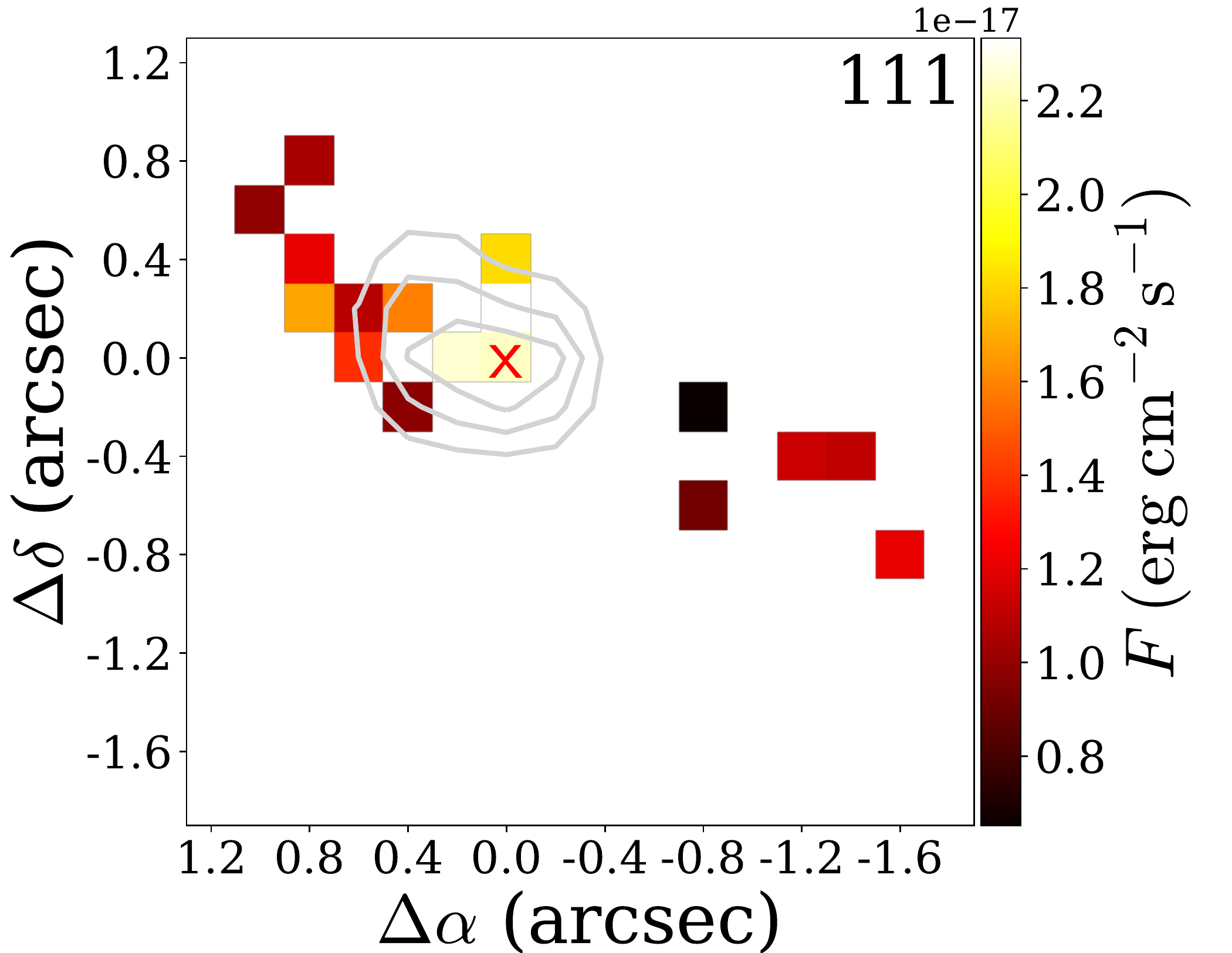}\hspace{-0.1cm}
\includegraphics[width=0.2\textwidth]{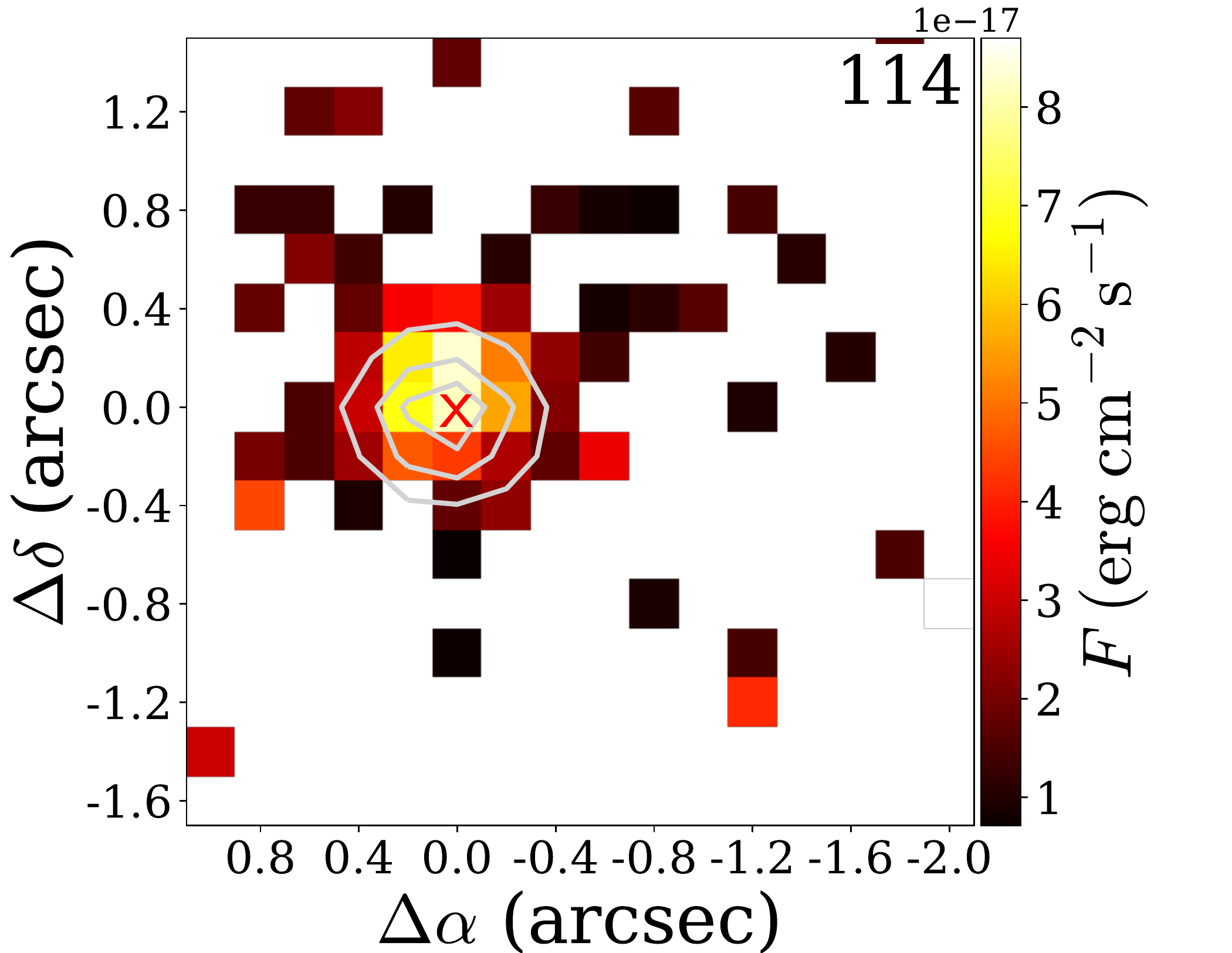}\hspace{-0.2cm}
\includegraphics[width=0.2\textwidth]{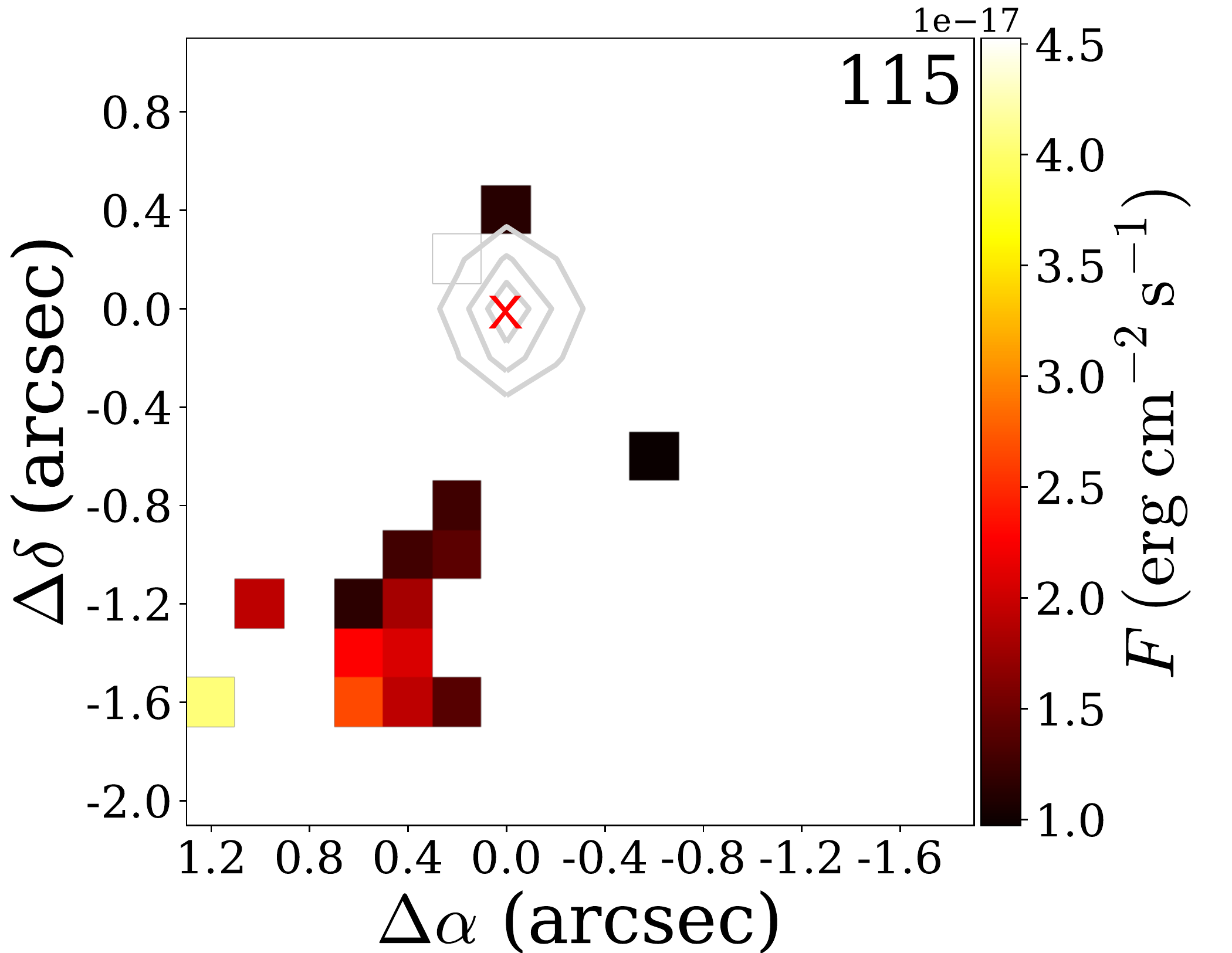}\hspace{-0.1cm}
\caption{Similar to Figure \ref{fig:emiss-2.1218}, but for the H$_2$ 1-0 Q(1) line at 2.4066 $\mu$m. {  Only emission above 2$\sigma$ is shown.}}
\label{fig:emiss-2.4066}
\end{figure*}

\begin{figure*}[h!]
\includegraphics[width=0.2\textwidth]{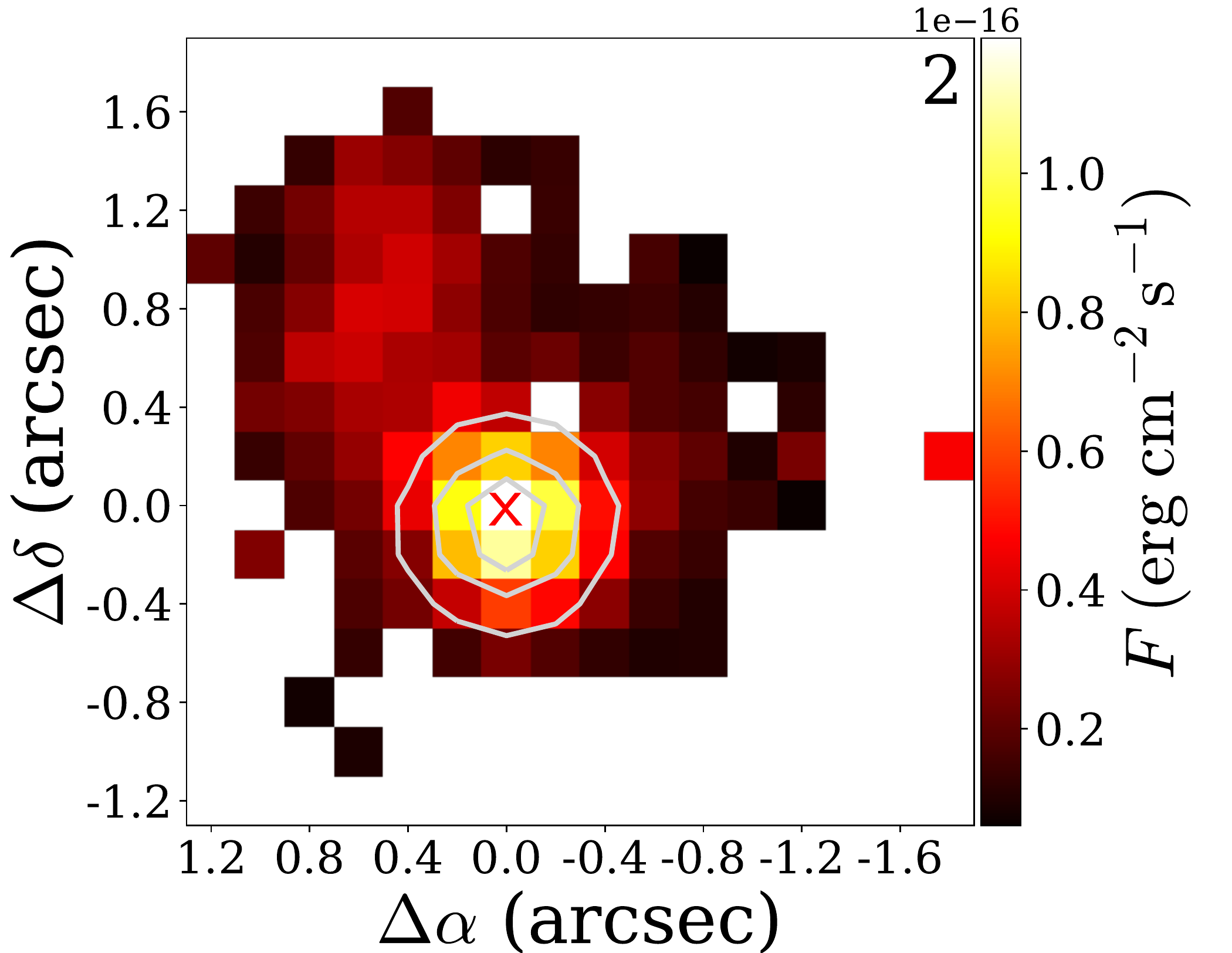}\hspace{-0.2cm}
\includegraphics[width=0.2\textwidth]{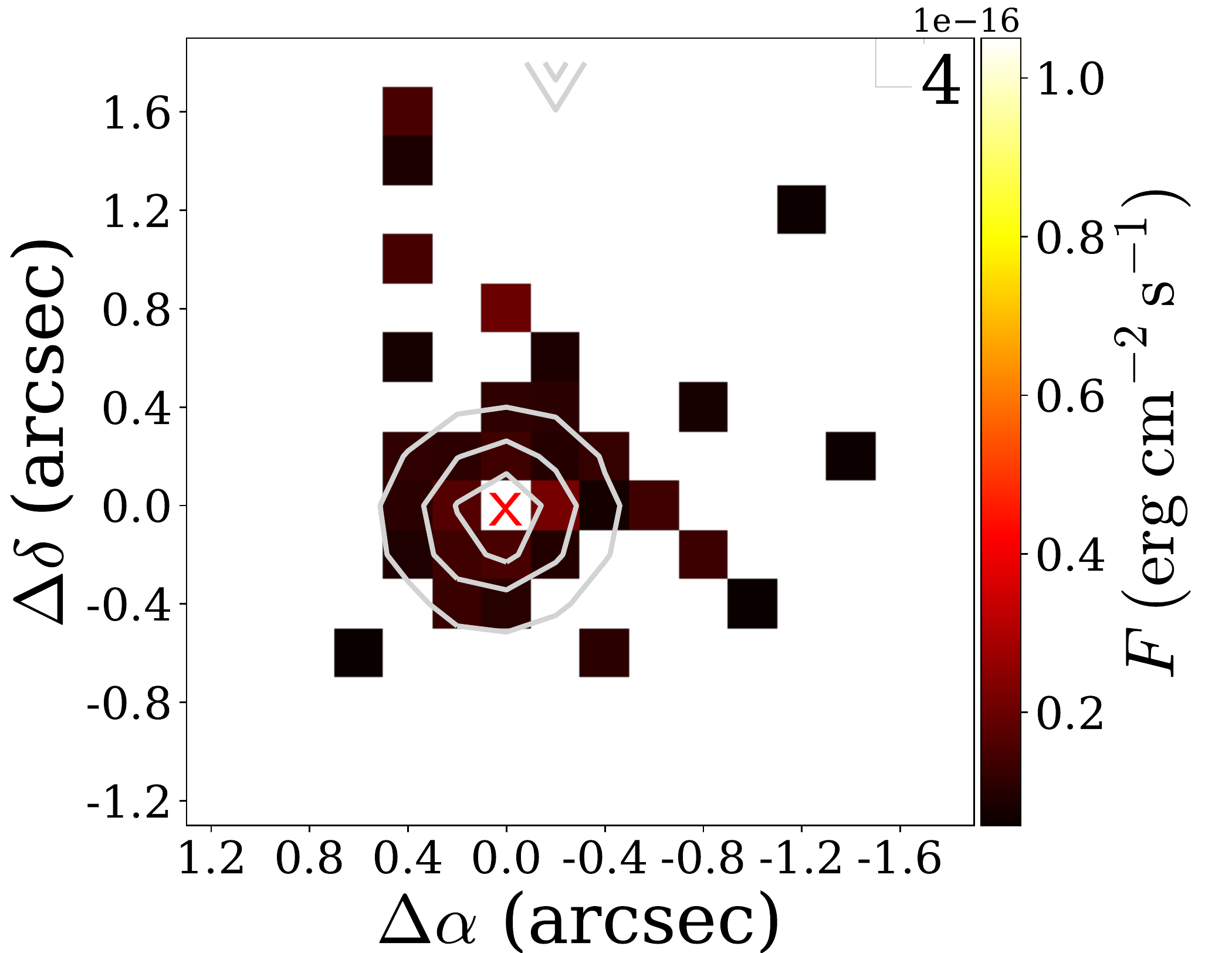}\hspace{-0.2cm}
\includegraphics[width=0.2\textwidth]{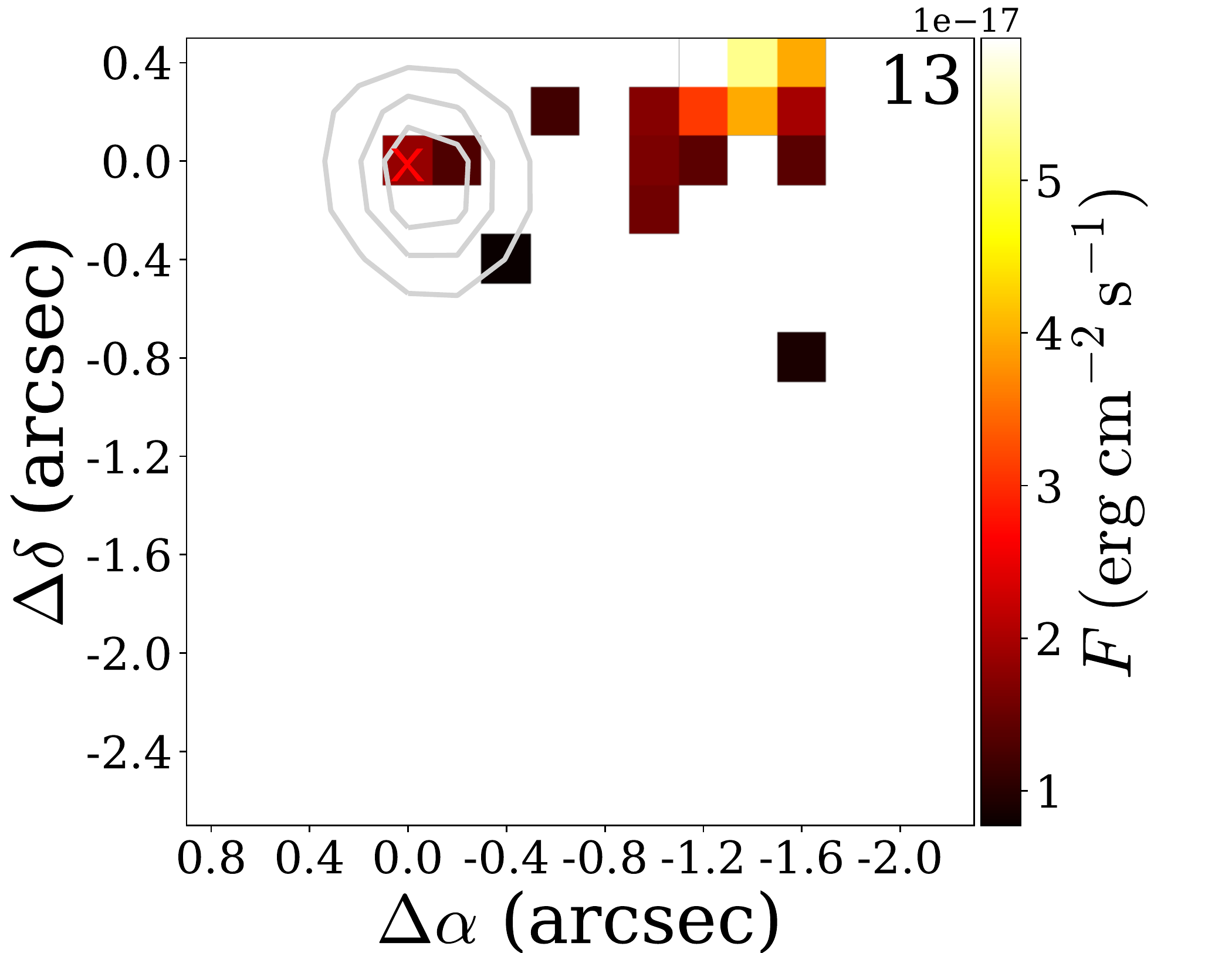}\hspace{-0.2cm}
\includegraphics[width=0.2\textwidth]{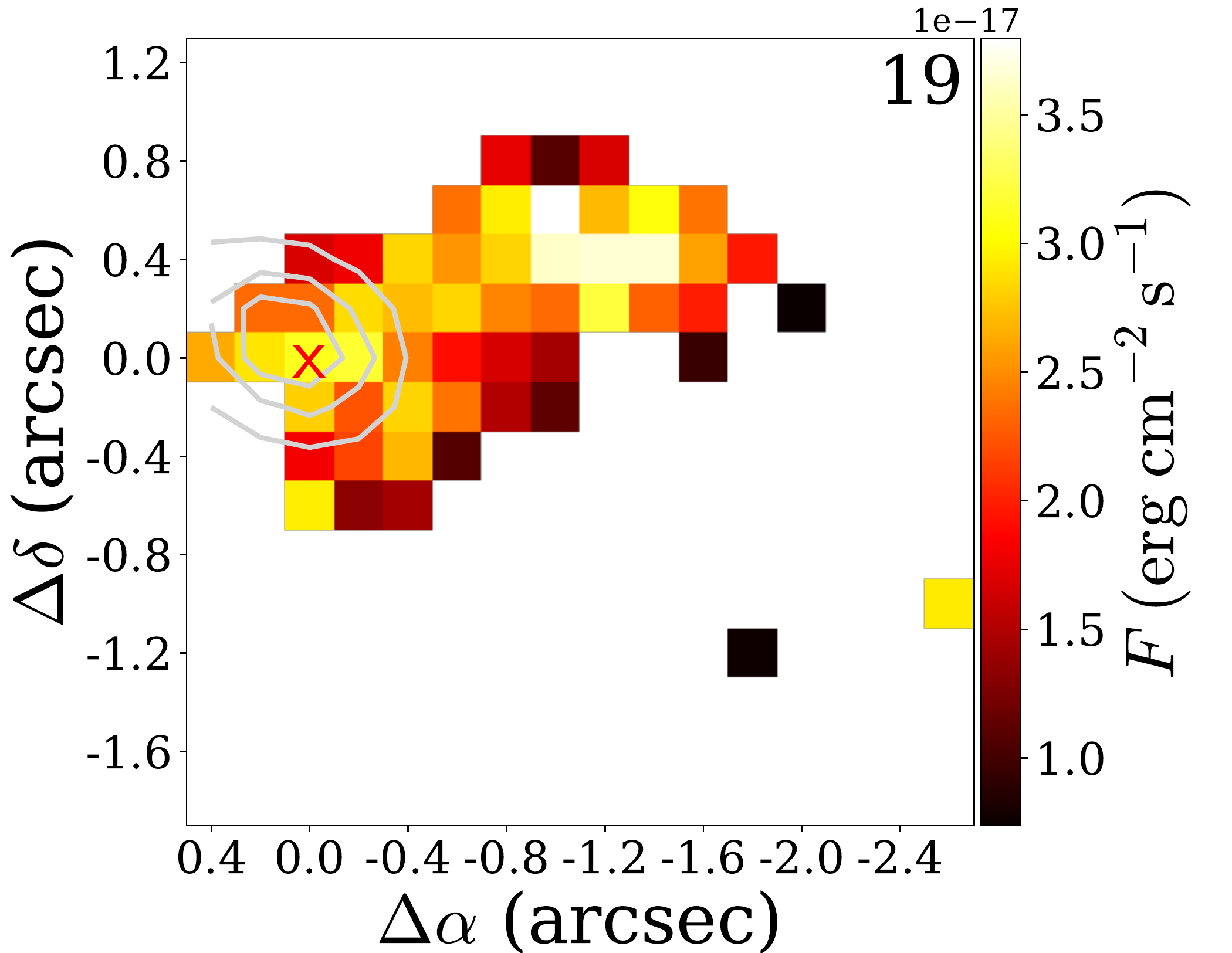}\hspace{-0.2cm}
\includegraphics[width=0.2\textwidth]{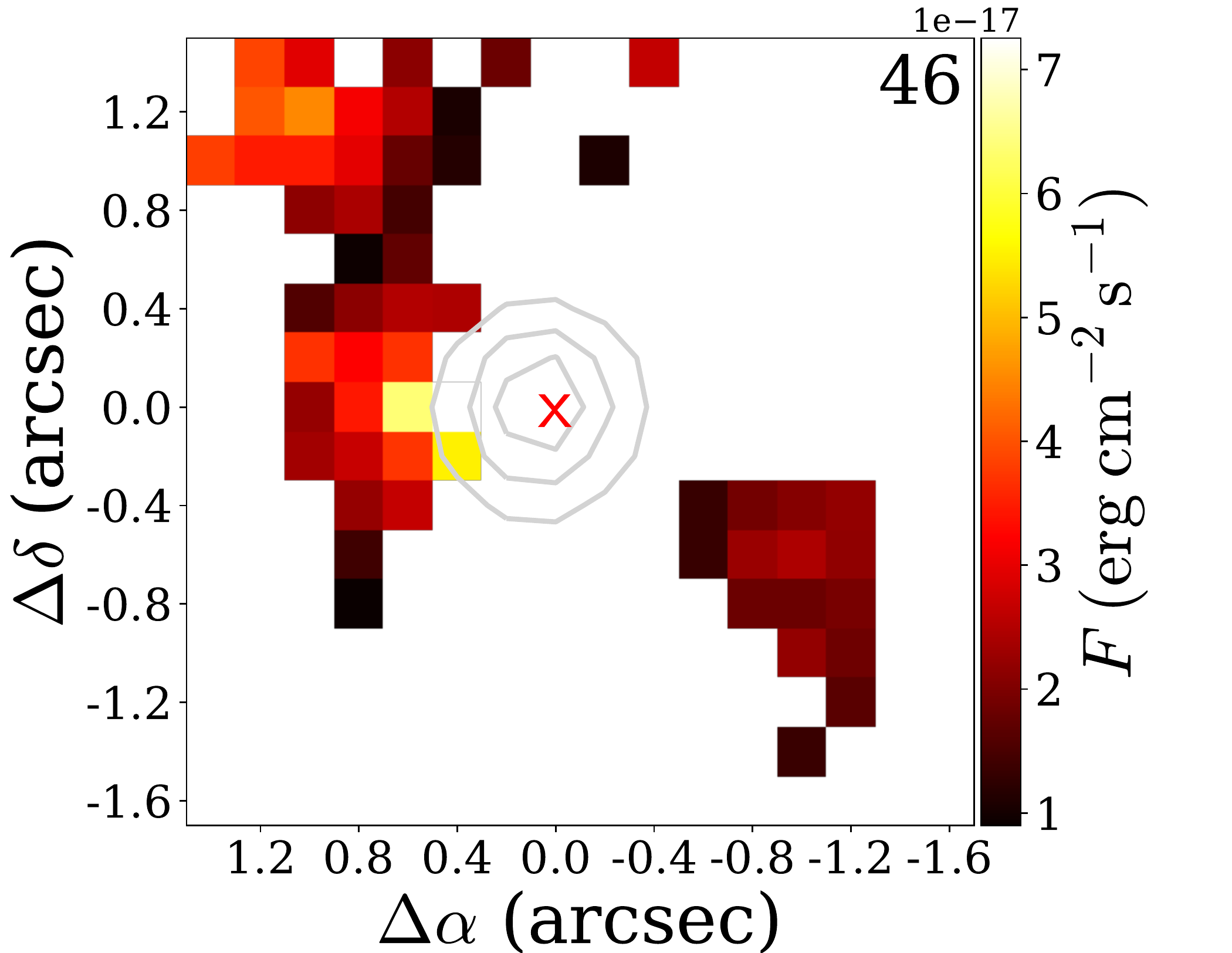}\hspace{-0.1cm}
\includegraphics[width=0.2\textwidth]{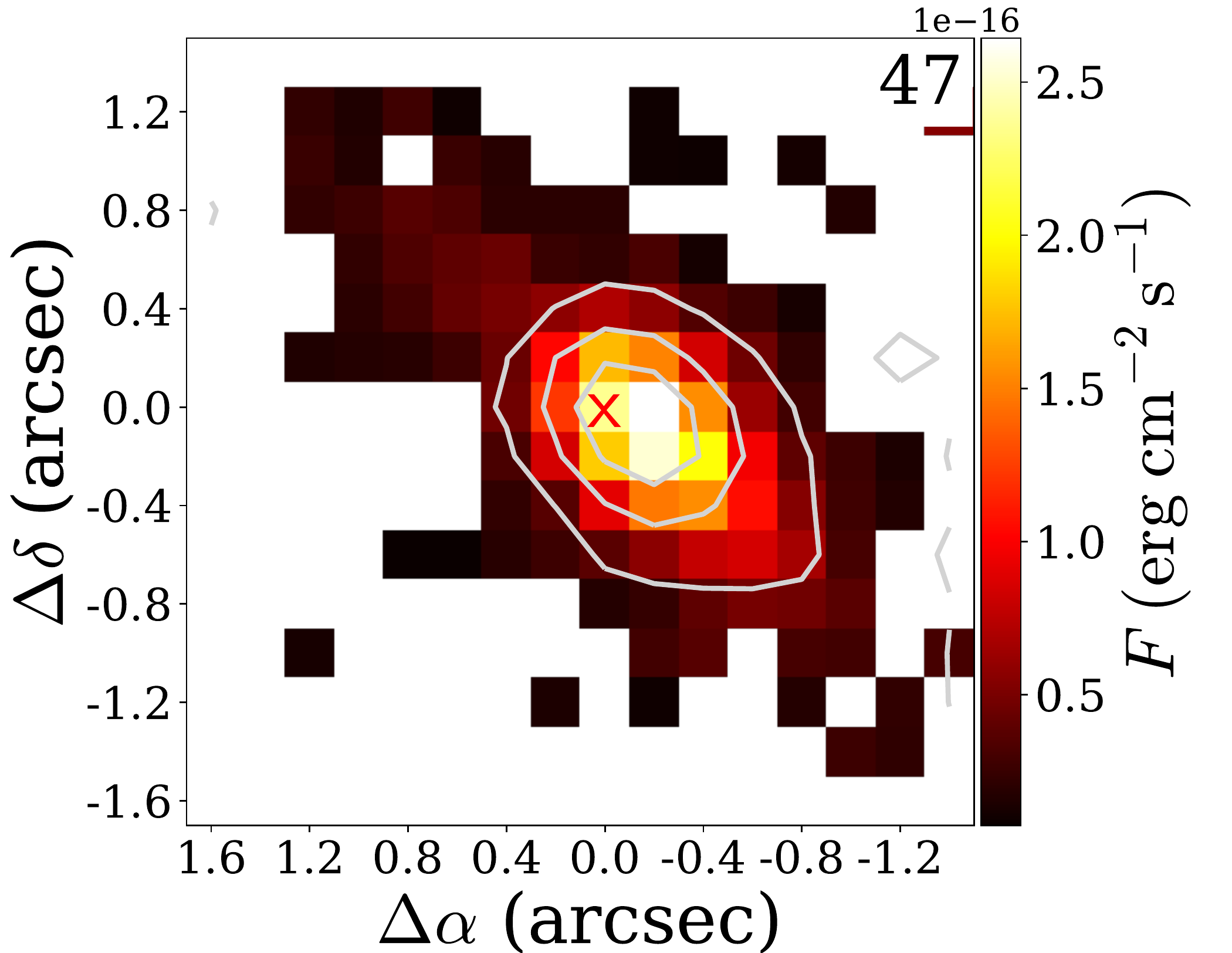}\hspace{-0.1cm}
\includegraphics[width=0.2\textwidth]{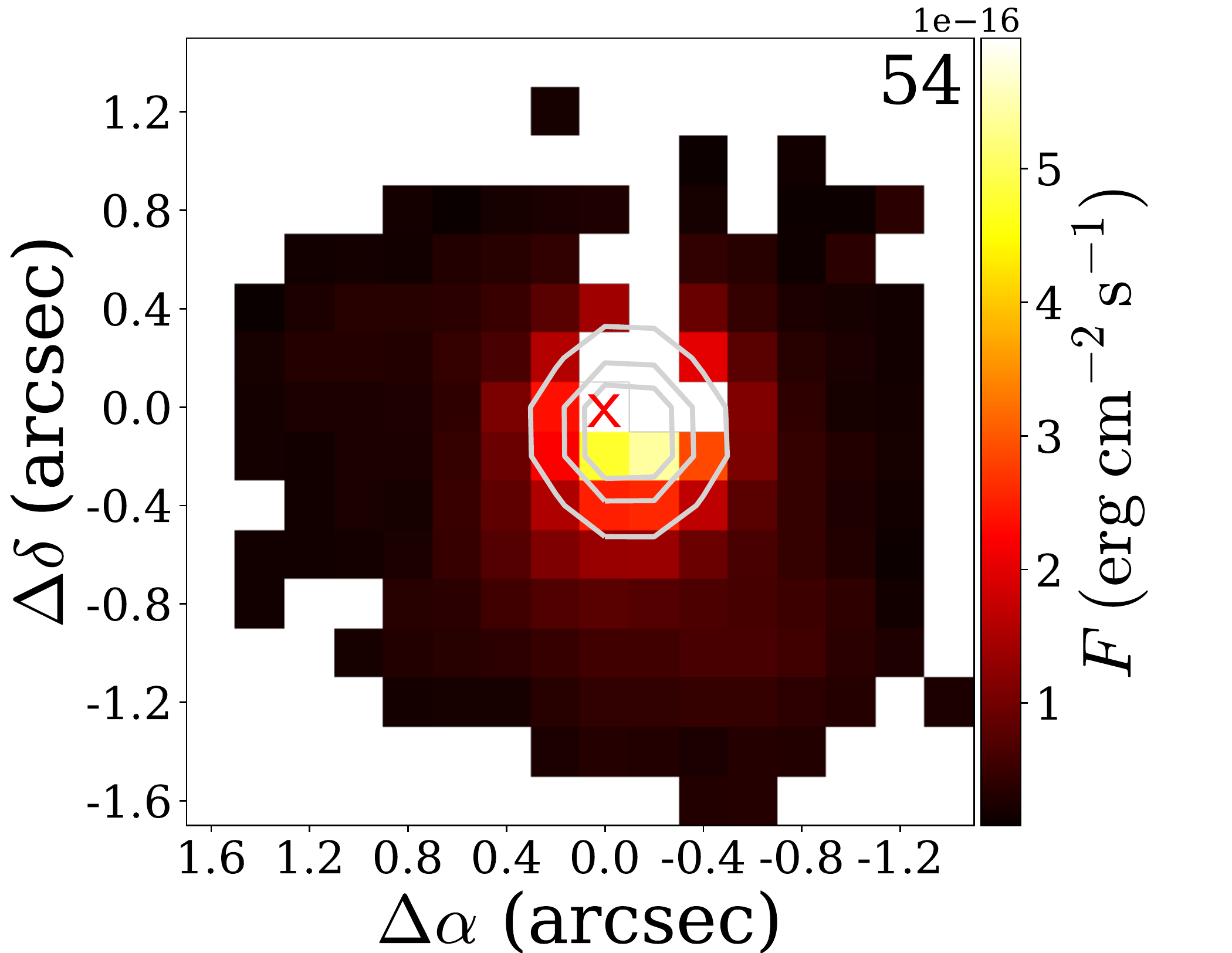}\hspace{-0.1cm}
\includegraphics[width=0.2\textwidth]{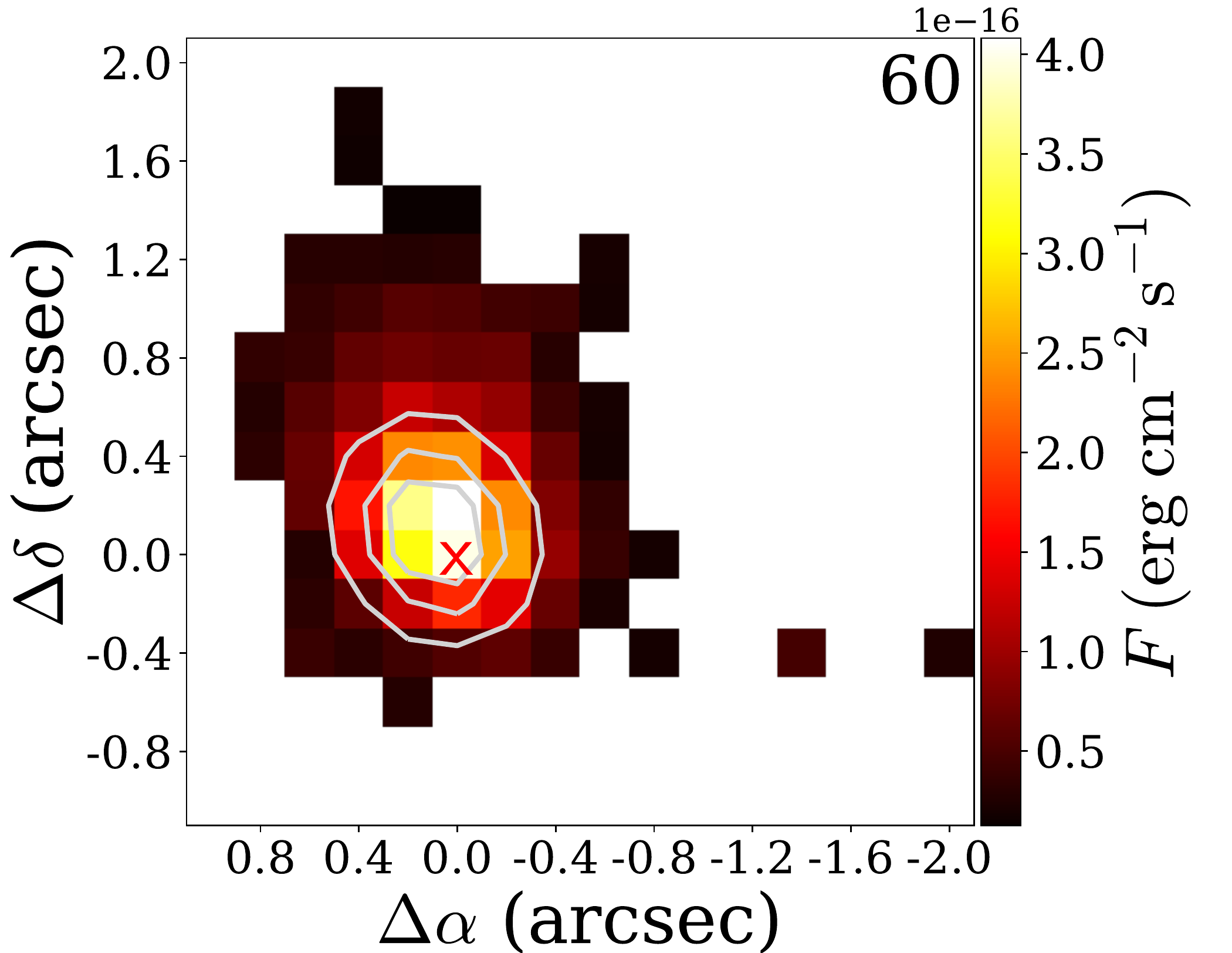}\hspace{-0.1cm}
\includegraphics[width=0.2\textwidth]{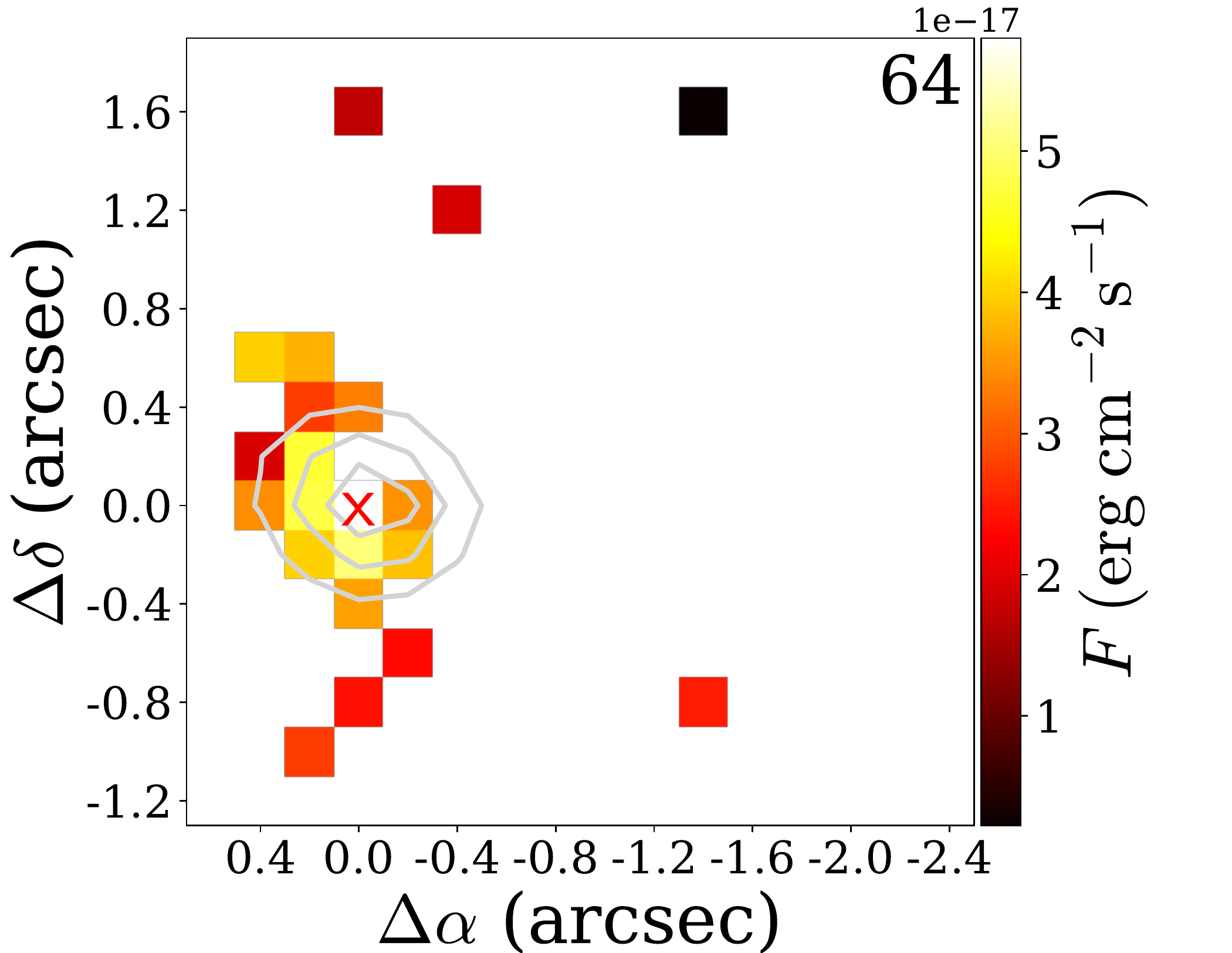}\hspace{-0.2cm}
\includegraphics[width=0.2\textwidth]{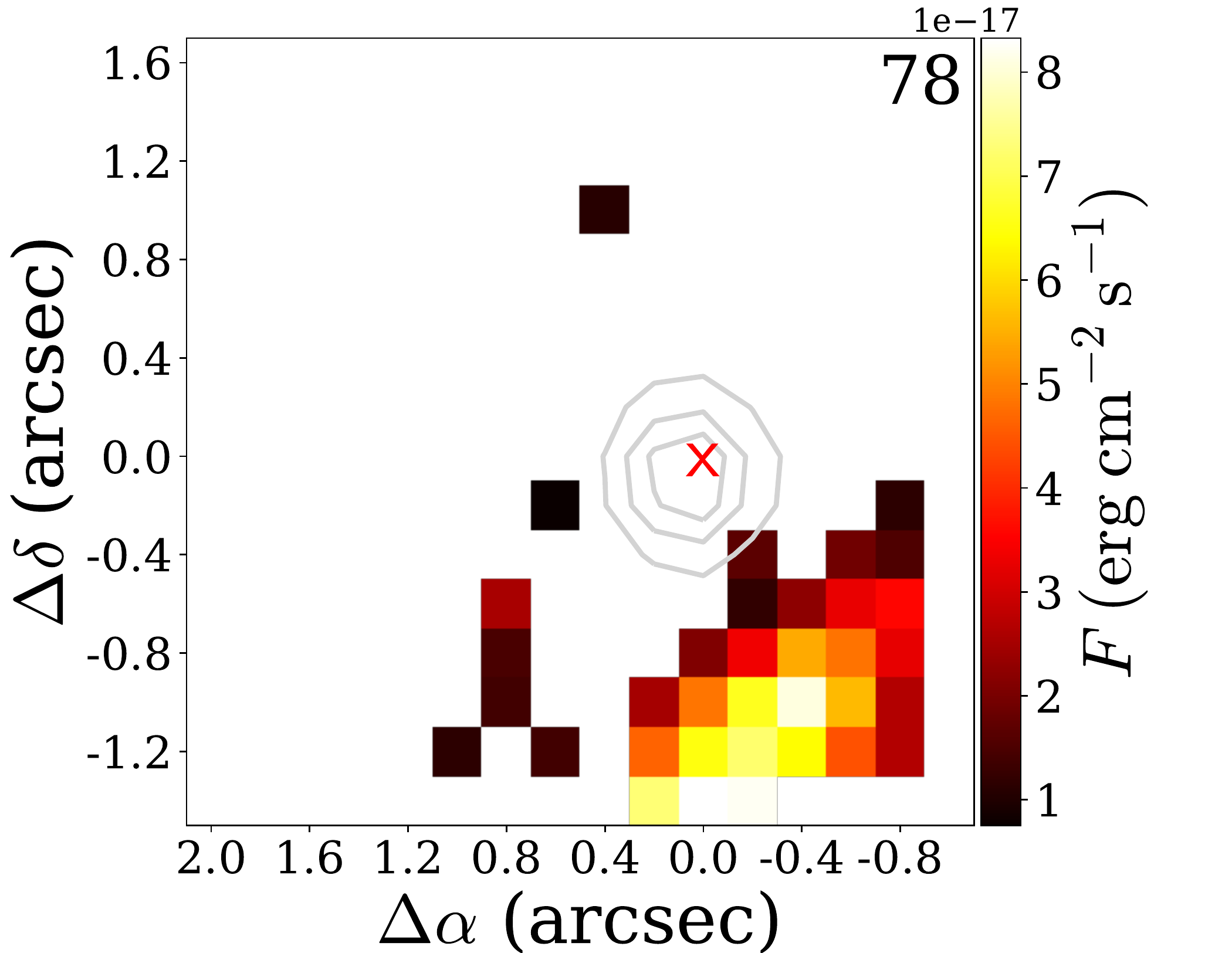}\hspace{-0.1cm}
\includegraphics[width=0.2\textwidth]{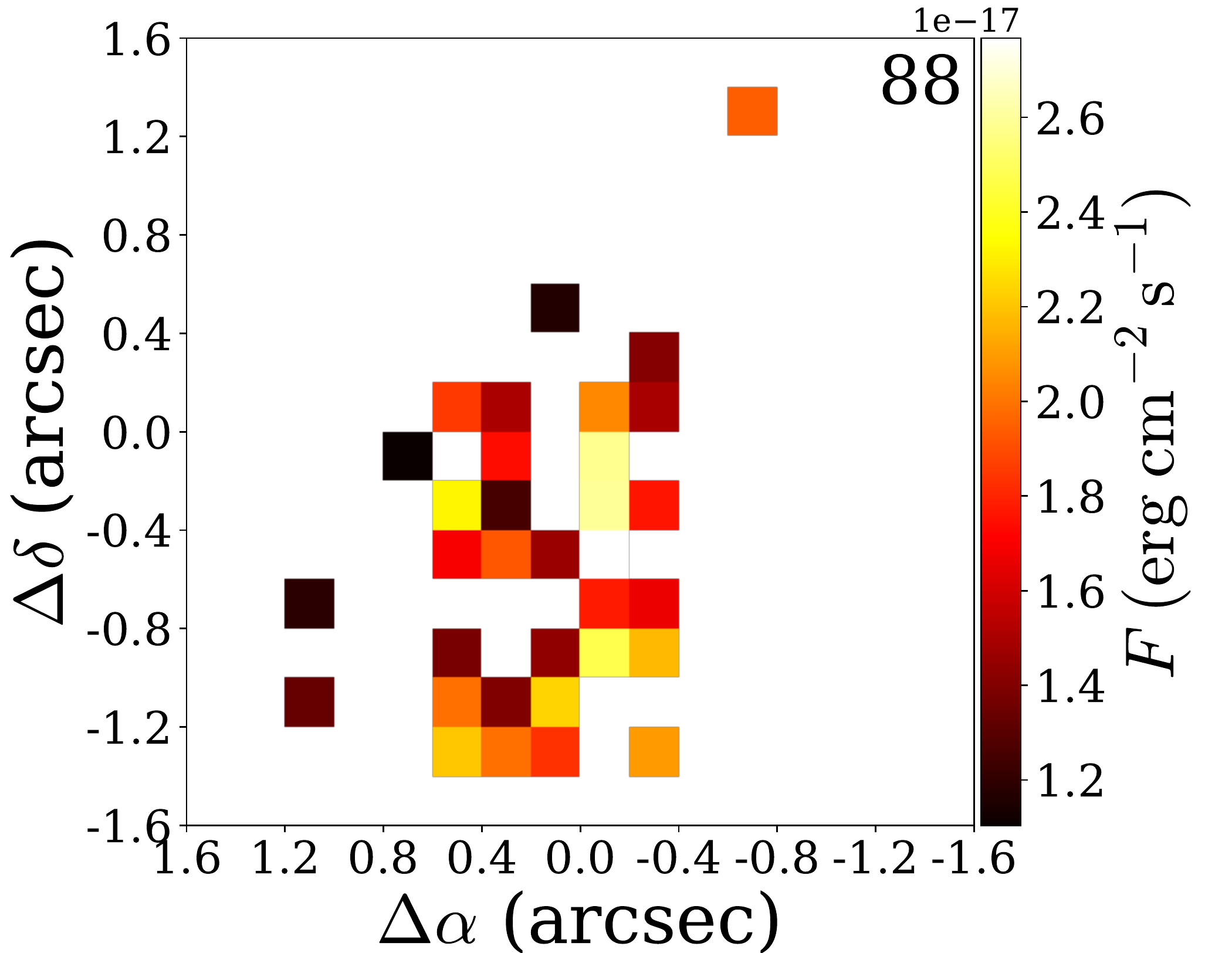}\hspace{-0.1cm}
\includegraphics[width=0.2\textwidth]{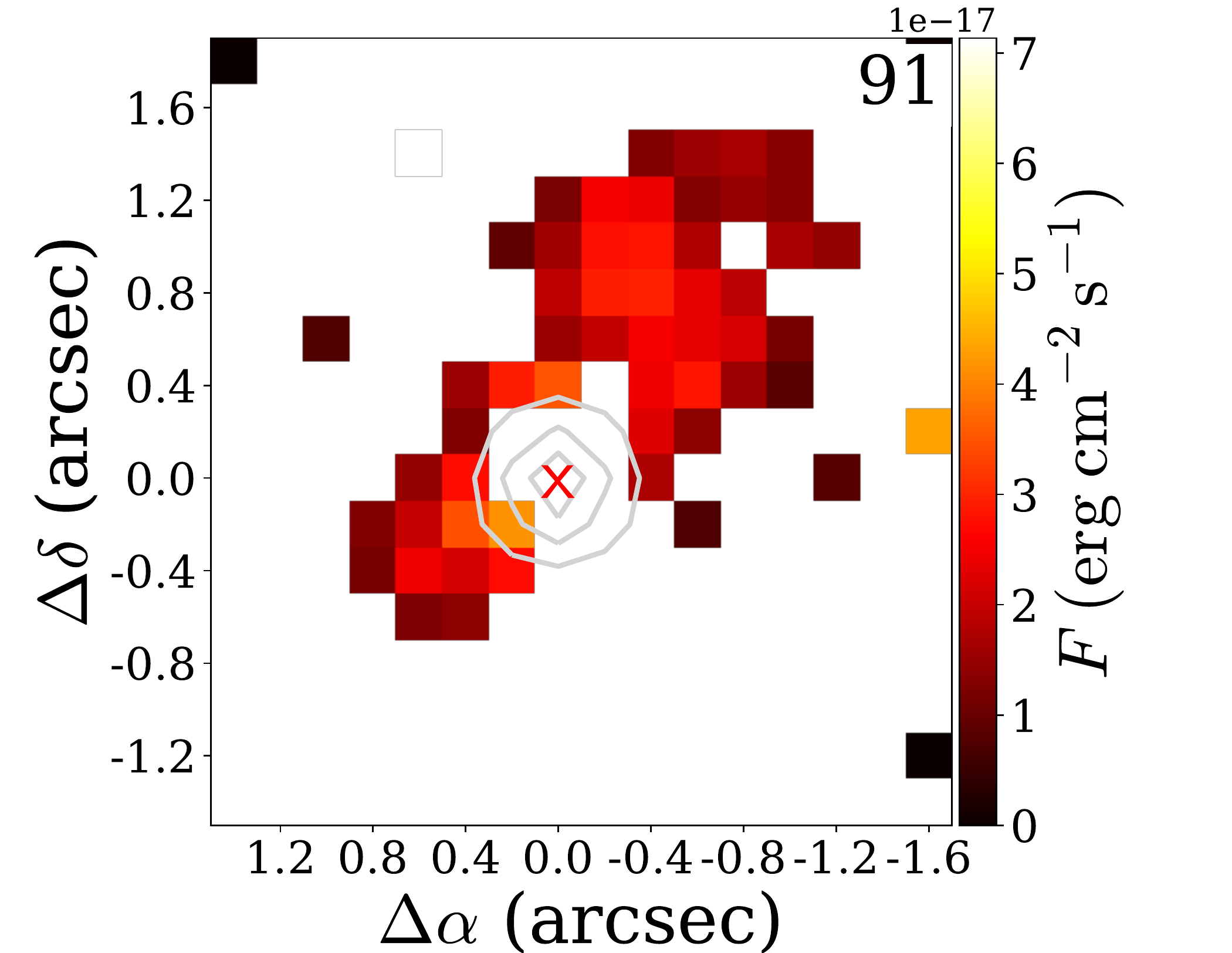}\hspace{-0.1cm}
\includegraphics[width=0.2\textwidth]{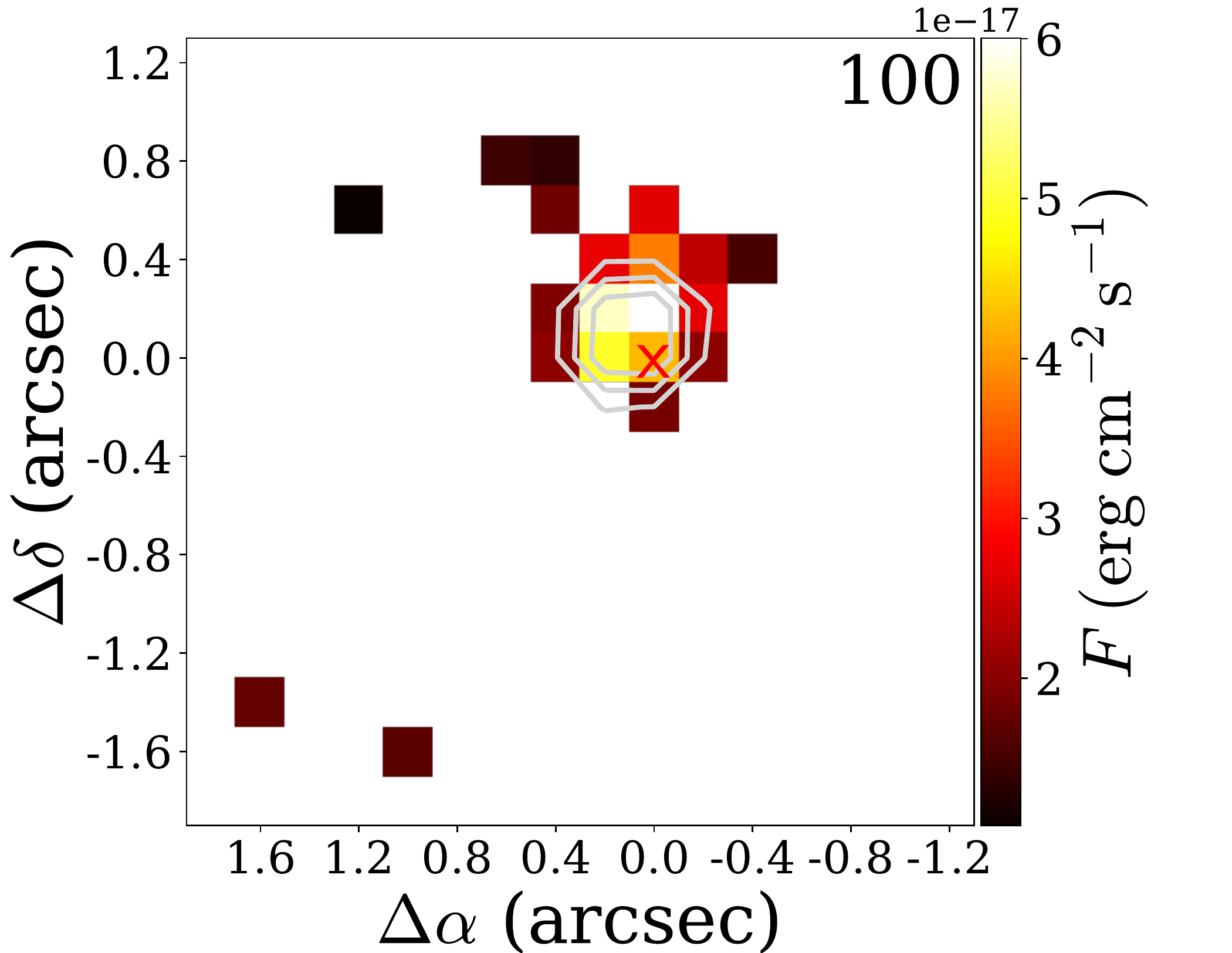}\hspace{-0.2cm}
\includegraphics[width=0.2\textwidth]{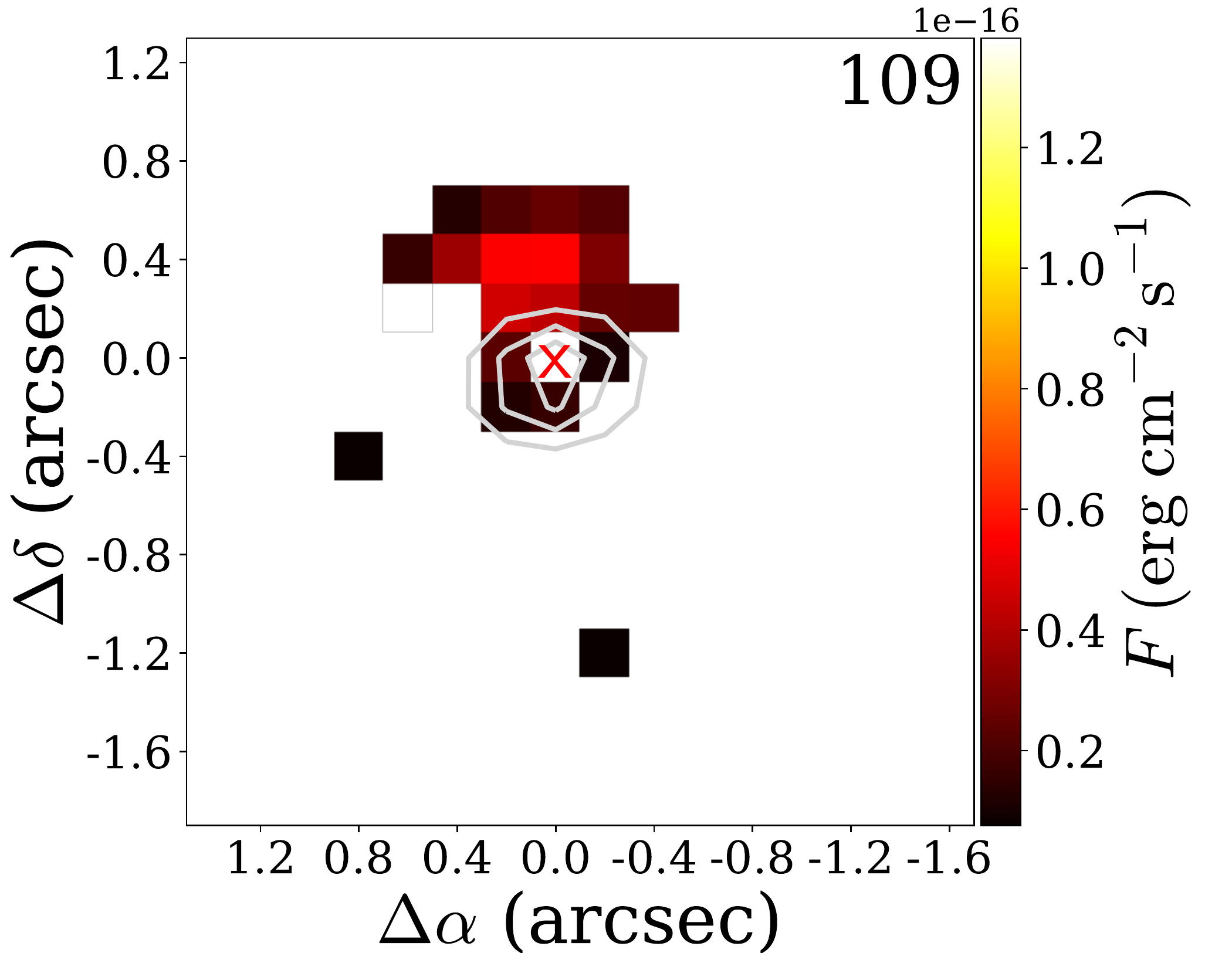}\hspace{-0.1cm}
\includegraphics[width=0.2\textwidth]{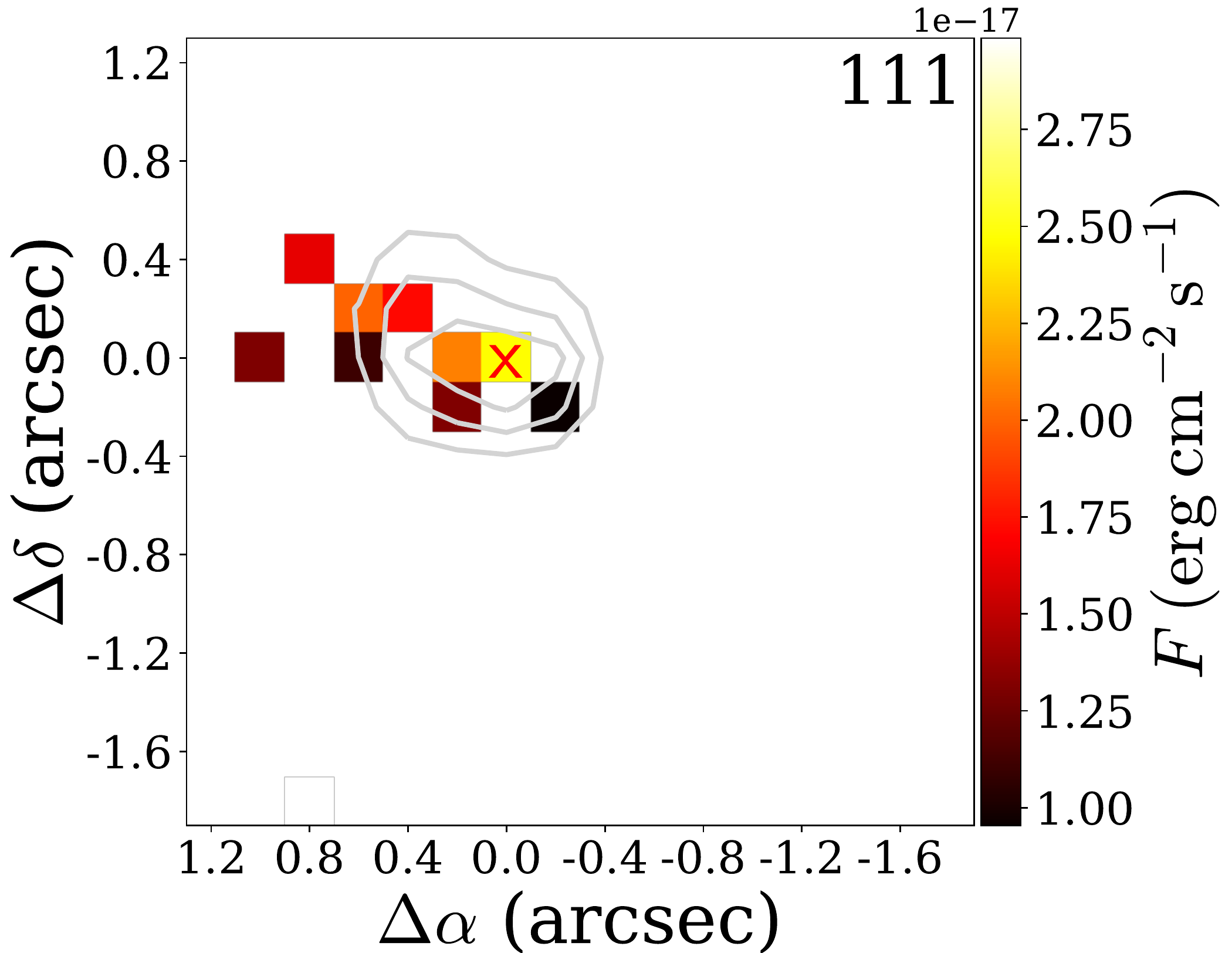}\hspace{-0.1cm}
\includegraphics[width=0.2\textwidth]{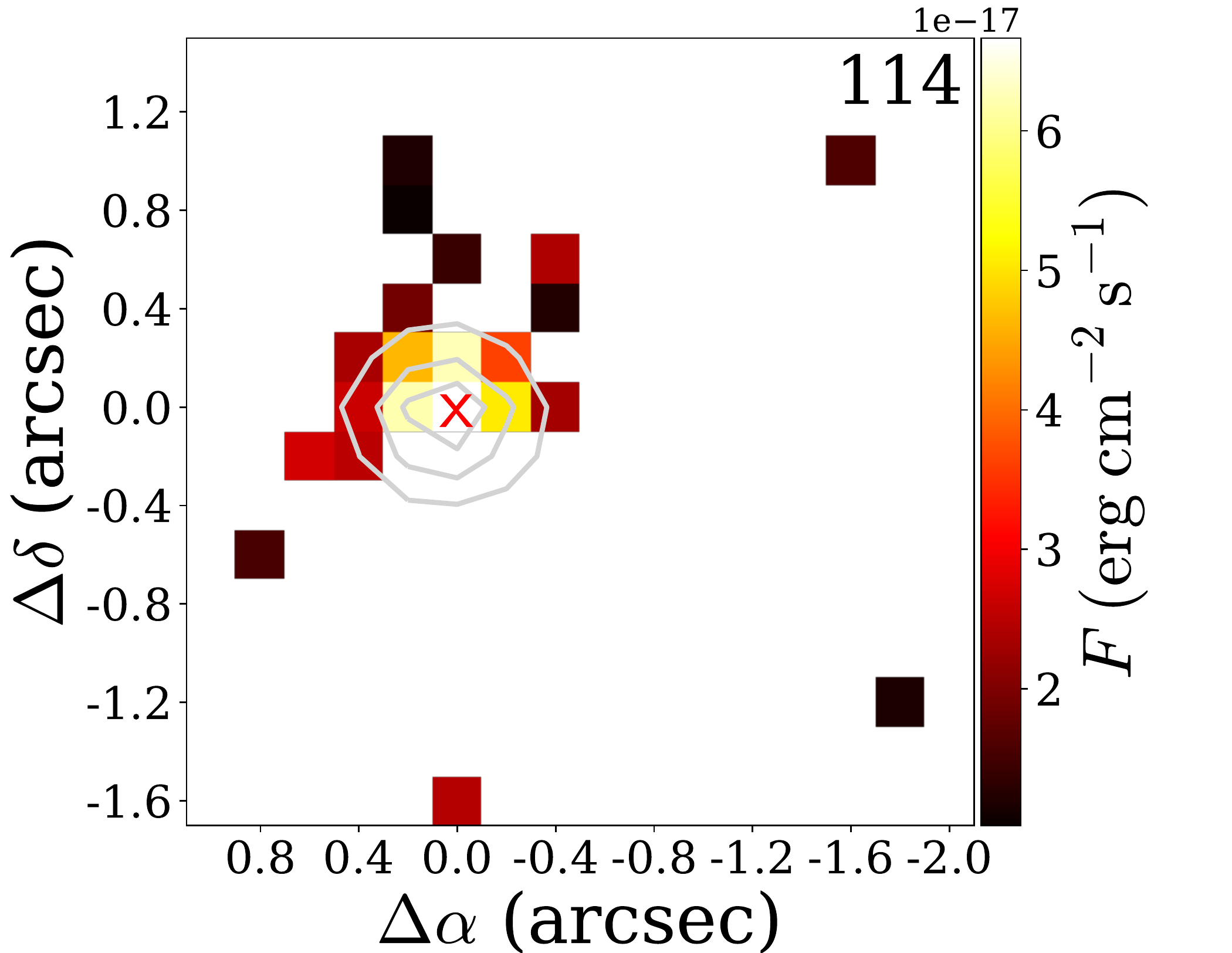}\hspace{-0.1cm}
\caption{Similar to Figure \ref{fig:emiss-2.1218}, but for the H$_2$ 1-0 Q(3) line at 2.4237 $\mu$m. {  Only emission above 2$\sigma$ is shown.}}
\label{fig:emiss-2.4237}
\end{figure*}

\begin{figure*}[h!]
\includegraphics[width=0.2\textwidth]{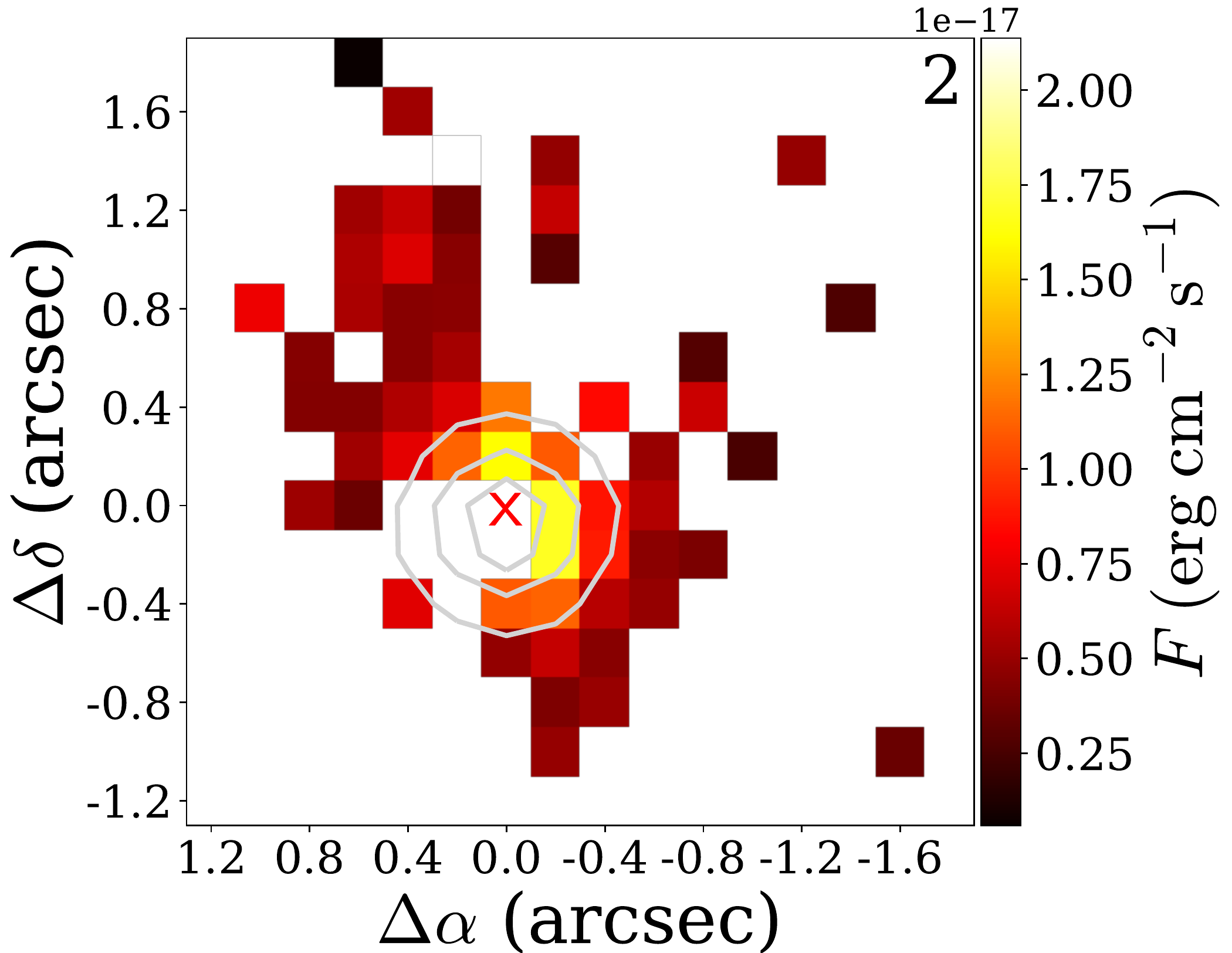}\hspace{-0.2cm}
\includegraphics[width=0.2\textwidth]{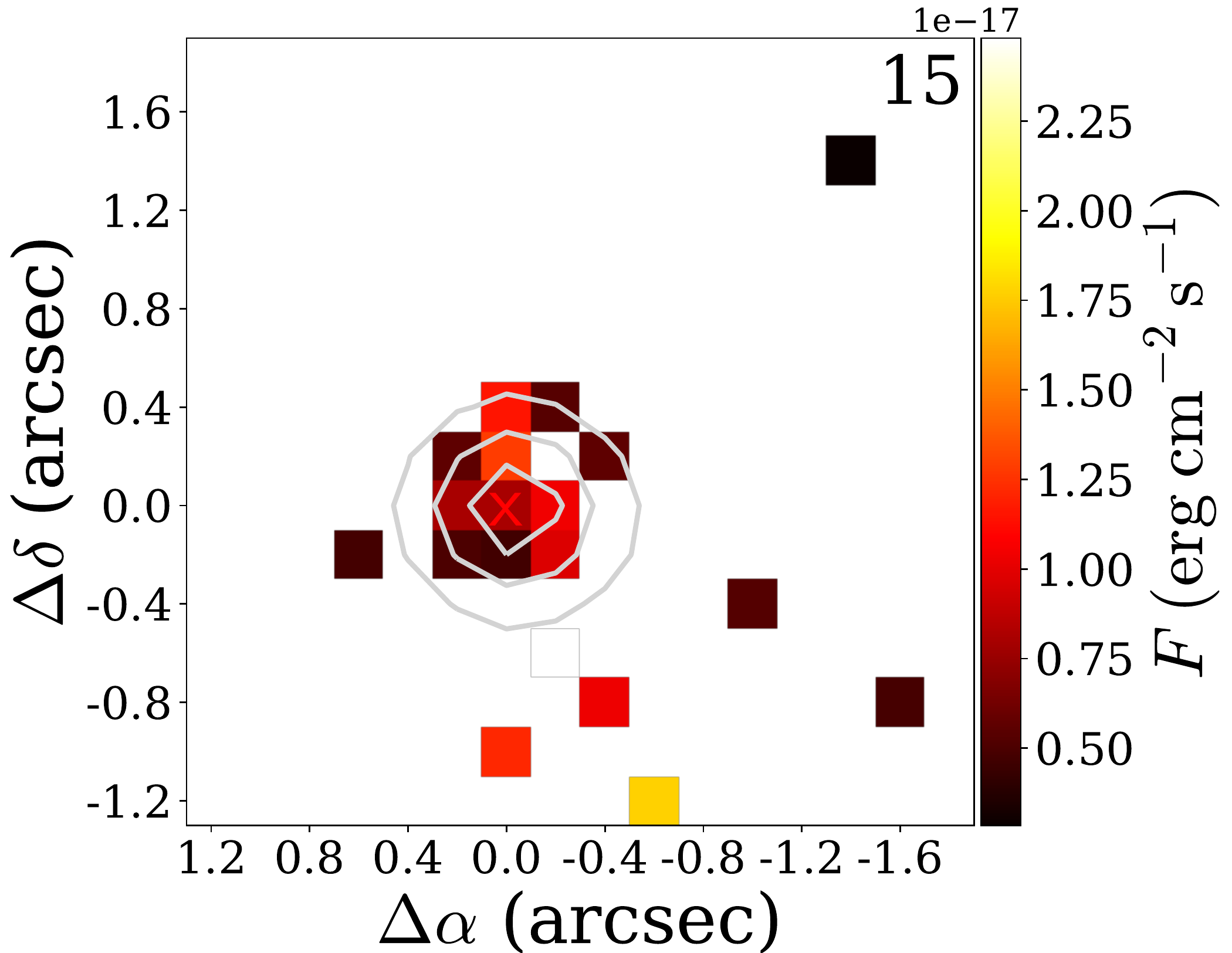}\hspace{-0.2cm}
\includegraphics[width=0.2\textwidth]{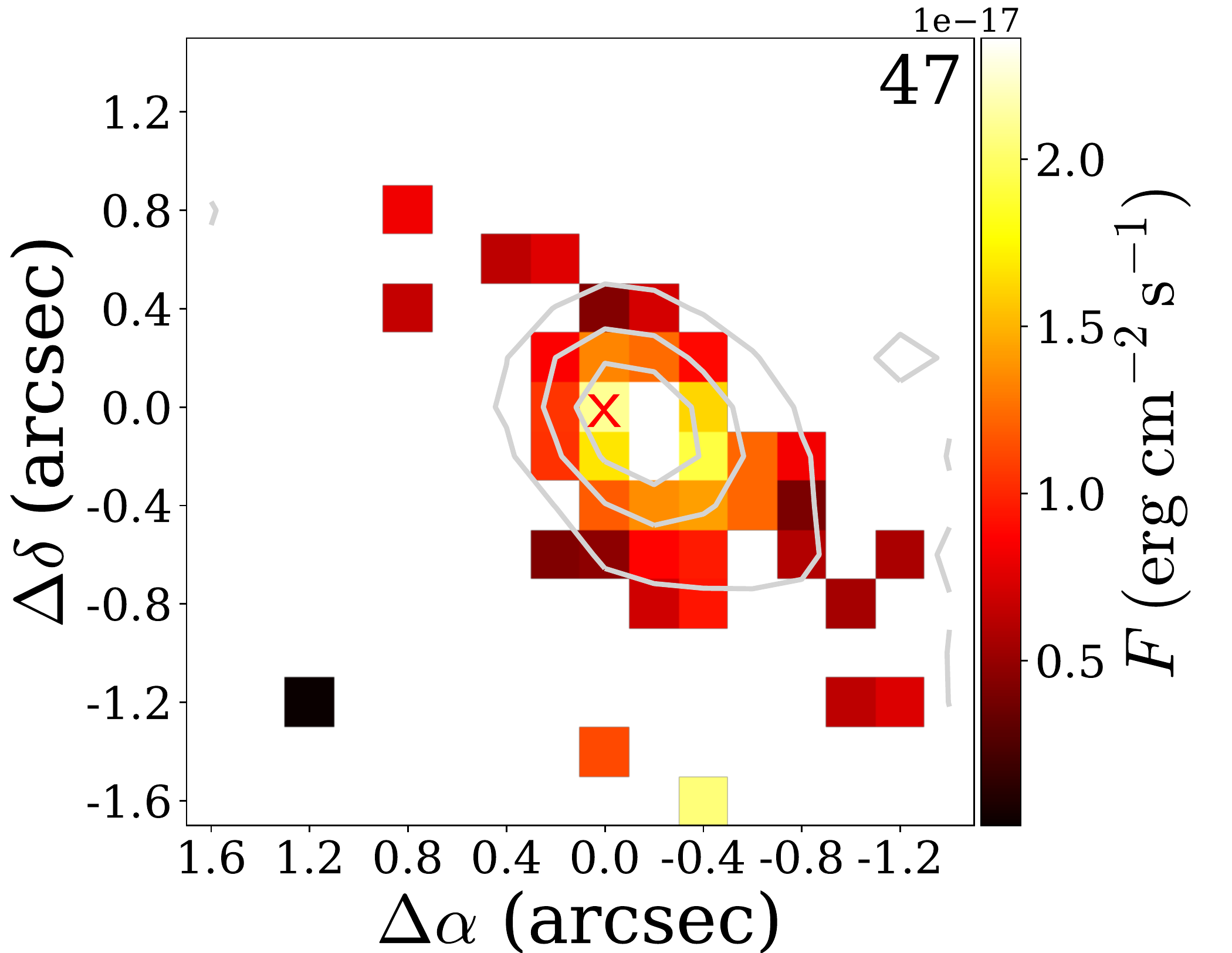}\hspace{-0.2cm}
\includegraphics[width=0.2\textwidth]{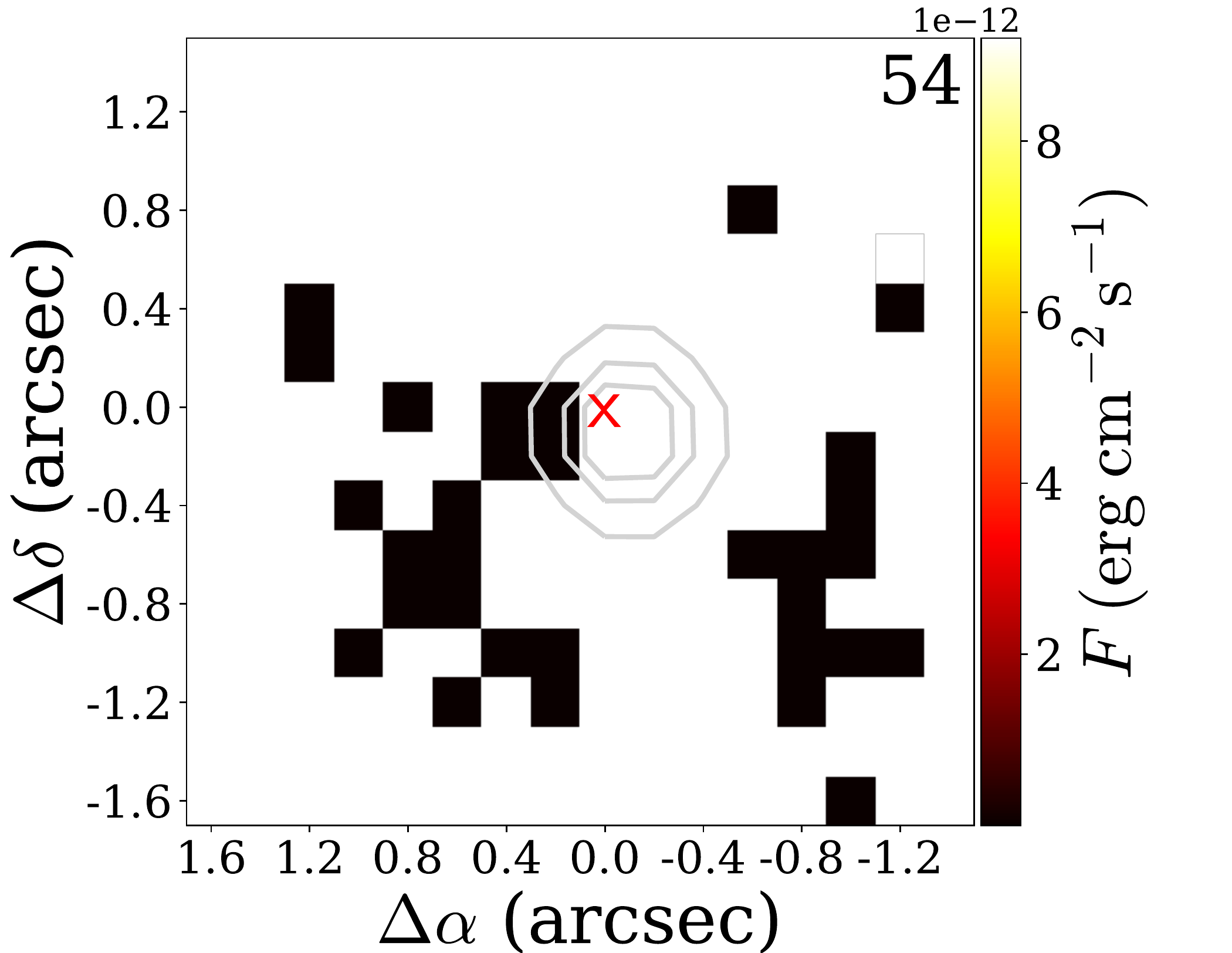}\hspace{-0.2cm}
\includegraphics[width=0.2\textwidth]{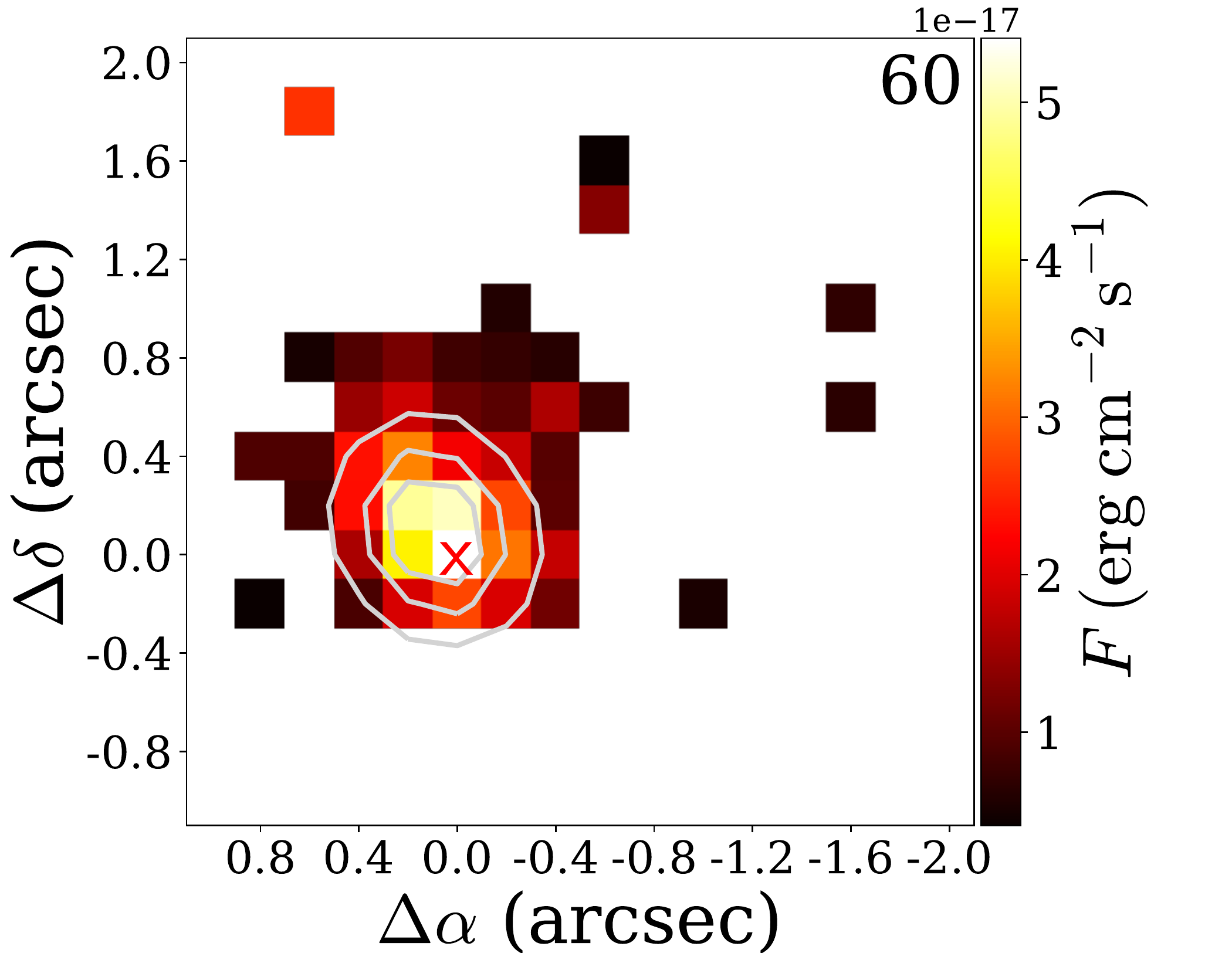}\hspace{-0.2cm}
\includegraphics[width=0.2\textwidth]{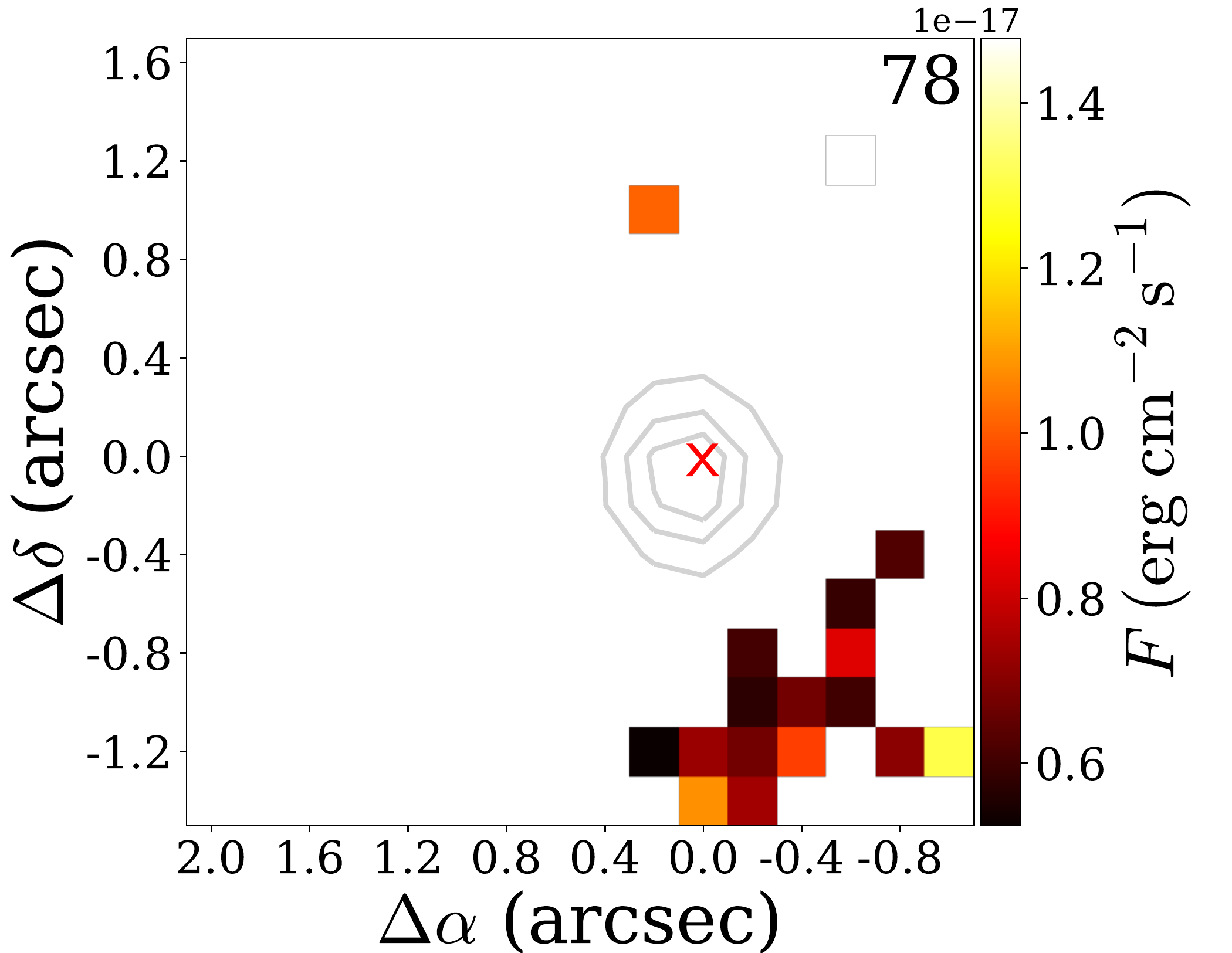}\hspace{-0.2cm}
\includegraphics[width=0.2\textwidth]{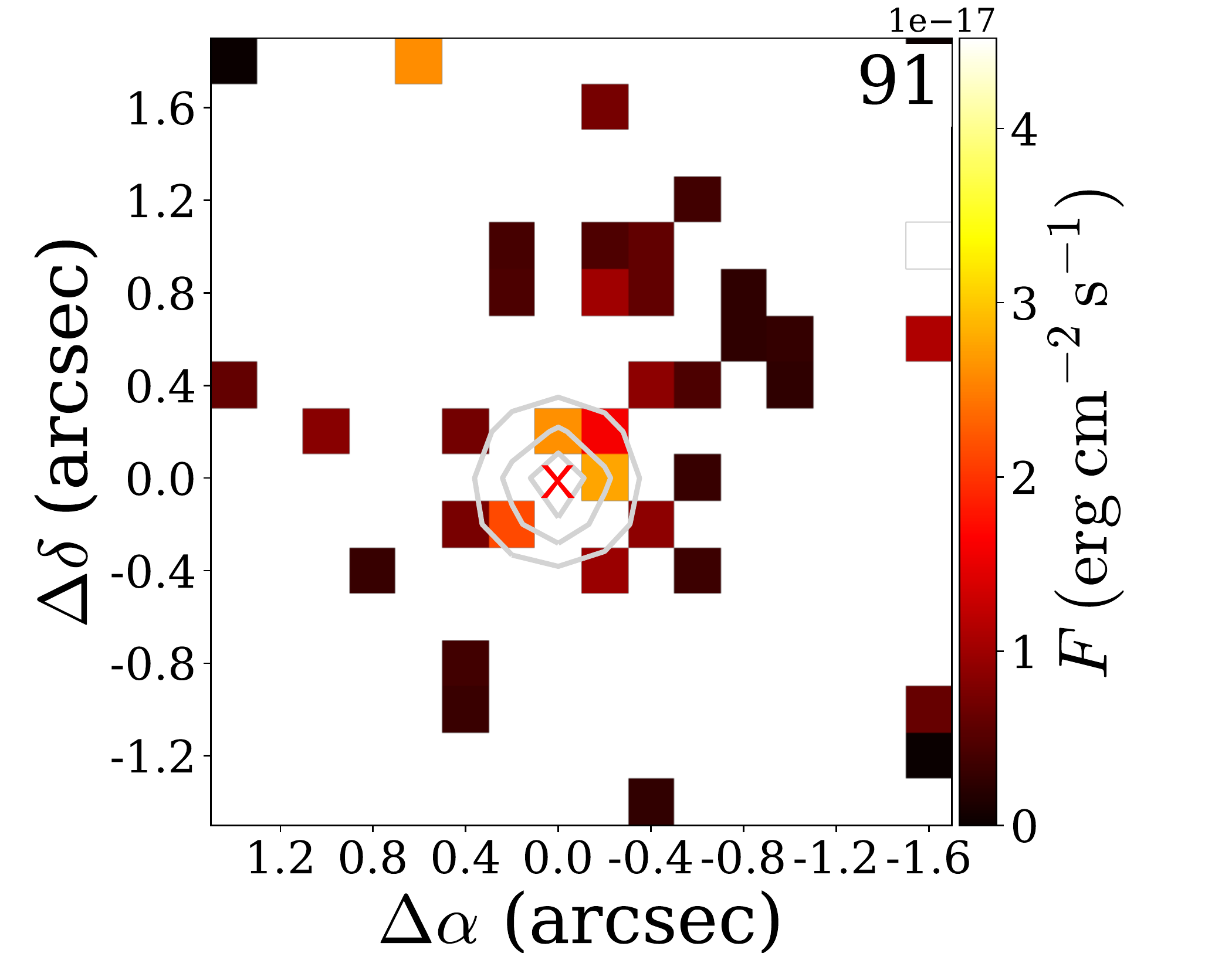}\hspace{-0.2cm}
\includegraphics[width=0.2\textwidth]{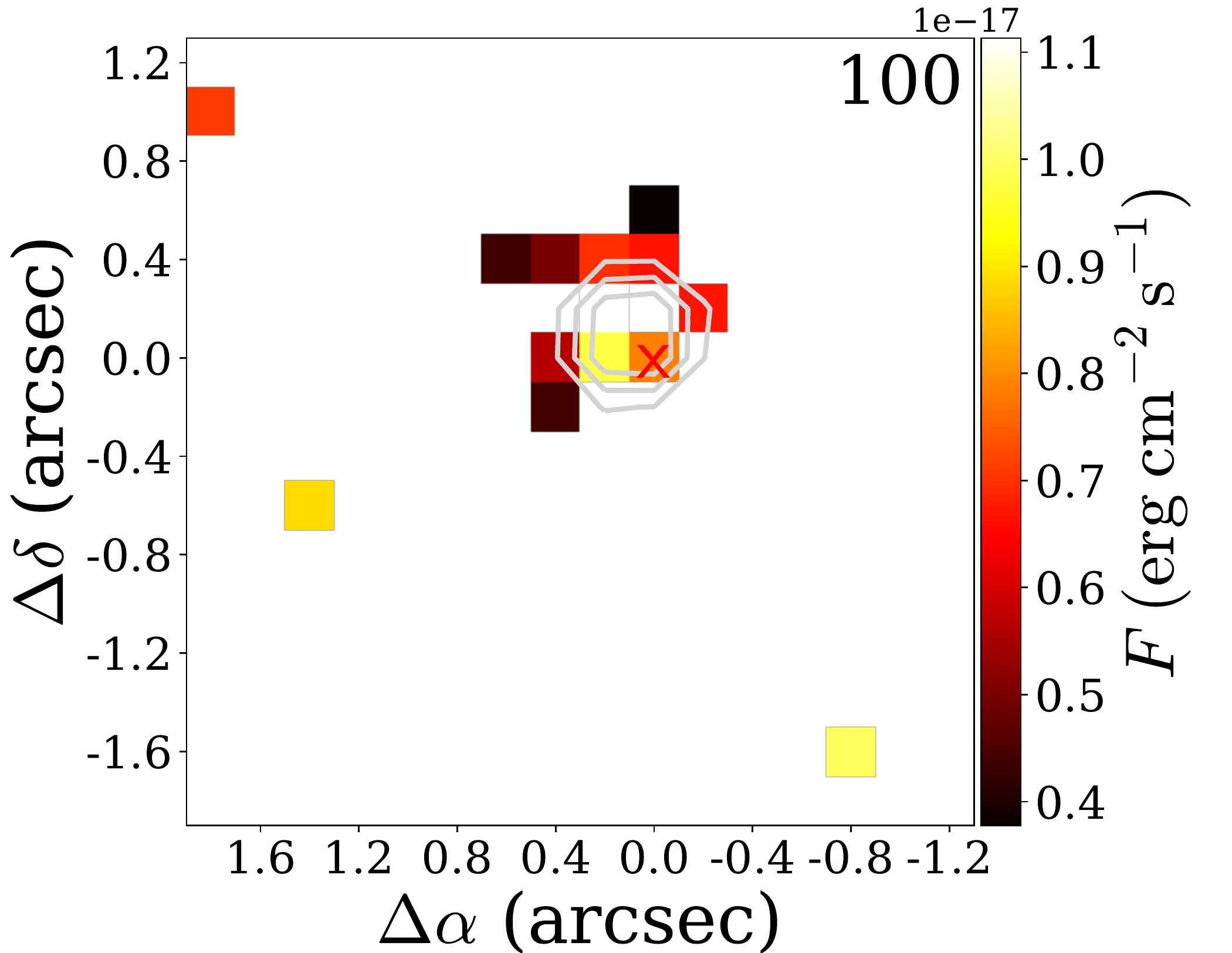}\hspace{-0.2cm}
\includegraphics[width=0.2\textwidth]{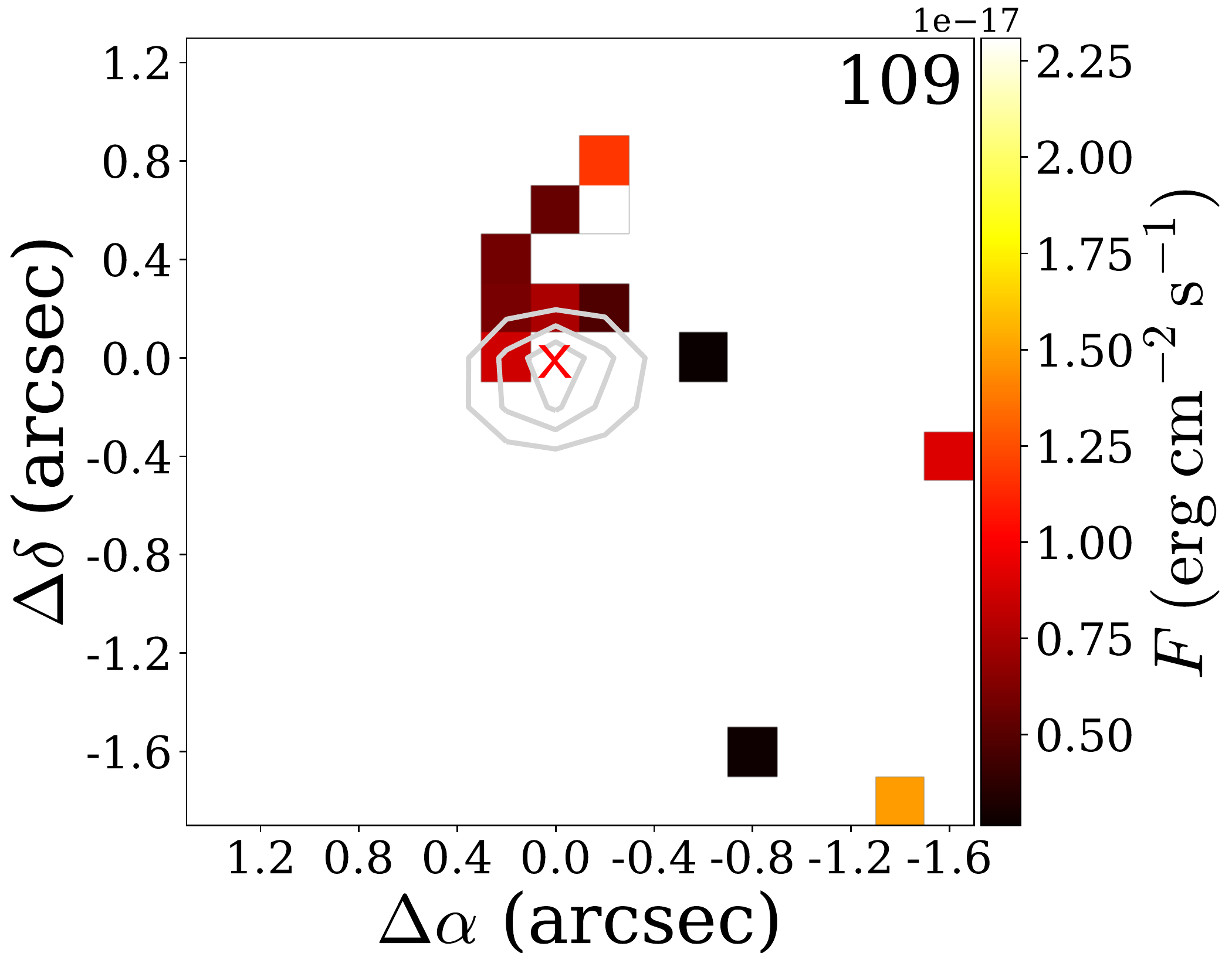}\hspace{-0.2cm}
\includegraphics[width=0.2\textwidth]{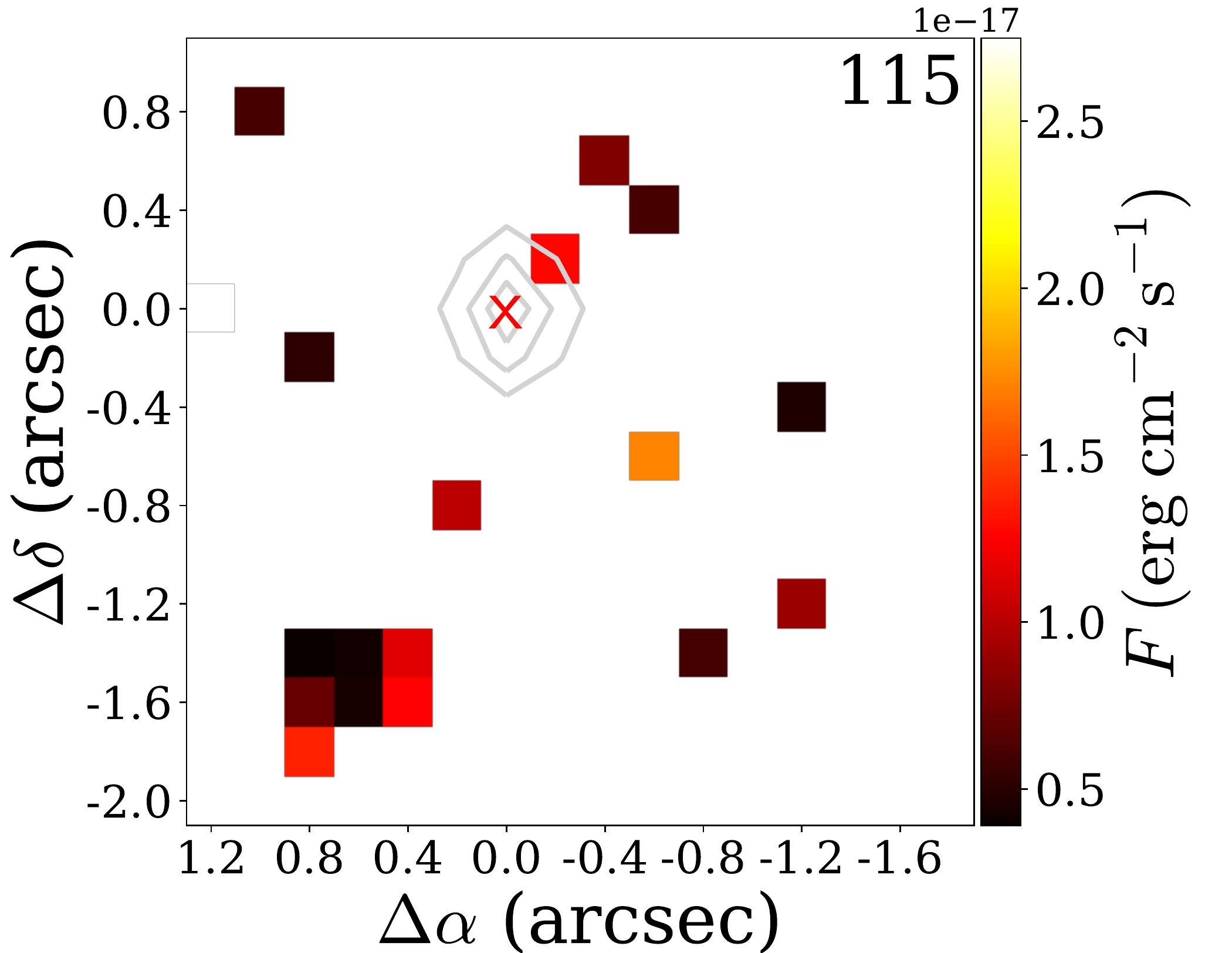}\hspace{-0.2cm}
\caption{Similar to Figure \ref{fig:emiss-2.1218}, but for the H$_2$ 2-1 S(1) line at 2.2477 $\mu$m. {  Only emission above 2$\sigma$ is shown.}}
\label{fig:emiss-2.2477}
\end{figure*}


\begin{figure*}[h!]
\includegraphics[width=0.2\textwidth]{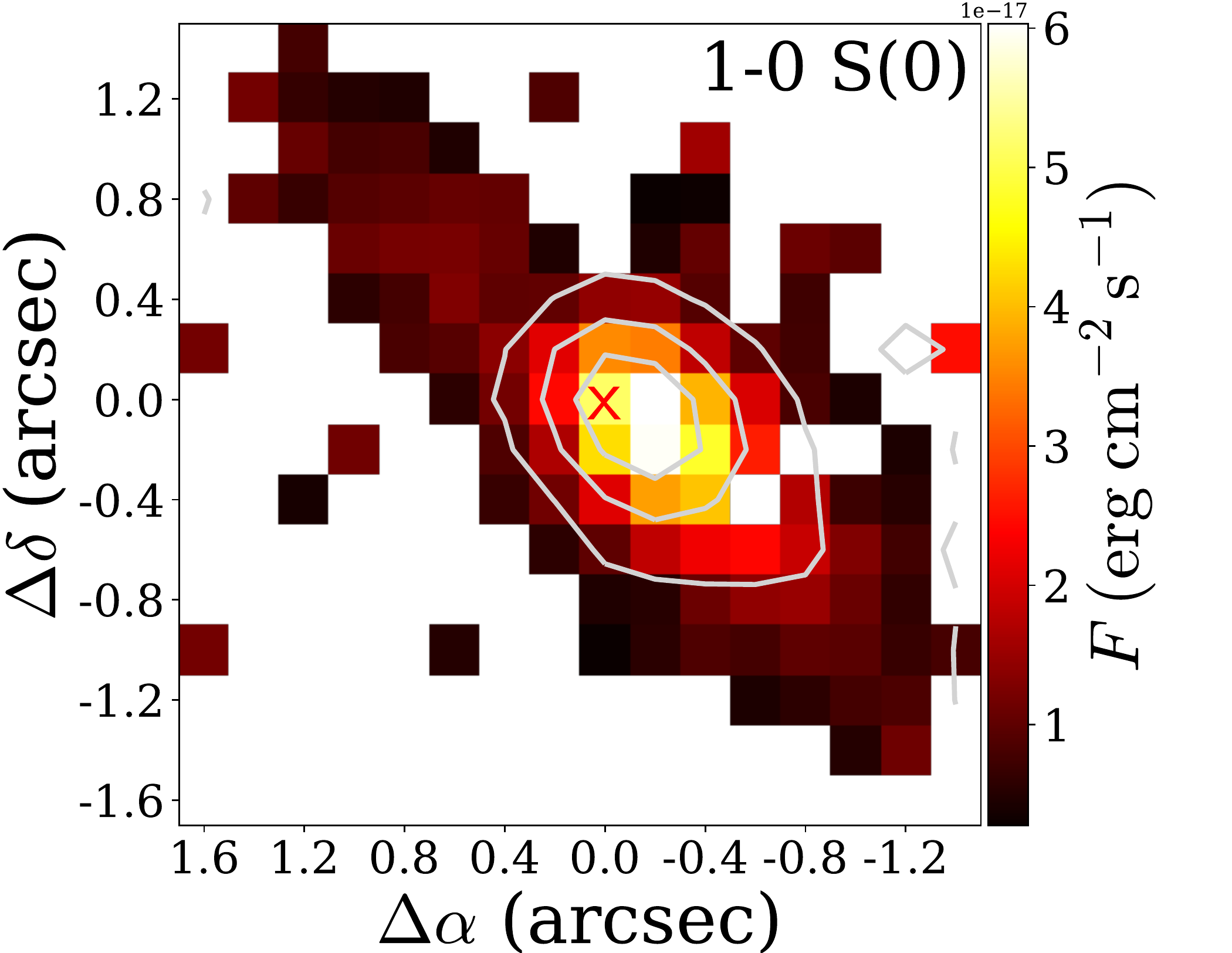}\hspace{-0.2cm}
\includegraphics[width=0.2\textwidth]{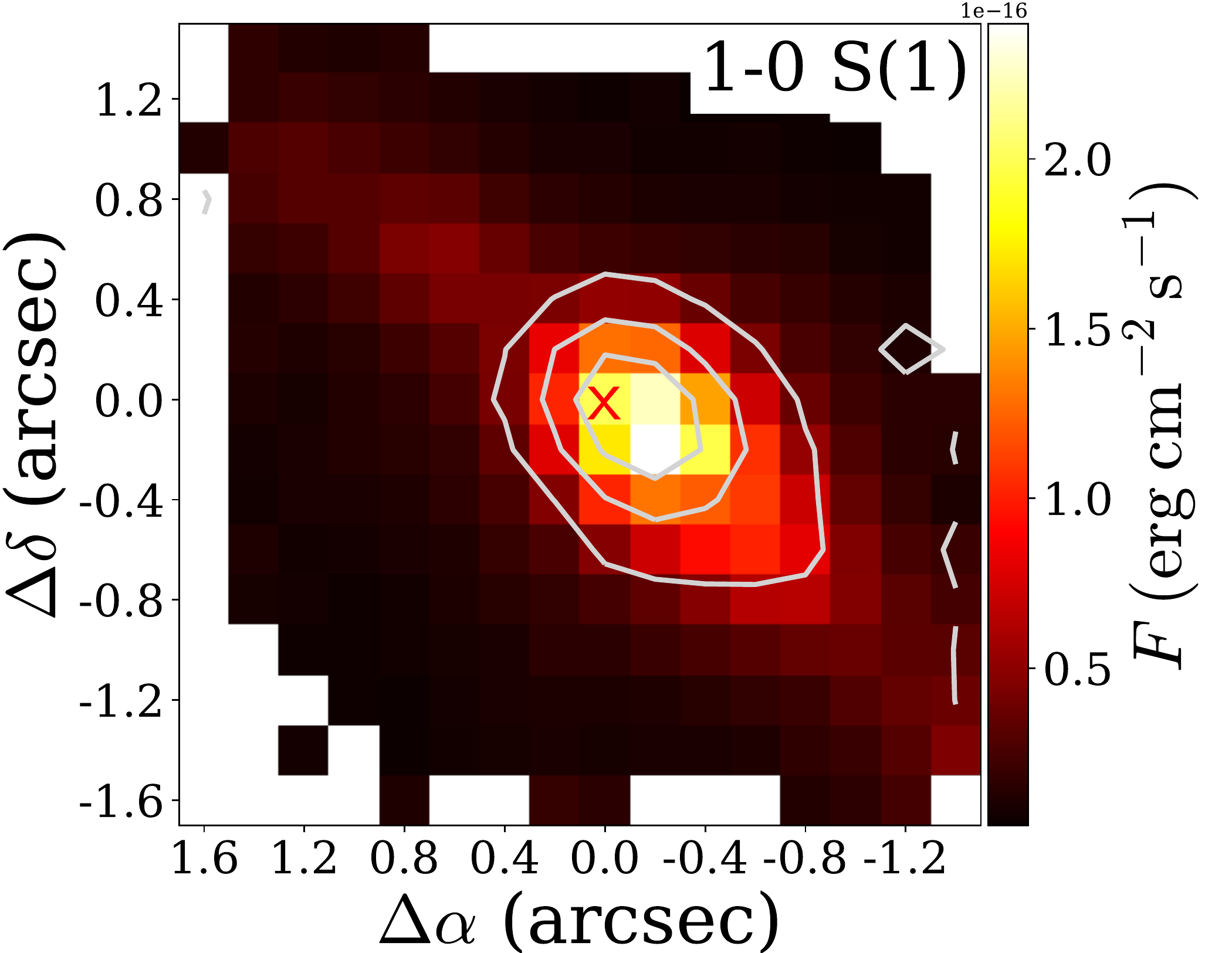}\hspace{-0.2cm}
\includegraphics[width=0.2\textwidth]{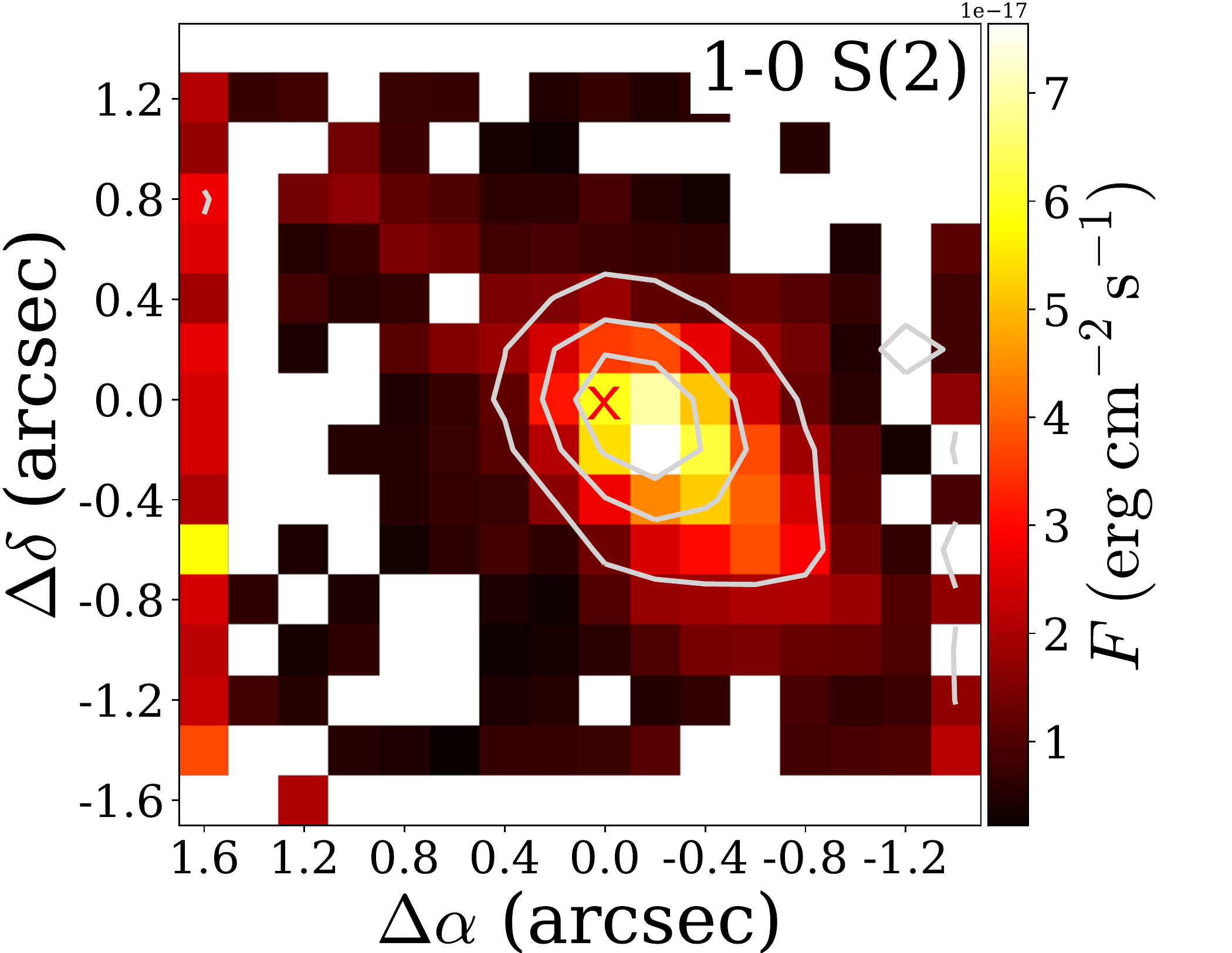}\hspace{-0.2cm}
\includegraphics[width=0.2\textwidth]{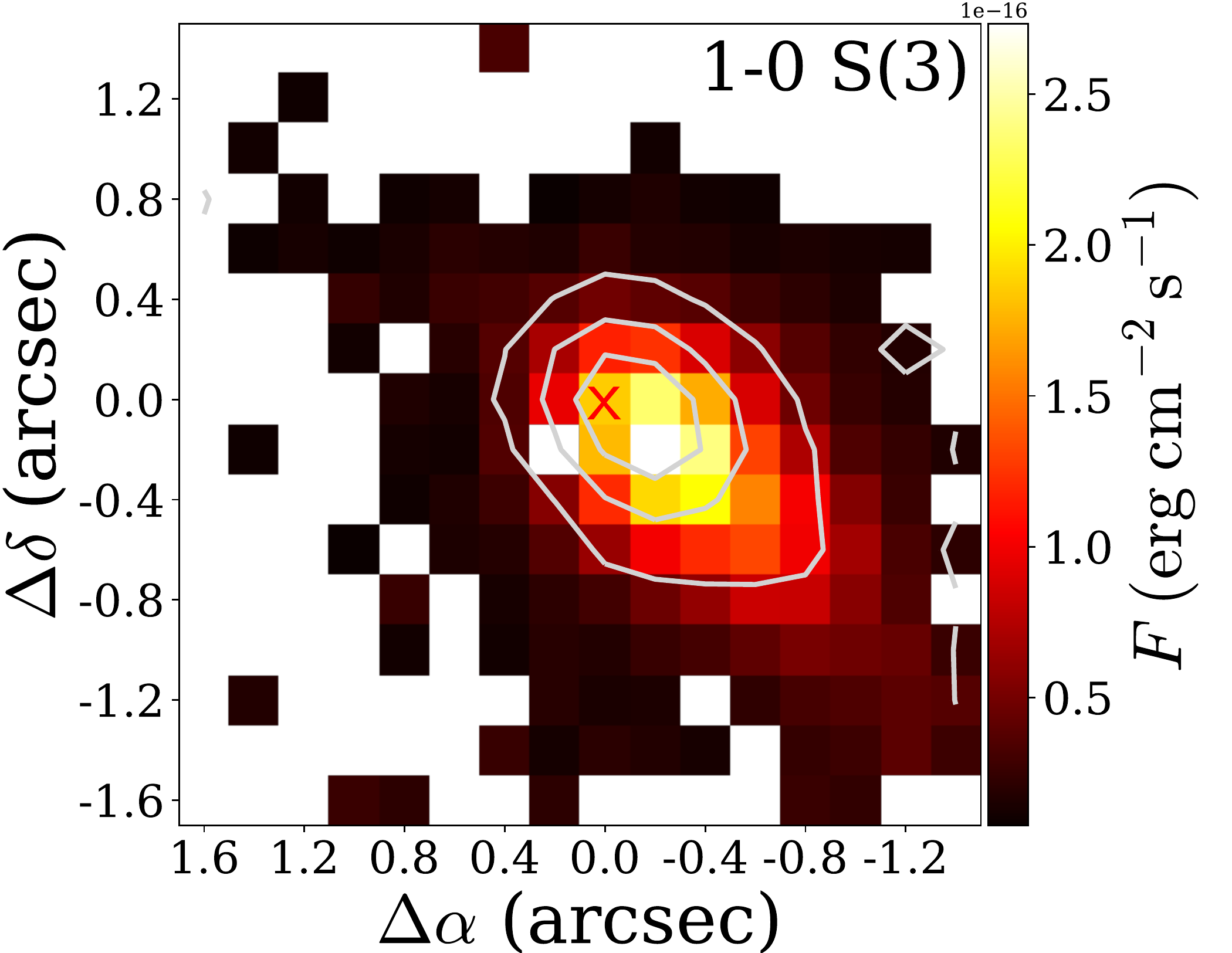}\hspace{-0.2cm}
\includegraphics[width=0.2\textwidth]{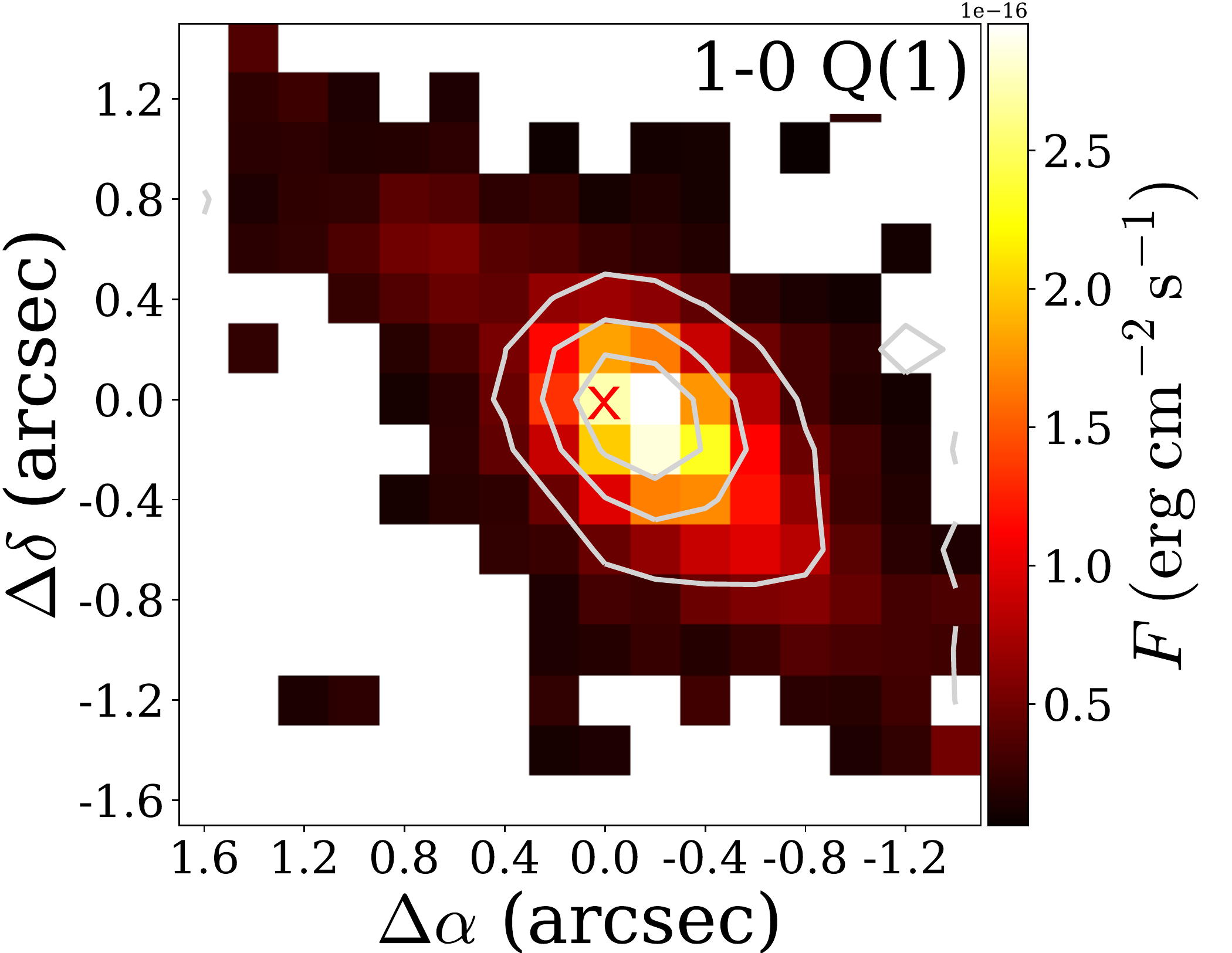}\hspace{-0.2cm}
\includegraphics[width=0.2\textwidth]{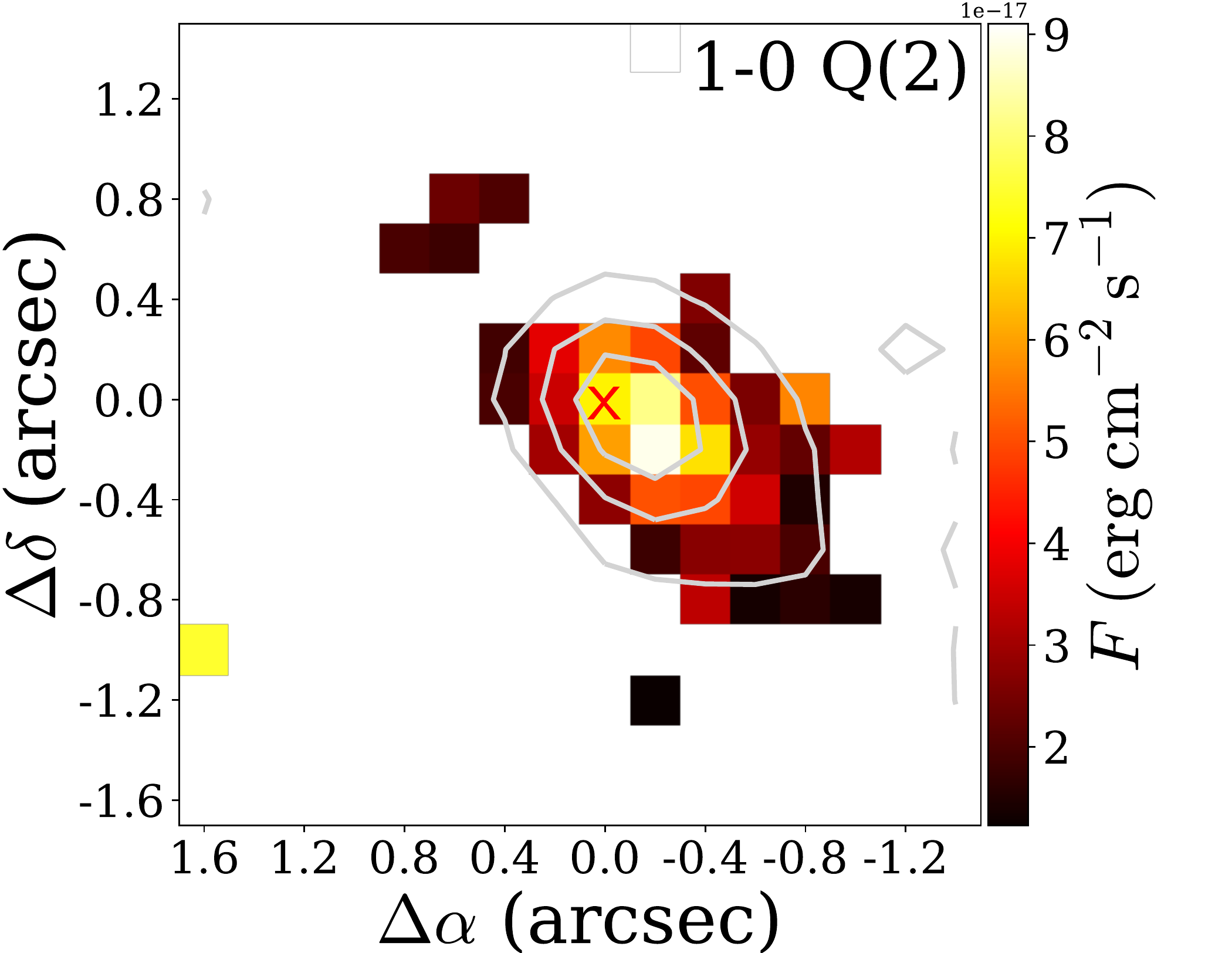}\hspace{-0.2cm}
\includegraphics[width=0.2\textwidth]{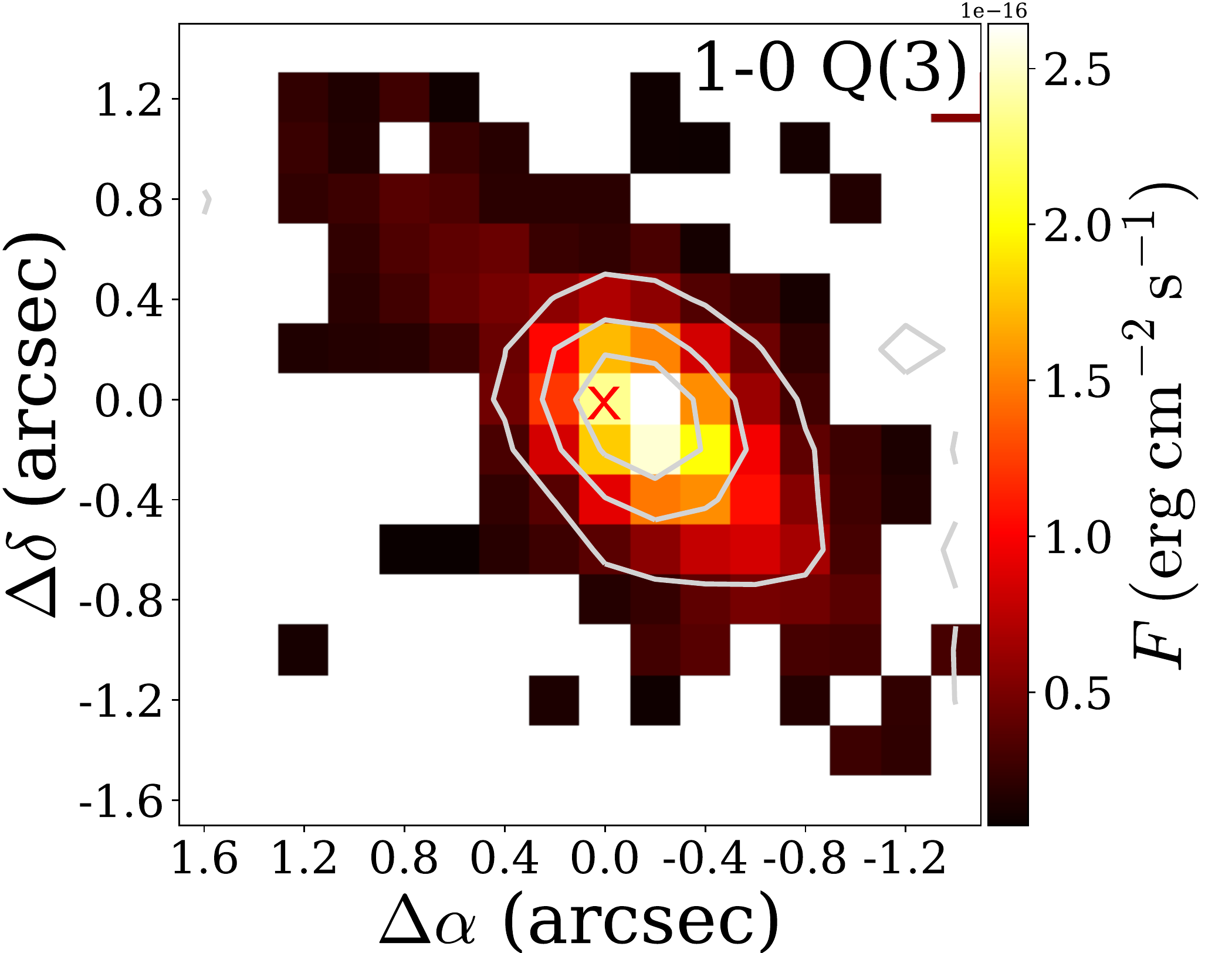}\hspace{-0.1cm}
\includegraphics[width=0.2\textwidth]{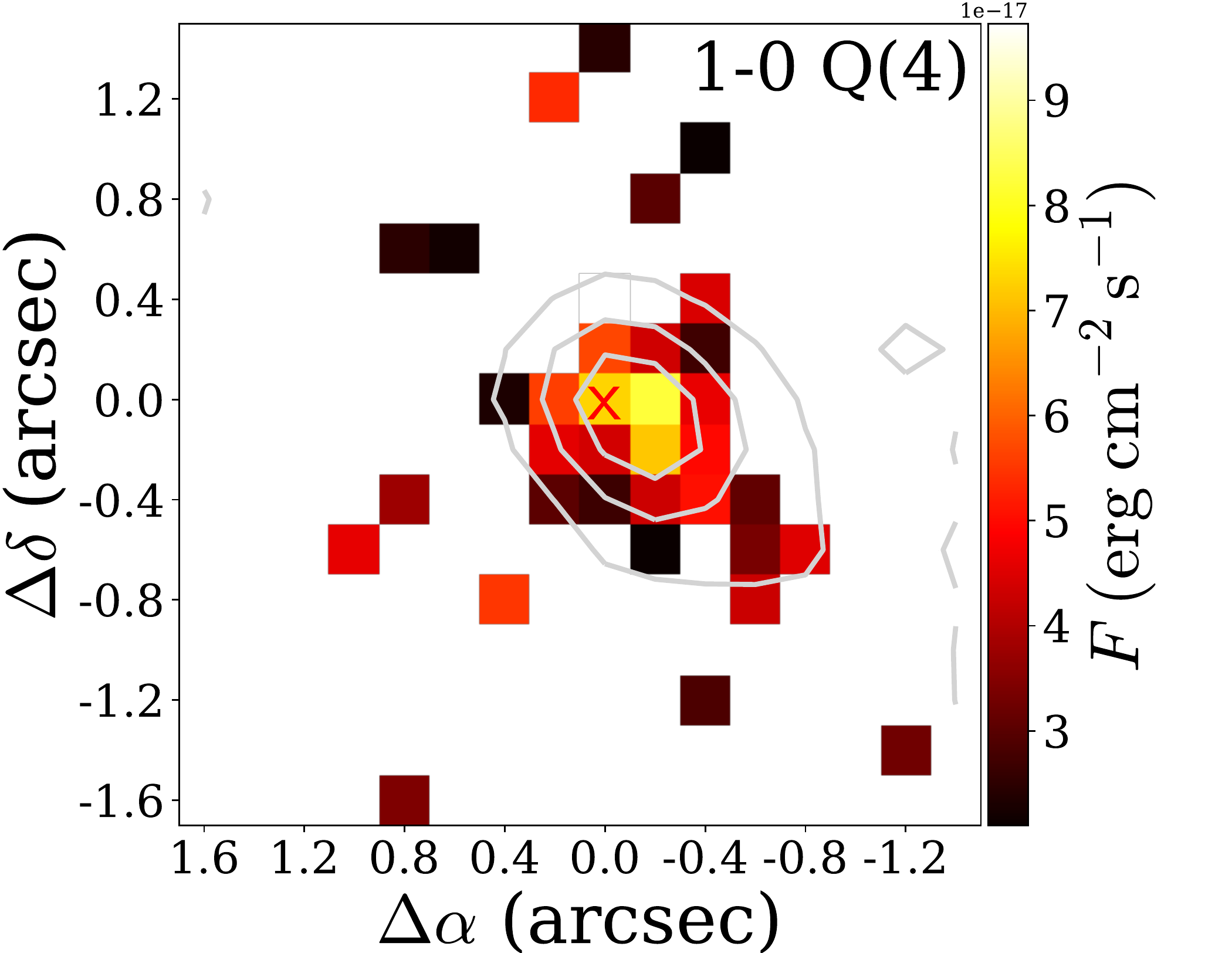}\hspace{-0.1cm}
\includegraphics[width=0.2\textwidth]{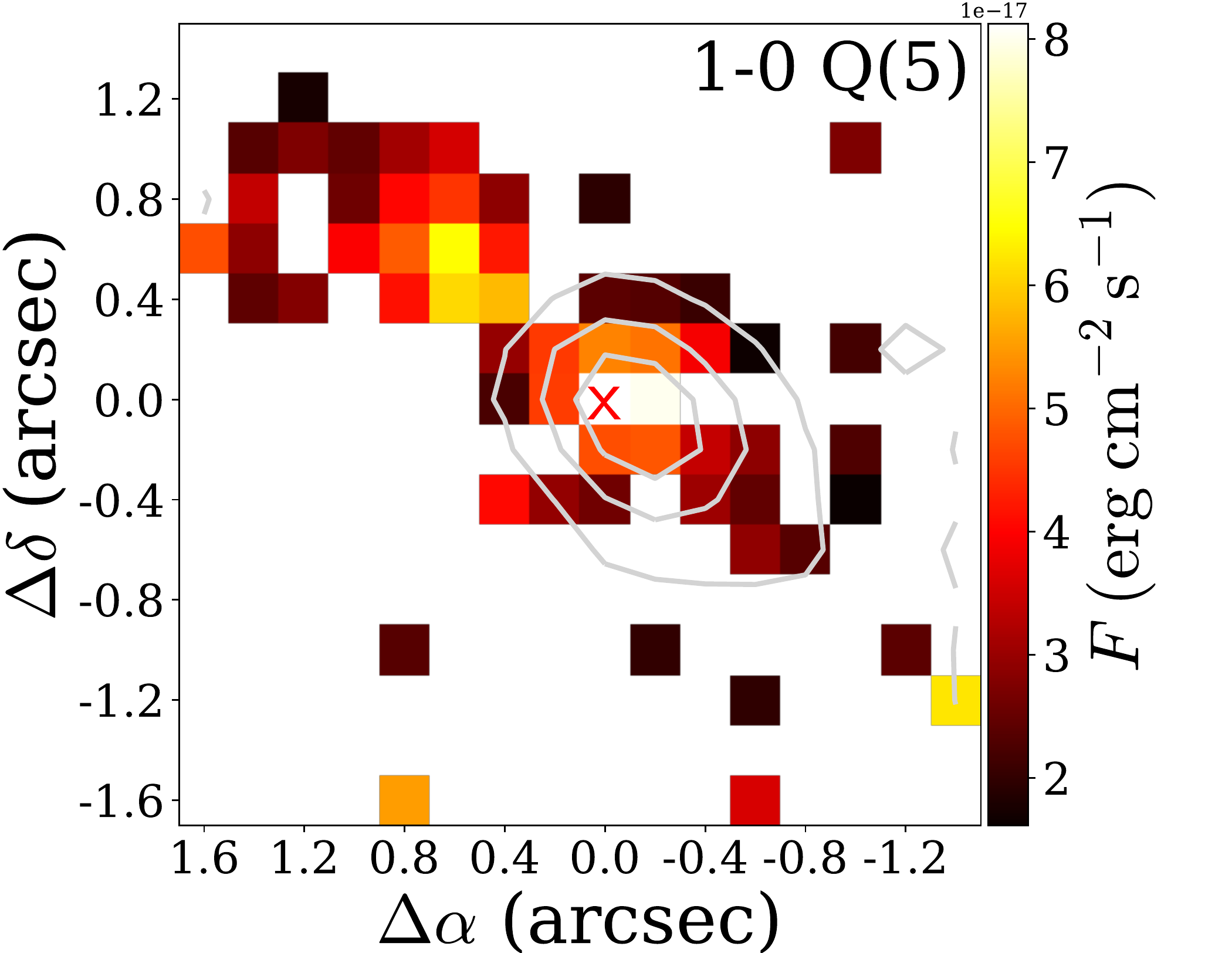}\hspace{-0.1cm}
\includegraphics[width=0.2\textwidth]{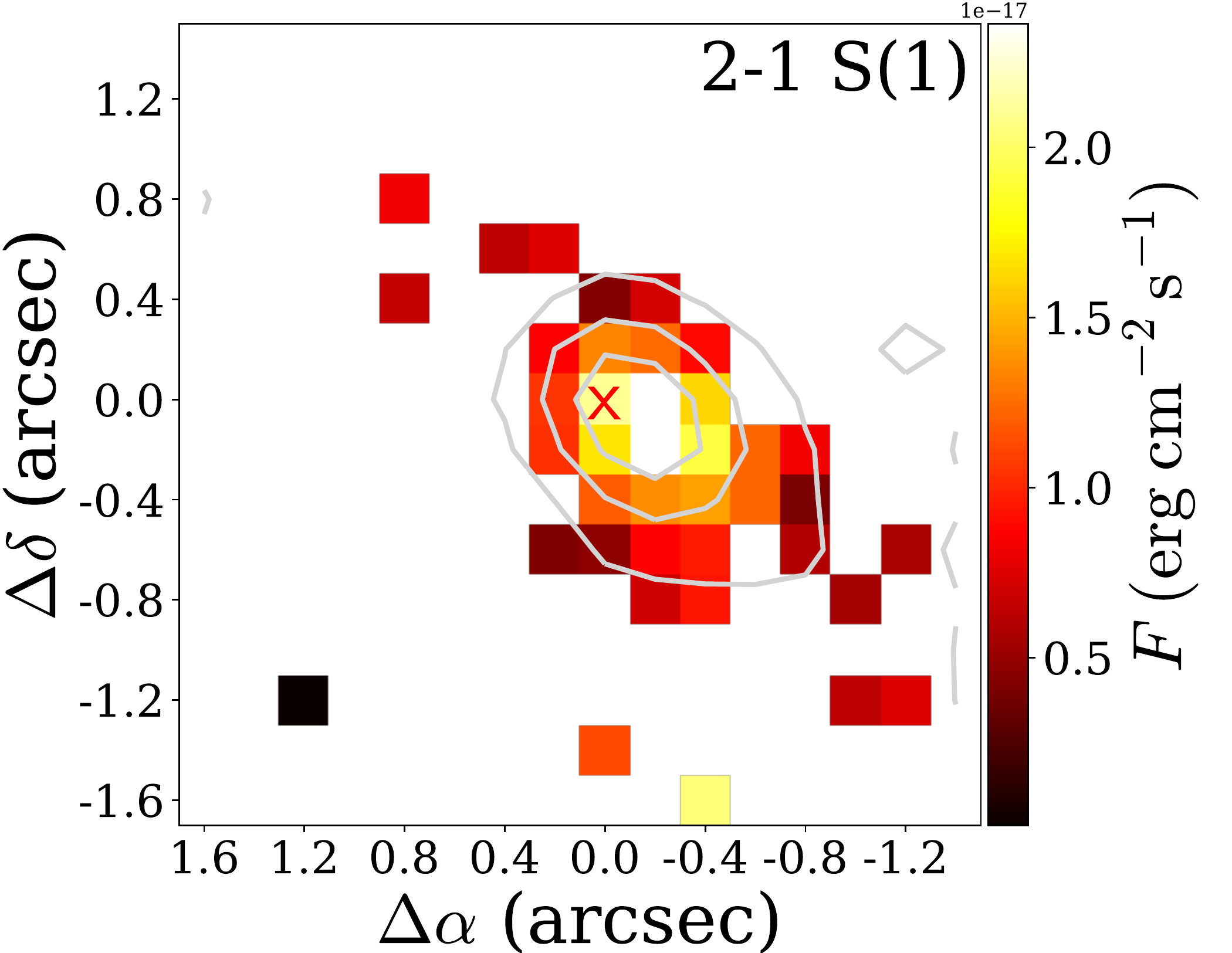}\hspace{-0.1cm}
\includegraphics[width=0.2\textwidth]{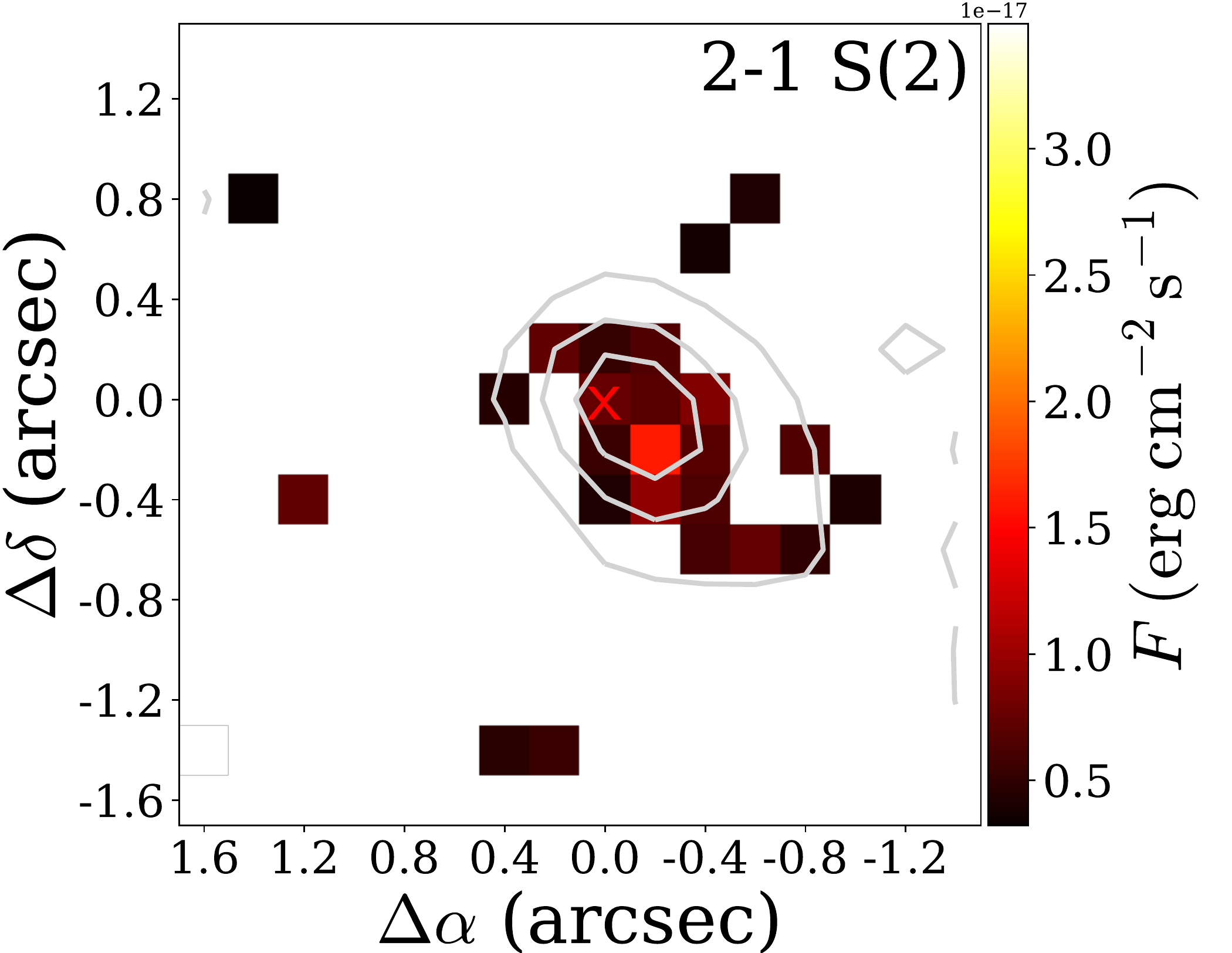}\hspace{-0.1cm}
\includegraphics[width=0.2\textwidth]{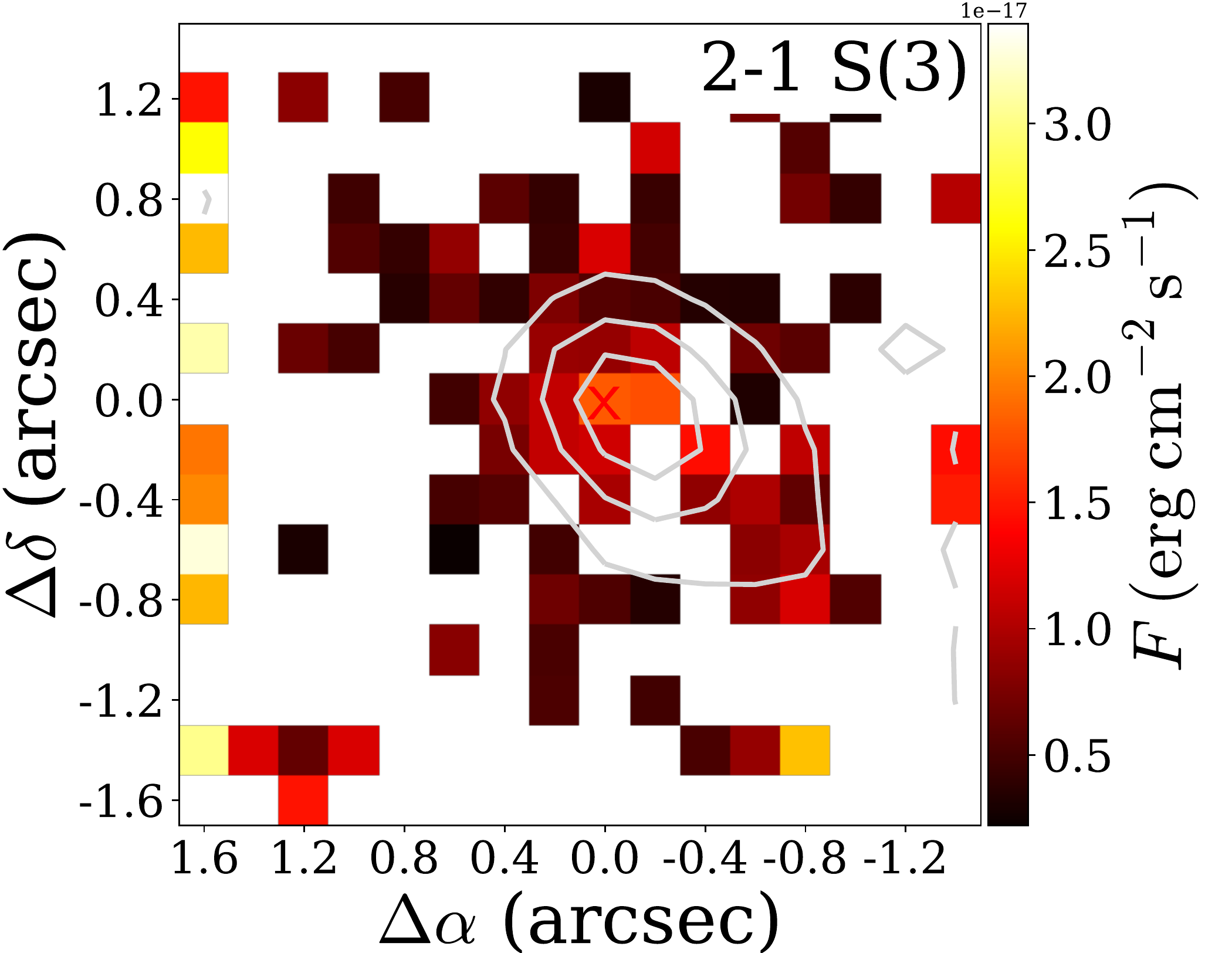}\hspace{-0.1cm}
\caption{Maps in several H$_2$ lines in source No. 47, which shows detections of the highest-excited H$_2$ lines in the entire sample (see Table \ref{tab:coordinates}).}
\label{fig:emiss-1512183}
\end{figure*}

\begin{figure*}[h!]
\includegraphics[width=0.2\textwidth]{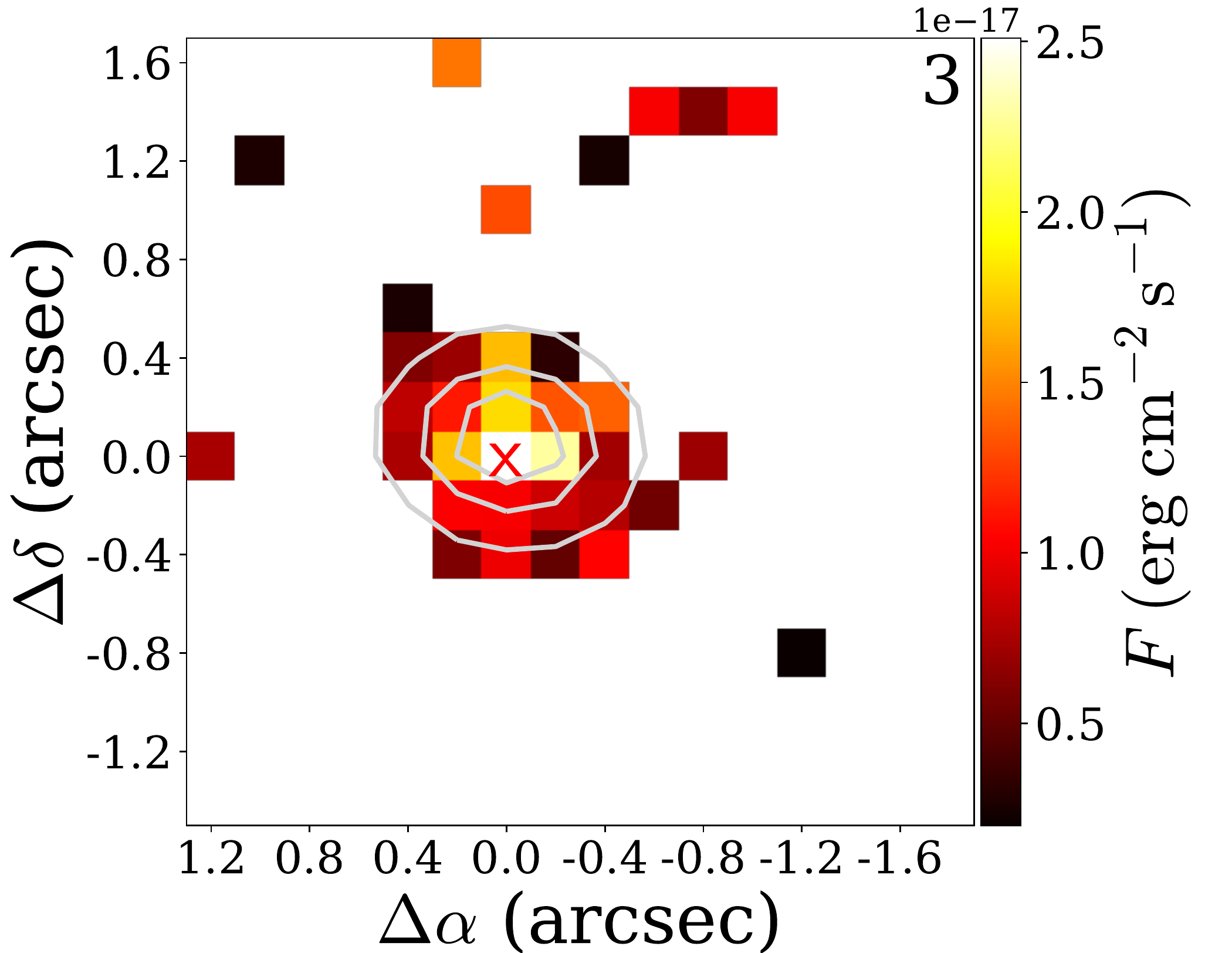}\hspace{-0.2cm}
\includegraphics[width=0.2\textwidth]{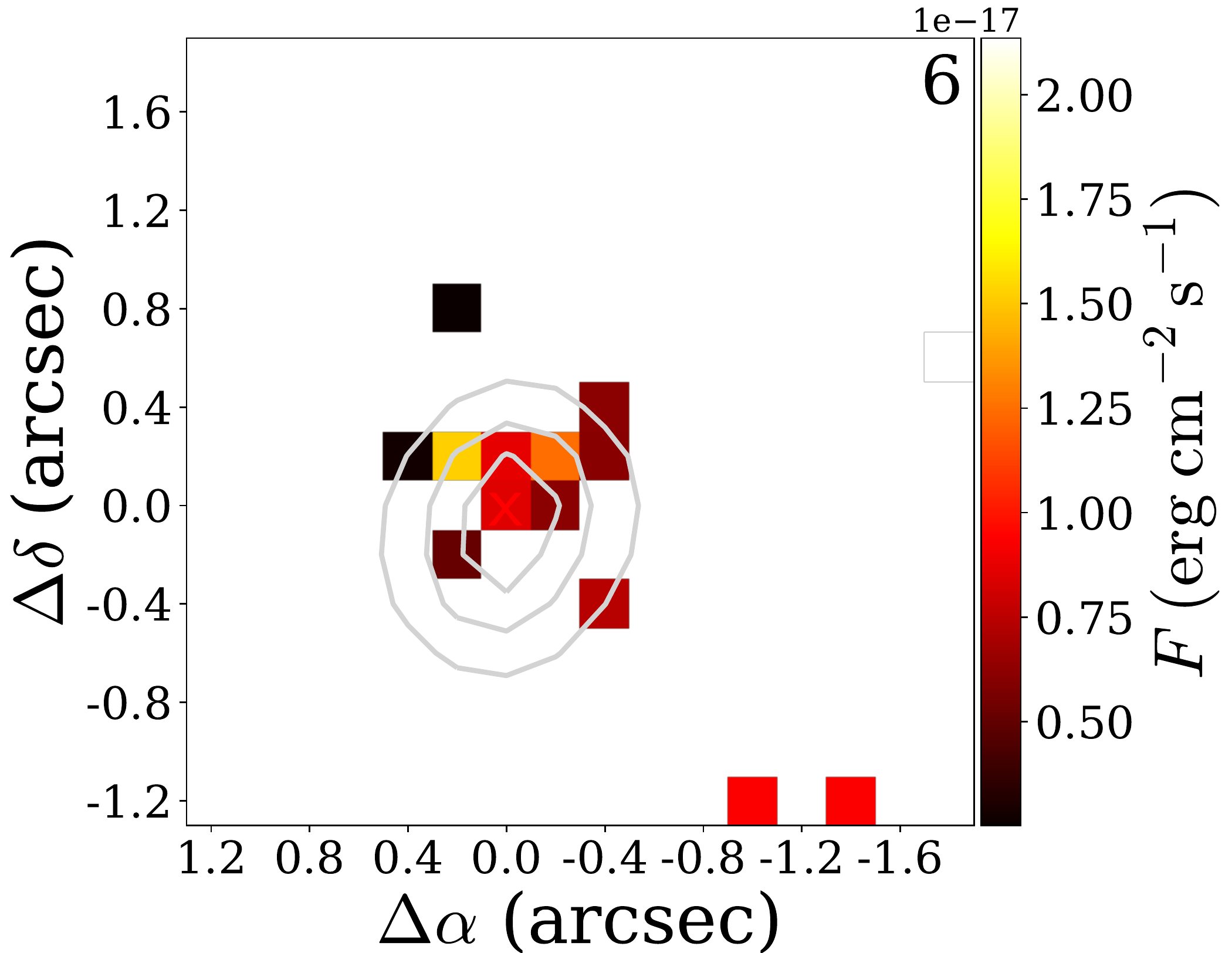}\hspace{-0.2cm}
\includegraphics[width=0.2\textwidth]{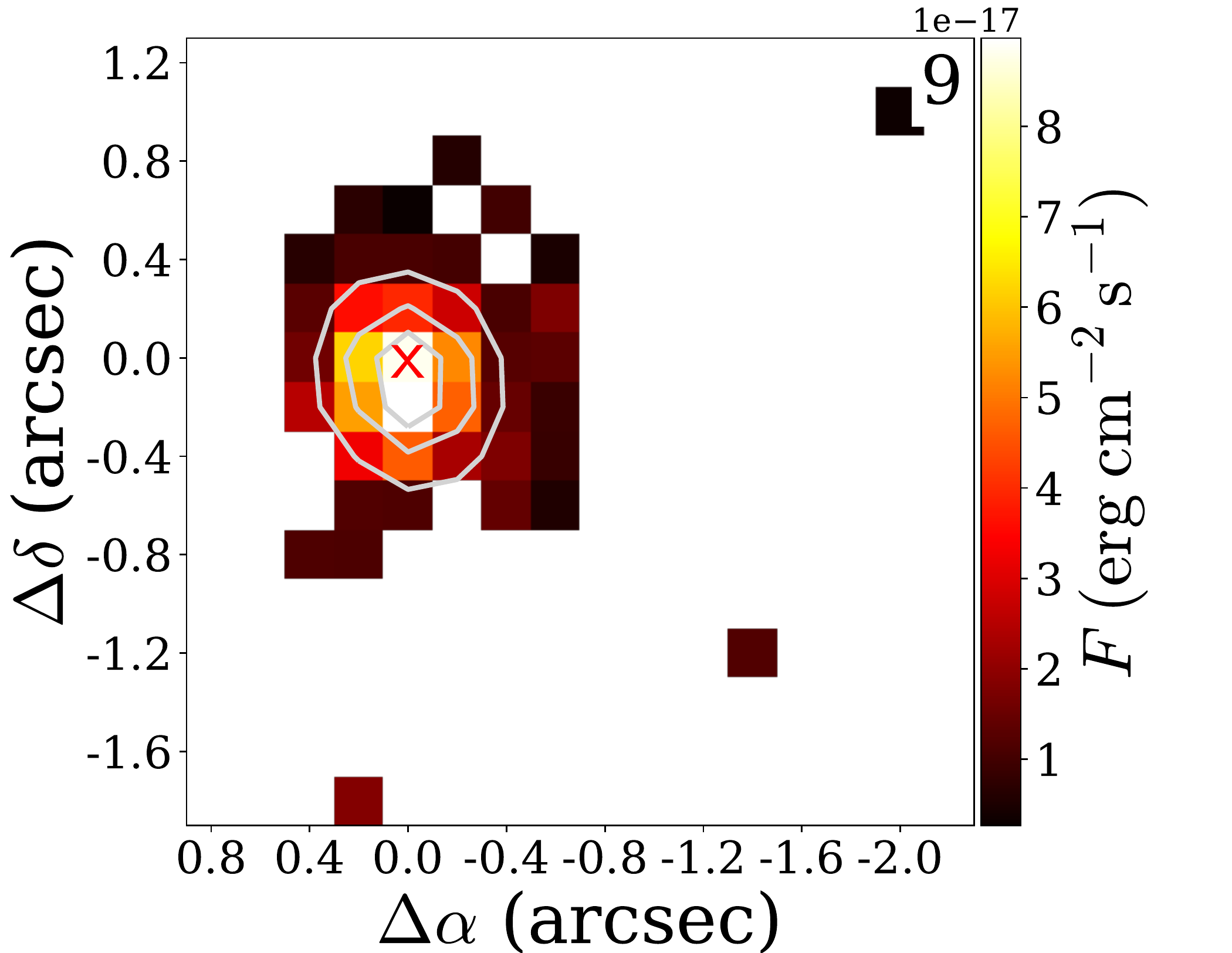}\hspace{-0.2cm}
\includegraphics[width=0.2\textwidth]{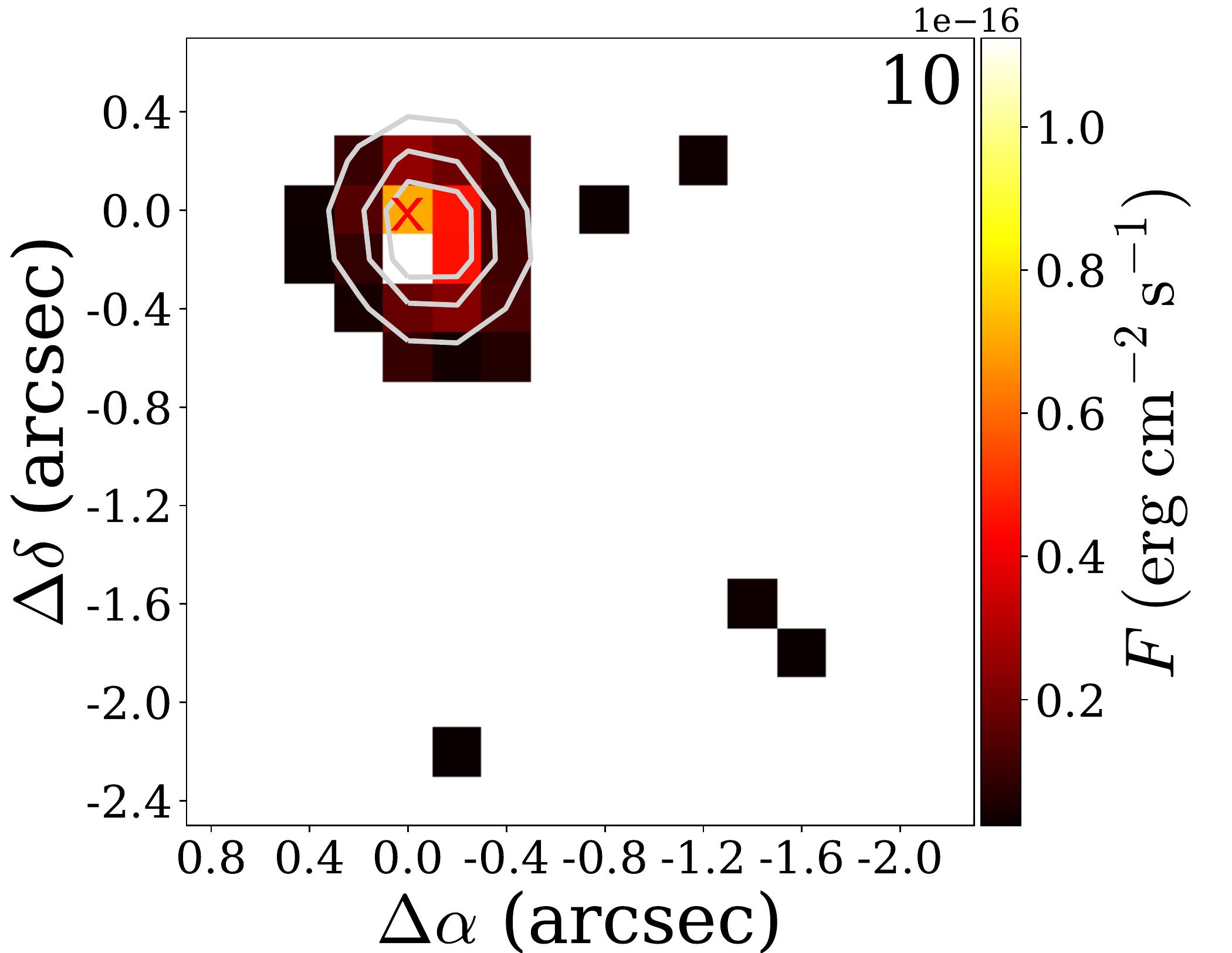}\hspace{-0.2cm}
\includegraphics[width=0.2\textwidth]{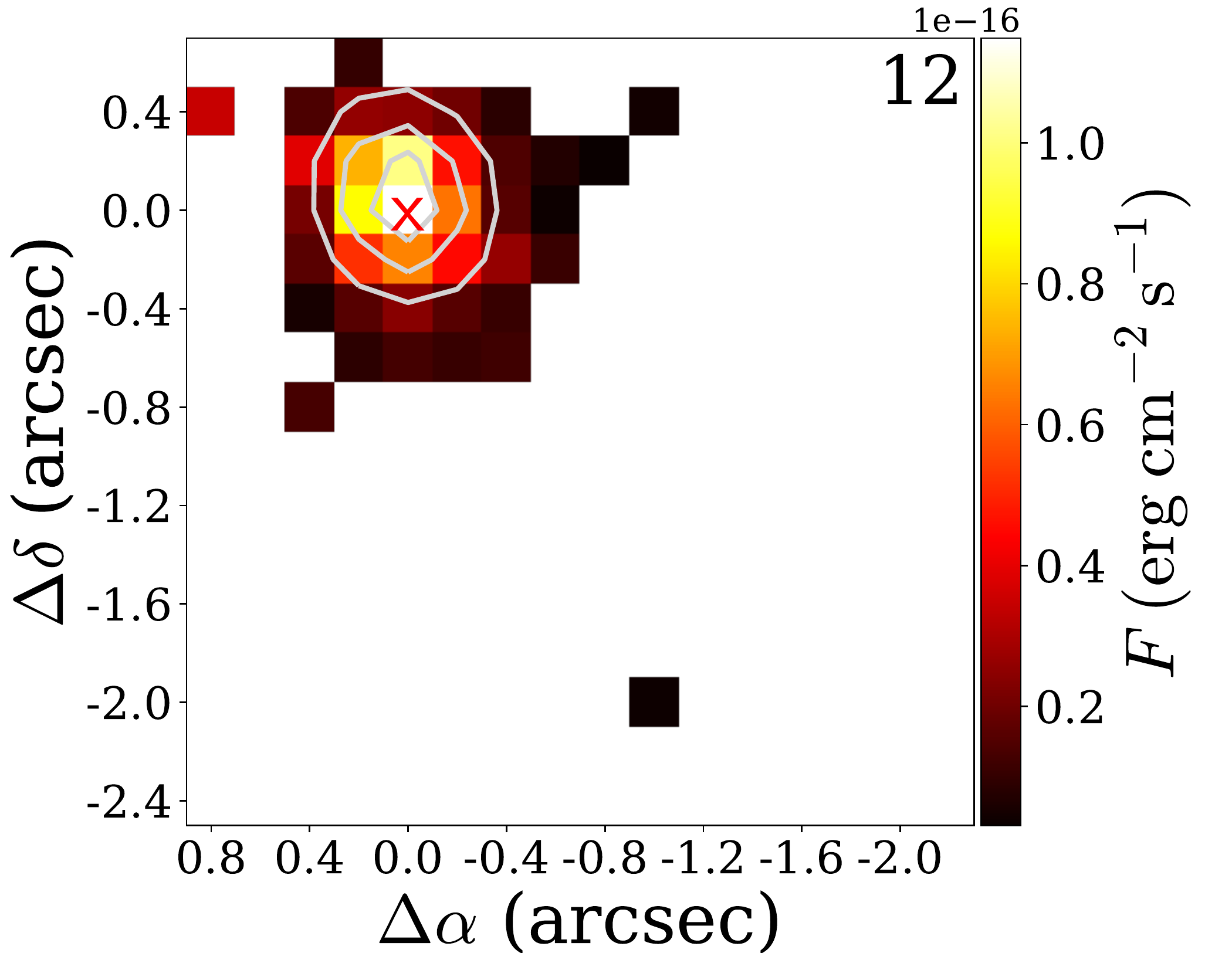}\hspace{-0.2cm}
\includegraphics[width=0.2\textwidth]{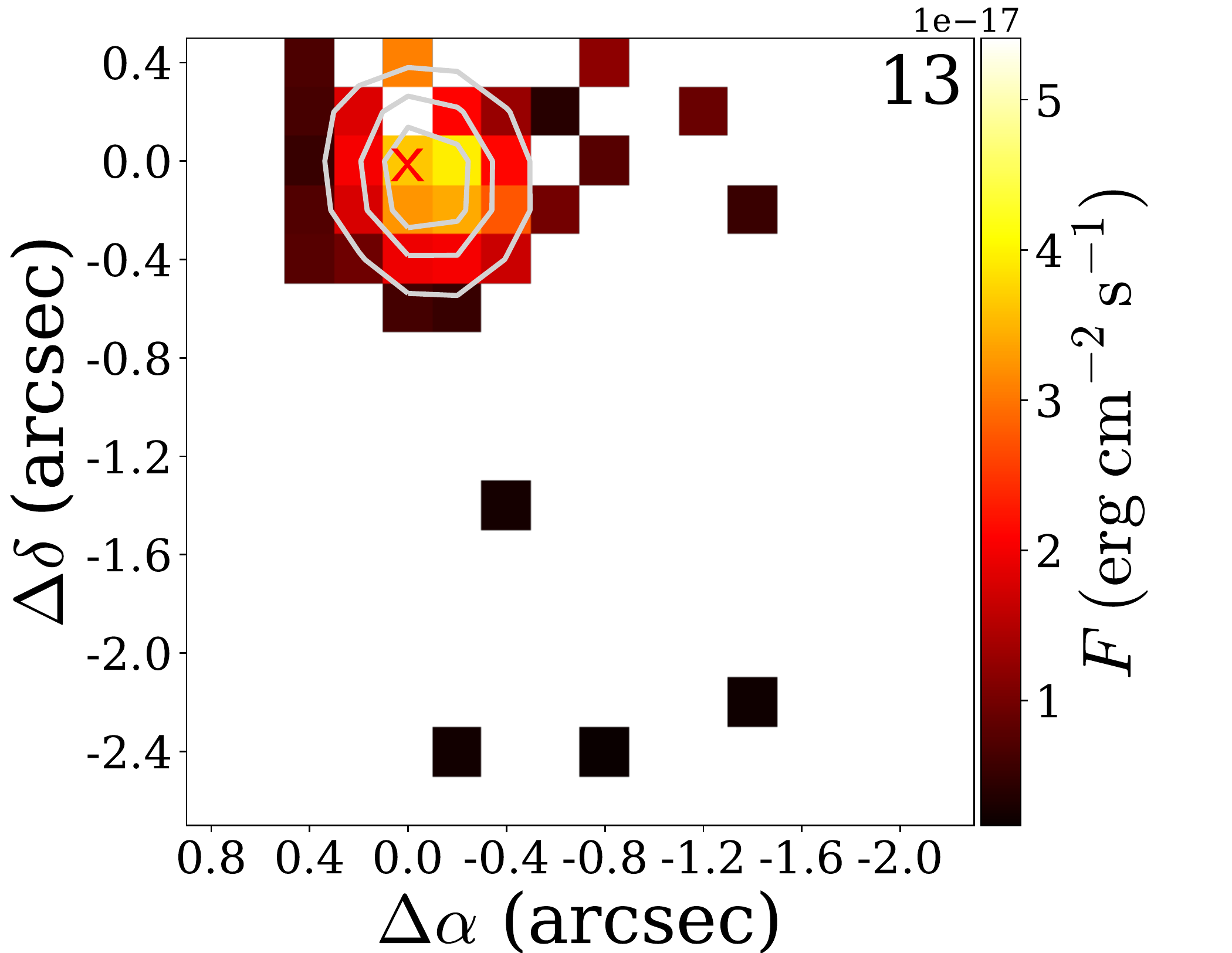}\hspace{-0.2cm}
\includegraphics[width=0.2\textwidth]{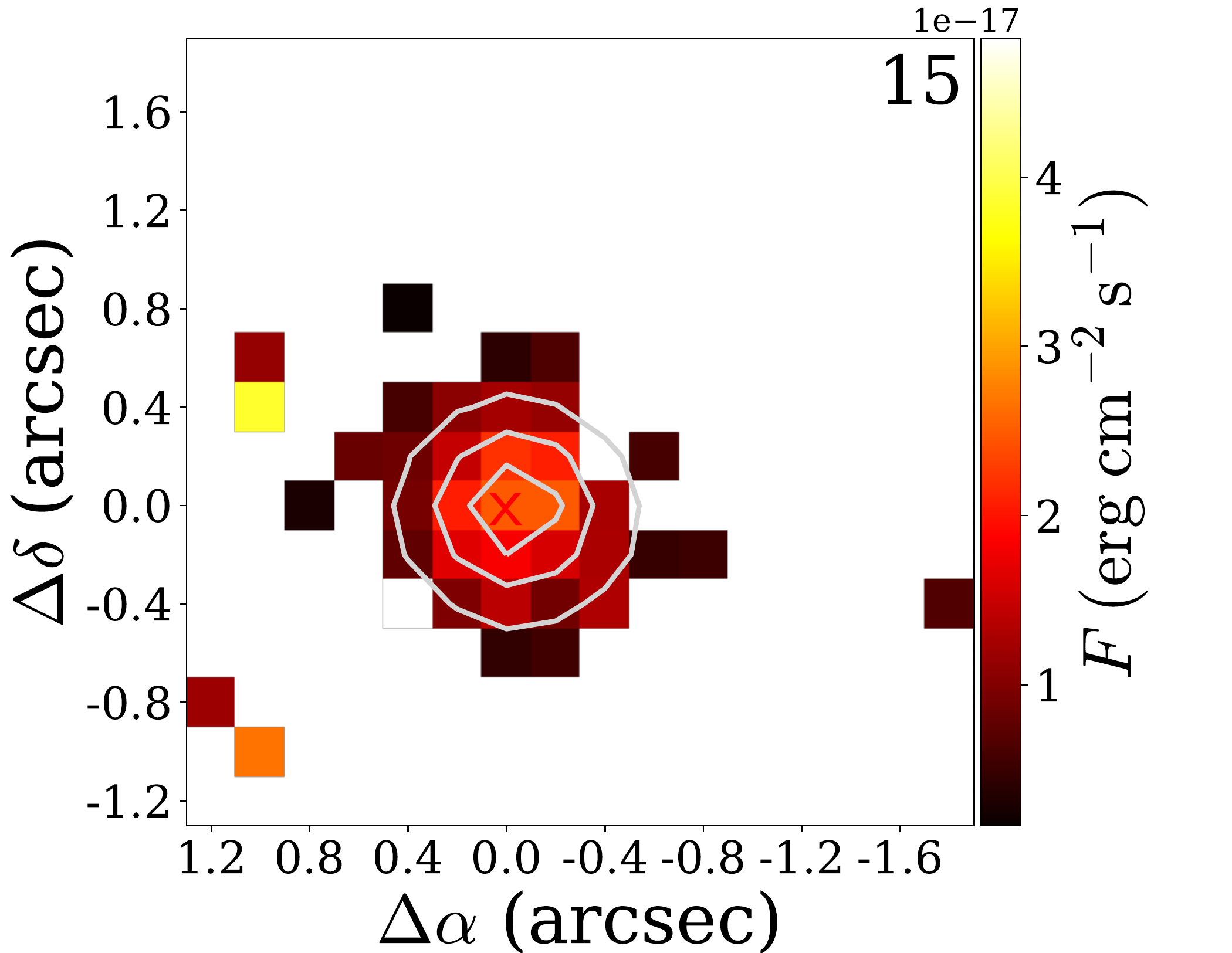}\hspace{-0.2cm}
\includegraphics[width=0.2\textwidth]{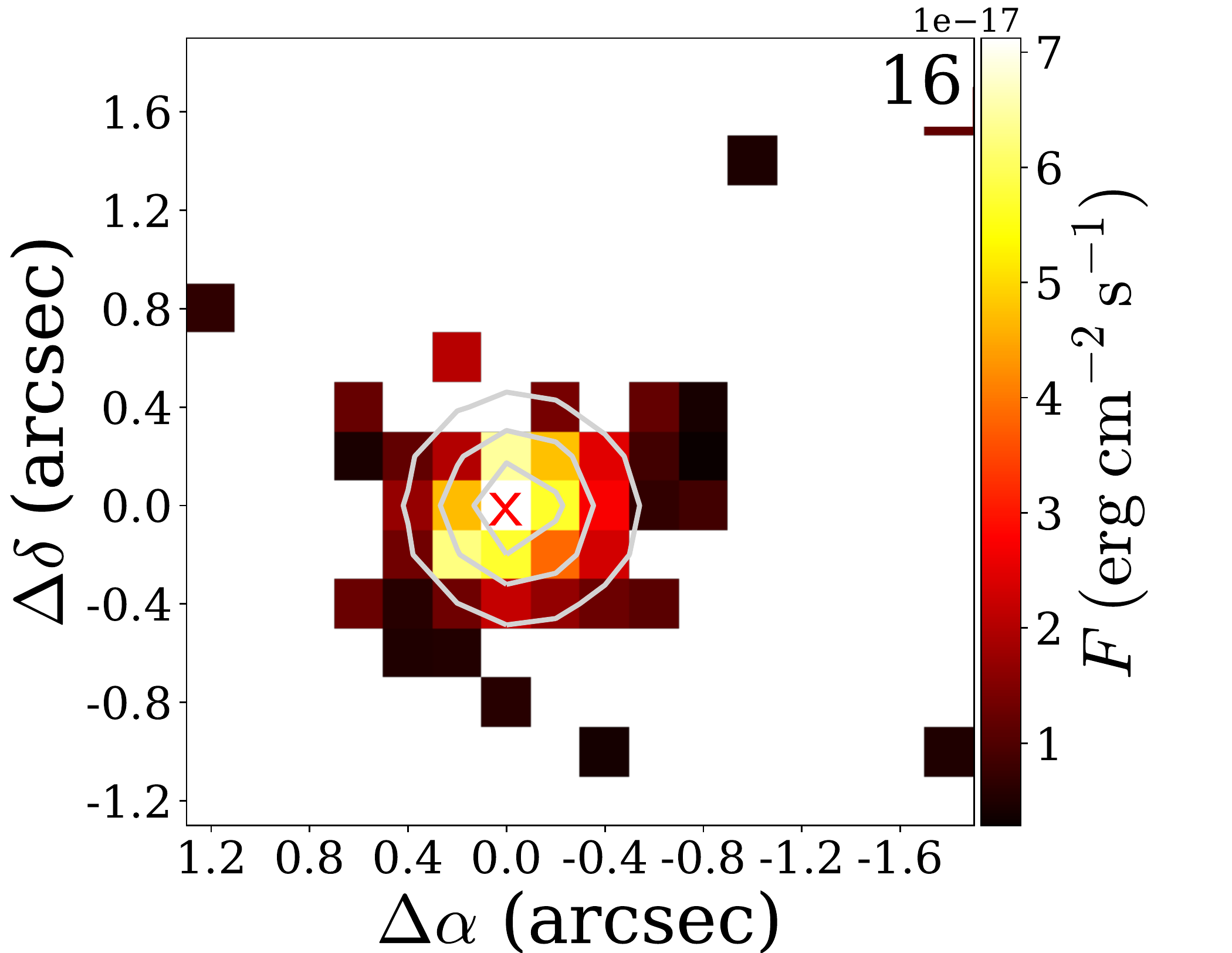}\hspace{-0.2cm}
\includegraphics[width=0.2\textwidth]{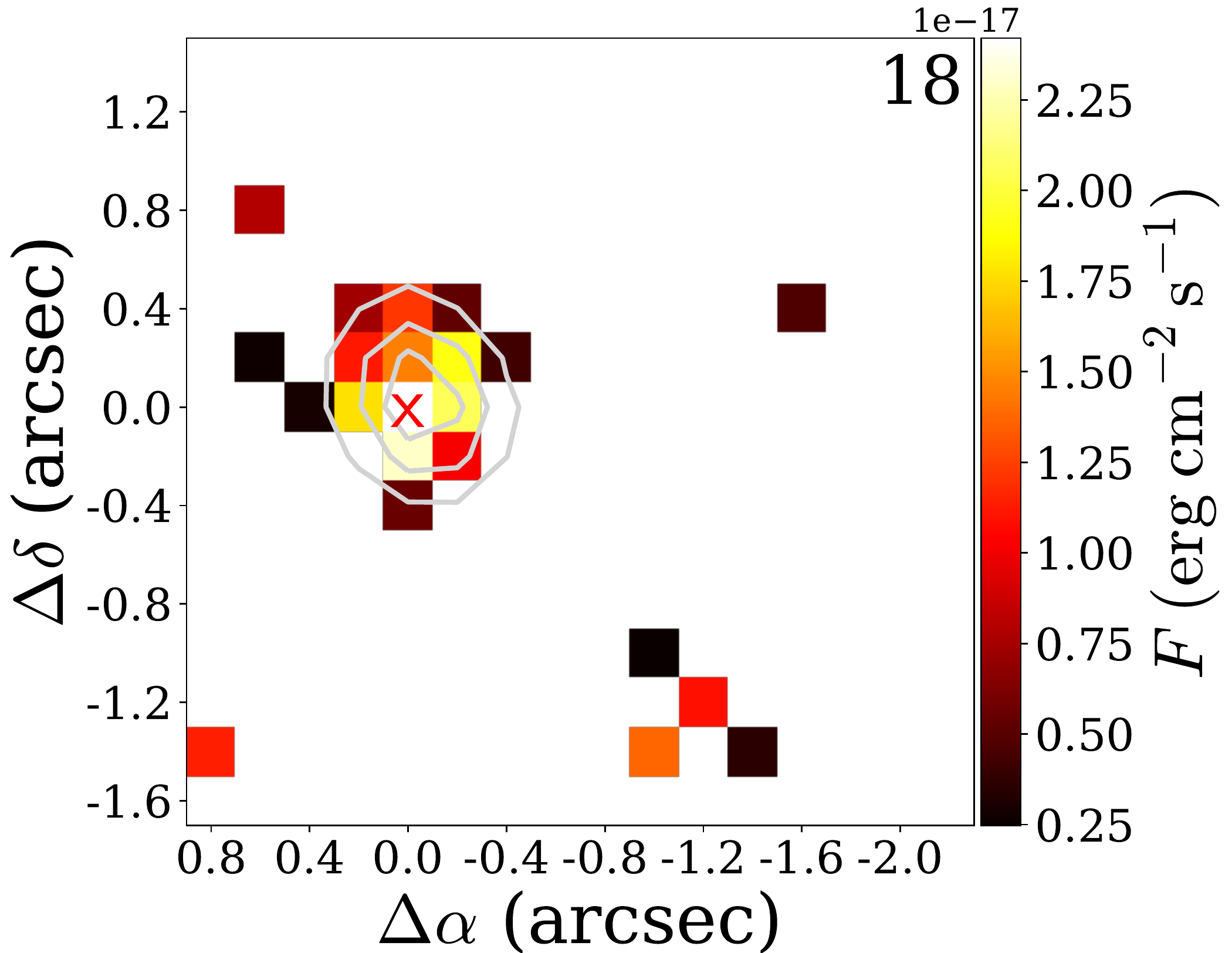}\hspace{-0.2cm}
\includegraphics[width=0.2\textwidth]{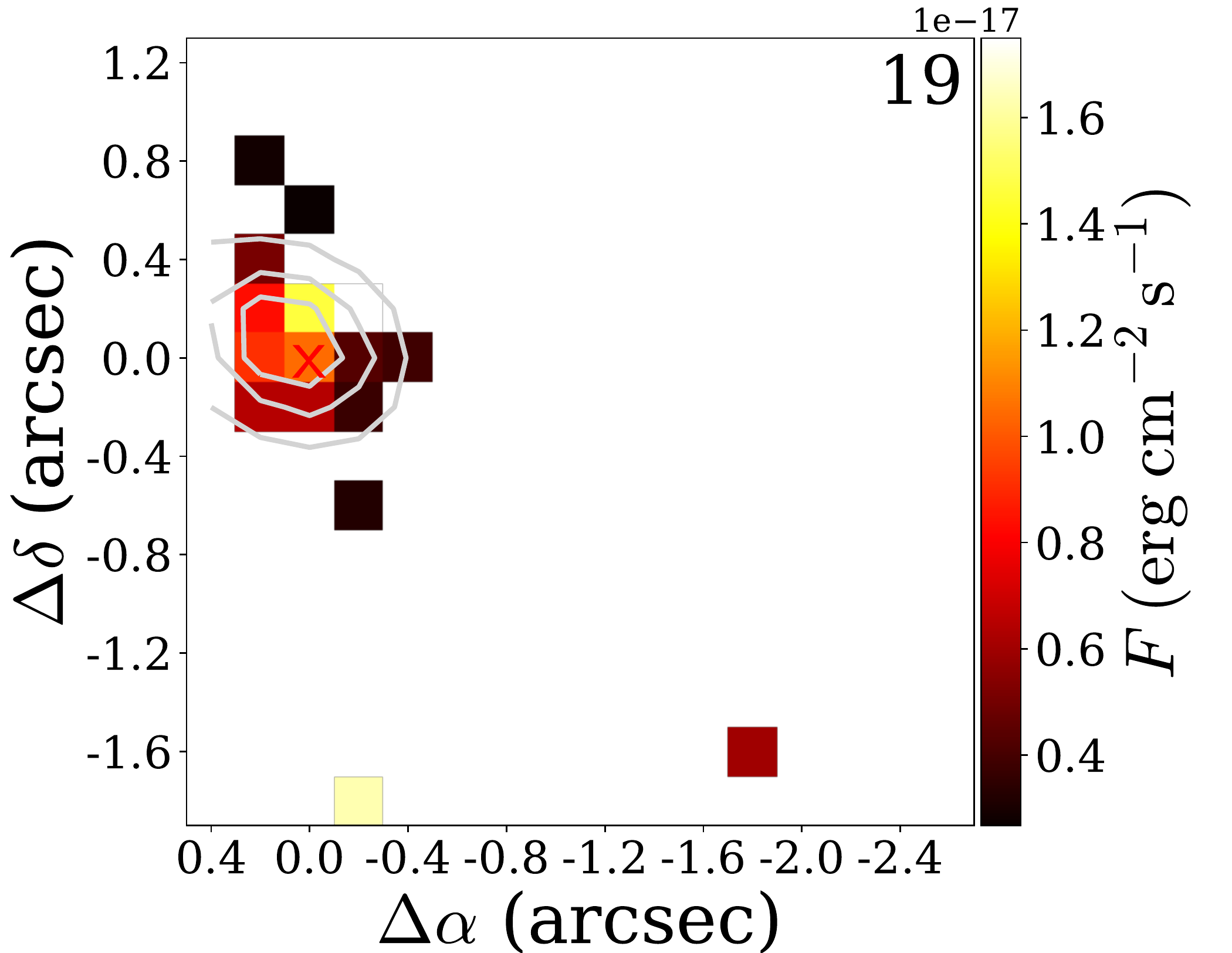}\hspace{-0.2cm}
\includegraphics[width=0.2\textwidth]{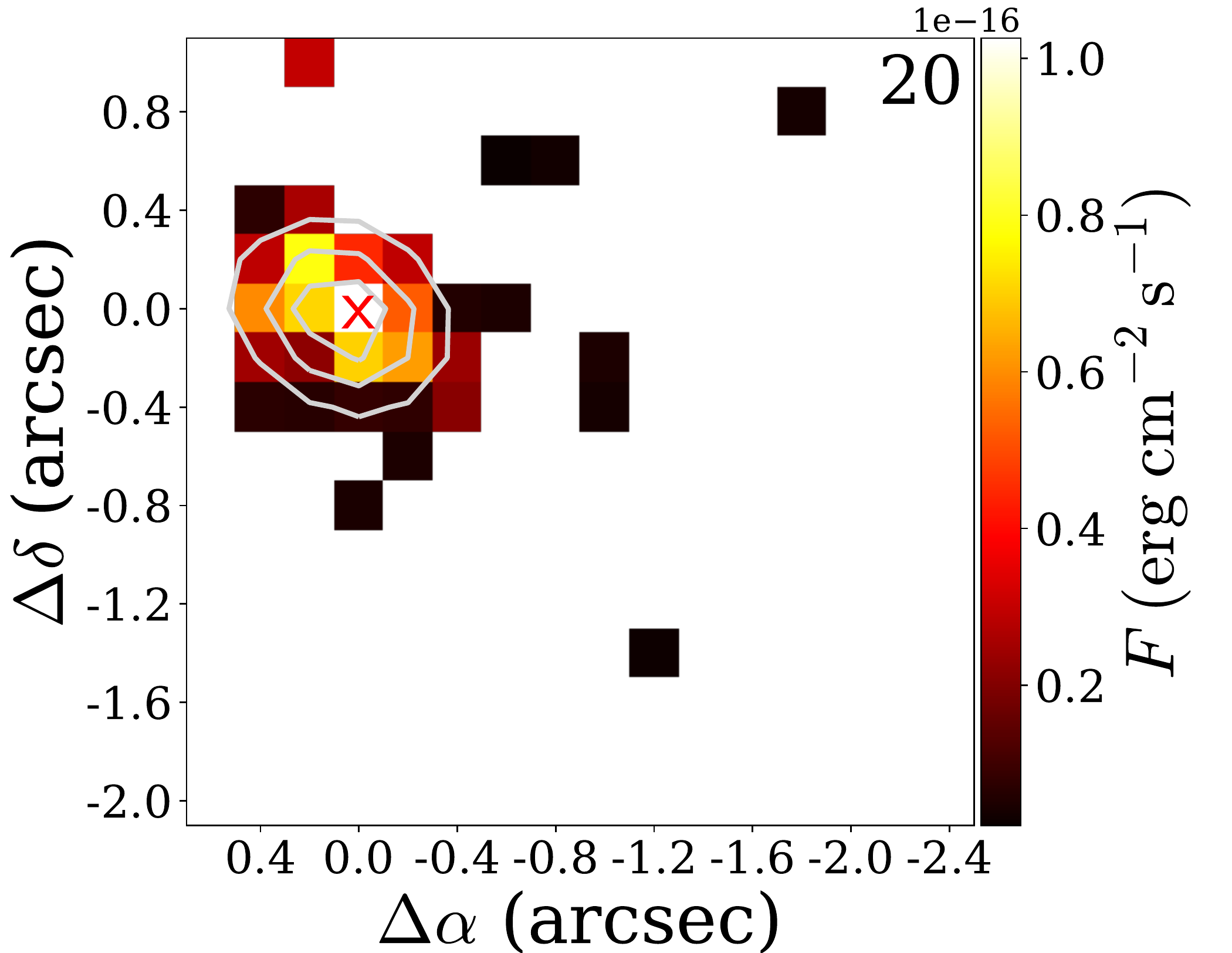}\hspace{-0.2cm}
\includegraphics[width=0.2\textwidth]{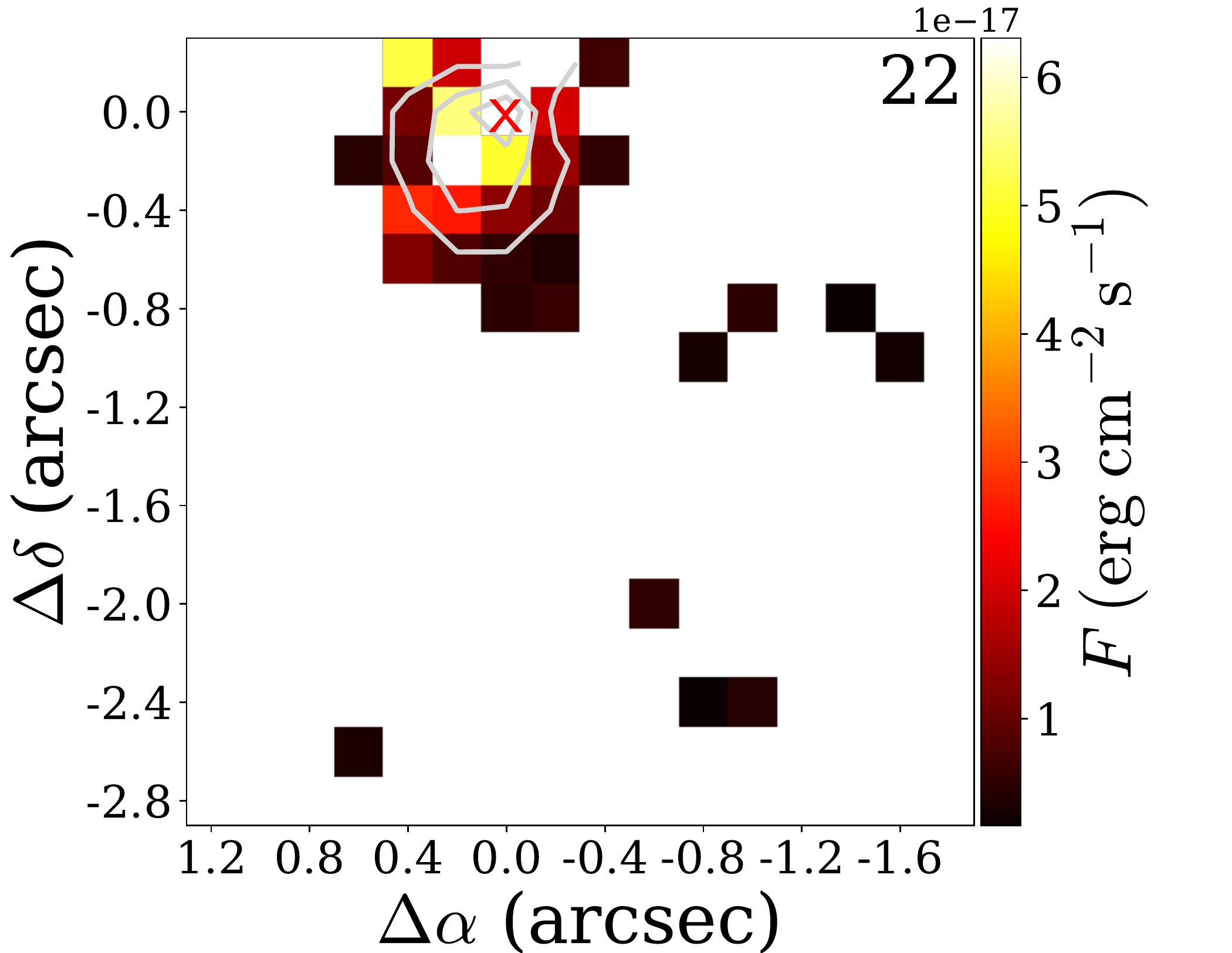}\hspace{-0.2cm}
\includegraphics[width=0.2\textwidth]{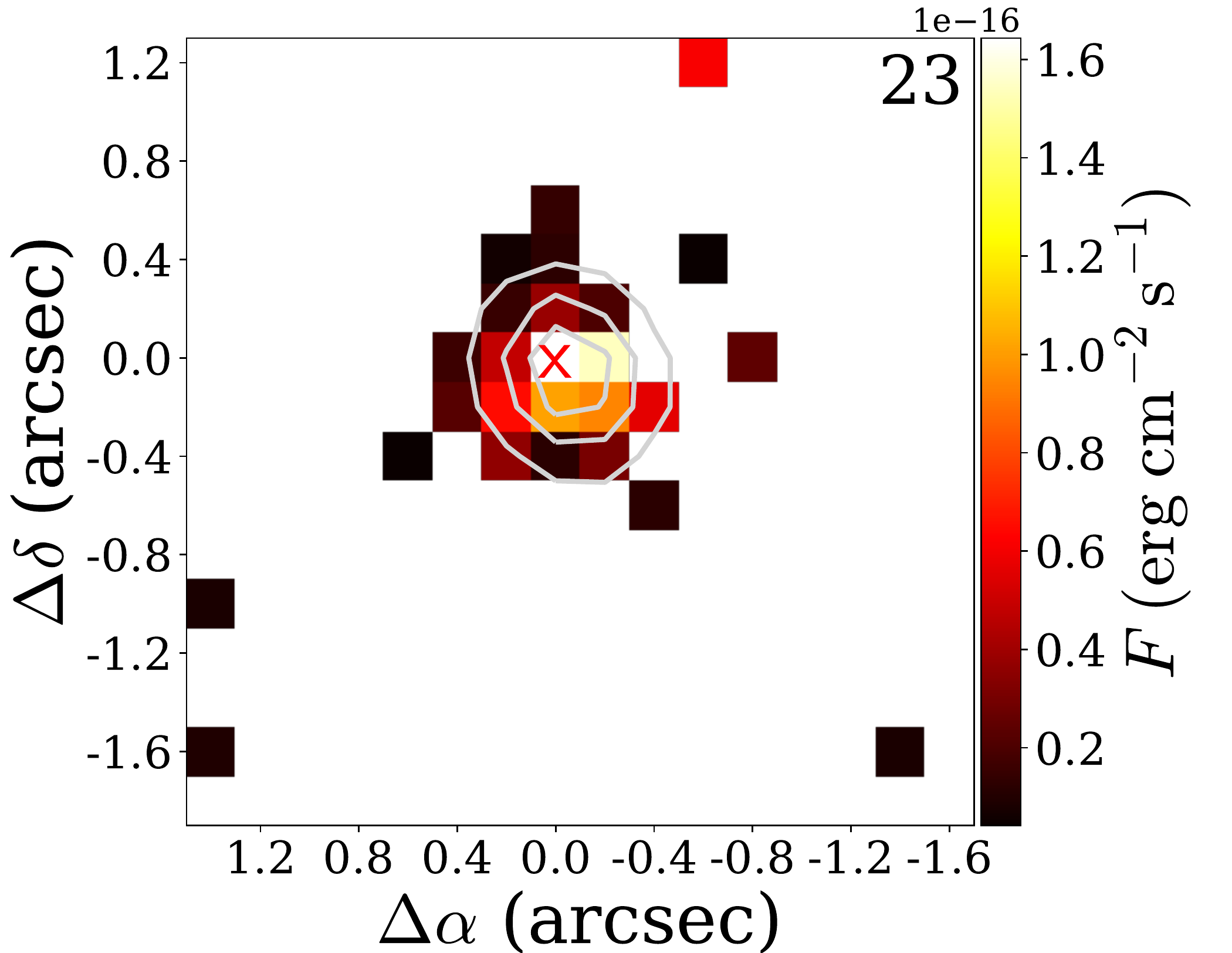}\hspace{-0.2cm}
\includegraphics[width=0.2\textwidth]{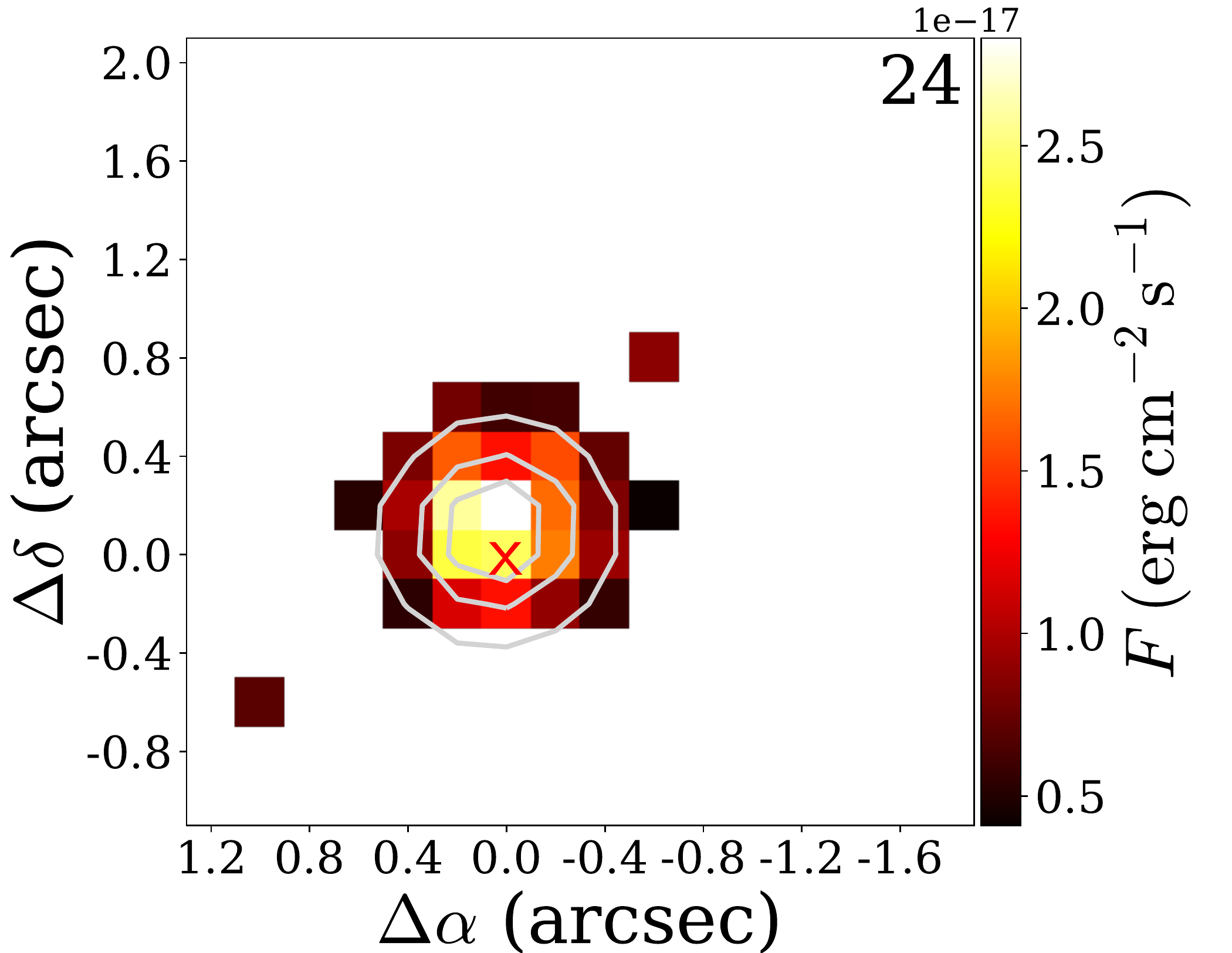}\hspace{-0.2cm}
\includegraphics[width=0.2\textwidth]{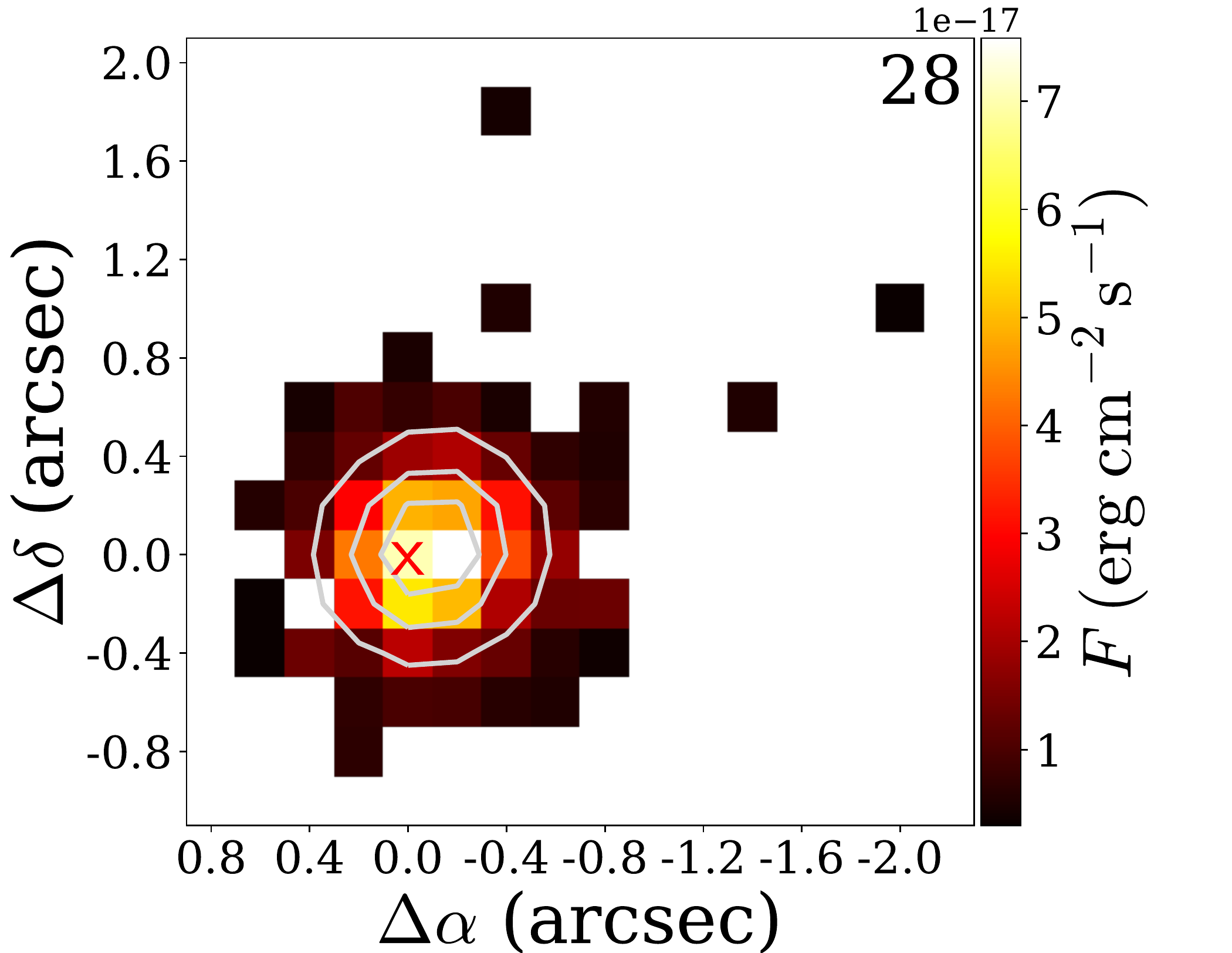}\hspace{-0.2cm}
\includegraphics[width=0.2\textwidth]{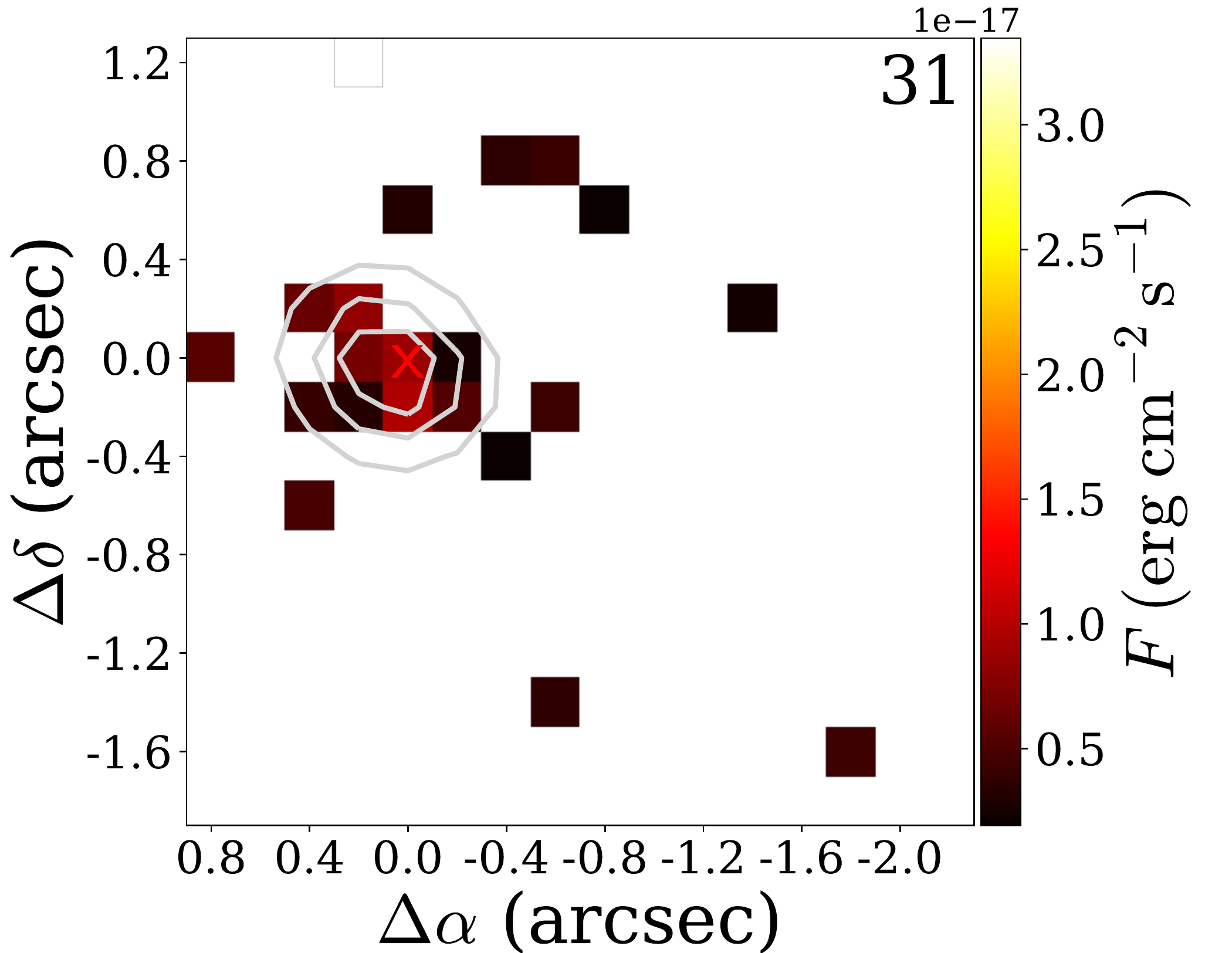}\hspace{-0.2cm}
\includegraphics[width=0.2\textwidth]{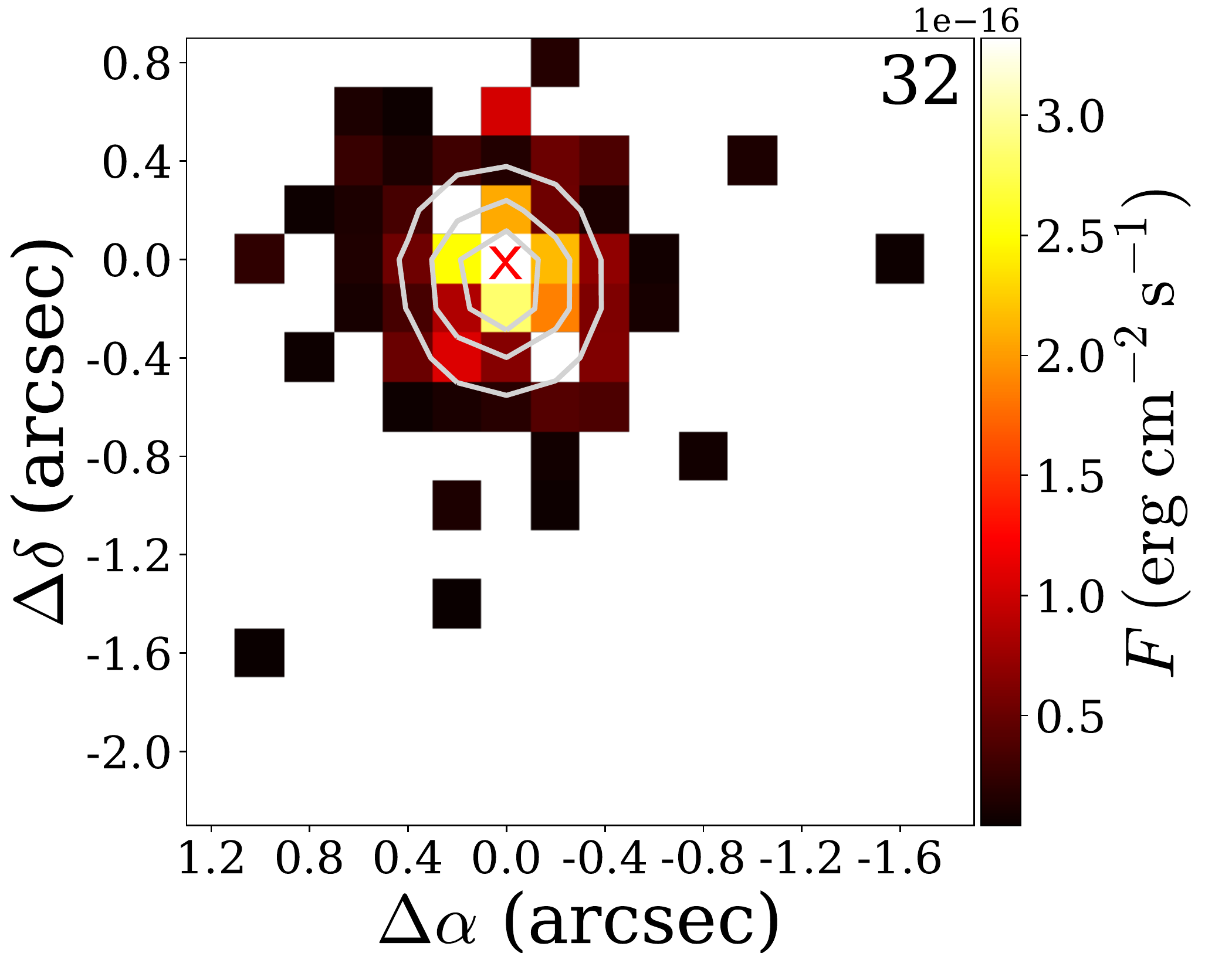}\hspace{-0.2cm} 
\includegraphics[width=0.2\textwidth]{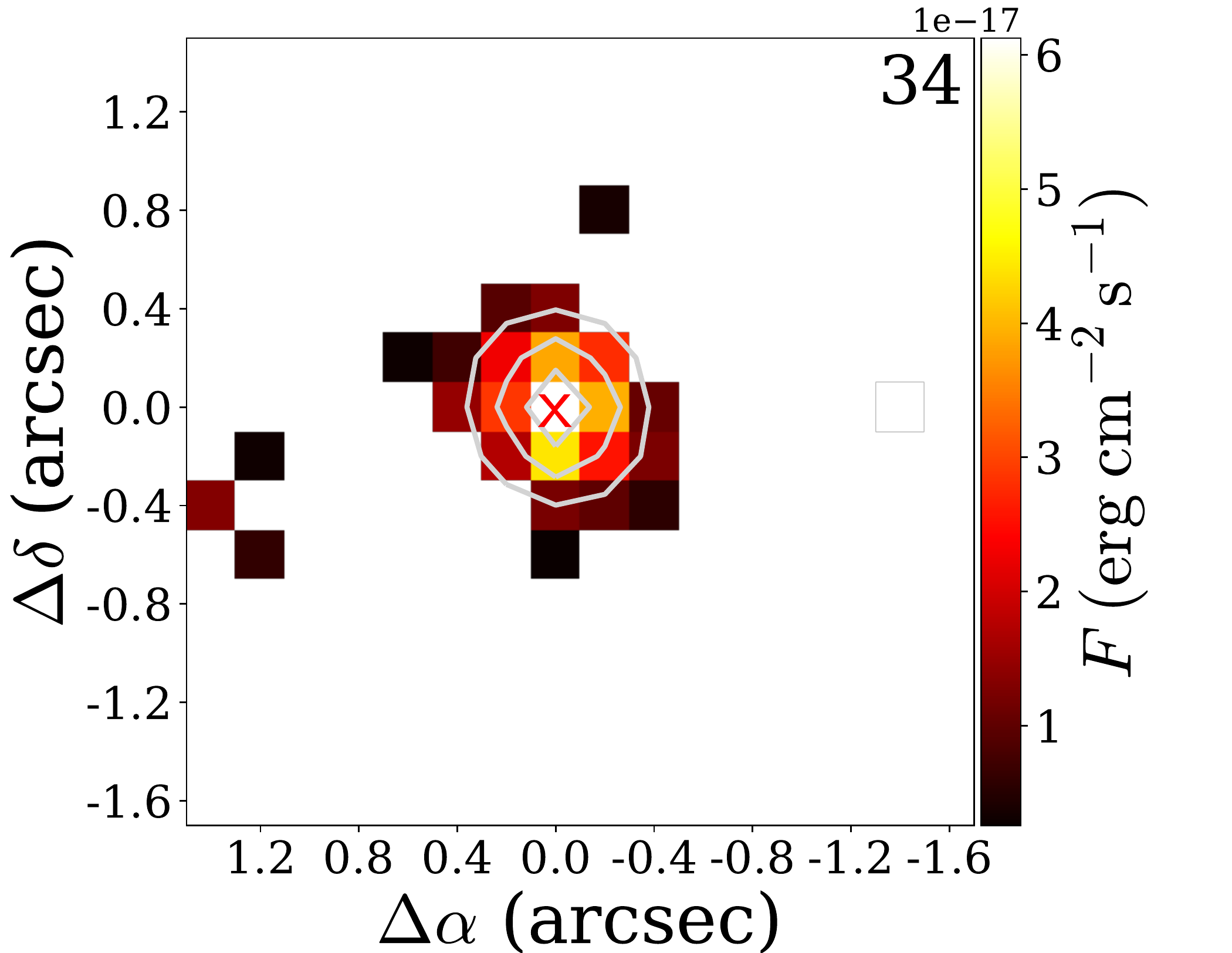}\hspace{-0.2cm}
\includegraphics[width=0.2\textwidth]{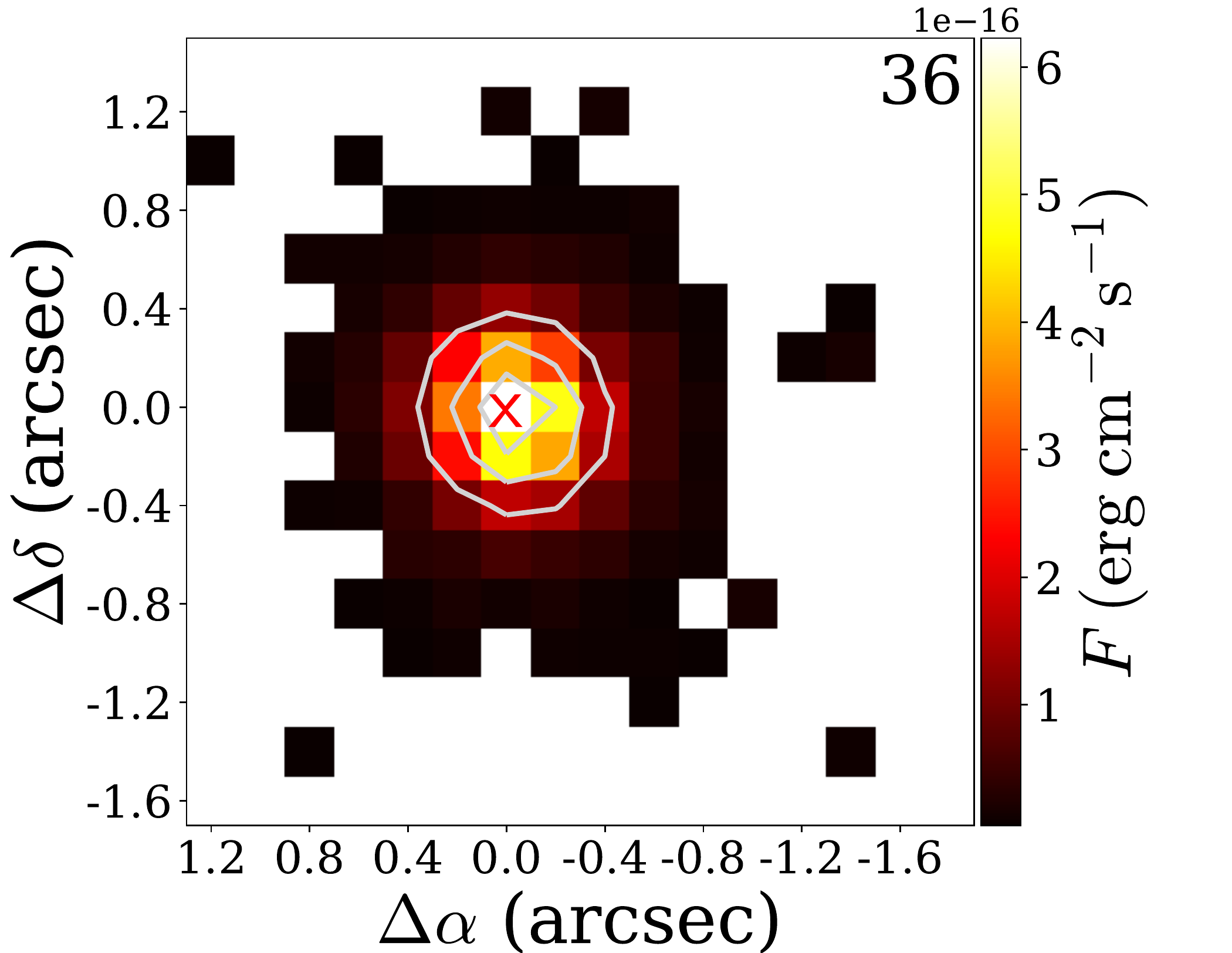}\hspace{-0.2cm}
\includegraphics[width=0.2\textwidth]{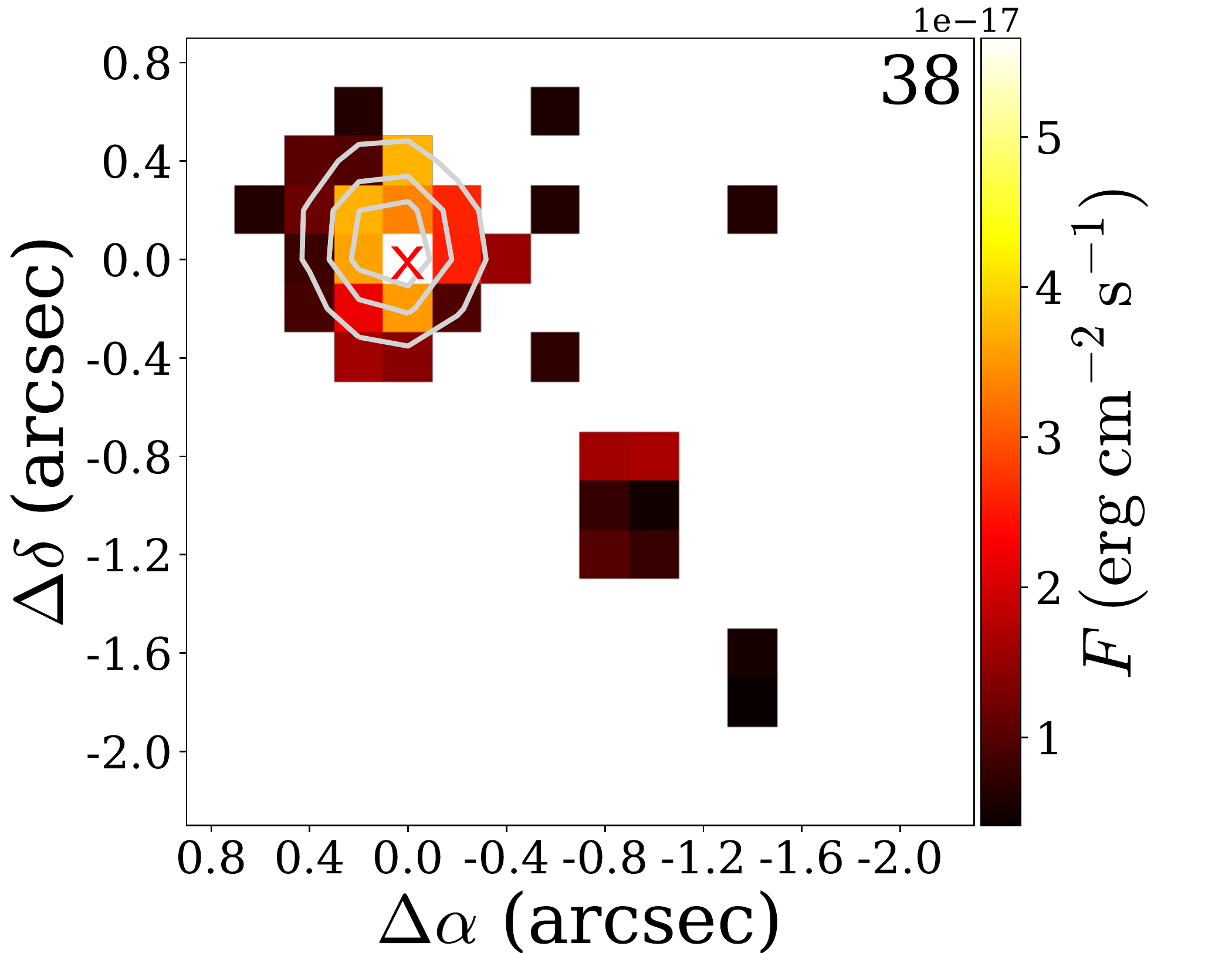}\hspace{-0.2cm}
\includegraphics[width=0.2\textwidth]{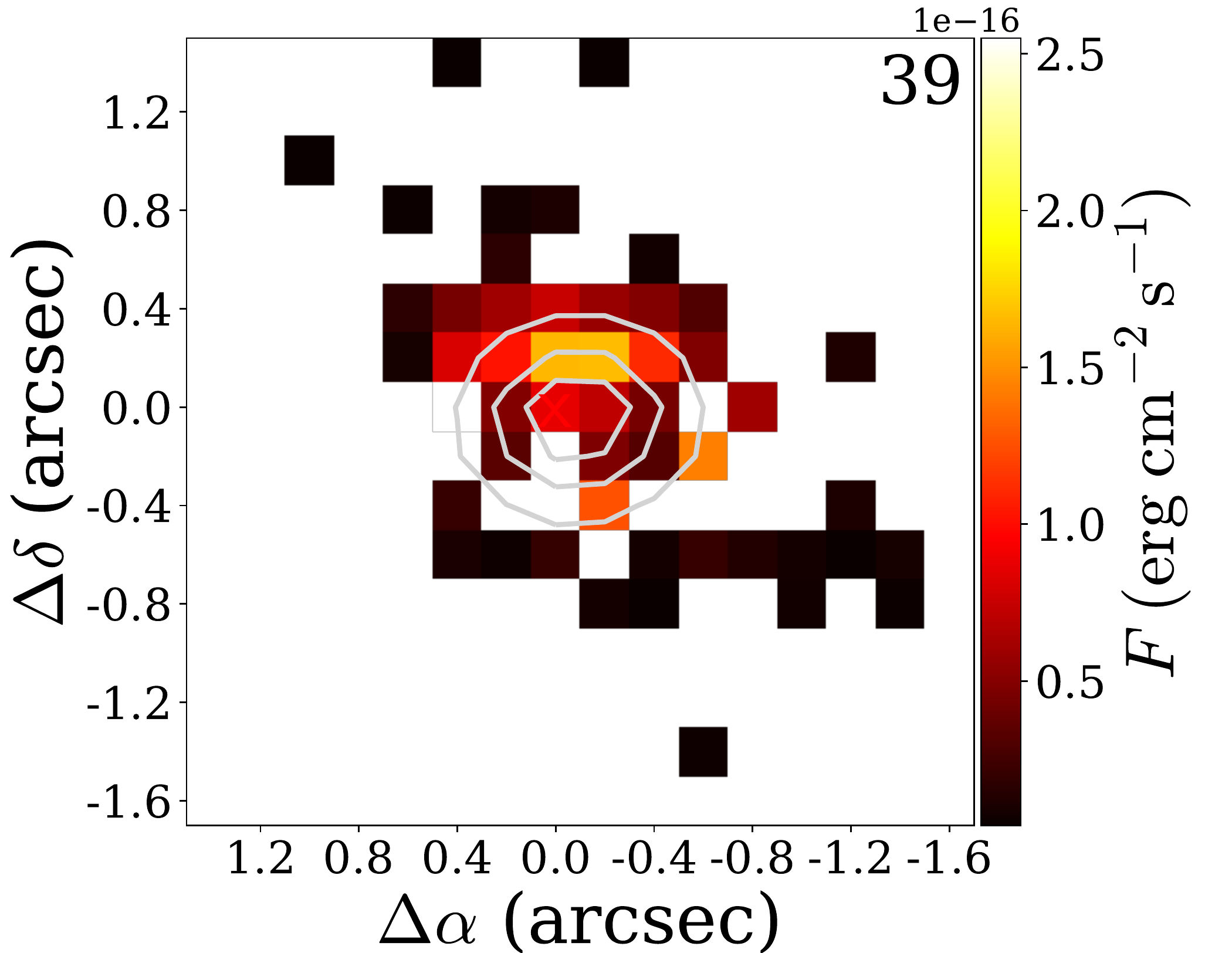}\hspace{-0.2cm}
\includegraphics[width=0.2\textwidth]{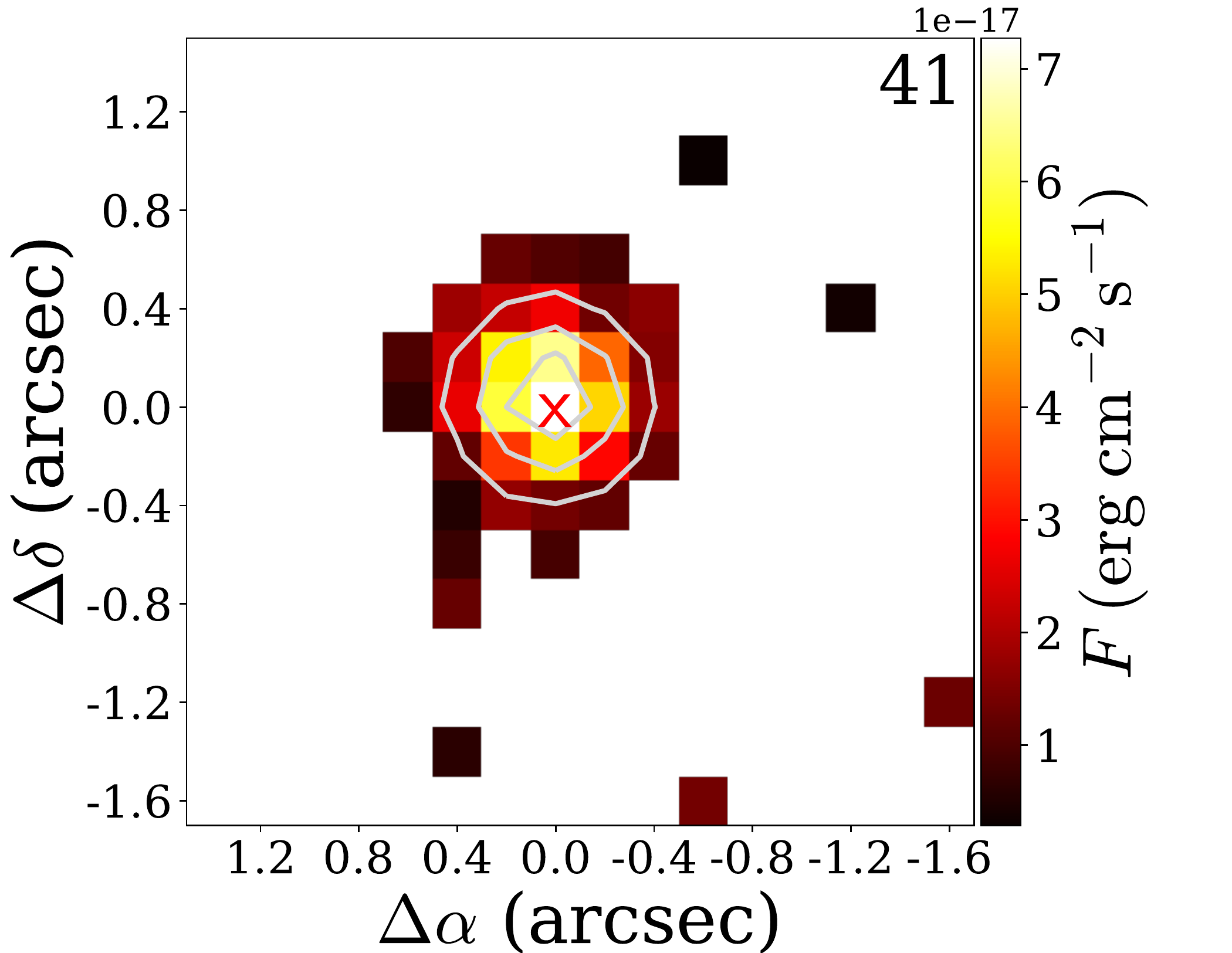}\hspace{-0.2cm}
\includegraphics[width=0.2\textwidth]{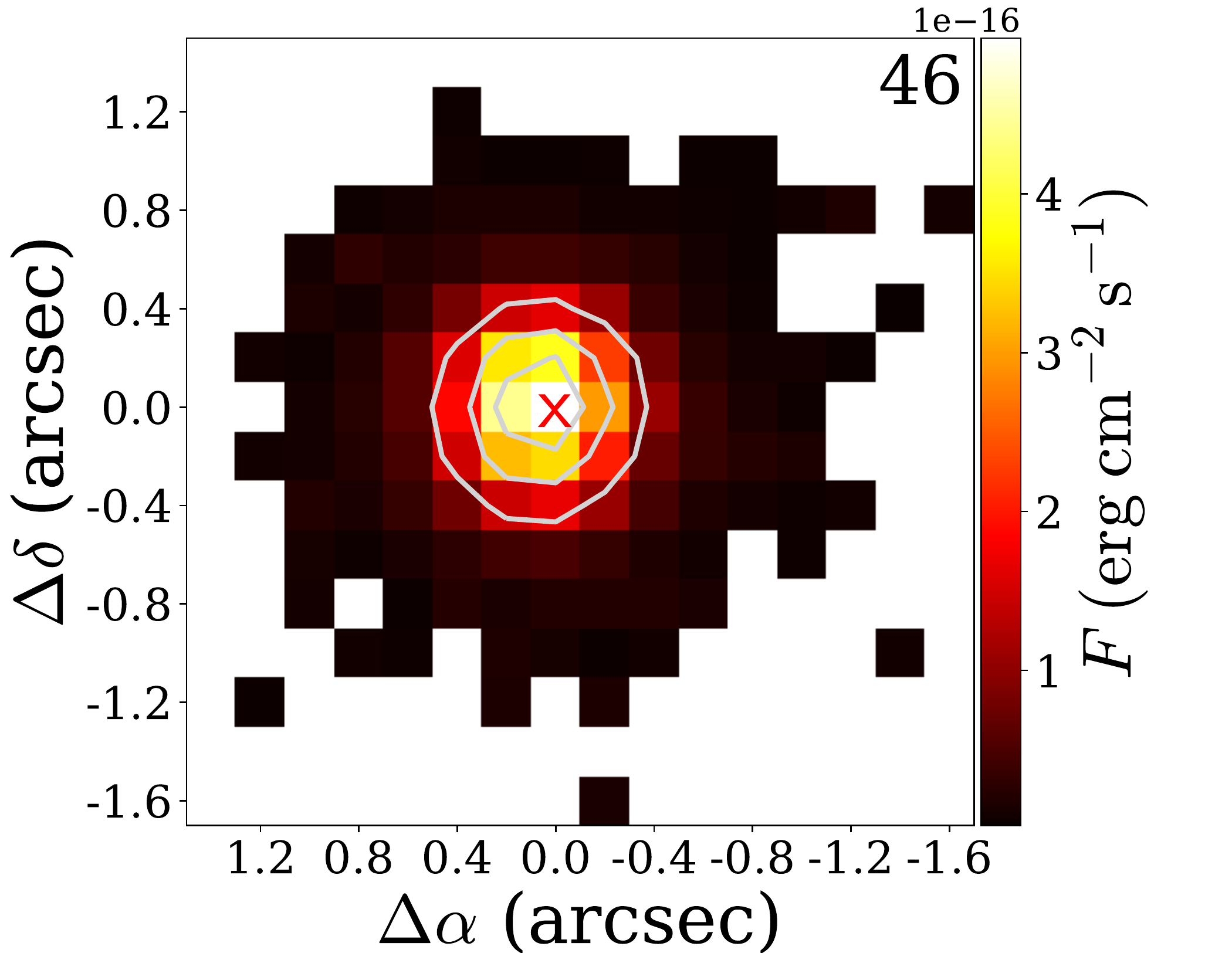}\hspace{-0.2cm}
\includegraphics[width=0.2\textwidth]{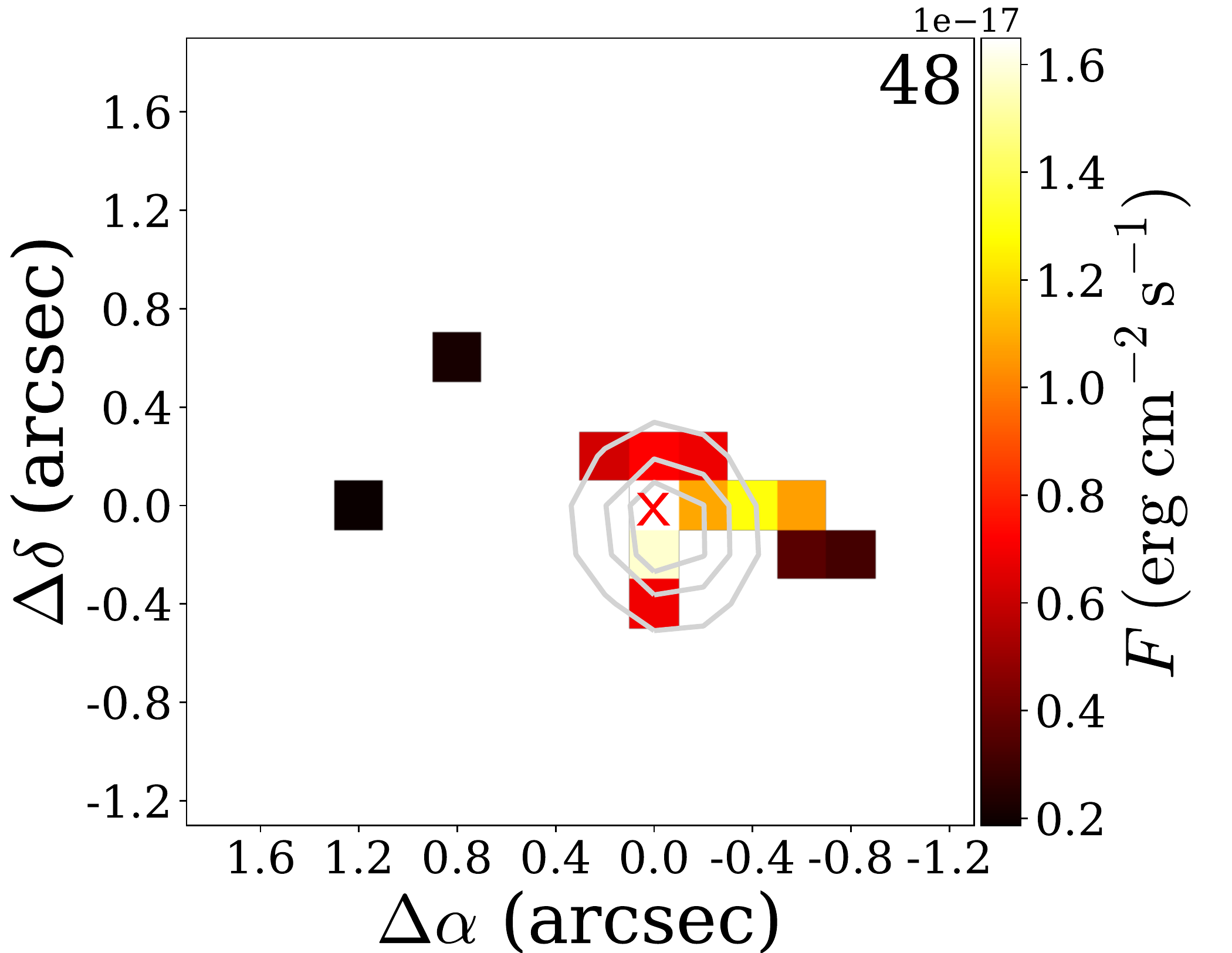}\hspace{-0.2cm} 
\includegraphics[width=0.2\textwidth]{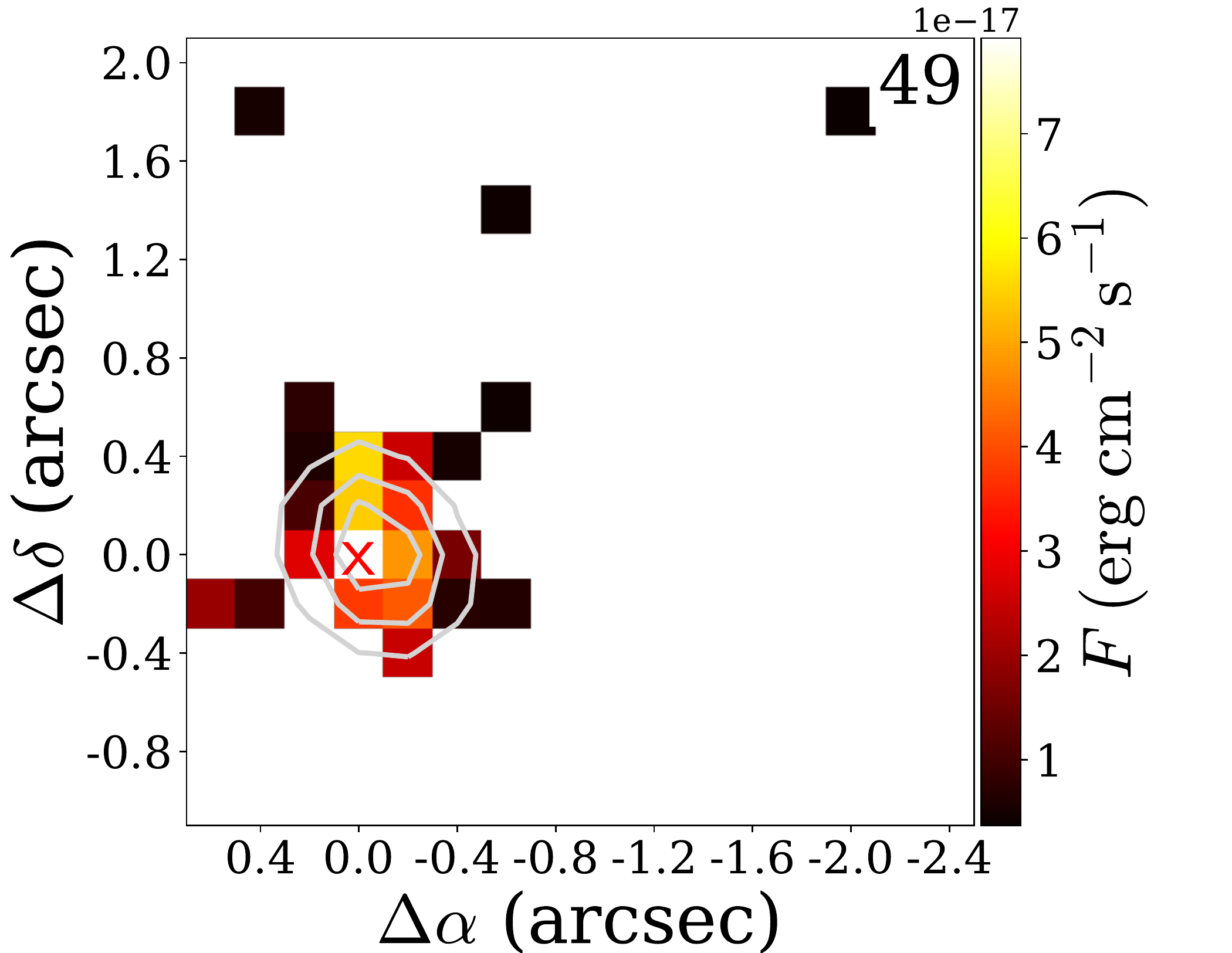}\hspace{-0.2cm}
\includegraphics[width=0.2\textwidth]{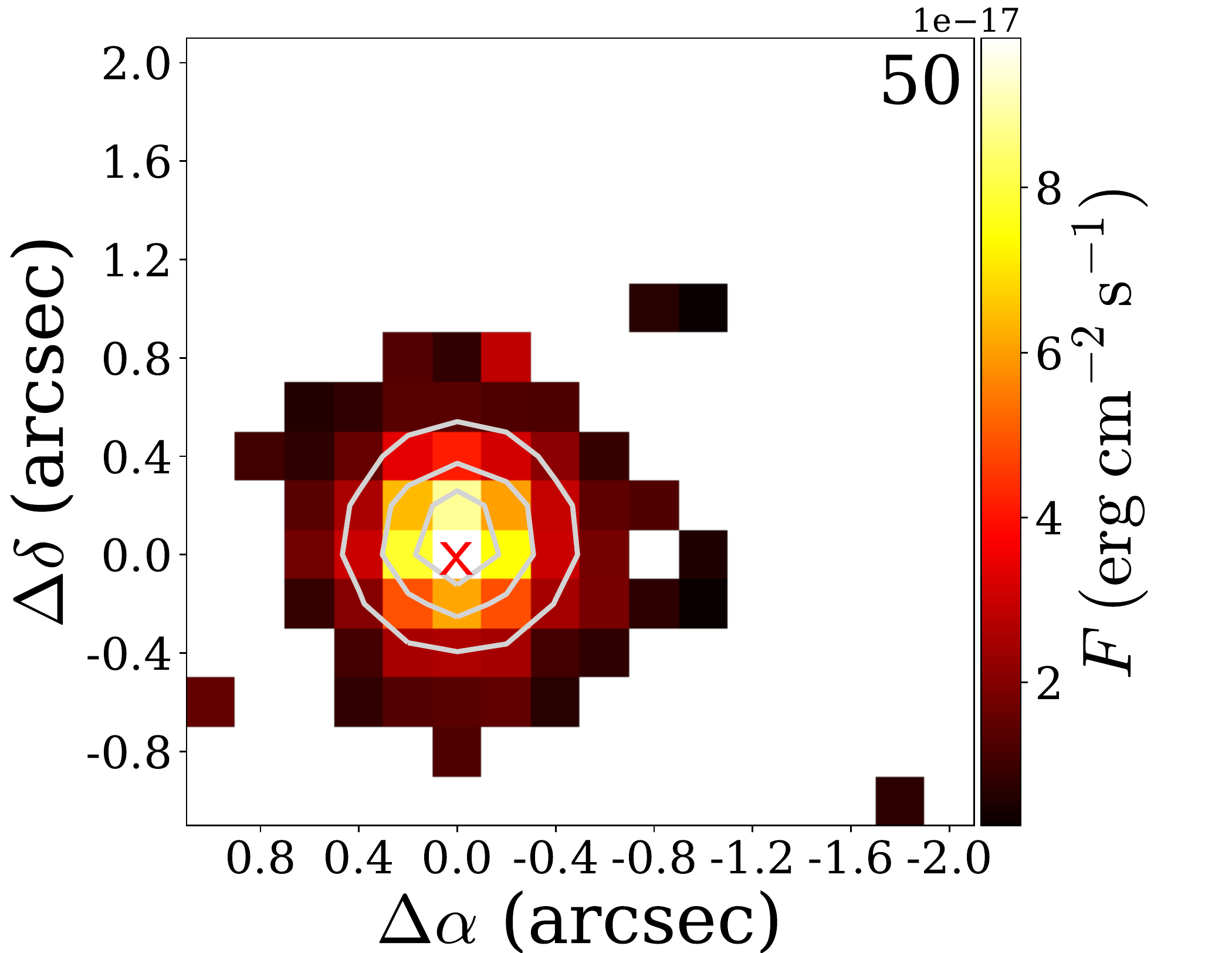}\hspace{-0.2cm}
\includegraphics[width=0.2\textwidth]{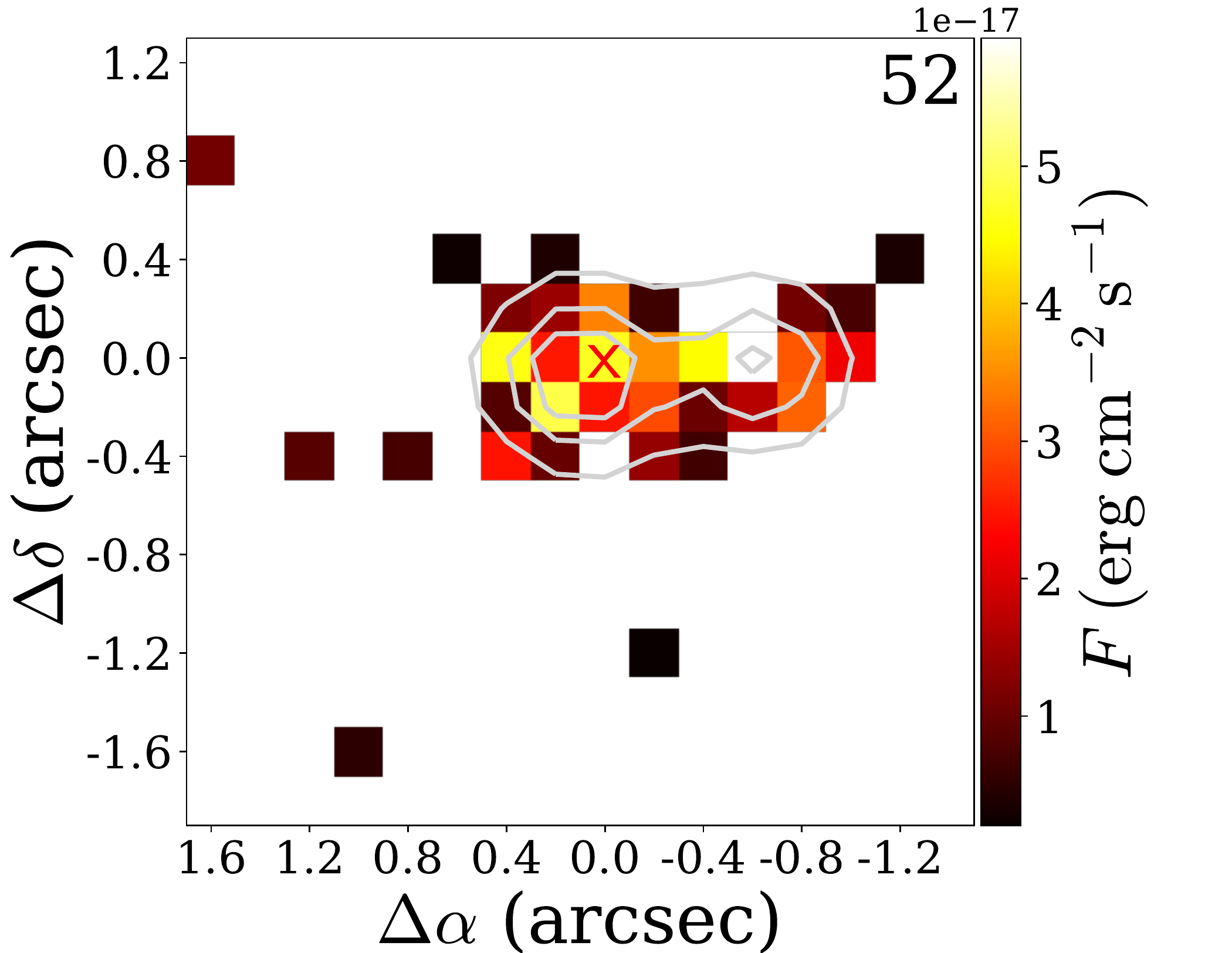}\hspace{-0.2cm}
\includegraphics[width=0.2\textwidth]{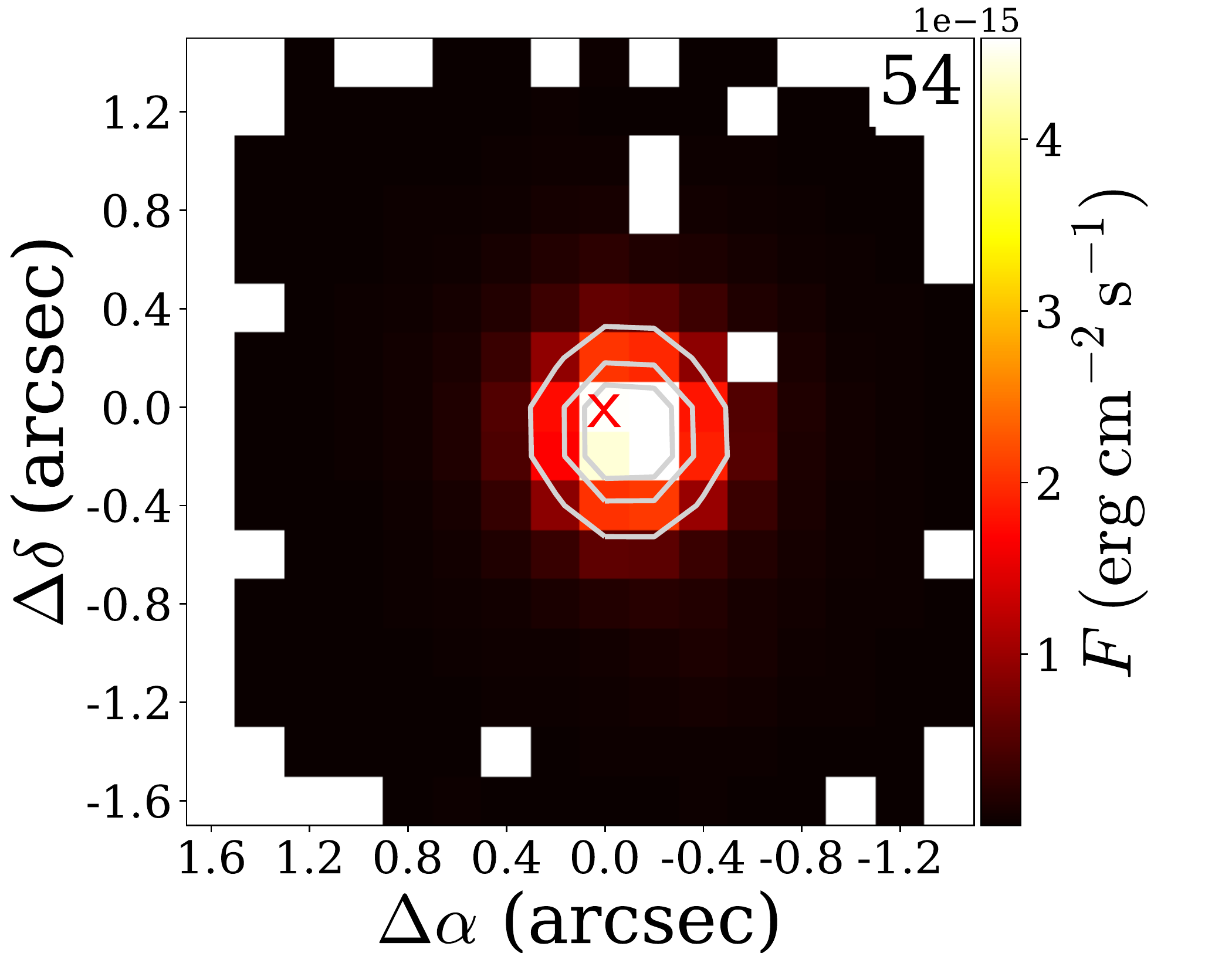}\hspace{-0.2cm}
\includegraphics[width=0.2\textwidth]{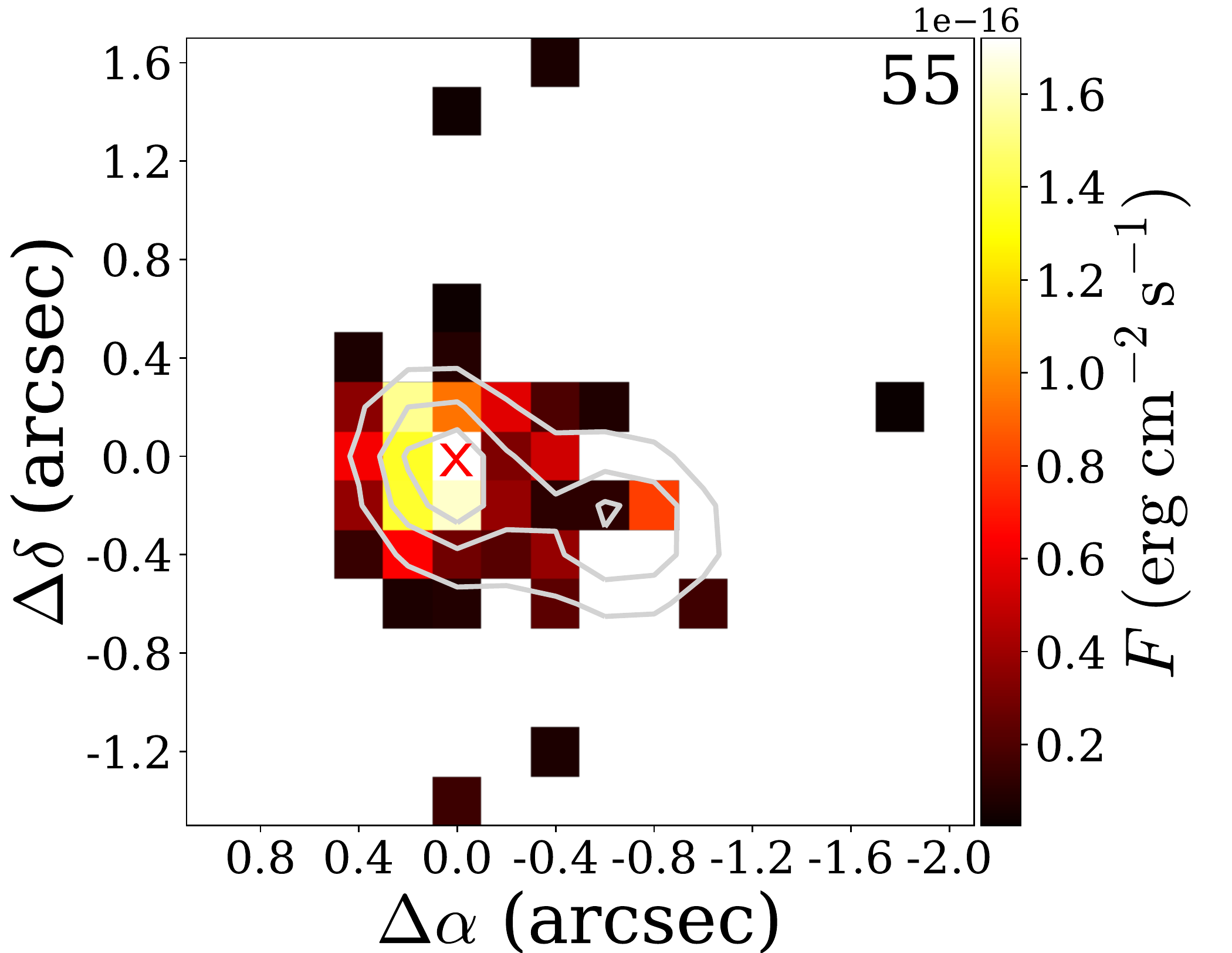}\hspace{-0.2cm}
\includegraphics[width=0.2\textwidth]{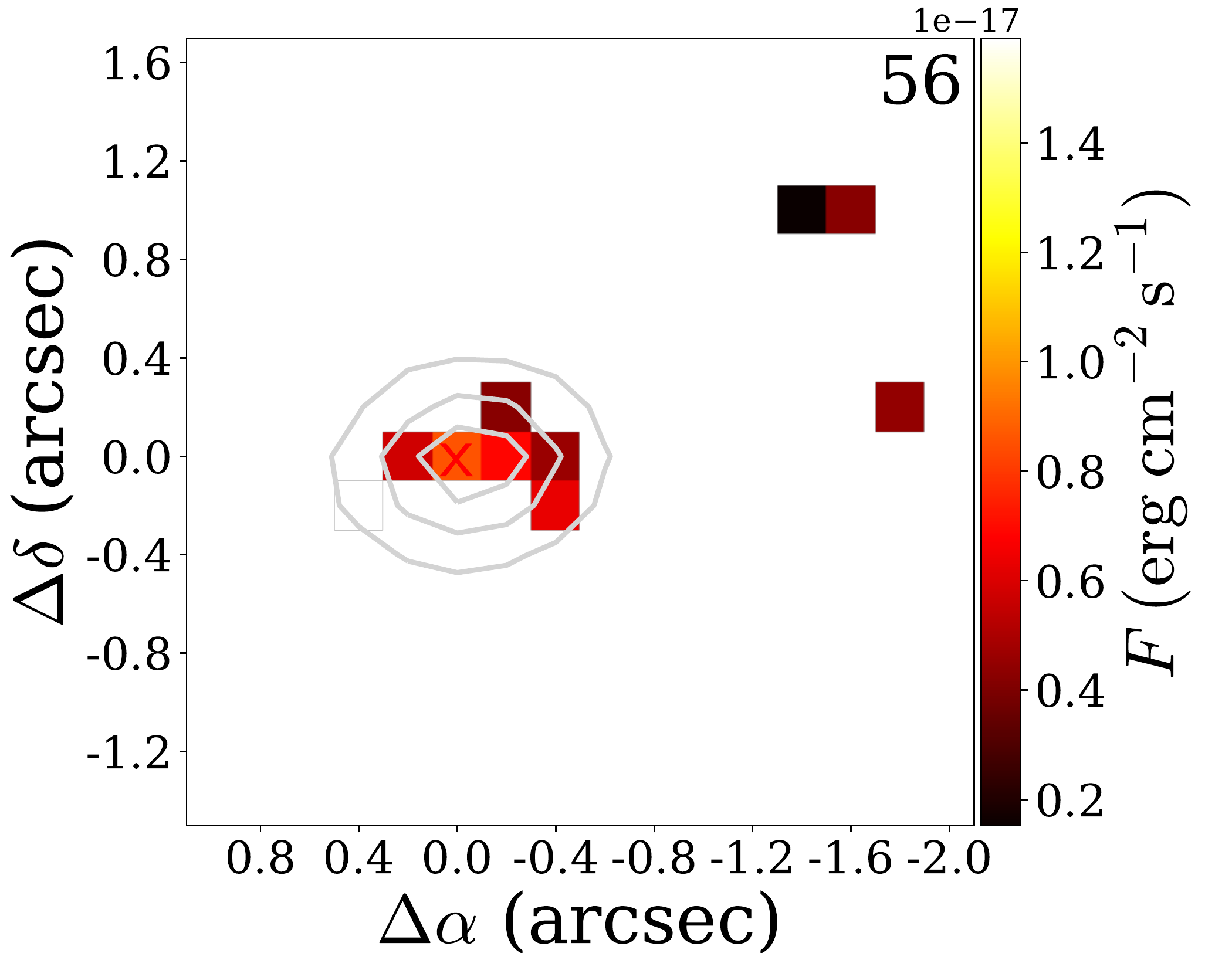}\hspace{-0.2cm} 
\includegraphics[width=0.2\textwidth]{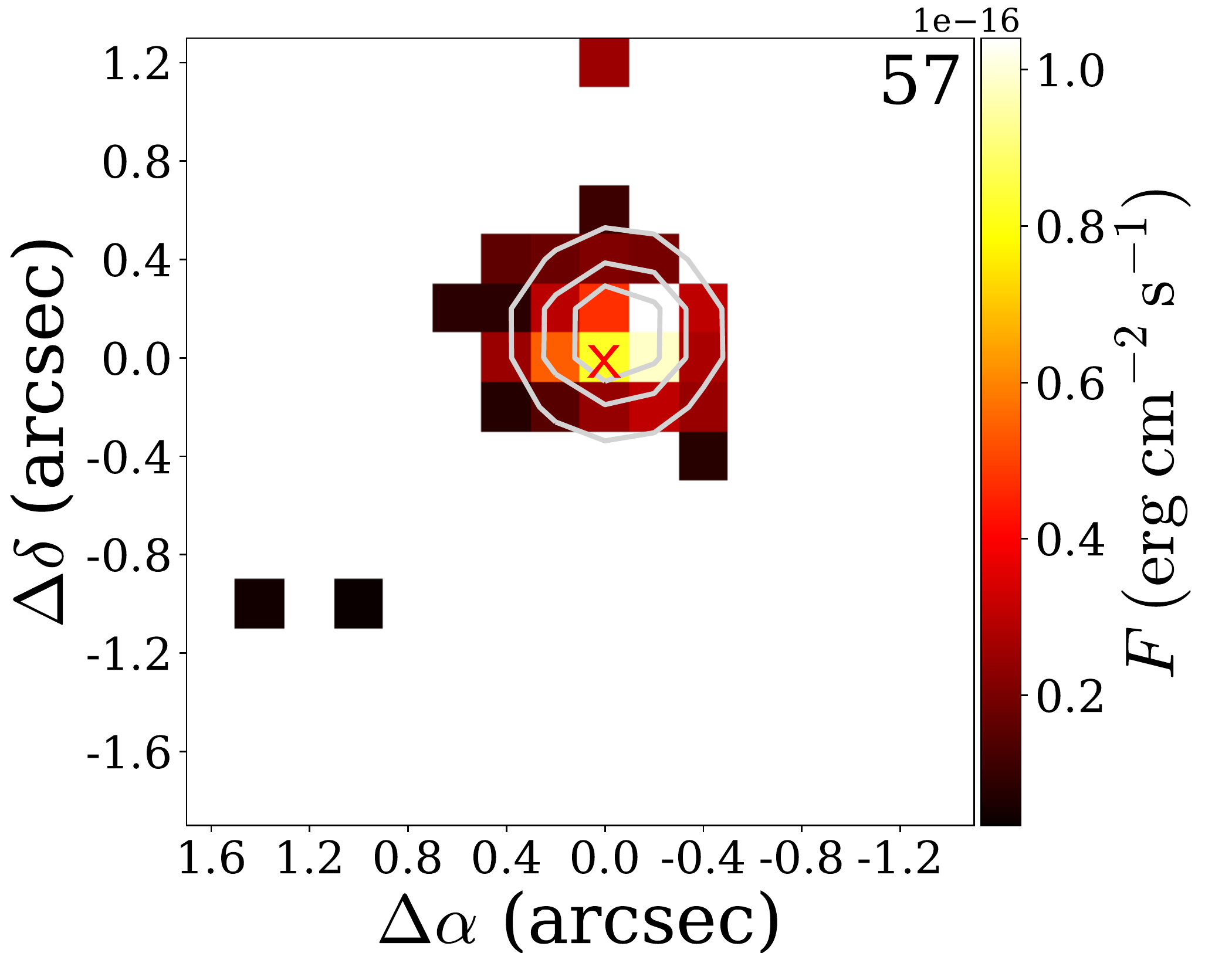}\hspace{-0.2cm}
\includegraphics[width=0.2\textwidth]{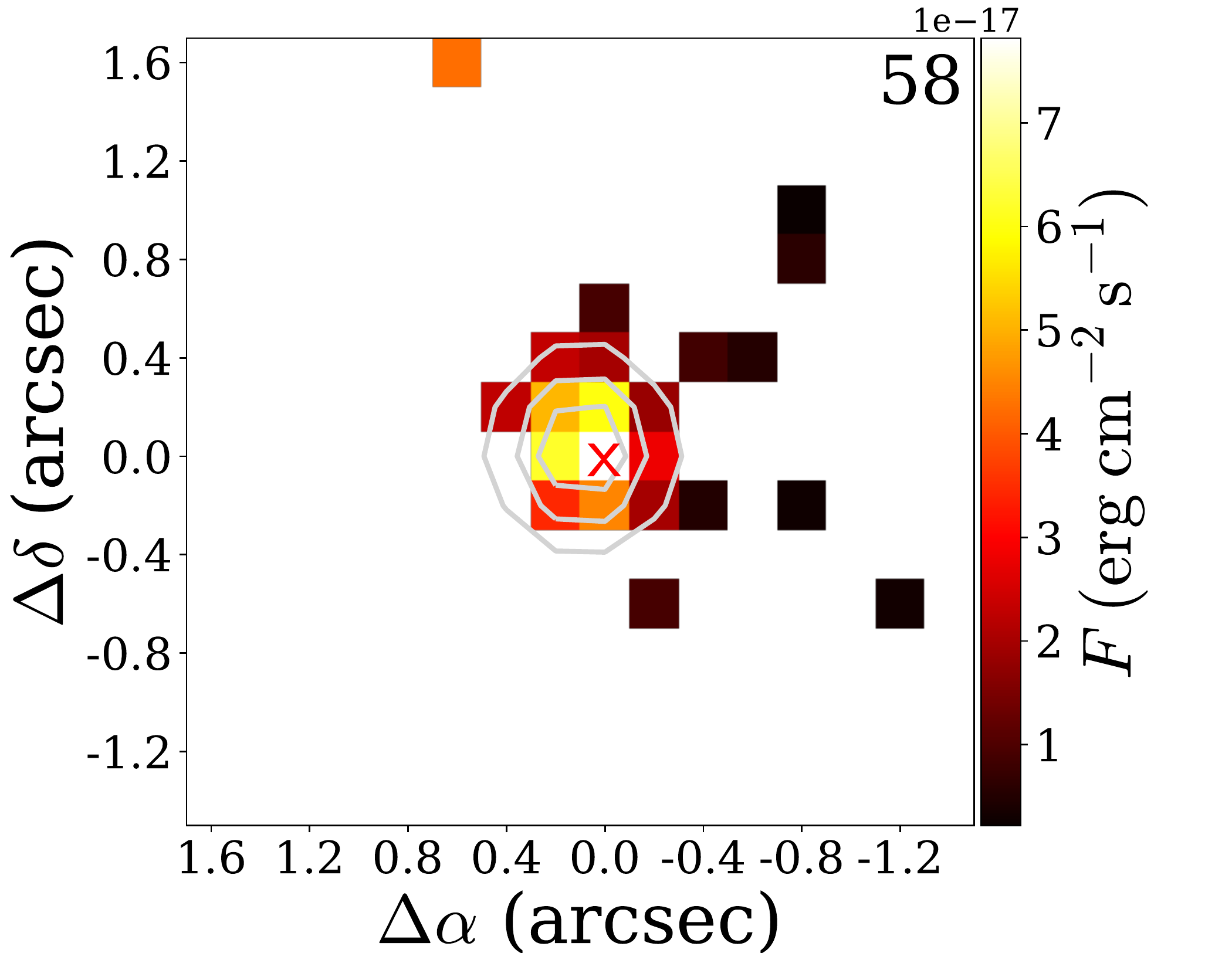}\hspace{-0.2cm}
\includegraphics[width=0.2\textwidth]{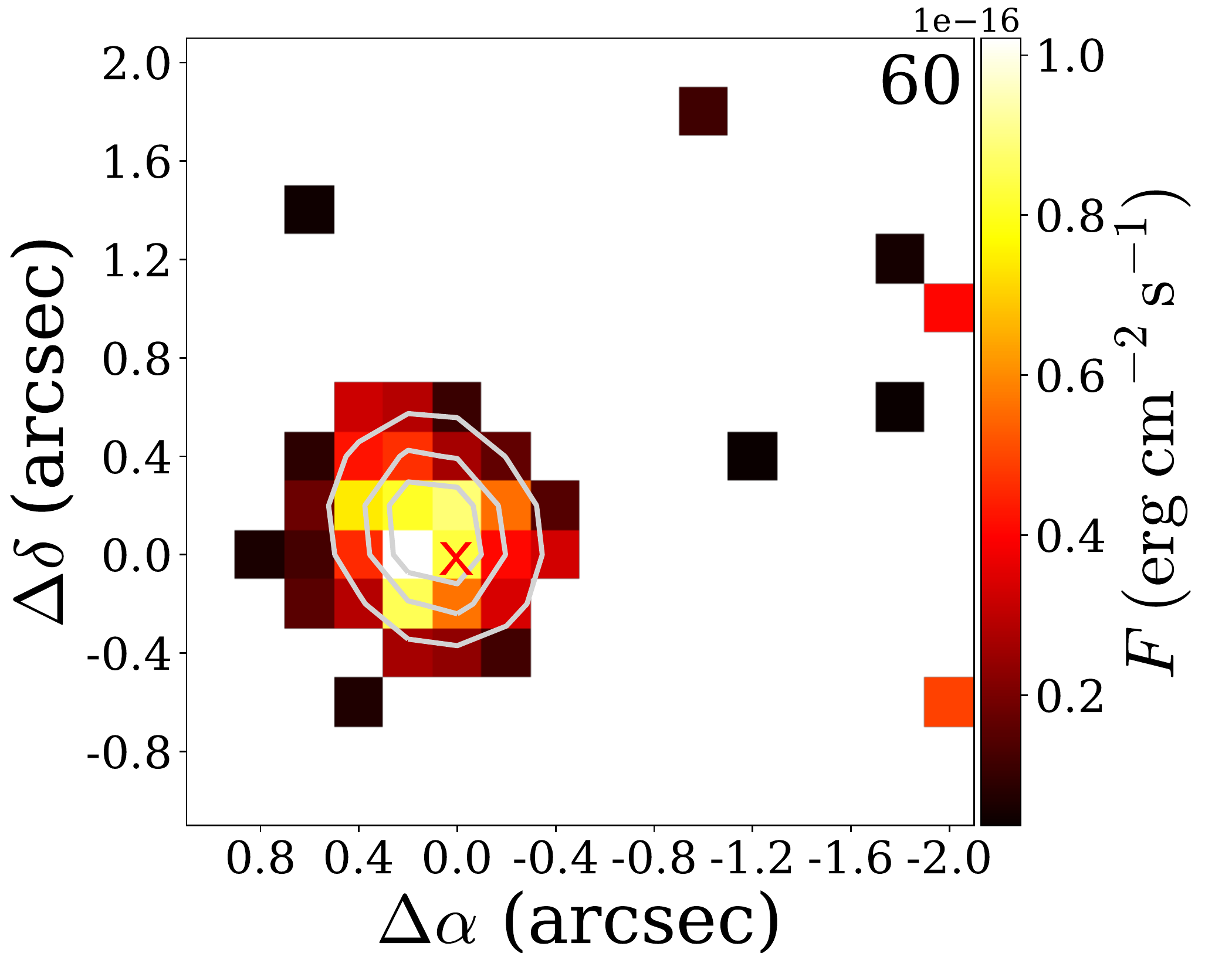}\hspace{-0.2cm}
\includegraphics[width=0.2\textwidth]{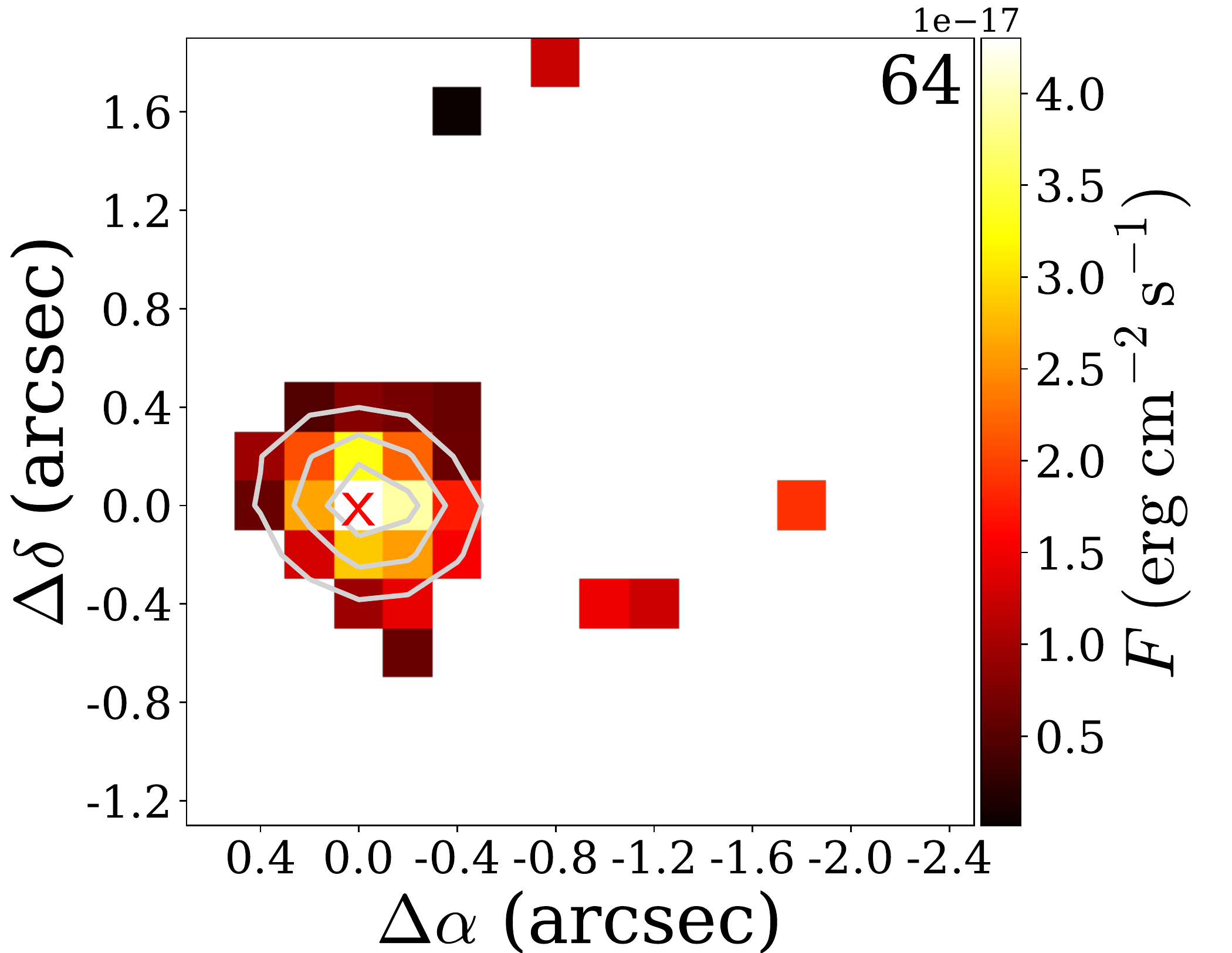}\hspace{-0.2cm}
\includegraphics[width=0.2\textwidth]{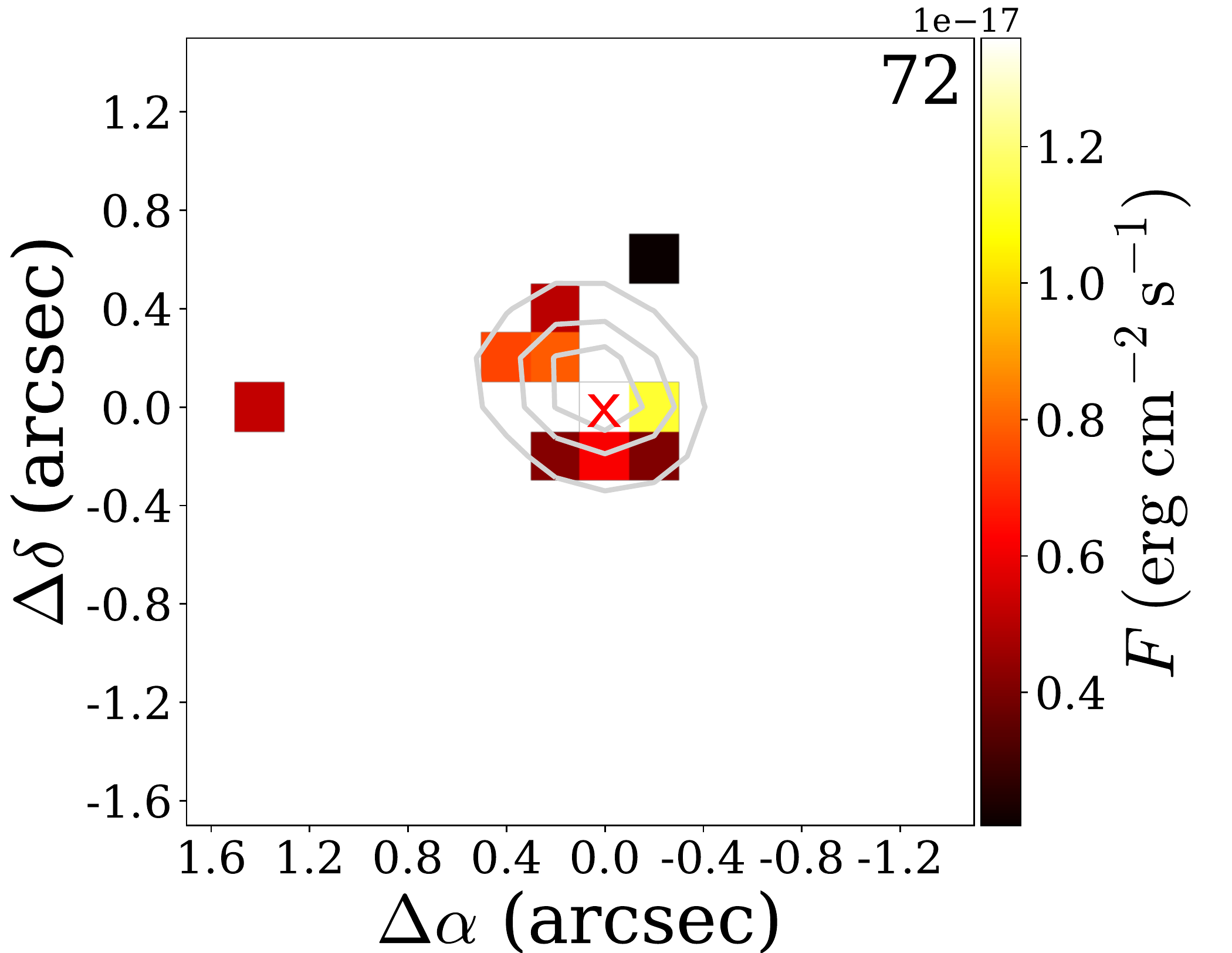}\hspace{-0.2cm} 
\caption{Integrated maps for the H Br$\gamma$ line at 2.1661 $\mu$m. Colors correspond to the line fluxes at each pixel (detections above 2$\sigma$) and white contours show the continuum emission in $K$-band (see also Appendix~\ref{app:cont}).}
\label{fig:emiss-2.1661}
\end{figure*}
\addtocounter{figure}{-1}
\begin{figure*}[h!]
\includegraphics[width=0.2\textwidth]{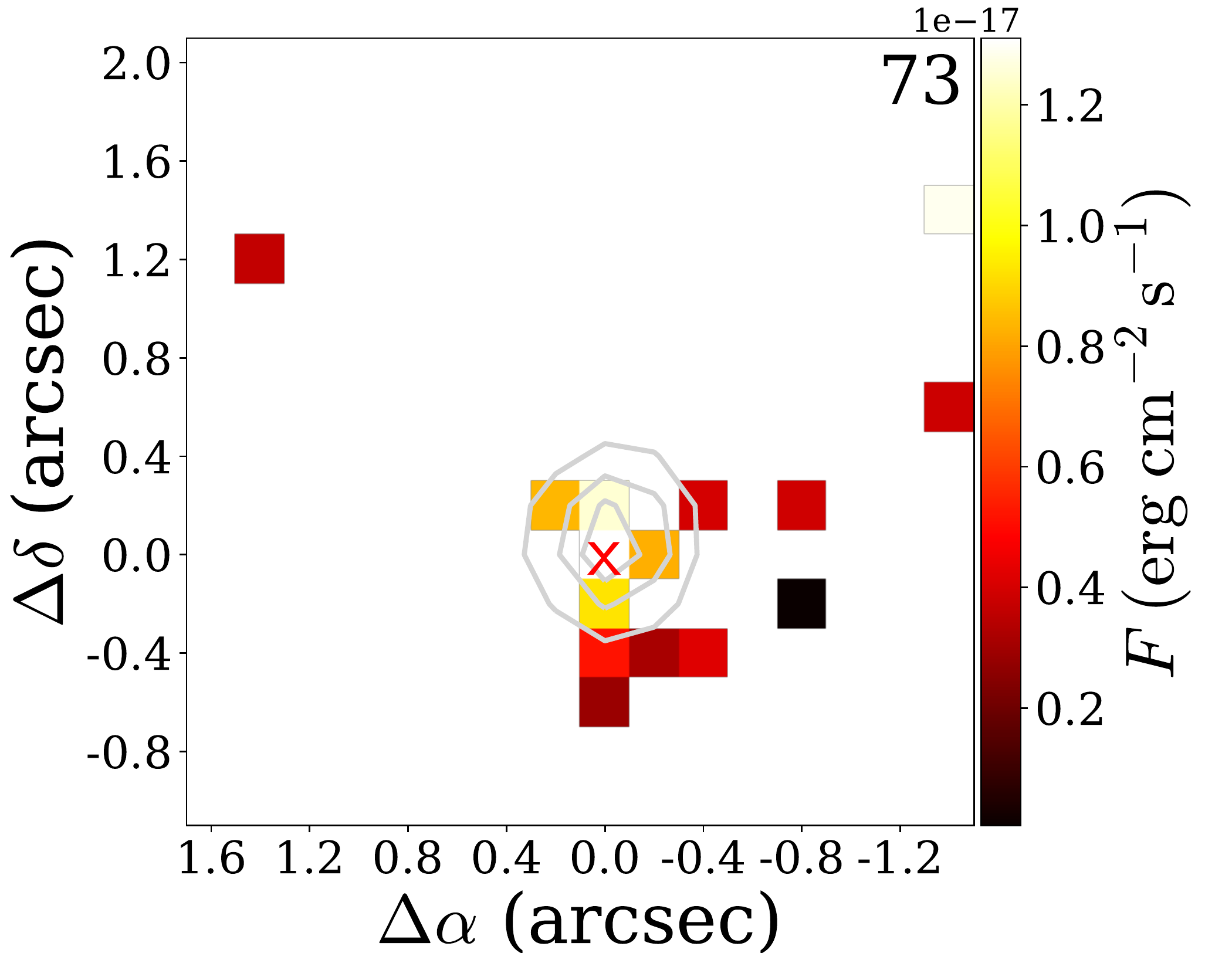}\hspace{-0.2cm} %
\includegraphics[width=0.2\textwidth]{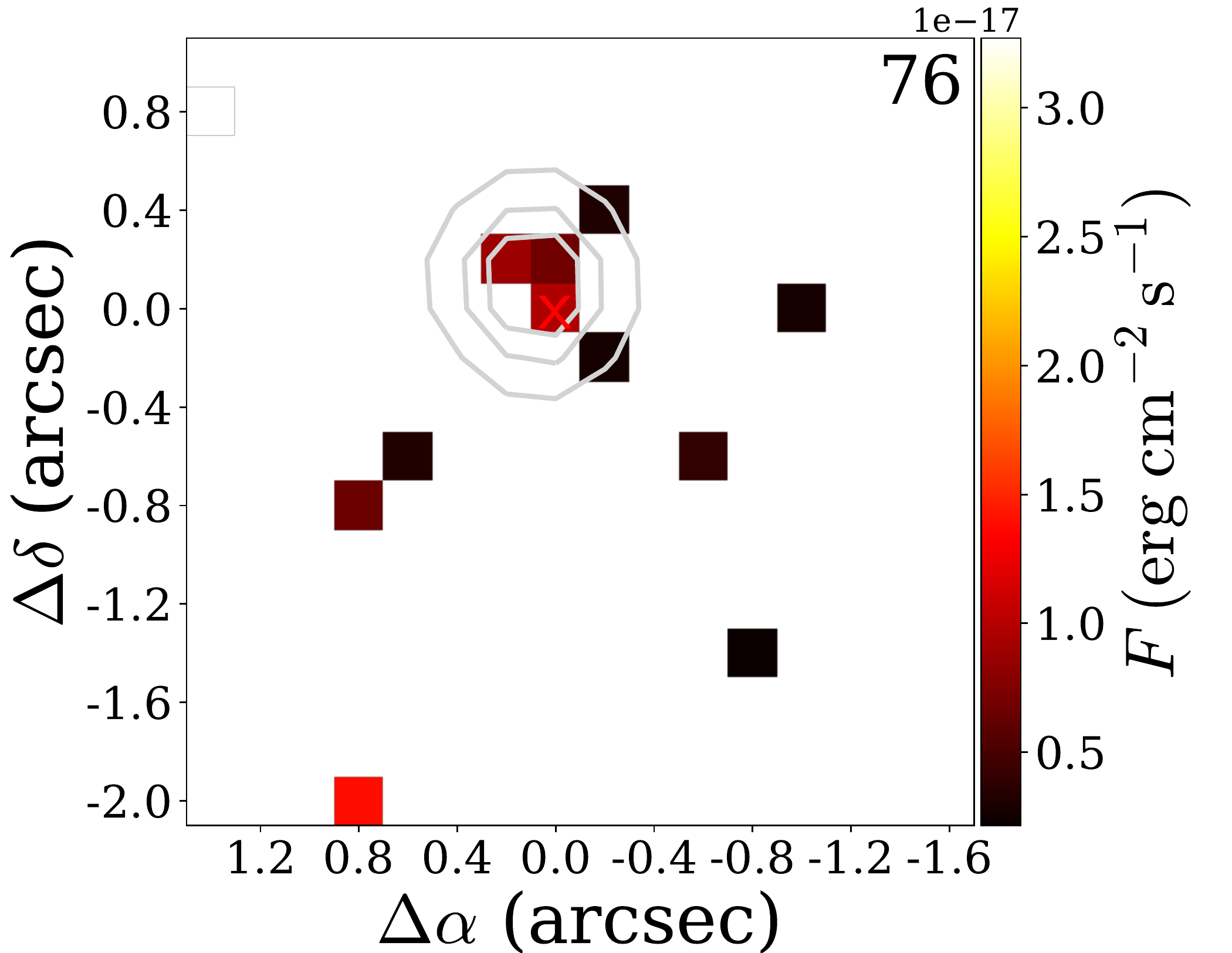}\hspace{-0.2cm} %
\includegraphics[width=0.2\textwidth]{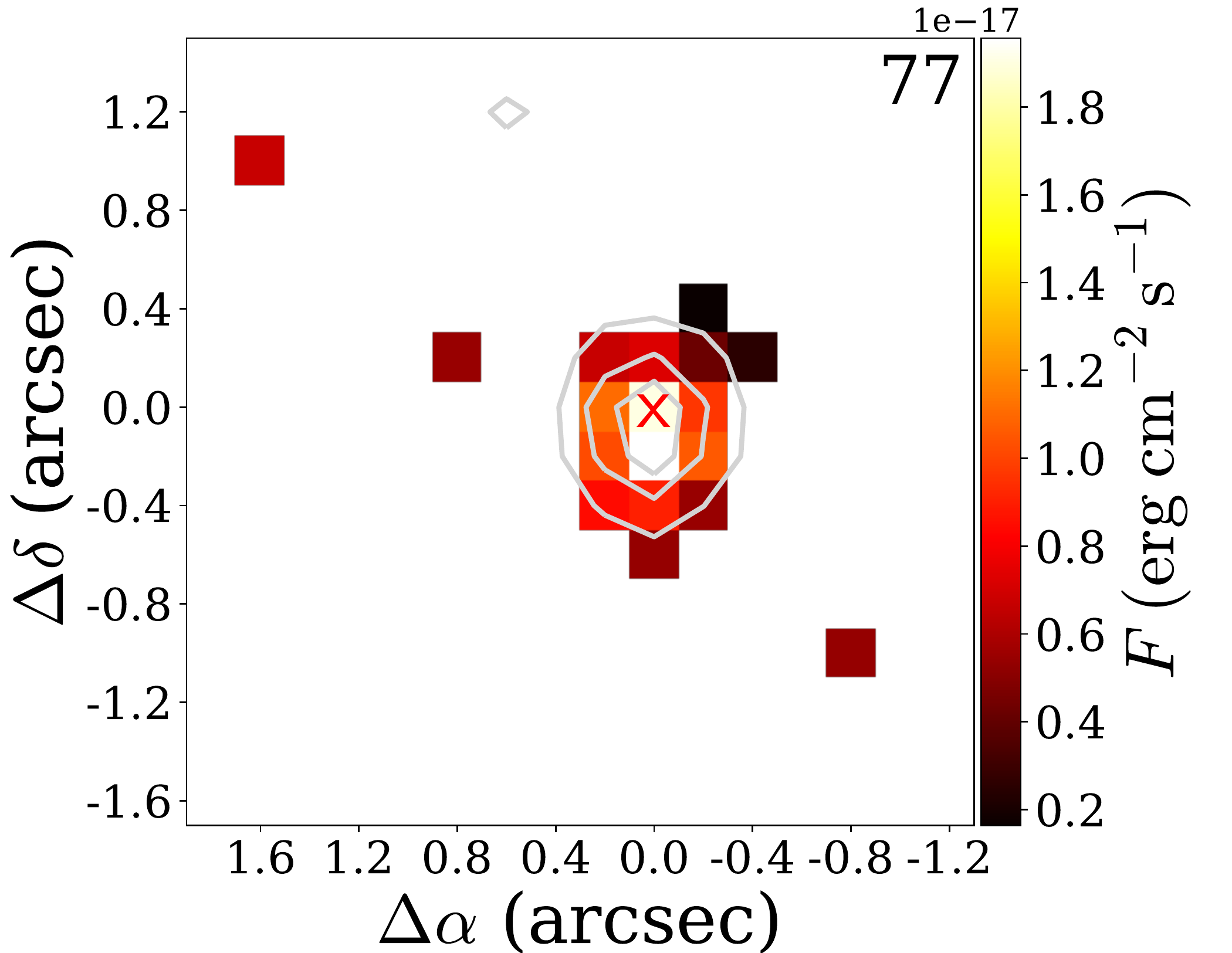}\hspace{-0.2cm}
\includegraphics[width=0.2\textwidth]{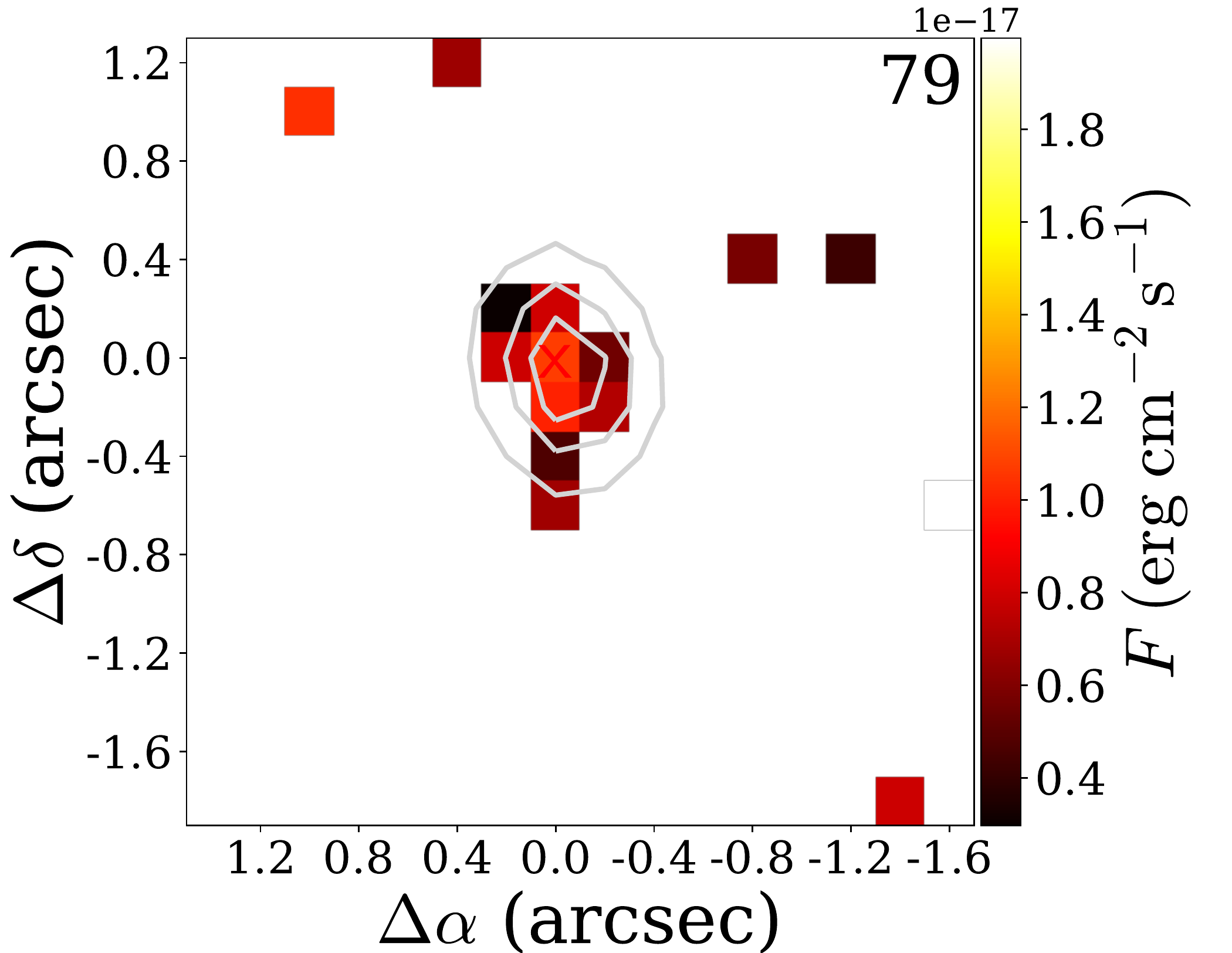}\hspace{-0.2cm}
\includegraphics[width=0.2\textwidth]{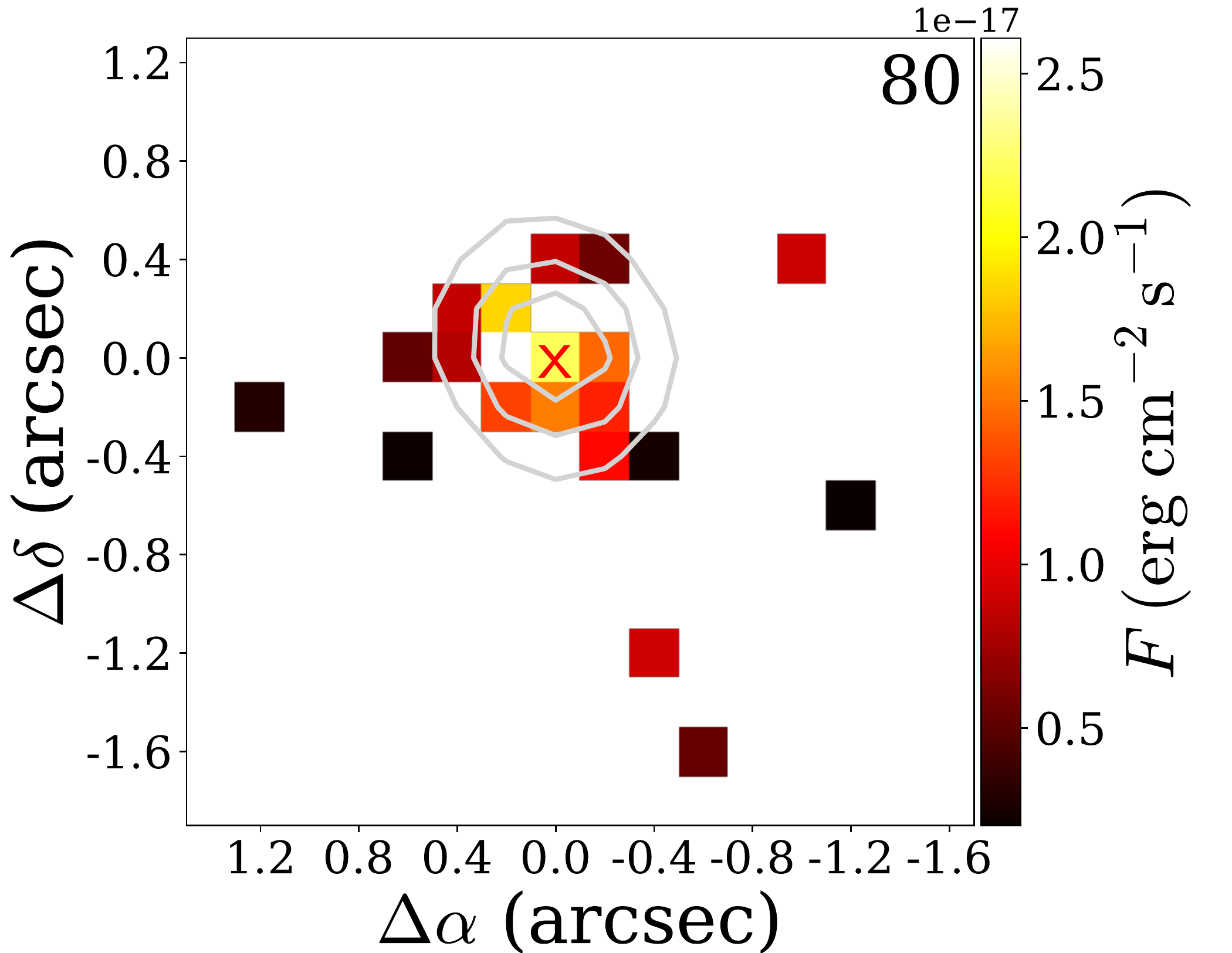}\hspace{-0.2cm}
\includegraphics[width=0.2\textwidth]{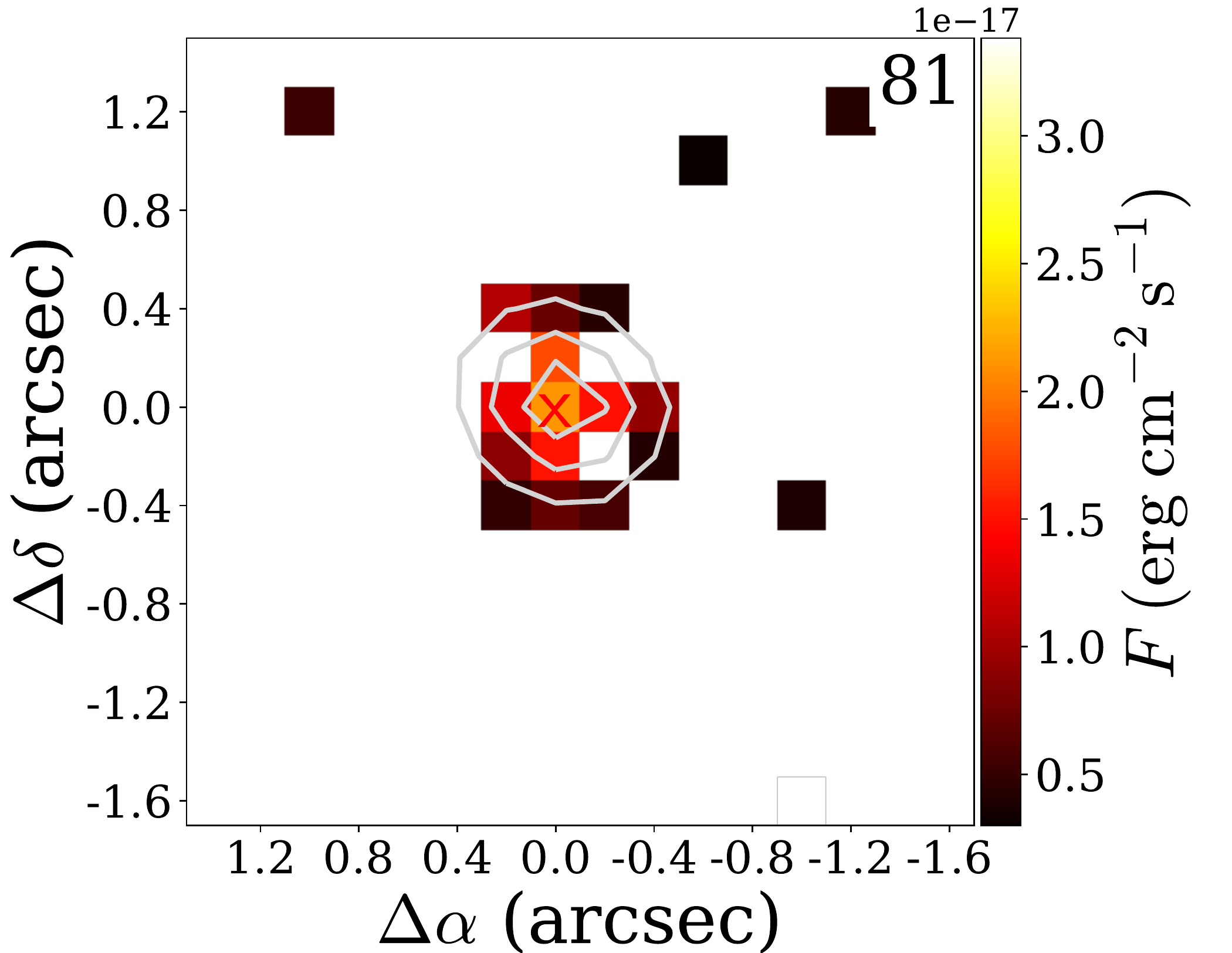}\hspace{-0.2cm}
\includegraphics[width=0.2\textwidth]{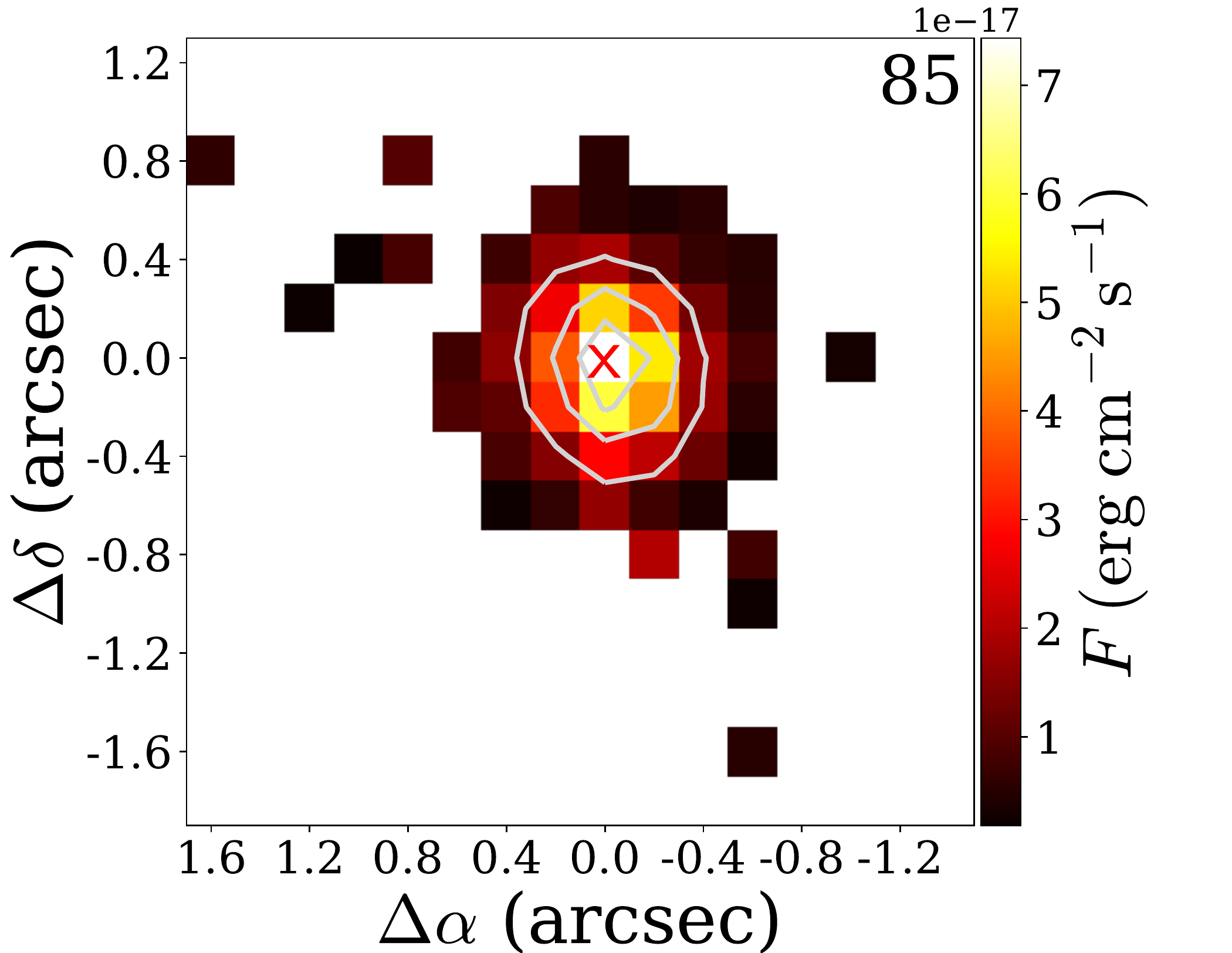}\hspace{-0.2cm}
\includegraphics[width=0.2\textwidth]{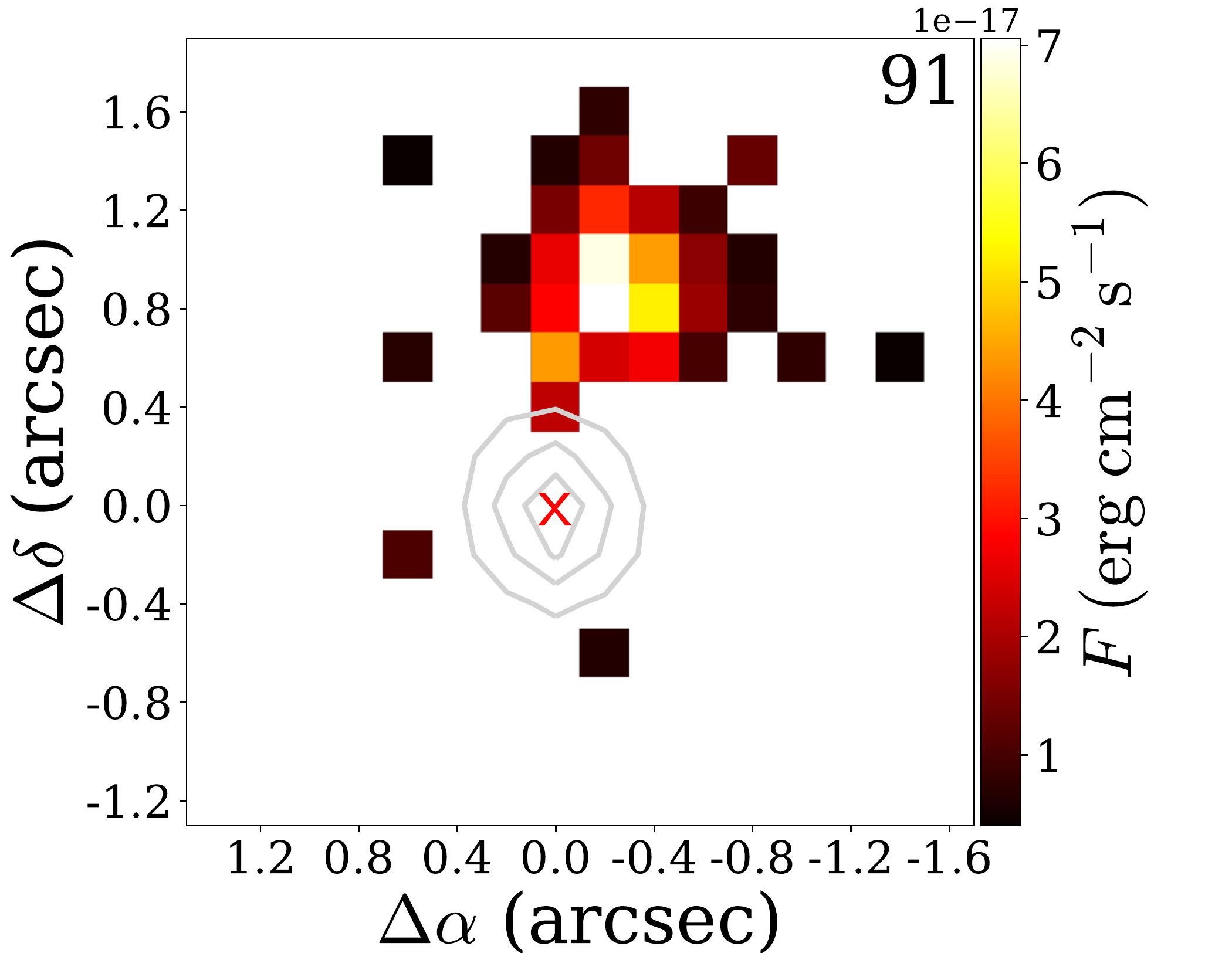}\hspace{-0.2cm}
\includegraphics[width=0.2\textwidth]{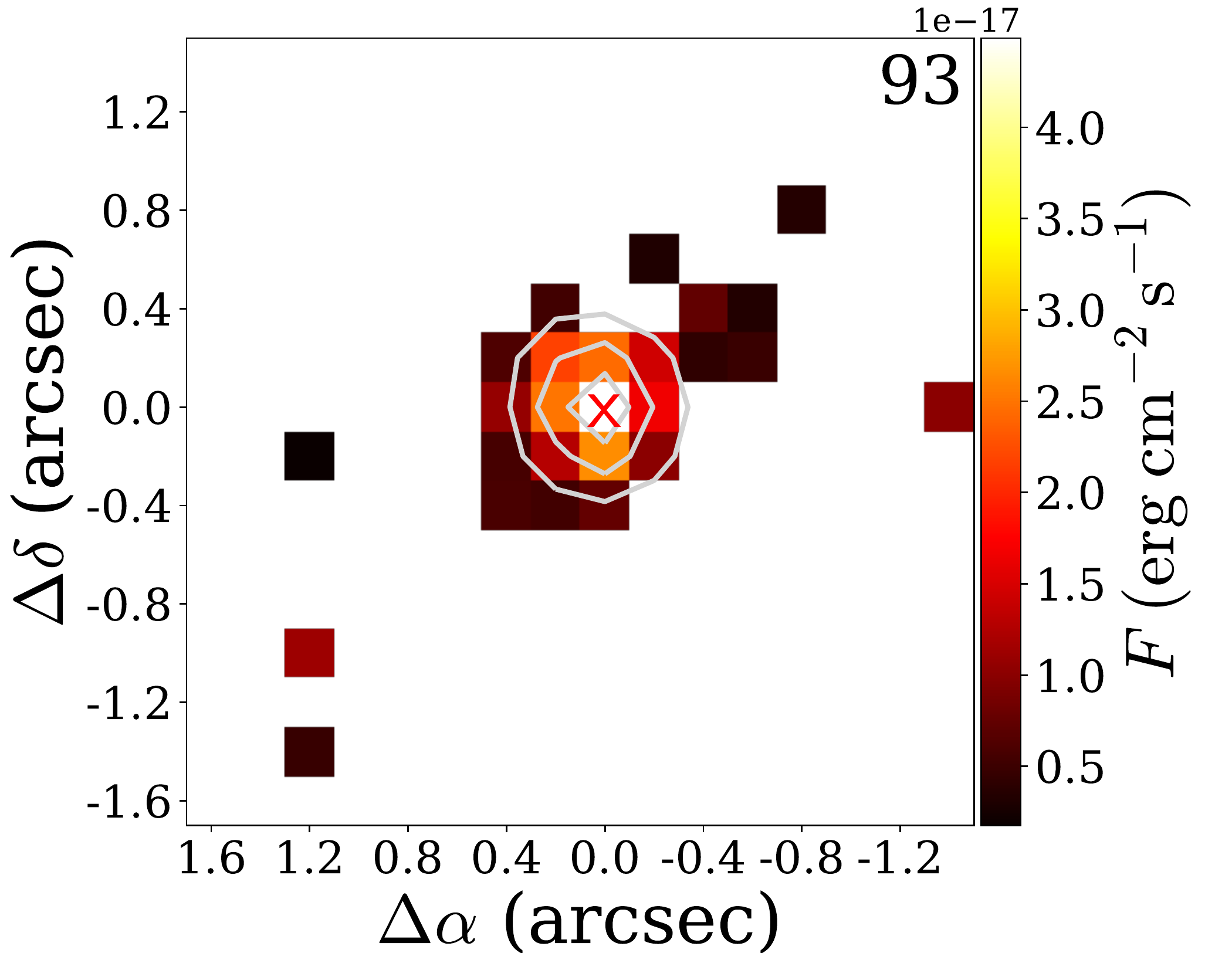}\hspace{-0.2cm}
\includegraphics[width=0.2\textwidth]{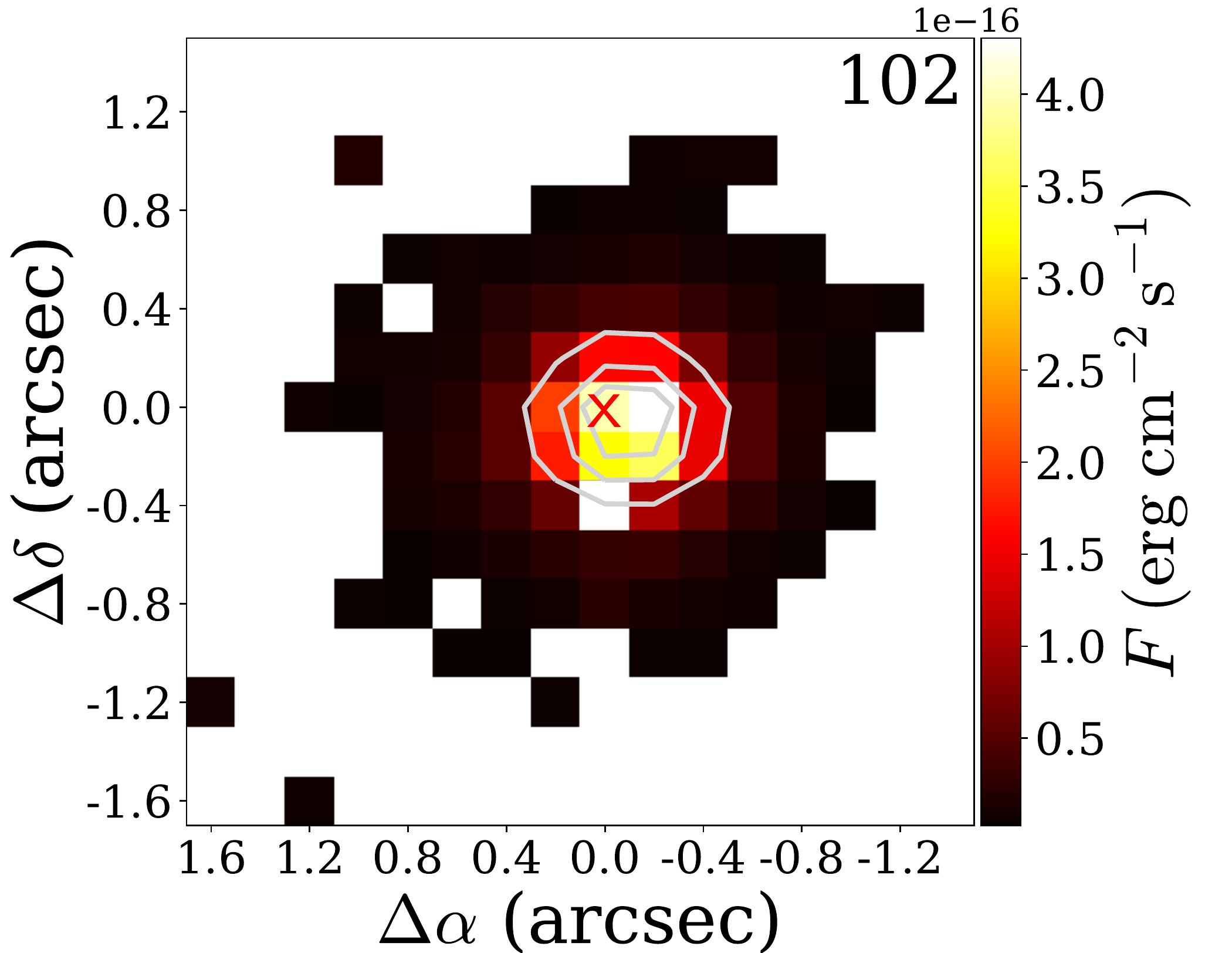}\hspace{-0.2cm}
\includegraphics[width=0.2\textwidth]{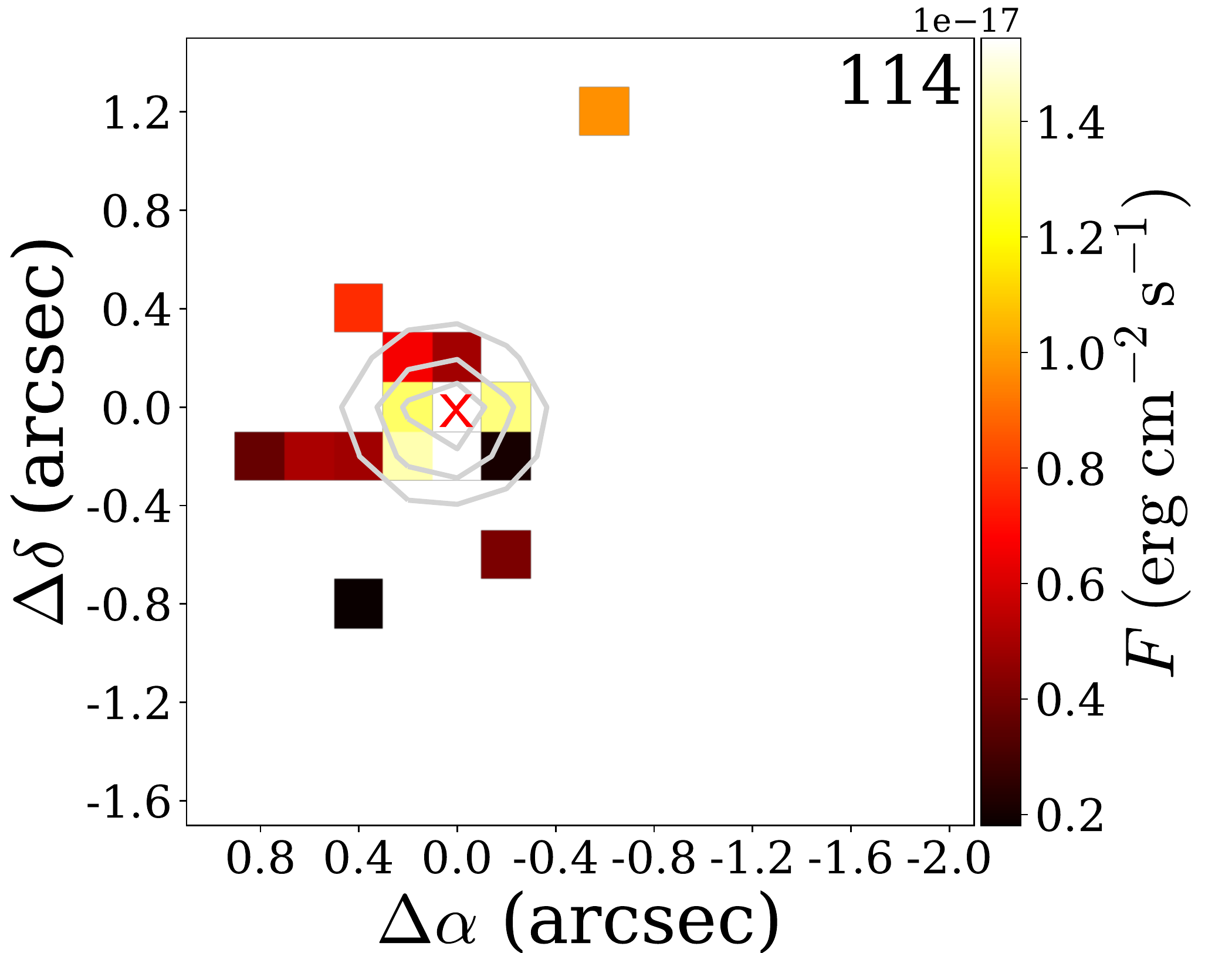}\hspace{-0.2cm} 
\includegraphics[width=0.2\textwidth]{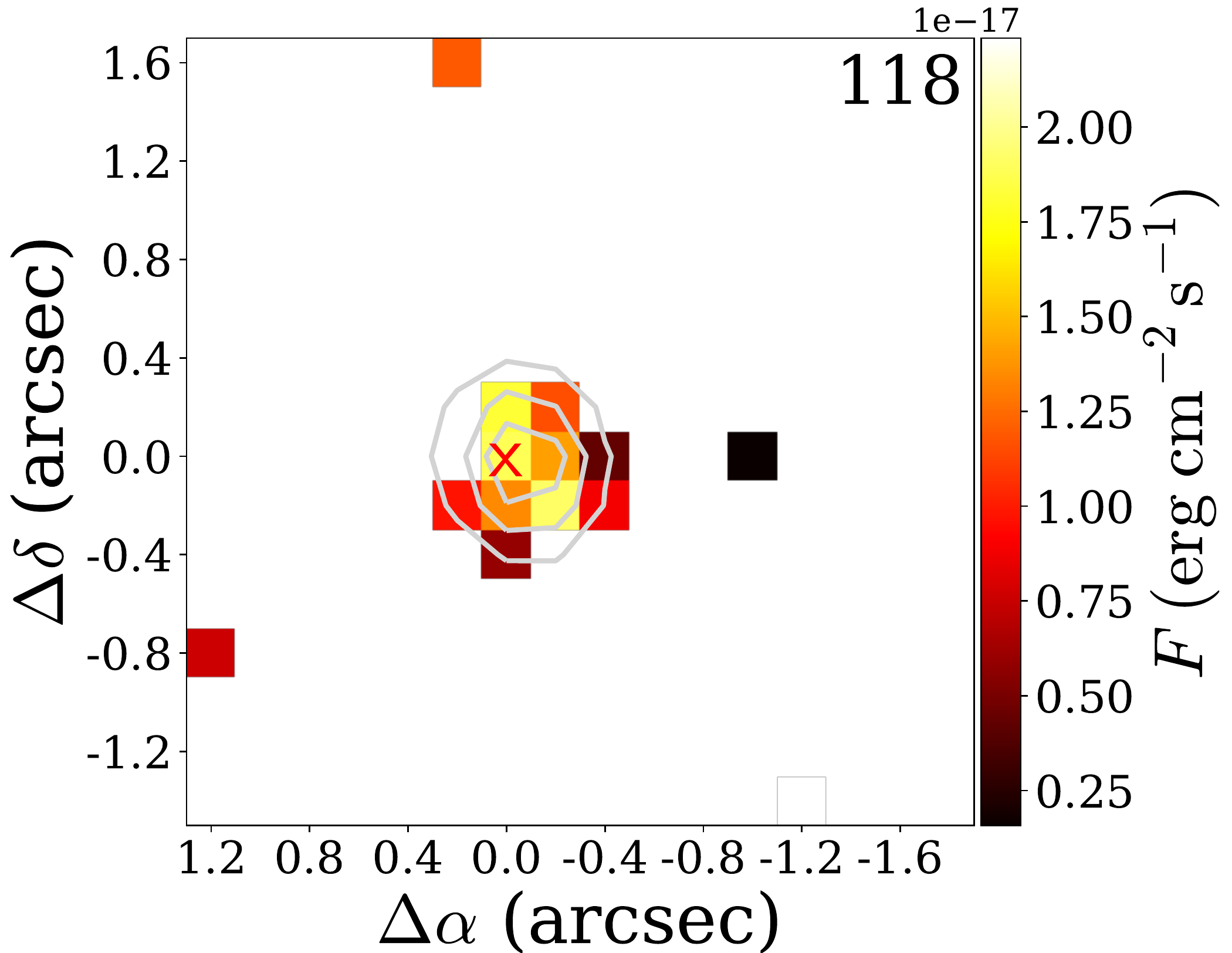}\hspace{-0.2cm} 
\caption{Continued} 
\end{figure*}

\begin{figure*}[h!]
\centering
\includegraphics[width=0.2\textwidth]{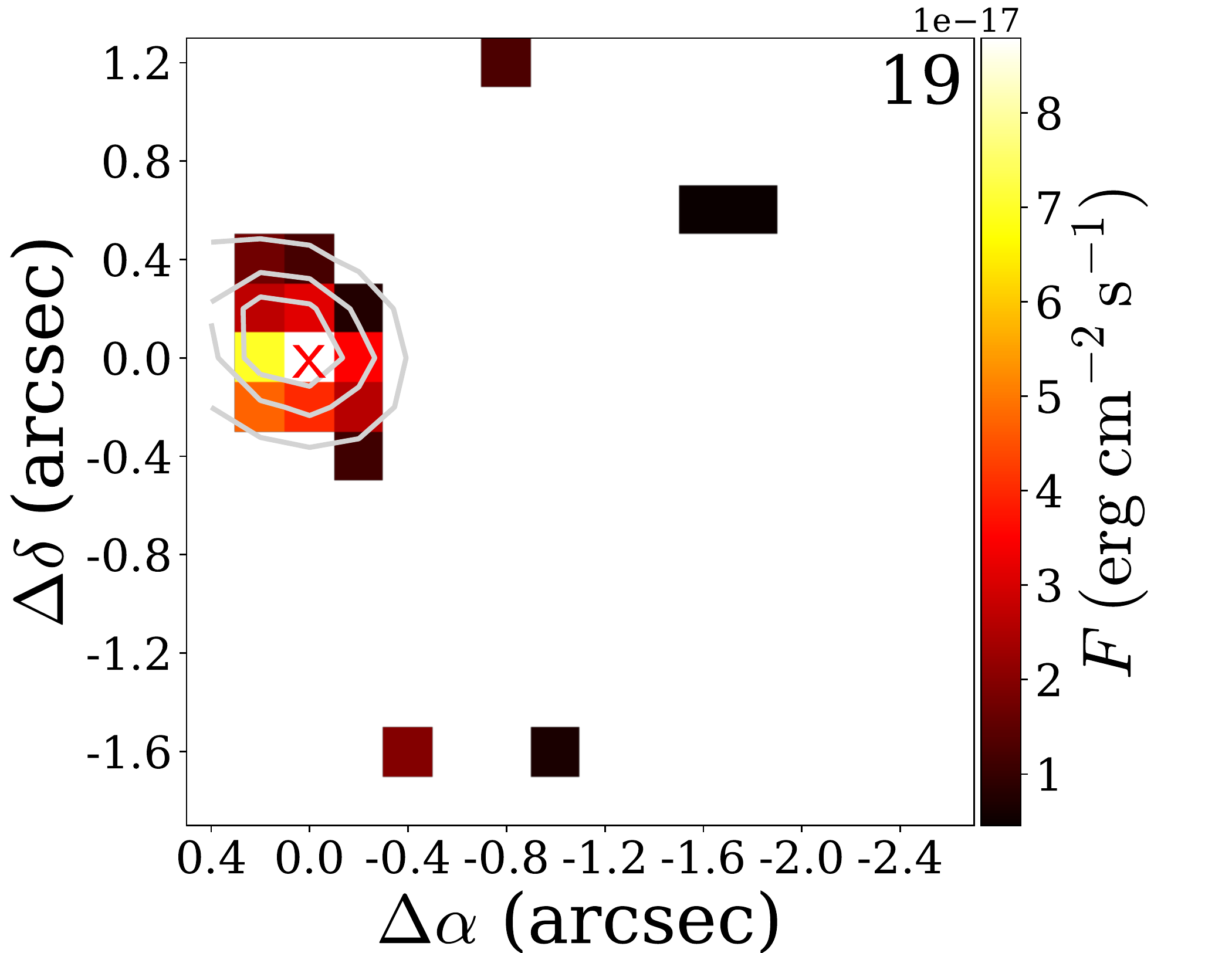}\hspace{-0.2cm}
\includegraphics[width=0.2\textwidth]{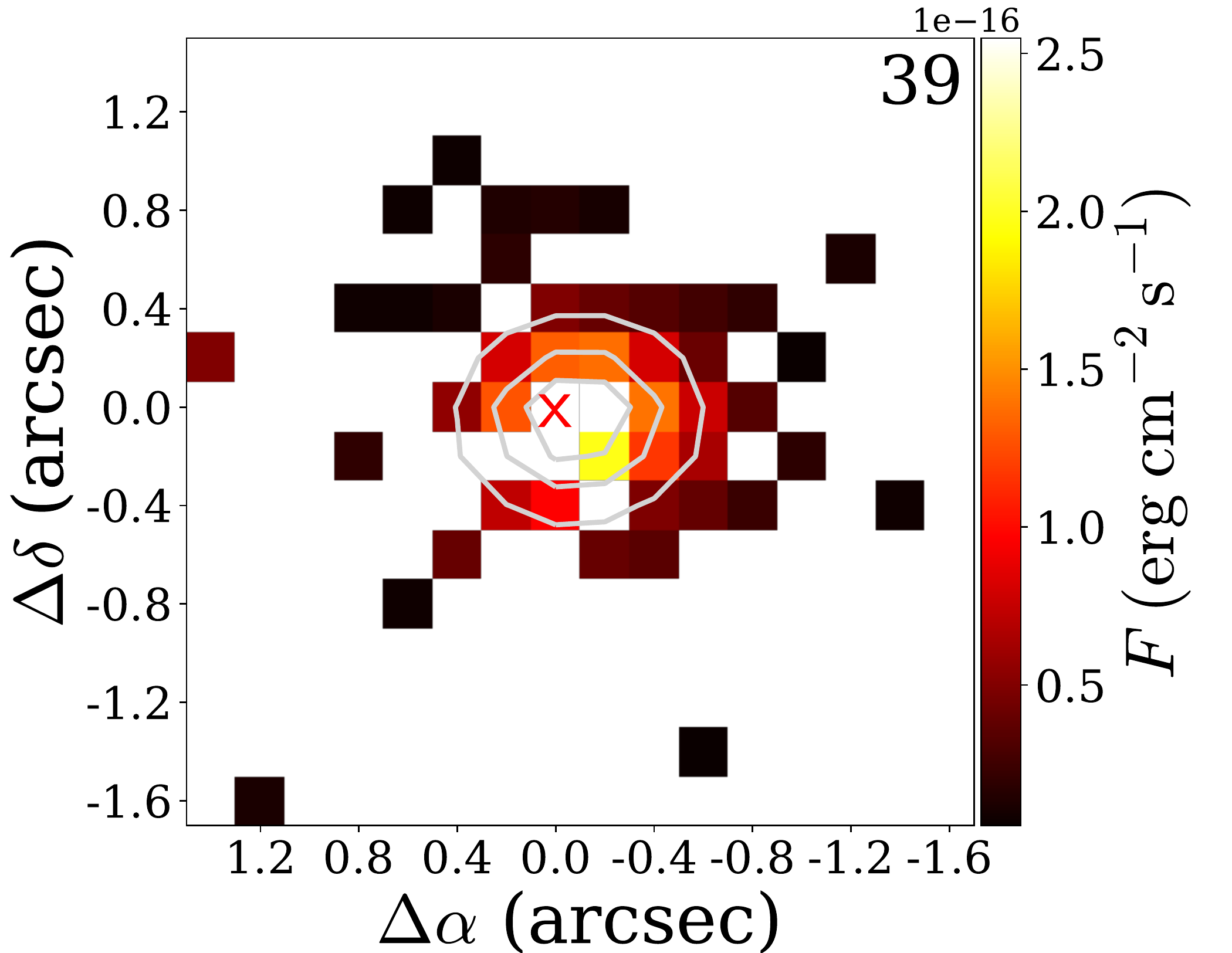}\hspace{-0.2cm}
\includegraphics[width=0.2\textwidth]{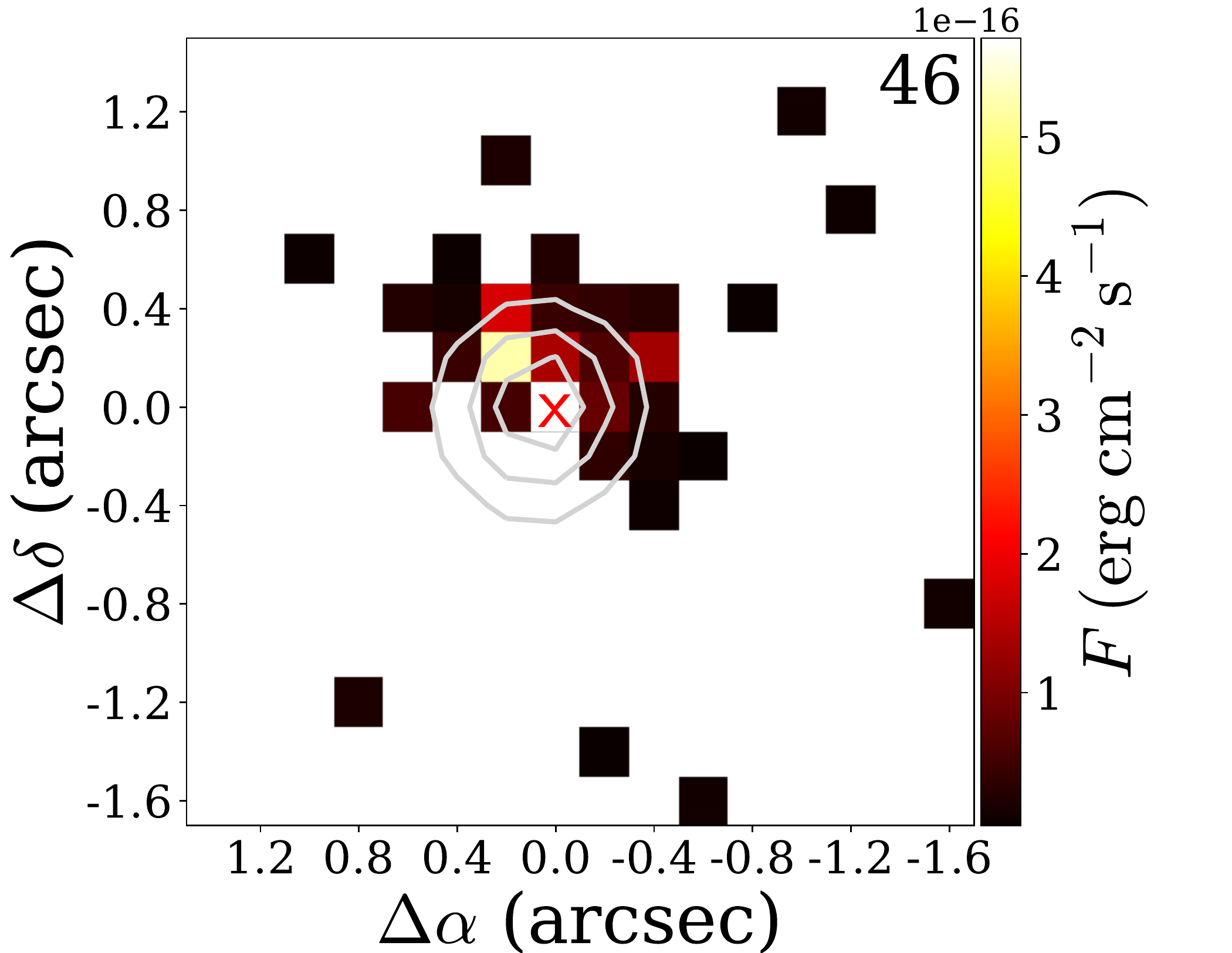}\hspace{-0.2cm}
\includegraphics[width=0.2\textwidth]{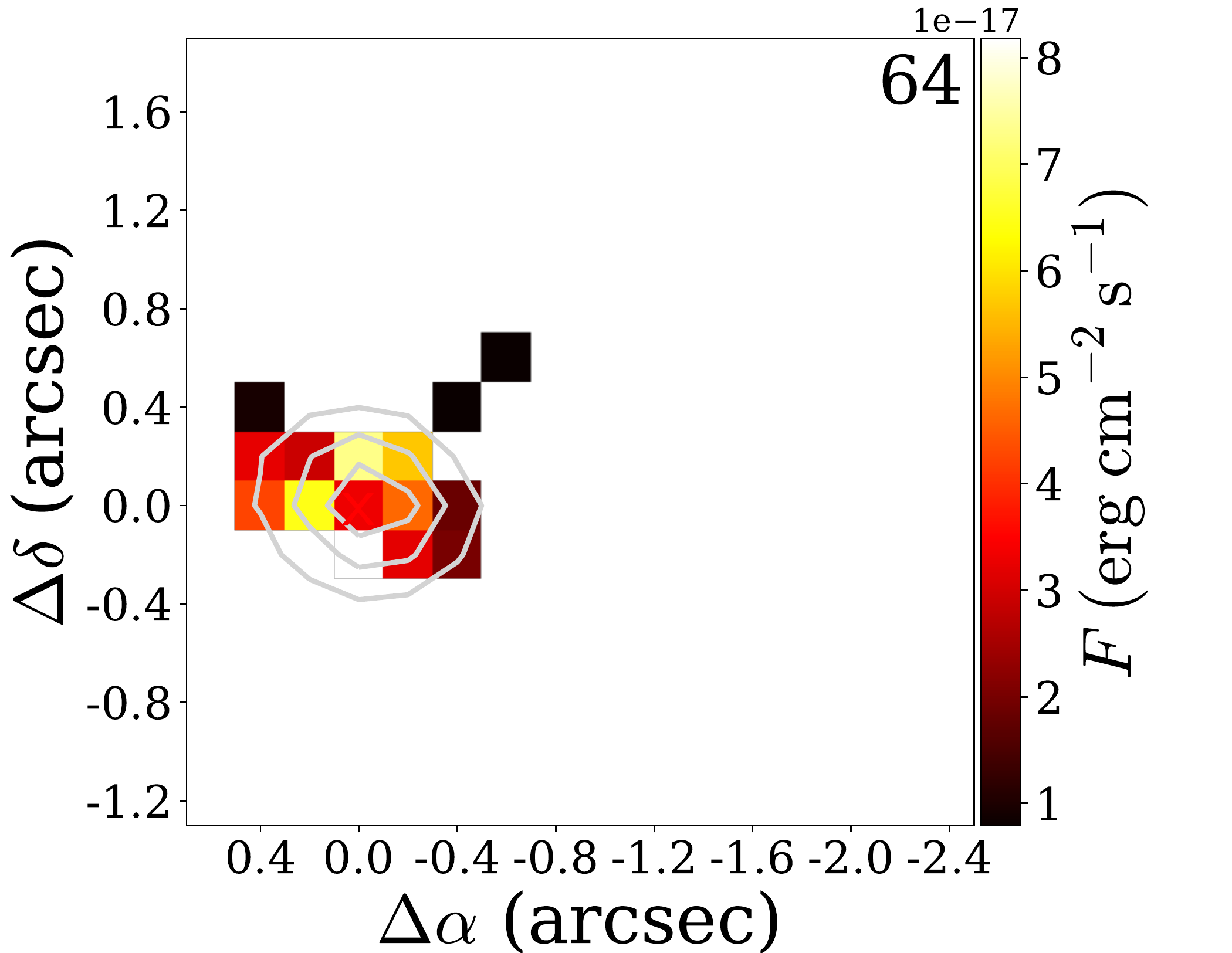}\hspace{-0.2cm}
\caption{Integrated maps of the CO $\nu$ = 2-0 line at 2.29 $\mu$m. Colors correspond to the line fluxes at each pixel (detections above 2$\sigma$) and white contours show the continuum emission in K-band (see also Appendix \ref{app:cont}).}
\label{fig:emiss-2.2940}
\end{figure*}
\begin{figure*}[h!]
\centering
\includegraphics[width=0.2\textwidth]{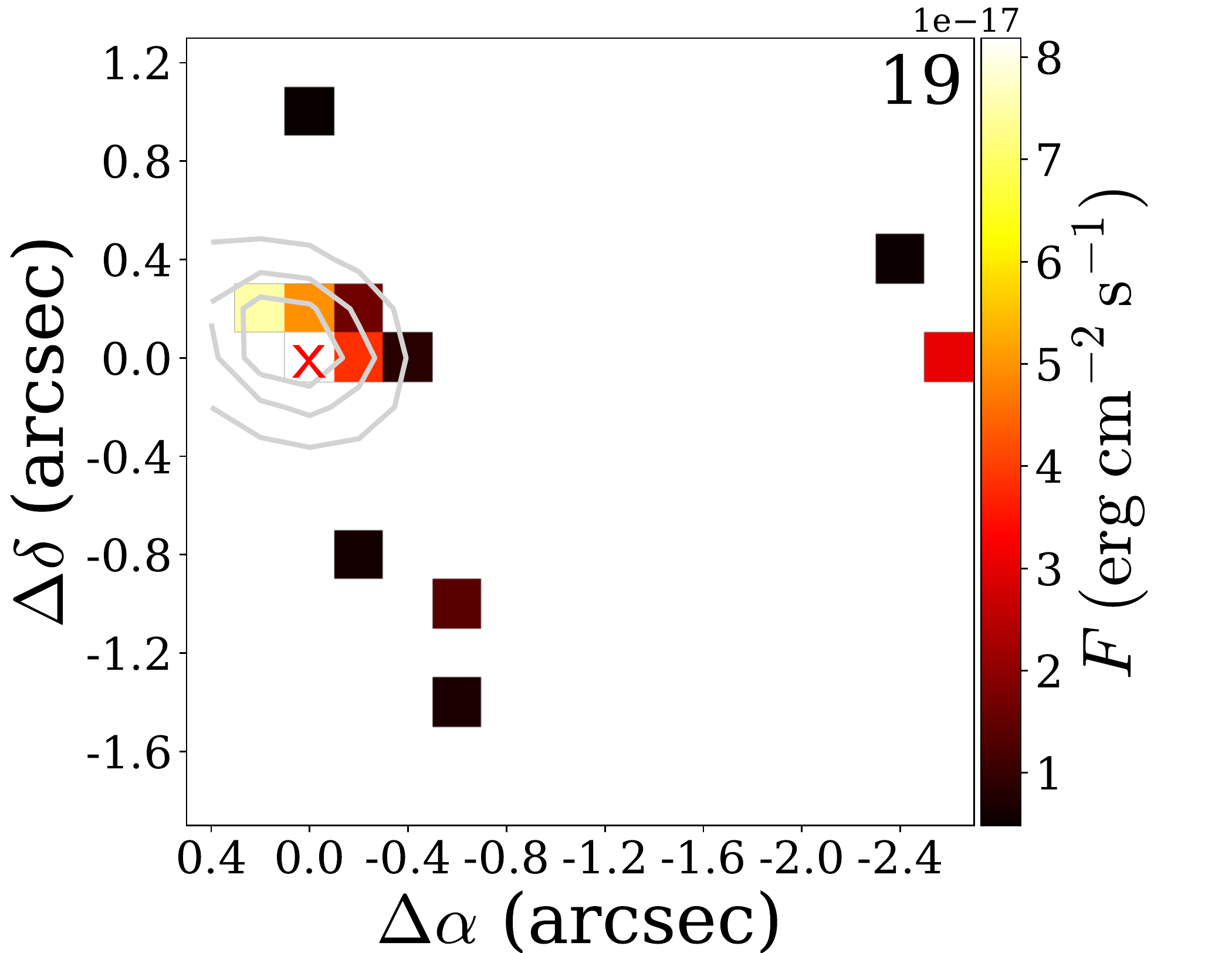}\hspace{-0.2cm} 
\includegraphics[width=0.2\textwidth]{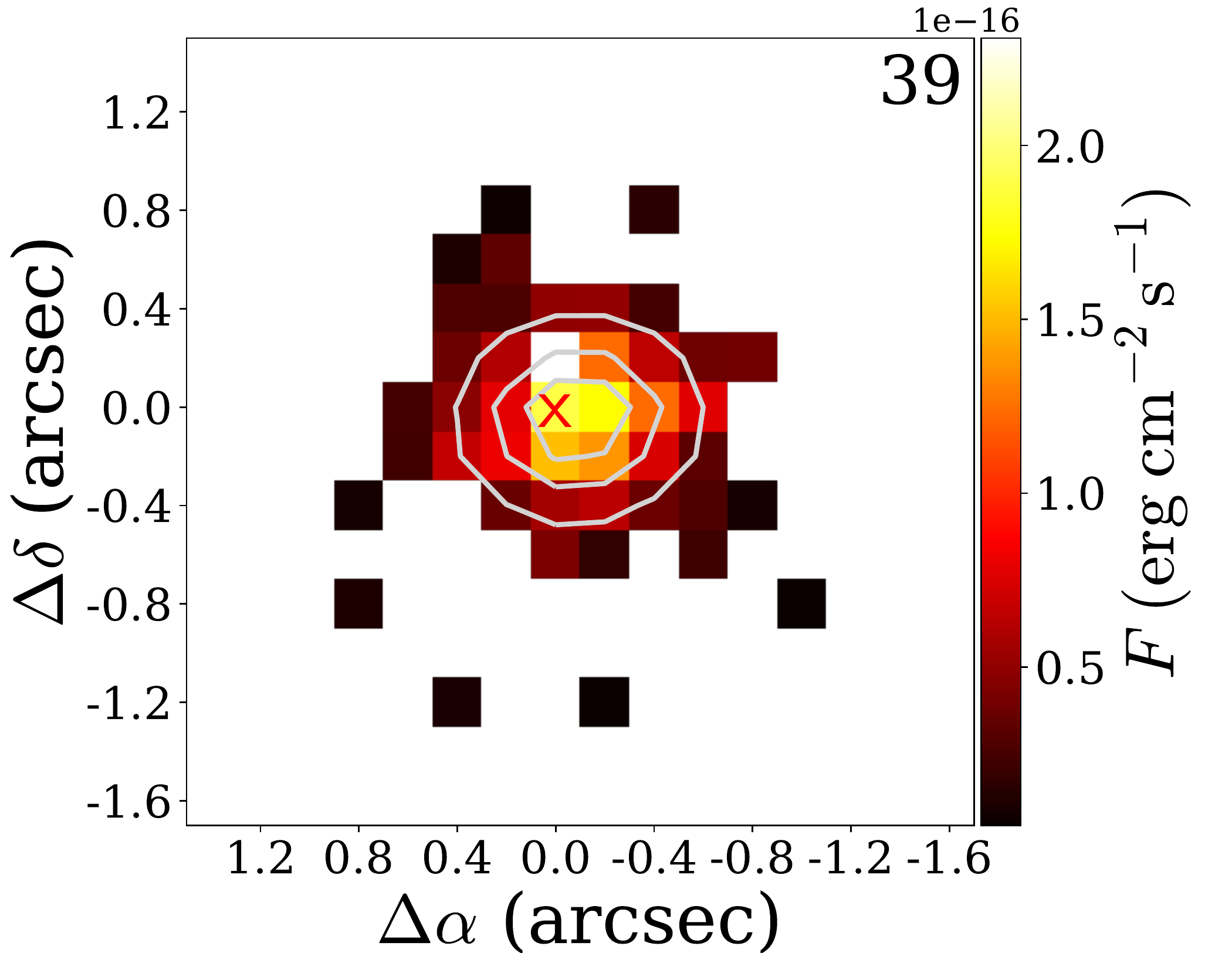}\hspace{-0.2cm}
\includegraphics[width=0.2\textwidth]{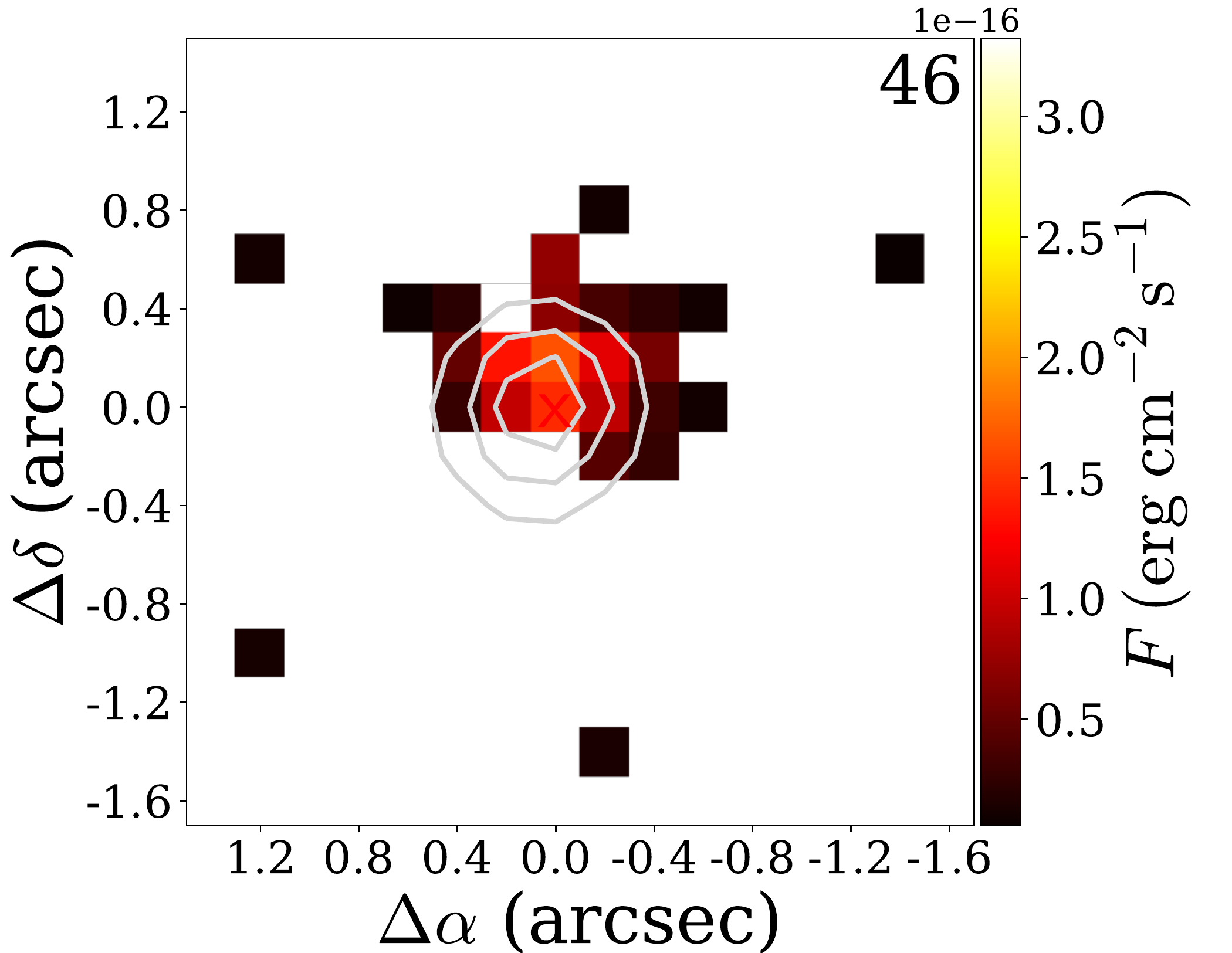}\hspace{-0.2cm}
\includegraphics[width=0.2\textwidth]{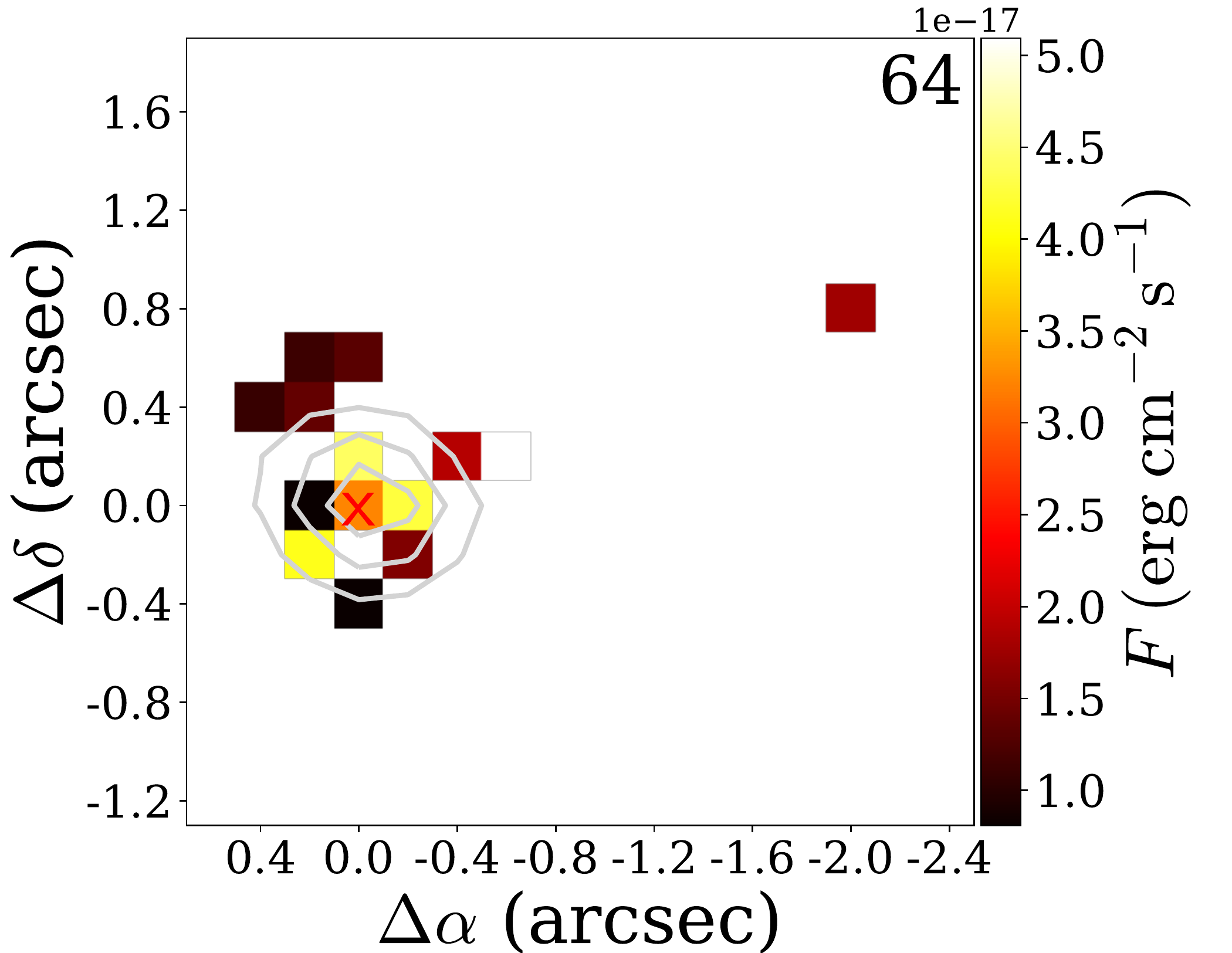}\hspace{-0.2cm} %
\caption{Similar to Figure \ref{fig:emiss-2.2940}, but the maps show the CO $\nu$ = 3-1 line at 2.32 $\mu$m.}
\label{fig:emiss-2.3230}
\end{figure*}
\begin{figure*}[h!]
\centering
\includegraphics[width=0.2\textwidth]{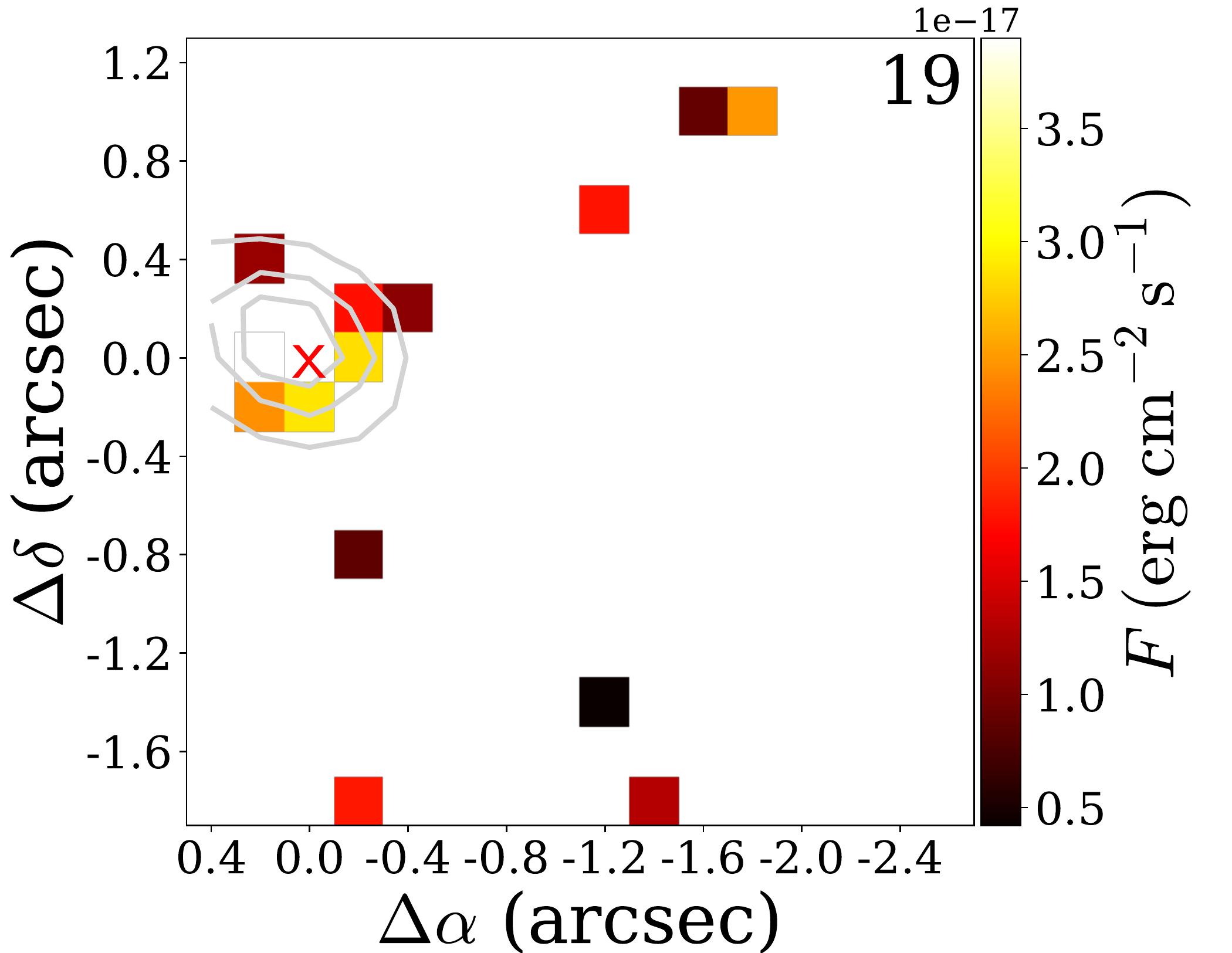}\hspace{-0.2cm} %
\includegraphics[width=0.2\textwidth]{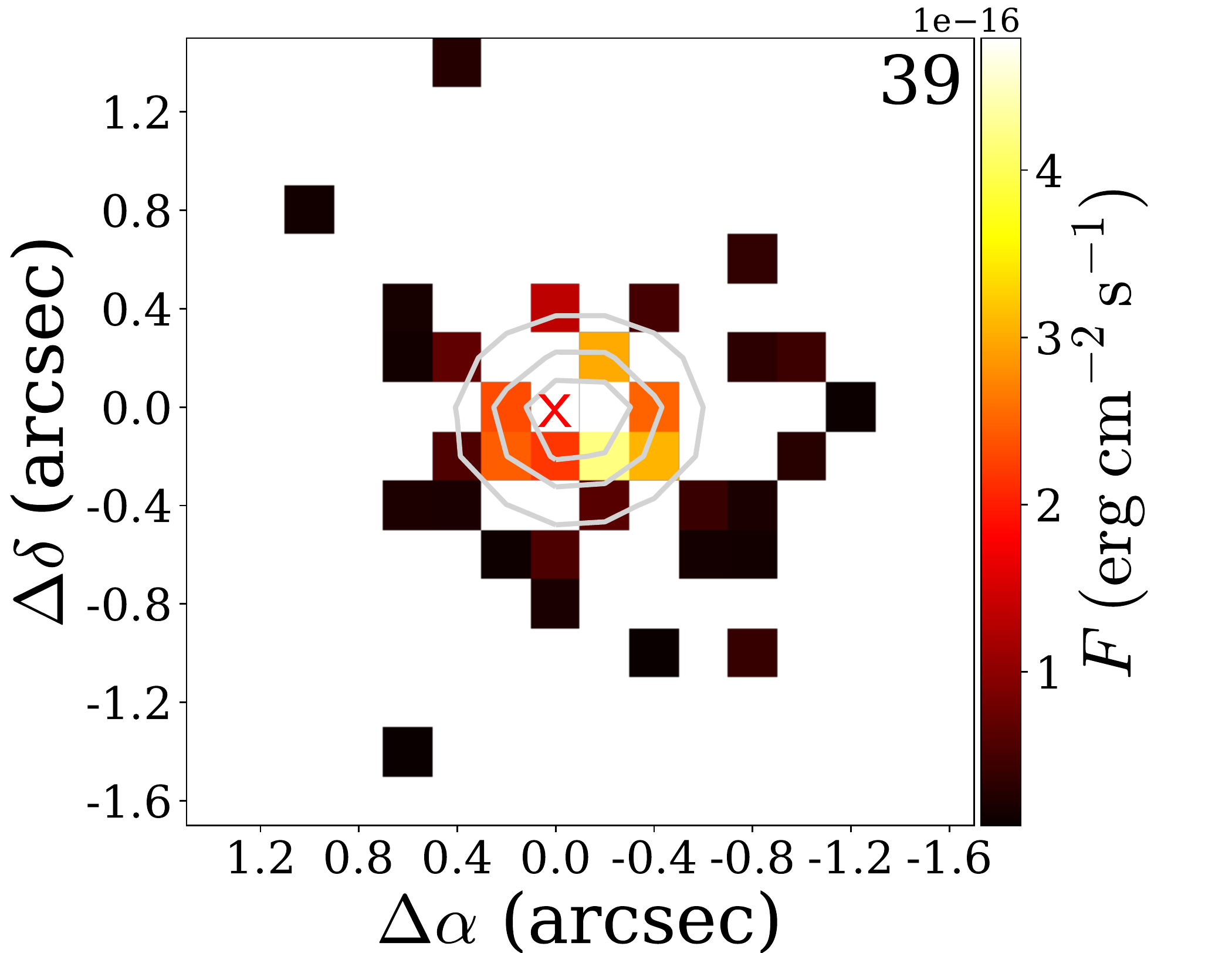}\hspace{-0.2cm}
\includegraphics[width=0.2\textwidth]{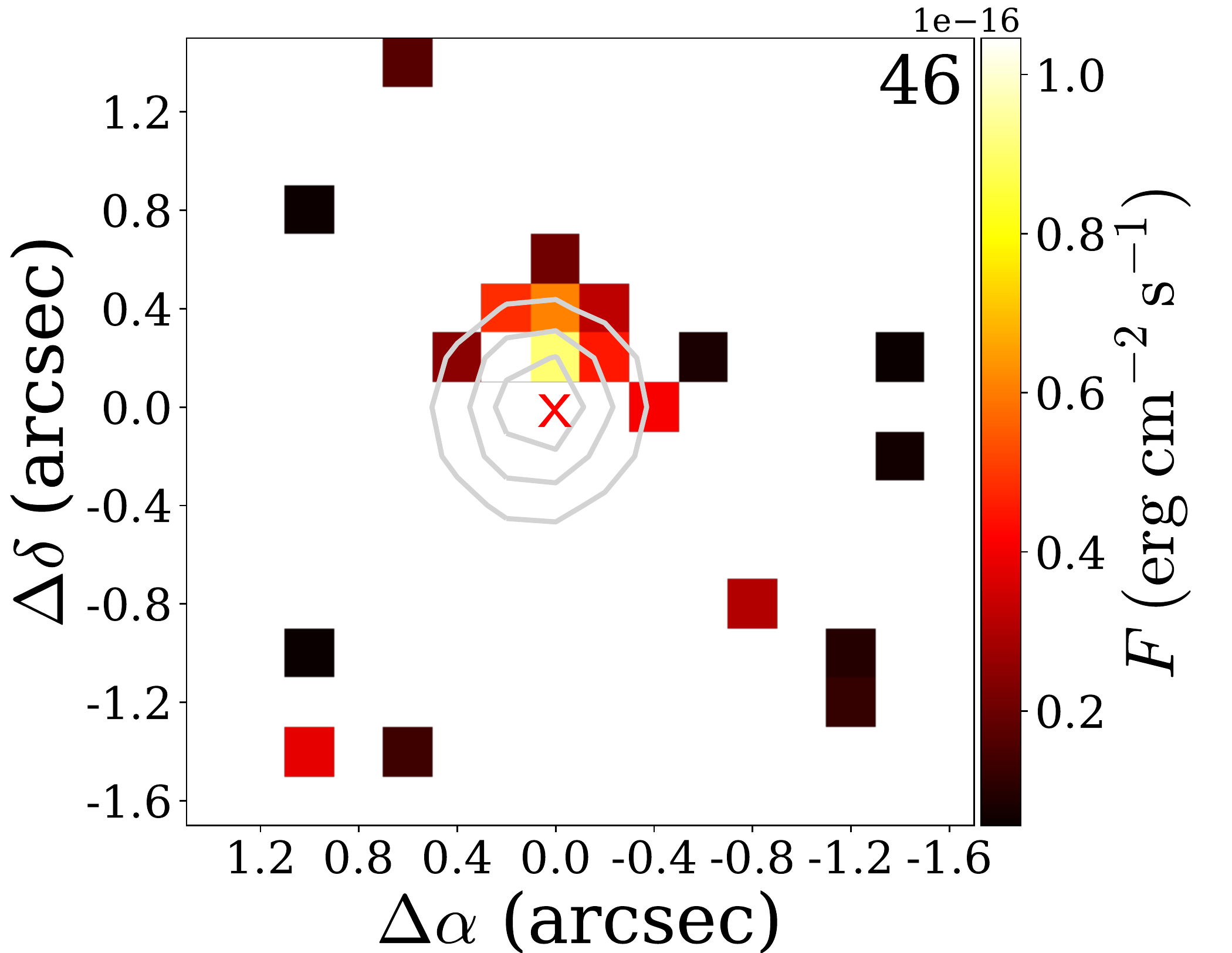}\hspace{-0.2cm} %
\caption{Similar to Figure \ref{fig:emiss-2.2940}, but the maps show the CO $\nu$ = 4-2 line at 2.35 $\mu$m.}
\label{fig:emiss-2.3540}
\end{figure*}
\begin{figure*}[h!]
\centering
\includegraphics[width=0.2\textwidth]{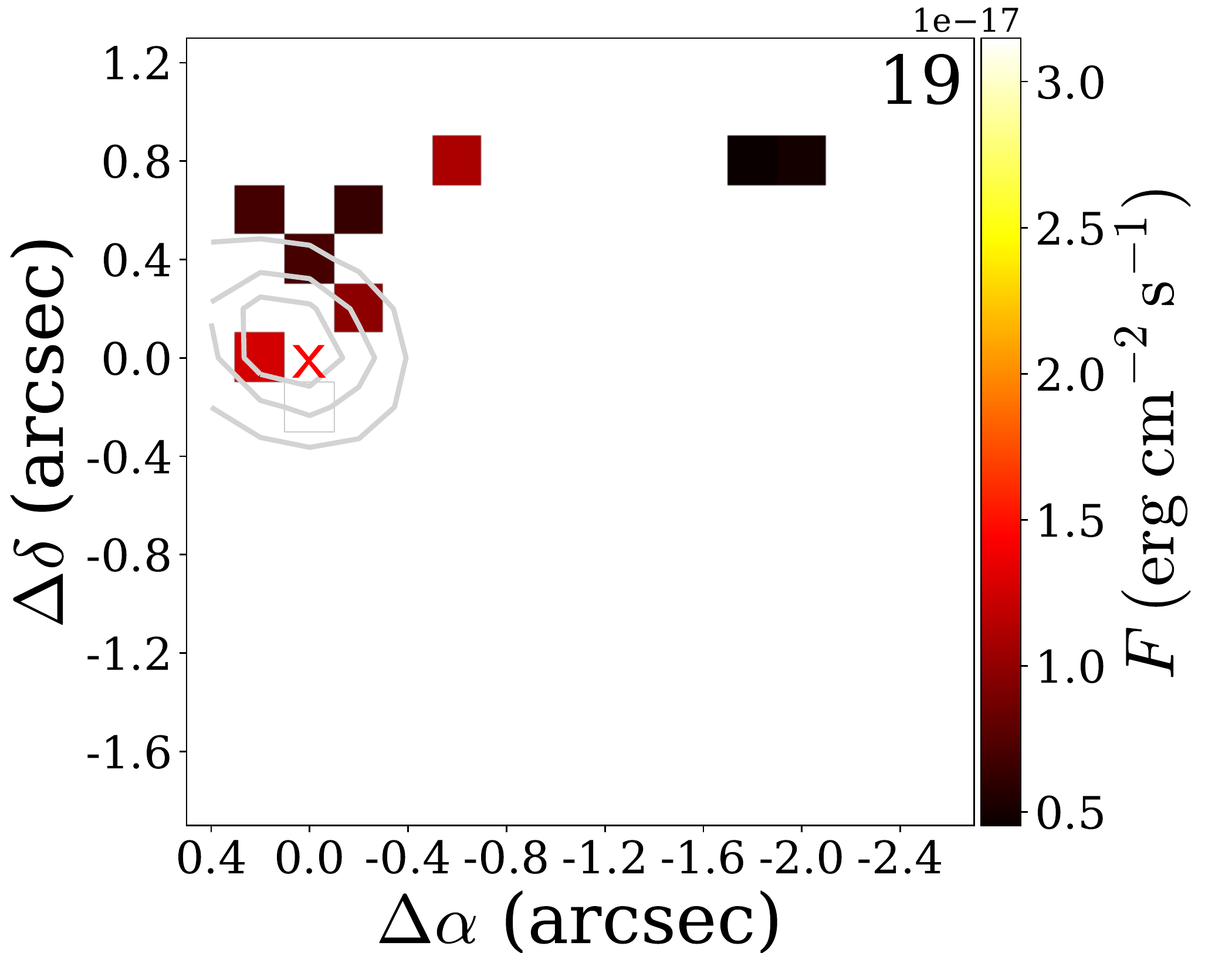}\hspace{-0.2cm} %
\includegraphics[width=0.2\textwidth]{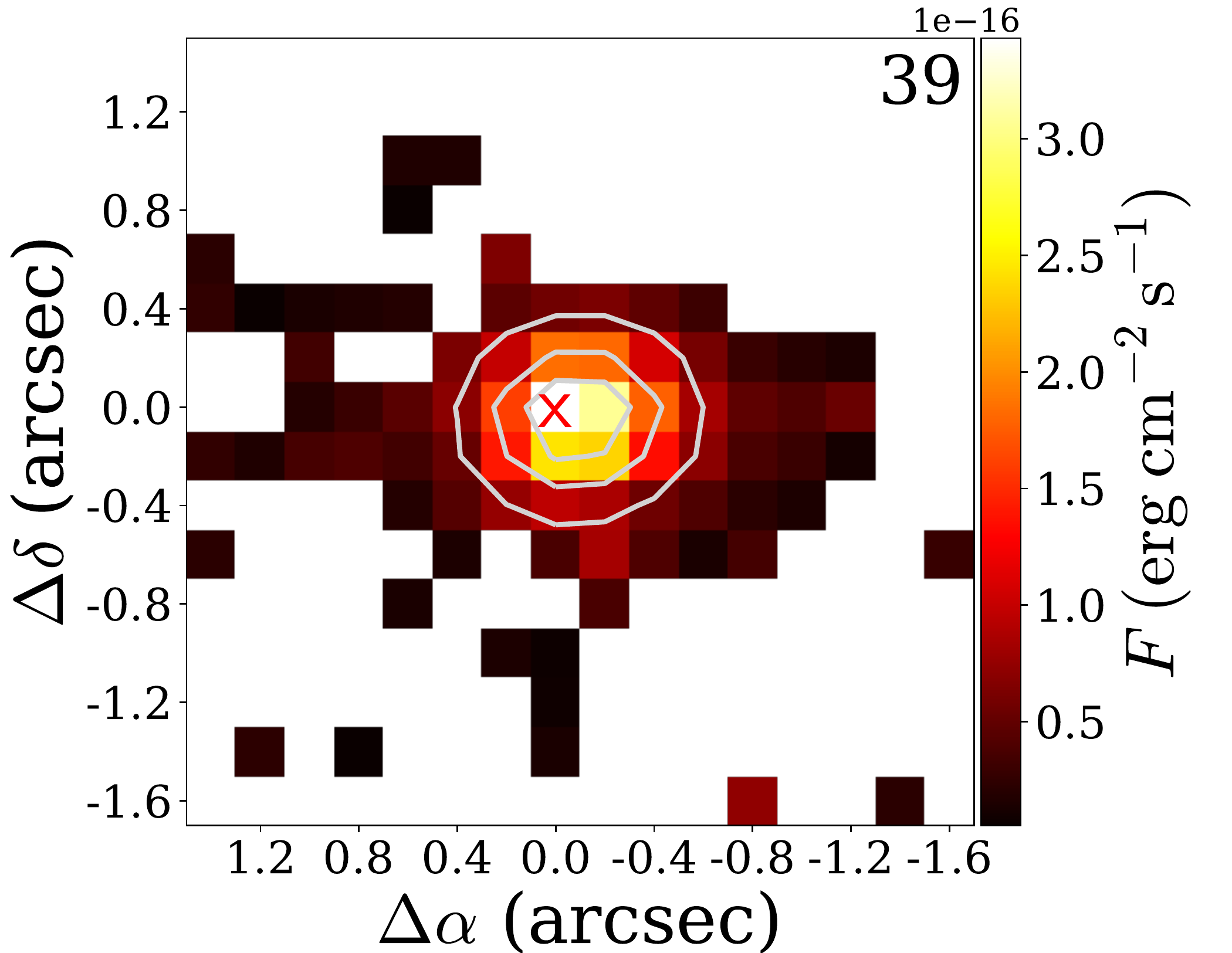}\hspace{-0.2cm}
\includegraphics[width=0.2\textwidth]{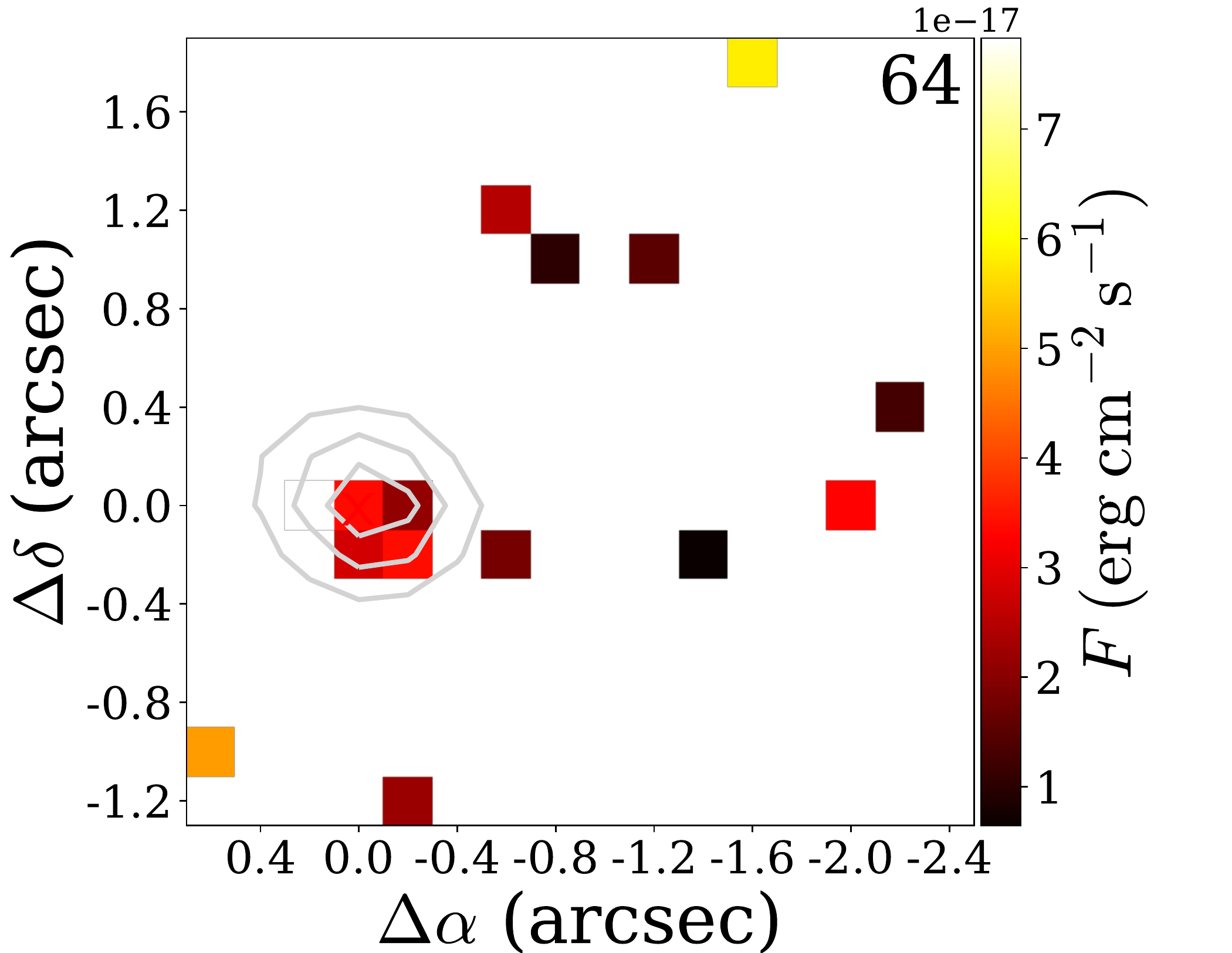}\hspace{-0.2cm}
\caption{Similar to Figure \ref{fig:emiss-2.2940}, but the maps show the CO $\nu$ = 5-3 line at 2.38 $\mu$m. 
}
\label{fig:emiss-2.3830}
\end{figure*}


\section{Extinction within stellar envelopes}
\label{app:Av}

The extinction values from SED modelling incorporate only the interstellar extinction to the source. Some of our stars were identified to be embedded in the envelope. Thus, to complete the estimations of extinction, 
we investigate the level of extinction due to the envelope. 
We consider only the targets, for which best-fit SED models included the envelope component \citep{sewilo2019}. For simplicity, we assume that the spherical envelope can be well described by the power-law density profile \citep[Eq. 3]{robitaille2017}:
\begin{equation}
\rho(r) = \rho_0 \cdot (r/R_0)^\gamma 
\end{equation}
where $\rho_0$ is a volume mass density at the radius $R_0$ and $\gamma$ represents the slope of the relation. We do not include outflow cavities, which were considered in the original models.

{To calculate column densities, we integrate the density profile along the line of sight, between the inner and outer radii of the envelope. We assume the mean molecular weight of 2.8 and the gas-to-dust ratio of 100. For calculations, we adopt parameter values from SED modelling of \cite{sewilo2019} and set the outer radius to be at 5000~au (the maximum value considered in models of \citealt{robitaille2017}). In cases where the inner radius was not available, we assume a value of 50~au. For target No. 47 we also assume the $R_0=100$~au, which is within the parameter range adopted in the SED modelling \citep{robitaille2017}. Some of our sources were originally fitted by a broken power-law density profile \citep[Eq. 4]{robitaille2017} and therefore they lack the value of $\gamma$. We assume a slope of -1/2 for $r<R_0$ and -3/2 for $r>R_0$, following the description in Sec. 3.2.3 of \cite{robitaille2017}. }

{We estimate the visual extinction due to the envelope of YSOs based on derived column densities. We follow the formula from \cite{predehl1995}:
\begin{equation}
A_{\rm V} = 1.086 \cdot N \cdot \pi a_g^2 \cdot Q_{\rm ext}
\end{equation}
where $N$ is a column density in 10$^{21}$ cm$^{-2}$, $a_g$ is a radius of a spherical grain in 0.1~$\mu$m, and $Q_{\rm ext}$ is an extinction efficiency. We adopted $a_g$ = 0.1~$\mu$m and $Q_{\rm ext}$ = 0.5. The resulting estimates are shown in Fig. \ref{fig:Av_env} as a function of a total $A_{\rm V}$ obtained from the SED models. $A_{\rm{V,env}}$  is higher for most Class I than for Class II sources, indicating that envelope is being dissipated with time, in line with evolutionary predictions. The only outlier, Class II star No. 65, is characterised by high $\rho_0 = 1.8\times10^{-18}$~g cm$^{-3}$, which leads to the very high value of the envelope extinction. This source is very bright, has relatively low $A_{\rm V}$, and has no emission lines, which are signposts of ongoing accretion or ejection. Clear detection of CO bandhead in absorption suggests that this target might be a foreground contamination, possibly a red giant. }

\clearpage

\startlongtable
\begin{deluxetable}{chhhhcc} 
\tablewidth{0pt}
\tablecaption{Interstellar and Envelope Extinction Values from SED Modeling. \label{tab:Av}}
\tablehead{
\colhead{No.} & \nocolhead{$A_{\rm{V}}$ (Q2/S0)} & \nocolhead{$A_{\rm{V}}$ (Q3/S1)} & \nocolhead{$A_{\rm{V}}$ (Q4/S2)} & \nocolhead{$A_{\rm{V}}$ (Q5/S3)} & \colhead{$A_{\rm{V,interst}}$} & \colhead{$A_{\rm{V,env}}$} 
}
\startdata
1 & \nodata & \nodata & \nodata & \nodata & 8.04 & \nodata \\
2 & 0.80 $\pm$ 0.26 & 9.75 $\pm$ 0.53 & \nodata & -1.51 $\pm$ 0.17 & 9.0 & 12.99 \\
3 & \nodata & \nodata & \nodata & \nodata & 11.9 & \nodata \\
5 & \nodata & \nodata & \nodata & \nodata & 34.9 & 3.98 \\
6 & \nodata & \nodata & \nodata & \nodata & 23.4 & \nodata \\
8 & \nodata & \nodata & \nodata & \nodata & 25.3 & \nodata \\
9 & 81.50 $\pm$ 49.80 & \nodata & \nodata & \nodata & 30.8 & \nodata \\
10 & \nodata & \nodata & \nodata & \nodata & 27.6 & \nodata \\
12 & \nodata & \nodata & \nodata & \nodata & 31.0 & \nodata \\
13 & \nodata & \nodata & \nodata & \nodata & 15.4 & \nodata \\
15 & 29.52 $\pm$ 18.11 & 32.04 $\pm$ 7.30 & \nodata & 2.52 $\pm$ 0.91 & 40.0 & \nodata \\
16 & \nodata & \nodata & \nodata & \nodata & 21.3 & \nodata \\
19 & -28.56 $\pm$ 27.09 & 25.06 $\pm$ 3.51 & \nodata & 8.92 $\pm$ 2.90 & 33.9 & \nodata \\
20 & \nodata & \nodata & \nodata & \nodata & 12.2 & \nodata \\
24 & \nodata & \nodata & \nodata & \nodata & 27.0 & \nodata \\
26 & \nodata & \nodata & \nodata & \nodata & 25.9 & \nodata \\
28 & \nodata & \nodata & \nodata & \nodata & 9.92 & \nodata \\
30 & \nodata & \nodata & \nodata & \nodata & 2.84 & \nodata \\
32 & \nodata & \nodata & \nodata & \nodata & 6.84 & \nodata \\
33 & \nodata & 32.84 $\pm$ 10.22 & \nodata & \nodata & 21.7 & \nodata \\
35 & \nodata & \nodata & \nodata & \nodata & 20.0 & 2.06 \\
39 & \nodata & 40.42 $\pm$ 7.08 & 49.49 $\pm$ 26.88 & \nodata & 10.2 & \nodata \\
40 & \nodata & \nodata & \nodata & \nodata & 28.6 & \nodata \\
42 & \nodata & \nodata & \nodata & \nodata & 26.2 & \nodata \\
43 & \nodata & \nodata & \nodata & \nodata & 2.29 & \nodata \\
45 & \nodata & \nodata & \nodata & \nodata & 19.3 & \nodata \\
46 & \nodata & \nodata & \nodata & \nodata & 14.0 & \nodata \\
47 & 17.83 $\pm$ 0.86 & 19.12 $\pm$ 0.74 & 15.79 $\pm$ 2.56 & -8.00 $\pm$ 0.46 & 33.3 & 12.65 \\
48 & \nodata & \nodata & \nodata & \nodata & 3.03 & \nodata \\
49 & \nodata & 17.03 $\pm$ 2.23 & 14.59 $\pm$ 14.10 & \nodata & 32.0 & \nodata \\
50 & \nodata & 30.12 $\pm$ 8.94 & \nodata & 20.18 $\pm$ 7.00 & 15.7 & \nodata \\
53 & \nodata & \nodata & \nodata & \nodata & 22.4 & \nodata \\
55A & \nodata & \nodata & \nodata & \nodata & 3.37 & \nodata \\
55B & \nodata & \nodata & \nodata & \nodata & 3.37 & \nodata \\
56 & \nodata & \nodata & \nodata & \nodata & 22.0 & \nodata \\
58 & \nodata & \nodata & \nodata & \nodata & 5.03 & \nodata \\
59 & \nodata & \nodata & \nodata & \nodata & 2.9 & \nodata \\
60 & -16.33 $\pm$ 2.42 & 13.64 $\pm$ 0.60 & 33.50 $\pm$ 8.31 & -15.68 $\pm$ 4.17 & 18.2 & 22.21 \\
61 & \nodata & \nodata & \nodata & \nodata & 8.51 & \nodata \\
62 & \nodata & \nodata & \nodata & \nodata & 17.3 & \nodata \\
63 & \nodata & \nodata & \nodata & \nodata & 22.6 & \nodata \\
64 & 24.99 $\pm$ 6.75 & 23.97 $\pm$ 1.67 & 38.78 $\pm$ 27.31 & \nodata & 7.15 & 28.74\\
65 & \nodata & \nodata & \nodata & \nodata & 3.84 & 109.96 \\
68 & \nodata & \nodata & \nodata & \nodata & 7.19 & \nodata \\
70 & \nodata & \nodata & \nodata & \nodata & 10.6 & \nodata \\
71 & \nodata & \nodata & \nodata & \nodata & 27.8 & \nodata \\
73 & \nodata & 32.84 $\pm$ 7.77 & \nodata & 30.29 $\pm$ 26.38 & 26.2 & \nodata \\
75 & \nodata & \nodata & \nodata & \nodata & 23.1 & \nodata \\
76 & \nodata & \nodata & \nodata & \nodata & 24.2 & \nodata \\
77 & \nodata & \nodata & \nodata & \nodata & 40.0 & 0.00 \\
79 & \nodata & \nodata & \nodata & \nodata & 15.1 & \nodata \\
81 & \nodata & \nodata & \nodata & \nodata & 25.7 & \nodata \\
82A & \nodata & \nodata & \nodata & \nodata & 11.8 & 0.01 \\
82B & \nodata & \nodata & \nodata & \nodata & 11.8 & 0.01 \\
83 & \nodata & -9.06 $\pm$ 7.47 & \nodata & \nodata & 12.8 & 41.15 \\
84 & \nodata & \nodata & \nodata & \nodata & 17.3 & 3.13 \\
86A & \nodata & \nodata & \nodata & \nodata & 6.14 & \nodata \\
86B & \nodata & \nodata & \nodata & \nodata & 6.14 & \nodata \\
87A & 94.83 $\pm$ 70.51 & -9.75 $\pm$ 2.87 & \nodata & \nodata & 30.3 & \nodata \\
87B & 77.67 $\pm$ 61.67 & \nodata & \nodata & 49.16 $\pm$ 43.48 & 30.3 & \nodata \\
88 & \nodata & \nodata & \nodata & \nodata & 0.0 & 0.23 \\
92 & \nodata & \nodata & \nodata & \nodata & 9.37 & \nodata \\
93 & \nodata & \nodata & \nodata & \nodata & 28.5 & \nodata \\
94 & \nodata & \nodata & \nodata & \nodata & 8.96 & \nodata \\
95 & \nodata & \nodata & \nodata & \nodata & 4.49 & \nodata \\
96 & \nodata & \nodata & \nodata & \nodata & 19.1 & 0.04 \\
97 & \nodata & \nodata & \nodata & \nodata & 13.9 & \nodata \\
98 & \nodata & \nodata & \nodata & \nodata & 7.47 & \nodata \\
102 & \nodata & \nodata & \nodata & \nodata & 15.4 & \nodata \\
104 & \nodata & \nodata & \nodata & \nodata & 7.78 & \nodata \\
105 & \nodata & \nodata & \nodata & \nodata & 28.1 & \nodata \\
108 & \nodata & \nodata & \nodata & \nodata & 7.74 & \nodata \\
109 & 15.66 $\pm$ 3.31 & 2.66 $\pm$ 0.29 & \nodata & \nodata & 15.0 & 29.33 \\
110 & \nodata & \nodata & \nodata & \nodata & 16.9 & \nodata \\
111 & 50.66 $\pm$ 45.27 & 16.41 $\pm$ 6.99 & \nodata & \nodata & 17.0 & \nodata \\
112A & \nodata & 48.09 $\pm$ 13.35 & \nodata & \nodata & 25.1 & \nodata \\
112B & \nodata & \nodata & \nodata & \nodata & 25.1 & \nodata \\
113 & \nodata & \nodata & \nodata & \nodata & 21.7 & \nodata \\
114 & -3.26 $\pm$ 1.76 & 12.61 $\pm$ 1.27 & \nodata & \nodata & 39.4 & \nodata \\
115 & \nodata & \nodata & \nodata & \nodata & 9.23 & 15.27 \\
116 & \nodata & \nodata & \nodata & \nodata & 2.43 & \nodata \\
117 & \nodata & \nodata & \nodata & \nodata & 5.11 & \nodata \\
118 & \nodata & \nodata & \nodata & \nodata & 12.7 & \nodata \\
\enddata
\tablecomments{All values are given in magnitudes. $A_{\rm V,interst}$ 
is the value adopted from \cite{sewilo2019} based on the SED modeling; we assume an uncertainty of 10~mag of their measurements (see text for details). The $A_{\rm V,env}$ is the extinction within the envelope estimated based on the results of the SED modeling. 
}
\end{deluxetable}

\begin{figure}
\includegraphics[width=\columnwidth, trim={0.4cm 0 1.5cm 1.4cm}, clip]{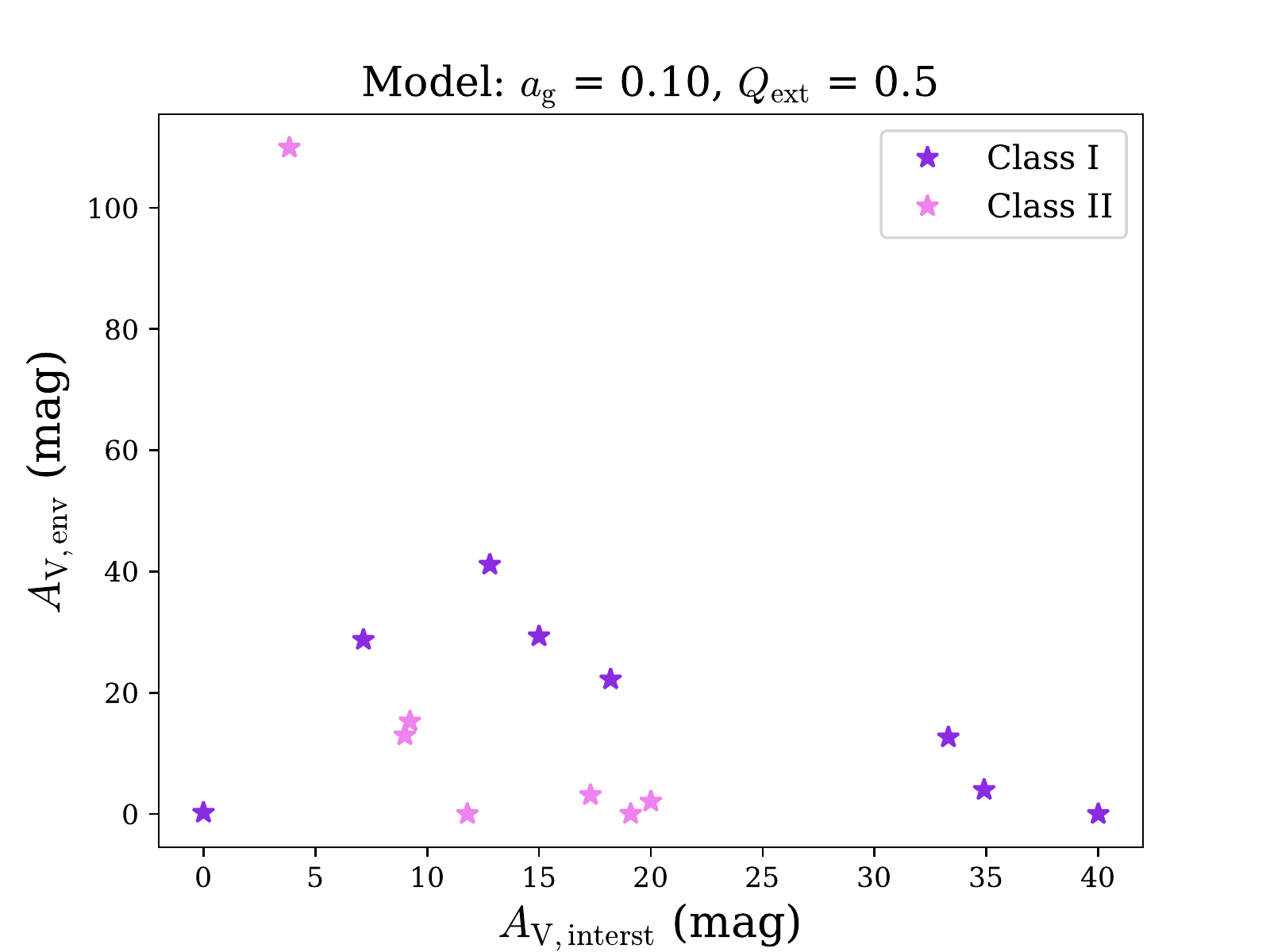}
\caption{Comparison between the total visual extinction $A_{\rm{V,total}}$ \citep{sewilo2019} and the extinction within the envelope $A_{\rm{V,env}}$. The latter was estimated assuming spherical envelope and a power-law density profile. See the text for details.}
\label{fig:Av_env}
\end{figure}

\end{document}